# I R I D E

## An **I**nterdisciplinary **R**esearch **I**nfrastructure based on **D**ual **E**lectron linacs&lasers

## A WHITE BOOK

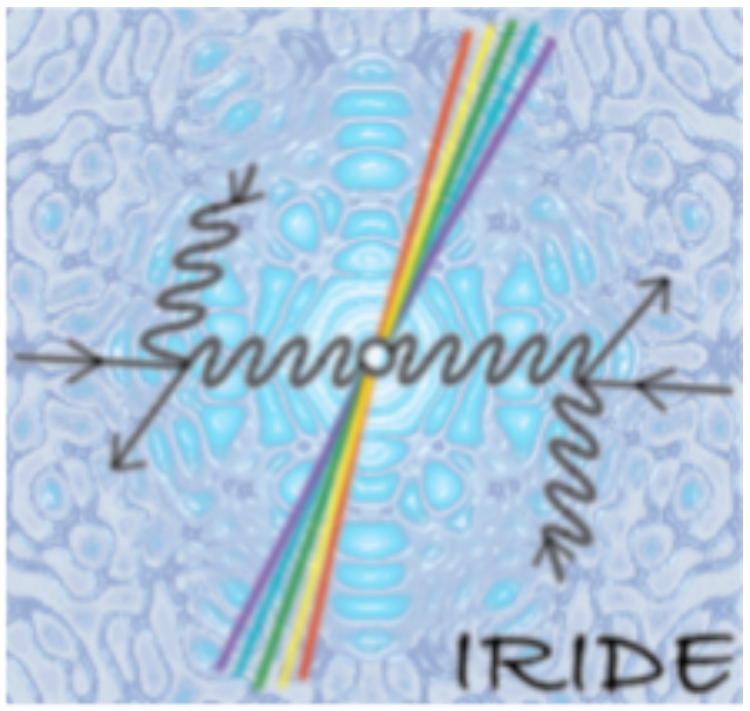

This report describes the scientific aims and potentials as well as the preliminary technical design of IRIDE, an innovative tool for multi-disciplinary investigations in a wide field of scientific, technological and industrial applications. IRIDE will be a high intensity "particle factory", based on a combination of a high duty cycle radio-frequency superconducting electron linac and of high energy lasers. Conceived to provide unique research possibilities for particle physics, for condensed matter physics, chemistry and material science, for structural biology and industrial applications, IRIDE will open completely new research possibilities and advance our knowledge in many branches of science and technology. IRIDE will contribute to open new avenues of discoveries and to address most important riddles: What does matter consist of? What is the structure of proteins that have a fundamental role in life processes? What can we learn from protein structure to improve the treatment of diseases and to design more efficient drugs? But also how does an electronic chip behave under the effect of radiations? How can the heat flow in a large heat exchanger be optimized?

The scientific potential of IRIDE is far reaching and justifies the construction of such a large facility in Italy in synergy with the national research institutes and companies and in the framework of the European and international research. It will impact also on R&D work for ILC, FEL, and will be complementarity to other large scale accelerator projects. IRIDE is also intended to be realized in subsequent stages of development depending on the assigned priorities.

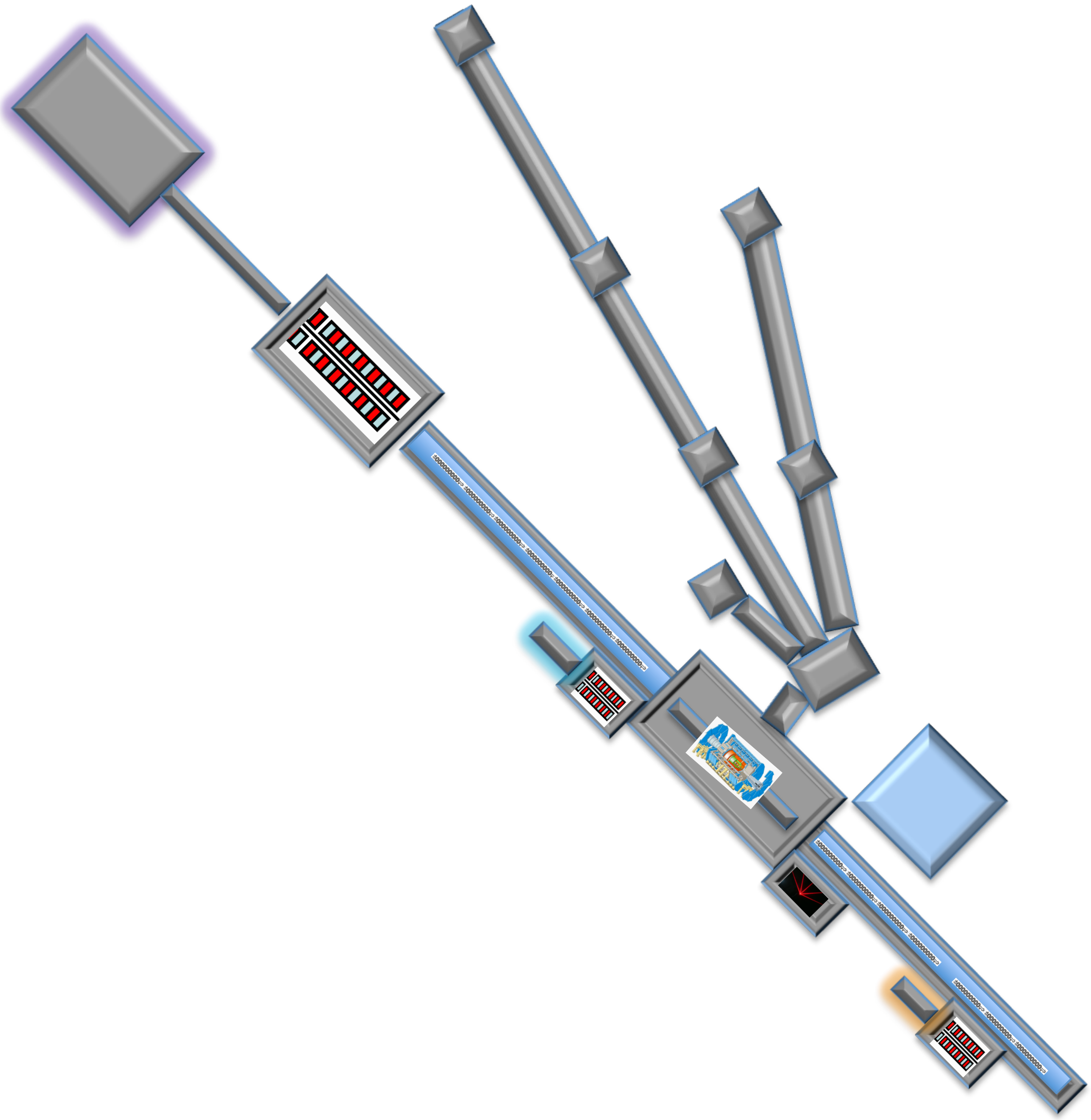

# Author's list

D. Alesini[1], M. Alessandroni[51], M. P. Anania[1], S. Andreas[56], M. Angelone[14], A. Arcovito [26] , F. Arnesano [25] , M. Artioli[14], L. Avaldi [40] , D. Babusci[1], A. Bacci[3], A. Balerna[1], S. Bartalucci[1], R. Bedogni[1], M. Bellaveglia[1], F. Bencivenga [50] , M. Benfatto[1,46], S. Biedron[66], V. Bocci[2], M. Bolognesi [27] , P. Bolognesi [40] , R. Boni[1], R. Bonifacio[7], M. Boscolo[1], F. Boscherini [43] , F. Bossi[1], F. Broggi[3], B. Buonomo[1], V. Calò [25] , D. Catone [41] , M. Capogni[14], M. Capone[14], K. Cassou[70],M. Castellano[1], A. Castoldi[17], L. Catani[4], G. Cavoto[2], N. Cherubini[14], G. Chirico [28] , M. Cestelli-Guidi[1], E. Chiadroni[1], V. Chiarella[1], A. Cianchi[4], M. Cianci [29] , R. Cimino[1], F. Ciocci[14], A. Clozza[1], M. Collini [28] , G. Colo'[3], A. Compagno[14], G. Contini [41] , M. Coreno [40] , R. Cucini [50] , C. Curceanu[1], S. Dabagov[1], E. Dainese [30] , I. Davoli[24] , G. Dattoli[14], L. De Caro [31] , P. De Felice[5], S. Della Longa [32] , G. Delle Monache[1], M. De Spirito [33] , A. Di Cicco [44] , C. Di Donato[57], D. Di Gioacchino[1], D. Di Giovenale[1], E. Di Palma[14], G. Di Pirro[1], A. Dodaro[14], A. Doria[14], U. Dosselli[1], A. Drago[1], K. Dupraz[70], R. Escrinabo[23], A. Esposito[1], R. Faccini[2], A. Ferrari[52], M. Ferrario[1], A. Filabozzi[4], D. Filippetto[18], F. Fiori[53], O. Frasciello[1], L. Fulgentini[15], G. P. Gallerano[14], A. Gallo[1], M. Gambaccini[11], C. Gatti[1], G. Gatti[1], P. Gauzzi[2], A. Ghigo[1], G. Ghiringhelli [49] , L. Giannessi[14], G. Giardina[54], C. Giannini [31] , F. Giorgianni[2], E. Giovenale[14], L. Gizzi[15], C. Guaraldo[1], C. Guazzoni[17], R. Gunnella [44] , K. Hatada [1,44] , M.Iannone[69], S. Ivashyn[12], F. Jegerlehner[58], P.O. Keeffe [40] , W. Kluge[58], A. Kupsc[60], L. Labate[15], P. Levi Sandri[1], V. Lombardi [34] , P. Londrillo[20], S. Loreti[5], A. Lorusso[9], M. Losacco [25] , S. Lupi[2], A. Macchi[15], S. Magazù[54], G. Mandaglio[54], A. Marcelli[1,46], G. Margutti[67], C. Mariani[16], P. Mariani [35] , G. Marzo[14], C. Masciovecchio [50], P. Masujan[61], M. Mattioli[2], G. Mazzitelli[1], N.P. Merenkov[21], P. Michelato[3], F. Migliardo[54], M. Migliorati[2], C. Milardi[1], E. Milotti[13], S. Milton[66],V. Minicozzi [24] , S. Mobilio [48] , S. Morante[24], D. Moricciani[4], A. Mostacci[2], V. Muccifora[1], F. Murtas[1], P. Musumeci[10], F. Nguyen[62], A. Orecchini[55], G. Organtini[2], P. L. Ottaviani[14], E. Pace[1], M. Paci [35] , C. Pagani[3], S. Pagnutti[14], V. Palmieri[6], L. Palumbo[2], G.C. Panaccione [45] , C. F. Papadopoulos[18], M. Papi [33] , M. Passera[63], L. Pasquini [43] , M. Pedio [45] , A. Perrone[9], A. Petralia[14], M. Petrarca[1], C. Petrillo[55], V. Petrillo[3], M. Pillon[14], P. Pierini[3], A. Pietropaolo[14], M. Pillon[14], A. D. Polosa[2], R. Pompili[4], J. Portoles[22], T. Prosperi [41] , C. Quaresima[41], L. Quintieri[5], J. V. Rau[15], M. Reconditi [34] , A. Ricci [38] , R. Ricci[1], G. Ricciardi[57], G. Ricco[71], M. Ripani[71], E. Ripiccini[2], S. Romeo[54], C. Ronsivalle[14], N. Rosato [37] , J. B. Rosenzweig[10], G. Rossi [24] , A. A. Rossi[6], A. R. Rossi[3], F. Rossi[20], D. Russo[15], A. Sabatucci [30] , E. Sabia[14], F. Sacchetti[55], S. Salducco[68], F. Sannibale[18], G. Sarri[19], T. Scopigno [35] , L. Serafini[3], D. Sertore[3], O. Shekhovtsova[64], I. Spassovsky[14], T. Spadaro[1], B. Spataro[1], F. Spinozzi [35] , A. Stecchi[1], F. Stellato [24,38] , V. Surrenti[14], A. Tenore[1], A. Torre[14], L. Trentadue[65], S. Turchini [41] , C. Vaccarezza[1], A. Vacchi[13], P. Valente[2], G. Venanzoni[1], S. Vescovi[1], F. Villa[1], G. Zanotti [39] , N. Zema[41], M. Zobov[1], F. Zomer[70].

1 INFN-Laboratori Nazionali di Frascati

2 INFN and Universita' di Roma"La Sapienza"

3 INFN and Universita' di Milano

4 INFN and Universita' di Roma"Tor Vergata"

5 Istituto Nazionale di Metrologia delle Radiazioni Ionizzanti, ENEA C R Casaccia.

6 INFN-Laboratori Nazionali di Legnaro

[7] INFN Universidade Federal da Paraiba, Brazil
[8] Universita' di Camerino
[9] INFN and Universita' del Salento
[10] UCLA, Los Angeles, USA
[11] INFN and Universita' di Ferrara
[12] TP, NSC "Kharkov Institute of Physics and Technology", Kharkov, Ukraine
[13] INFN and Universita' di Trieste
[14] ENEA
[15] CNR
[16] CNISM and Universita' di Roma"La Sapienza"
[17] Politecnico di Milano and INFN sez. Milano
[18] LBNL
[19]The Queen's University of Belfast, Belfast, UK
[20]INFN-Sezione di Bologna
[21]NSC KIPT, Kharkov, Ukraine
[22]instituto De Física Corpuscular, Spain
[23]Universitat Autonoma de Barcelona
[24] Dipartimento di Fisica Università di Roma *"Tor Vergata"* and INFN Sezione di Roma *"Tor Vergata"* - 00133 Roma, Italia.
[25] Dipartimento Farmaco-Chimico, Università di Bari "A. Moro" - 70125 Bari, Italia
[26] Istituto di Biochimica e Biochimica Clinica, Università Cattolica del Sacro Cuore - 00167 Roma, Italia.
[27] Dipartimento di Scienze Biomolecolari e Biotecnologia, Università di Milano - 20131 Milano, Italia.
[28] Dipartimento di Fisica, Università di Milano Bicocca - 20126 Milano, Italia.
[29] European Molecular Biology Laboratory, Hamburg Outstation - 22603, Hamburg, Germany
[30] Dipartimento di Scienze Biomediche, Università di Teramo - 64100 Teramo, Italia.
[31] Istituto di Cristallografia, CNR – 70125 Bari, Italia.
[32] Dipartimento di Medicina Sperimentale, Università dell' Aquila - 67100 L' Aquila, Italia.
[33] Istituto di Fisica, Università Cattolica del sacro Cuore - 00168 – Roma, Italia.
[34] Laboratorio di Fisiologia, Dipartimento di Biologia Evoluzionistica, Università di Firenze - 50019 Sesto Fiorentino, Italia.
[35] Dipartimento SAIFET, Sezione di Scienze Fisiche Università Politecnica delle Marche, Ancona, Italia.
[36] Dipartimento di Chimica Università di Roma *"Tor Vergata"* - 00133 Roma, Italia.
[37] Centro NAST, Nanoscienze & Nanotecnologie & Strumentazione - 00133 Roma, Italia, and Dipartimento di Medicina Sperimentale e Scienze Biochimiche, Università di Roma "Tor Vergata" - 00133 Roma, Italia.
[38] Center for Free Electron Laser Science c/o DESY - 22607, Hamburg, Germany
[39] Dipartimento di Chimica Biologica, Università di Padova - 35121 Padova, Italia.
[40] CNR-Istituto di Metodologie Inorganiche e dei Plasmi Area della Ricerca di Roma 1 - 00016 Monterotondo Scalo, Italia
[41] CNR-Istituto di Struttura della Materia Area della Ricerca di Roma 2 – Roma, Italia
[42] Dipartimento di Fisica, Universita' Roma "La Sapienza" – 00185 Roma, Italia
[43] Department of Physics and Astronomy, University of Bologna - 40127 Bologna, Italy
[44] CNISM, Scuola di Scienze e Tecnologie, Sezione di Fisica, Università di Camerino, 62032 Camerino Italy
[45] CNR-Istituto Officina Molecolare Lab –TASC area science park - 34149 Basovizza Trieste Trieste
[46] University of Science and Technology of China, Chinese Academy of Science, Hefei 230026, P.R. China.
[48] Department of Science, University of Roma Tre - 10046 Roma, Italy
[49] CNR/SPIN and Dipartimento di Fisica - Politecnico di Milano, Italy
[50] Sincrotrone Trieste area science park Basovizza – 34149 Trieste Italia
[51]RMP Srl
[52] Helmholtz-Zentrum Dresden-Rossendorf

[53] Università Politecnica delle Marche - Di.S.C.O.
[54] Dipartimento di Fisica e di Scienze della Terra dell'Università di Messina
[55] Dipartimento di Fisica, Università di Perugia
[56] Deutsches Elektronen-Synchrotron DESY, Hamburg, Germany
[57]Dipartimento di Fisica dell’Università di Napoli “FedericoII” and INFN Sezione di Napoli, Napoli, Italy
[58]Humboldt-Universität zu Berlin, Institut für Physik, Berlin and DESY, Zeuthen, Germany
[59]Institut für Experimentelle Kernphysik, Universität Karlsruhe, Germany
[60]Department of Physics and Astronomy, Uppsala University, Uppsala, Sweden
[61]Institut f□ur Kernphysik, Johannes Gutenberg-Universit□at, Mainz, Germany
[62]Laboratório de Instrumentação e Física Experimental de Partículas, Lisbon, Portugal
[63]INFN Sezione di Padova, Padova, Italy
[64]Institute of Nuclear Physics, Cracow, Poland
[65]Dipartimento di Fisica e Scienze della Terra “Macedonio Melloni”, Universit di Parma and INFN, Sezione di Milano Bicocca, Milano, Italy
[66]Colorado State University’s
[67]LFoundry, Avezzano (AQ), Italy
[68]Menikini s.r.l. , Albairate (MI), Italy
[69]Alenia  Aermacchi, Pomogliano d’Arco (NA), Italy
[70]CNRS-IN2P3, Université Paris-Sud, Orsay, France
[71]INFN, Sezione di Genova

## INDEX

# 1. EXECUTIVE SUMMARY

## 1.1. The IRIDE concept: technological breakthroughs as a basis for new research in fundamental and applied science

*The proposed IRIDE infrastructure (Interdisciplinary Research Infrastructure with Dual Electron linacs&lasers) will enable new, very promising synergies between fundamental-physics-oriented research and high-social-impact applications.* Conceived as an innovative and evolutionary tool for multi-disciplinary investigations in a wide field of scientific, technological and industrial applications, it will be a high intensity "particle beams factory", based on a combination of a high duty cycle radio-frequency superconducting electron linac (SC RF LINAC) and of high energy lasers. It will be able to produce a high flux of electrons, photons (from infrared to γ-rays), neutrons, protons and eventually positrons, that will be available for a wide national and international scientific community interested to take profit of the most worldwide advanced particle and radiation sources.

We can foresee a large number of possible activities, among them:

- Science with Free Electron Lasers (FEL) from infrared to X-rays,
- Nuclear photonics with Compton back-scattering γ-rays sources,
- Fundamental physics investigations with low energy linear colliders
- Advanced neutron sources by photo-production,
- Science with THz radiation sources,
- Physics with high power/intensity lasers,
- R&D on advanced accelerator concepts including plasma accelerators and polarized positron sources
- International Linear Collider (ILC) technology implementation
- Detector development for X-ray FEL and Linear Colliders
- R&D in accelerator physics and industrial spin – off

The main feature of a SC linac relevant for our facility is the possibility to operate the machine in continuous (CW) or quasi-continuous wave (qCW) mode with high average beam power (~1 MW) and high average current (~300 μA). The CW or qCW choice, combined with a proper bunch distribution scheme, offers the most versatile solution to provide bunches to a number of different experiments, as could be envisaged in a multi-purpose facility.

Europe is in a strategic position in the SC RF technology, mainly due to the strong contribution of European countries to the TESLA Collaboration. *In particular Italy is in a leading position, with knowledge and strong capabilities in the design, engineering and industrial realization of all the main component of a superconducting radiofrequency accelerator,* see Figure 1.1. INFN strongly participated since the early design stages through the final engineering and shares the know-how and has the recognized intellectual property of several main components (one of which is the cryo-module concept and its evolution).

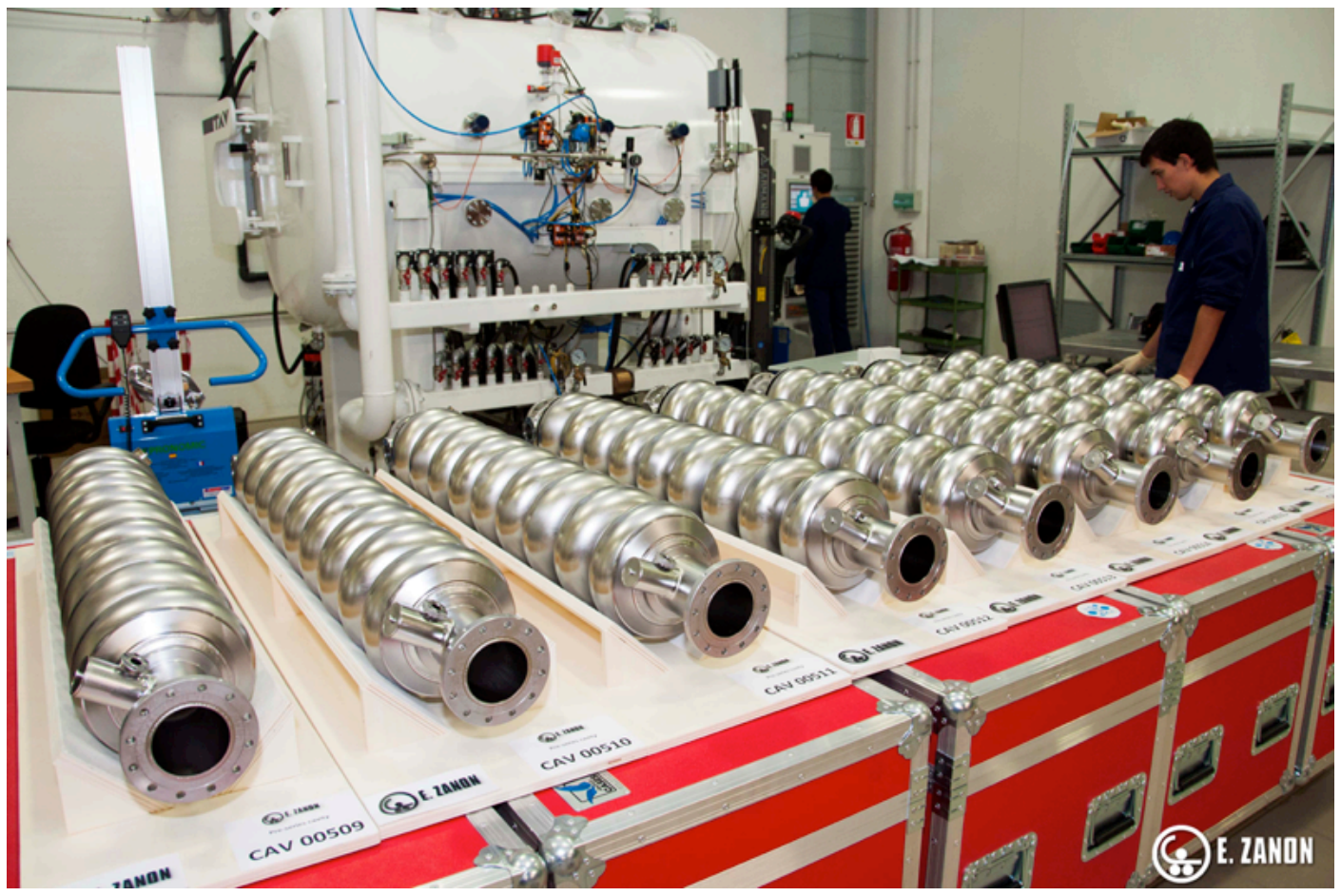

**Figure 1.1:** A set of pre-series XFEL cavities at the company E.Zanon.

As specific example of IRIDE applications the FEL radiation in the soft X-ray spectrum open possibilities for novel imaging methodologies and time-resolved studies in material science, in biology and in medicine, along with non-linear optics applications. It will allow studying crystals of size of the order of 100 nm edge outside the capability of any micro focus beam line in SR facilities with diffraction measurements of streams of nano-crystals and even of single molecules.

Another example is the high-luminosity Compton back-scattering source would be able to drive both a low-energy $\gamma-\gamma$ collider for fundamental studies on QED, i.e. to observe and measure photon-photon scattering, and would be of great relevance to pursue strategic applications in the fields of nuclear waste diagnosis and treatment, as well as national security applications, not to mention materials science and advanced medical and radio-biological researches made possible by such a new photon source. In this way $\gamma$–rays with energies of 10-20 MeV can also be produced and directed head-on against electrons of ~700 MeV. A rich physics program can be studied, which includes – among others – the precise measurement of the $\pi^0$ width through the process $e^- \gamma \rightarrow \pi^0 e^-$ (Primakoff effect), and precision tests of QED in the MeV range. These measurements, which provide important tests of the Standard Model, are not possible at present electron-photon colliders due to the low photon intensities of the machines.

The scientific activities foreseen at IRIDE lead to specific and challenging requirements with respect to x-ray instrumentation and in particular to x-ray detectors for FEL and Compton radiation. These requirements ask for detectors that cover a wide energy range, from soft to hard x-ray energies, with specification in some cases exceeding the existing technology. *Successful exploitation of the unprecedented features of the x-ray radiation that will be available at IRIDE calls for a dedicated and substantial detector R&D program which also opens the opportunity to enhance the*

*Italian scientific and technological leadership as well as the international links in this cutting-edge field.*

The electron beam can also be used to produce neutrons by photo-production. The neutron spectrum produced in this way, peaked around 1 MeV, has important implications both in fundamental (nuclear X section and decay measurements) and applied physics (Neutron tomography and material studies with neutron scattering).

Last but not least, advanced accelerators concepts could be also tested, like Polarised Positron Source in the context of ILC design effort, and Plasma acceleration or Dielectric wake field acceleration to increase the final beam energy giving new opportunities for basic and applied scientific cases.

*The realization of such a large facility at the Tor Vergata site in close proximity to the LNF laboratories will allow INFN to consolidate a strong scientific, technological and industrial role in a competing international context both to deploy a national multi-purpose facility along the scientific applications discussed in the following sections, and to prepare a strong role for the contribution to possible future large international HEP projects as the International Linear Collider.*

The aim of this report is to illustrate the wide range of applications of a multi-purposes linac as the one foreseen for IRIDE, providing also a consistent set of machine parameters and a preliminary project cost evaluation.

### 1.2. IRIDE layout: staging and upgrade potentials

The backbone of the IRIDE facility is a superconducting high duty cycle electron linear accelerator, with the required 2 K cryogenic plant, based on the L-band standing wave RF (1.3 GHz) cavities, developed by the TESLA collaboration, which currently drive the FLASH FEL facility in DESY and which, with minimal improvements of the cryo-module cooling system, could be upgraded to CW or qCW operation.

The second core device of the facility is the high energy cryogenically cooled Yb:YAG Laser system operating in a chirped pulse amplification architecture followed by a frequency conversion stage to achieve 515 nm wavelength. This technology allowed achieving recently 1 J at 100 Hz in the picosecond regime with a bandwidth of 0.1%.

By using standing wave SC accelerating structures, that can accelerate beams in both longitudinal directions, one can see an attractive scheme based on two linacs operating at a maximum energy of 2 GeV each, when working in the collider mode, or used in cascade, as a single longer linac, to boost the electron energy up to 4 GeV for higher energy electron beam applications. *In addition when operating in the collider mode both linacs may partially recover the electron kinetic energy of the beam leaving the opposite linac after the interaction, thus increasing the overall efficiency of the system and simplifying the beam dump design.*

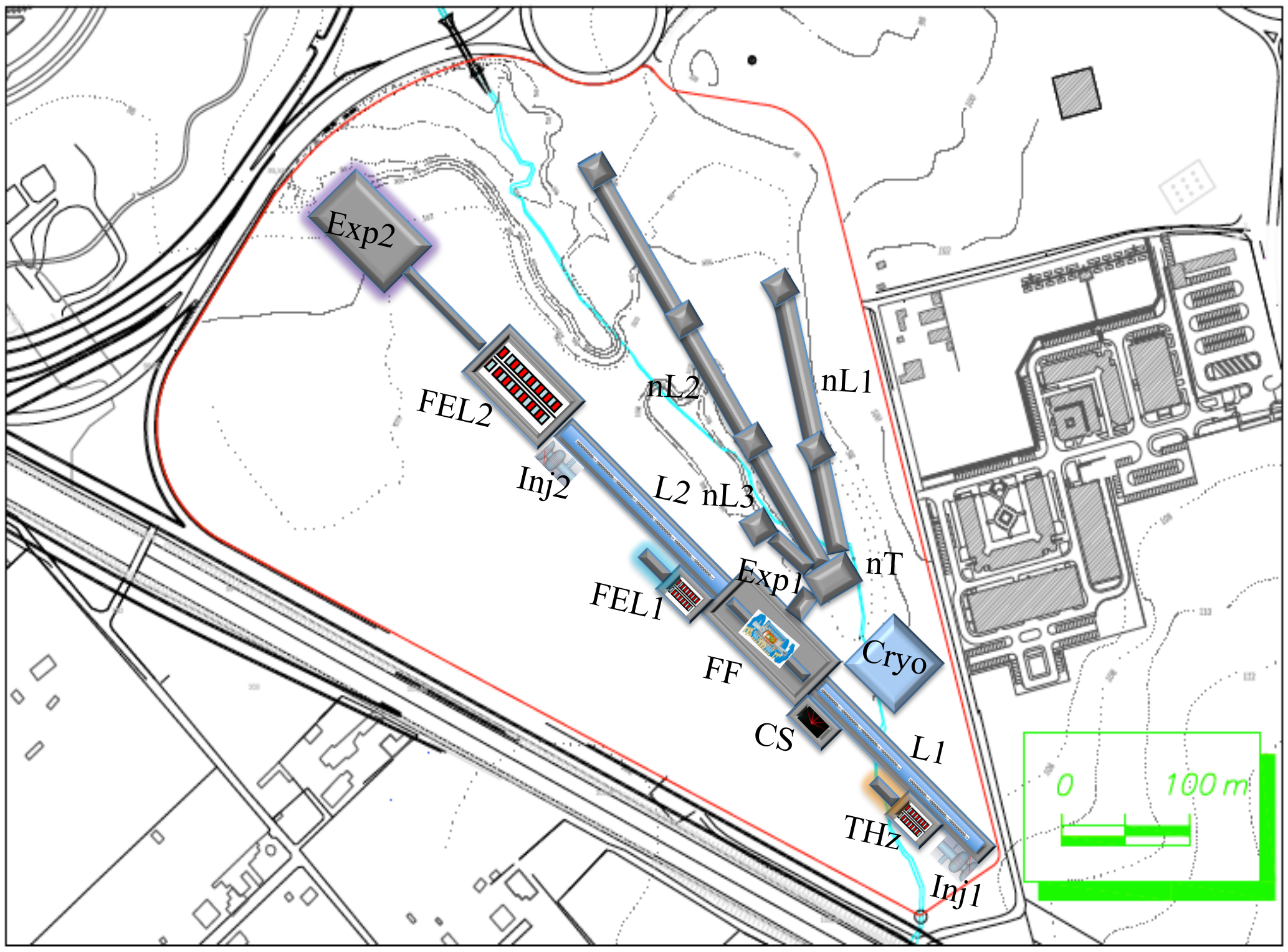

**Figure 1.2:** Schematic layout of the IRIDE accelerators and radiation sources complex. Some relevant components are indicated: the cryogenic plant hall (Cryo), the electron injectors (Inj1, Inj2), the SC linacs (L1, L2), the FEL devices (THz, FEL1, FEL2) and the user experimental hall (Exp1,Exp2), the laser system for Compton source (CS), the Final Focus beam line (FF) in the experimental hall Exp1 for colliders and nuclear photonics, the neutron source target (nT) with experimental beam lines (nL1, nL2, nL3).

As indicated in Figure 1.2 the first 2 GeV linac system (L1) can drive FEL, Neutron and THz radiation sources, electron-on-target physics experiments and, in combination with the high energy laser, a γ-ray Compton source is also possible with many applications in the field of nuclear-photonics and as a possible tool for developing a polarized positron source as the one required for ILC.

With also the second 2 GeV linac installed (L2) one can envisage a low energy linear collider scheme for electron-electron, electron-photon, photon-photon and eventually electron-positron scattering studies. The combination of the two linacs, boosting the electrons up to 4 GeV, could also drive a short wavelength FEL user facility (Exp2).

The **Neutron source (nT)** requires a medium energy electron beam as the one extracted from L1 driven to impinge on bulk of high Z cooled target (nT), where it loose energy mainly by bremsstrahlung, producing an electromagnetic shower cascade. The photons of the shower can excite the nuclei of the target with which they interact and these excited nuclei go back into the fundamental state by emitting one or more nucleons. At the state of the art of the project, we have mainly focused on the Tungsten as possible choice for the target of IRIDE: the estimated rate emissions of neutrons (up $10^{15}$ n/s) and other secondary particles, that are described in more details in chapter 6, have been obtained for a Tungsten cylindrical cooled target with 7 cm diameter and 6 cm height. A 3m thick Iron shielding is also foreseen. The beamlines of interest are of three types: short beamlines for Chip Irradiation and Imaging (**nL3**), long beamlines for applications requiring time of flight measurements, like Bragg Edge Transmission, Diffraction and Nuclear Resonance Capture Analysis (**nL1**), and even longer beamlines (~200 m) needed for neutron oscillation studies (**nL2**). Each beam line needs to be equipped with shielding, diagnostics and detectors that are described in detail in chapter 6.

The main components of the **Compton Source** (**CS**), see chapter 7, at the exit of L1 are:

a) a high brighntess *GeV*-class electron Linac (L1) capable to deliver multi-bunch trains, *i.e.* working at 100 *Hz* rep rate with at least 50 electron bunches distributed over the RF pulse duration (from 0.5 up to 1 microsec), carrying a fraction of a *nC* bunch charge at very low rms normalized emittances (< 1 *mm·mrad*) and energy spreads (<0.1%)

b) a high energy, high quality, high repetition rate laser system, delivering pulses carrying at least 1 *J* of energy (in the fundamental), psec pulse duration, 100 *Hz* repetition rate, high quality ($M^2$<1.2), such to be focused down to typical spot sizes of 10 *μm* at collision with the electron bunch

c) a laser recirculator consisting of a two parabolic confocal mirror set, capable to recirculate the laser pulse a number of times equal to the electron bunches within the train (<50), by focusing it down to the collision point, recollimating and reflecting it back to the other mirror which in turns refocuses it down back to the interaction.

The expected performances for the *γ*-ray beam delivered are: tunability between 1 and 50 *MeV*, bandwidth smaller than 0.3%, full control of polarization (linear, larger than 99.8%), spectral density larger than $10^4$ *photons/s·eV*, and peak brilliance larger than $10^{22}$ ($photons/s \cdot mm^2 mrad^2 \cdot 0.1\%$).

Several **FEL source** configurations are possible at IRIDE ranging from IR to X ray wavelength radiation, as discussed in chapter 3.

A Seeded configuration is possible at the exit of L1, by using an externally injected laser signal. In this case the maximum operating energy is fixed by the source exploited as seeding. If we consider the 27-th harmonics of the Ti-Sa (26.9 nm), the beam energy is constrained below 1GeV and the FEL tunability could range from 27 nm to 1.65 nm (**FEL1**). User beam lines can be accommodated in the first experimental hall (**Exp1**).

At the highest energy end of the linac (**FEL2**) a combination of Oscillator, SASE and Seeded operational mode offer an attractive and unique possibility. As shown in Figure 1.3 an oscillator operating in the VUV region is used to produce bunching in the e-beam, which is successively injected into the downstream sections of the undulator chain tuned at higher harmonics of the oscillator.

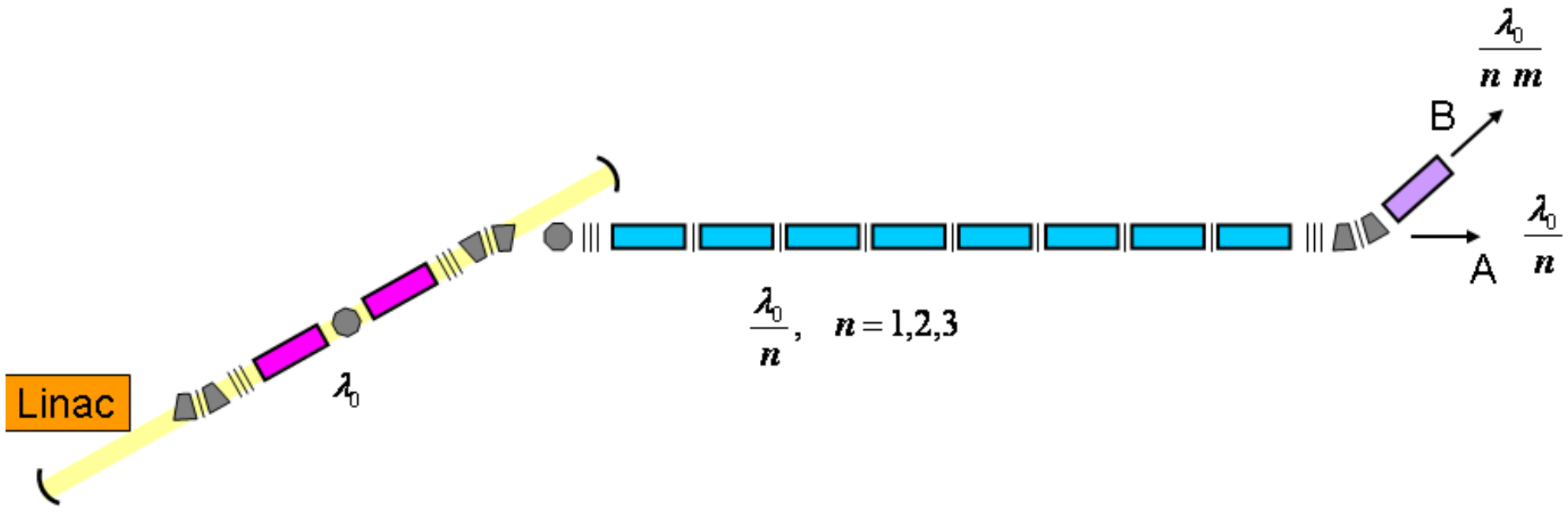

**Figure 1.3:** Undulator Chain at the S-C LINAC output, the first component is provided by an oscillator

The rather narrow bandwidth of operation of the cavity provides a constraint for the energy of the e-beam at an energy around $2.28\,GeV$. In this configuration a significant amount of third and fifth harmonics allows its use in the successive section to get pre-bunched SASE operation at $4.5\,nm, 2.7\,nm$. Removing the cavity mirrors and operating at full linacs energy (up to 4 GeV) we get an output wavelength of 0.6 nm, which can be extended to 0.2 nm, if the choice of a segmented undulator is foreseen and the last sections are replaced by a super-conducting undulator with $\lambda_u = 1.cm, K = 1$. User beam lines can be accommodated in the second experimental hall (**Exp2**).

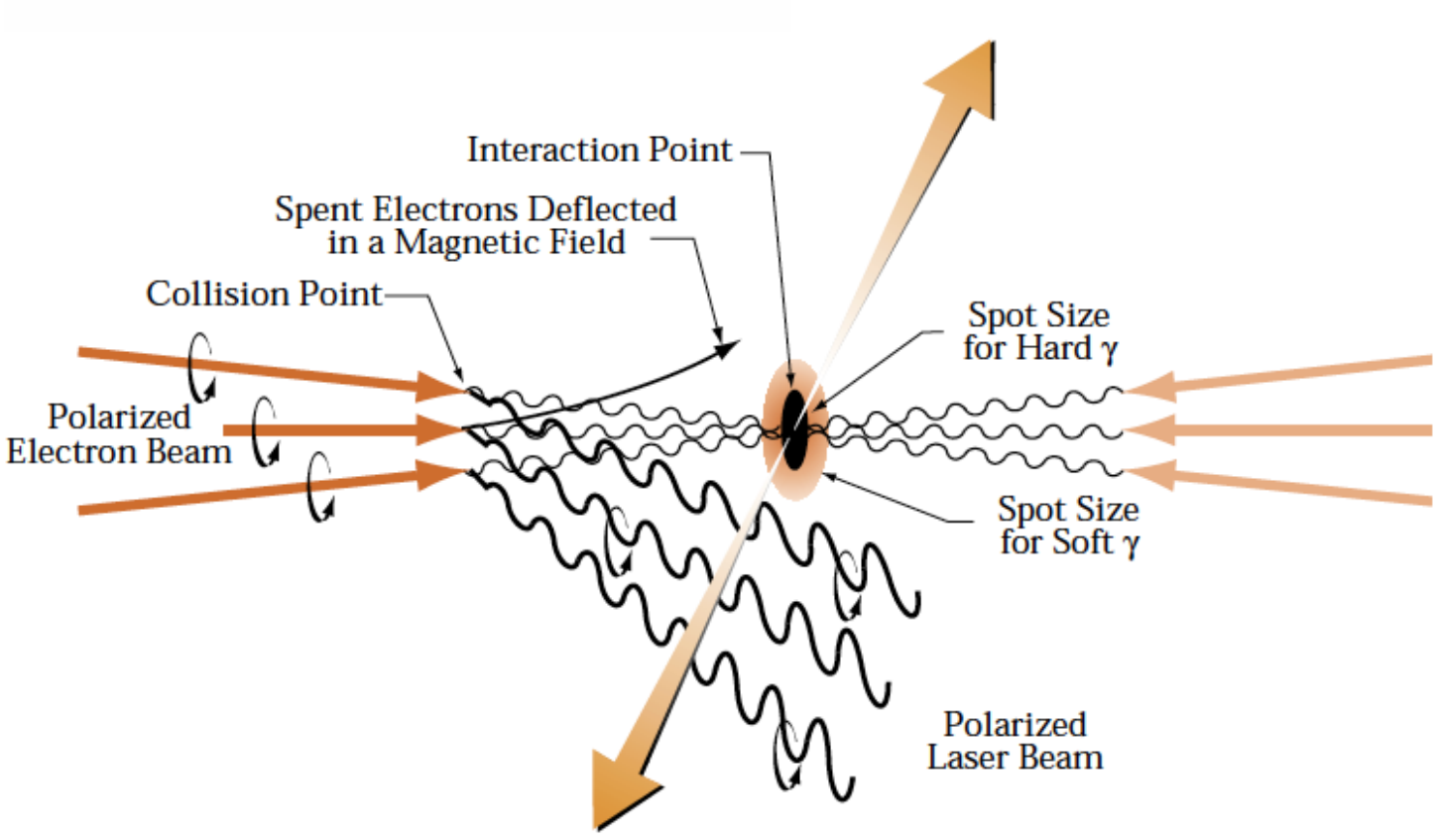

**Figure 1.4:** Scheme of principle of a photon collider. High-energy electrons scatter on laser photons and produce high-energy photon beam which collides with a similar photon or electron beam at the interaction point IP.

Experimental hall 1 (**Exp1**) will host also the most challenging components of the entire project. The **electron-electron collider option** will be essentially based on the final focus (**FF**) system already operating at ATF2, based on the recently proposed compact final focus optics with local chromaticity correction and where 100 nm electron beam spot sizes have been already achieved. The feasibility of the **electron positron collider** is strongly dependent on our capability to produce low emittance positrons. In the IRIDE R&D program is comprised the development of a positron source based on direct conversion of a gamma ray beam in a solid target as discussed in chapter 9. The **gamma-electron and gamma-gamma options** will require a careful design and development of the interaction region. In a photon collider in fact two high energy electron beams after the final focus system travel towards the interaction point (IP) and at adistance of about 1- 5 mm from the IP collide with a focused laser beam. After scattering, the photons follow their direction to the interaction point (IP) where they collide with a similar opposite beam of high energy photons or electrons. Such a new collider configuration has never been realized so far and is the subject of many design studies around the world. A dedicate design for the IRIDE facility is under way.

Advanced accelerators techniques could be also investigated in the large experimental hall (**Exp1**). The success of the advanced accelerator activity as a vigorous and intense R&D program focused on the enabling technologies of plasma accelerators, Compton converters, gamma beam focusing, polarized positron source, superconducting RF gun and the associated advanced diagnostics instrumentation, could allow envisage a convenient energy upgrade of the facility to tens GeV level in a higher energy range of scientific applications.

The IRIDE facility can be hosted in the 30 hectares area on the University of Rome Tor Vergata campus site a few km southeast of the city of Rome. It will be disposed along a linear geometry, approximately 700 m long. The interested area is just alongside of the CNR territory and it is approximately a couple of km away from the ENEA and INFN sites in Frascati.

Highlights extracted from the IRIDE scientific cases are reported in the next paragraphs, more details are available in the following chapters of this report.

### 1.3. Science with photons: new insights into the facets of nature and life

Photons are the most important probe to investigate our environment. From radio frequencies to hard X-ray photons are used since many years to get both electronic and structural information on virtually any materials, from biology to condensed matter systems to nuclear physics. Among the different types of sources, storage rings to emit Synchrotron Radiation (SR) have a special place because of the capability to produce radiation with very high flux and brightness in an energy range from infrared to hard X-ray. The development of storage rings with special magnetic components over the past thirty years has led to third generation machines, especially designed for SR research, with the peculiarity to have radiation with unprecedented degree of brilliance.

Last progresses in the physics of the linear accelerators (LINAC) opened the possibility to a new jump in photon sources quality especially in the X-ray energy domain. The LINAC based sources are called Free Electron Laser (FEL). These machines are sources of coherent radiation presenting similar optical properties as conventional lasers. The main difference is in the lasing medium, being in this case an electron beam moving through a magnetic structure made by a very long undulator. *The X-ray FEL has the possibility to generate X-ray beams with a peak brilliance of several order of magnitude* *higher that the third generation SR sources for both the spontaneous and coherent emission. It also offers an extremely short pulse length, typically less than $10^{-13}$ s (femtosecond (fs) time domain), with some degree of tunability and a high degree of either linear or circular polarization.*

The advent of such short and powerful pulses of fs laser sources has disclosed the opportunity to make real time experiments in the fs time domain using many different techniques, to study the interaction of radiation with matter in a non-linear regime and to use nano-crystals in protein-crystallography. Actually there are four operating FEL facilities (Flash, Fermi, LCLS, SACLA) and three in construction (XFEL, LCLSII, SwissFel), covering the spectral range from soft to hard X-ray. The first experiments performed in the existing facilities begin to enlighten the strong potentialities in many different fields, see Figure 1.5, and at the same time several critical points that should be resolved to fully exploit the possibilities of these new machines.

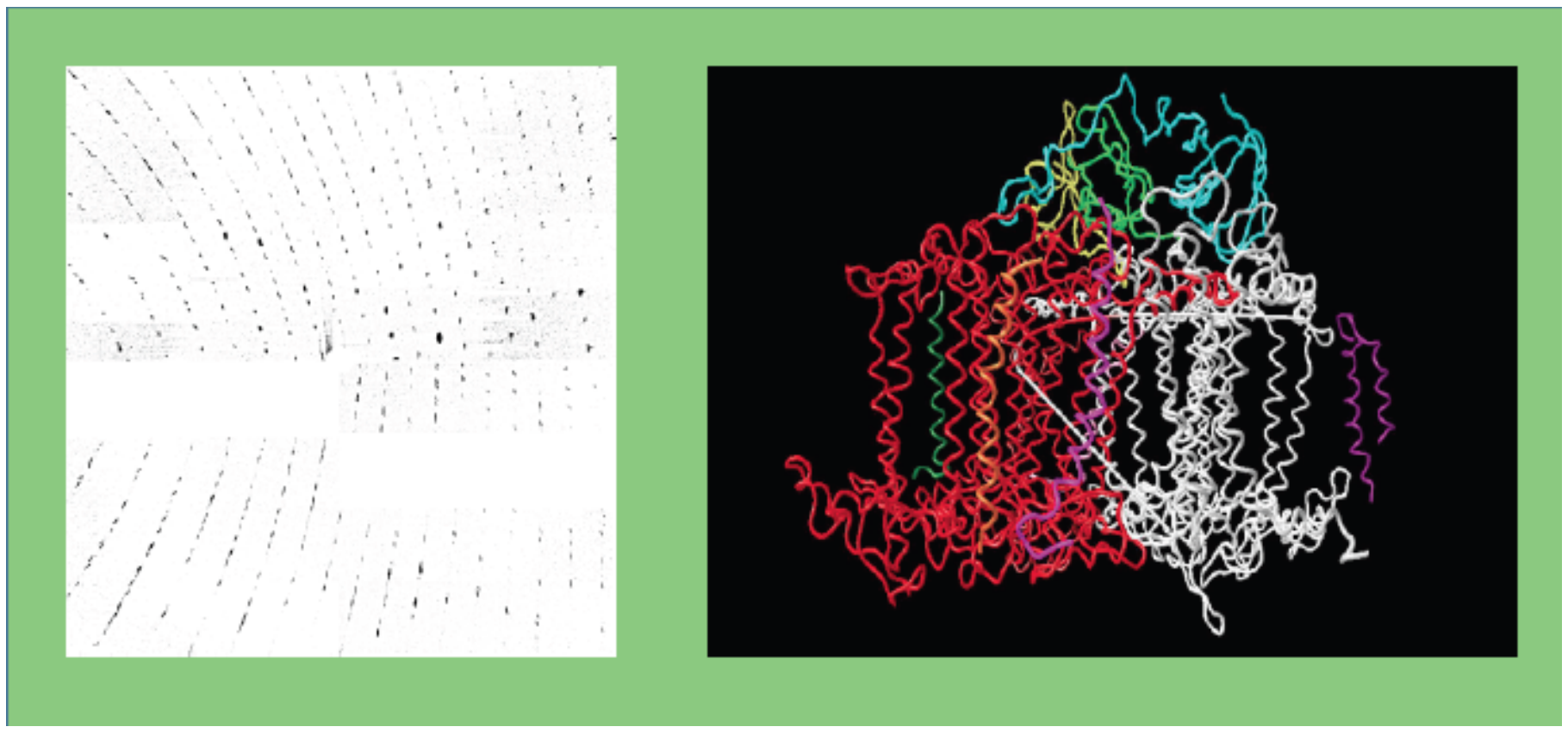

**Figure 1.5** Left: diffraction pattern of the Photosystem I complex measured at LCLS with a photon beam of 1.8 KeV and a time length of 80 fs. Right: Electronic density reconstruction at 7Å of structural resolution.

One of the FEL options at the IRIDE facility is a source of coherent X-rays, covering the range of wavelengths (λ) ranging from 0.5 to 4 nm at fundamental harmonics, depending on the electron beam energy, and it will be also able to reach the Ångström region using the third and fifth harmonics. *In some way it covers a radiation region complementary to those of other existing or in construction facilities, and will be provided also with an ancillary equipment to produce radiation down to the infrared and THz region.* The IRIDE FEL has a reasonable wavelengths overlap to satisfy users in many different fields of science and to ensure at the same time a beneficial level of competiveness.

At the same time the IRIDE FEL will provide radiation with the unprecedented characteristics like:

- *Self-seeding: narrow bandwidth, wavelength stability, higher brightness and energy tunability.*
- *Polarization control: tunable linear and circular polarization*
- *Two color pulses: simultaneous delivery of independent wavelength pulses*
- *Delayed pulses: independent delay of two pulses up to a few ps*

*An X-ray source with these specifications will allow the investigation of photon-matter interaction in a new regime, and will become a high competitive facility in the worldwide growing FEL framework.*

The considered scientific case for IRIDE FEL reports experimental proposals ranging from time dependent spectroscopies in condensed matter to imaging for biological applications.

Among the highlights of the IRIDE Scientific Case, see chapter 4 for more details, of paramount importance is the *Spectroscopy of flying proteins*. Proteins, long linear chains composed by L-α-amino acids joined together by peptide bonds, rule fundamental functions in life processes. To achieve their functional properties the interplay between electronic properties and structural properties is crucial. Proteins take their shape spontaneously after the synthesis in the cell, but the structure depends also on the environment properties (solvent, salt concentration, pH, temperatures and molecular chaperones). The chirality influences the assembly, folding and activity of biological molecules: amino acids that form proteins are all in the L configuration, with the exception of (non-chiral) Glycine. So far the task to determine the structure of proteins, and subsequently their electronic properties, was carried out by X-ray diffraction and nuclear magnetic resonance in crystals and in solution. *The characterization of structural and electronic properties of proteins in the gas phase, as is possible at IRIDE, would provide valuable information to understand the folding.* The absence of solvent interaction can reveal the balance of the molecular weak forces that determine the shape of the protein.

To bring un-fragmented proteins into the gaseous phase, the state of the art technique is represented by Electro Spray Ionization (ESI). The ESI technique solved the problem of how to study large molecules in solution by mass spectrometry that needs a high vacuum environment. Pioneering works combining ESI with laser spectroscopies revealed the possible application in the characterization of electronic and structural processes. *These preliminary results envisage the application of ion-spectroscopy to proteomics, but this research field is still completely unexplored.* The low density of the target (space charge limits the maximum ion density $10^6$ ions/cm$^3$) prevents to extend such experiments in the VUV-soft X-ray wavelength range with the present sources. *IRIDE with its high flux and focusing represents an ideal source for these studies. Moreover the IRIDE wavelength range can cover the excitations to the core states, focusing on bond character and local environment, and providing a rich and detailed description.* These results would be complementary with respect to those obtained in crystals or solution, and give insight into the influence of the solvent on the protein shape.

In addition the possibility of reaching wavelength of the order of 1 Å will allow to measure protein crystals of very small dimension, typically of the order of 100 nm edge, which are those produced most of the time in the crystallization trials. The measurement of such samples is outside the capability of any micro focus beam line in the standard Synchrotron Radiation facilities. The fluence of IRIDE in the hard X-ray energy range is still enough to enable structure determination from diffraction measurements of streams of nano-crystals, as recently done at LCLS-FEL facility where diffraction patterns from a large photosystem membrane complex that crystallizes in samples of about 250 nm size have been recorded.

The very broad energy range available at IRIDE will allows a complete structural characterization of protein samples in many different physical-chemical conditions. This is crucial to understand the structure/function relationships from a molecular to a cellular level opening new perspective for the treatment of genetic disabilities and diseases and for the design of more effective drugs.

Powerful THz radiation sources have been also considered in the IRIDE design. The accepted paradigm of condensed matter physics in fact is that the high-energy short-time dynamics affects the low-energy long-living degrees of freedom. Actually, pushing a system out-of-equilibrium, this hierarchy could be reversed. This determines a non-linear coupling among several degrees of freedom providing the possibility to coherently manipulate different states of matter. In this scenario, one can cite for instance the possibility to coherent induce a conformation transition in macromolecules, selectively pumping a low-energy collective mode; or even inducing a coherent structural phase transition through a phonon pumping. THz radiation could be coupled to x-ray light for THz-pump x-ray diffraction probe experiments.

### 1.4. Science with γ-rays: a deep view of exotic nuclear structures

The technology of producing *γ*-ray beams by Thomson/Compton back-scattering of high brightness electron beams with high energy lasers is rapidly progressing: in the last decade significant advancements in designing and commissioning Thomson sources is leading this technology to opening the Nuclear Photonics era, see chapter 8.

Thanks to several initiatives world-wide (mainly LLNL in the US, Japan-JAEA and the European ELI-NP) the state of the art in producing high brilliance/spectral density mono-chromatic *γ*-ray beams will be soon enhanced, stepping up from the present performances (*γ*-ray beams with bandwidth nearly 3% and spectral density of about 100 *photons/s·eV*) up to what is considered the threshold for Nuclear Photonics, *i.e.* a bandwidth of the *γ*-ray beam lower than 0.3% and a spectral density larger than $10^4$ *photons/s·eV*. *γ*-ray beams of these characteristics will enable Nuclear Physics studies and Nuclear Applications based on exciting nuclear states not accessible with present machines.

Radiation at short wavelength as $\gamma$ rays is used to excite the nuclear resonant fluorescence (NRF), so that different nuclei can be identified by the distinct pattern of NRF emission peaks.

In nuclear physics there is large interest at present for the neutron-rich systems. On the one hand, existing and planned radioactive beams facilities aim to locate the position of the neutron and proton drip-lines (i.e., the limits defining whether a nuclear system is bound), and to study the properties of the isotopes in which the neutron/proton ratio differs from the values that characterize stable nuclei. On the other hand, another example of nuclear matter under extreme conditions is the matter that compose the compact astrophysical objects like the neutron stars. From a general point of view, an understanding of these “exotic” nuclear systems cannot be reached without a better assessment of the isoscalar (T = 0) and isovector (T = 1) components of the nuclear Hamiltonian. Even in stable nuclei the isovector properties are poorly known: the neutron radii, the systematics of isovector collective states, the pairing interaction in the T = 0 and T = 1 channel - to name only a few - are still object of strong debate. The possibility of new experiments that clarify these issues is consequently of great importance for the development of nuclear science.

In particular, Nuclear Physics will benefit from the availability of these new generation $\gamma$-ray beams for:

A) studies of the nucleus structure at the Pigmy and Giant Dipole Resonance excitation (to probe the structure and isospin properties of nuclear systems) with unprecedented resolution in reconstructing the nuclear states: this is crucial also to understand some unknown processes in the stellar nucleo-sinthesys

B) studies of two level barionc states in the high energy resonance of the nuclei, above 20 *MeV* and up to 60 *MeV*, crucial to reconstruct the equation of state of the nuclear matter

While new Nuclear Applications will be pursued thanks to these $\gamma$-ray beams in several fields:

A) detection and imaging of fissile and strategic material with isotopic reconstruction of the components (*e.g.* detection of fissile materials hidden in metallic containers), with large impact on the national security scenario

B) remote sensing and diagnosis of nuclear wastes in containers, with reconstruction of the isotope and nuclear composition of the waste material, with large impact on the atomic energy scenario

C) Medical imaging and therapy

It is interesting to explore also the possibility of generating positrons by direct conversion of a gamma ray beam in a solid target. There are two main advantages in this scheme:

- If the gamma photons are circularly polarized it is possible to produce a positron beam with a high degree of polarization.

- By taking advantage of a thinner target, a reduction in the opening angle of the positron beam is possible. This in conjunction with a smaller source size could bring an order of magnitude improvement to the emittance of the positron beam.

Positron production by photon conversion has *already* been shown superior to the traditional electron bombarding because the high average intensities of the beams required to maximize the collider luminosity would be otherwise seriously limited by the target damage consideration. For the same number of positron generated, a gamma ray based positron source reduces the thermal load problems because one is able to use a shorter target with lower Z material with a higher heat capacity. These advantages quantify in a two orders of magnitude increase of the thermal damage threshold in the positron production.

### 1.5. Science with neutrons: from fundamental physics to industrial applications

Neutrons represent a unique probe for studying matter on the molecular scale, thus opening a wide range of applications: from material science to life science, from engineering and industrial applications to fundamental physics experiments, see also chapter 6. *They cannot compete with electromagnetic radiation in intensity, but they are complementary with it because they penetrate substances that block the electromagnetic radiation (like metals) and are stopped by long radiation length materials, in particular hydrogenated and deuterated ones*. Highest intensity neutron sources in the world are based essentially on three different production mechanisms: some fission processes (such as that of $^{235}U$) in nuclear reactors, proton-driven spallation sources, and electron-driven (photo-production) sources. With a large community using neutrons and due to the high specialization of instruments for the different classes of neutron experiments, photo-production sources demonstrated to be very useful, as the GELINA facility based on a 70-140 MeV electron beam in Geel, Belgium. Another interesting example is the n-Elbe source, driven by a 40 MeV electron beam, at FZ Dresden-Rossendorf. Indeed, photo-production facilities can be more cost-effective than spallation sources for neutron fluxes up to $10^{15}$-$10^{16}$ n/s at the target, even though the neutron yield per primary electron is (depending on the primary beam energy) at least 10-20 times lower with respect to proton-induced spallation. *The opportunity of using an intense, high-energy, electron beam – produced in the case of the IRIDE project by superconducting high average current linacs – opens the possibility of having a new photo-production neutron source that can have a significant place in the European panorama in the coming decade and can be a very useful facility*. A source able to produce up to **$10^{15}$ n/s** at the target, i.e. like a medium size reactor, would indeed not only enlarge the access capacity for the large community of European scientists, but also could provide a complementary and original approach with respect to the existing neutron facilities, thus enriching the experimental and application opportunities, and profiting of the large wealth of expertise in the field.

Regarding fundamental physics investigation that are possible with IRIDE neutron facility, **neutron-antineutron oscillations** are very important since they would allow precision testing of the fundamental CPT-symmetry, where C is the charge conjugation, P is the inversion of space and T is the time-reversal, very closely connected to the quantum field theory through the CPT-theorem. After A. D. Sakharov has clearly put in connection the violation of CP symmetry and the baryon asymmetry of universe and V. A. Kuzmin noticed that baryon number violation could lead to n→nbar oscillations, more recently it was concluded that these oscillations would represent one of the most accurate test of the CPT symmetry. Of course, the observation of such an oscillation also would make the neutron a Majorana fermion (with a tiny Majorana component). *If discovered, n →nbar oscillation will establish a new force of nature and a new phenomenon leading to the physics beyond the SM at the energy scale above TeV. In addition, will help to provide understanding of matter-antimatter asymmetry and origin of neutrino mass.*
An experiment aimed to improve the present limit on the n→nbar oscillation lifetime (i.e. $\tau > 10^8$ s) obtained at Institut Laue-Langevin, would require, in addition to the possibility of producing cold neutrons with a cryogenic moderator, a dedicated long beam-line with high-vacuum and terrestrial magnetic field shielding, and a detector placed around a thin target in order to reconstruct the anti-neutron annihilation products.
In addition impinging the electron beam on a target produces also charged pions that decaying produce muons. High intensity $\mu^+$ beams are used for the search of the lepton violating deacy μ-->e γ. Preliminary simulations show that with a not yet optimized carbon target a rate of 50 times the beam on which the Muon Electron Gamma (MEG) experiment at PSI is currently operated can be achieved. *This solution opens for a huge potential in the search for lepton-flavor-violation.*

As far as applied physics is concerned, the interest from the scientific community (in particular the Societa' Italiana di Spettroscopia Nucleare) and the industrial community. The former stresses the absence of devoted neutron beamlines in Italy and the long time needed to satisfy beam time requests. As far as industries are concerned, several companies have expressed interest in utilizing neutron beams in research and control activities.

Possible applications of the IRIDE neutron source in the field of applied physics are:

**Neutron Resonance Capture Analysis (NRCA)**: Each resonance is the fingerprint of a nuclear specie (isotopical recognition) thus allowing for the elemental material analysis (qualitative and quantitative) especially on metallic samples (e.g. cultural heritages).
**Bragg Edge Transmission (BET):** By means of this technique, stresses and strain in bulky samples can be analysed. This analysis is very important for both industrial as well as cultural heritages applications.
**Chip irradiation** : In order to test the robustness of electronic devices to neutron field in a few minutes, neutron beams produced at facilities are desirable as the may provide an almost atmospheric-like neutron spectrum but several order of magnitude more intense.
**Radiography and Tomography (NR, NT)**: By means of radiography it is possible to obtain an image of a object that evidences the internal structure, by rotating the sample with respect to the incident beam and collecting images for each angular position a 3D image of the object is obtained (tomography).
**Neutron metrology**: In this context, the Italian National Institute of Ionization radiation Metrology (INMRI) is interested in having in Italy (and especially in Roma area) a high energy neutron source in order to develop primary standards for neutron emission rate and energy spectrum calibration

The characteristics of the neutrons that make them of interest for applied research can of course be used in industrial research. Examples of industrial field with known applications with neutrons are:

- Efficient and cost-effective fabrication of a variety of **advanced materials**. Neutron scattering techniques can be used for the development of novel transformation-induced plasticity steels or precipitation-hardening Al and Ni alloys, to be used, for example, in aeronautics.
- **Pharmaceutical products**. The development of new drugs and drug delivery systems, which is strictly related to the detailed understanding of the mechanisms of disease, as well as the improvement of the product shelf-life can be carried out by the employment of neutron techniques.
- **Thermoelectric materials**. Here, neutrons allows to identify efficient and non-pollutant systems for the development of innovative thermoelectric devices combining low thermal conductivity with high electrical conductivity, to be employed for waste-heat recovery and in the refrigeration industry.
- **Renewable energy** sources. In such a field, more and more effective engines, materials for lower heat loss and less energy spill and greener processes for industry are requested. Novel materials for solar and fuel cells, as well as hydrogen storage materials can be developed thanks to neutrons.
- **Agro-food** systems. Plant strategies and metabolism in resistance to drought can be characterized by neutron methods.

*Finally, a neutron facility of the kind we are proposing, can have a positive impact under other two important points: training and education of young scientists, and development of new detection techniques*. An interesting example in this respect is the field of neutron detectors, that has gained importance in the last years due to the growing problem of $^3$He-replacement, and can clearly benefit of an easier access and usability, lower intensity facility.

### 1.6. Particle physics opportunities: assembling the Standard Model puzzles

It is commonly accepted that the Standard Model (SM) of elementary particles interactions is the model, which describes the visible part of the nature and of the Universe. Recently the experimental results from the Large Hadron Collider at CERN have provided us with very important information on the mass of the SM higgs-like particle. However, the existence of this particle with a given mass does not solve, by itself, all the long-standing puzzles of the SM, such as a problem of the SM hierarchy, the naturalness of the higgs boson and the electroweak (EW) symmetry breaking. Even though all the SM parameters are now measured to a high accuracy, the necessity of the New Physics (NP) existence for explaining the SM puzzles is still an open question. From a theoretical point of view, precise and complicated calculations are required to answer these questions, and high-precision input information on the SM parameters is a must. Due to the intrinsic complexity of the calculations, as one needs to study the running of the non-abelian gauge theory parameters over a dozen of orders of magnitude up to the Planck scale, even small experimental uncertainties in the SM parameters have a drastic impact on the conclusions, which can be drawn from such computations. The implications affect our understanding of the fundamental issues of the “conspiracy” between the SM couplings, the EW phase transition, Universe inflation, the cosmological constant, and also the nature of the Dark Matter (DM).

*It is important to stress that the precise values of the SM parameters, due to the renormalization group evolution, can be obtained only by simultaneous studies at high-energy and low-energy scales.* The former point highly motivates the International Linear Collider (ILC) initiative, while the IRIDE project can pursue the latter one and serve as an accelerator-technology test installation and a research facility, see also chapter 9. The latter point motivates the possible use of the IRIDE facility as a precision tool for the SM exploration at low- and medium-energy scales, with a high priority on the information about the EW couplings of SM, which drives the evolution of the electromagnetic running coupling and the squared sine of the weak angle. Also a rich hadron phenomenology is accessible at these scales, which allows to study issues of the QCD confinement, where the ordinary perturbation theory approaches fail to work. While the technological aspects are discussed in another Chapter of this White Book, the present Chapter deals with an overview of the fundamental particle physics opportunities of the IRIDE project.

It is anticipated that the construction of the IRIDE facility will be realized step-by-step. We review the particle physics goals of the full accelerator complex according to the order in which one can launch the various steps of the facility. We start with the physics program that can be pursued with an electron beam on target, further we investigate that of the electron-photon collider, of photon-photon collisions and finally of the electron-positron and electron-electron collider. It is important to stress that a synergy of all the proposed measurements can lead to a very reliable and cross-checked experimental exploration of the SM. *In addition with the expected luminosity of IRIDE, in the electron-positron mode the operational time required for the physics program would be limited and well in accordance with the beam requests for the other functioning modes (e.g, FEL) of the machine.*

The **electron-on-target** physics program makes IRIDE a discovery and also a precision physics machine. Among the searched candidates there are the hypothetical particles, like the very-weakly interacting massive U(1) gauge boson (U-boson) as a DM particle candidate and the non-hypothetical, well investigated theoretically, but yet undiscovered, “true muonium” states (TM), which are the bound states of muon and anti-muon with the lifetime of an order of a picosecond. Utilizing the polarized electron beam dumped onto the proton target, one can measure the left-right parity violating asymmetry of electron-proton scattering at the per cent level, and thereby extract precisely the electroweak mixing angle.

The **electron-photon collider** allows to utilize the elementary Primakoff process to produce the light pseudoscalar (and scalar) mesons in order to precisely measure their two-photon decay widths and thus to tackle the triangle anomaly of QCD. In addition, one can perform the U-boson search in the lepton triplet production channel. A special feature here is the availability of the highly-polarized photon beam. This allows to use the lepton triplet production at IRIDE as a research laboratory for development of the methods of polarimetry to be used in astrophysics to measure the polarization directions of incoming high energy γ-rays. Finally, triple Compton effect can be used to study the properties of entangled states. *These measurements, which provide important tests of the SM, are not possible at present electron-photon colliders due to the low photon intensities of the machines.*

Low-energy **photon-photon collisions** give a direct view into the vacuum properties of Quantum Electrodynamics (QED), allowing for precision tests of QED in the MeV range, and more generally of Quantum Field Theory (QFT). *The IRIDE accelerator complex can generate for the first time colliding photon-photon beams by Compton backscattering, and this opens the fascinating field of low-energy photon-photon physics.* The technology needed to carry out a photon-photon physics program at energies close to 1 MeV would disclose new developments at higher energies, where a photon-photon Higgs factory could be a nearly ideal discovery machine.

The high-luminosity **electron-positron** and **electron-electron collider** with variable energy would be an extremely useful tool for the study of hadronic vacuum polarisation effects, measurements of the effective electroweak mixing angle and contributing to the description of the muon anomalous magnetic moment and the running QED coupling constant by providing the hadronic cross sections with high accuracy. In addition, these measurements can contribute to the extraction of the light quark masses, flavour symmetry breaking pattern in the light meson sector and allow to study precisely the meson mixing phenomenology through the various meson decays produced with high statistics in lepton collisions. The gamma-gamma fusion sub-processes in the positron/electron-electron inelastic scattering gives us the opportunity to investigate the two-photon couplings and form-factors of the various hadronic resonances (and also the many-particles states, like $\pi^+\pi^-$ or $\pi^0\pi^0$), which is important for the understanding of the quark contents of these resonances, of hadron phenomenology and for improvement in the estimate of the hadronic light-by-light scattering contribution to the anomalous magnetic moment of the muon. *The LHC, or a future $e^+e^-$ International Linear Collider (ILC), will answer already many questions. However, their discovery potential may be substantially improved if combined with more precise low energy tests of the SM. In this framework an electron-positron collider such as IRIDE with luminosity of $10^{32}$ $cm^{-2}s^{-1}$ with centre of mass energy ranging from the mass of the ϕ-resonance (1 GeV) up to ~ 3.0 GeV, would complement high-energy experiments at the LHC and a future linear collider (ILC).* The direct competing project is VEPP-2000 at Novosibirsk which will cover the center-of-mass energy range between 1 and 2 GeV with two experiments. This collider has started first operations in 2009 and is expected to provide a luminosity ranging between $10^{31}$ $cm^{-2}$ $s^{-1}$ at 1 GeV and $10^{32}$ $cm^{-2}$ $s^{-1}$ at 2 GeV. Other “indirect” competitors are the higher energy $e^+e^-$ colliders (τ-charm and B-factories) that can cover the low energy region of interest by means of radiative return (ISR). However, due to the photon emission the “equivalent” luminosity produced by these machines in the region between 1 and 3 GeV is much less than the one expected in the collider discussed here.

### 1.7. IRIDE Parameter list

Table 1.1 summarizes the main beam parameter requirements arising from the analysis of the IRIDE physics cases (described in the subsequent chapters of this WhiteBook) and that have been used to assess and propose the RF configuration options for the IRIDE multipurpose linac.

**Table 1.1**: Beam parameter requirements for the IRIDE physics cases.

| | Neutron source | Neutron source with TOF | Nuclear photonics | Thomson source, $\gamma$-$\gamma$, $e^-$-$\gamma$ | FEL (SASE, seeded) | FEL (oscillator) | THz radiation | $e^+$-$e^-$, $e^-$-$e^-$ |
|---|---|---|---|---|---|---|---|---|
| $E$ [GeV] | > 0.8 | > 0.8 | 0.1 ÷ 1 | 0.1 ÷ 1 | 0.75 ÷ 4 | 0.03/2.3 | 0.1/1.5 | 0.5 ÷ 2 |
| $q_b$ bunch charge [pC] | 1000 | 1000 | 10 | 500 | 150÷600 | 50÷200 | < 500 | <350 |
| $I_{peak}$ bunch peak current [kA] | any | any | any | Any | 0.75 ÷ 4 | 1 | 2 | < 500 |
| $\sigma_t$ bunch length [fs] | any | any | any | Any | 150-200 | 20÷80 | ≈100 | <1000 |
| $\sigma_{x,y}$ bunch transv. size [μm] | any | any | 10 | 10 | 50 | 50 | 100 | <1.5 |
| $\sigma_E/E$ bunch energy spread [$10^{-3}$] | any | any | 0.1 | 0.1 | 0.1 | 0.1 | 0.1 | 0.1 |
| $\varepsilon_n$ bunch norm. emittance [μm] | any | any | 1 | 1 | 1 | 1 | < 5 | <5 |
| $f_b$ bunch rep. rate [MHz] | 1 | 1000 | 100 | 65 | < 1 | 5.0 | < 1 | 1 |
| $T_p$ pulse duration [μs] | CW | 0.5 | CW | 0.9 | Any | > 100 | any | CW |
| $f_{pulse}$ pulse rep. rate [Hz] | CW | 50 | CW | 100 | Any | any | any | CW |
| $I_p$ current in the pulse [mA] | 1 | 1000 | 1 | 32.5 | < 1 | 0.25÷1 | < 5 | CW |
| $I_{ave}$ average current [μA] | 1000 | 25 | 1000 | 3 | < 250 | < 300 | < 300 | 350 |
| $(\Delta E/E)_{train}$ energy spread along the train [$10^{-3}$] | | | | | | 0.5 | | 0.1 |
| <$P_b$> ave. beam power [kW] | | | | | | | | 500 |
| $L$ luminosity [1/cm$^2$ s] | | | | | | | | $10^{32}$ |

The above requirements can be largely satisfied by a linac configuration either based on **high duty cycle pulsed operation** or on a **pure CW operation**, providing various beam parameters in the energy span of 1-2 GeV (per linac) with the same machine layout (i.e. number of cavities, modules and gradient settings), but different RF power systems. In table 1.2 a few cases (mainly at different energies and currents) are shown for the two RF architecture choices (pulsed or CW). For each option any of the beam parameter listed within its cases can be delivered by the same machine, operating at different accelerating field values with different current settings. Besides the fact that the two RF options require different RF power systems and a different power coupler, the linac design for each energy range in terms of choice of number of modules and cavities, and cavity accelerating field, are the same. A minimal increase in the linac length for the CW case is due to the necessity of increasing the number of cryogenic connections in order to accommodate the larger cryogenic loads caused by CW operation.

The two options can cover the same energy range with (approximately) the same machine length of about 150 m, with a different RF system (either tubes of SSA) and a different size of cryoplant. The high energy operation at 2 GeV for the CW version can be safely guaranteed by limiting the duty cycle of the RF sources at 60% over long times (~ scale of seconds) in order to decrease the cryogenic losses. This marginal limitation, purely given by a conservative assumption on the cryoplant capacity, is based on the use of a single LHC like cryoplant (18 kW @ 4.5 K) feeding both IRIDE identical linacs. It takes into account all the static losses, as experienced at FLASH, the 2 linac fluid distribution, and a 50% overhead for off-nominal operation, uncertainties on heat load estimations and transients.

**Table 1.2**: Possible SC linac parameters for the two RF options (pulsed/CW).

| | **Pulsed Option** | | | **CW/qCW Options** | | |
|---|---|---|---|---|---|---|
| | **Pulse 1** | **Pulse 2** | **Pulse 3** | **CW 1** | **CW 2** | **qCW** |
| $E$ [GeV] | 1 | 1.5 | 2 | 1 | 1.5 | 2 |
| $I$ (within pulse) [mA] | 5 | 3.5 | 2.5 | 0.5 | 0.35 | 0.260 |
| $I$ (average) [mA] | 0.58 | 0.33 | 0.17 | | | 0.155 |
| RF pulse duration [ms] | 1.5 | | | CW | CW | 1000 |
| Beam pulse duration [ms] | 1.17 | 0.95 | 0.67 | | | 990.74 |
| RF Duty cycle [%] | 15 | | | 100 | 100 | 60 |
| Beam Duty cycle [%] | 11.7 | 9.5 | 6.7 | | | 59.4 |
| $f_{RF}$ [MHz] | 1300 | | | | | |
| $E_{acc}$ [MV/m] | 10.04 | 15.05 | 20.07 | 10.04 | 15.05 | 20.07 |
| $Q_0$ | 2.00E+10 | | | | | |
| $Q_{ext}$, Design coupling factor | 4.00E+06 | | | 4.00E+07 | | |
| Cavity rise time [ms] | 0.98 | | | 9.79 | | |
| # of cavities | 96 | | | | | |
| # of modules | 12 | | | | | |
| $P_{beam}$/cavity (ave) [kW] | 6.09 | 5.20 | 3.48 | | | 3.22 |
| $P_{RF}$/cavity (ave) [kW] | 8.77 | 8.21 | 8.81 | | | 3.62 |
| Available $P_{RF}$/cavity (pulse) [kW] | 80.00 | | | 7.00 | | |
| $P_{cryo}$ (@ 4.5 K) [kW] Cryogenic 4.5 K equivalent power, accounting all loads and 50% margin | 1.4 | 1.9 | 2.6 | 4.0 | 7.3 | 7.7 |
| Total $P_{beam}$ (peak) [kW] | 5000 | 5250 | 5000 | 500 | 525 | 520 |
| Total $P_{beam}$ (ave) [kW] | 585 | 499 | 334 | | | 309 |
| Linac length [m] | 149 | | | 159 | | |

In the following table we summarize the main characteristic FEL@IRIDE SASE emission calculated for three possible scenarios of electron beam energy. The pulse duration will be in the range 200-70 fs.

**Table 1.3**: FEL performances at 1.5 GeV electron beam energy

| | Fundamental | 3° harmonic | 5° harmonic |
|---|---|---|---|
| λ(nm/KeV) | 4/0.413 | 1.33/1.23 | 0.8/2.07 |
| peak flux (n/s/- 0.1%BW) | $2.7*10^{26}$ | $2.5*10^{24}$ | $1.9*10^{23}$ |
| Peak brilliance | $1.56*10^{30}$ | $1.4*10^{28}$ | $1.1*10^{27}$ |
| photon/bunch | $5.94*10^{13}$ | $5.5*10^{11}$ | $4.18*10^{10}$ |

**Table 1.4**: FEL performances at 3 GeV electron beam energy

| | Fundamental | 3° harmonic | 5° harmonic |
|---|---|---|---|
| λ(nm/KeV) | 1/1.24 | 0.3/3.72 | 0.2/6.2 |
| peak flux (n/s/- 0.1%BW) | $4.6*10^{25}$ | $4.1*10^{23}$ | $3.4*10^{22}$ |
| Peak brilliance | $6.4*10^{31}$ | $5.7*10^{29}$ | $4.7*10^{28}$ |
| photon/bunch | $1.01*10^{13}$ | $9.02*10^{10}$ | $7.48*10^{9}$ |

**Table 1.4**: FEL performances at 4 GeV electron beam energy

| | Fundamental | 3° harmonic | 5° harmonic |
|---|---|---|---|
| λ(nm/KeV) | 0.563/2.2 | 0.188/6.5 | 0.113/10.9 |
| peak flux (n/s/- 0.1%BW) | $1.2*10^{25}$ | $5.9*10^{22}$ | $2.8*10^{21}$ |
| Peak Brilliance | $1.92*10^{31}$ | $1.8*10^{29}$ | $1.2*10^{28}$ |
| photon/bunch | $2.1*10^{12}$ | $1.06*10^{10}$ | $5.0*10^{8}$ |

Table 1.5 summarizes the performances of THz/MIR coherent radiation source from the undulator. Beam and undulator parameters also reported.

**Table 1.5:** Parameters of the THz/MIR coherent undulator radiation (CUR) source.

| Undulator | |
|---|---|
| Period (cm) | 40 |
| Number of periods | 10 |
| Magnetic field (T) | 0.1 -1 |
| Coherent Radiation parameters | |
| Wavelength (μm) | 100 (with K = 6, i.e. B ≈ 0.2 T)-10 (K=1.4, i.e 0.04T) |
| Peak power (MW) | > 100 |
| Micropulse energy (mJ) | ≈ 10 |
| Micropulse duration (fs) | 200 |

Table 1.6 summarizes the performances of THz/MIR coherent radiation source from a diffraction radiation (DR) target.

**Table 1.6:** Performances of coherent diffraction radiation (CDR) source.

| Coherent Radiation parameters | |
|---|---|
| Wavelength (μm) | > 50 |
| Peak power (MW) | > 100 |
| Micropulse energy (μJ) | ≈ 100 |
| Micropulse duration (fs) | 200 |

Table 1.7 summarize the performances of the IRIDE neutron source compared to other existing sources.

**Table 1.7:** performances of the IRIDE neutron source

| **Facility Parameters** | **nElbe** | **Gelina** | **nToF** | **ISIS** | **IRIDE** |
|---|---|---|---|---|---|
| **Source** | | e-Linac | p spallation | p spallation | SC e- Linac |
| **Part E (MeV)** | 40 | 120 | 20000 | 800 | 1000 (2000) |
| **Max Power (kW)** | 18 | 11 | 45 | 160 | 32/1000 (pulsed/continuous) |
| **Neutrons/s** | 3.4E+13 | 3.2E+13 | 8.1E+14 | 1E+16 | 7E+14/3E+15 |

**Table 1.8:** parameters at the IP of the IRIDE linear collider for 3 GeV c.m. energy

| **Parameters** | **Units** | **Electrons >< Electrons** | **Electrons >< Positrons** |
|---|---|---|---|
| | | | |
| Beam energy | [GeV] | 1.5 | 1.5 |
| Beam power | [MW] | 0.45 | 0.53 |
| Charge | [nC] | 0.3 | 0.35 |
| Bunch length rms | [μm] | 270 | 150 |
| Peak current | [A] | 333 | 700 |
| Rep. rate | [MHz] | 1 | 1 |
| Average current | [mA] | 0.3 | 0.35 |
| Transverse rms spot at IR | [μm] | 0.5 | 0.5 |
| Norm. emittance | [μm] | 2 | 5 |
| Beta at IR | [mm] | 0.4 | 0.15 |
| Disruption parameter | D | 2.6 | 1.23 |
| Beam-strahlung parameter | $\delta_E$ | $\sim 10^{-7}$ | $\sim 10^{-6}$ |
| Luminosity enhancement factor | $H_D$ | 1.6 | 1.2 |
| Luminosity | $cm^{-2}s^{-1}$ | $1.1\ 10^{32}$ | $1.3\ 10^{32}$ |

For the Linear Collider option the parameters listed in Tab. 1.8 have to be obtained at the Interaction Point (IP) in order to achieve the required luminosity, assuming the SC linacs are both operating in CW mode and both electron and positron beams are round.

The third column shows the required beam parameters in the case of electron-electron collider mode. In this case the reasonable low emittance of both beams should allow operating at 500 nm spot size at IP.

The forth column illustrate an example of electron-positron collider in which both beams have the same parameters. The parameters are within the state of the art for low energy electron beams but not yet for positrons. The positron source is still an open problem that might partially overcome by using a Damping Ring or by a new positron source concept based of direct photon conversion.

## 1.8. Preliminary project cost estimate

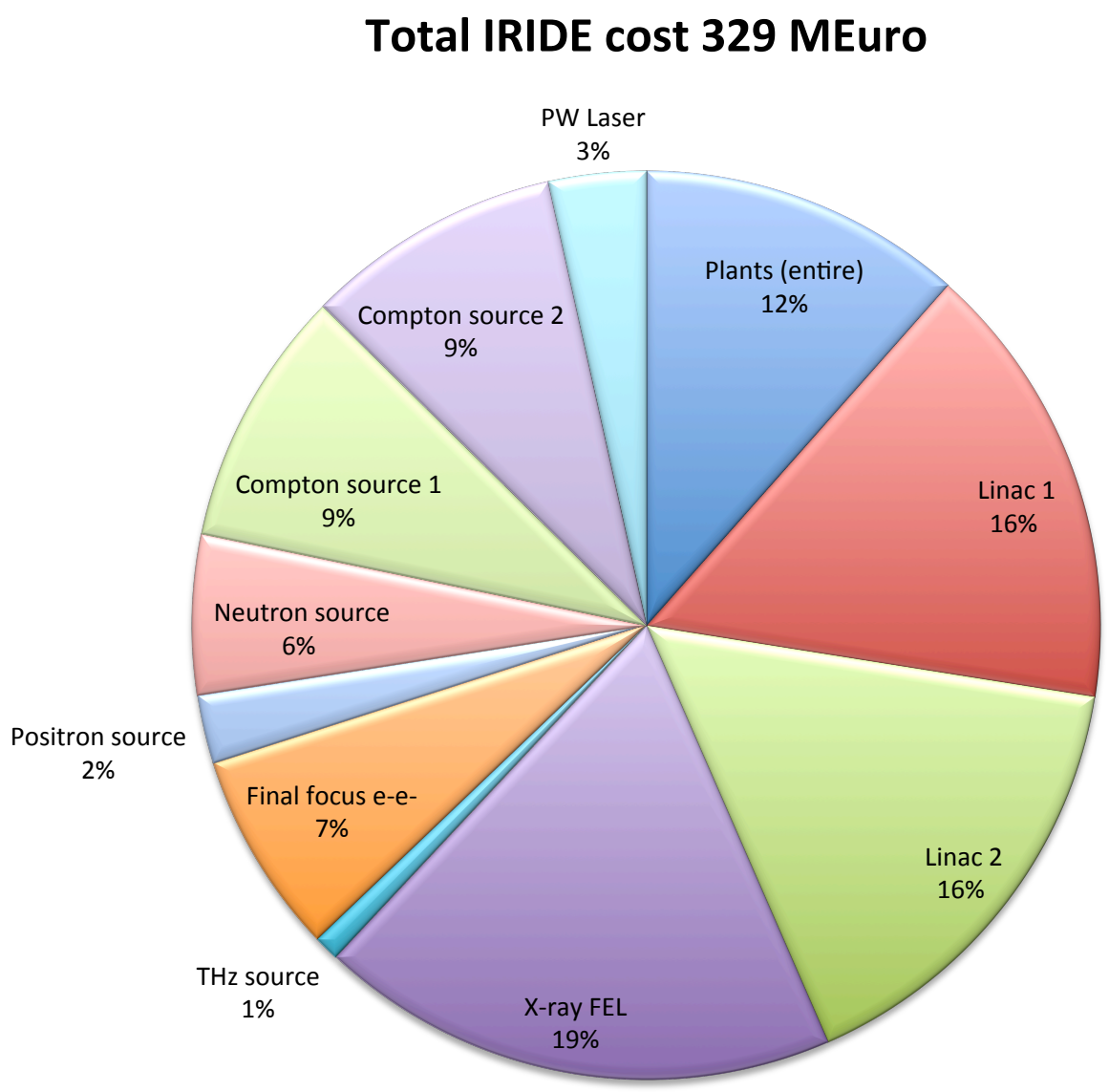

The costing for the civil engineering are based on the assumption of a cost per cubic meter of 350 €/m$^3$ excluding VAT.

The costing for the superconducting linac components is largely based on the XFEL Project, reviewed for the different RF distribution options foreseen for the two operation options. Costs for the RF systems have been assessed through contact with companies potentially able to provide the klystron or Solid State Amplifier systems. For the cryogenic system experience and scaling laws published for LHC has been used.

The costing for the other IRIDE components are based on the experience with similar facilities: NGLS (CW RF injector), SPARX (FEL), ATF2 (Final focus), GELINA (Neutron source), ELI_NP (Compton source), SPARC_LAB (PW laser and THz source).

The total cost of the entire IRIDE facility amounts to **329 M€** as detailed in the following table where the third column show also the incremental cost. In the previuos pie-chart the relative cost of each component is compared.

| **Components** | **Cost [M€]** | **Incremental Cost [M€]** |
|---|---|---|
| | | |
| Cryogenic plant | 20 | |
| LHe transfer lines | 5 | |
| Building (50m x 20m x 8m x350 Euro/m3) | 3 | |
| **Total Cryogenic plant** | **28** | **28** |
| | | |
| **Electrical eng. for 8 MW distribution** | **4** | **32** |
| **Fluids eng.** | **6** | **38** |
| | | |
| Injector 1 (RF gun + laser) | 6 | |
| Linac 1 modules including CW RF and LLRF | 32 | |
| Beam diagnostics | 1 | |
| Bunch compressor system | 2 | |
| Transfer Lines | 5 | |
| Synchronization | 1 | |
| Control system | 2 | |
| Linac 1 building (170mx5mx5mx350Euro/m3) | 1,5 | |
| Additional plants + cabling | 2 | |
| **Total Linac 1** | **52,5** | **90,5** |
| | | |
| Injector 2 (RF gun + laser) | 6 | |
| Linac 2 modules including CW RF and LLRF | 32 | |
| Beam diagnostics | 1 | |
| Bunch compressor system | 2 | |
| Transfer Lines | 5 | |
| Synchronization | 1 | |
| Control system | 2 | |
| Linac 2 building (170x5x5x350) | 1,5 | |
| Additional plants + cabling | 2 | |
| **Total Linac 2** | **52,5** | **143** |
| | | |
| FEL undulators and optical cavity | 25 | |
| X-ray Transport optics | 5 | |
| User end station | 5 | |
| X-ray detector | 8 | |
| Undulator hall (100x30x8x350) | 8,4 | |
| X-ray transport tunnel (80x5x5x350) | 0,7 | |
| Experimental hall (80x30x8x350) | 6,8 | |
| Additional plants + cabling | 2 | |
| **Total X-ray FEL** | **60,9** | **203,9** |
| | | |
| **Total THz source** | **3** | **206,9** |
| | | |
| ATF2 Final focus system e-e- (thanks to Seryi) | 11,3 | |
| Diagnostics | 2,1 | |
| Experimental hall (100x30x8x350) | 8,4 | |
| Additional plants + cabling | 2 | |
| **Total final focus e-e-** | **23,8** | **230,7** |
| | | |
| Neutron Target area | 8 | |
| Beam lines 4 | 8 | |
| Neutron transport tunnel (150x5x5x350) | 1,3 | |
| Experimental station 5 (10x10x8x350) | 1,4 | |
| **Total Neutron source** | **18,7** | **249,4** |
| | | |
| Dedicated injector | 5 | |
| high energy Yb:AG laser | 7 | |
| Interaction region and recirculator | 4 | |
| g-ray beam collimation and diagnostics | 3 | |
| User beam line | 10 | |
| Laser and users exp hall (20x20x8x350) | 1,1 | |
| **Total Compton source** | **30,1** | **279,5** |
| | | |
| Target area | 5 | |
| Positron Capture | 2 | |
| Diagnostics | 1 | |
| **Total test positron source** | **8** | **287,5** |
| | | |
| **Second Compton source for gamma-gamma** | **30,1** | **317,6** |
| | | |
| Laser PW (TiSa) | 8,7 | |
| Laser-Plasma diagnostics and control | 1,2 | |
| PW target area | 1,5 | |
| **Total PW Laser for AAC** | **11,4** | **329** |
| | | |
| **Total IRIDE facility** | **329** | |

Two possible investment profiles are shown in the next plots as examples of the flexibility of the project development. Other solutions downstream the first linac are possible depending on the assigned priorities.

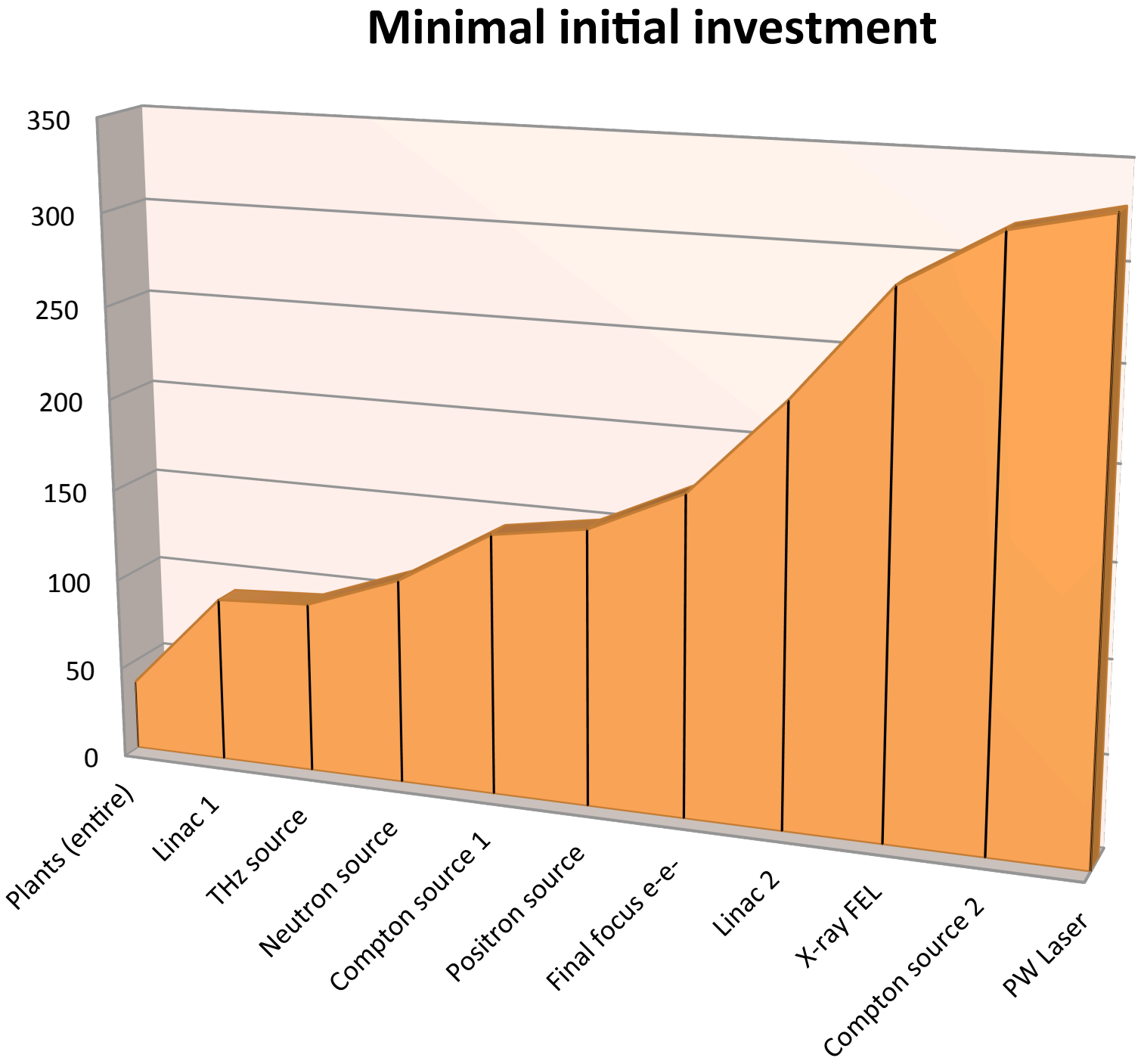

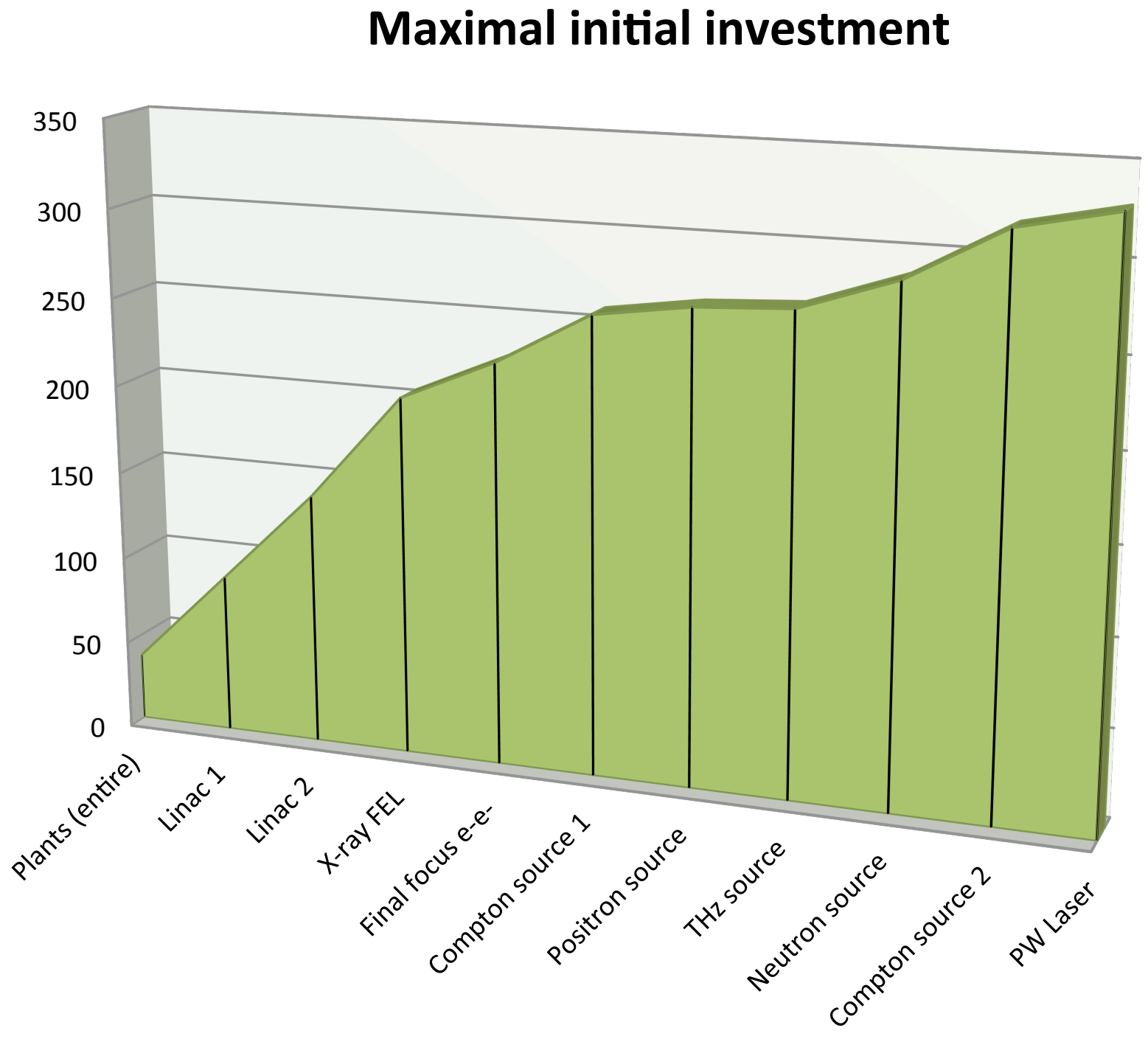

## 2. THE SUPERCONDUCTING LINAC COMPLEX

### 2.1. Introduction

IRIDE is a multipurpose facility with an evolutionary approach. Physics requirements and temporal staging opportunities of the Project foresee the realization of two SRF linacs, separated by an experimental hall, each with an energy reach in the 1 to 2 GeV range. Two possible operation options can be conceived to serve the requirements of the physics cases which are described in the following chapters:

- **a high-duty cycle pulsed operation**
- **a CW operation**.

Although the linac length and machine layout are almost the same in these two operation options, which are exclusive, as they lead to a substantially different RF generation and distribution infrastructure (large klystrons distributed over many cavities vs SSA in a one-to-one configuration) and to a different scale for the cryogenic plant.

In the following we provide a summary of the state of the art of the technology, outline the main design criteria and describe a few important technical components needed for the IRIDE linacs.

### 2.2. Superconducting L-band cavities and accelerator modules

The workhorse of the SRF technology is represented by the L-band (1.3 GHz) cavities [1] and modules, developed by the TESLA Collaboration [2,3], which currently drive the FLASH FEL facility in DESY and which form the main accelerator complex of the ~1.6 km European XFEL accelerator facility under construction, see Fig. 2. The **TESLA/XFEL cavities and module concepts are well proven and understood and** furthermore:

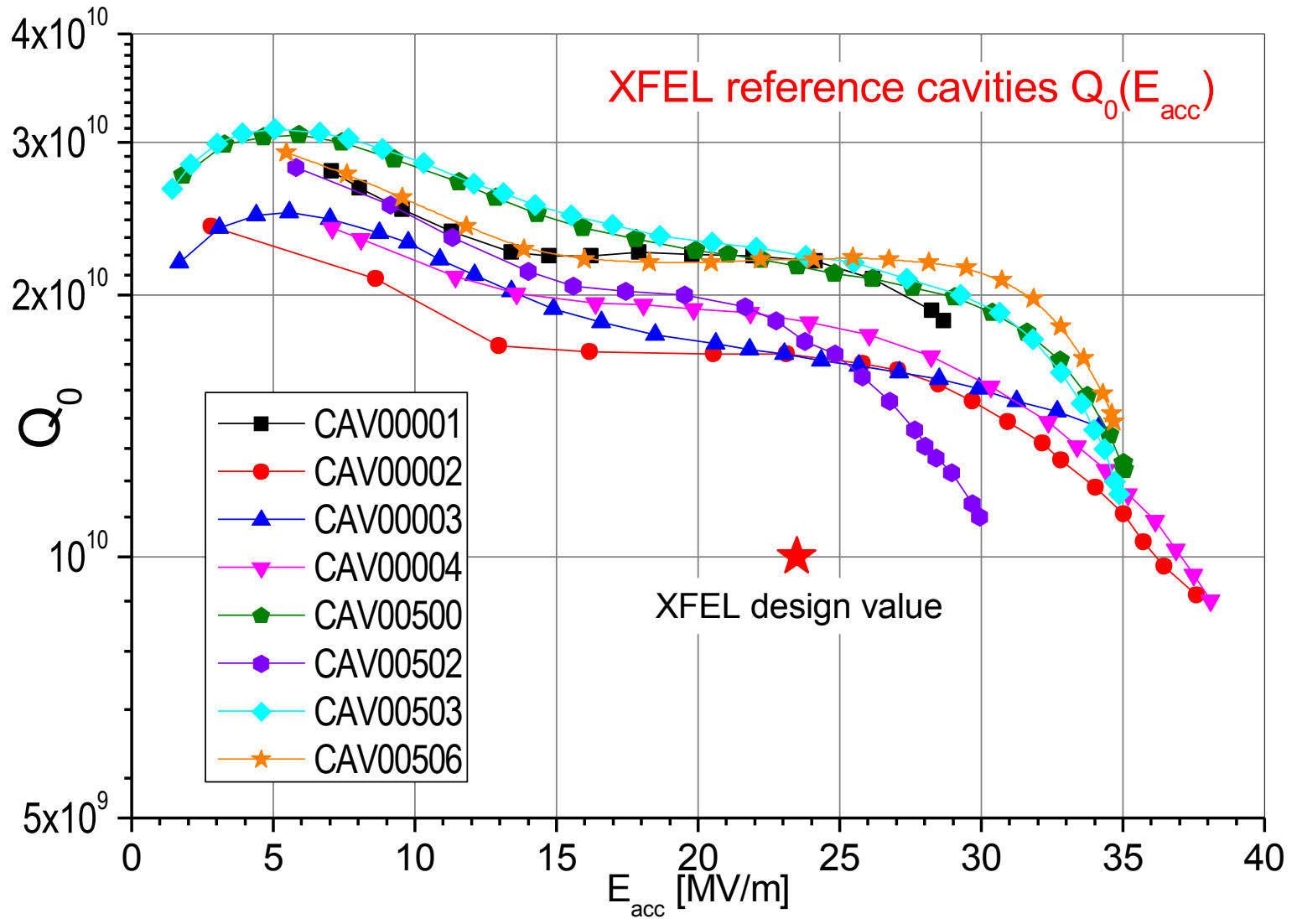

**Figure 2.1:** 2 K Test results of the eight cavities used for the qualification of the XFEL mechanical production infrastructure at the two vendors.

- The series production of the 800 XFEL cavities and 100 cryomodules is running at the nominal production stage of 1 module per week. More than 50 cavities have been successfully tested outperforming the results of the qualification pre-series, shown in Figure 2.1. $Q_0$ values at 2 K close to 2 $10^{10}$ at moderate accelerating fields are shown in vertical tests, strengthening the path towards high duty cycle operation. Operating the linac at a temperature between 1.8 to 1.9 K (taking profits of the further decrease in surface resistance) would imply that a reference $Q_0$ of 2 $10^{10}$ is a safe design value.
- Variations of the concepts for almost all components and subcomponents have been explored and tested for a variety of operating conditions (e.g. modified cavity and module geometries to deal with the larger heat loads associated to CW operation, longer modules to increase the real estate gradients, ERL configurations, etc.). The ILC S1-Global collaborative experiment at KEK proved that the concept can easily be adapted to the case of a combination of substantially different design variants for important components like cavities, tuners and couplers.
- A large industrial effort is currently ongoing in order to manufacture in Europe a large number of the European XFEL components (all 800 cavities, 50% of the ~100 cryomodules, all power couplers and RF stations). Half of the cavities and all the modules produced in Europe are procured in Italy, see Fig. 2.2. Industries have deployed large production facilities and infrastructures, taking over steps in the cavity preparation process that were previously performed at research institutions (e.g. the needed surface treatments for the bulk niobium cavities to prepare them for operation). This capability and infrastructures exist and will contribute towards a potential decrease of the main accelerator components procurement costs and to rapid procurement times (the XFEL procurement reaches a cavity production rate of 4 cavities per week per vendor).

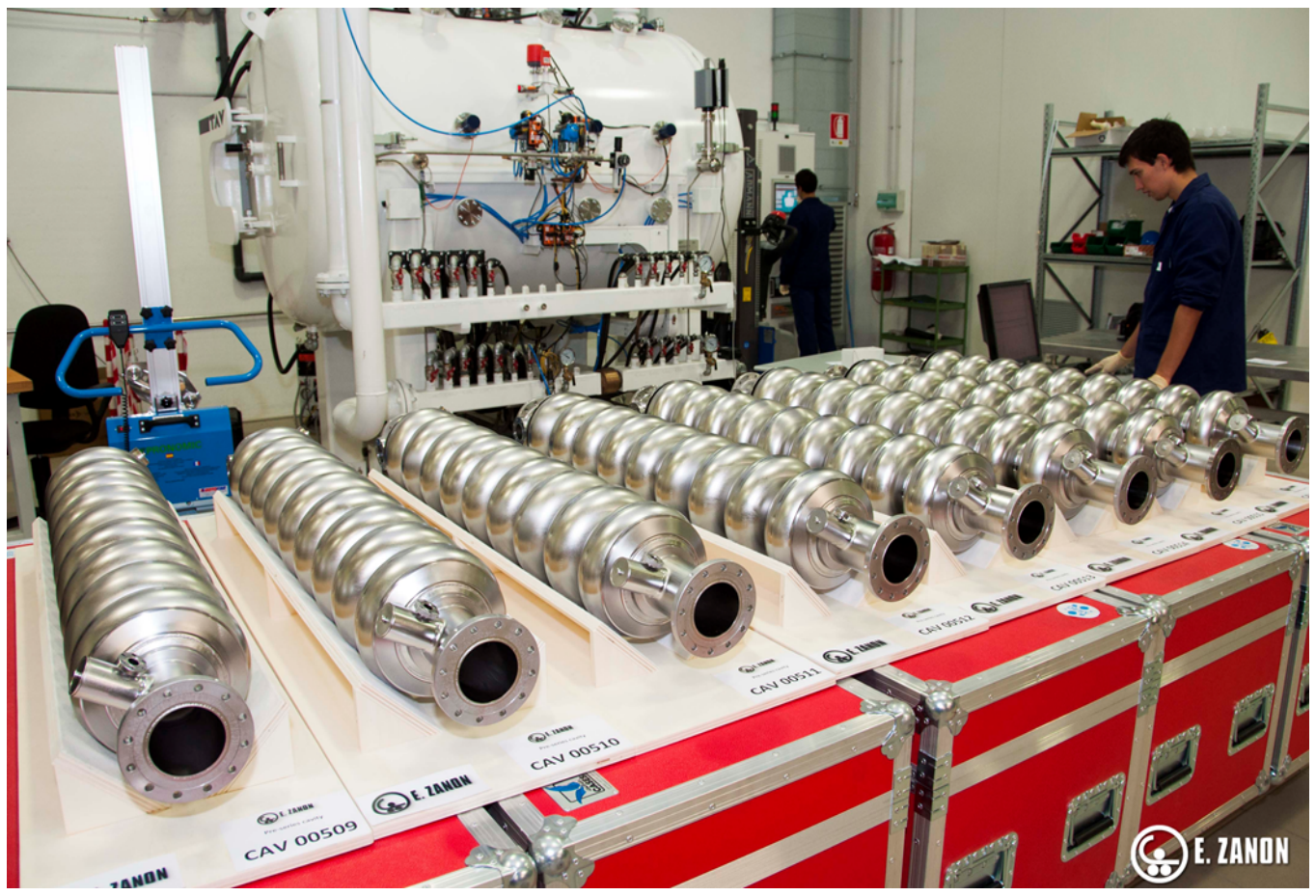

**Figure 2.2:** A set of pre-series XFEL cavities at the company E.Zanon.

Table 2.1 resumes the fundamental parameter and features for the state-of-the-art L-Band cavity and module technology.

**Table 2.1**: Main parameters or SRF L-Band components

| **Cavity** |
|---|

| | |
|---|---|
| Frequency | 1300 MHz |
| Active length, L | 1038 mm |
| Number of cells | 9 |
| Beam aperture | 70 mm |
| R/Q | 1036 Ohm |
| Geometric Factor | 270 Ohm |
| Achieved gradient in vertical tests (electropolished), $E_{acc}$ | > 40 MV/m |
| Achieved operation in module conditions, $E_{acc}$ | ~ 35 MV/m |
| Quality factor ($Q_0$) at moderate gradients (20 MV/m), 2 K | $\sim 2 \times 10^{10}$ |
| **Cryomodule** | |
| Number of cavities in module | 8 |
| Overall module slot (length + interconnection space) | 12000 mm |
| Outer vessel diameter | 965.2 mm (38”) |

### 2.3. CW versus pulsed RF operation

The choice of accelerating gradient for the superconducting linac cavities is strongly affected by the operating mode, time structure and current required to the beam.
The reference TESLA Technology was originally developed for the realization of an efficient linear collider. The SRF Technology was chosen for the International Linear Collider, a machine designed in a multi-parameters optimization process with the following major goals: minimize the cost per MV, maximize the machine real estate gradient (i.e. minimum length) and maximize the beam to plug power conversion efficiency. High gradients above 30 MV/m (needed to reach high energies) impose the choice of a pulsed beam structure at percent level RF duty cycle in order to limit power dissipation on the cavities, thus reducing the cryogenic requirements. RF pulses of ms duration at ~10 Hz in the TESLA/XFEL cavities can deliver trains of thousands high-charge electron bunches (with ~ MHz repetition within the RF pulses) to the experiments. The RF sources and RF distribution system need to provide the necessary RF power to the cavity through the main couplers in order to provide beam acceleration. Typically, klystrons with power rated in the MW range are used to feed simultaneously many cavities in a number of subsequent cryomodules.

The following Fig. 2.3 illustrate the main typical parameters for a model case, for a 1% duty cycle linac at 1 GeV.

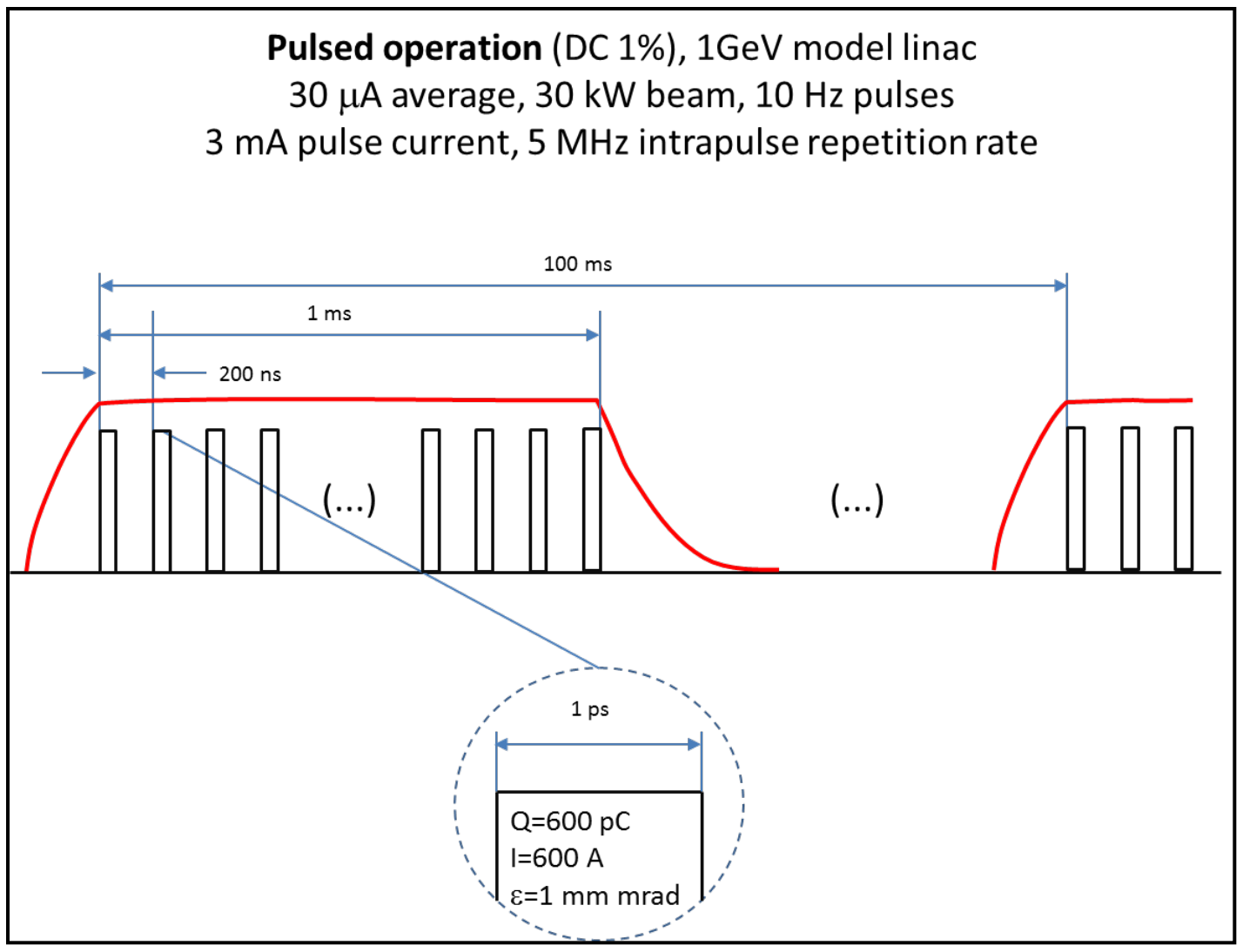

**Figure 2.3:** Typical time structure for 1% duty cycle pulsed operation .

The pulsed option can be extended to larger RF and beam duty cycles (up to 15%) for moderate gradient operation in the 10-20 MV/m by the use of existing RF sources delivering 1.5 ms RF pulses at 100 Hz.

The existing SRF technology allows also to operate the accelerator in CW RF mode, delivering an uninterrupted train of moderate charge bunches at ~ MHz repetition rates to the experimental areas. As RF dynamic losses on the cavity walls depend quadratically on the accelerating gradient and inversely on the cavity $Q_0$, gradients need to be kept to moderate levels (~10-20 MV/m) to limit cryogenic consumption. Nevertheless the factor of 2 gained on the available $Q_0$ demonstrated by the industrially produced cavities for the XFEL allows increasing the optimum gradient by square root of 2 for the same cryogenic load. That makes CW operation very attractive and possibly the reference choice for the future 4$^{th}$ generation light sources under study, which add to facilities already existing or near to completion, such as LCLS, XFEL and SACLA. Moderate power CW sources (ideally solid state transmitters in the <10 kW power range) can be used to power a single cavity, with significant benefits on cost and complexity of the RF system. Figure 2.4 illustrates the temporal structure of a model case in the parameter space, for a CW case at 1 GeV.

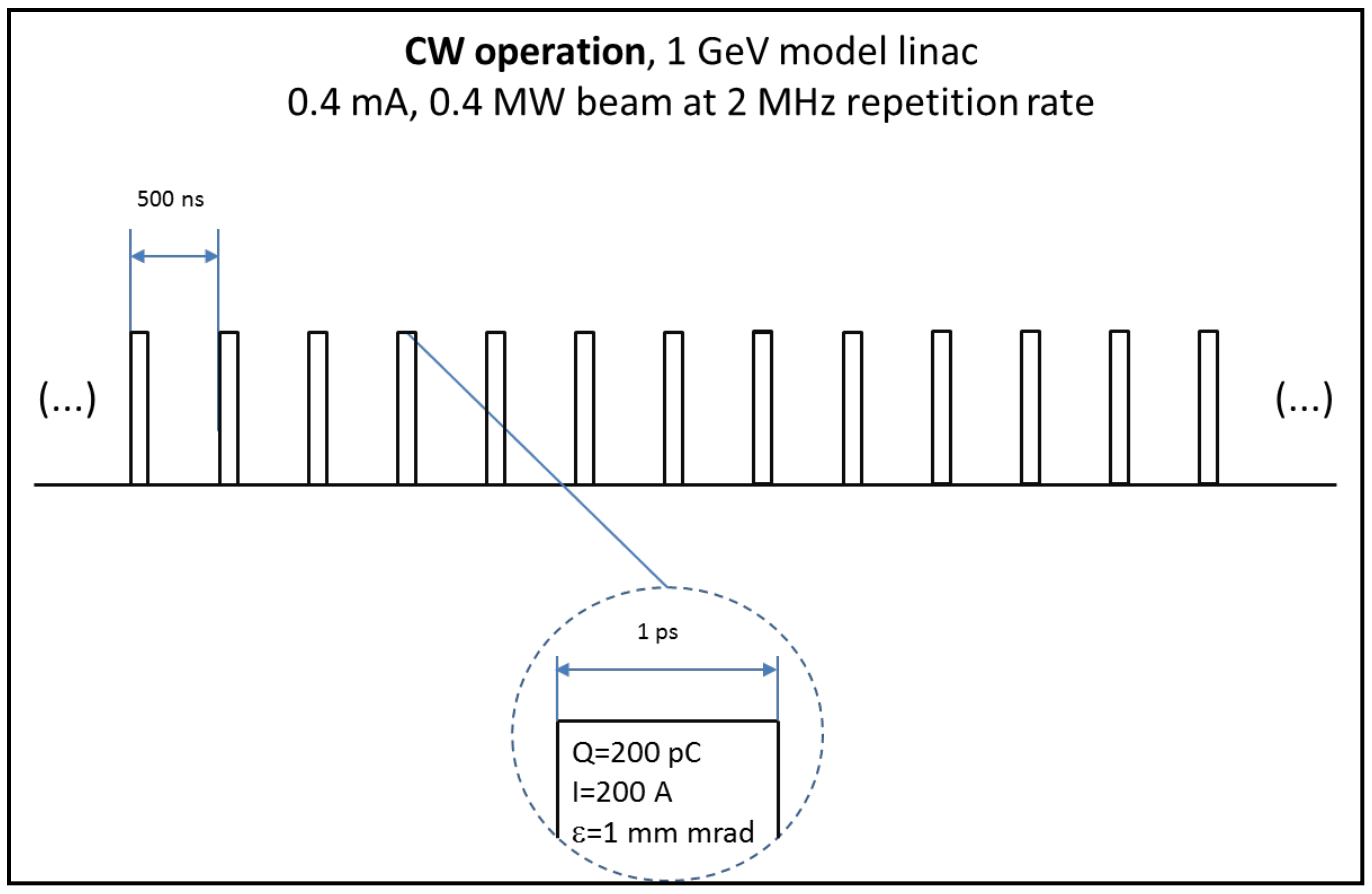

**Figure 2.4:** Typical time structure for CW operation .

The numbers in the two figures are intended only to show the different possible scenarios for the linac parameter space as a guideline towards the finalization of a conceptual design meeting the combined requirements arising from the proposed experiments at the IRIDE multipurpose facility. A further discussion on the beam parameter flexibility in the two options is outlined later. The CW choice, combined with a bunch distribution scheme, would offer the most versatile solution to provide bunches to a number of different experiments, as could be envisaged in the development of the multipurpose facility.

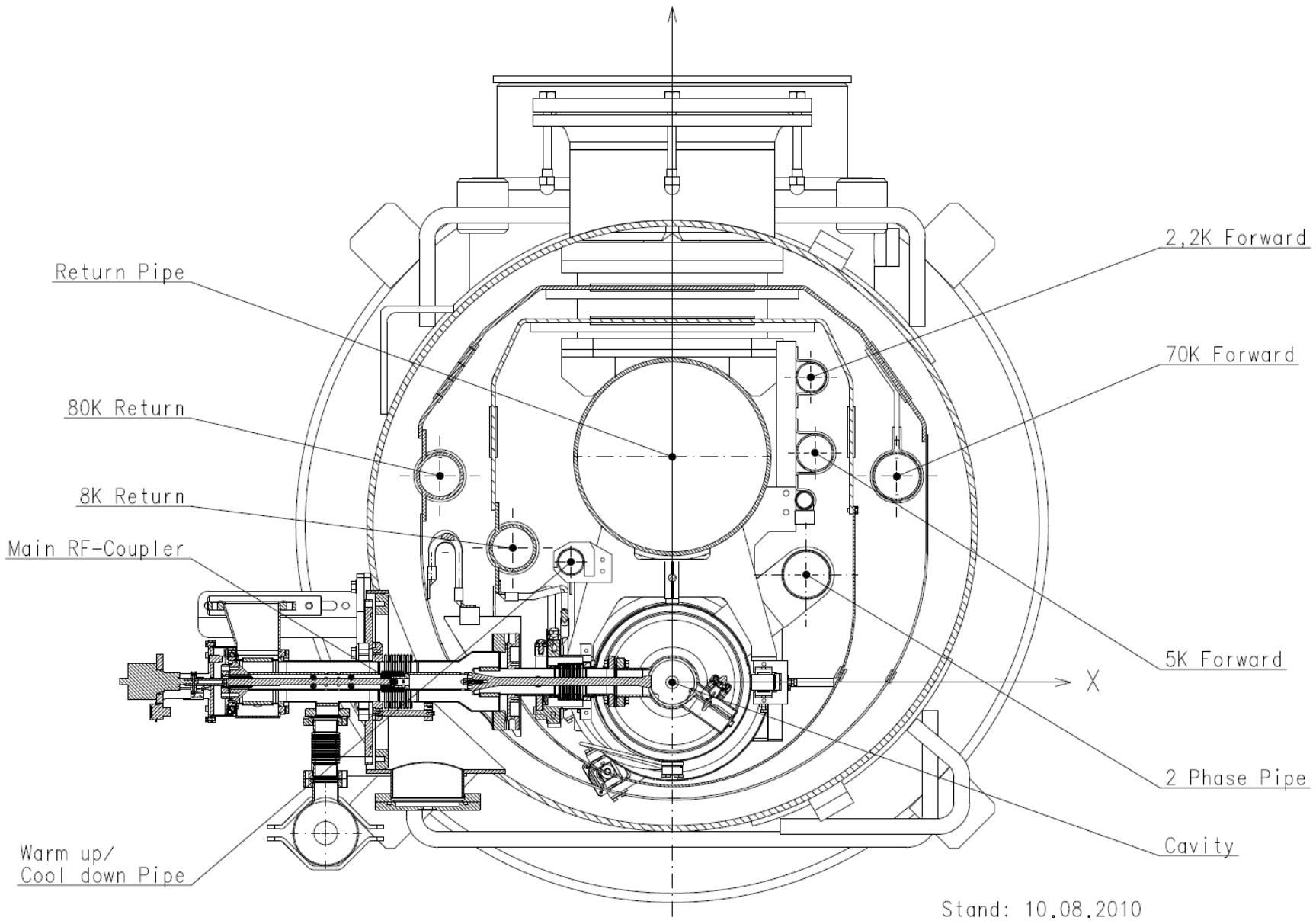

**Figure 2.5:** The TESLA/XFEL cryomodule transverse cross-section, showing the integration of the cryogenic lines.

The TESLA/XFEL cryomodule design is driven by the combined requirements of a good filling factor (in terms of ratio of nominal accelerating length with respect to the physical beamline length), moderate capital and operation costs (mainly focused on design simplicity and low static losses) and provisions for effective assembly and alignment of the beamline components. In this scheme (see Fig. 2.5) the linac cryogenic lines are embedded in the cryomodules and strings of cryomodules are efficiently packed in strings of approximately 10-12 units fed by a single J-T supply valve able to provide the necessary low pressure 2 K helium for operation. In the ILC design [3], which shares this concept, a single standard LHC-unit cryoplant would be able to provide He for a ~2.5 km string of nearly 200 modules. These design requirements, originating from the linear collider application, imply that the concept, with few minor adaptations can serve both the scope of a high-gradient long linear accelerator at the percent duty cycle (as the case of the linear collider concept) and a short, moderate gradient facility delivering continuous (CW) beams. All piping is sized to provide minimal pressure drop in long (km size) circuits with low mass flow and can therefore be operated with larger mass flows for smaller lengths.

This versatility is demonstrated by the fact that the original TESLA Linear Collider concept has been chosen by XFEL (approximately 4% of the LC size), and is proposed for the CW schemes as the BESSY FEL, LBNL NGLS and even the Cornell ERL.
The adaptations required to fine tune the module concept to the operation mode concern mainly the cryogenic aspects:

- Set the frequency of cryogenic boxed for LHe supply (J-T valves) in the string to provide the low pressure superfluid liquid He for the removal of the heat deposited by the RF accelerating field on the cavity walls. The LHe supply should be provided approximately after 200 W of low temperature load on the circuit, requiring an extra 2.5 m on the beamline for each cryogenic connection box.
- Review the cryo piping dimensions to ensure the correct pressure drop at the nominal mass flow in the cryogenic circuits, in order to provide the necessary temperature stability and flow conditions for safe cavity operation.

### 2.4. RF system and couplers

Both pulsed and CW options rely on existing technology, available on the market by several vendors. Pulsed klystrons at 1.7 MW peak delivering 1.5 ms pulses at 100 Hz repetition rate are for instance available from Thales (TH2104D). Solid State Amplifiers delivering 5-10 kW CW power are available from several Italian or European vendors. XFEL/LC couplers exceed the IRIDE parameter requirements in term of both peak and average power.

The RF system configuration is greatly different for the pulsed and CW operating options. The input coupling factor of the cavities in both cases is chosen to optimally match the beam loading. In pulsed mode the beam loading depends on the beam current value within the RF pulse, while in CW operation depends on the average beam current value. As a consequence, in pulse modality the beam loading is larger by about one order of magnitude, which requires external Q-factors of the cavity input coupler in the range of $\approx 10^6$, one order of magnitude smaller of those required for CW operation. In this respect SC cavities configured for pulsed operation show larger loaded bandwidths (in the ≈100 Hz range) and faster filling times (in the ≈1 ms range) with respect to those configured for CW operation (≈10 Hz bandwidths, ≈10 ms filling times). Narrowband cavities are more sensitive to microphonic noise and require more sophisticated active (piezoelectric based) tuning systems to keep their resonant frequencies close to the RF excitation.

In pulsed modality the use of high power klystrons require the realization of a waveguide network distribution to power many cavities (typically 16÷32, i.e. 2÷4 full modules) in parallel. The number of the needed Low Level RF control units (typically one per power source) for the whole linac is reduced with respect to the CW operation, but their complexity is greater since they have to deal with vector combination of RF probe signals coming from many cavities. As the cavities are powered in large groups, the RF regulations have a limited degree of flexibility. This scenario is reversed for the CW operation modality. As the CW RF power required by each cavity is in the < 10 kW range, a one-to-one amplifier to cavity architecture is conceivable and cost effective, leading to a much simpler distribution network. Cavities can be individually controlled for a fully flexible RF regulation. Individual RF cavity control is also beneficial to reduce the effects of microphonic noise by applying proper amplitude and phase modulations to compensate the microphonic detuning.

## 2.5. Injector system

Both IRIDE design options (CW at MHz repetition rate or pulsed operation at high duty cycle with 100 MHz bunch trains) have fundamental implications on the electron injector and on the electron gun. Indeed, the injector accelerating sections must be of the superconducting type, as the rest of the electron linac, while the gun can be in both cases a sub-harmonic, low frequency RF gun, operating at room temperature. The results recently obtained at Berkeley with the CW APEX gun [4-6], in the framework of NGLS [7], are consistent with the CW option and can also be extrapolated with the required brightness in case the high charge bunch, high repetition rate option is chosen.

The outstanding results of the relatively low repetition rate FEL guns, as the ones of the LCLS at Stanford or of PITZ at DESY, cannot be directly scaled to higher repetition rates or CW. Moreover the need of keeping dark current within acceptable values pushes the accelerating gradient in the electron gun down to values significantly smaller than in the lower repetition rate case. All together we consider the APEX gun as the best reference candidate for IRIDE. Beam dynamics in this "low gradient" regime assumes quite different characteristics and although simulations indicate the capability of achieving the required results, a complete experimental demonstration has still to be performed and the successful work ongoing at Berkeley is a significant step forward.
Similarly to the large majority of high brightness gun schemes, also in IRIDE photocathodes are used for the flexibility they offer in controlling the electron bunch distribution. The high-repetition rate and the available laser technology require the use of high quantum efficiency (QE) photocathodes. As for APEX the reference cathode system will be the one developed by INFN for FLASH and XFEL [8-10].

Table 2.2 shows the injector and electron gun requirements for operating IRIDE for various users and in both CW and pulsed modes of operation. In the following part of this section, some of the Table requirements will be analyzed and addressed in some detail.

### 2.5.1. **RF Gun**

In the proposed configuration the electron gun is based on a RF technology capable to operate in either continuous wave (CW) or in high duty cycle pulsed mode. This choice is compatible with both the proposed linac operational options, so that the final choice between the 2 possible linac configurations will not be driven by considerations on the injector characteristics and requirements.

**Table 2.2:** IRIDE Injector and Gun Requirements

| **Parameter** | **Unit** | **Values** |
|---|---|---|
| Bunch repetition rate | MHz | ≤ 186 |
| Charge per bunch | pC | ≤ 2000 |
| Average current | mA | ≤ 4 |
| Normalized transverse emittance (slice, rms) | μm | <1 |
| Beam energy at the gun exit | MeV | >~0.75 |
| Beam energy at injector exit | MeV | ~100 |
| Electr. field at the cathode during photoemission | MV/m | ~20 |
| Bunch length at the cathode (rms) | ps | ~10 |
| Peak current at the injector/main linac exit | A | 60/3000 (@ $q_b$=250 pC) |
| Compression factor (injector/main linac) | | ~6/50 (@ $q_b$=250 pC) |
| Compatibility with magnetic field (gun/cathode area) | | yes |
| Operational vacuum pressure | nTorr | ~0.1 |
| 'Easy & fast' cathode replacement//regeneration | | yes |

Small normalized transverse emittances $\varepsilon_n$ are extremely important in X-ray FELs because of the required matching with the small X-ray photon emittance $\lambda/4\pi$. Small emittances allow also for shorter FEL gain lengths and thus for shorter undulators. The described benefits are particularly important because they permit reducing the size and cost of large and expensive X-ray FEL facilities. For different applications, such as the Linear Collider operation mode, the optimization tradeoff requires higher charge per bunch and a somehow more relaxed emittance requirement.

More details are given in ref [4-6] where APEX, the proposed reference injector, is described and the first experimental data are presented. That is enough for the CW FEL mode, including the laser system and the very effective control of the time jitter that in the pulsed machine is a critical issue. As in the NGLS proposal, two color, pump and probe, experiments will be available with sub-fs control of the delay between the two photon pulses.

The scheme developed and commissioned at LBNL has been designed to conservatively satisfy requirements similar to those of IRIDE. The core of the gun is a normal-conducting copper RF cavity resonating at 186 MHz in the VHF band, i.e. the 7$^{th}$ sub-harmonic frequency of the 1.3 GHz that feed the superconducting linac . Figure 2.6 shows a cross section of the VHF cavity with the main components.

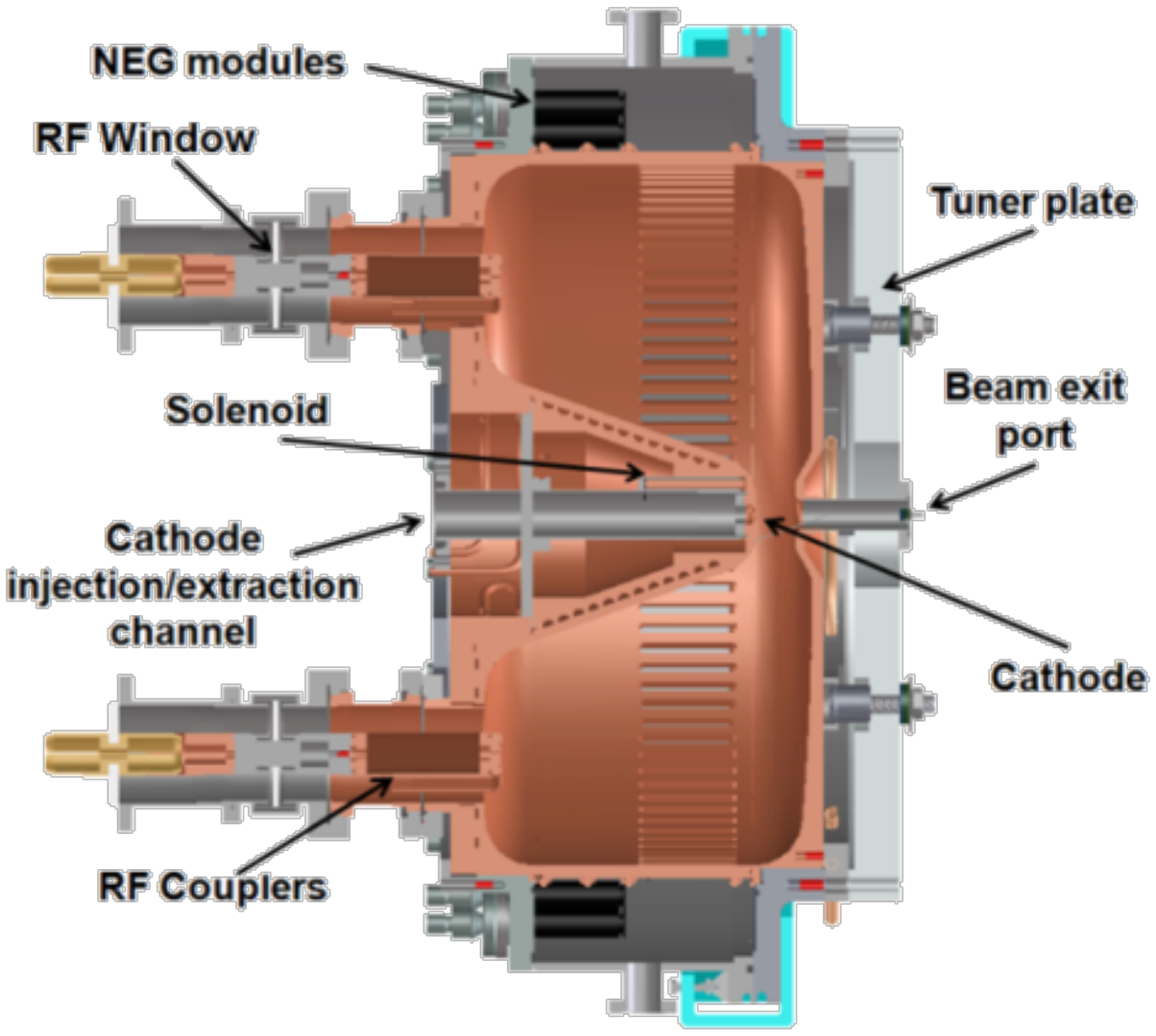

**Figure 2.6:** VHF cavity cross-section, showing the cavity main components.

The pulsed mode is performed with the same RF gun still operating in CW, but with a different laser system, in which, as for XFEL and FLASH, the full ≈1ms bunch train is amplified all together. The expected beam quality is somehow lower in term of emittance and time jitter but still sufficient for wide variety of experiments. For the gun thermal stability and the consequent effect on the beam time jitter, a CW operation at constant RF field is envisaged.

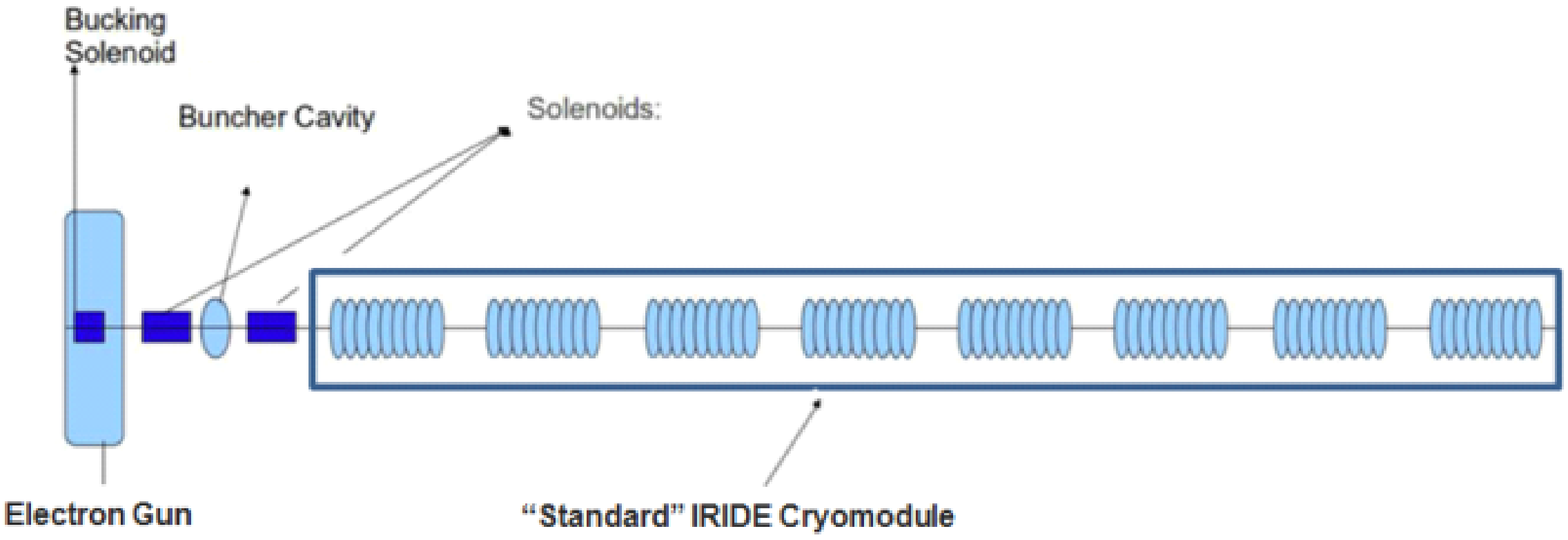

**Figure 2.7:** IRIDE Injector block schematic diagram.

A block schematics of the IRIDE electron injector is shown in Figure 2.7. The gun cavity can be seen in extreme left of figure, followed by solenoids, buncher and diagnostic system before the "RF booster" consisting of a "standard" IRIDE cryostat with nine-cell Tesla-like 1.3 GHz cavities.

**Table 2.3:** LBNL VHF Gun Parameter List.

| Parameter | Symbol | Unit | Value |
|---|---|---|---|
| Frequency | $f_G$ | MHz | 186 |
| Operation mode | | | CW |
| Beam Energy at the gun exit | $E_G$ | keV | 750 |
| Field at the cathode at photoemission | $E_C$ | MV/m | 19.5 |
| Quality factor (ideal conductor) | $Q_0$ | | 30900 |
| Shunt impedance | $R$ | MΩ | 6.5 |
| Nominal RF power for $Q_0$ | $P_0$ | kW | 88 |
| Stored energy | $E_S$ | J | 2.3 |
| Peak surface field | $E_P$ | MV/m | < 24.1 |
| Peak wall power density | $P_W$ | W/cm$^2$ | < 25 |
| Accelerating gap | | mm | 40 |
| Diameter/Length | | mm | 694/350 |
| Design operating pressure | | Torr | ~$10^{-10}$ $10^{-9}$ |

### 2.5.2. **Photocathodes**

One of the main characteristics of the IRIDE injector design is the MHz-class repetition rate with charges up to 2 nC per bunch. This precludes the use of traditional metallic photocathodes as used in other low repetition rate facilities due to their low Quantum Efficiency (QE) and hence high demand for laser power. Indeed, running a metal cathode with its typical 5 $10^{-5}$ QE at the IRIDE beam conditions would require a laser with a power in the IR of many kW, well beyond the capability of the present short pulse laser technology.

Cathodes with high QE in the $10^{-2}$ range exist and can be potentially used for IRIDE with laser power well within the limits of the present laser technology (see next section for more details). These cathodes are represented by semiconductor materials such as for example $Cs_2Te$, multi-alkali antimonides, and Cs:GaAs (a complete review of cathodes can be found [11]). Unfortunately, all these materials are chemically reactive and require very low operational vacuum pressures in the UHV/XHV range in order to achieve acceptable QE lifetimes. The good news is that new gun schemes, such as the gun here proposed for IRIDE, have already demonstrated vacuum pressures compatible with such cathodes.

In what follows some possible cathode candidates for IRIDE are briefly discussed.
In the category of positive electron affinity materials (PEA), special attention will be placed to $Cs_2Te$ and multi-alkali antimonides. $Cs_2Te$ is a relatively robust material that has been extensively used in the FLASH RF gun [10, 12]. QEs of up to 10% have been routinely achieved with month-class lifetime at the FLASH operational conditions. A minor drawback is that $Cs_2Te$ requires UV light in order to achieve high QE, hence imposing some additional load on the laser system compared to cathodes emitting in the visible.

Multi-alkali antimonides, and $K_2CsSb$ and Na2KSb in particular, are materials with very interesting characteristics. Due to their small band gap and small electron affinity, they photo-emit in the visible, and at this energy, QEs > 5 % can be readily achieved. Recent results on $K_2CsSb$ [11, 12] have demonstrated promising lifetimes and appealing small intrinsic emittances..

Cs:GaAs has been used extensively in DC guns achieving an overall good performance but its compatibility with RF guns still need to be experimentally demonstrated.

### 2.5.3. Drive Laser Systems

The photoinjector drive laser is a key component of an injector system because its properties directly determine the quality of the electron beams.

The pulse-to-pulse stability specifications are stringent in terms of both longitudinal and transverse parameters: timing, pointing, energy stability are all key parameters for IRIDE.

Transverse and longitudinal laser pulse shaping are required to deliver flat-top distribution with transverse hard edges and sharp rise and fall time. This is required in order to minimize non-linear self-forces in the beam that would act during the electron beam transport increasing the beam RMS emittance.

The IRIDE machine will provide a variable repetition rate, up to values in the 100 MHz range. Among the various operating scenarios considered for IRIDE, the most demanding one in terms of photocathode laser intensity is the pulsed RF option where a beam current as high as 4 mA in the 1 ms long macropulse is required at 100Hz repetition rate. This sets the needed average power of the photo-injector laser. High quantum efficiency semiconductor photocathodes will be used (i.e $Cs_2Te$) emitting in the UV (266 nm) with a quantum efficiency exceeding 10%. If one takes 5% as a conservative value for the quantum efficiency, than the laser energy budget calculation lead to an initial IR average power out of a MOPA (Master Oscillator-Power Amplifier) system of about 40W. This was calculated taking into account 10% IR-to-UV conversion efficiency and 90% energy loss during laser shaping and transport to the cathode.

Given the tight requirements on stability and transverse beam quality, the preferred choice for the IRIDE photo-injector laser is represented by fiber based MOPA systems. The oscillator frequency has to be sub-harmonic of the electron Gun and accelerating RF frequency. A possible choice is to set the laser oscillator cavity at the same frequency of the RF gun, choosing the 186 MHz APEX gun frequency for both. The synchronization between the accelerating cavity frequency and the laser is assured by the presence in the cavity of a piezo actuator with few KHz bandwidth controlling the cavity length.

Single pass Ytterbium-doped silica fibers can be used as amplifying media in CPA (Chirped Pulse Amplification) schemes, taking the average power up to the desired level (tents of Watts). An acousto-optic pulse picker used at low power will lower the repetition rate as needed (1-100 MHz).

The control on the macropulse duration will be done at high power, when a pulse picking system based on a Pockels cell plus a waveplate and a polarizing beam splitter can be used. High voltage power supplies with rise and fall times as fast as 2 ns are commercially available, assuring complete discrimination between 100 MHz laser pulses.

## 2.6. Beam parameters flexibility

Table 2 summarizes the main beam parameter requirements arising from the analysis of the IRIDE physics cases (described in the subsequent chapters of this WhiteBook) and that have been used to assess and propose the RF configuration options for the IRIDE multipurpose linac.

**Table 2.4**: Beam parameter requirements for the IRIDE physics cases.

| | Neutron source | Neutron source with TOF | Nuclear photonics | Thomson source, $\gamma$-$\gamma$, $e^-$-$\gamma$ | FEL (SASE, seeded) | FEL (oscillator) | THz radiation | $e^+$-$e^-$ , $e^-$-$e^-$ |
|---|---|---|---|---|---|---|---|---|
| $E$ [GeV] | > 0.8 | > 0.8 | 0.1 ÷ 1 | 0.1 ÷ 1 | 0.75 ÷ 4 | 0.03/2.3 | 0.1/1.5 | 0.5 ÷ 2 |
| $q_b$ bunch charge [pC] | 1000 | 1000 | 10 | 500 | 150÷600 | 50÷200 | < 500 | <350 |
| $I_{peak}$ bunch peak current [kA] | any | any | any | Any | 0.75 ÷ 4 | 1 | 2 | < 500 |
| $\sigma_t$ bunch length [fs] | any | any | any | Any | 150-200 | 20÷80 | ≈100 | <1000 |
| $\sigma_{x,y}$ bunch transv. size [μm] | any | any | 10 | 10 | 50 | 50 | 100 | <1.5 |
| $\sigma_E/E$ bunch energy spread [$10^{-3}$] | any | any | 0.1 | 0.1 | 0.1 | 0.1 | 0.1 | 0.1 |
| $\varepsilon_n$ bunch norm. emittance [μm] | any | any | 1 | 1 | 1 | 1 | < 5 | <5 |
| $f_b$ bunch rep. rate [MHz] | 1 | 1000 | 100 | 65 | < 1 | 5.0 | < 1 | 1 |
| $T_p$ pulse duration [μs] | CW | 0.5 | CW | 0.9 | Any | > 100 | any | CW |
| $f_{pulse}$ pulse rep. rate [Hz] | CW | 50 | CW | 100 | Any | any | any | CW |
| $I_p$ current in the pulse [mA] | 1 | 1000 | 1 | 32.5 | < 1 | 0.25÷1 | < 5 | CW |
| $I_{ave}$ average current [μA] | 1000 | 25 | 1000 | 3 | < 250 | < 300 | < 300 | 350 |
| $(\Delta E/E)_{train}$ energy spread along the train [$10^{-3}$] | | | | | | 0.5 | | 0.1 |
| $<P_b>$ ave. beam power [kW] | | | | | | | | 500 |
| $L$ luminosity [1/cm$^2$ s] | | | | | | | | $10^{32}$ |

The above requirements can be largely satisfied by a linac configuration either based on **high duty cycle pulsed operation** or on a **pure CW operation**, providing various beam parameters in the energy span of 1-2 GeV (per linac) with the same machine layout (i.e. number of cavities, modules and gradient settings), but different RF power systems. In table 2.5 a few cases (mainly at different energies and currents) are shown for the two RF architecture choices (pulsed or CW). For each option any of the beam parameter listed within its cases can be delivered by the same machine, operating at different accelerating field values with different current settings. Besides the fact that the two RF options require different RF power systems and a different power coupler, the linac design for each energy range in terms of choice of number of modules and cavities, and cavity accelerating field, are the same. A minimal increase in the linac length for the CW case is due to the necessity of increasing the number of cryogenic connections in order to accommodate the larger cryogenic loads caused by CW operation.

The two options can cover the same energy range with (approximately) the same machine length of about 150 m, with a different RF system (either tubes of SSA) and a different size of cryoplant. The high energy operation at 2 GeV for the CW version can be safely guaranteed by limiting the duty cycle of the RF sources at 60% over long times (~ scale of seconds) in order to decrease the cryogenic losses. This marginal limitation, purely given by a conservative assumption on the cryoplant capacity, is based on the use of a single LHC like cryoplant (18 kW @ 4.5 K) feeding both IRIDE identical linacs. It takes into account all the static losses, as experienced at FLASH, the 2 linac fluid distribution, and a 50% overhead for off-nominal operation, uncertainties on heat load estimations and transients.

**Table 2.5**: Possible machine parameters for the two RF options (pulsed/CW).

| | **Pulsed Option** | | | **CW/qCW Option** | | |
|---|---|---|---|---|---|---|
| | **Pulse 1** | **Pulse 2** | **Pulse 3** | **CW 1** | **CW 2** | **Quasi CW** |
| $E$ [GeV] | 1 | 1.5 | 2 | 1 | 1.5 | 2 |
| $I$ (within pulse) [mA] | 5 | 3.5 | 2.5 | 0.5 | 0.35 | 0.260 |
| $I$ (average) [mA] | 0.58 | 0.33 | 0.17 | | | 0.155 |
| RF pulse duration [ms] | 1.5 | | | CW | CW | 1000 |
| Beam pulse duration [ms] | 1.17 | 0.95 | 0.67 | | | 990.74 |
| RF Duty cycle [%] | 15 | | | 100 | 100 | 60 |
| Beam Duty cycle [%] | 11.7 | 9.5 | 6.7 | | | 59.4 |
| $f_{RF}$ [MHz] | 1300 | | | | | |
| $E_{acc}$ [MV/m] | 10.04 | 15.05 | 20.07 | 10.04 | 15.05 | 20.07 |
| $L$, Cavity length [m] | 1.038 | | | | | |
| $R/Q$ [Ohm] | 1036 | | | | | |
| $Q_0$ | 2.00E+10 | | | | | |
| $Q_{ext\ opt}$, Optimal coupling | 2.01E+06 | 4.31E+06 | 8.04E+06 | 2.01E+07 | 4.31E+07 | 7.73E+07 |
| $Q_{ext}$, Design coupling factor | 4.00E+06 | | | 4.00E+07 | | |
| Reflected RF power at design | 1.09E-01 | 1.38E-03 | 1.13E-01 | 1.09E-01 | 1.38E-03 | 1.01E-01 |
| Cavity BW [Hz] | 162.50 | | | 16.25 | | |
| Cavity rise time [ms] | 0.98 | | | 9.79 | | |
| Rise-time to target $V_{acc}$ [ms] | 0.33 | 0.55 | 0.83 | 3.57 | 6.01 | 9.26 |
| # of cavities | 96 | | | | | |
| # of modules | 12 | | | | | |
| $P_{beam}$/cavity (pulse) [kW] | 52.08 | 54.69 | 52.08 | 5.21 | 5.47 | 5.42 |
| $P_{beam}$/cavity (ave) [kW] | 6.09 | 5.20 | 3.48 | | | 3.22 |
| $P_{RF}$/cavity (pulse) [kW] | 58.49 | 54.76 | 58.70 | 5.85 | 5.48 | 6.03 |
| $P_{RF}$/cavity (ave) [kW] | 8.77 | 8.21 | 8.81 | | | 3.62 |
| Available $P_{RF}$/cavity (pulse) [kW] | 80.00 | | | 7.00 | | |
| Unloaded asyntotic voltage [MV] | 36.42 | 36.42 | 36.42 | 34.06 | 34.06 | 34.06 |
| $P_{diss\ RF}$ (@ 2 K), RF dissipation [W] | 75.41 | 169.67 | 301.64 | 502.73 | 1131.15 | 1206.56 |
| $P_{static}$ (@ 2 K) [W] | 42.00 | | | | | |
| $P_{cryo}$ (@ 4.5 K) [kW] Cryogenic 4.5 K equivalent power, accounting all loads and 50% margin | 1.4 | 1.9 | 2.6 | 4.0 | 7.3 | 7.7 |
| Total $P_{beam}$ (peak) [kW] | 5000 | 5250 | 5000 | 500 | 525 | 520 |
| Total $P_{beam}$ (ave) [kW] | 585 | 499 | 334 | | | 309 |
| Linac length [m] | 149 | | | 159 | | |

## 2.7. Timing and Synchronization

The Timing and Synchronization central system is a very crucial part in a multipurpose, multiuser facility like IRIDE that has to serve a large variety of experiments such as various FEL (SASE, seeded, oscillator), Linear colliders, Compton sources, Neutron sources, THz radiation sources, etc. Most of these experiments require an extremely accurate synchronization, at the fs scale, among the various lasers of the facility (photocathode, heater, seeding, high intensity photon source for Compton scattering …), the RF accelerating fields in the cavities and, ultimately, the linac electron beam. Moreover, a relevant part of the beam diagnostics needs to be synchronized at the same level (bunch arrival monitors, streak cameras,...) as well as pump lasers whenever FEL radiation users perform pump and probe class experiments.

Start-to-end simulations probing the sensitivity to the synchronization errors of each of the processes expected from IRIDE will be needed to set the synchronization specifications for each machine subsystem. However, in this early stage of the proposal, on the base of consolidated results from similar facilities already in operation or in a more advanced state of design/construction, we are aware that IRIDE will need a state-of-the-art synchronization system. In the last decade, mainly pushed by the requirements of the FEL radiation production physics [16], this field has impressively progressed and global facility synchronization at the fs scale is presently an ambitious but still realistic goal.

The design and development of the IRIDE synchronization system will benefit the studies and the operational experience of FLASH [13] in this field, as well as the work done for XFEL [14]. Similarly to NGLS and other proposed facilities, IRIDE will have to extend the concepts already developed for pulsed machine to CW operation.

### 2.7.1. Timing and Synchronization General Architecture

The timing clients, i.e. all sub-systems that need to be kept tightly synchronized, will be positioned all over the entire facility, on a length scale of ≈ 800 m. Therefore the timing and synchronization system will consists of three main parts:

- **Timing generation and distribution**. An ultra-stable reference signal generated in a central timing station will be distributed to the various clients through actively stabilized links. Due to the remarkable link lengths, an optical reference will be distributed to exploit the fiber-link low attenuation and the large sensitivity obtainable by optical based timing detection. The required stability of each link is <10 fs over any time scale. The link stability is part of the synchronization budget assigned to each client.
- **Client Locking**. Each individual client (laser systems, RF power stations, beam diagnostics hardware...) has to be locked to the local reference provided by the timing distribution systems. The lock technique depends on the particular client (laser PLLs, laser direct seeding, RF pulse-to-pulse or intra-pulse phase feedbacks...) while the needed lock accuracy for each sub-system will be ultimately set by a systematic study of the effects of its timing error on the machine performances.
- **Client triggering**. Together with a continuous reference signal, low repetition rate trigger signals must be provided to some clients, which contain essentially all the information related to the beam time structure that is needed to prepare all the systems to produce and monitor the beam itself as required (laser amplification pumps, klystron HV video pulses, beam and FEL diagnostics ...). This task is especially delicate here, being IRIDE proposed as a multitasking machine that may require complex time structures in the beam to serve many different users simultaneously. The triggering system is a coarser timing line (≈10 ps required stability) that can be distributed either optically (through fiber-links) or electrically (through coaxial cables).

The proposed architecture for the IRIDE timing and synchronization system to implement the illustrated general approach is sketched in Figure 2.8.

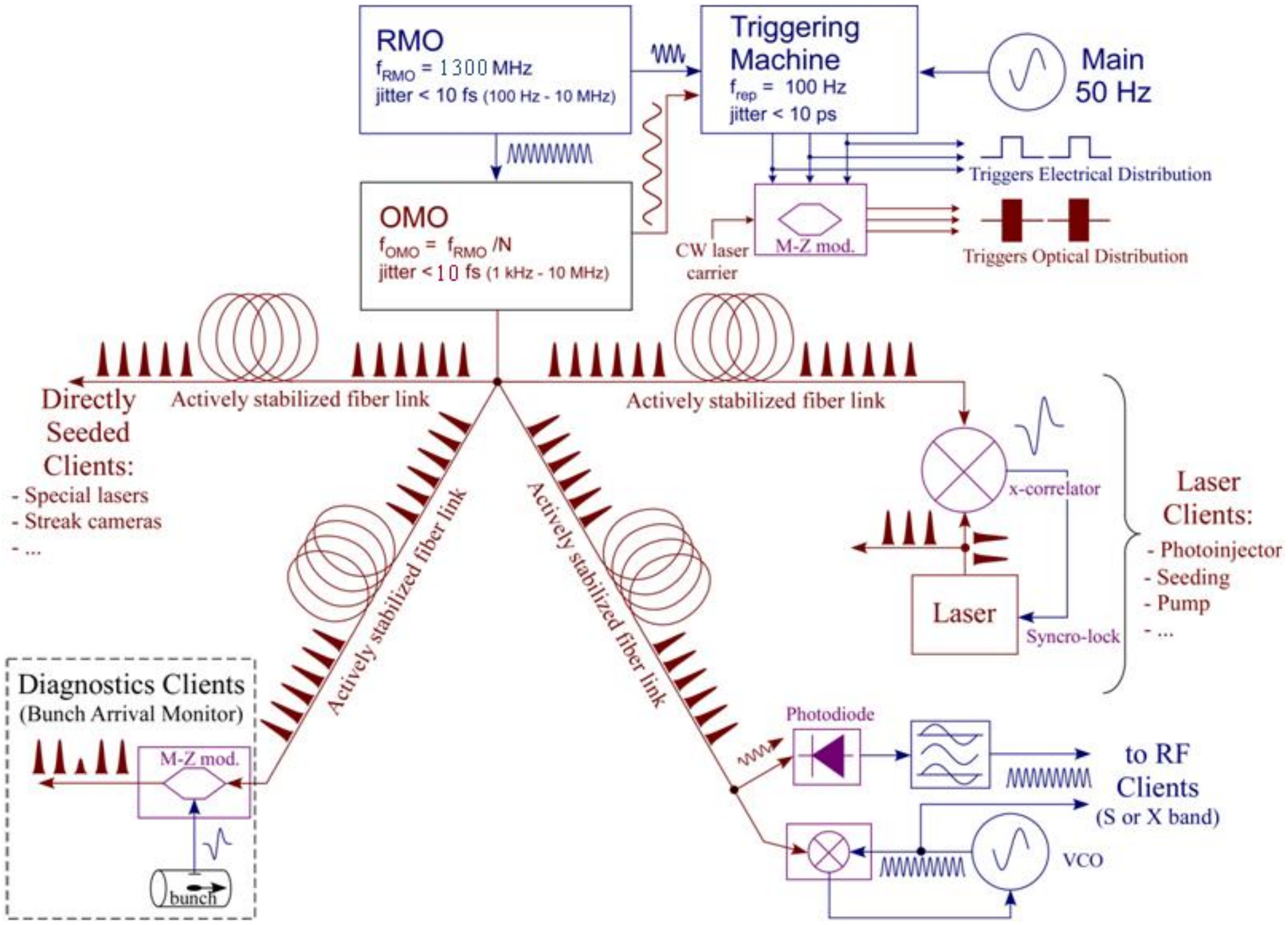

**Figure 2.8:** IRIDE Timing and Synchronization system schematics.

### 2.7.2. Timing and Synchronization General Architecture

The reference signal is originated by a Reference Master Oscillator (RMO) which is a μ-wave crystal oscillator with ultra-low phase noise characteristics. The role of this device is to provide a reliable reference tone to an Optical Master Oscillator (OMO) which is a highly stable fiber-laser that encodes the reference timing information in the repetition rate of short optical pulse in the IR spectrum.

The RMO guarantees the long term stability of the OMO, and, through the OMO phase lock system, imprints its low-frequency noise figure to the whole facility timing line.

The state of the art low-noise μ-wave oscillators can provide pure sine tones with phase jitter at few fs level over a spectral range from 10 Hz to 10 MHz. As an example, the phase noise SSB spectra of sapphire oscillators commercially available from Poseidon Scientific Instrumentation (PSI) are reported in Figure 2.9.

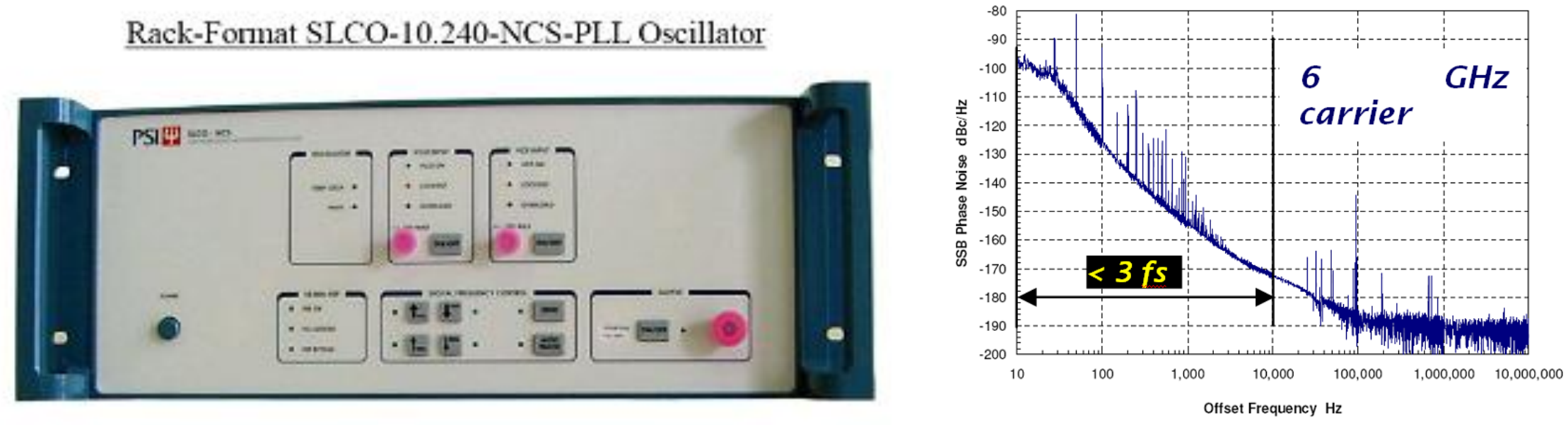

**Figure 2.9:** Ultra-low phase noise sapphire oscillator from PSI.

The timing reference will be encoded in an optical signal before being distributed over the whole facility through a glass fiber network. The μ-wave to optical conversion is accomplished by mode-locking a low noise fiber laser (the facility Optical Master Oscillator – OMO) to the RMO. The OMO to RMO synchronization is obtained by a PLL scheme controlling the path length of the fiber laser cavity by stretching the fiber with piezo-controllers driven by the relative phase error between the two oscillators. This is a standard technique to synchronize also in-air laser oscillators to external references, with the piezos controlling the position of one or more mirrors in this case. Due to the limited frequency response of piezo-controllers, the loop gain rolls off typically around 5 kHz. Above this cut-off frequency the OMO retains its typical noise spectral properties, while below the cut-off frequency the OMO phase follows the RMO one, and the spectra of the two oscillators result to be very similar. However, the intrinsic phase noise spectrum of a good fiber laser oscillator above the piezo based PLL cut-off frequency is comparable or even better respect to that of a μ-wave reference oscillator.

Fiber lasers with characteristics similar to the IRIDE OMO specifications are available on the market. A commercially available unit is shown in Figure 2.10.

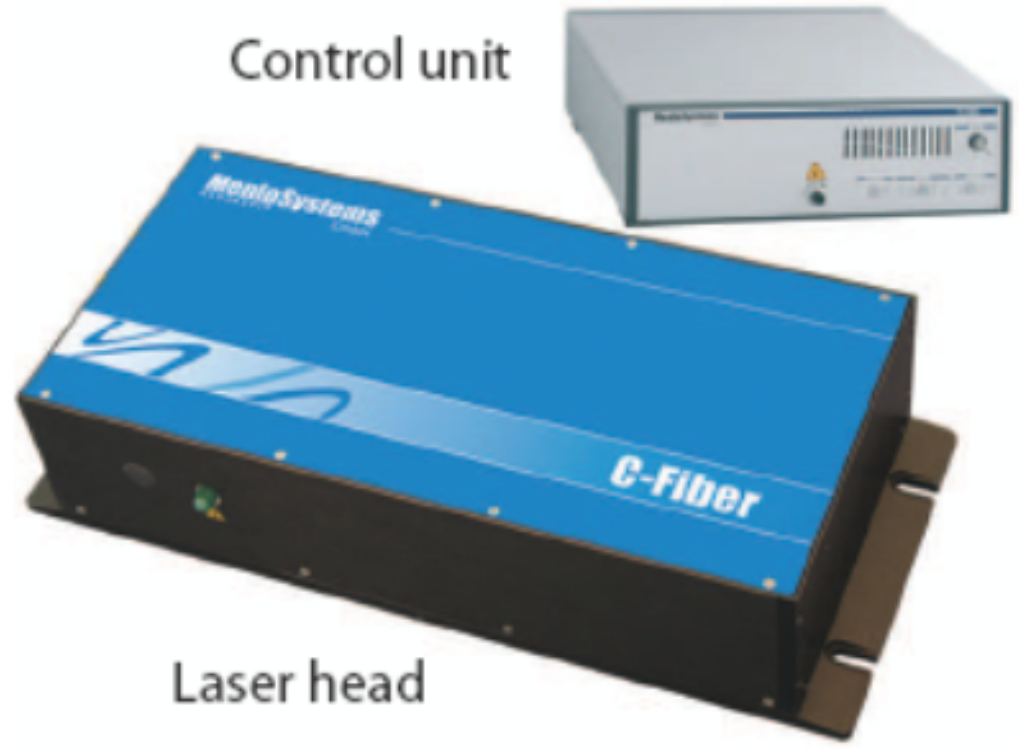

**Figure 2.10:** MENLO Sytem GmbH fibre lase C-Fiber 780.

### 2.7.3. Optical Distribution of the Synchronization Reference

The distribution of the synchronization reference for IRIDE will be realized using multi fiber optic channels. The source of the temporal reference signals for all the sub-systems is foreseen to be generated by an optical master oscillator (OMO) situated at a median position of the facility infrastructure. The channels extend from the source to the several end users up to 400 meters distance. The choice of optical waveguide channel to distribute the synchronization signals is motivated by a number of advantages of the optical link respect to the electrical cables. In fact the optical fibers offer THz bandwidth, immunity to electromagnetic interference and very low attenuation. These properties make the optical fiber a mandatory technology when one aims to synchronize a large scale facility at sub 10 fs level.

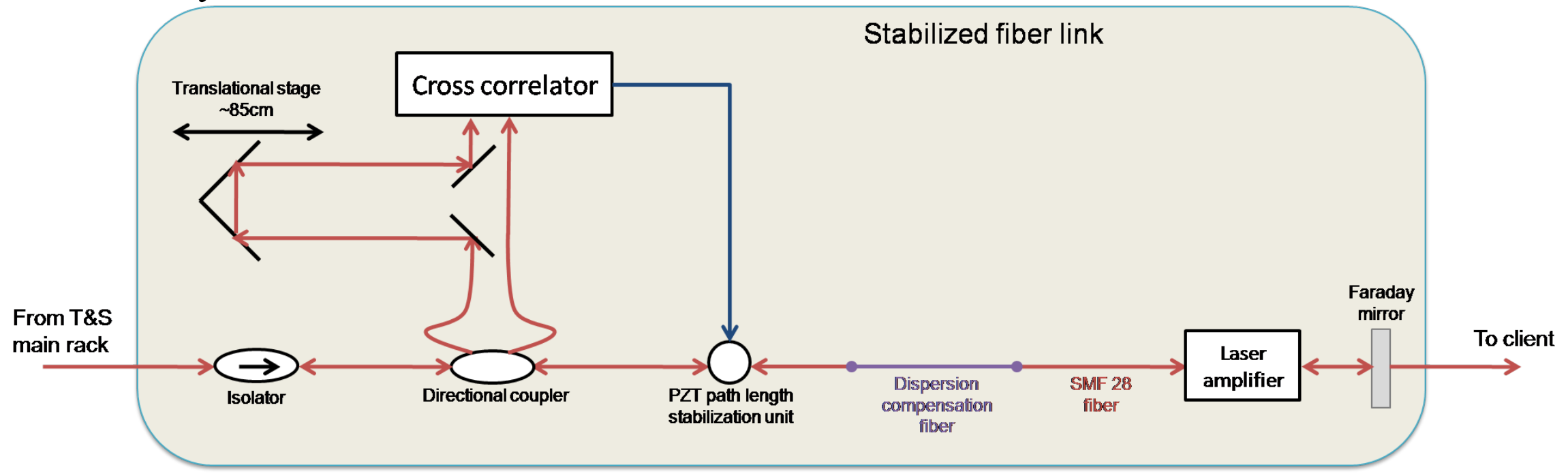

**Figure 2.11:** Layout of a path-length stabilized optical link.

The optical signal will be sent to several fiber links with equalized optical length. The fiber optics will to be length-stabilized in order to deliver synchronous pulse to the end users. In fact the time of propagation in the fiber is in general affected by temperature drift and acoustic noise. The stabilization is achieved sending back a pulse from the end of the fiber and optically compare it with another pulse from the oscillator as schematized in Fig. 2.11. To reflect part of the power a Faraday mirror can be used. In details the pulse to send through the fiber is divided with a polarization beam splitter: part of the pulse is sent to the fiber and the other is used as a reference for the optical mixing with the retro-reflected pulse. The comparison between the reference and the other pulse is carried out with cross-correlator. The error signal will be used to drive a piezo path length stabilization unit and a stepper actuator to move trombone. The feedback aims to keep constant the pulse propagation length.

### 2.8. Cryogenic plant

Besides the RF distribution architecture, the CW and pulsed linac differ on the size of the cryogenic plant required for operation. A cryogenic heat load assessment which takes into account the TTF/FLASH/XFEL/ILC experience and studies [15] has been developed, resulting in Table 2.6 summarizing the needed cryoplant capacity for the different options and the same beam parameters assumed in Table 2.5.

**Table 2.6:** Cryogenic plant requirements for IRIDE Options.

| | **Pulsed Option** | | | **CW/qCW Option** | | |
|---|---|---|---|---|---|---|
| | **Pulse 1** | **Pulse 2** | **Pulse 3** | **CW1** | **CW2** | **Quasi CW** |
| E (GeV) | 1 | 1.5 | 2 | 1 | 1.5 | 2 |
| RF Duty Cycle (%) | 15% | | | 100% | | 60% |
| Total needed 4.5 K equivalent power per linac (kW) | 1.43 | 1.93 | 2.63 | 3.99 | 7.32 | 7.72 |
| Contribution of 2 K load to total (%) | 51% | 64% | 73% | 80% | 89% | 90% |
| Required 4.5 K equivalent power for both linacs (kW) | 2.85 | 3.85 | 5.25 | 7.98 | 14.64 | 15.45 |

The cryogenic consumption is dominated by the dynamic RF loads in the CW linac version, where the low temperature 2 K load provide 80-90% of the total requirement. In the high duty cycle pulsed version the cryogenic power is decreased, but not linearly with the RF dynamic load as all static contributions need to be taken into account.

The assessment summarized in Table 2.6 leads to the conclusion that the CW machine can be operated in pure CW for most of its energy reach, and in quasi-CW (with a 60% duty cycle) at its highest energy reach, to be able to conservatively fit in a standard cryogenic LHC unit of 18 kW 4.5 K equivalent power, which can be provided by independent vendors in Europe. The 60% duty cycle, in addition to a 50% overcapacity on the heat load assessment described previously, has been conservatively assumed to limit the required cryogenic power to ≈16 kW, below the 18 kW specifications, but these margins give confidence for raising it further towards a full CW mode, as experience in all large SRF installations (as in LEP) shows that the overcapacity margins can be later on used to cope with additional heat dissipation resulting from increased performance goals.

In the pulsed option the required cryogenic power can be delivered by a smaller size plant, well within the 8 kW XFEL cryoplant sizes, to cover all beam parameter modes. A maximum of ≈5 kW (1/3 of the CW option) is required in this case.

### 2.9. Cost analysis

The costing for the superconducting linac components reported in the cost assessment table is largely based on the costs budgeted and assessed by the XFEL Project, reviewed for the different RF distribution options foreseen for the two RF operation options. Costs for the RF systems have been assessed through contact with companies potentially able to provide the klystron or SSA systems. For the cryogenic system experience and scaling laws published for LHC and widely available in literature has been used.

# 3. THE FREE ELECTRON LASER SOURCE

## 3.1. Introduction

The use of superconducting (S-C) LINACs to drive a Free Electron Laser (FEL) or a chain of FEL's offers significant advantages in terms of flexibility and opens new possibilities for large scale facilities. The breakthrough of the architecture of future FEL devices will be the development of an integrated facility embedding SASE sources combined with FEL oscillators, having different functions, including those of seeding and radiation damping (beam heater).

The first operating FEL exploited the Stanford S-C LINAC as electron beam source, being at that time the most reliable accelerator in terms of beam qualities (in particular peak current and energy spread), RF bunch structure and stability.

In the subsequent development of the FEL devices the choice of more conventional accelerators (namely Microtrons, conventional Linacs, Storage Ring…) was dictated by the intrinsic complexity of the underlying technology and by the associated high cost of operation.

The technological evolution, occurred since the seventies of the last century, has significantly reduced the costs and the operating conditions of a S-C accelerating device.

The use of a S-C LINAC would offer significant advantages for short wavelength FEL oscillators, would increase the average power of fourth generation synchrotron radiation sources and would expand the relevant capabilities in terms of beam-lines and, therefore, of users.

In Fig. 3.1 we report the existing FEL sources characterized by the energy of the driving e-beam and by the output photon energy. The short wavelength-high electron beam energy region is getting more and more crowded.

The scaling of the photon beam energy vs. the e-beam energy displays an almost linear behavior in a log-log plot, as it must be since $e_p \propto E_e^2$. However we note that some deviations occurs and in some energy regions the slope tends to become flat.

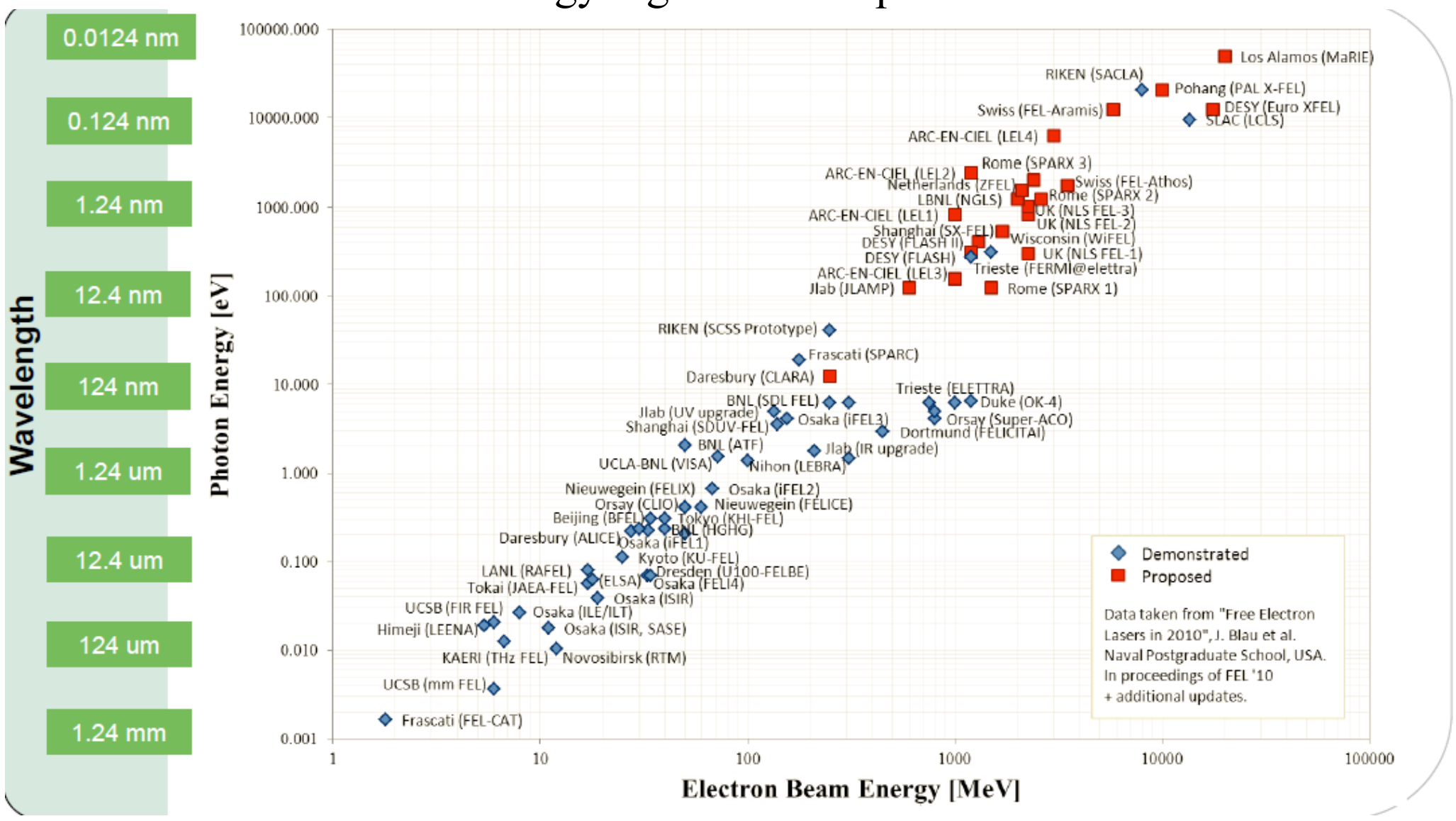

**Figure 3.1:** FEL photon energy-electron beam energy chart

Such a fact has quite a deep meaning and corresponds to changes in FEL technologies. To reach shorter wavelengths, conventional Linacs have been replaced by Storage Rings, which have been later substituted by the fourth generation synchrotron radiation sources. This last technology has provided a new conception for the production of FEL radiation (mirror-less single passage).
An analogous trend can also be inferred from the Livingston chart in Fig. 3.2a, where we have shown the evolution during the years of the accelerator energy and from Fig. 3.2b where we have reported the same for the X-ray brightness. In both cases the figures display the logistic paradigm: exponential growth, saturation, innovation, exponential growth…

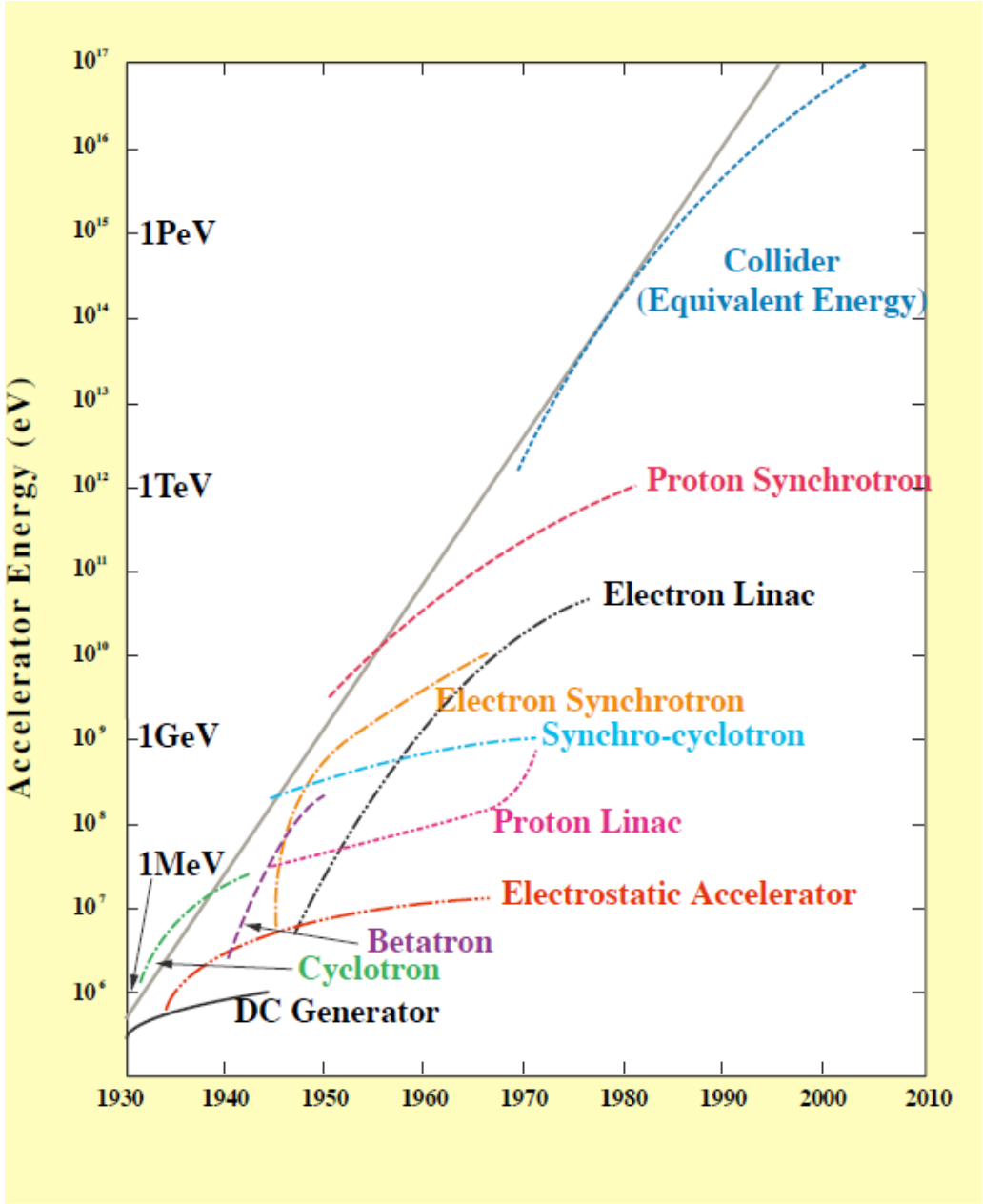

**Figure 3.2a:** Livingston Chart accelerators

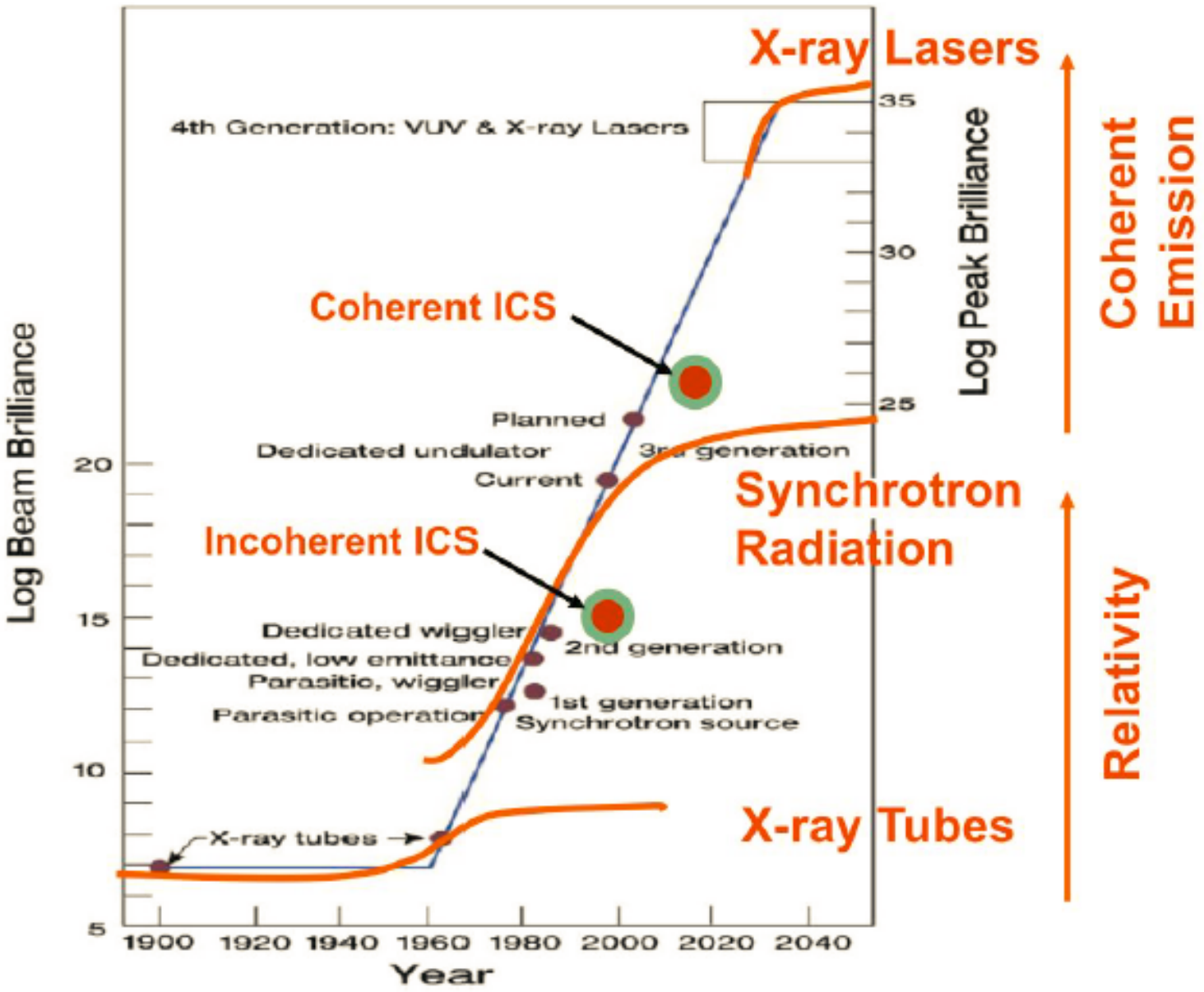

**Figure 3.2b:** X-ray brightness and relevant technological developments

It is evident that in the development of X-ray production the great impact have been provided, during the years, by the use of conventional SR sources, whose technology “saturated” with third generation devices, which were followed by the “invention” of the so (improperly) called 4-th generation sources (4GS), whose propulsive action is expected to be exhausted in three decades from now.
Even though not reported in the previous figures, the evolution scenario should be completed with FEL-Oscillators. The relevant temporal chart can be comprised in two decades, their technological limiting factor has been the non-availability of mirrors in the VUV-X region of the spectrum and this has paved the way to 4GS devices.
A significant breakthrough in FEL technology and science could be provided by new conceptions capable of merging high gain devices and oscillators, as illustrated in Fig. 3.3, where the possibility of developing X-FEL oscillators is combined with other promising innovations like Energy Recovering Linac and the possibility of exploiting FEL triggered guns.

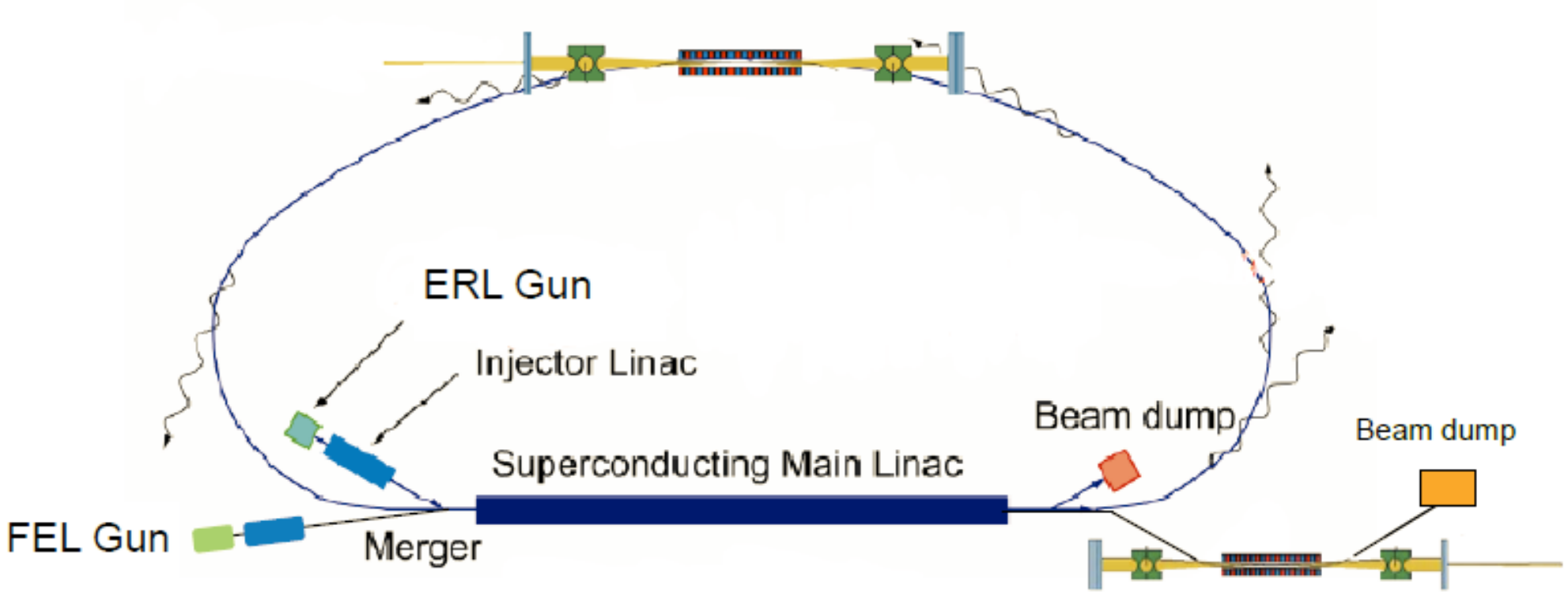

**Figure 3.3:** ERL superconducting LINAC and X-FEL-O

The interest for large scale FEL facilities is growing, but, since the world wide scenario is getting more and more crowded (see Fig. 3.4), distinctive features are mandatory to make it competitive

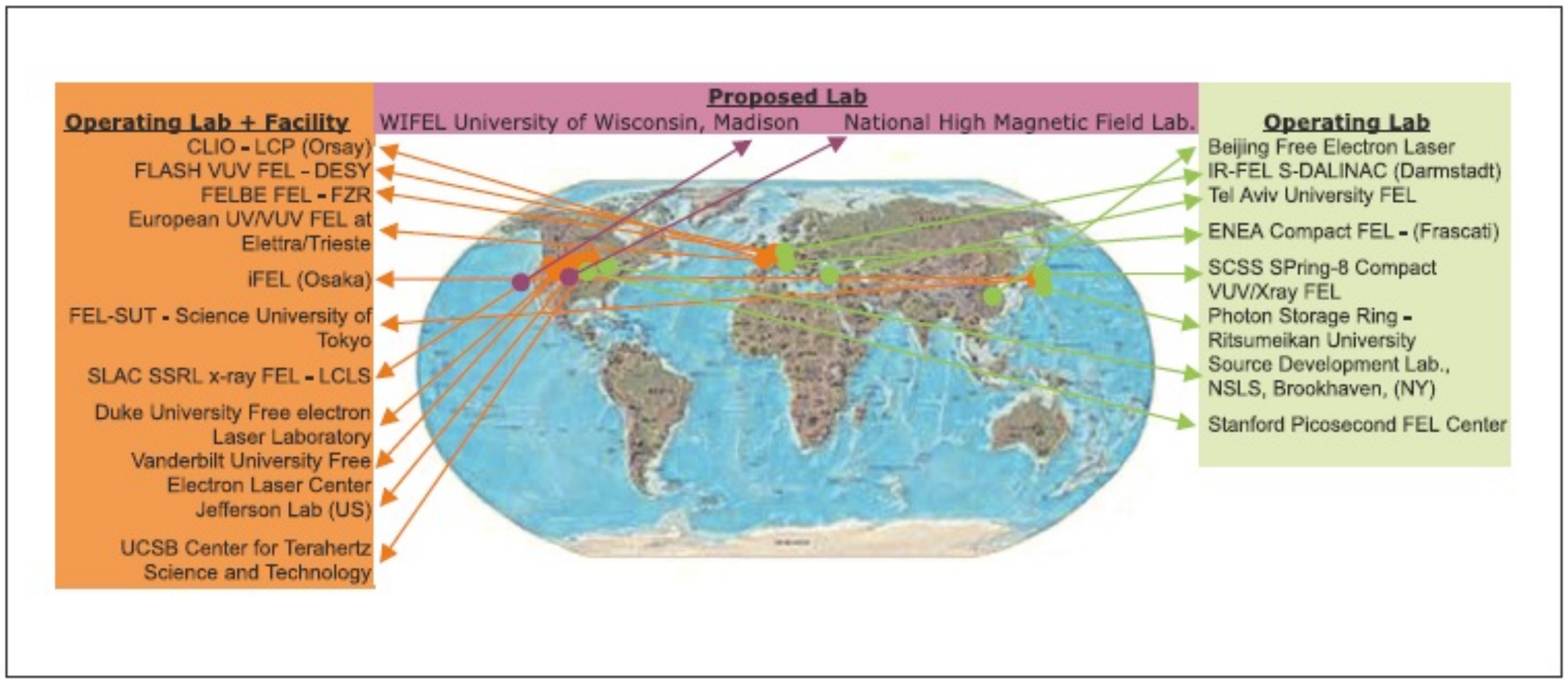

**Figure 3.4:** World-Wide FEL scenario

The question we will address before entering into the specific design details of this proposal is the following

*"What should be the architecture of a FEL device in order to be a large scale facility, going beyond the fourth synchrotron radiation paradigm?"*

A non-conventional design should provide the possibility of covering all the enormous range of energy and characteristic times necessary to probe characteristic phenomena in matter. An interesting and very effective technique is the pump-probe method, in which a strong long wave-length laser is used together with a short wavelength coherent source to study a variety of effects in matter (see Figs. 3.5)

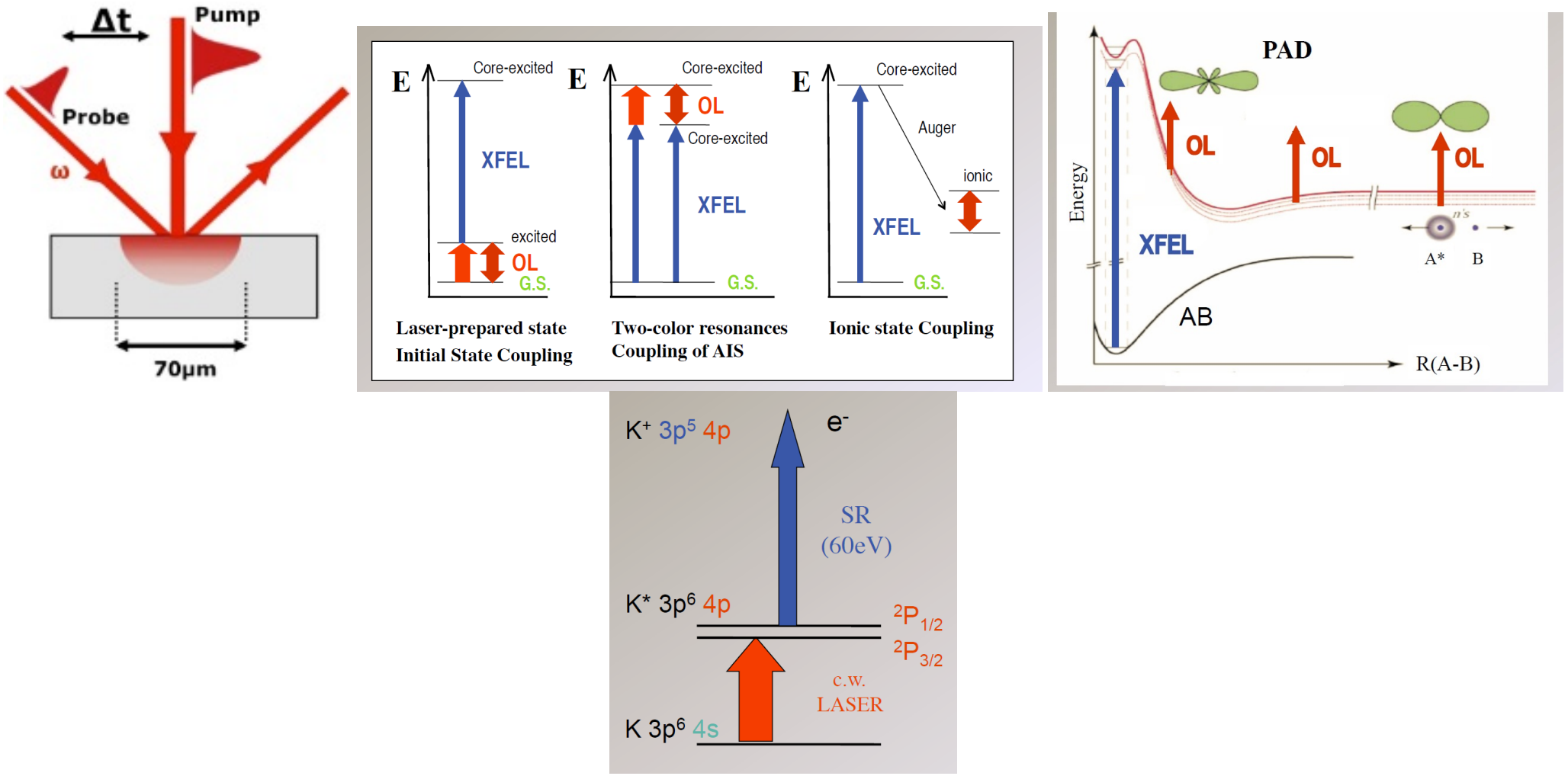

**Figure 3.5a:** Pump-Probe arrangement and possible applications

An idea in terms of photon energies and pulse length involved is given in Fig. 3.5b.

The FEL architecture we are considering is tailor-suited for these type of experiments, since "Optical-Laser" and X-UV beams will be provided by naturally synchronized Free Electron Laser sources.

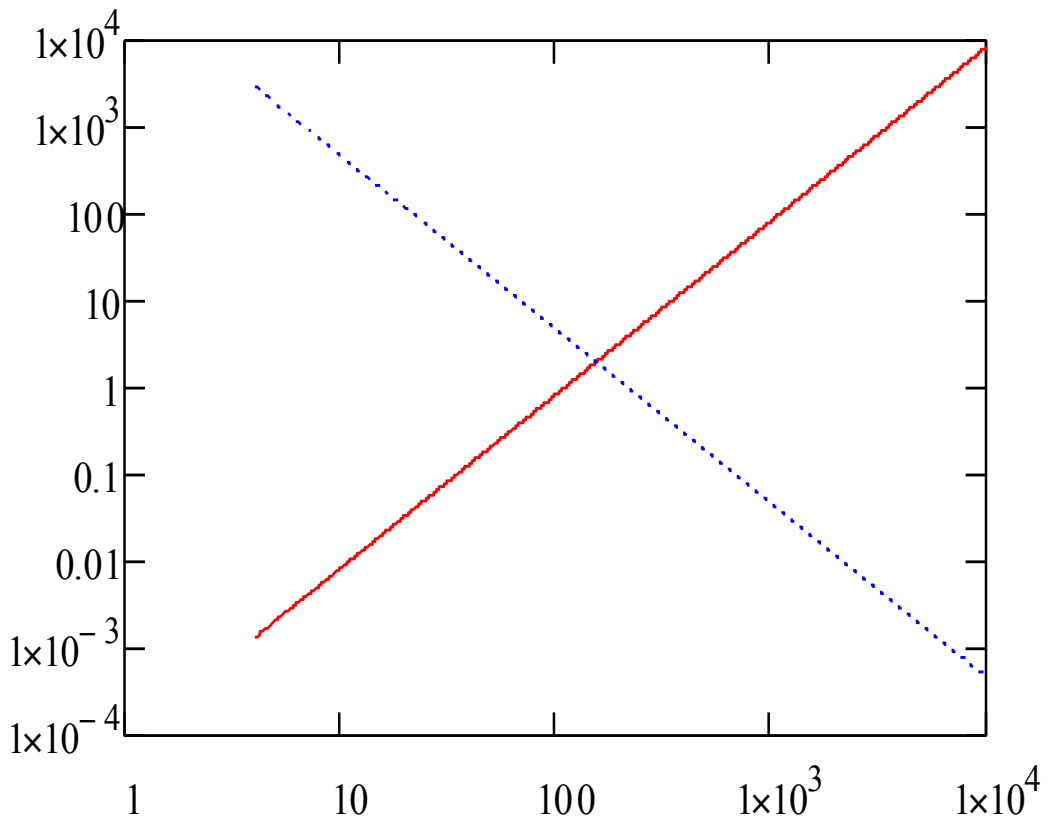

**Figure 3.5b:** Photon energy-pulse length Scenario of pump-probe experiments

The problem of conceiving the design of a FEL device, capable of providing all the needs summarized in Figs. 3.5, is very challenging since it would require an effort aimed at combining different aspects of FEL technology ranging from oscillators, to SASE, SEEDED and SELF SEEDED configurations. As already stressed, not a secondary challenge would be the possibility of driving the various FEL components with an e-beam provided by the same accelerators, thus automatically satisfying all the issues associated with time synchronization, necessary to exploit the different laser beams in e.g. pump-and-probe experiments.
Furthermore, a FEL source with such a flexibility should allocate part of its time for experiments aimed at testing fundamental problems in electrodynamics, to develop new schemes for FEL operation and for the production of tools for nuclear physics studies and the production of polarized positron beams to be eventually injected into a damping ring.
The conception of the Free Electron Laser line is such that it provides both a FEL facility and a complement to the architecture of a multi-facility laboratory.
The various components are listed below and sketched in Fig. 3.6.

a) *Two S-C-LINACS with 1.5 GeV maximum energy*

b) *Between the two Linacs a double FEL oscillator, with a manifold role, is inserted*

c) *The undulator chain can be powered by the beam operating at full energy (3-GeV) or less*

d) *A second FEL oscillator is added for the operation in the UV region and for intra-cavity backscattering for the realization of a gamma source to be exploited for Nuclear Physics studies and the production of polarized electrons*

e) *The third FEL section may operate in SASE or SEDEED mode*

f) *The seeding will be achieved by exploiting a conventional seeding procedure or by using the self- seeding scheme based on a kind of oscillator-amplifier device, according to the scheme developed in [3]*

g) *An oscillator with mirrors at* $13.5\,nm$ *can eventually be considered for the operation at short wavelength seeding*

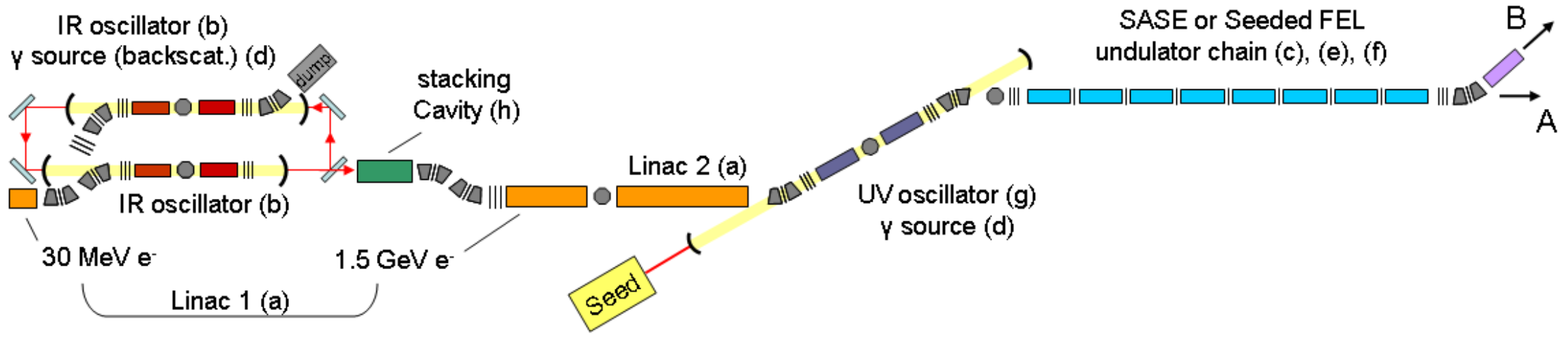

**Figure 3.6:** FEL SPARC Architecture

The realization of the whole device should be considered evolutionary and each step should provide a well-defined item with a well-specified application.
The S-C LINAC sections can be developed along with the Infra-LINAC FEL, whose role is strategic for the rest of the facility.
The two sections may also be operated with a low number of accelerating cavities to produce a beam with the energy necessary for the realization of an IR-FEL and for QED test, as discussed in the forthcoming section.
The second oscillator in the UV can be operated when the LINACs are not at the full energy, and the performances of the first oscillator as beam heater can be tested.
The construction of the undulators can be undertaken in parallel to that of the accelerators, and techniques of beam sharing on different undulator lines can be tested. Two "parallel" FEL oscillators can be constructed by using different portion of the e-beam, as shown in Fig. 3.7, the lasers, if operating at closer frequencies, can be exploited for two colors experiments.

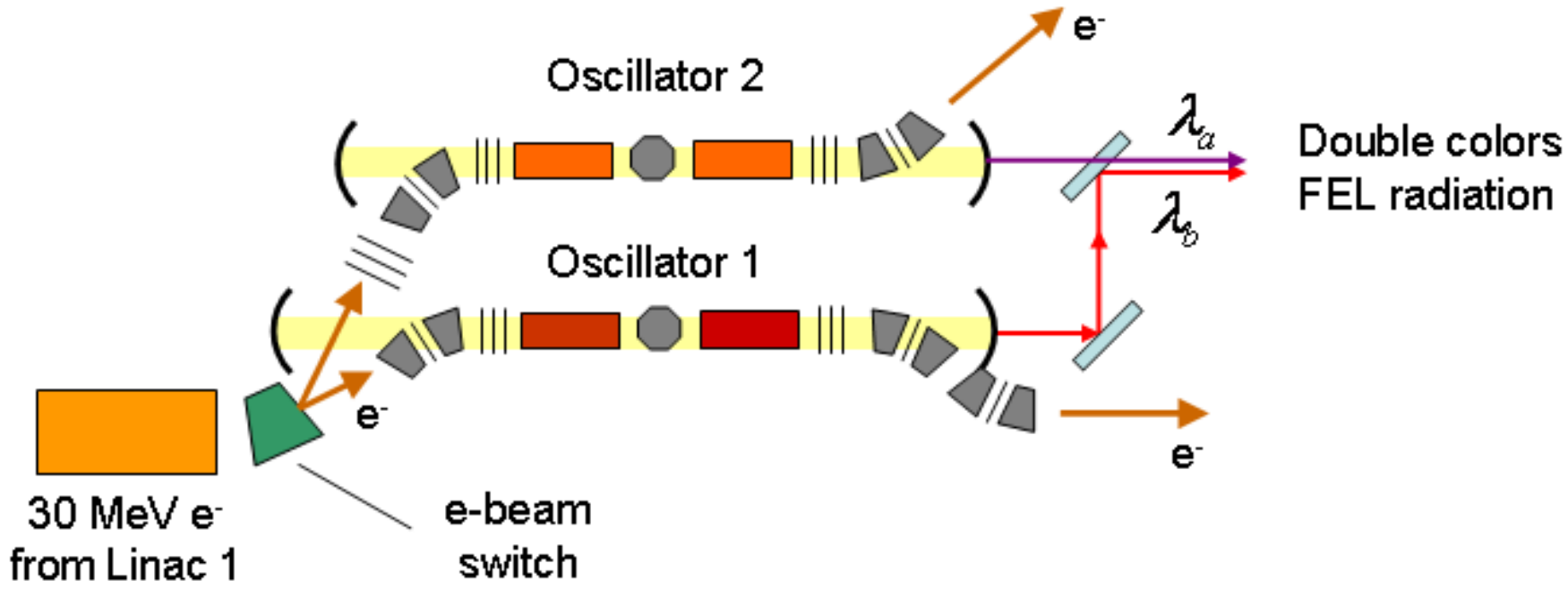

**Figure 3.7:** Two colour configuration for the two parallel low energy Oscillators.

### 3.2. ILIO the first IR oscillator

The oscillator within the LINACS (Fig. 3.8) is designed to play a multipurpose task, which will be summarized below.
ILIO (Infra-Linac-Interface-Oscillator) is conceived to provide an oscillator operating in the IR region, with the first LINAC kept at low energy (below 30 MeV) to drive FEL radiation in the region below $10\,\mu m$, using a SPARC type undulator ($\lambda_u = 2.8\,cm, K = 1$). The necessary electron peak beam current to ensure a successful operation is fairly modest, on the order of $50\,A$, an optical cavity with a total length of 4 m is sufficient (this space can however be reduced). In these conditions, a peak power around 5 MW can be obtained with a corresponding average power larger than 100 W, if a pulse train duration around $1\,ms$ is used with $1\,Hz$ repetition frequency (see the All. 1a for the details of the design).

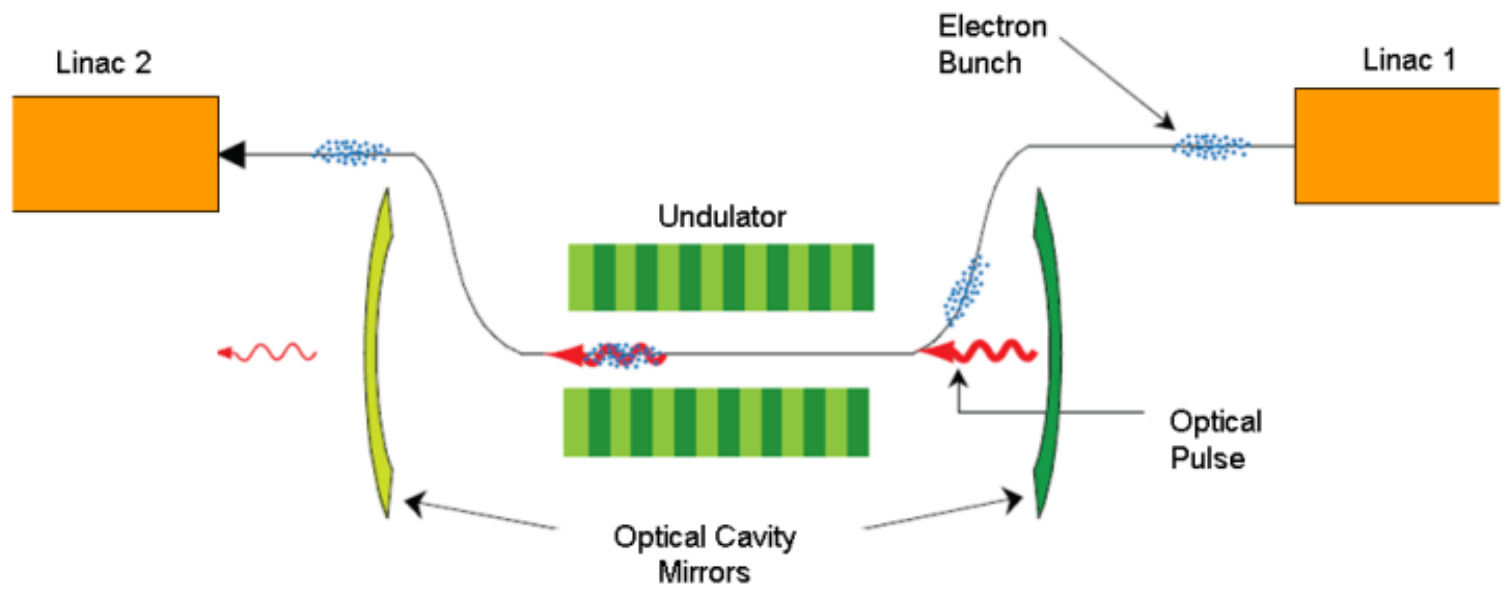

**Figure 3.8:** ILIO-layout

The long electron pulse and the associated number of electron bunches inside the optical cavity allows a substantive emission of intra-cavity backscattered X-ray photons Fig. 3.9 $(10^{11}\#\,peak\ flux/0.1\%bw)$ at $1\,nm$.

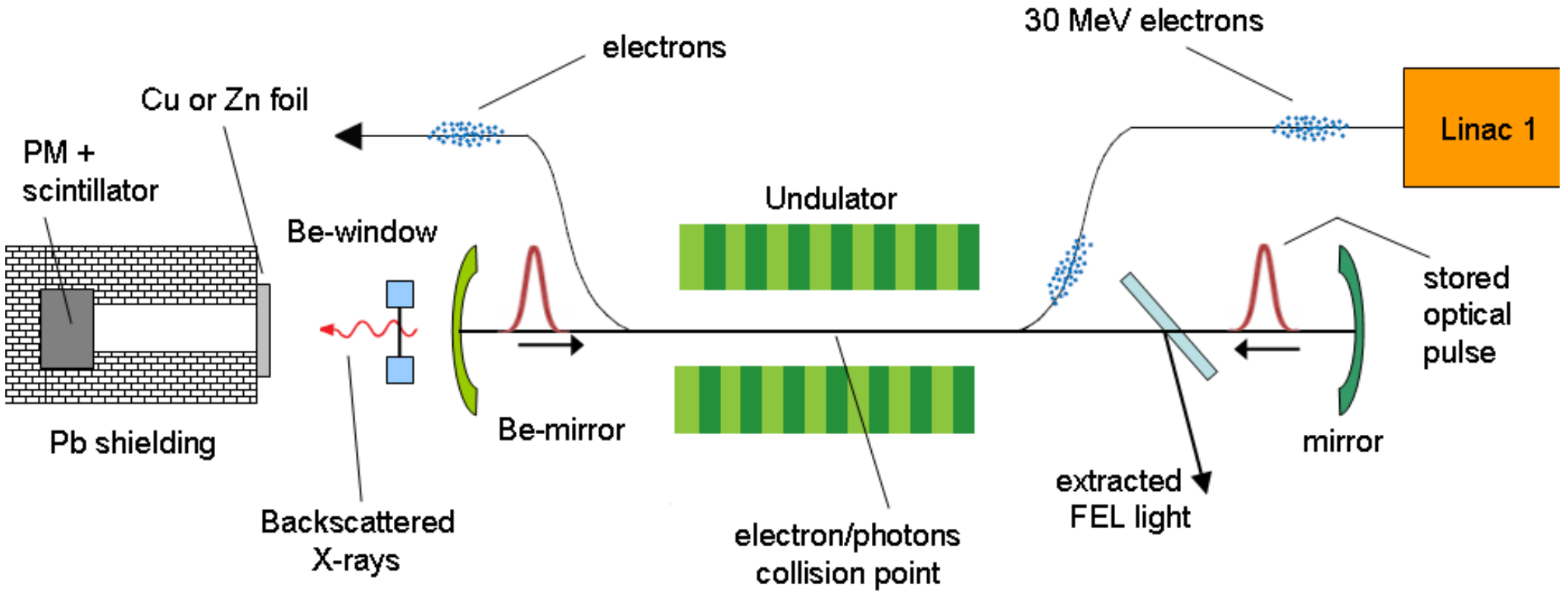

**Figure 3.9:** Intra-cavity backscattering and X-ray production

The design of the backscattering device can also be made less conventional by taking advantage from the structure of the e-beam. Indeed, we can split the e-beam and use it as indicated in Fig. 3.10, in which the laser beam is recirculated in a ring cavity, in which the upper part is exploited to produce the backscattering.

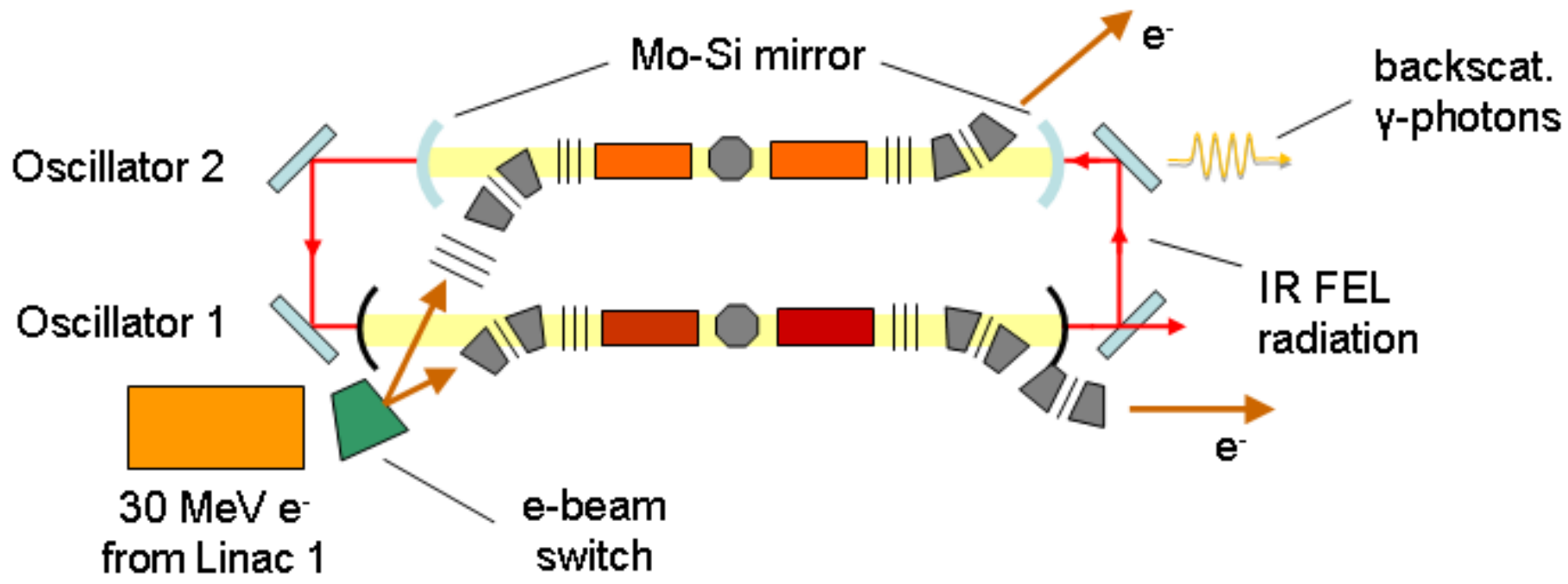

**Figure 3.10:** Backscattering configuration for the two parallel low energy Oscillators

In the following, we will discuss the application of the X-ray beam for the test of Molybdenum-Silicon (Mo-Si) mirrors, to be installed for the operation of the second oscillator CUPIDO (Coherent Ultimate Pulsed Infrared Double Oscillator).

A further use is that of the double FEL configuration, which can be exploited for QED test.
This is a fairly new conception of the FEL oscillator, which foresees the production of two laser beams in the same cavity, by two counter-propagating e-beams, which, in turn, are exploited to produce counter-propagating gamma photons (see Fig. 3.11).

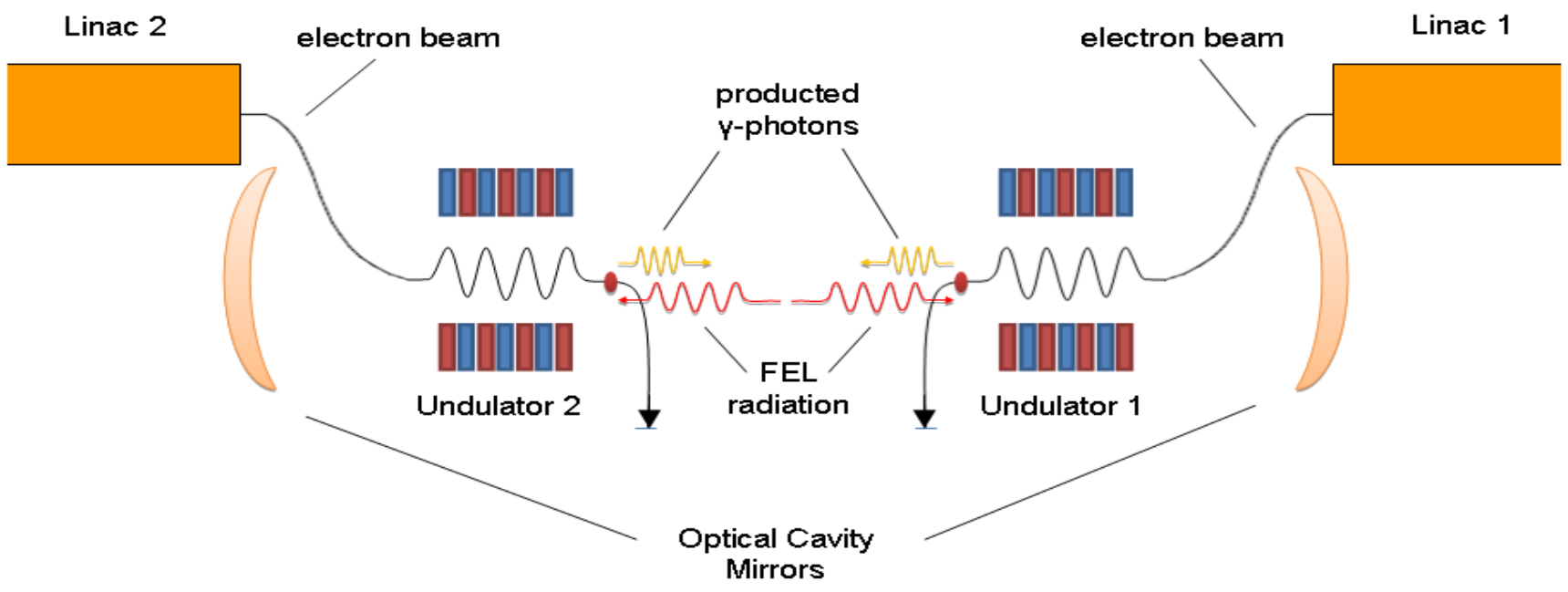

**Figure 3.11:** Double FEL configuration

In this configuration both LINACS are active and are used to produce two identical laser beams, propagating in opposite directions. The FEL intracavity photons are then backscattered, to produce counter-propagating gamma photons of energy below 1 MeV.
Also in this case a particular relevance is played by the macropulse duration, which would allow the possibility of observing a number collisions, sufficient to test the QED predictions of the photon- photon scattering effect ***(see All. 1a for the details of the design).***
Finally, the role of this FEL as laser heater should be considered carefully, especially in connection with the use of bunch compression or velocity bunching or of other operation modes aimed to enhance the peak current. For this option the infra-FEL is used as a ***"Landau damper"*** to produce sufficient additional energy spread, enough to inhibit the on-set of coherent synchrotron radiation contribution, in the bending magnets. The FEL oscillators are inserted between the sections of the LINAC, in a region allowing the injection inside the undulator at low energy, in order to drive the oscillator in the IR region. The beam is then injected in the successive accelerating sections to be brought to larger energies.
The role of ILIO and CUPIDO is not limited only to the first part of the facility, but it can be embedded to the high energy part for the use of its radiation in pump-and-probe experiments (see the concluding section).

### 3.3. FILOS the second undulator section and the associated intracavity backscattering for the production of gamma photons

The system is flexible enough to allow different sol utions, which will be explored in the following. In Fig. 3.12, we have reported a possible option, in which the oscillator operating in the VUV region is used to produce bunching in the e-beam, which is successively injected into the last sections of the undulators tuned at higher harmonics of the oscillator.

The oscillator will be called FILOS (FEL Intracavity Light Oscillator Scattering) to stress its role in the production of intra-cavity backscattered photons.

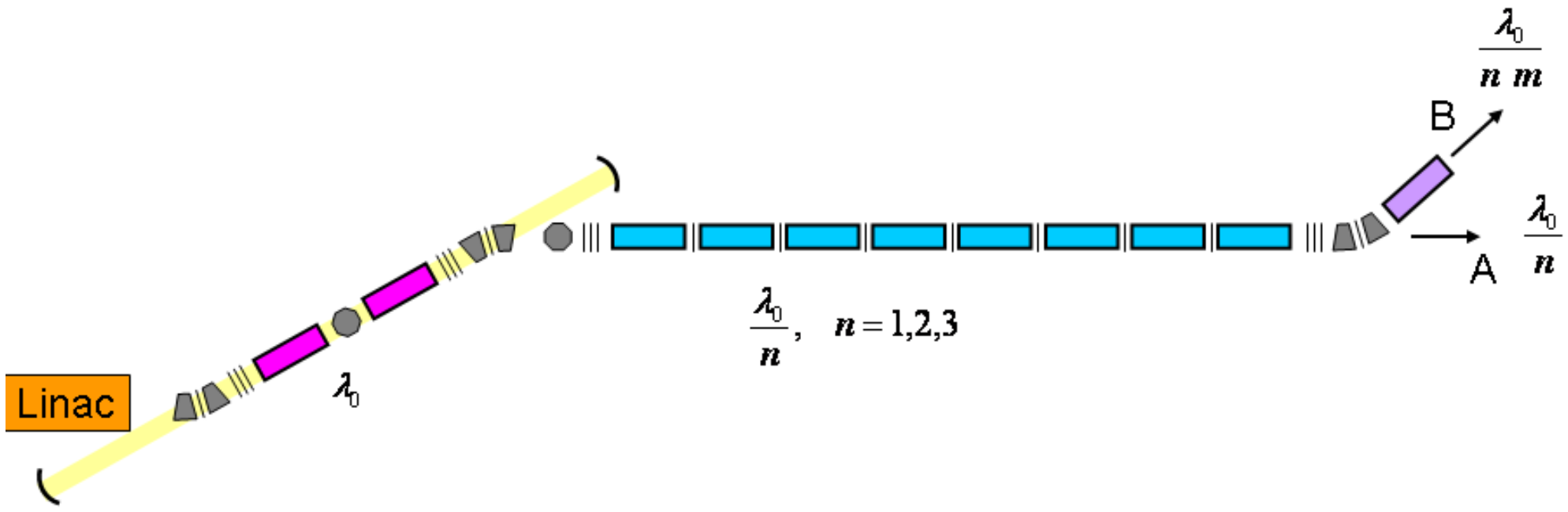

**Figure 3.12:** Undulator Chain at the S-C LINAC output, the first component is provided by an oscillator

The details of the oscillator design are reported in All. 1b, here we report the main characteristics.
The use of the Linac operating at 750 MeV and an undulator with a period of $\lambda_u = 4.4\,cm$ and parameter strength $K \cong 4.643$ allows the emission of radiation in a region around $\lambda \cong 120\,nm$ and higher order harmonics at ( $\lambda_3 \cong 40\,nm, \lambda_5 \cong 25\,nm$ ). The bunching produced by the FEL operation can eventually drive the second section for a pre-bunched SASE operation. The undulators of the second section are in this case tuned by changing the relevant gaps.
The intra-cavity radiation can be backscattered to produce a high flux $(10^9 / s)$ of monochromatic gamma rays with maximum energy of 70 MeV.
The Mo-Si cavity mirrors option may provide a significant step forward for FILOS, the oscillator can, indeed, be operated at $\lambda \cong 13.5\,nm$.
The rather narrow bandwidth of operation of the cavity provides a constraint for the energy of the e-beam at an energy around $2.28\,GeV$ (for the same undulator characteristics as before). In this configuration a significant amount of third and fifth harmonics allows its use in the successive section to get pre-bunched SASE operation at $4.5\,nm, 2.7\,nm$.
It is worth stressing that the fairly long e-beam pulse allows the possibility of splitting the beam, emerging from the oscillator cavity in two beams to be injected into two parallel undulator chains, tuned at the different harmonics of the ***"master –oscillator"***, in order to provide radiation for different users stations.
It should be finally noted that the energy of the backscattered intra-cavity radiation could, in this configuration, be larger than 600 MeV (see All. 1b for the details of the design, oscillator+intracavity backscattering source) and it may provide a fairly interesting source for Nuclear Physics studies.

### 3.4. The SASE Option

In Fig. 3.13, we have reported the SASE configuration, which is quite a natural evolution of SPARC, albeit in a region of the spectrum very much appealing for users.

In this case the undulators of the second section are SPARC like ($\lambda_u = 2.8\,cm$) and the last undulator with $\lambda_u = 1.7\,cm$ can be tuned, by varying K, at different harmonics to produce radiation at shorter wavelength.

If the device is exploited as in the case of Fig. 13 in the mirror-less configuration, with the LINAC at an energy of 750 MeV, coherent (SASE) FEL radiation is obtained at 27.4 nm at the fundamental harmonic, while at the higher harmonics, with the first undulator section used as modulator. The last undulator can be tuned at higher harmonics of the fundamental, and coherent radiation up to 1.83 nm can be obtained.

The operation at full energy of one LINAC (1.5 GeV) would allow SASE radiation in first harmonic up to 6.8 nm, but the number of undulators of the main section should be brought to 14, since for the operation at 3 GeV a significantly longer undulator chain is required (30-35 SPARC like undulators). This number can be reduced according the peak values of the current, which can be obtained. The operating wavelength lies in the sub-nanometer region and depends on the choice of the K values. For SPARC like parameter and K=1, we get an operating wavelength of 0.6 nm, which can be extended to 0.2 nm, if the choice of a segmented undulator is foreseen and the last sections are replaced by a super-conducting undulator with $\lambda_u = 1.cm, K = 1$. The present considerations have been developed for a peak current below 1 kA, further details of the design are reported in All. 2.

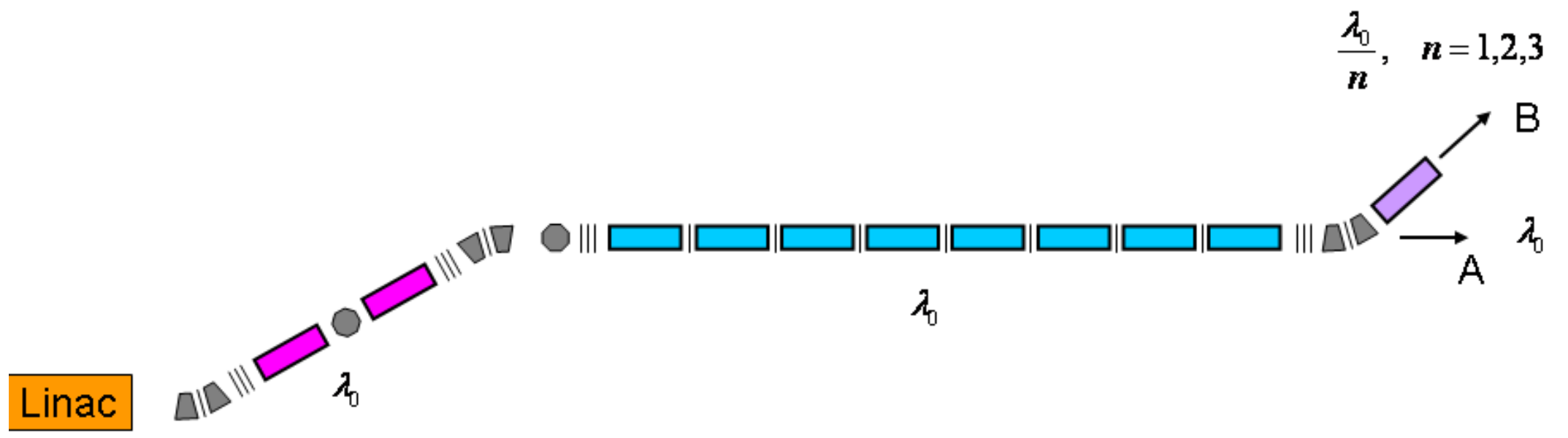

**Figure 3.13:** SASE configuration

### 3.5. The Seeded Operation

The philosophy of the seeded mode is reported in Fig. 3.14, it is not dissimilar from the conceptual point of view developed for the oscillator assisted configuration and allows some margin of flexibility. In this case, however, the maximum operating energy is fixed by the source exploited as seeding. If we consider the 27-th harmonics of the Ti-Sa (26.9 nm), the beam energy is constrained below 1GeV. The tunability could range from 27nm to 1.65 nm. (The details of the design are reported in All. 1,2)

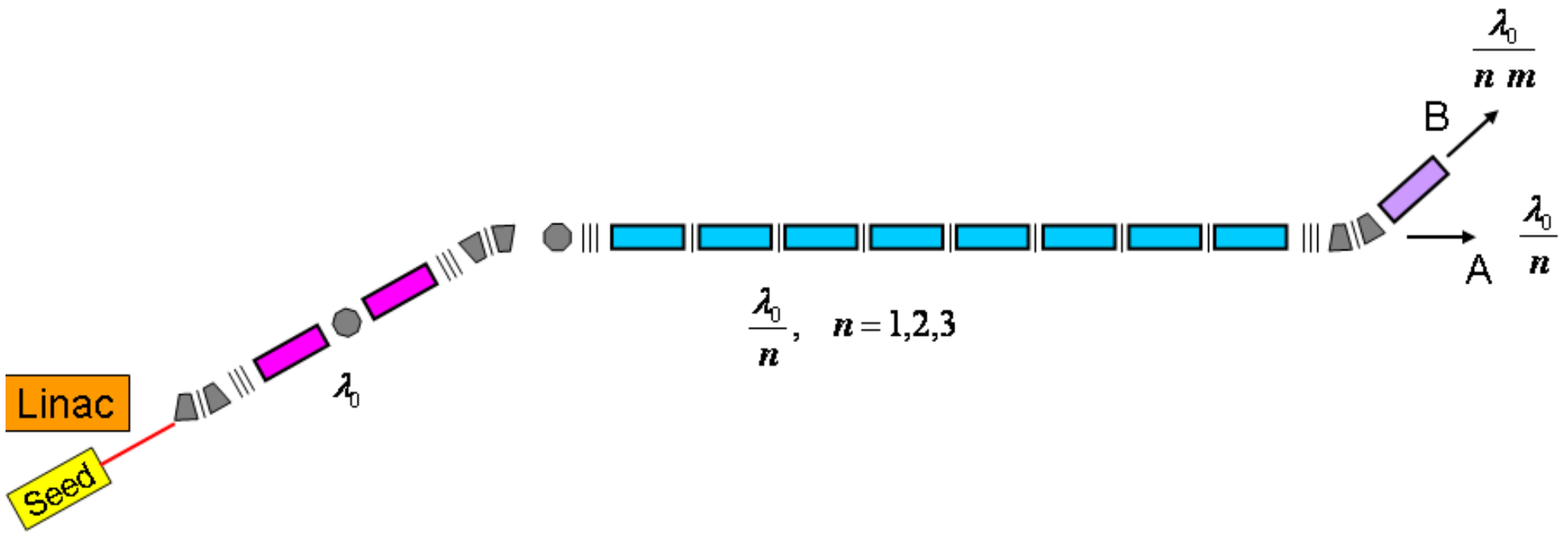

**Figure 3.14:** The seeded mode configuration

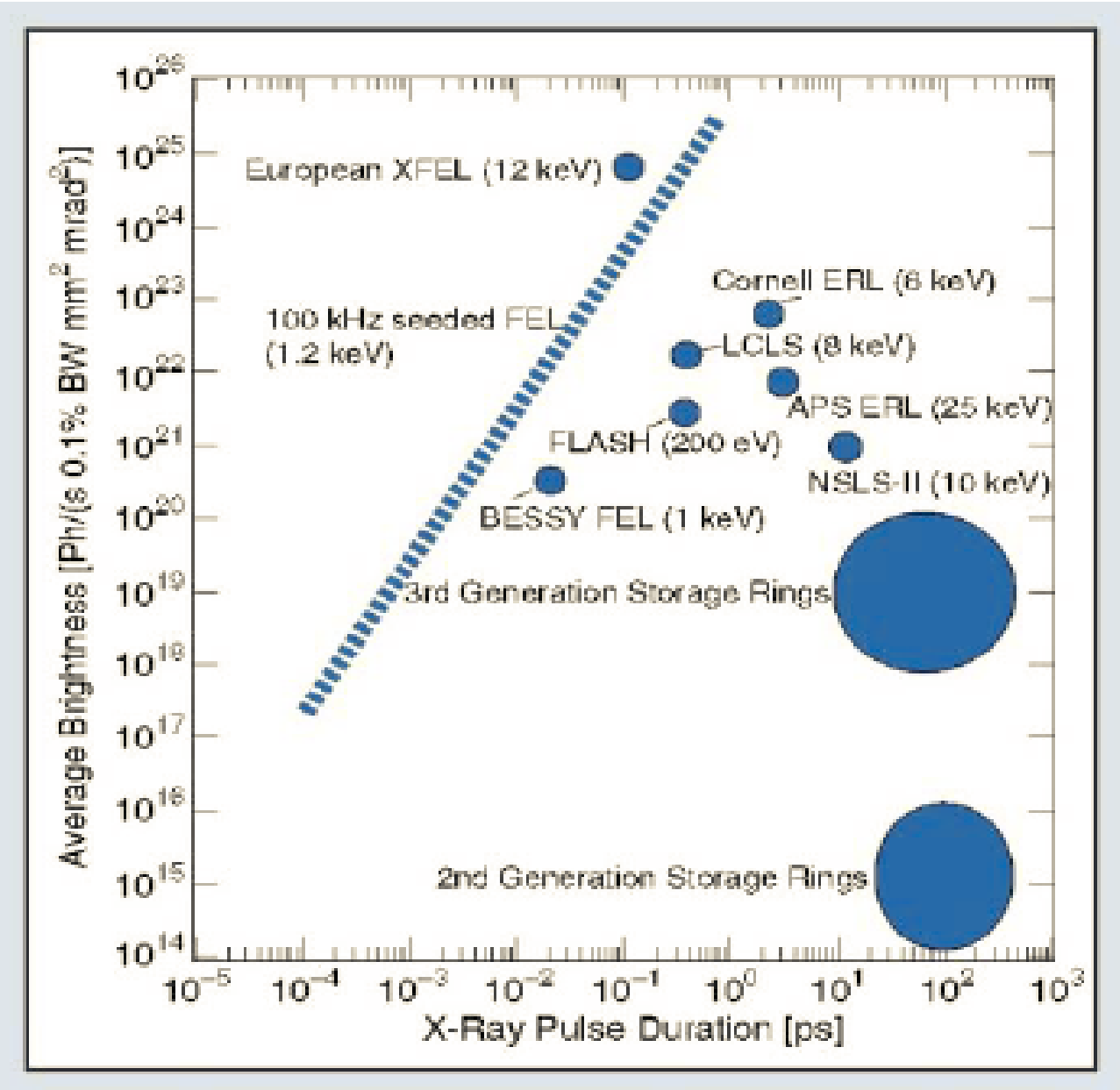

**Figure 3.15:** FEL-Average Brightness and pulse duration

To summarize, we can say that the proposed device should be accordingly capable of providing radiation, from IR to X-Ray, in particular we have:

A) In the first double oscillator ILIO and CUPIDO, with tunable wavelength of $5-20\,\mu m$, the coherent harmonic generation could ensure the emission in the visible range of spectrum, the intracavity backscattering the generation of an intense EUV-X flux of photons.

B) The FILOS oscillator spans in the UV region, the present mirror technology allows operation up to $120\,nm$, higher order harmonics and bunching allows the emission in the region of tens of $nm$.

C) The Use of Mo-Si mirrors allows the generation of radiation at 13.5 nm and in the nm region using the mechanism of harmonics generation.

D) In both B) and C) intra-cavity back-scattered gamma photons (with energies 70 and 600 MeV, respectively) are produced.

E) The successive undulator chain can be designed at further stages of the project in order to produce radiation from tens of nm to a few Angstrom.

In Fig. 3.15 we have reported the various FEL sources characterized in terms of average brightness and pulse duration, IRIDE even though not shifted towards the very hard X-ray region, will be capable of providing bright and short pulses of X-ray bursts in a region of noticeable interest for applications

### 3.6. Concluding remarks

The construction of a device of this type involves the development of different technological components, which, apart from the accelerators, demands for a significant effort. In particular, different types of undulator lines can be developed, including those of superconducting type allowing high field and short period.
The undulators of the main line can be designed to provide radiation with variable polarization and optimized to enhance the emission at higher harmonics. This solution becomes particularly important when the system is operating at high energies to produce radiation in the sub-nanometer region.
The design of the optical cavity of the second oscillator requires particular care specially for the case of the operation with Molybdenum-Silicon mirrors.
The first oscillator may accordingly be exploited to produce radiation through backscattering at 13.5 nm in a pulsed stacked mode and accumulated in the Mo-Si cavity (see Fig. 3.16), which in this way can be tested and can also be exploited to produce a substantive source of E-UV photons.

In the following table are summarize the FEL performances.

| | ILIO - CUPIDO | FILOS | Seeded at 13 *nm* | SASE | | | |
|---|---|---|---|---|---|---|---|
| e.beam energy | 30 MeV | 2.28 GeV | 2.28 GeV | 750 MeV | 1.5 GeV | 3 GeV | 4 GeV |
| Peak current | 50 A | 600 A | 600 A | 750A | 1.5 kA | 3 kA | 4 kA |
| Initial microbunch time length | 6 ps | 1 ps | 1 ps | 1 ps | 1 ps | 1 ps | 1 ps |
| compressed microbunch time length | | 330 fs | 330fs | 200 fs | 200 fs | 200 fs*** | 150 fs*** |
| Charge per Bunch | | 200 pC | 200 pC | 150 pC | 300 pC | 600 pC | 600 pC |
| Micropulse reperition rate | 5 MHz | 5 MHz | | | | | |
| Resonant wavelength (1th harmonic) | 10μm<br>0.124 eV | 13.5nm<br>91.84 eV | | ≈ 16 nm<br>77.5 eV | ≈ 4 nm<br>310 eV | ≈ 1 nm<br>1.24 keV | ≈ 0.3 nm<br>3.76 keV |
| Resonant wavelength | | | | ≈ 5.3 nm | ≈ 1.3 nm | ≈ 0.3 nm | ≈ 1 Å |
| (3th harmonic) | | | | 234 eV | 0.95 keV | 3.76 keV | 12.4 keV |
| Peak Power | 5 MW | 150 MW | | | | | |
| Undulator length | 2 m | 18 m | 18 m | 21 m | 31 m | 34 m | 58 m |
| Undulator period $\lambda_u$ | 2.8 cm | 4.4 cm | 2.8 cm - 1.4 cm | 2.8 cm | 2.8 cm | 2.8 cm | 2.8 cm |
| Undulator Parameter $K$ | 1 | 4.736 | | 1.71 | 1.71 | 1.71 | 1. |
| Cavity Length $L_c$ | 4 m | 30 m | | | | | |
| **Cost estimate** | **2 M€** | **5-10 M€** | **20 M€** | | | | |
| | | | | | | | |
| **Note** | Energy and Peak current are the constrains for the different FEL operations. The green cells represent only indicative values for the accelerator operation modes. | | | | | | |

The FILOS configuration can be also exploited in the seeded mode too. In this hypothesis the radiation from the oscillator should be injected in the undulator section using grazing incident mirrors.
The use of the radiation from the low energy oscillator in two colors or pump-and-probe configuration is shown in Fig. 3.17.

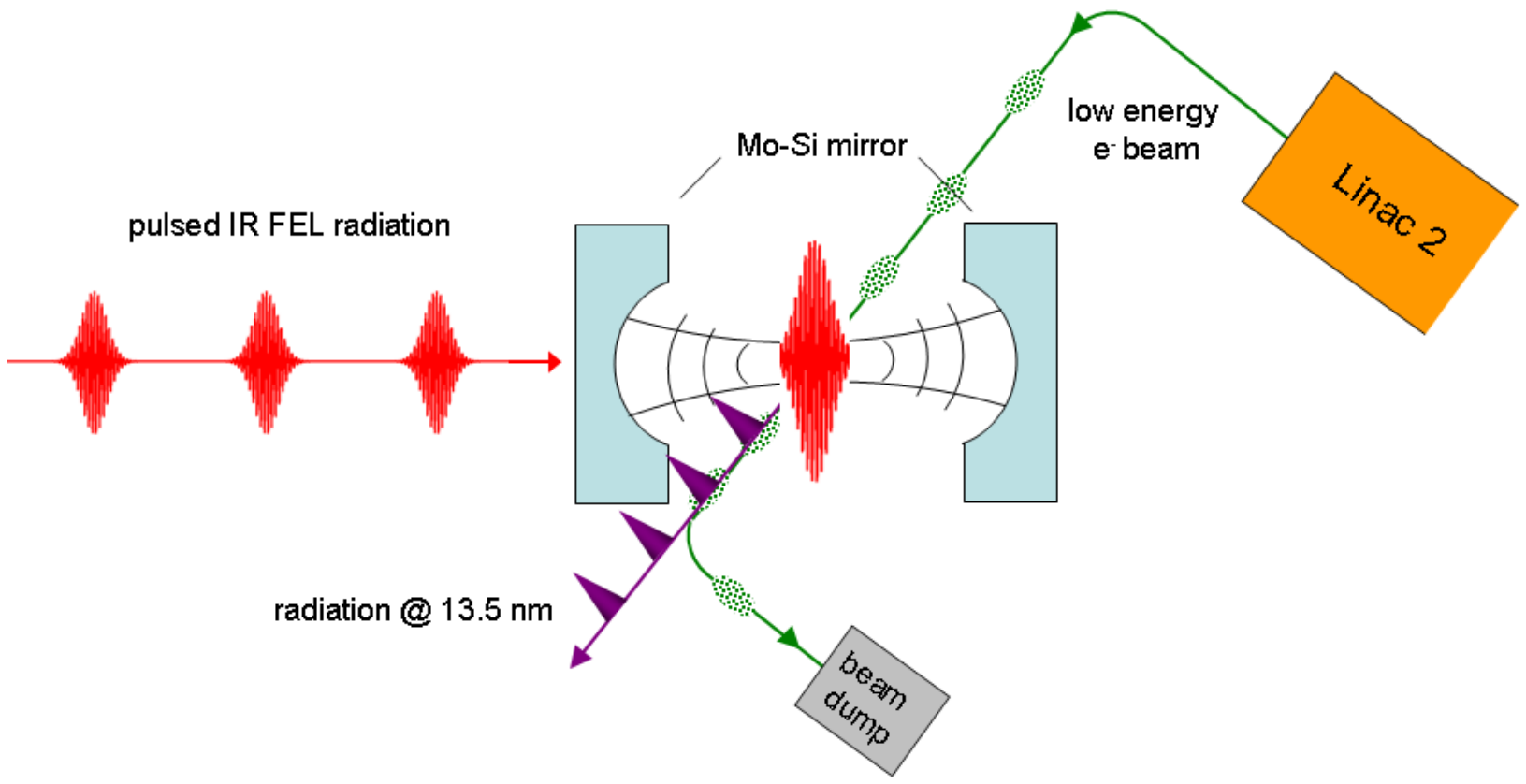

**Figure 3.16:** Pulse stacking cavity

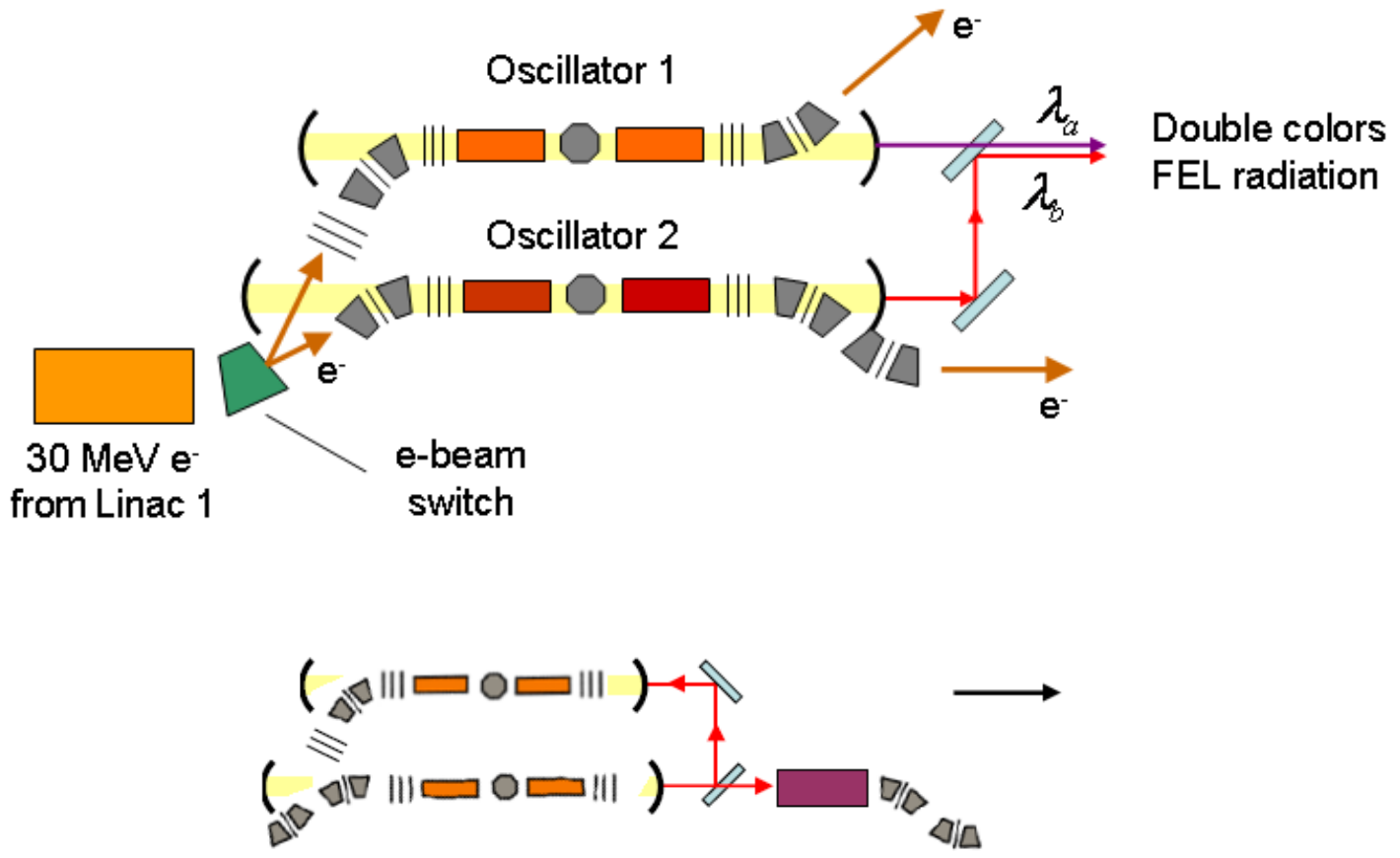

**Figure 3.17:** (a) Two color configuration of low energy FEL operation, b) Pump-and-probe

As already remarked, FILOS is aimed at producing intra-cavity gamma photons to be exploited as a probe in nuclear physics experiments. However, they can also be a tool to produce polarized positron beams to be eventually injected into a damping ring. This possible scheme of operation is not dissimilar from the CLIC proposal and is shown in Fig. 3.18. The photons from FILOS are send on a target and converted into electrons and positrons, the positrons are then injected into a damping ring.

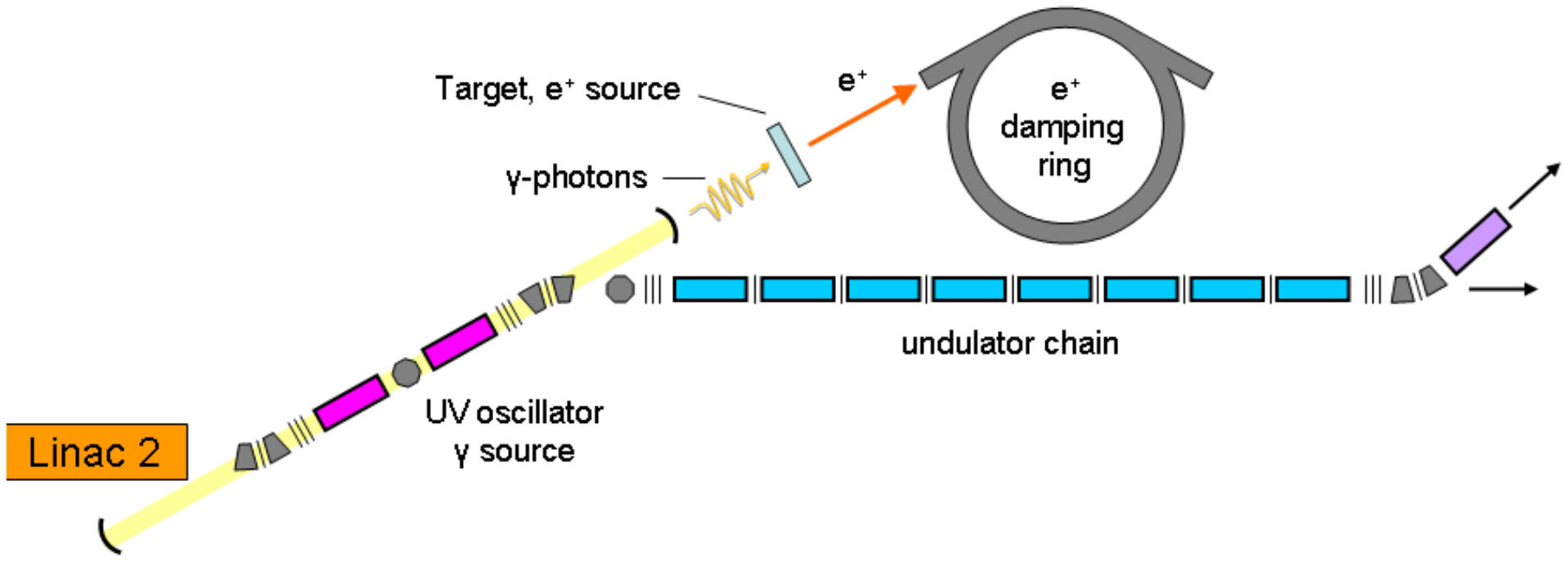

**Figure 3.18:** FILOS gamma photons on a target and production of electrons on a target and relevant injection into a damping ring

The advantage of this configuration stems from the fact that the backscattering occurs in the cavity at a very high repetition rate. The rate of produced photons can in turn be modulated by "chopping" the cavity via a suitable cavity damping consisting in a periodic mismatching of the cavity mirrors.

(This aspect of the problem has not been properly treated and is a mere possibility, corroborated by the analysis already developed by the CLIC team design. No simulation has been done for this specific case). The design of the damping ring demands for precise statement on the required performances in terms of emittances and of damping times. A very preliminary analysis to define the main parameters and costs can be afforded in reasonably short times.

### 3.7. Cost analysis

The cost of the FEL device, as described in this report, is dominated by that of the undulators. We have considered essentially SPARC like type, and, therefore, we can estimate a financial effort of 400.000 Euros per undulator, inclusive of control and diagnostics. If we assume a total of 45 undulators a cost of 18 M Euros can be safely considered. A further amount of 7 M Euros for the realization of the optical cavities, relevant control and diagnostics and electron beam transport brings the financial effort to 25 M Euros.

A practical formula to estimate the costs of the undulators is the following

$$C_U[\varepsilon] \cong 4 \cdot 10^5 L_s[m]$$

The contingency cannot be calculated if not correlated to the uncertainties on the e-beam qualities. The inhomogeneous broadenings provide an increase of the saturation length and, for example, the energy spread induces a variation of the saturation length reported in Fig. 19, quantified according to the relationship

$$L_s[m] \cong L_s^0[m]\left(1 + 0.185\frac{\sqrt{3}}{2}\tilde{\mu}_\varepsilon^2\right),$$

$$\tilde{\mu}_\varepsilon = 2\frac{\sigma_\varepsilon}{\rho}$$

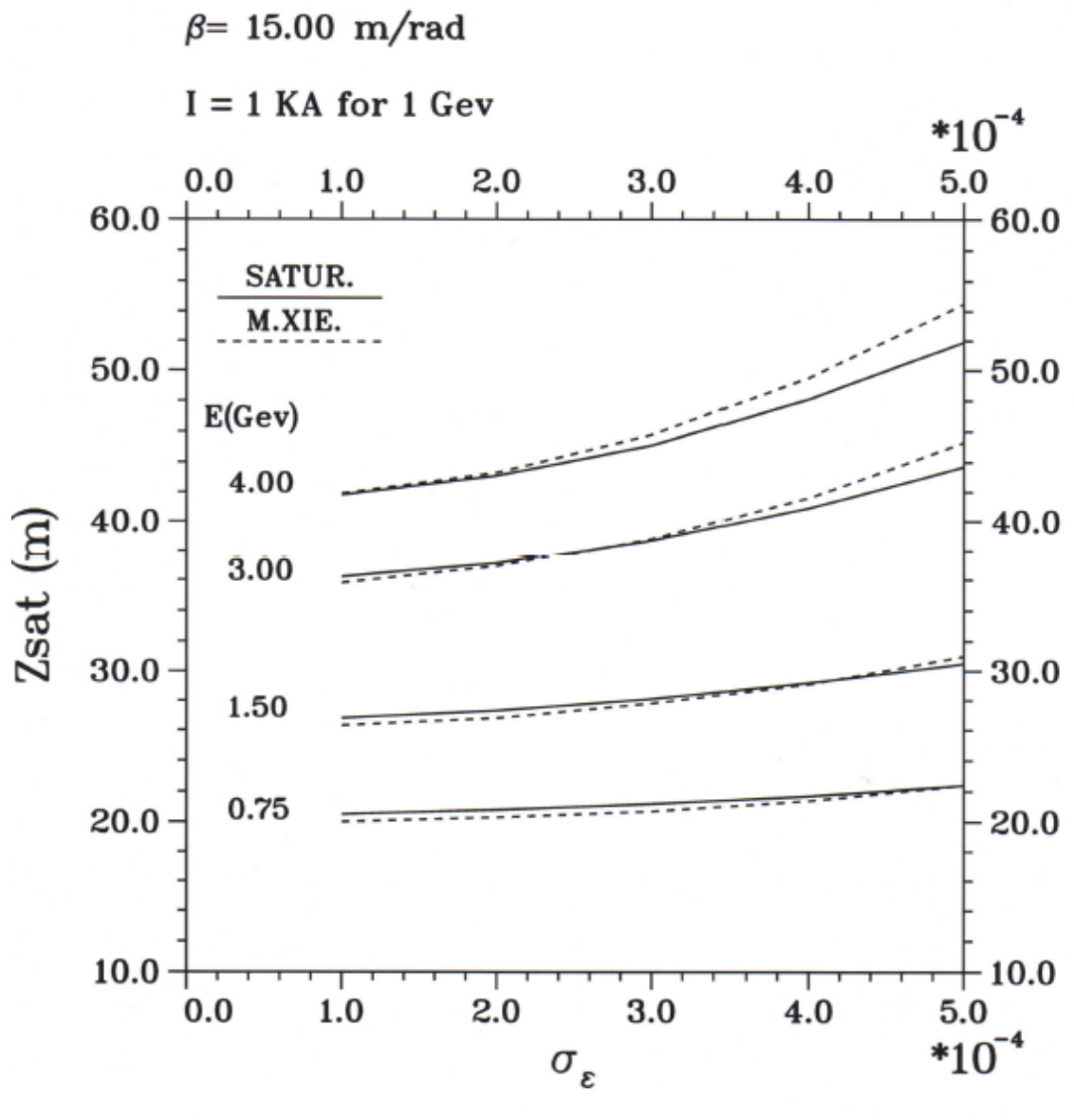

**Figure 19:** Saturation Length Vs. the relative energy spread for different energies

An increase of the energy spread from $10^{-4}$ to $5 \cdot 10^{-4}$ might be responsible for an increase of the saturation length of 6 m (at 3 GeV), which amounts to an increase of costs of about 6- million euros.

# 4. THE SCIENTIFIC CASE WITH THE FREE ELECTRON LASER SOURCE

## 4.1. Introduction

Photons are the most important probe to investigate our environment. From radio frequencies to hard X-ray photons are used since many years to get both electronic and structural information on virtually any materials, form biology to condensed matter systems. Among the different types of sources, storage rings to emit Synchrotron Radiation (SR) have a special place because of the capability to produce radiation with very high flux and brightness in an energy range from infrared to hard X-ray. The development of storage rings with special magnetic components over the past thirty years has led to third generation machines, especially designed for SR research, with the peculiarity to have radiation with unprecedented degree of brilliance, defined as photons/(sec*mm$^2$*mrad$^2$*bandwidth ($\Delta E/E=0.1\%$)).

This important property makes possible to routinely perform experiments which were previously only conceivable, like X-ray diffraction (XRD) from surfaces, huge biological molecules or magnetically ordered structures. In parallel, new experimental techniques like X-ray absorption spectroscopy, which allows measuring the local structural and electronic properties of matter or inelastic X-ray scattering, to probe the lattice dynamics in solids, were developed. Today, in the SR facilities thousands of researchers, coming from many different countries perform their experiments, contributing to the advance of scientific knowledge in many different fields and bridging the traditional dichotomy between applications and fundamental science.

Last progresses in the physics of the linear accelerators (LINAC) opened the possibility to a new jump in photon sources quality especially in the X-ray energy domain. The LINAC based sources are called Free Electron Laser (FEL). These machines are sources of coherent radiation presenting similar optical properties as conventional lasers. The main difference is in the lasing medium, being in this case an electron beam moving through a magnetic structure made by a very long undulator. This magnetic structure causes transverse electron oscillations that produce electromagnetic radiation. The coupling of this radiation with the electron transverse velocity produces micro-bunching of the electron beam depending of the radiation wavelength and leads to coherent emission. The X-ray FEL has the possibility to generate X-ray beams with a peak brilliance of several order of magnitude higher that the third generation SR sources for both the spontaneous and coherent emission. It also offers an extremely short pulse length, typically less than $10^{-13}$ s (femtosecond (fs) time domain), with some degree of tunability and a high degree of either linear or circular polarization.

The advent of such short and powerful pulses of fs laser sources has disclosed the opportunity to make real time experiments in the fs time domain using many different techniques, to study the interaction of radiation with matter in a non-linear regime and to use nano-crystals in protein-crystallography. Actually there are four operating FEL facilities (Flash, Fermi, LCLS, SACLA) and three in construction (XFEL, LCLSII, SwissFel), covering the spectral range from soft to hard X-ray. The first experiments performed in the existing facilities begin to enlighten the strong potentialities in many different fields and at the same time several critical points that should be resolved to fully exploit the possibilities of these new machines.

The FEL@IRIDE project is a source of coherent X-rays, covering the range of wavelengths (λ) ranging from 0.5 to 4 nm at fundamental harmonics, depending on the electron beam energy, and it will be also able to reach the Ångström region using the third and fifth harmonics. In some way it covers a radiation region complementary to those of other existing or in construction facilities, and will be provided also with an ancillary equipment to produce radiation down to the THz region. FEL@IRIDE has a reasonable wavelengths overlap to satisfy users in many different fields of science and to ensure at the same time a beneficial level of competiveness. The IRIDE project is planned as an evolutionary research infrastructure exploiting the large site available at the Tor Vergata University campus, an area about 1.5 km long. We strongly support establishing a close collaboration between INFN, the University of Rome Tor Vergata and major national research institutes (CNR, ENEA and CNISM, the latter representing the Condensed Matter network of Italian Universities) in order to guarantee strong ties with the user community. The outcome of the FEL@IRIDE project should be a user facility providing beam-time on the basis of a peer reviewed proposals systems in the tradition of the international synchrotron radiation and free electron laser community.

In the following table we summarize the main characteristic FEL@IRIDE SASE emission calculated for three possible scenarios of electron beam energy. The pulse duration will be in the range 200-70 fs.

**1.5 Gev electron beam energy**

| | Fundamental | 3° harmonic | 5° harmonic |
|---|---|---|---|
| λ(nm/KeV) | 4/0.413 | 1.33/1.23 | 0.8/2.07 |
| peak flux (n/s/- 0.1%BW) | $2.7*10^{26}$ | $2.5*10^{24}$ | $1.9*10^{23}$ |
| Peak brilliance | $1.56*10^{30}$ | $1.4*10^{28}$ | $1.1*10^{27}$ |
| photon/bunch | $5.94*10^{13}$ | $5.5*10^{11}$ | $4.18*10^{10}$ |

**3.0 Gev electron beam energy**

| | Fundamental | 3° harmonic | 5° harmonic |
|---|---|---|---|
| λ(nm/KeV) | 1/1.24 | 0.3/3.72 | 0.2/6.2 |
| peak flux (n/s/- 0.1%BW) | $4.6*10^{25}$ | $4.1*10^{23}$ | $3.4*10^{22}$ |
| Peak brilliance | $6.4*10^{31}$ | $5.7*10^{29}$ | $4.7*10^{28}$ |
| photon/bunch | $1.01*10^{13}$ | $9.02*10^{10}$ | $7.48*10^{9}$ |

**4.0 Gev electron beam energy**

| | Fundamental | 3° harmonic | 5° harmonic |
|---|---|---|---|
| λ(nm/KeV) | 0.563/2.2 | 0.188/6.5 | 0.113/10.9 |
| peak flux (n/s/- 0.1%BW) | $1.2*10^{25}$ | $5.9*10^{22}$ | $2.8*10^{21}$ |
| Peak Brilliance | $1.92*10^{31}$ | $1.8*10^{29}$ | $1.2*10^{28}$ |
| photon/bunch | $2.1*10^{12}$ | $1.06*10^{10}$ | $5.0*10^{8}$ |

At the same time the FEL@IRIDE will provide radiation with the following and unprecedented characteristics

- *Self-seeding: narrow bandwidth, wavelength stability, higher brightness and energy tunability*
- *Polarization control: tunable linear and circular polarization*
- *Two color pulses: simultaneous delivery of independent wavelength pulses*
- *Delayed pulses: independent delay of two pulses up to a few ps*

An X-ray source with these specifications will allow the investigation of photon-matter interaction in a new regime, and will become an high competitive facility in the worldwide growing FEL framework.

This Scientific Case reports experimental proposals ranging from time dependent spectroscopies in condensed matter to imaging for biological applications. It has been organized in three paragraphs: atomic and molecular process, biological applications, time-dependent experiments plus a paragraph with some very general considerations about beam lines. Each of the paragraphs contains the contribution of several groups working mainly in the Italian research institutions and already strongly involved in FEL experiments mainly in the European facilities.

Actually each paragraph is susceptible of revisions during the time due to the new advances coming from the recent experiments at the existing FEL facilities. It is also clear that it is impossible to evaluate all the possible applications of this new source, also due to the unprecedented extreme beam conditions, but these proposals represent a well-defined road map for the future experiments at FEL@IRIDE.

### 4.2. Atomic and molecular processes

An X-ray source with these specifications of the FEL@IRIDE will allow the investigation of photon-matter interaction in a new regime, where inner shell electrons are the dominant “mediator” of the interaction, as well as to drive chemical transformations by controlled optical or infrared pulses and understanding the atomic and electronic transformation by means of x-ray spectroscopies. Such a source will offer opportunities and challenges to the atomic physics communities

The opportunities include the application of established methods and techniques to tackle established topics with unprecedented accuracy and to investigate low density targets (clusters, radical and metastable species), due to the high flux of the new source. The challenges are represented by the possibility to perform two (multi)-photon and "pump-probe" experiments on core orbitals making use of the characteristics of the new source. To achieve this ad-hoc methods and instrumentation have to be developed. In the following some examples are reported.

### *4.2.1. Non-linear processes*

Non-linear spectroscopy is a standard technique with the high-power visible and UV lasers. Multiphoton absorption spectra (single colour, 3-4 photons) have been reported in atoms and molecules [1,2]. The number of photons per pulse in the FEL sources will make the investigation of two-photon processes possible in the soft X-ray region. The Keldysh parameter $\gamma$ describes the two regimes of high field ionization: for values of $\gamma \ll 1$, ionization is due to barrier suppression and tunneling, while for values $\gamma > 1$, multiphoton absorption is the mechanism of emission. From another point of view $\gamma < 1$ corresponds to a system whose behaviour is dominated by the applied electric field, and $\gamma > 1$ corresponds to a system dominated by the nuclear electric field. While high power and visible lasers have mainly operated in the tunnelling regime, IRIDE will offer the possibility to explore the physics of multiphoton ionization in the x-ray region.

#### *4.2.1.1. Two-photon inner shell excitation in atoms*

The observable excited states of atoms by optical absorption are governed by selection rules. For single photon absorption, dipole selection rules apply while for two photon absorption, monopole and quadrupole rules apply. Thus different excited states are probed for these two cases. For example in the rare gases the np and nf states can be accessed by two-photon excitation of an outermost (n-1)p electron. Two-photon absorption is a non-linear process and thus requires a high power laser to function. This will be the case at new FEL sources, where at least $10^{13}$ photons in pulse of few fs are foreseen.

Considering the available range of wavelength, the inner shell excitations from the Ne 1s (870 eV) and Ar 1s (about 3200 eV) states appear to be the most suited candidates for the first studies. Recently some calculations have been done in the case of the Ne 1s excited states by Novikov and Hopersky [3]. The calculations of these authors shown in figure 1 clearly display a) the suppression of the dipole $1s^{-1}np$ transitions, b) an enhancement of the two-photon absorption cross section of at least one order of magnitude in correspondence of the $1s^{-1}$ns and nd transitions and c) the sensitivity to the polarization state of the incident radiation. Using as a reference the values of the cross section calculated for those cases, the typical conditions of target density of a gas phase experiment and the expected 100 fs duration of the pulses one can estimate about a few thousand events per laser pulse. This appears to make the experiment feasible. However the main experimental issue will be the separation of signal produced by two-photon inner shell excited states from the direct ionization of the valence shell by a single photon, which is expected to produce a few orders of magnitude higher signal. This leads to the exclusion of conventional methods based on the measurement of the total ion or electron yield. The detection of energy selected photoelectrons is therefore a most suited approach. The addition of angular selectivity in the detection will also help in disentangling the emission due to two-photon processes from the that due to second order radiation. Indeed in the last case the electron angular distribution will be determined by dipole selection rules and characterized by the angular asymmetry parameter $\beta_2$, while in the case of two-photon processes different angular distributions characterized by two asymmetry parameters $\beta_2$ and $\beta_4$ are expected. For example the 1-photon ionization of a s shell of a rare gas by linearly polarized radiation will lead to a vanishing photoelectron yield at 90° with respect to the direction of the polarization of the incident radiation. Thus a detector placed in that direction would be a sensitive probe of non-linear processes.

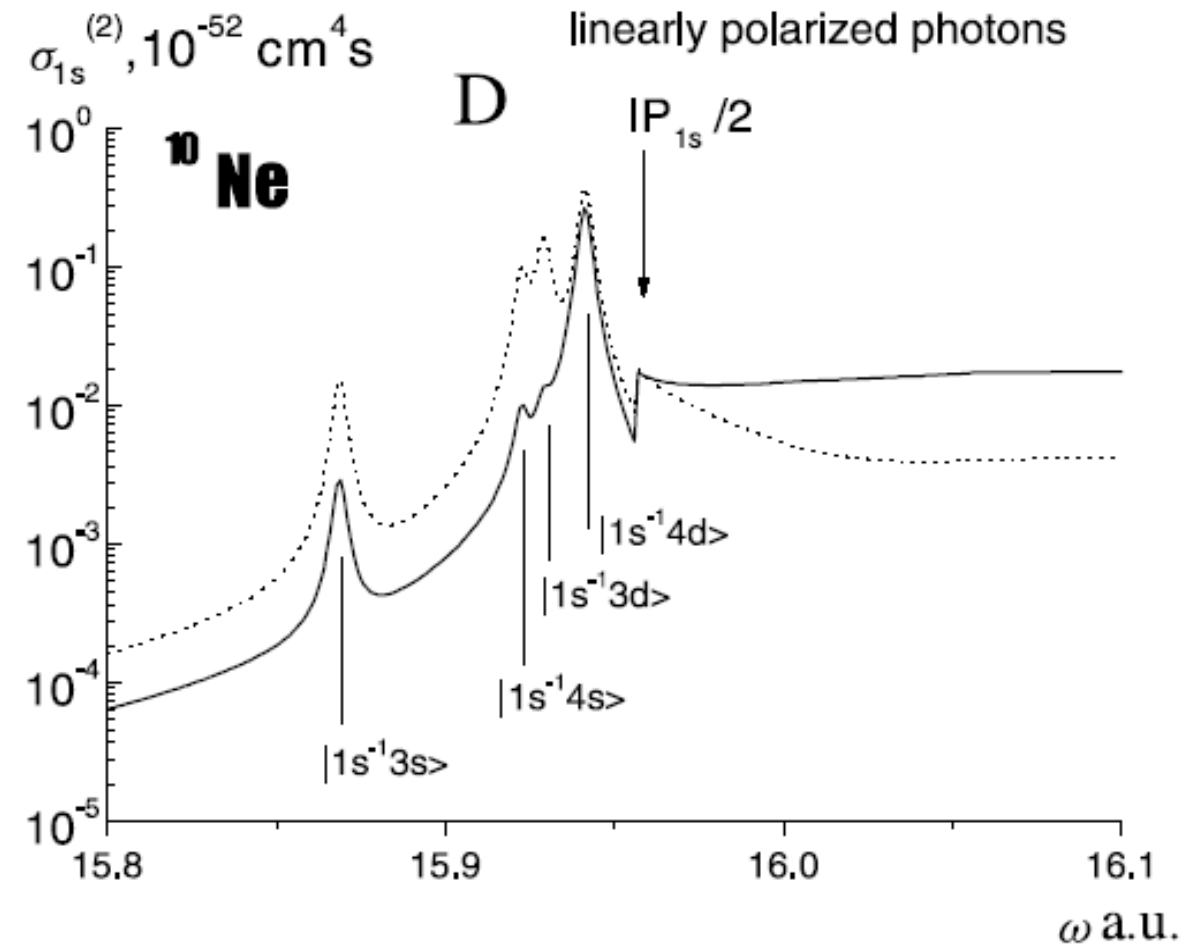

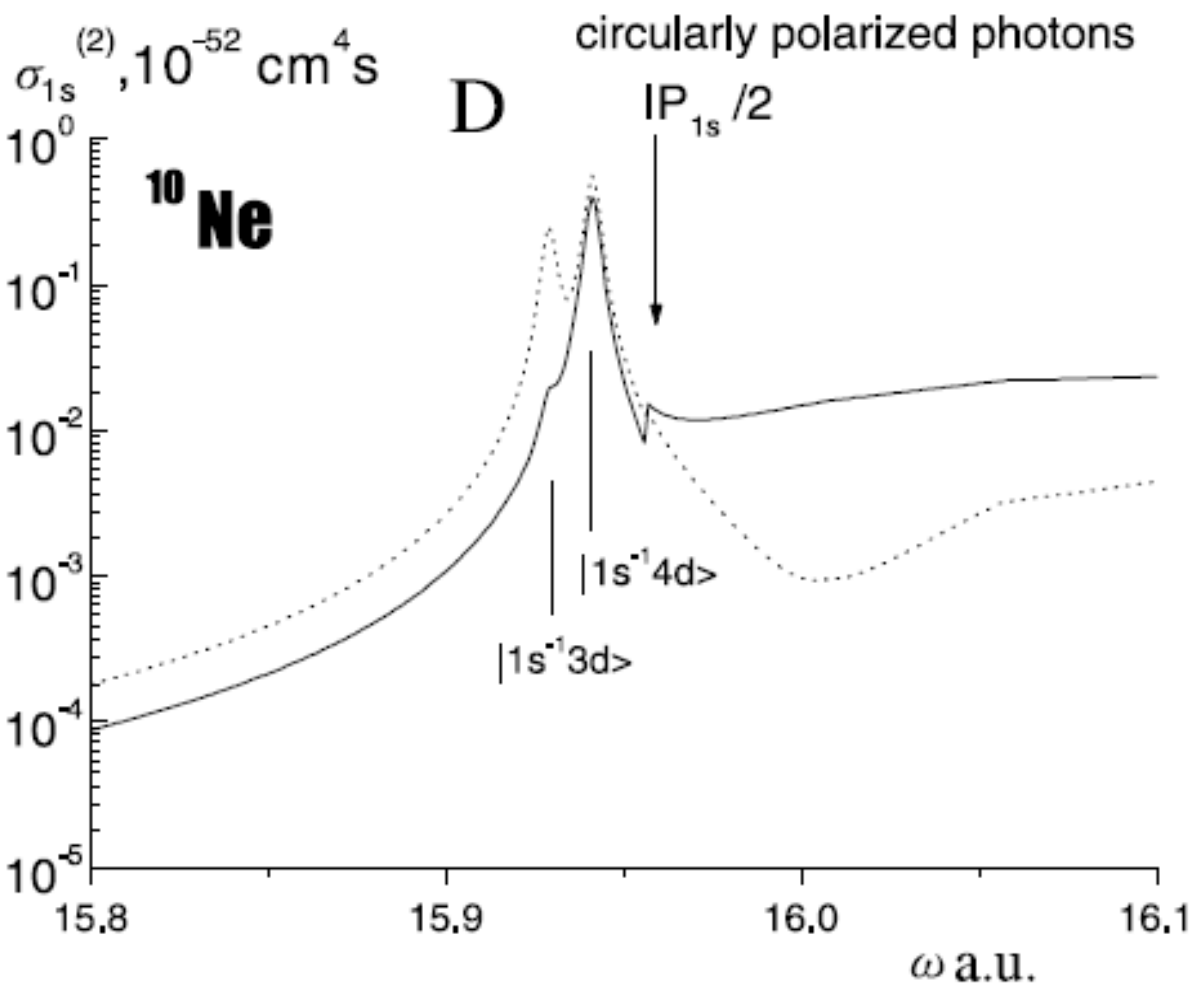

**Figure 1.** The two-photon cross section $\sigma^{(2)}$ of the process of excitation/ionization of the Ne 1s shell by two linearly (top panel) and circularly (bottom panel) polarized photons calculated taking into account the effects of relaxation of the atomic residue in the field of the creating vacancies with (full curve) and without accounting for this effect (dotted curve). In the figure the discrete final states are indicated.

Due to the limited energy tunability of the existing FEL facilities only one example of two-photon excitation has been reported [4] at Fermi recently, but some two-photon ionization [5,6] on He, Ne and $N_2$ have been achieved. These results prove the feasibility of the proposed experiments, once FELs will allow an easier tuning of the incident energy.

#### *4.2.1.2. Double core hole spectroscopy*

Conventional X-ray photoelectron spectroscopy based on one-photon ionization is one of the most used spectroscopic techniques to investigate the electronic structure and the chemical environment of a sample. More than two decades ago [7], Cederbaum et al. investigated double core-hole states in molecules such as $C_2H_2$, $C_2H_4$, $C_2H_6$, C6H6, $SiH_4$, and $SiF_4$. They found that the double core-hole states, especially with two holes at different sites, are very sensitive to the chemical environment of the two holes. The one-photon ionization rates to the two-site double-hole states are vanishing. This hampers the access of the double core-hole states by conventional X-ray photoelectron spectroscopy. These double-core states are accessed very efficiently by sequential X-ray two-photon absorption, leading to X-ray two-photon photoelectron spectroscopy (XTPPS), as demonstrated by recent work at LCLS [8]. Inspired by the advent of the X-ray FEL at LCLS, Santra et al. [9]. demonstrated theoretically that two-site double core-hole states can indeed be probed by XTPPS. In the proof-of-principle simulations, they used the organic molecule para-aminophenol, which contains one O, one N, and six C atoms, and assumed that the molecule is irradiated by a 1.2-fs, 1 keV X-ray pulse containing 1.0 $10^{11}$photons, focused to a width of 1 μm. For XTPPS, it is very helpful to have such a pulse significantly shorter than the core-hole lifetimes, because it guarantees that the second x-ray photon is absorbed by the singly-ionized core-hole state as prepared by the first X-ray photon absorbed by the molecule, before any Auger decay occurs.

The energy relations between electron kinetic energies and the single and double ionization potentials, SIP(Sj) and DIP(Si,Sj), for the Sj single core-hole and SiSj double core-hole states, respectively, can be summarized as follows

$$\Delta E = KE(Sj) - KE(Si,Sj) = DIP(Si,Sj) - [SIP(Si) + SIP(Sj)],$$

where KE(Sj) and KE(Si,Sj) are the kinetic energies of electrons ejected via the ionization leading to the Sj single core hole state and via the ionization from the Si single core-hole state to the SiSj double core-hole state. Si and Sj can be at one site (i=j) or at two sites (i≠j). The values of $\Delta E_1$ = DIP(S1,S1) – [SIP(S1) + SIP(S1)] for the one-site double core-hole states depends on the potential, V, experienced by the two electrons in the same atom and the relaxation energy [10]. The dominant term V is roughly proportional to Z, thus it is very important to find experimentally the deviation of $\Delta E_1$ from V, because the deviation includes the direct information about the change of the electron density via core-hole creation, which is very sensitive to chemical environments of the double core holes.

The values of $\Delta E_2$ = DIP(S1,S2) – [SIP(S1) + SIP(S2)] for the two-site double core-hole states can be interpreted as the difference of the repulsion energy $V \cong 1/R_{12}$(a.u.) between two holes located at atom 1 and atom 2 minus the inter-core relaxation energy [10]. Thus, knowing the distance $R_{12}$ between atom 1 and atom 2, we can experimentally determine inter-core relaxation energy via XTPPS. Studies of simple organic molecules have been undertaken as for the C,O and N atoms at LCLS. The energy range of IRIDE will open the possibility to apply XTPPS to systems with higher Z and provide new information on inter- and intra-atomic correlations.

### *4.2.2. Dissociation and isomerisation of polyatomic molecules*

#### *4.2.2.1. Dissociation due to multi-photon ionization*

A topic which attracted a lot of interest in the strong field community is whether in the case of double/multiple ionization the electrons escape from the target “sequentially” or “non-sequentially”, i.e. if each electron absorbs photons independently, or one electron absorbs all of energy from the field and then internally shares it with the other electrons via electron-electron correlation. In the molecular case our interest is the investigation of the role that the mechanism of the double/multiple ionization plays in the dissociation of a molecule. In a molecule the HOMO electronic state determines the characteristic of the bond and of the Potential Energy Surface (PES) experienced by the nuclei. The effect of the external laser field is to modify the PESs of both bonding and dissociating states, affecting the motion of the nuclei. It is known that each cation/dication state correlates with one or more dissociating states. Thus different fragmentation channels can be followed, depending on the way the PESs are modified by the laser field. Among the studies on single and double ionization of polyatomic molecules the work of Cornaggia et al. [11]. regarding $CO_2$, $C_2H_2$ (acetylene) and $C_3H_4$ (propyne) molecules is a milestone: it investigates the correlation between the single and double ionization rates as a function of the peak intensity of the laser. They established the regime of sequential (S) or non - sequential (NS) ionisation comparing the ion yield with the different theoretical models. The FEL allows to extend these investigations to shorter wavelengths and to compare one and two-photon processes directly in the same experiment. Varying the intensity regime ($10^{13}$ W/cm$^2$), and scanning the energy range one can follow the double/multiple ionization and dissociation of these molecules from a regime dominated by the NS to a regime dominated by the S mechanism. The crossing from the NS to the S regimes will modify the population of the final ionic states and the path on the PES leading to the dissociation. This kind of study may explain why several dissociation channels identified with other techniques (synchrotron radiation, IR low intensity lasers) have not been recorded [11].

##### *4.2.2.2. Isomerisation of polyatomic molecules*

Isomerisation, induced by light absorption, corresponds to changes of the nuclear geometry which results in a reorganization of the electrons, thus to an interwined motion of nuclei and electrons. In order to perform a certain function the motion has to be directed into a particular channel [12]. This for example plays a key role in molecular machines that transform light energy in chemical energy which can be stored and transformed more easily. Time resolving the structure during the ultrafast photo-isomerization will have a tremendous effect on our understanding and potentially control of direct chemical conversion. An X-Ray FEL with high intensity and short pulses will allow the combination of diffraction studies of these molecules with time resolved spectroscopic studies of species both isolated in the gas and in liquid phases.

As a benchmark study in the gas phase one may consider acetylene molecule Indeed the isomerization time in acetylene dication (the time needed for the proton to migrate from one to the other end of the molecule) has been estimated to be around 60 fs [13]. The investigation of the isomerization in small molecules (such as acetylene and the propyne) where the dication state is populated by NS or S ionization. Indeed, beside the modification of the dication energy level, the new shape of the PES might speed up (or prevent!) the isomerization to take place.

#### *4.2.3. "Pump-probe" experiments*

The ideal study of a chemical process involves the transfer of a known amount of energy on a specific molecular site and then the observation of the evolution of the target towards the final products. Typical vibrational periods are in the range 20-100 fs, thus spectroscopy with fs time resolution allows the observation of the evolution of the electronic structure of a molecule due to the motion of the nuclei. In the “pump-probe” experiments the time evolution of a dissociation process can be followed, obtaining detailed information on the transition states and the nature of the barriers determining the pathways for the chemical reaction. These experiments using a X-FEL source in combination with external synchronised lasers will extend to higher photon energy the pioneering work that lead A.H. Zewail to the Nobel Prize for femtochemistry studies [14]. The advantage of using X-ray is that the inner shell of a particular atom of the molecule can be selected; adding to the previous studies localisation and “site selectivity”.

An interesting and valuable application is represented for example by the study of the photostability of biomolecules, DNA, to UV exposure. It is known all the basic molecules of the DNA alphabets (thymine, adenine, cytosine and guanine) strongly absorb in the region of the 250 nm. This UV radiation contains sufficient energy to damage DNA by either breaking the DNA double strand or by fragmenting the nucleobases. To avoid permanent damage nature has provided nucleobases with ultrafast mechanisms to redistribute the excess energy and transfer population from the excited states down into the ground state. The dynamic involved is non-adiabatic and the relaxation from the excited state to the ground state may be possible with a violation of the Born-Oppenheimer approximation. Our understanding of these mechanisms is based on extensive theoretical studies, sometime contradictory, and ultrafast absorption and valence shell ion/phototelectron spectroscopy. The X-ray from the new FEL sources provides the opportunity to learn about nature’s photoprotection mechanisms. Proposals to use LCLS pulse for this purpose have been recently presented and the first experiments performed in 2012. A new seeded facility which provides high intensity, monochromatic pulses with well determined and stable spectrum shot-to-shot guarantees a better and more complete insight in the process and possibly lead to the understanding of how different nucleobases were selected in the early bionic ages.

#### *4.2.4. Spectroscopy of flying proteins*

Proteins, long linear chains composed by L-α-amino acids joined together by peptide bonds, rule fundamental functions in life processes. To achieve their functional properties the interplay between electronic properties and structural properties is crucial [15]. Proteins take their shape spontaneously after the synthesis in the cell, but the structure depends also on the environment properties (solvent, salt concentration, pH, temperatures and molecular chaperones). The chirality influences the assembly, folding and activity of biological molecules: amino acids that form proteins are all in the L configuration, with the exception of (non-chiral) Glycine. So far the task to determine the structure of proteins, and subsequently their electronic properties, was carried out by X-ray diffraction and nuclear magnetic resonance in crystals and in solution. The characterization of structural and electronic properties of proteins in the gas phase would provide valuable information to understand the folding. The absence of solvent interaction can reveal the balance of the molecular weak forces that determine the shape of the protein. These results would be complementary with respect to those obtained in crystals or solution, and give insight into the influence of the solvent on the protein shape. In addition, the application of "top down" techniques [16] in characterization of protein sequences and post-translational modifications by MS shows an increasing popularity in ongoing biological research [17]. To bring un-fragmented proteins into the gaseous phase, the state of the art technique is represented by Electro Spray Ionization (ESI) [18,19]. The ESI technique solved the problem of how to study large molecules in solution by mass spectrometry, that needs a high vacuum environment. After the commercial development of ESI sources, mass spectrometry has become a powerful tool for the investigation of very large molecules or as John Fenn described his invention in his Nobel prize address "Electrospray wings for molecular elephants".

The exploitation of ESI sources allows to extend the state of the art gas phase spectroscopies to protein ions and van der Waals clusters of one molecular system surrounded by a well-defined number of water molecules.

Pioneering works combining ESI with laser spectroscopies revealed [20,21] the possible application in the characterization of electronic and structural processes. These preliminary results envisage the application of ion-spectroscopy to top-down proteomics, but this research field is still completely unexplored. The low density of the target (space charge limits the maximum ion density $10^6$ ions/cm$^3$) prevents to extend such experiments in the VUV-soft X-ray wavelength range with the present sources. IRIDE with its high flux and focusing represents an ideal source for these studies. Moreover the IRIDE wavelength range can cover the excitations to the core states, focusing on bond character and local environment, and providing a rich and detailed description. The bulk of the experimental apparatus will be an ESI source coupled with suitable ion optics and ion traps. The spectroscopic techniques will be photo-absorption, photo-fragmentation and photoelectron spectroscopy. Photo-fragmentation will allow the understanding of protein structure as well as the study of charge transfer processes and the unraveling of reaction mechanisms.

This activity is complementary to the imaging of biomolecules via single-shot diffraction and could be also applied to characterize structures for organic nano-device deposition.

### *4.2.5.* ***References.***

### 4.3. Biological Applications

Owing to today's availability of genome sequences for a very large number of organisms, including humans, and the need of identifying, correlating, and understanding, the function of hundreds of thousands proteins, as well as the inter-relations among themselves and with other biomolecules (membranes, DNA, RNA,...), the interest of present investigations has moved from "molecular" biology to "modular" biology, where the specific biological process of interest is modeled as a complex system of functionally interacting macromolecules. This quite recent change of perspective opened a new and inspiring era in biology the so-called '*omic era*'.

Despite the enormous amount of knowledge acquired through the genome sequencing of thousands of living organisms, our instruments to interpret DNA functioning are still quite limited. Among the many processes that are waiting for a satisfactory explanation there is, for example, the orchestrated activation and silencing of the many thousands of genes in a living cell that establishes and maintains differences among cells and ensures the development and functioning of a complex multicellular organism.

It is clear that the functional role of biological molecules is very often dependent on the ability of their tertiary or quaternary structure to respond to the interaction with other molecules by conformational changes. However, in consideration of the large number (thousands) of genes, which have been and are being characterized through genomic research (implying access to the corresponding translated proteins), structural genomics initiatives have been started throughout the world, aiming at a massive/parallel study of protein 3D structures along with their genomic and functional characterization. At the present time, X-ray structure analysis of protein crystals and NMR are the only methods by which detailed structural information can be obtained at atomic resolution. However, obtaining crystals of suitable size and quality for a single-crystal diffraction experiment is the major drawback for many proteins, especially those with a high predominance of hydrophobic interactions (e.g. membrane proteins, or large protein complexes). Therefore, the number of experimentally determined protein structures is lagging far behind the output of sequenced proteins.

There are many open questions in biology and biophysics, among them we want to mention here the case of membrane proteins. Membrane proteins continue to be among the most challenging targets of structural biology, but the determination of their structure has proven to be difficult to study owing to their partially hydrophobic surfaces, flexibility and lack of stability[1]. These proteins are involved in a wide variety of biological processes including photosynthesis, respiration, signal transduction, molecular transport and catalysis. Precisely because they are notoriously more difficult to work with than soluble proteins (crystallization remains the most significant bottleneck in the process), researchers know little about their structures: for example the human genome encodes an estimated 10,000 membrane proteins, but we know the three-dimensional structure of less than 1% of them. Further technical development is clearly necessary to facilitate membrane protein structural analysis by the use of either membrane integral fragments, micelles or 2D nano-crystals. A more fundamental understanding of the structure-function relationships of membrane proteins would make invaluable contributions to structural biology, pharmacology and medicine.

The advent and the development of “very brilliant” synchrotron radiation sources in the energy range of hard X-ray has already led to very fruitful investigations in many different fields of biological interest, from crystallography to in-solution structural studies, from diffraction to absorption experiments. The new paradigm of omic-biology is asking for new experimental approaches and tools because the role of molecular complexes and molecular interrelationships are especially important. In this context the availability of intense and coherent sources of (soft to hard) X-rays will add other dimensions of information.

In particular soft X-rays, focused in the range of a few nanometers, will allow microscopy with nanometer resolution. Most importantly, monochromatic soft X-rays will enable us to reach an energy resolution in the milli-eV range, thus allowing spectroscopic investigations leading to an unprecedented richness of chemical and physical information.

There are at least two characteristic features of FEL spectroscopy that are very promising for experiments of biological interest: high brilliance and the simultaneous spatial and temporal coherence of the produced radiation.

It should be recalled, however, that high brilliance may lead to important radiation damage on biological specimens. It has been shown, in fact, that cooling cannot eliminate radiation damages that accumulate during the time needed for conventional measurements.

This strong limitation can be overcome by taking advantage of the very short exposure time needed in a FEL experiment. Ultrashort (of the order of the femtosecond, fs) high intensity X-ray pulses in combination with container-free sample handling methods based on spray techniques, may provide a viable approach to structural investigations.

A very interesting and promising domain of investigation in structural biology and functional genomics is the study of dynamical properties in the short time scale range. Biological processes of interest span over different time scales, ranging from millisecond (ms) to fs, and their study can benefit of the third outstanding property of X-FELs, namely its extremely short pulses. A FEL tuned in the wavelength range 0.1-13 nm can ideally combine its very high time resolution with either high spatial resolution, yielding structural information, or with high energy resolution, providing detailed chemical information, or both, thus opening the way for a wealth of new biological experiments.

We now wish to examine and discuss a number of possible experiments aimed at answering some of the many biological questions that are emerging owing to the analysis of the enormous amount of biochemical and structural data today available. We will group the proposed experiments according to the FEL radiation characteristic (coherence, intensity, time resolution) that is “mostly wanted” for achieving the desired goal.

### 4.3.1. *Coherence*

#### *4.3.1.1. X-ray microscopy of large macromolecular complexes or smaller samples at medium resolution*

Soft X-rays are a useful tool to bridge the gap between the experimental resolution reachable by light microscopy in the visible, which yields insight on the gross structural organization of the cell and its organelles, and the atomic resolution provided by hard X-ray spectroscopy.

In the wavelength region between the K-absorption edges of oxygen and carbon, i.e. in the so-called “water window”, which lies around 2.7 nm,the absorption by organic carbon and nitrogen containing material is much higher than that by oxygen. For this reason imaging measurements can be carried out on biological specimens up to about 10 μm thickness. This means, for example, that a whole cell can be imaged in an aqueous environment close to its native state. As already mentioned in the introduction, none of the methods so far devised to limit radiation damage of biological sample in experiments where the necessary exposure time is on a much too long time scale, are really effective. The only way out to possibly maintain structural stability during exposure is to take an image within one very short pulse, during which no appreciable structural changes will occur. The very high peak brilliance required for such experiments is only available at FEL facilities.

It is possible to exploit FEL radiation for refractive or scattering methods in microscopy. X-ray microscopy with FEL could clearly take great advantage of light coherence properties. A number of interferometric microscopy methods, such as Schlieren Microscopy, based on detecting the scattered radiation, could be implemented in this context. An alternative and even more attractive possibility is to perform single molecule X-ray imaging by means of holographic microscopy. In this case we could imagine, as recently proposed [2], to have a very dim radiation coherently scattered by a biomolecule beam with a reference X-ray wave (Fig.1). In this heterodyning holographic acquisition method we could reach much better signal-to-noise ratio than in the equivalent homodyning method (without reference wave). The experiment could be done using a single shot irradiation of molecules labeled with nanometer size amorphous gold particles. Coherent imaging at FELs has already allowed reconstructing the structure of big viruses [3] even in the absence of holographic references.

Concretely X-ray microscopy offers the possibility of performing a number of important experimental investigations on topics of large impact for the research in biology and medicine, a non-exhaustive list of which is given here below:

a) diffusion processes of small molecules in biological membranes;

b) changes in membrane structure induced by electroporation and sonoporation;

c) monitoring of protein folding and protein aggregation in living cells;

d) monitoring of molecular events induced by the presence of drugs.

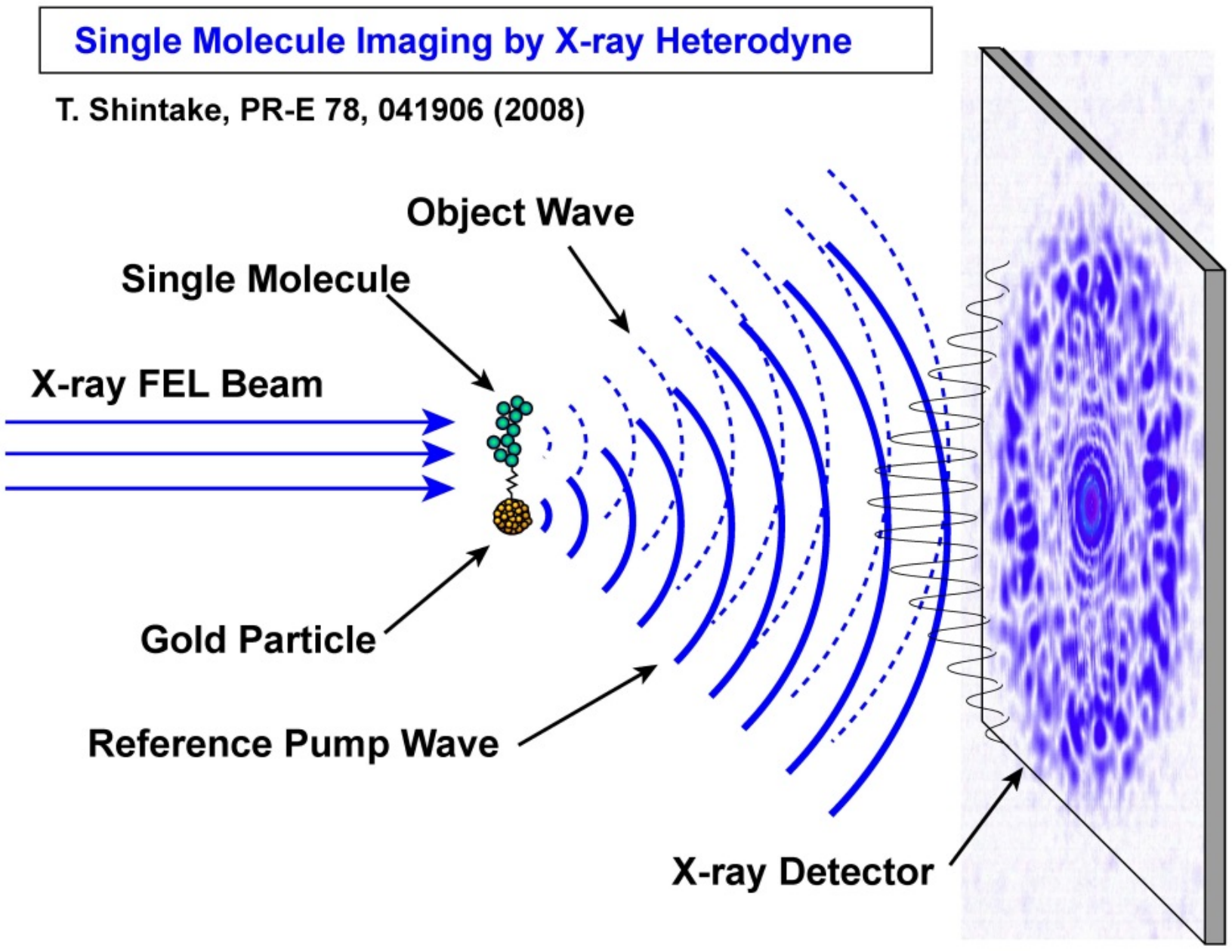

Figure 1. Principle of holographic microscopy of single peptides. The protein is linked to an amorphous gold particle that provide a reference spherical wave for the holographic recording of the diffracted X-ray FEL beam (with kind permission of T.Shintake).

In recent years neurodegenerative diseases (e.g. Parkinson's, Alzheimer's) have seen a large spread. They are believed to share a common molecular mechanism involving protein misfolding and aggregation [4]. In particular, misfolding of well-defined proteins (or proteins fragments) leads to their insolubleness and consequent accumulation in vital organs, such as brain or liver. The capability to prevent the formation of toxic aggregates and to eliminate them is indeed a key feature to avoid cell death and neurodegenerative diseases. Moreover there is an increasing number of observations indicating that metal ions are capable of accelerating these processes [5;6].

The high coherence and brilliance of the IRIDE FEL source will be used to image, by Coherent diffraction imaging (CDI), different stages of the aggregation process of these proteins in presence metal ions. CDI is a "lensless" imaging technique for the 2D or 3D reconstruction of nanoscale structures [7]. In CDI, a highly coherent nanosized beam of x-rays is incident onto an object. This beam is scattered by the object and the diffraction pattern is collected by a detector. This recorded pattern is then used to reconstruct an image via an iterative feedback algorithm. Effectively, the objective lens in a typical microscope is replaced with software to convert from the reciprocal space diffraction pattern into a real space image. The advantage in using no lenses is that the final image is aberration –free and so resolution is only diffraction and dose limited (dependent on wavelength, aperture size and exposure). CDI experiments on biological samples have been already demonstrated in the past [8], with limitations imposed by the radiation damage [9]. A very crucial point is to perform an effective data treatment to phase recover the object image at the best allowed spatial resolution [10;11].

#### *4.3.1.2. Diffraction ultra-microscopy*

An important problem that has to be solved in order to exploit most efficiently the FEL potentiality in the field of biological investigations is the possibility of orienting proteins in a controlled way. The idea that we wish to put forward is to use a method similar to that employed for alignment of isotopically labeled macro-molecules in a viscous medium as it is done in high resolution NMR spectroscopy.

A valuable test of the efficiency of the method can be carried out using the well-known Lysozyme protein[1]. This soluble protein is commercially available at a high level of purity and may be employed, with different orienting media, either alone or in association with other molecules so as to reach the desired level of immobilization. It is expected that the percentage of three-dimensionally oriented proteins in the medium can be pushed up to a quite high level. Something of the order of 70-80% is not inconceivable.

Different orienting media and tools may be exploited, choosing among detergents, liquid crystals in different phases or different chemical gels, like the ones that are currently used in biochemistry for proteins manipulation and separation.

The main goal of the test proposed here is to check the possibility of obtaining a sufficiently good diffraction pattern to be eventually compared to the three-dimensional structure of the protein, which is known both for the protein in solution (by NMR experiments) and for the crystallized protein (by X-Ray diffraction experiments). Experimental structural data at different resolutions and in different physico-chemical conditions are available in the PDB (Protein Data Bank).

We are confident that the measured diffraction pattern (possibly improved by the use of the "oversampling method" suggested by Miao [12]) can be analyzed to reconstruct the three-dimensional external profile of the object. The reconstructed profile can then be used to check whether the resolution that can be achieved for each of the selected orientation techniques, is good enough for reconstructing the characteristic structural features of the protein, like helix assembling and similar secondary structure arrangements.

The feasibility study we are proposing here should be performed with the aim of verifying the possibility of scaling up this innovative strategy to larger proteins or complexes which are out of the range of investigation of NMR spectroscopy and/or cannot be crystallized (like the afore-mentioned membrane proteins or naturally unfolded proteins)

---

[1] Lysozyme is a 14.4 kilodalton enzyme that damages bacterial cell walls. It is abundant in a number of secretions, such as tears, saliva, and is also present in egg whites.

#### *4.3.1.3. Structural studies of proteins using nanocrystals*

X-ray crystallography provides the vast majority of macromolecular structures, but it relies on growing crystals of sufficient size. In conventional measurements, the necessary increase in X-ray dose to record data from crystals that are too small leads to extensive damage before a diffraction signal can be recorded.

The technique of serial crystallography has been recently developed at FELs [13;14] and can overcome this limitation thanks to the high peak brilliance of the source. The FEL delivers enough photons on the sample to provide an usable diffraction patterns before the radiation-induced damage changes the structure of the sample. This methods allows to study very small crystals. FEL experiments performed at the LCLS [13] have shown that it is possible to measure useful diffraction data from Photosystem I crystals as small as 300 nm, corresponding to about unit cells.

The outline of the experiment performed at the LCLS is depicted in Fig.2 A gas focused liquid microjet was used to inject randomly oriented photosystem I microcrystals and nanocrystals into the X-FEL interaction region. There, they intersected with 300 fs long X-ray pulses with 1.8 keV photon energy. Since the crystals were destroyed upon exposure to a single femtosecond FEL pulse, the data were collected in a serial fashion, with the liquid jet providing a fast, gentle, and convenient means of replenishment. Serial femtosecond crystallography resulted in interpretable electron density of photosystem I, with a resolution limited to 8.5 Å.

Diffraction data are routinely measured on crystals of the size of 10-100 μm using microfocus beamlines. Very recently, experiments have allowed the measurement of diffraction data using a beam size of 1 μm [15]. It is not known yet what would be the lower limit achievable by this technique, i.e. how small a crystal can be in order to still obtain a discrete spectrum characterized by Bragg peaks. Simulations suggest that a crystal composed of 10x10x10 unit cells can still give rise to a discrete spectrum. Thus, assuming a unit cell of 10 nm (which is a reasonable size for a protein crystal), 10 unit cells would yield a crystal of 100 nm edge. It should be recalled that very small crystals may be easier to grow than large ones: most of the precipitates obtained during the crystallization trials contain crystals of extremely small size.

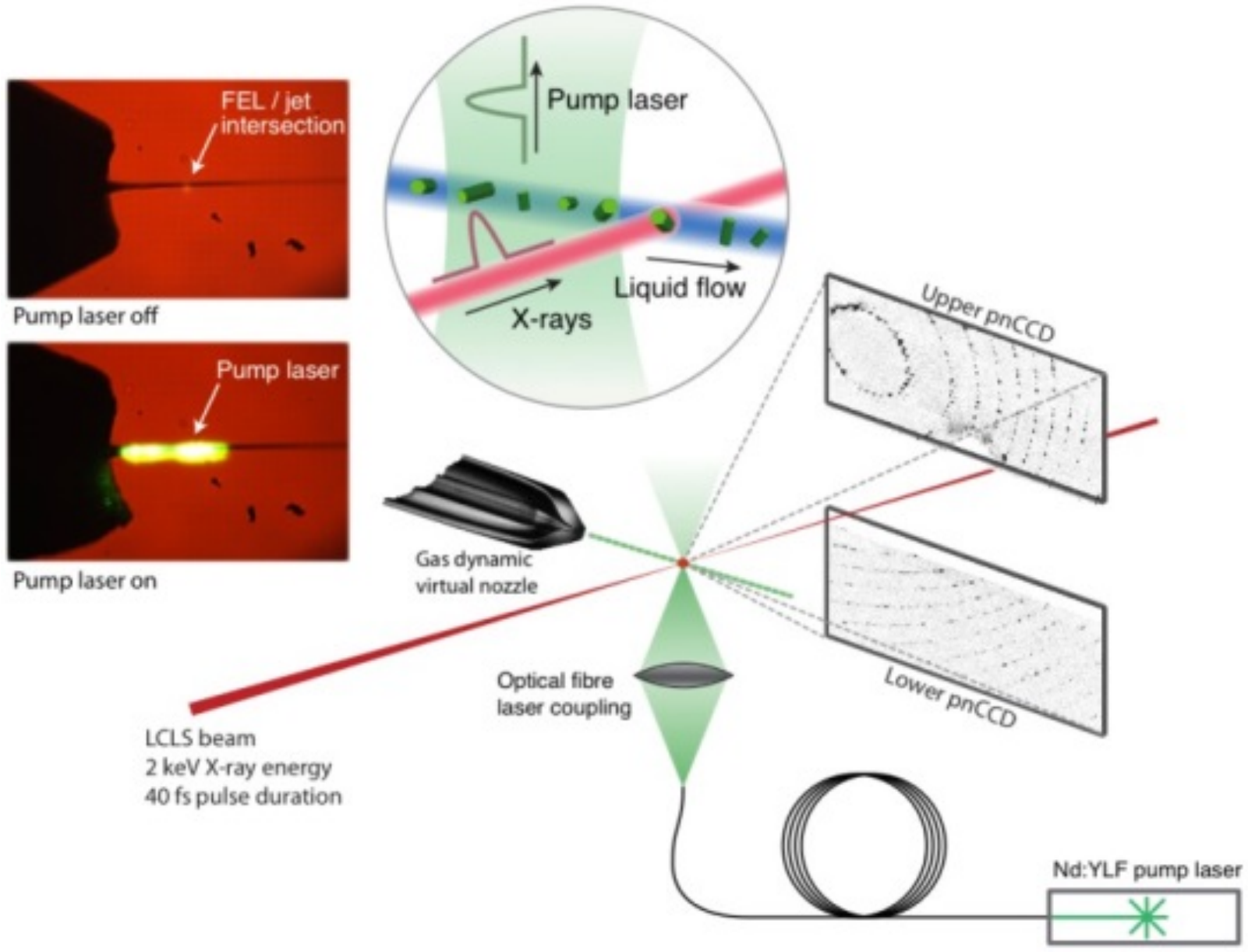

Figure 2 Serial femtosecond crystallography

The use of nanocrystals with a monochromatic source does not allow measuring the entire diffraction dataset from a single crystal. A serial approach, in which a new crystal is put into the FEL beam at every shot is therefore required. For that purpose, random portions of the reciprocal space, collected from several crystals, will be assembled in order to achieve the complete data set [16]. To obtain resolution at atomic level, the third harmonic of the beam should be used. Finally, nanocrystals can be used to exploit another interesting feature of the diffraction spectrum. Because of the presence of a very limited number of cells, the Laue conditions are not fulfilled and a continuous signal in between the Bragg reflections will be present. This continuous signal can be exploited to recover the phases of the reflections, for example making use of the Hilbert transform technique. Several papers have been published to derive relationships that, using non-integer reflections, allow the reconstruction of the phase [17]. The issue of radiation damage on such small crystalline samples remains however the main issue that will require further simulation studies as well as experimental proof of principle.

The IRIDE light will be ideal to perform serial crystallography experiments on protein crystals. Thanks to the short wavelength provided by the 5$^{th}$ harmonic it will allow to collect diffraction data up to atomic resolution.

#### *4.3.2. Intensity*

##### *4.3.2.1. Use of high fluxes for atomic analysis*

The high spatial resolution and high fluxes of the FEL beam will open the way to the possibility of atomic or molecular composition analysis. One of the possibilities that has not been explored up to now is that of exploiting the ionization efficiency of the FEL beam to perform Electron Analysis Microscopy (EAM), such as in commercial TEM microscopes. The use of a coherent beam would allow achieving much higher spatial resolution and intensity while opening the way to an analysis on biological samples at an atomic scale.

An additional example in this direction is the mass spectrometry of biomolecule by UV/FEL matrix-assisted laser desorption/ionization [18]. In this case we actually make use of one of the drawbacks that FEL has in biological applications, viz. the radiation damage induced on the sample, and use it to induce desorption of biomolecules or ionization in order to perform mass spectrometry. The combination of MALDI techniques and FEL high fluxes will offer the possibility of enhancing the resolution and sensitivity of the mass spectroscopy analysis [18]. A second possibility to exploit the high flux available at IRIDE would be to perform hole-burning spectroscopy.

#### *4.3.2.2. Small Angle X-ray Scattering of Biological Molecules*

Small angle X-ray scattering (SAXS) is a universal tool for probing structures on the nanometer scale for a wide variety of non-crystalline materials and represents an emerging technique in the framework of structural and functional studies of biological molecules [19; 20; 21]. SAXS data are collected at low scattering angles, typically much smaller than those used for diffraction patterns from most crystalline biomolecules, so that the achievable experimental resolution is typically much lower than 'atomic' roughly being located between the spatial resolutions typical of EM and crystallography. SAXS nevertheless is an important tool whose chief advantage resides in the possibility of performing measurements on sample s in diluted solutions thus providing easy access to the study of structural transitions ranging from conformational changes to assembly processes triggered by a variety of perturbations. In the case of biological macromolecules SAXS immediately provides information on particle average size (radius of gyration), aggregation state, molecular weight and volume, and a shape function from which low resolution 3D models of the biomolecules under study can be obtained [22; 21]. Moreover, while X-ray crystallographic studies are essentially static, SAXS can be carried out in a time-resolved manner [23] providing information about dynamic processes. Due to the very low scattering signal relative to background and the sensitivity to parasitic scattering, synchrotron X-rays have proven to be essential for biological time-resolved studies (currently with time resolution in the few millisecond or slightly shorter time range) [20]. In this context, applications range from the study of the number, shape and organization of subunits or domains in complex proteins [24] to the analysis of protein conformational changes after addition of small ligands (e.g. drugs, effectors, substrate analogues) in solution [25] or at the folding-unfolding transition, allowing to investigate the molecular basis of "protein misfolding" diseases, such as mad-cow and Alzheimer diseases [26], and to analyze protein-protein interaction mechanisms [20; 27; 28]

Another application of SAXS at IRIDE is represented by X-ray fiber diffraction studies of muscle contraction that have developed in the direction of investigating a single muscle cell (Fig.3), where the mechanical events can be associated with the structural changes of the molecular motor with sub-millisecond time resolution [29]. X-ray fiber diffraction studies at the level of the single cell have been possible thanks to the improved brilliance of the third generation synchrotrons, and allowed to characterize the mechanical and structural working stroke of the myosin motors in situ [30].

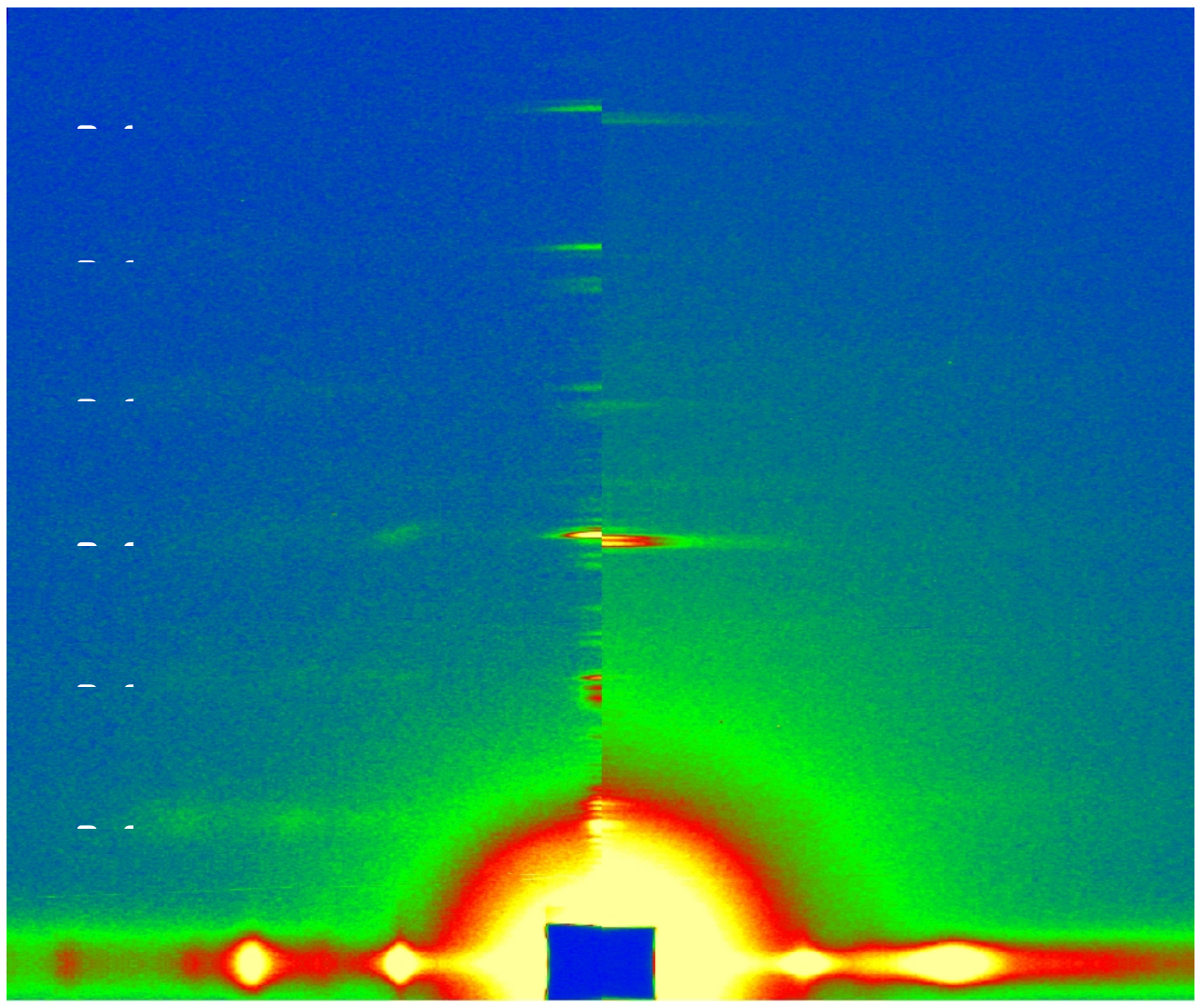

Figure 3. High resolution X- ray diffraction patterns from a single fibre from frog skeletal muscle at rest (left) and during isometric contraction (right). The patterns have been collected on a CCD detector at the BioCAT beamline of the Advanced Photon Source, Argonne IL, USA. Each pattern is the sum of 5 exposures 100 ms each, taken on different positions along the fibre to spread radiation damage. Flux on sample was about $10^{12}$ ph/s. A $10^{3}$ times higher flux would be necessary to collect patterns with the same signal/noise ratio and ca 0.1 ms time exposure adequate to follow structural dynamics of the fastest events in the myosin motors .

The internal architecture of fibrin fibers, the basic constituent of blood cloths and thrombi, has also been intensively investigated [31]. Indeed one of the therapeutic approaches in the treatment of patients suffering from acute myocardial infarction (and, in general, with thromboembolic diseases) is based on the use of thrombolitic drugs that promote clot lysis. Even if the mechanisms leading to clot lysis are not completely understood, it is widely accepted that they depend their ready diffusion into the fibers.

Highly ordered distribution of protofibrillar repeats have been found to form a unit cell, within which each basic protofibrillar entity is separated by a distance of 8 nm and half-staggered in relation to adjacent protofibrils. A relevant advantage would consist in the possibility of following the frustrated dynamic of selected, labeled, drug molecules inside fibrin fibers by using X-ray photon correlation spectroscopy at the high spatial resolution achievable with IRIDE. More in general, biocompatible polymer gels are employed as nano-carriers for low release drug delivery, and even in this case the drug release kinetics can be followed.

Since radiation damage is already a limiting factor for measurements at state-of-the-art third-generation synchrotron beam lines, higher X-ray flux will be primarily beneficial for time-resolved measurements. Indeed, SAXS experiments can in principle access very short time scales (from nano- to pico-seconds) thanks to sophisticated approaches to sample handling (e.g., faster mixing systems, more efficient means to trigger fast initiation of the event to be studied [32]). Increased average brightness that can be provided by IRIDE source could play a unique role in stretching even further this time domain. With the extreme brightness of a X-FEL, and controlling radiation damage by approaches like rapid flowing of sample solutions, it will be possible to imagine recording a two-dimensional scattering pattern on the ultra-fast time scale provided by the pulses of the accelerator (in an experiment analogous to that of single-molecule imaging discussed elsewhere in this document). If the challenging problems posed by sample manipulation and background scattering can be solved, it may be feasible to study very fast structural changes at low resolution - like time-dependent aggregation during amyloid fibrillation [26] and time-modification of the shape function - on a much faster time scale than the current millisecond regime limit.

In addition, structural biology needs to uncover the extremely fast steps in biochemical reactions such as enzymatic processes and drug–receptor interactions using time-resolved SAXS. Thus, an obvious evolutionary development from the current success obtained by micro-focus beam-lines at third generation synchrotron sources is to build nano-focus beam-lines at a X-FEL source for ultra-fast time-resolved in-solution SAXS measurements.

A non-exhaustive list of the scientific topics that could be attacked at IRIDE is given here below:

a) low resolution structure of non-crystallizable large and complex assemblies in very dilute solutions in conditions that are at or near those normally found in vivo (e.g. membrane fragments or bicelles containing functional membrane proteins);

b) low resolution structure of proteins interacting with membranes analyzing different soluble mutants obtained by varying the length and composition of domains by which the proteins are anchored to the lipid bilayer (e.g. integral membrane proteins or cytosolic proteins translocating to membranes);

c) time-resolved analysis of the ultra-fast changes in the structure of biological materials in very dilute solutions, ranging from single protein to large and complex assemblies;

d) high-throughput low-resolution structure of proteins in very dilute solutions to provide complementary information useful in optimizing the structure determination pipelines to estimate phases for crystallographic data from complexes that can be crystallized and determine the mutual disposition of subunit structures that are obtained from high resolution methods such as crystallography and NMR [33; 32];

e) analysis of the structure and ultra-fast dynamics of the aggregation of amyloid protofibrils and unstructured aggregates that precede their formation in dilute solutions. Ultra-fast time resolved structural analysis of the amyloid “nucleation growth” process;

f) ultra-fast time resolved structural analysis of protein folding-unfolding mechanisms in very dilute solutions;

g) ultra-fast time resolved structural analysis of intrinsically unstructured proteins. Indeed, it has recently been recognized that a large number of proteins contain non-globular, disordered segments, which become structured in certain environments or when interacting with ligands or other proteins;

h) diffraction analysis of small size and low-diffracting demembranated fibres from mammalian muscle, to collect the signal from the weaker reflections originating from the regulatory and citoskeletal proteins associated with myofilaments. These experiments require a higher intensity than that available with present X-ray sources.

The improved brilliance of FEL X-ray sources will not only allow measurements of this kind, but also will open the possibility to investigate the simplest level of organization the contractile machinery, the myofibril. Myofibril is the substructure of the 1 μm wide muscle fibre where the reduced dimension will allow stepwise changes of the chemical composition of the medium and replacement of native regulatory proteins with exogenous isoforms or disease-associated and genetically engineered mutants. All this will make possible in situ studies of the actomyosin ATPase cycle responsible for energy conversion and its regulation, offering a new powerful tool to the investigation of the molecular basis of the related pathologies.

Very similar considerations hold about the possibility of performing Wide Angle X-ray Scattering (WAXS) experiments at IRIDE. Structural changes of proteins have been already successfully tracked at synchrotrons by time-resolved WAXS with nanosecond time resolution [34]. Thanks to the high brilliance of the IRIDE source similar experiments can be performed at IRIDE aiming at even better time resolution.

### *4.3.3. Temporal structure*

#### *4.3.3.1. Pump-probe and time resolved spectroscopy*

The characteristic features of the FEL beam allow the generation of secondary laser radiation in the range from tens to few hundreds of nm. This is a critical range in which most of the absorption of the molecular transitions of protein and DNA backbone lies. The existence of these molecular transitions has never been fully exploited due to the need of laser radiation in this region of the spectrum. A way to circumvent this problem is to have an external laser focused in the region of the first undulator. The interaction between the electron beam and the external laser leads to a density modulation resulting in coherent emission with harmonic content in the next undulator. Experiments in Japan have shown that a seeding 800 nm laser beam can provide easily up to the fourth harmonic (400, 266 and 200 nm). Being the absorption of DNA at 260 nm and that of the amidic bond around 190 nm, this radiation would be extremely useful for time resolved far UV spectroscopy. In addition, ligands in proteins can be efficiently photo-lysed using the Soret band at ~400 nm, and this will open new possibilities of highly sensitive study of photolysis of prosthetic groups in proteins.

The most obvious application of FEL would then be to make use of very short pulses in pump probe experiments. The obvious requirement is to reduce the FEL beam intensity to levels at which the samples can be followed over time. This can be achieved by expanding the FEL beam and collecting only the central part of it (reducing the power used and enhancing the time resolution of the laser pulses), thanks to the intrinsic inhomogeneity of the pulse width and the jittering of laser shots on the wave surface. Pump-probe is a time resolved spectroscopic technique aimed at measuring the population dynamics or the time evolution of transient species. It allows in principle to follow temporal changes in chemical composition or in structure of biomaterials or cells with sub-picosecond time resolution. It consists typically in the ionization, isomerization or photolysis of the molecule (or the aggregate) under study by two ultra-fast laser pulses (pump and probe) delayed in time one with respect to the other.

Femtosecond pump-probe spectroscopy enables to follow in real time vibrational motions coupled to electronic transitions. Indeed, when the system is excited by a pulse shorter than its vibrational period, vibrational coherence is initiated, providing in this way information on both the ground and the excited electronic states, on nuclear dynamics and, eventually, on the coupling between different degrees of freedom (multidimensional spectroscopy).

In a pump-probe experiment, the output pulse train from an ultra-fast laser is divided into two beams: the sample is excited by one pulse train (pump) and the changes it induces in the sample are probed by the second pulse train (probe), which is appropriately delayed with respect to the pump. A number of properties related to the probe (reflectivity, electronic and vibrational spectrum, luminescence, atomic structure) are then monitored in order to investigate the changes induced by the pump on the sample.

A further application of FEL radiation to life science would be probing transient molecular structures in photo-biological processes by time-resolved X-ray Absorption Spectroscopy (XAS) [35]. Although the time course of biological reactions could be as long as minutes or longer, the underlying elemental chemical events, such as charge transfer, spin crossover, bond breaking or bond forming, take place on a timescale of about 100 fs or shorter [36]. Ultrafast pump-probe laser spectroscopy is used in such studies, obtaining both the coherence of excited states as well as the kinetics of photo-induced biochemical reactions. For example, flash photolysis of hemoglobin derivatives ($O_2$, NO, CO) in the femtosecond time scale has revealed that two short-lived species, Hb I* and Hb II*, are formed on the absorption of a visible photon by the heme [37]. The lifetime of Hb II* is 2.5 ps (before fast recombination), and this species is significantly populated when the ligand is NO or $O_2$, but not when the ligand is CO. Thus the apparent low quantum yield of photo-dissociation for the former two ligands could be in part explained by the fast recombination of the ligand from the highly reactive state Hb II*, a physiologically relevant information.

However, the understanding of such biochemical events will not be complete without solving molecular structures along the reaction path from the initially photo-excited Frank-Condon state to a thermally equilibrated excited state. Moreover, structural changes underlying fast reactions have to be solved at atomic resolution, at least at the level of the reaction site, and XAS spectroscopy can in principle in many cases accomplish this difficult task [38]. Very recently, picosecond and femtosecond time-resolved pilot experiments done at Synchrotron radiation sources have been shown XAS to be uniquely suited to this purpose [39; 40]. Structural changes smaller than 0.1 Å accompanying the spin crossover dynamics of iron(II) tris(bipyridine) and nickel(II)-porphyrin in solution have been provided by theoretical analysis of the XAS spectrum in both the high-energy [41] and the low-energy range [40; 42]. The interest in extending such experiments to system of biological interest or useful models by means of the new light sources seems obvious.

A final example of application of FEL radiation could be, for instance, the study of the dynamics of the important processes that occur when photosynthesis organisms adapt to different illumination conditions [43] Ultraviolet radiation (UV-A and UV-B) and possibly soft X-rays are in fact well known damaging factors of plant photosynthesis.

Thanks to the short wavelength accessible with the $5^{th}$ harmonic, the IRIDE source is also suitable for X-ray Absorption and Emission experiments [44] of proteins and peptides in complex with metal ions. This approach has already been proven to be successful at the LCLS. The high peak brilliance of the source will allow monitoring with high time resolution chemical reactions.

#### *4.3.4. Theoretical and Numerical Tools*

There is a last important point that is worth mentioning in devising a bio-scientific case. We are deeply convinced that whilst a challenging new machine like FEL is being designed and constructed, at the same time computer simulation tools and know-how should be promoted and developed according to the needs of this emerging field of investigation. Numerical methods will obviously play a pivotal role in designing experiments and analyzing data.

Owing to the enormous power of today computers and the development of sophisticated mathematical algorithms and simulation strategies, we are in position of carrying out accurate *in silico* studies of biologically realistic systems with the aim of investigating their dynamical and thermodynamic properties and following at atomic level (or, if needed, at a quantum level) their physico-chemical behavior under the action of different kinds of stimuli (from temperature and pH changes to photon and electronic excitations).

Three main applications of numerical simulations are easily foreseeable. The first one consists in the simulation of "gedanken" experiments in order to test their feasibility and foretaste the information obtainable from them. When challenging experiments, like the ones proposed at FEL, are devised this kind of tests are of fundamental importance [9; 45; 46] to avoid miscalculation of risks and ensure the achievement of the expected scientific goal. The two remaining applications, namely data analysis and modeling, may at first sight look more hackneyed, but given the amount of uncertainty inherent in this new research field, it would be quite possible that a lot of original efforts will be required to invent and project new theoretical instruments for high-throughput data collection and analysis. Numerical simulations have been recently also used to shed light on the damage processes in FEL serial crystallography and imaging experiments [47].

In the field of biophysics there is a well-established and recognized experience on the many theoretical and numerical approaches today available to deal with systems of biological interest. It is enough to recall the extraordinary large number of applications of *ab initio* calculations to biosystems that have followed the development of the Density Functional Theory [48]. The latter (under the form of Quantum Chemistry or disguised as Car-Parrinello [49] molecular dynamics) has been used in many important areas at the frontier of today's research activity (macromolecule interactions, immunological recognition, proto-fibril formation, etc.) [50; 51]. Thanks to the recent unprecedented improvements in computer architectural design and algorithmic developments, systems with up to a few hundreds atoms and some thousand electrons can be comfortably simulated. This knowledge and experience will certainly be very precious to investigate the output of FEL experiments where often single molecule events are involved.

#### *4.3.5.* **References**

### 4.4. Time resolved X-ray spectroscopy in condensed matter

The extremely high fields that x-ray FELs produce are widely exploited in studies of condensed matter. The pulses delivered by IRIDE will result in nonlinear effects becoming significant or even dominating the interaction with the sample. They might make it possible to create and probe plasmas as dense as solids and the underlying fundamental physical processes are in many respects unknown. This is due not only to the extension of the photon energy range with the atom selectivity in the inner shell of the elements forming the studied systems, but also to new processes related to the enormous peak field strengths generated by the IRIDE. Therefore, it is essential to study in detail the interaction of the high peak-power, x-ray pulse with matter so that correct interpretation of experimental results can be made with confidence.

In condensed matter excitation of electrons in unoccupied states radically alters the effective potential energy which determines the nuclear motion in matter. This leads to a variety of phenomena which can include dissociation and appearance of short lived transient structures and new effects [1]. The operation of the new pulsed FEL facilities [1;2] made access to explore dynamics at previously unavailable time and length scales. X-ray time-resolved (TR) scattering and direct TR methods ( electron spectroscopies and absorption) are complementary approaches to temporal information and, together, provide precious tools of investigating dynamics in condensed matter.

The generation of short (1 – 100 fs) and intense ($10^9$ – $10^{12}$ photons) pulses is one of the most notable features of FEL radiation [2;3]. In fact for pump-probe experiments, the duration and intensity of the pulses will allow significant advancements in the study of the transient states of matter which follow (optical) photoexcitation and of the accompanying relaxation processes due to the interaction of the excited electrons with structural degrees of freedom (structural distortions / phonons, typical times 0.1 – 10 ps) and with other excitations (e.g. plasmons and excitons, typical times 1 ps). In many cases, use of FEL radiation will allow radical or even revolutionary improvements with respect to what can be achieved using currently available pulsed sources (e.g. storage rings, plasma x-ray sources, harmonics of fs lasers and others).

In the soft to hard X-ray region the photon wavelengths match very closely the periodicity of the dynamical fluctuations, thus, providing an ideal probe. All this opens new horizons to dynamics and out of equilibrium characterization of systems, offering new opportunities to research areas such as vibrational control of phase transitions, coherent excitation of phonons, magnons and orbital waves, femtochemistry, ultrafast surface melting. Our objective is to establish new classes of experiments, able to determine energy time and spin, dependent properties of electronic correlations where ultrafast probes (together with external parameters) are intentionally used to drive the system out of equilibrium. This is complementary with the information on collective dynamics in equilibrium or non-equilibrium provided by transient grating techniques.

The exciting opportunity offered by IRIDE to probe transient states and relaxation processes with spectroscopic techniques such as x-ray absorption (XAS), photoelectron spectroscopy (PES) and related techniques is driving significant research efforts worldwide. In the x-ray region, studies have been performed exploiting the TR-XAS and TR-x-ray diffraction (XRD) [4]; these studies are suggestive of the kind of studies which will be possible using the advanced characteristics of IRIDE. The possibility to mixing different radiations beams and to synchronize the FEL pulse with external laser sources or to split and delay the FEL pulse itself is directly related to the time structure, to perform a rich variety of pump-probe (PP) experiments involving X rays. This, together with the photon energy range, provides the possibility to explore and resolve ultrafast dynamical processes rather than the ground state of matter, as electronic relaxation and scattering processes inherent to any energy transport or signal transduction process, from meso- to nano- scale.

The FEL of IRIDE is foreseen as a medium-energy high repetition rate pulsed x-ray source. In the soft x-rays (0.5-1.5 keV) this puts the peak brilliance and pulse intensity somehow below the European XFEL, SCS and FXE (Hamburg) beam lines, but with comparable average flux, i.e about 2 orders of magnitude higher than synchrotron radiation sources. The development of photon-in photon-out spectroscopy at IRIDE should be made in consideration of these parameters. Less photon hungry techniques (x-ray absorption, resonant elastic scattering) can be fully developed up to single pulse pump-probe experiments. Moreover the more photon hungry resonant inelastic x-ray scattering (RIXS) can become competitive in standard operation with respect to synchrotron sources, with as an optional development towards pump-probe time resolved experiments.

The challenge is to develop methods and protocols which will enable extensions of optical pump – probe techniques, established since the availability of fs lasers, to the (soft and hard) x-ray region, in which spectroscopic techniques possess a unique sensitivity to the local atomic and electronic structure.

Condensed matter systems, from liquid to amorphous and to crystals, show an huge variety of configuration in the inter-relation of the degree-of-freedom. In condensed matter dynamics, charge, lattice and spin degrees of freedom are coupled in characteristic time scales of which excitations of one particular subsystem interact and equilibrate with the remaining subsystems.

In Fig. 1 we show the characteristic time-energy relation of fundamental interaction in solids, reporting schematically our present knowledge about typical times involved in the interactions of atoms, electrons, and spins. The corresponding quantum-mechanical interaction energies are estimated from the energy–time correlation DE Dt = h ~ 4 fs·eV. The characteristic time scales can differ from tenth of fs to hundreds of ps. TR spectroscopy facilitates direct and independent access of these degrees of freedom, providing novel insight into coupling mechanisms.

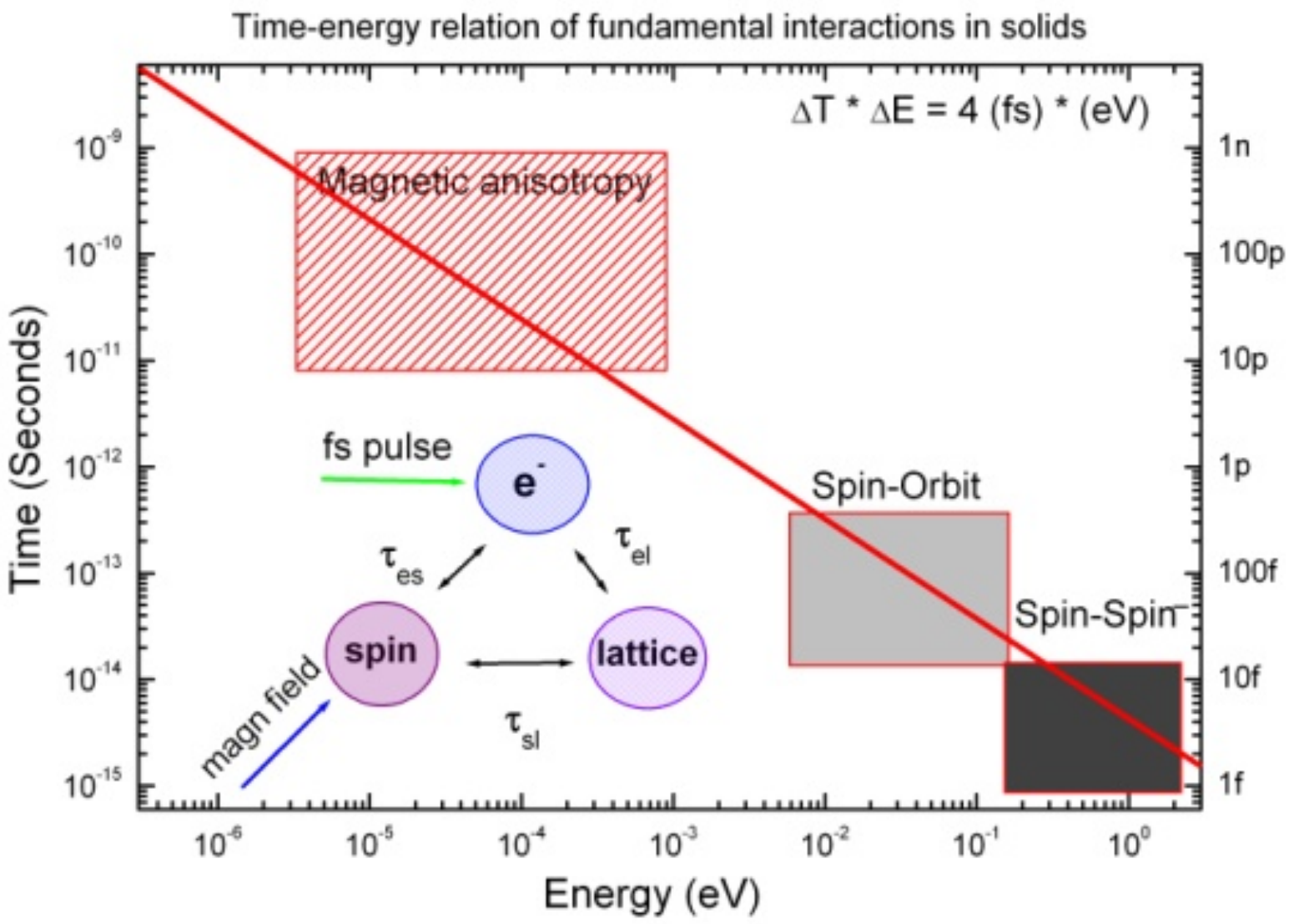

**Figure 1**. Characteristic time-energy relation of fundamental interactions in solids, as deduced from Heisenberg's uncertainty principle DE*Dt ~ h ~ 4 fs*eV . Inset: The model assumes that the electron motion (e), the electron spins (spin) and lattice phonons (lattice) represent three mutually-interacting thermal reservoirs. Furthermore, the reservoirs interact with one another via the electron-phonon interaction (el), the spinlattice relaxation rate (sl) and the spin-orbit interaction (es).

Fluctuations are intimately related to the complexity of the phase diagram in materials.

For this reason the  multiphase systems where fs excitations could be studied by different novel techniques. A large variety of systems with competing short and long range interactions self organizes in domain patterns [5]. Examples range from ferromagnetic systems to diblock copolymers and, at least theoretically, neutron star matter. A wide variety of experimental results and theoretical investigations in recent years have singled out multiphase complex systems category for which several physical interactions—spin, charge, lattice, and/or orbital—are simultaneously active leading to different dominant states are not spatially homogeneous [6]. Most correlated electronic systems exhibit exotic glassy behavior with notoriously slow dynamics, establishing one of the prime connections between traditional biological or soft systems and the complex states described here. Phase separation is also been revealed in thin films, surfaces and interfaces. Notable examples are the high-temperature superconductors (HTc), where magnetic order and, depending on the doping level, charge order compete with superconductivity. Fig. 2 shows the clear similar images taken on frustrated phase separation in different magnetic and organic film systems.  For details on the domain scales see ref 5.

Our ultimate goal is to get dynamic information and to relate that with the microscopic local properties to the meso-nano scales properties of condensed systems, such as liquid, amorphous, glassy, films, interfaces and highly electron correlated systems. These tasks can be achieved with the complementary use of ultra-fast X-rays spectroscopies and non-linear wave mixing techniques. The achievement of these goals would correspond to a clear breakthrough in experimental physics and spectroscopy, i.e. to go beyond the simple observation of collective phenomena in correlated systems and step towards the control of fundamental interactions in solids. We project to bring into operation not only a number of conceptually novel experimental tools but also will develop a highly interdisciplinary approach, where the a priori different knowledge in theory, spectroscopy, fast magnetization dynamics, strongly correlated oxides, new materials, will efficiently glue together as an added value of the proposed research.

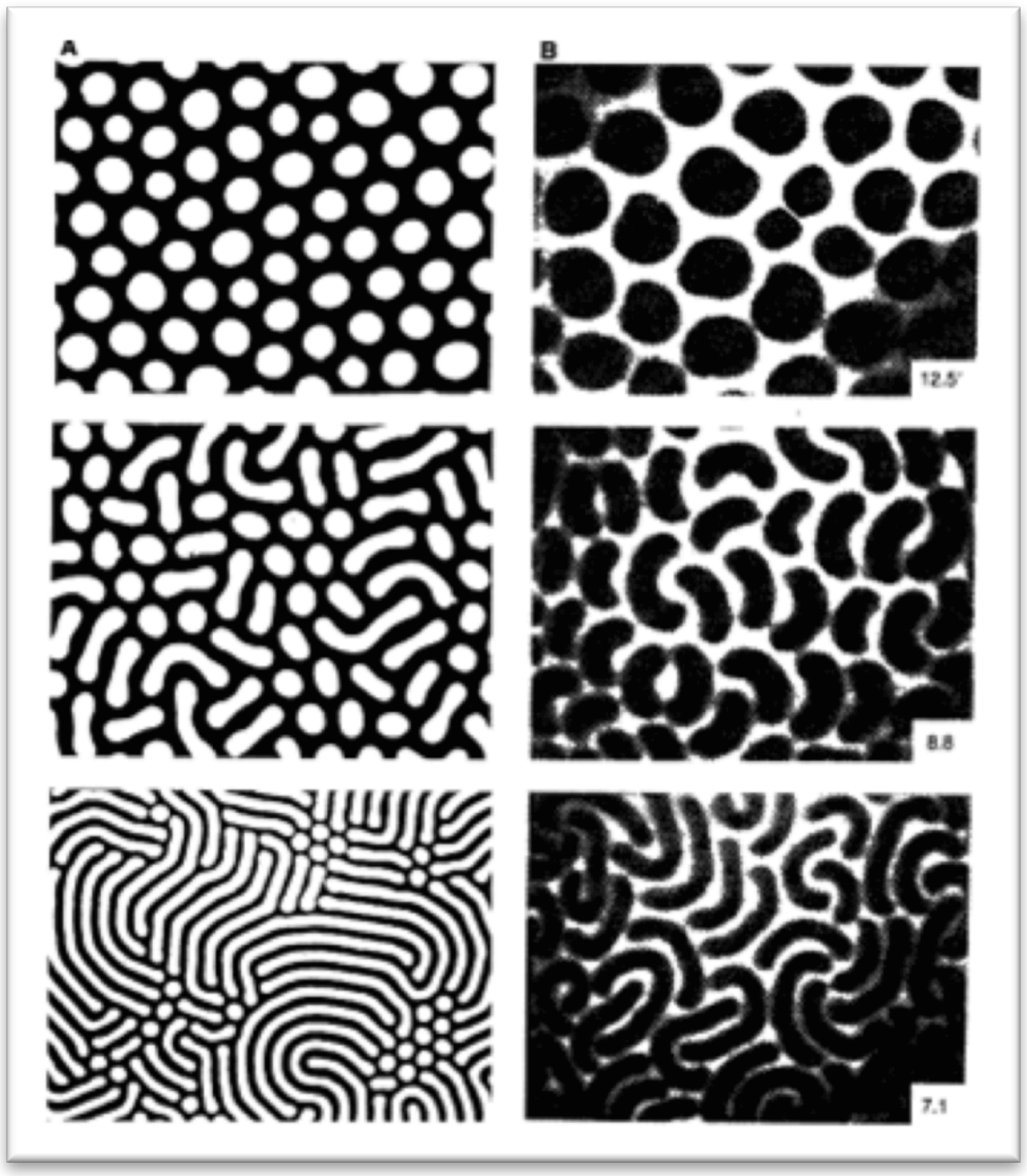

**Figure 2**: Frustrated phase separation in different systems magnetic and organic thin films (From ref.5)

In order to enlighten the interconnection between the different phases present in a systems and the detail of the microscopic (example molecule-molecule bonds) scales to macroscopic properties, the different time and spatial scales must be taken into account. IRIDE will offer this unprecedented possibility, allowing to probe fluctuations by virtue of the excellent match between the relevant length scales and the soft X-ray wavelength as well as the ability to tune the photon energy into a core or valence threshold. The scientific goals can be summarized in the following points:

a) Enlightening the interconnection between the different phases present in a system and relate this info with the dynamics of the local bonds.

b) Exploit new tools for electronic studies of condensed matter by combining element specificity of x-ray techniques with the selection rules and symmetries typical of nonlinear optics.

c) Investigate time-dependent phenomena in solids at surfaces, exotic excited states, ultrafast phase transitions.

d) Assess the validity of present theoretical description in nonlinear properties of condensed matter in the XUV regime and receive specific guidance for the experimental design. Explore both with experiments and with theory regimes of optical nonlinearity in condensed matter in the XUV domain.

These future developments lye in the combination of high temporal resolution in the fs range with the elemental and chemical state selectivity of X-rays. The advance in developing research this project represents, will act as a bridge between the present and the future experiments in condensed matter systems.

Substantial developments in experimental techniques, control of experimental conditions and end-station instrumentation are required for effective utilization of the advanced features of the FEL. Major areas include:

*Sample damage control*

The high power density of the x-ray pulses can instantaneously damage the sample, and refocusing the beam onto the sample down to micron sized spots is often impractical [7]. The ablation threshold for a large variety of systems (metal, semiconductors or insulators) is less than 0.5 $J/cm^2$. For example for pulses of hv=100 eV, photon density about $10^{13}$ photon/pulse and spot size of 100x100 $\mu m^2$, the fluence corresponds to 2 $J/cm^2$ .

More accurately the temperature increasing ΔT, assuming all the pulse energy absorbed on the sample illuminated volume, results:

$$\Delta T = \frac{\Delta N}{\left(\frac{3K_B}{E}\right) * \left(\frac{d^2}{\sigma_{ABS}}\right)} = 4 * \sigma_{ABS}[cm^2] * 10^{23} K$$

where N and E are the photon flux and energy, respectively, $k_B$ is the Boltzmann constant, $d^2$ is the illuminated area and $\sigma_{ABS}$ is the absorption cross section. In the previous example σ values lye in the range of $10^{-19}$ – $10^{-17}$ Mb for almost all materials and considering the spot size, ΔT results 4 $10^4$- 4 $10^6$ K. In other words, a pulse energy of 1 mJ corresponding to $10^{13}$ photons/pulse could raise the sample temperature by up to 100 K after several 10 ps. This will decay again on a ns time scale. An effective sample cooling, however, avoid a significant increase of the average sample temperature. Pioneering experiments at FLASH demonstrated that relatively unfocused spots and high repetition rates limited the sample damage and the perturbation on the measured spectra. Alternatively single shot and refresh of sample can be designed: jet for liquids, feasible movement (piezomotors) of solid samples etc. Moreover a high repetition rate strongly increases the feasibility of specific techniques, as Inelastic X-ray Scattering, RIXS or Valence-band PES.

*Source requests*

The baseline specifications to fully exploit the IRIDE source are here summarized:

1) High brightness pulsed coherent light up to 2 KeV

2) Repetition rate at least of the order of 1-10 Khz: in order to reduce the beam damage and make possible performing Inelastic Scattering, RIXS and other spectroscopies.
3) Tunability *is highly advisable*: smooth tuning across most of the spectral range
4) To allow, for example TR XAS (even dichroic) using the coherent part of the output.
5) Tunability of FEL first harmonic is feasible. Possible PP experiments either by dispersive detection (single shot experiment). In some cases, as in absorption for inner shell C, N, O K-edges and $L_{2,3}$ transition metals, the time structured spontaneous emission could also been used.
6) Fully control of the polarization, from linear to circular: Circularly Polarized FEL pulses are requested to perform dichroic measurements.
7) Pulse duration < ≈100 fs, possibly down 20 fs for ultrafast spectroscopy-Synchronized to short pulsed lasers
8) Two colour capability for Pump Probe experiments with synchronization jitter better than 10 fs, for example: Colour 1: IR - Colour 2: 1000 eV
9) For a future instrumentation layout a common optical configuration could be designed for X-ray/X-ray PP techniques, in addition to a synchronized optical laser system. In this way two colour PP and VIS/X-ray experiments will be feasible. High degree of transverse beam coherence. High degree of temporal coherence up to > 1keV

### *4.4.1.* ***Pilot experiments at FEL@IRIDE in condensed matter***

After the recent starting operation of FELs facilities the techniques improvements and the scientific results in this field are rapidly evolving. In the following we highlight some specific topics and experiments.

**Non-thermal melting**

Melting is generally due to the increase of the mean temperature of the lattice degrees of freedom in equilibrium with the other degrees of freedom, which induce an increase in the atomic motions; according to the Lindemann criterion, melting occurs when the mean square displacements of the lattice atoms reaches 10-20% of the lattice constant. An increase of the atomic motions can be induced also by using fs laser pulses, through different mechanisms; for example, an increase of a few percent of the carrier density with respect to the valence band electron density induces first a reduction and then a disappearance of the shear restoring force, eventually resulting in melting [8]; also the resonant excitation of particular phonon modes may induce coupled and large atomic movements [9]. In addition, following a short and intense excitation, the electrons, lattice and spin degrees of freedom may no longer be in equilibrium; their temperatures equilibrate with different time constants and through different interaction mechanisms. Time resolved studies may monitor such phenomena and measure the different relaxation times; for example, in the case of half metallic ferromagnetic $Sr_2FeMoO_6$, time resolved magneto optical Kerr effect has shown that the electron temperature increases immediately after laser optical excitation, transferring heat to and thermalizing with the lattice within few ps; on the other hand, the spin temperature takes much longer to reach again equilibrium. The strong coupling existing between lattice and spin degrees of freedom in these materials makes it very interesting to look also at the structural evolution after the laser excitation, which happens through thermalization first with the electrons and then with the spins. Time resolved X-ray diffraction is the proper tool, but it was used only in few cases to investigate such phenomena; in a thin crystalline layer of germanium a fs laser-induced solid-to-liquid phase transition was observed, with non-thermal melting taking place within 300–500 fs and followed by strong acoustic perturbations evolving on a picosecond time scale [10]. In InSb a delayed onset of diffusive atomic motion was observed, signalling the appearance of liquid-like dynamics; it was observed that the root-mean-squared displacement in the [111] direction increases faster than in the [110] direction after the first 500 fs, indicating the existence of an initially generated anisotropic fluid phase significantly different from the equilibrium isotropic liquid. In InSb an abrupt softening of the lattice potential as a function of carrier density and at very high carrier density and also the appearance of accelerated atomic disordering were observed [11].

**Carrier dynamics in semiconductors and oxides**

Absorption of light in semiconductors is responsible for a rich variety of phenomena at the heart of important technological applications. Absorption of photons leads to the creation of charge carriers (electrons and holes) which subsequently thermalize and recombine following a complex series of interactions. Typical interactions and their time scales include inter – carrier scattering and emission of optical phonons (0.1 – 10 ps), emission of acoustic phonons (1 – 10 ps), creation of free or bound excitons (often centered on dopant sites) and finally electron – hole recombination (radiative and non radiative). Analysis of the radiative recombination leads to Photo-luminesce (PL) spectroscopy, one of the workhorses of experimental semiconductor physics. Femtosecond time – resolved PL has developed into a powerful tool to study the rich panorama of carrier dynamics; despite its success it lacks structural sensitivity and it would be of great interest to develop a tool to study the transient structural changes associated to carrier dynamics, especially those localized on dopant atoms. For example, in GaAs/AlGaAs structures the free to bound exciton transition is faster in doped samples because the bound exciton is created by trapping the free exciton on the dopant [12].

Limited work has been performed in this direction. For example, Adams et al.[13] were able to detect, on the 100 ps time scale following laser excitation, a transient increase of the X-ray absorption cross – section near the low energy tail of the near edge region; this increase was interpreted as due to the additional final states created in the valence band by laser absorption and available for core electron transitions. In the future, with FEL sources, it might be possible to probe structural relaxations associated with carrier excitation, relaxation and recombination around dopant or minority atoms using resonant XRD or XAS.

Since the pioneering work of Honda and Fujishima [14] and of Grätzel $TiO_2$, especially in the form of nanoparticles or nanostructured materials, has been studied for photo-assisted electrochemical splitting of water ("photocatalysis") and for photovoltaic energy production; there are also significant environmental applications [15]. A complex series of elementary excitation and relaxation processes, the time scales of which span many orders of magnitude, are responsible for the functioning of these photo assisted processes. Injection of carriers in the conduction and valence band may be direct via absorption of UV light on the sub fs time scale or via the dye molecule (able to absorb visible light) in characteristic times of the order of 100 fs – 1 ps. Carriers may be trapped and released by shallow traps in ~ 100 ps – 10 ns or fall into deep traps in ~ 10 ns. To reduce or eliminate carrier recombination, which occurs on the order of 1 – 10 ns, metal nanoparticles deposited on the oxide can act as "scavengers" which are able to supply reducing electrons to the exterior; electron transfer to the scavenger occurs in ~ 100 fs. Slower interfacial charge transfer processes also occur on μs or ms time scales.

Time resolved X-ray absorption and/or X-ray emission spectroscopy can be used to study the modifications of local atomic and electronic structure which accompany photoassisted processes in $TiO_2$ nanoparticles. In particular, availability of x-ray pulses of duration of the order of 10 fs could allow to study changes in the charge state of Ti and / or O related to the evolution of occupancy of localized deep trap states and to the associated structural relaxations; charge transfer from the dye molecules to the nanoparticle and from the latter to electron scavengers could be monitored. Given the time scales described above, storage ring sources (~ 100 ps pulses) allow only to probe the slower processes involved. A unique opportunity at FEL@IRIDE will be to exploit the "two color" x-ray beam using time delayed photon pulses tuned to different absorption edges (e.g. Ti and the scavenger metal) to "follow" excitation and relaxation processes in real time with an original approach.

**Advanced RIXS experiments**

We propose a pathway towards the use of the FEL X-ray source at IRIDE for advanced resonant spectroscopies of solid samples, with the ultimate goal of performing time-resolved experiments. In a broader vision we include under the RIXS denomination also resonant absorption/reflectivity and resonant elastic scattering (diffraction) in the soft x rays, which are scientifically complementary and prerequisite to RIXS itself, and share with it part of the technical setup. The possible availability at IRIDE of circularly polarized photons could allow performing x-ray magnetic circular dichroism (XMCD) on ferromagnetic materials in the femtosecond time scale, providing remarkable new insight in the field of ultrafast spin dynamics of magnetically ordered materials. Ultimately the aim is to realize high resolution RIXS, possibly in a time resolved mode, in order to study the electronic and magnetic structure of (mainly) 3*d* transition metal compounds. This last task is particularly challenging given the requirements in terms of integrated/average flux, energy resolution and beam stability.

The future availability of sub-picosecond, ultra intense X-ray pulses from the IRIDE Free Electron Laser facility stimulates the interest for time-resolved spectroscopic applications. In the most recent years, photon in – photon out spectroscopic techniques have been greatly profiting from the fast improvement of synchrotron radiation facilities and are among the most promising methods to be based also on FEL radiation. On the other hand, the applicability of FEL-based techniques to solid samples is at present under investigation at the LCLS x-ray FEL in Stanford due to damage effects. The goal is thus to explore the experimental conditions that best allow to exploit the exceptional characteristics of the FEL pulses in resonant spectroscopies of solid samples. In consideration for the recent momentous development of RIXS, which has lead several storage ring facilities to realize high resolution RIXS end-stations in the next future (ERIXS at the ESRF, 2014; Centurion at NSLS II, 2016; IXS at Diamond, 2016; Veritas at MAX IV, 2016; Metrix at Bessy II, 2017). An important project is foreseen at the European XFEL in Hamburg (hRIXS), with the ambition of performing single puls pump-probe RIXS experiments on solid, liquid and gaseous samples. Thus IRIDE cannot ignore RIXS and related techniques: specific strengths of IRIDE in this connection will have to be carefully considered in the future, but it seems evident that high repetition rate and (relatively) low pulse intensity are the properties that differentiate IRIDE from the European XFEL. RIXS at IRIDE might cross the gap between storage rings and XFEL-EU. We propose a pathway towards the development of time-resolved resonant inelastic X-ray scattering in the 100-1500 eV energy range to be realized at the FEL@IRIDE facility.

The energy range covered by the FEL@IRIDE includes the L-edges of 3*d* transition metals (from Ti to Cu). A variety of interesting compounds could be studied with resonant spectroscopies going from high $T_c$ superconductors to giant magneto-resistance manganites, from Mott insulators to technological semiconductors, from metallic ferromagnetic films to Kondo systems. In all those cases the possibility of resonant time-resolved studies in a pump-probe setup pertains almost uniquely to FEL facilities. The 20-200 fs X-ray pulse duration combined with femtosecond laser pulses or fast pulsed magnetic fields serving as pump will allow the exploration of the dynamics of the electron interaction (that determines the electronic and magnetic properties of materials) with lattice-related degrees of freedom, of the electron-electron interaction and of the quasi-particle formation. Moreover, non-linear phenomena in the X-ray absorption and emission will become naturally at hand at the FEL source: their study does not require a pump-probe set up and would inaugurate a totally unexplored field for the solids.

Standard RIXS experiments require the photon fluxes of 3$^{rd}$ generation synchrotron radiation beam lines, whereas XAS experiments are much less photon hungry. Modern RIXS necessitates average monochromatic flux of at least $10^{12}$ ph/s concentrated on $10^3$ micron$^2$ sample area, whereas XAS and reflectivity can be performed with 2 orders of magnitude lower flux possibly spread over 100 times bigger area. These requirements will have to be confronted with the target performances of IRIDE. In fact, judging from the numbers available today, we think that in the "1.5 GeV mode" IRIDE will deliver a flux (peak and average) sufficient for all the experiments mentioned in this section. On the other hand the short X-ray pulse duration (<100 fs), as compared to standard synchrotron light pulses, makes the FEL extremely attractive for time-resolved experiments. The drawback, however, is that working at the high repetition rate, the synchronization of the X-ray pulse with an external laser source acting as a pump is a non-trivial problem. However, this well-known issue has been recently addressed, suggesting a possible workout [16].

RIXS is unaffected by macroscopic charging and compatible with strong magnetic fields, unlike electron spectroscopies, and it has momentum resolution, unlike optical spectroscopies [17]. RIXS with soft x-rays is especially interesting because the transitions that directly probe the valence states of the most interesting materials fall in this domain, namely the $L_{2,3}$ ($2p \rightarrow 3d$) edges of the $3d$ TMs, the $M_{4,5}$ ($3d \rightarrow 4f$) edges of the lanthanides, and the oxygen K ($1s \rightarrow 2p$) edge. While pioneering RIXS experiments have been performed over more than a decade [18;19], recent drastic improvements in the experimental conditions have considerably enhanced the general relevance and impact of the technique. High-resolution data from the SAXES instrument at the SLS have clearly demonstrated that in a RIXS spectrum it is possible to distinguish different charge states and to determine with high accuracy the local crystal field parameters [20]. They have also shown that the q dependence of RIXS can be exploited to map the dispersion of collective excitations [21]. In particular, this has enabled the complete determination of the spin wave dispersion in selected insulating and superconducting layered cuprates [22;23;24]. Moreover RIXS, with fair energy resolution (0.2-0.5 eV), has been very successfully used to measure the crystal field (*dd*) excitations in strongly correlated electron systems like cuprates and simple oxides [25;26;27;28]. Those excitations, ranging from few 100s of meV's to some eV's in $3d$ transition metals, are electric dipole forbidden for optical absorption but not in the double photon process of RIXS. A special effort has to be devoted to the design of the X-ray spectrometer that analyzes the spectral distribution of the X-rays emitted by the sample. The optimization of the optical matching between beamline and spectrometer also plays a crucial role (a small focus on the sample helps in getting high resolution without loss in the efficiency) [29].

**X-ray Absorption Spectroscopy, X-ray Magnetic Circular Dichroism and Resonant X-ray Reflectivity ( RXR) experiments**

These methods, widely used at the $L_{2,3}$ edges of 3*d* transition metals (450-1500 eV), have proved to be highly sensitive to the electronic structure of the samples [30;31;32]. XMCD in particular can provide element selective information on the magnetic status of the sample [33;34]. Very recently, a pioneering time-resolved XMCD experiment with femtosecond resolution has been conducted on thin nickel film [35], using the slicing technique at the BESSY synchrotron radiation facility. Although this method provides extremely short (100 fs) X-ray pulses naturally synchronized with an external laser source, the photon flux is still exiguous for routine operations. The high brilliance of the FEL would therefore overcome this limitation. The study of magnetization dynamics on the femtosecond time scale has received considerable attention in the last decade. Most of the investigations are conducted with lasers in the near-IR to near-UV spectral region. XMCD would add chemical sensitivity to these measurements opening extraordinary capabilities and providing a considerable leap in the understanding of ultrafast spin dynamics. Time-resolved XAS and XMCD on longer timescales (100 ps to 100 ns) have been measured at the $L_{2,3}$ edges [36;37;38]. Accessing the sup-picosecond time window, however, seems to be the natural prerogative of the FEL. These techniques are broadly applicable to any materials containing 3d transition metals and/or lanthanides, metals as well as insulators, strongly or weakly correlated.

**Resonant Elastic X-ray Scattering (REXS)**.

In the recent years soft x-ray diffraction performed at the Tm L edges and O K edge has been used several times to study the orbital and spin ordering and the charge density modulation in strongly correlated systems [39;40;41;42]. This technique can be ideally extended to pump-probe, where the pump melts the relevant order parameter and x-rays can probe the time-scale of the loss and recovery of the long and mid-range correlation. This kind of experiments can be performed jointly with XAS, in order to make the link between the ordering and the local properties usually probed by absorption spectroscopy. REXS is a difficult technique because it requires an in-vacuum diffractometer, but it is much less photon demanding than RIXS because it does not require the energy analysis of the scattered photons. The first results of time resolved REXS from LCLS are being published right now, such as the interesting case of striped $La_{1.75}Sr_{0.25}NiO_4$, where the spin and charge ordering re found to follow different time-scales [43].

For the realization of XAS, REXS and RIXS experiments, mainly in the pump-probe mode, a series of technical issues will have to be solved. The most important are the following.

1) A femtosecond pulsed laser source will be needed as pump for pump-probe experiments. The main difficulty will be in the synchronization to the x-ray pulse.

2) The intrinsic band width and the pulse-to-pulse stability of the central energy of the FEL energy peak will determine the need and the characteristics of a monochromator at the beam line.
3) For XAS and REXS experiments it will be important to scan the energy of the FEL source. The design of the beam line should be compatible with these requirements.
4) For XAS and reflectivity a relatively big (200x500 micron$^2$) spot on the sample is ideal, whereas for RIXS a small spot (5-10 micron), at least in one direction, is necessary. The optics of the beam line should allow some flexibility in the focusing.

**Surface Science and Femtochemistry**

Many scientific cases and experiment on new X-ray FEL facilities [FLASH, LCLS, Fermi] are focussed on ultrafast spectroscopies applied to surfaces and interfaces, and TR- PES experimental tools have been developed recently, providing unique information on the dynamics in surface systems. Recently it has been shown that ultrafast pump-probe x-ray fluorescence spectroscopic techniques based on an x-ray free-electron laser, the Linac Coherent Light Source (LCLS), can be used to probe the electronic structure of a transiently populated, weakly adsorbed state in CO desorption from Ru(0001) [44]. The observed electronic structure changes are consistent with a weakening of the CO interaction with the substrate but without notable desorption and a phonon-mediated transition into a weakly adsorbed precursor state occurring on a time scale of >2 ps prior to desorption, while within the first picosecond after laser excitation the metal-adsorbate coordination is initially increased due to hot-electron-driven vibrational excitations. This process is faster than, but occurs in parallel with, the transition into the precursor state. With resonant x-ray emission spectroscopy, each of these states was selectively proved and determine the respective transient populations depending on optical laser fluence.

Femtosecond-chemistry and the understanding the dynamics and formation of a chemical bond at interfaces strongly benefit of FEL facilities. Thus the dynamics of the complete electronic structure can be observed using PES as a tool, providing a direct time domain approach to the study of electron correlation in atoms, molecules and solids. These studies enlighten the paramount interest of this topic.

**Ultrafast Photoemission Spectroscopy and solid dynamics**

Photoemission is the most direct method to understand the electron dynamics of surfaces, condensed matter (both solids and liquids), clusters, molecules and atoms. This is so because it utilizes the photoionization process itself and the related relaxation processes in the most direct way: by measuring the properties of the electrons themselves. A photoemission experiment at IRIDE will exploit both the high energies, the possibility of a defined polarization state of the X-rays and the ultra-short pulses. Several groups are active in TR-PES even with laser sources. Though the induced by spatial charge leads to severe experimental problems and an intense theoretical effort is active in this field. The high X-ray energies available at IRIDE will - generally speaking - lead to lower cross sections. This will make the problem of so called space charge broadening less destructive. The space charge broadening is caused by the high charge density of the plume of electrons leaving the sample after each FEL pulse. This effect has recently been studied in great detail [45]. It should be noticed that also high kinetic energies reduces the effect of space charge.

The high average flux of photons from IRIDE will also enable photoemission studies of diluted samples. Also here the lower cross sections and the weaker signal - which per se is no advantage but will reduce the space charge problems.

With the high photon energy possible at the European XFEL it is of course possible to study deeper core shells. These are often naturally broadened, relaxing the demands on the resolution of the analyzer.

The efficient detection of medium and high kinetic energy electrons implies a development effort on both energy and angular selection of photoelectrons, of efficient spin detection and ultrafast spin-polarimetry, and on electron detectors that must provide adequate space resolution (to exploit angular resolution in parallel detection mode), multi-hit linearity in order to provide a high dynamical range for photoemission signals of very different cross sections, and a high read-out refresh rate to cope with the 4.5 MHz clock of the XFEL. The photoemission community is also interested in the low-energy mode based on 10-GeV operation of the XFEL, provided that the beam-line optica actually allow for the efficient use of near C K-edge energies.

The energy and angular analysis calls for the further development of the new generation of angular resolving time of flight spectrometers that have recently become commercially available (ARTOF) [46]. These instruments are capable of analyzing the three components of the momentum of the photoelectrons simultaneously. If such an instrument can be combined with spin resolved detectors it creates a very complete experiment. There are still some development steps to be taken on this roadmap, but they are most probably possible to overcome, provided that at sufficient funding can be obtained.

As presented in Fig.1, ultrafast time and spin resolved spectroscopies need a direct access to both the relevant energy (1-100 meV) and time (1ps-10 fs) domain, where electron, spin and lattice relaxation time constants are separately accessible. At present, it is an open question whether and how the electronic bands, the exchange splitting, and the spin-orbit coupling change in the laser-excited state.

The inset of Figure 1 represents the three temperature model for magnetization dynamics, showing the various processes occurring after a femtosecond laser pulse has interacted with a ferromagnetic material. From early ultrafast demagnetization experiments following an optical stimulus performed in metallic systems the indication that electron thermalization occurs on the time scale of 10 fs and that electron phonon equilibrium is reached on the 1 ps timescale were retrieved. The 3-temperature model shown in Fig. 2 is of purely thermodynamic origin and hence fails to account for the wavelength of the excitation. With a tunable source, from FEL sources, the dependence of bath temperature on wavelength can be investigated. Other open questions concern the spin-lattice coupling times with predictions ranging from some 100 fs to 100 ps. All optical as well as time resolved PES and fs-slicing X-ray experiments support an ultrafast electron-spin coupling of some 100 fs [47]. However, there seem to be indications that a transition from a ferromagnetic d-shell transition metal to a half-metallic system increases this equilibration time drastically. In fact, the microscopic mechanisms are heavily disputed. Eliott-Yafet spin scattering, super-diffusive spin transport with and without phonon contributions, etc. More experiments are needed to detail the mechanisms and their interplay.

In the case of strongly correlated systems, the investigation of non-equilibrium states by fs-resolved spin analysis will address the *e*-ph scattering in the proper time domain. Indeed, an ultra-short light pulse excites the electrons of a metal on a shorter time scale than the perturbation of the atomic lattice. After the spin and charge fluctuations of the photo-excited electrons have thermalized, the rate of energy transfer from hot electrons to the lattice is governed by the electron-phonon interaction. In magnetic systems, ultrafast Kerr spectroscopy and electron spectroscopy with spin analysis are able to follow the resulting loss of the magnetization, related to the transfer of energy between the electron, phonon and spin lattice.

It thus becomes clear that joint efforts are mandatory to shed some light on the fundamental physics underlying ultrafast demagnetization and ultra-fast generation of magnetic order. Clearly electronic excitations are involved. Thus, ultrafast electron spectroscopy in the vicinity of the Fermi energy with high temporal or energy resolution in combination with laser excitation is a key ingredient. To probe the transient dielectric function, laser based photon in /photon out spectroscopy techniques (linear reflectivity and linear transmission) are fast and reliable experimental tools. Together with laser-based time resolved Kerr effect, these experiments probe a collective state of the system, the magnetization. Photoemission-based Spin Polarization measures the evolution of the band structure in a single electron picture and may elucidate whether this picture still holds on ultra-short time scales after photo-excitation in the nonequilibrium regime. Adding spin detection to other spectroscopies allows one to disentangle the spin and orbital contribution in ultrafast magnetic processes.

The class of experiments outlined above should clarify: a) how tools probing the electronic and lattice properties at equilibrium and out of equilibrium are necessary to tackle fundamental scientific challenges in solid state physics, b) the presence of a common lack of technology in ultrafast spectroscopies, spanning over a-priori different domains, c) the strategic perspective of creating a common, highly interdisciplinary, community focused on the investigation of electronic correlation in the out of equilibrium regime and d) the basic materials science parameters to tailor magnetic anisotropy, blocking, Curie temperature.

Our focus is the fundamental understanding of charge, spin and lattice dynamics, i.e. the direct prerequisite to understand magnetization dynamics and the peculiarities of nanostructured matter, direct spin injection, current injection as well as the study of ground state properties of spin-stabilized long range structures in non-magnetic materials and other exotic phases of matter in special and/or extreme conditions.

**Ultra-fast dynamics in graphene and purely 2D systems**

Carbon in the purely two-dimensional (2D) lattice of graphene, since its discovery has been shown to be an actual prototype system for fundamental science as well as appealing applications, thanks to the unprecedented characteristics of graphene for its ponential use in nanotechnology applications. A purely 2D graphene sheet can be prepared by chemical vapour deposition (CVD) on hexagonal-terminated surfaces, among which transition metals that offer a well-suited support for epitaxial growth. In particular, the Ir(111) surface has revealed to be a good template for minimizing the carbon–substrate interaction [48], thus leading to a quasi-free standing graphene layer.

Upon electromagnetic excitation, several time-scales can be found in the de-excitation in C-based systems, due to the coupling of the electron-hole excitations with optical phonons and eventually lattice thermalization. These processes of ultra-fast dynamics still present intriguing problems related to the actual nature of the C-lattice.

On-campus techniques have been applied on highly-oriented pyrolitic-graphite (HOPG) by using both pump and probe in the optical energy range, and the analysis of the variation in the optical constants obtained upon measuring transmission and reflectivity as a function of time delay between pump and probe allowed to single out the main sources of de-excitation following the electron-hole excitation across the pi-bands of HOPG: electron-electron fast scattering (order of 30 fs), carrier-optical phonon scattering (order of 200 fs), finally phonon thermalization in the ps-time scale. Recent on-campus experiments have been carried-out at 3-55 layers grapheme [49], where the 1.6 eV photon energy allowed a photo-excited carrier density of the order of $10^{11}$ $cm^{-2}$, obtaining quasi-equilibrium conditions in an analogous time-scale. Only very recently, on-campus experiments have been performed by applying variable photon energy pump in the UV spectral region, while probing in the visible photon energy region and the de-excitation in HOPG has been accompanied by a rather strong band-gap renormalization [50]. This result was achieved thanks to the variable pumping photon energy, tuned to the highest joint-density of states for graphite, close to the M point of the surface Brillouin zone (SBZ).

Still, there are open questions concerning the fast de-excitation processes in actual 2D graphene single sheet. Thus, the technology for obtaining actual graphene and the photon-beam tuning with appropriate polarization are becoming a necessary pre-requisite, which can be solved by using the FEL radiation from a machine like IRIDE. In particular, we can envisage to apply excitation in the UV range, or adjusted to other poin ts of the SBZ, even involving sigma-like bands, and probe the C 1s core-levels (about 284 eV) to the first π*-empty resonances excitation by x-ray absortpion spectroscopy (XAS), and follow the processes of de-excitation as a function of time. Furthermore, there are other 2D systems with different electronic properties, like BN and BCN layers, presenting analogous purely 2D nature, with clear symmetry reduction and band-gap opening. Several study on hybrid nanostructures consisting of patchwork of graphene and BN 2D layers have been recently considered for tailoring of physical properties of graphene, and we are exploiting the technology for in-situ preparation of these hetero-systems in 2D. The ultra-fast dynamics can be probed also in the de-excitation processes at these hetero 2D systems, opening a new path for fundamental physics knowledge and potential applications in nanotechnology.

**Scattering and Wave Mixing non-linear optics techniques extended to FEL**

Recently the wave mixing have been tested on diamond at LCLS [51]. Glover et al. directed optical laser light and a beam of X-ray light from the Linac Coherent Light Source into a diamond sample. Most of the beam's X-rays are scattered elastically (their wavelength is preserved) and produce the standard Bragg intensity peak on the detector. The rest of the X-rays undergo inelastic scattering, resulting in optically modulated X-rays which have a frequency that is the sum of the X-ray frequency and the laser frequency. These rays generate an intensity peak that is slightly shifted from the standard peak. In standard X-ray crystallography, X-ray illumination and scattering at different crystal orientations yield information about a crystal's three-dimensional charge density. In much the same way, the optically modulated, scattered X-rays can allow the three-dimensional, optically induced charge-density changes to be reconstructed. The optically modulated X-ray diffraction was the first proof of principle of sum-frequency signals recorded for one crystal orientation. Extending the experiment to several crystal orientations, will allow determination of the three-dimensional, induced charge-density variation, should be straightforward. X-ray and optical wave mixing in crystals enables optical polarization to be measured on a microscopic scale.

One of the most important goals of condensed matter physics today is the experimental determination of the collective dynamical properties due to atomic, electronic and magnetic density fluctuations, responsible for specific vibrational, electronic and magnetic elementary excitations, and often related to many macroscopic properties characteristic of the considered system. TR x-ray scattering can be measured to examine localized disturbances resulting from photon excitation. Information about t-dependent sizes and shapes of electronic excitations (e.g. excitons, charge transfer states) and the molecular distortions surrounding them, localized structural responses (e.g. solvent rearrangements around photoexcited species), and collective structural responses (e.g. nucleation or domain evolution) can be extracted and correlations between response time scales and distance scales are elucidated. The S(Q,E) of disordered systems has been measured in the mesoscopic region only recently. TG provides information on dynamics in condensed matter in terms of the Fourier transform of the dynamic structure factor, F(Q, t), i.e. information on the collective dynamics is provided, associated with density (lattice, charge or spin) fluctuations, whose base quantity is the dynamical structure factor S(Q,E). This is of extreme interest for the study of disordered systems since it will make accessible the mesoscopic kinematics region that cannot otherwise be explored by the use of visible laser based instruments. In Fig.3 we report a sketch of the available kinematics regions.

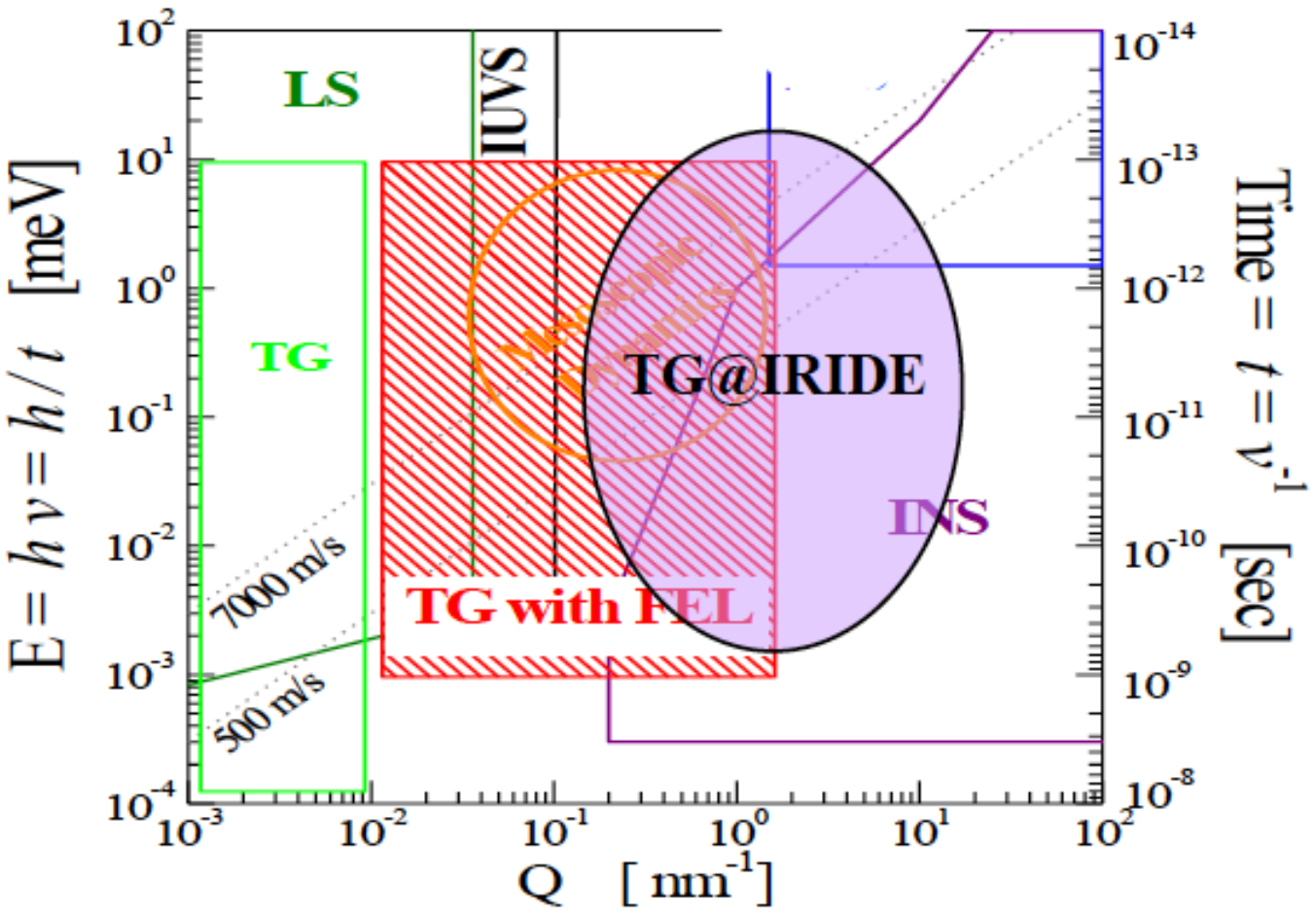

**Figure 3** Kinematic regions accessible by existing techniques: Transient Grating (TG), Light Scattering (LS), Inelastic UV Scattering (IUVS), Inelastic X-ray Scattering (IXS) and Inelastic Neutron Scattering (INS). In red we display the region that will be made accessible by the present project (TG with FEL). The two dotted lines indicate energies of collective excitations with characteristic sound speeds of 500 m/s and 7000 m/s. The oval shadowed area is related to the regions accessible to the SparX pulses, with a slight superposition with the Fermi (indicated as TG with FEL).

In this field Transient Gratings Spectroscopy (TG) is a coherent mixing (x-ray excitation, x-ray probe) non-linear technique to study the dynamics in condensed matter; it is usually carried out with pulsed laser sources. TG is based on the splitting of the beam in three components (PP in the different photon range (soft X-soft X in our case) and a Δt varying the delay time. This will be achieved by using two identical pulses impinging at the same time on the sample to create a standing wave that imposes a transient density modulation in the sample. A third, delayed pulse at the third harmonic is then scattered by the TG. The scattering amplitude is related to the collective dynamics present in the system and can serve as a time-dependent monitor.

The feasibility of TG in the EUV and soft X pulse energy range has been explored some of us and a facility at the new Fermi FEL (Trieste) is in commissiong. The setup is expected to be able to study a wide variety of materials, at it would permit to extend the actual knowledge of the dynamics at nano-metric scale allowing sensitive probing of interfaces, extremely thin films, heat transport and correlations in nanostructured materials.

Novel transient grating spectroscopy setups have been recently developed to produce charge grating or spin grating in complex materials. The importance of extending TG to IRIDE source is, first of all, due to the possibility to have larger penetration depth that will allow to use, for many materials, transmission geometry (i.e. to fulfil the tick TG conditions) thus having a simpler detection condition. At the same time the constraint on the sample environment may be realised being not any more necessary the precautions typical of VUV experiments. Moreover the maximum accessible momentum transfer (Q) will be higher and, namely, in the range of 10 $nm^{-1}$ making the proposed technique also complementary to well established IXS and INS techniques

**Probing matter under extreme conditions**

The expected performances of the IRIDE FEL sources, delivering 10-100 fs pulses in the 300-3000 eV range open the way to investigate fundamental properties of condensed matter under non equilibrium thermodynamic conditions with an unprecedented temporal resolution. In particular, the new machine is expected to represent a unique and optimized tool for accessing ultrafast dynamical processes at the nanometric scale, including investigation of materials under extreme and excited or metastable conditions.

At the time of writing, several FEL sources with different characteristics are operating or under development. Matter under extreme conditions is part of the scientific case of the LCLS (MEC beamline, Stanford) and XFEL (European X-ray FEL, Hamburg) facilities. In typical single-shot FEL experiments, a large fraction of the electrons of the specimen are excited within the pulse duration, raising the temperature of the specimen. A typical sample equilibrates its temperatures within a few picoseconds and can reach very high temperatures (up to $10^3$-$10^5$ K) still maintaining typical densities of condensed matter (warm dense matter -WDM- regime). The ultrafast isocore heating induced by the interaction of the FEL pulses with condensed matter, depending on the intensity of the pulses, is able to raise pressures up to the Mbar range. Matter can be measured at high-pressure/temperature conditions for a few ps, before the hydrodynamic expansion of the ion lattice takes place.

These states of matter is poorly known and exceedingly difficult to study, while its knowledge is of basic interest because such disordered states are those found in the interior of large planets and in stars. As an example, intense and ultrashort FEL pulses were used for creating and investigating matter in, or near to, the warm-dense-matter (WDM) regime at FLASH (Hamburg) [52]. More generally, the availability of ultrashort and intense FEL pulses, in a pump-probe configuration allows us to probe excited and metastable short-living systems. The new FEL facilities are continuously upgrading their performances and the user community is expanding, including that of beamlines dedicated to matter under extreme conditions (see for example TIMEX [53] at Fermi@ELETTRA).

However, new perspectives are opened by the possibility of having a new x-ray FEL facility with the unique characteristics prospected for IRIDE in terms of pulses wavelengths, duration, intensity, tunability and with the possibility of an intrinsic jitter free two-color configuration. In fact, ultrafast pump -probe experiments performed under extreme and metastable or non-equilibrium conditions are relevant to a variety of physical and chemical phenomena for which full control and understanding are presently beyond our experimental capabilities. This includes forefront research in high-pressure and high-temperature physics, non-equilibrium and metastable states of matter, applied material studies, understanding of chemical reaction and catalysis paths under extreme conditions, planetary interiors, inertial fusion, and various forms of plasma production in which energy is rapidly deposited into a solid.

Our previous experience with currently available FEL sources indicates that several techniques and pump-probe configurations can be successfully used for the study of matter under such non-equilibrium and extreme conditions, including XAS which is sensitive to both electronic and structural ultrafast changes. A FEL two-color pump-probe configuration could be extremely useful to study various fundamental phenomena (dynamics of non-thermal and thermal melting, for instance). A typical experimental configuration possible with a new machine includes a first FEL pulse used as a pump (for example in the soft x-ray range) and a second FEL pulse tuned near an absorption edge, that can be used to detect the dynamics of phase transitions within timescales of the order of picoseconds. Similar configurations can be used for XRD and other techniques, allowing us full control and time-resolved imaging of phase transitions.

Furthermore, those experiments include studies of phase transitions involving metastable states occurring in regions which are presently un-accessible. An important example is the occurrence of polyamorphism and the hypothesis about the existence of a coexistence line and a critical point separating low-density and high-density fluids in a class of substances which include water, C, Ge, Si and their oxides.

In spite of the fundamental importance of these phenomena, the transition region often is located in a no man's land where crystallization of the undercooled liquid or of the glassy phase prevents observation of the fluid structure transformation.

Ultrafast heating using a FEL pulse will also allow probing of presently un-accessible metastable states in glassy and liquid matter with extremely high nucleation rates. In particular, ultrafast heating of amorphous germanium (the same holds for other glassy systems including a-GeO2, a-SiO2, a-Si, water and so on) will allow us to study the existence of a two-fluid regime with a transition line and of a hypothetic critical point between the low-density semiconductor and high-density metallic phases. Photon energy, sample thickness, beam dimensions can be adjusted to reach selected metastable states so a wide region of the phase diagram can be probed. Under suitable conditions, metastable configurations possibly useful for applications (hard materials, special energy gap requirements and so on) can be obtained and frozen in their states for successive exploitation.

As outlined in this overview, the class of experiments possible at the new IRIDE facility is extremely important for probing states of matter under extreme conditions that are presently un-accessible, and the new knowledge is extremely useful for progresses in basic science and for studying new materials for advanced applications.

**Theoretical tools**

In recent years the role of computational spectroscopy, that is the *ab-initio* simulation of spectra, has become increasingly important. Thanks to increasingly refined codes and to more easily available computing power the interpretation of x-ray spectroscopic data is increasingly performed with the aid of computational tools. In the field of fs x-ray spectroscopy important new applications of computational spectroscopy will occur, in consideration of the need to predict the spectra of short lived transient states for which it will often be impossible to record static experimental spectra of reference compounds. In the specific field of XAS, it is to be expected that initially spectra in a narrow energy range (~ 40 eV) might be recorded, highlighting the usefulness of *ab – initio* simulations of this spectral region. In this context, a very useful approach to simulations is based on the real – space multiple scattering (MS) formalism, which has been especially applied to XAS but can usefully be extended to the interpretation of PES and related techniques.

The short available energy range points to a MS scheme going beyond the muffin-tin (MT) approximation for the shape of the potential and the effective one-particle scheme. This can be done following some general MS schemes proposed in the literature that must be implemented in a practical and easy to use computer code. At the moment MXAN [54], the code used in the low-energy XAS data analysis, works with the MT approximation within a one-electron quasi-particle scheme. An interesting perspective for the analysis of short lived states is to take advantage of Molecular Dynamics calculations to predict the atomic structure.

### *4.4.2.* References

## 4.5. Radiation Beamlines

### *4.5.1. General Considerations*

The FEL@IRIDE will produce radiation from soft to hard X-ray, arriving at the Angstrom with the higher harmonics with extremely narrow bandwidths in very short time pulses, the fs regime. As a consequence of the high number of photons available in the short duration pulse emission an unprecedented high peak power up to GW imposes severe constrains to the radiation handling in the beam-line design, in particular special attention is required to avoid radiation damage on optics (mirrors, gratings, slit blades, etc.). Also the keeping of temporal duration of FEL pulses after photon transport and monochromatization is a new challenge imposed by these new sources.

Some experiments are feasible by using the intrinsic energy resolution provided by the FEL source allowing keeping the pulse length at the shortest possible value. When a higher degree of monochromatization is needed, a lengthening of the pulse duration is to be paid. Consequently, the fundamental conflict between higher energy resolution and simultaneously short pulse availability should be faced in the beam line design according to the specific needs.

Here we plan to have at least two different beam lines in order to maximize the users' accessibility. The two beam lines will be optimized for hard X-ray short-pulse and soft X-ray high energy resolution handling. Depending to the repetition rate we are thinking to design the optics to use the two beam lines at the same time. If this it will not be possible, the switching between them is accomplished by means of a moveable optics.

This concept has been already used successfully at insertion device beam lines in many Synchrotron Radiation Facilities.

In comparison to conventional SR beam lines, FEL beam lines stand for their length, of the order of 100 meters, especially those for sub-nanometer wavelengths. Physical reasons make such length compulsory.

The scarce reflectivity of all the optical elements in the range of tenths of nanometer imposes very high incidence angles for the radiation onto the optical surfaces (90°›α›89,5°). This means that, in order to have sufficient room to independently operate, the distance between two beams lines from the switching mirror must be, at least, of about 100 m.Very important is also the need of spatial separation of the FEL radiation direction from the Bremsstrahlung γ radiation beam generated inside the undulator by the interaction of highly energetic electrons with the residual gas. Such a radiation, in addition to the natural hazards to the health safety, might be harmful for all the apparatus along the beam line from those of radiation diagnostic to the experimental station.

Many experiments such as nanometric imaging, single shot coherent scattering on macromolecules and so on, can be realized only if the FEL radiation is concentrate in a few tens of $nm^2$. In a 100÷120m long beam line this can be simply achieved with a proper choose of the magnification ratio, $f_{source}/f_{image}$ where the first term is of the order of 100 m while the second one is some millimeter.

A further consideration that leads to prefer long FEL beam lines concerns the needs of limiting the exceptionally high power that hits the investigated objects and perform a strong reduction of the contribution of higher harmonics, when needed. A noble gas filter is an optimal solution but, in order to avoid light scattering phenomena that might degrade the quality of the beam.it has to be very long (80÷ 100m) with a very low pressure inside ($10^{-4}$ ÷$10^{-1}$ mbar).

The principal characteristic of the FEL radiation is its very short pulsed structure that should be preserved as much as possible passing through the optical elements along the whole beam line. In fact when this radiation pulses are energy dispersed by a crystal monochromator its time distribution results modified. This is a consequence of two contributions: a) the different optical paths travelled by the individual photons within the same pulse and b) the mutual dependence of pulse length and energy bandwidth through the Fourier transform limit relation $\Delta T \Delta E \geq h/4\pi$ In the case of mirrors the optical path differences are close to zero and no energy dispersion is introduced in the photon pulse leaving almost unperturbed the pulse width. In the case of diffractive elements the light of wavelength λ illuminating a crystal with N planes density at an angle of incidence α is diffracted at angle β at first order. The diffraction pattern is determined by the interferences of rays which experienced a delay resulting in a path difference given by the equation $N\lambda=(\sin\alpha-\sin\beta)$. A pulse time width variation is roughly estimated to be equal to the path difference divided by the speed of light. According to beam divergences and the angles of incidence needed, the pulse width broadening will range from 0.01 to 10 ps depending on the requested photon energy resolution [1].

The very high peak energy reaching the optical elements together with the extremely short pulse length make different the design of FEL beams lines with respect to standard synchrotron radiation (SR) beam lines. This difference requires modifications of the usual SR beam line designs as well as the choice of proper materials for optical coating and bulk. The peak energy density from the FEL is four to five orders of magnitude higher than what produced at 3rd generation synchrotron undulator beams. Such energy, delivered in sub-picosecond pulses, creates a large number of excitations and ionizations at surface atoms of the optical elements, inducing desorption (ablation) of surface layer atoms [2]. Such behavior suggests avoiding strong focusing of the undispersed FEL radiation onto optical elements (slits included) and makes use of grazing incidence angles on the optical surfaces more suitable for photon transport. Finally, surface materials with as low as possible ionization cross section is preferred [3].

The typical integral power load delivered on the first optical element by a FEL with a 30 Hz repetition rates ranges around some hundredths of watt. Also taking into account the power loads due to the spontaneous emission of the undulator, as well as, the radiation emitted by the bending magnet used to deflect the electron beam out of the photon trajectory very low values result for the average thermal load, values much smaller than what expected in typical SR beam lines, making conventional side cooling sufficient to remove the average heat load from optical elements.

On the other hand the peak power on the mirrors may be in the GW range. At these values, the principal problem may no longer be thermal distortion, but rather the possibility of plasma formation at the optical element surfaces, which affects the reflection efficiency and may even cause permanent damage. Formation of plasma will depend on the radiation dose per pulse, described as the energy absorbed per atom per pulse. In X-ray range, tenths of degrees grazing angles are commonly used to obtain good reflection efficiency and to reduce power per unit area on the mirror surface. The induced heat load, however, is related to the deposited energy per unit volume, i.e. the dose. This quantity becomes small only at incidence angles significantly larger than the critical angle $\alpha_0$ for two reasons:

1) the penetration of the field into the mirror becomes roughly independent of the angle and no longer compensates the increasing size of the beam footprint,

2) the mirror also starts to reflect efficiently. In some cases in X-ray region the index of refraction of matter is smaller than its vacuum value allowing taking advantage of total external reflection from optical surfaces at incident angles higher than the critical angle $\theta c$. In this regime the penetration depth of radiation into materials is nearly independent on the incidence angle.

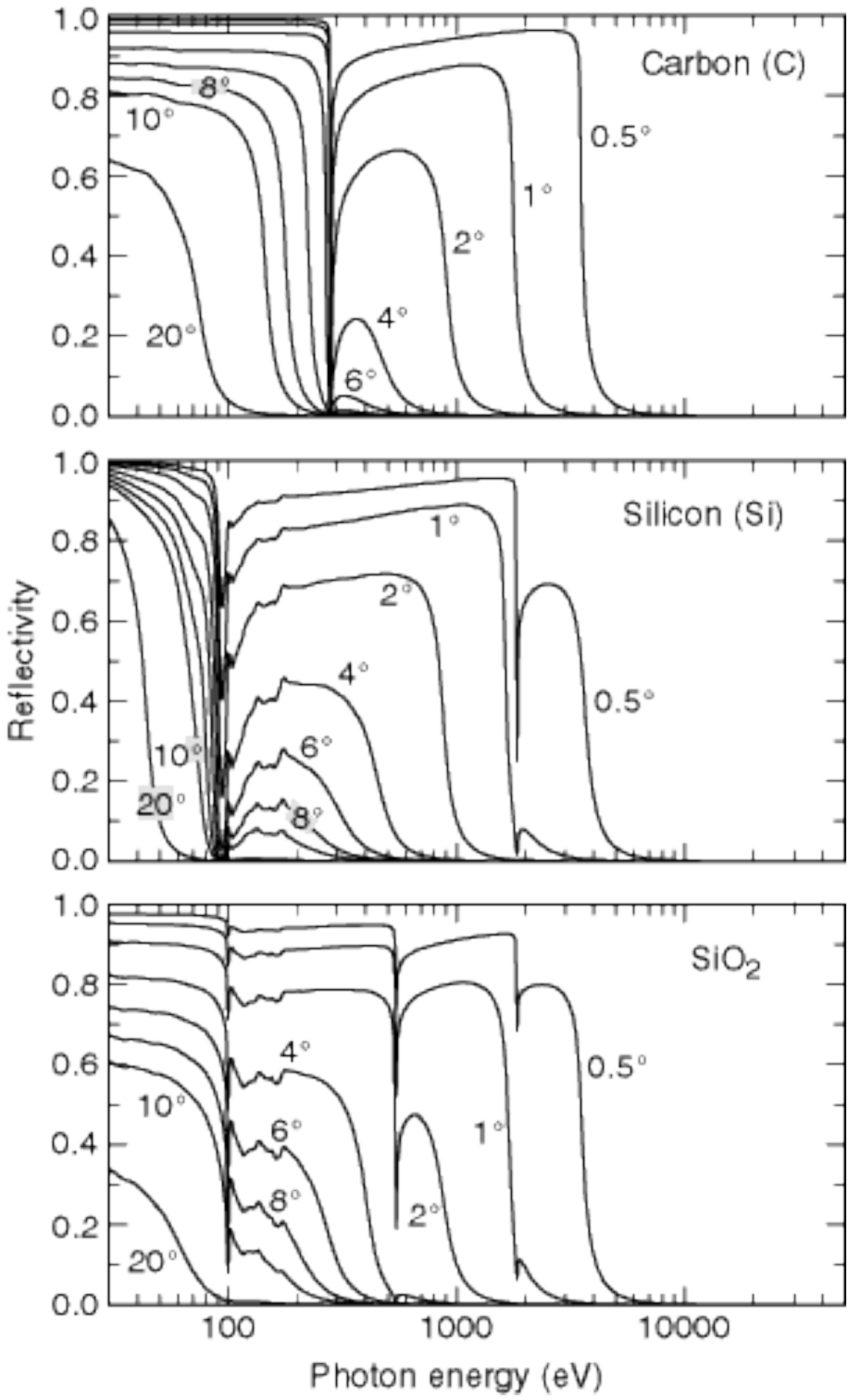

Figure 1: Specular reflectivity of carbon (ρ = 2.2 g/cm$^3$), silicon (ρ = 2.33 g/cm$^3$), and silicon dioxide (ρ = 2.2 g/cm$^3$). The reflectivity is calculated for s-polarization at grazing angles of 0.5, 1, 2, 4, 6, 8, 10, and 20 degrees.

The previous considerations indicate that in order to avoid damage of reflecting optical surfaces, smaller grazing angles are preferred, possibly in total reflection conditions.

The reflectivity behavior is influenced by the inner shell absorption edges that introduce strong changes in the values of the optical constants of the surface mirror coating. Optical elements for VUV and soft X-ray radiation are, usually, coated with a material different from the substrate to enhance the reflectivity in a specific spectral range. Very strong structures in the reflectivity spectrum are evident in correspondence of the excitation energies at the K- and L-edges of the optical material due to the strong increase of the absorption cross section by more than one order of magnitude [4], enhancing the risk for surface damage. The reduced reflectance also precludes their use in these energy ranges; as a consequence the choice of the mirror coatings is dependent on the spectral interval. Coating materials with small atomic numbers are more suitable to withstand the intense FEL beam. Figure 1 shows reflectivity with respect to photon energy of some low Z materials for X-ray optics at several grazing incidence angles [5]. The absorption cross section of the low Z elements such as carbon are up to two orders of magnitude lower, for example, than high Z element like gold. Due to its damage resistance and to the absence of absorption edges and high reflectivity, carbon is a very good coating material. Moreover, carbon has excellent thermal properties when used in its diamond structure. Assuming a grazing angle on the mirror as small as $\approx 0.2°$, carbon coating operate in total reflectivity mode in the energy range 5-8 keV.

Due to the low number of possible beam-lines with respect to 3$^{rd}$ generation synchrotrons different possible optical layouts will be considered and designed with multiple and multipurpose end stations to support highly diverse set of experimental configurations. The main characteristics of beam lines are summarized here:

1) Beam-line supporting the end stations should have "removable" monochromator to filter, if necessary the impinging photon pulses energy, the resolving power can be in average of the order of 2000 at least in soft X-ray energy region.
2) The monochromatization spoils the time structure. The optics layout should provide, for some experiments, a spot onto the sample with temporal resolution $\Delta t$ in the tenths of fs.
3) The focus onto the sample must be variable, from totally unfocussed spot, up to 50 microns
4) The FEL/FEL wave mixing need special delay lines.
5) For non-linear and wave mixing a well-defined (Gaussian-like) time-space profiles of the photon pulses is a mandatory requirement.
6) For TR spectroscopies: energy range soft and mid hard X rays. Pulse width 10 fs, 100 fs are sufficient for start-up phase. The synchronization 10 fs with external laser/or FEL output at different photon Energy, repetition rate pulse structure 1-10 kHz, Mode seeded, Energy resolution close to the Fourier-transform limit ($<10^4$ for O1s).

**Front-end**

Before the FEL radiation passes the shielding wall, to feed the beam lines, a front-end section is met. The front-end is the interface between the photon source and the user beam line. It is designed for radiation safety purposes and for vacuum separation. It is the last equipment met before the FEL radiation enters the experimental hall and the beam line, a sketch is reported in figure 2. A front-end must fulfill the following functions [6]:

1. Vacuum separation of the machine from the beam line. This is obtained by means of narrow pipelines equipped with remotely controlled valves.

2. Safety vacuum interlock system to regulate the manual opening and manual/automatic closing of separation valves.

3. Vacuum protection of the machine from sudden ventings that may happen somewhere in the beam line. This is made by means of a 6m vacuum delay pipe to break the pressure shock-wave and allow the action of a fast closing valve set in between the front–end and the undulator.

4. Radiation safety separation of the machine area from the experimental hall and the beam line. This is obtained by means of a beam shutter made of a tungsten alloy of suitable dimensions to strongly reduce the bremsstrahlung radiation from the undulator vacuum chamber; it will be placed before the radiation safety wall.

5. Definition of the optical path and suppression of scattered light between the machine and the beam line. A precision moveable diaphragm allows determining the propagation direction of the laser beam and defining the angular acceptance of the beam line. Variable acceptance angle is crucial in the suppression of scattered light from the source and in reducing the spontaneous emission content collected from the FEL source.

6. Diagnostics for determining the position and the emission angle of the radiation source (photon beam position monitors) should be also present in the frontend; these devices should be developed and must be able to perform 10µm spatial and 0.1 µrad angular resolution. After passing the front-end, the photon beam passes through shielding wall that separate the undulator hall from the experimental area. After introduced into the experimental hall, it is deflected and energy characterized by a nondestructive in-line spectrometer which acts also as the first deflecting mirror to remove the FEL photon beam from the direct observation of the undulator axes.

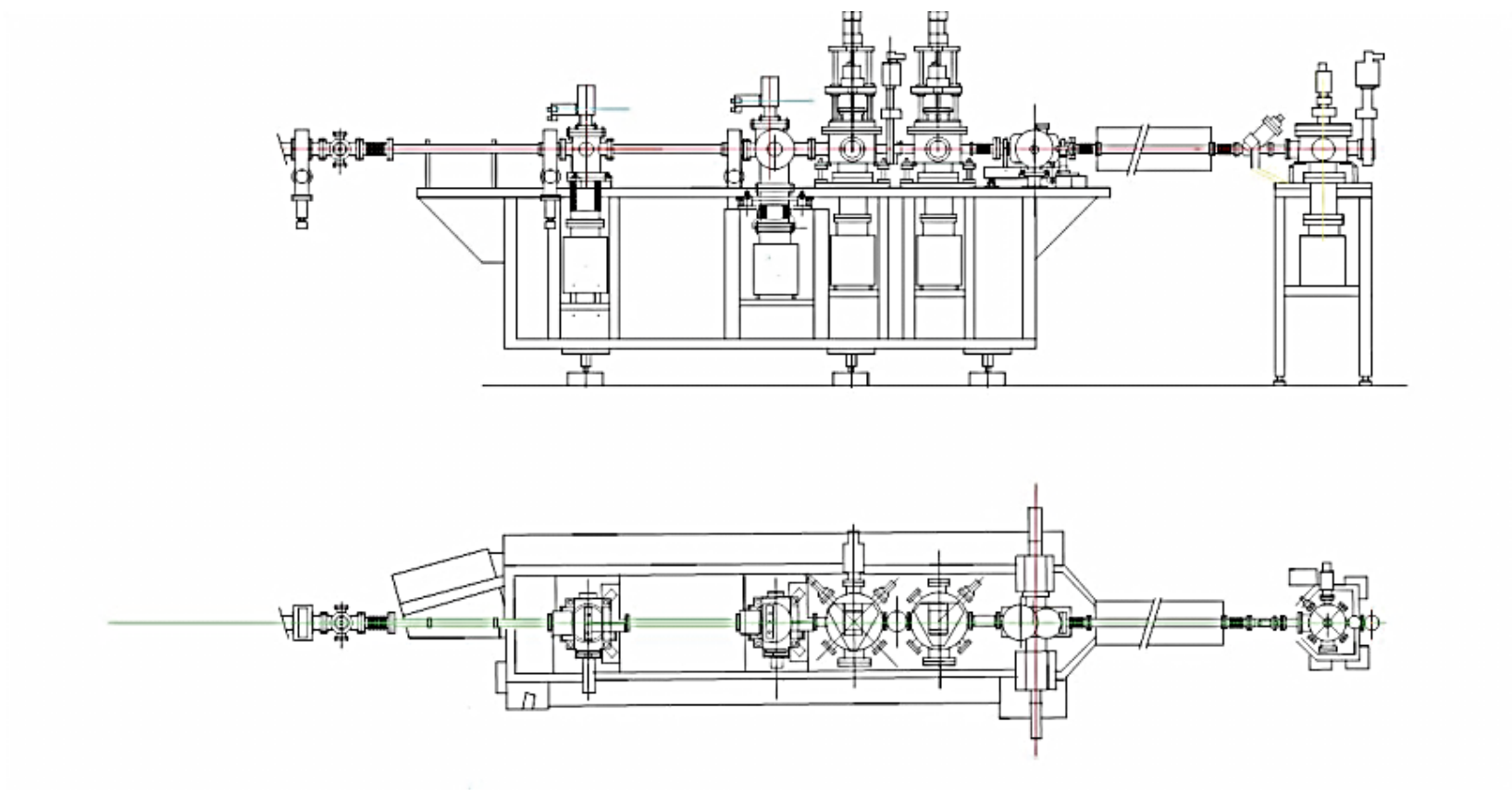

Figure 2: Sketch of a possible vacuum front-end for the FEL beam lines.

The spectrometer is made by a fixed angle crystal monochromator who's zero order acts as a mirror while the first internal diffraction order is used for energy distribution analysis.

**Switching mirror**

After passing the front-end the radiation meets, in the experimental hall, a mirror, plane or concave, that distribute the radiation between two different sections with different monochromator performances, one with large bandpass, used to "clean" the pulse from any photons other than the selected harmonic, and one that allows to select narrow bandwidths. The switching consists of two mirrors placed into the same chamber. The two mirrors are plane in shape and are placed in series in such a way to deflect the incoming light of a total angle that is shared between the two, this allows to use on each mirror a smaller grazing angle which ensure: a reduced power density shining the surface, to maintain the possibility of working in total reflection conditions and, especially for high photon energy to take advantage that, at fixed photon energy, the square of the reflectivity at θ is larger than the value of the reflectivity at 2θ. The twin mirrors may be moved in and out the radiation beam by means of a mechanism that displace the mirror vacuum chamber of a few centimeters.

The optical scheme of this mirror chamber is reported in figure 3. After the switching mirror, the beam enters a long pipe (80-100m) equipped, at the two ends, with differential pumping stage. This is the gas absorber stage that allows controlling the intensity of FEL pulses and the high harmonic content of the "white" FEL photon beam, when working without monochromator spectral selection.

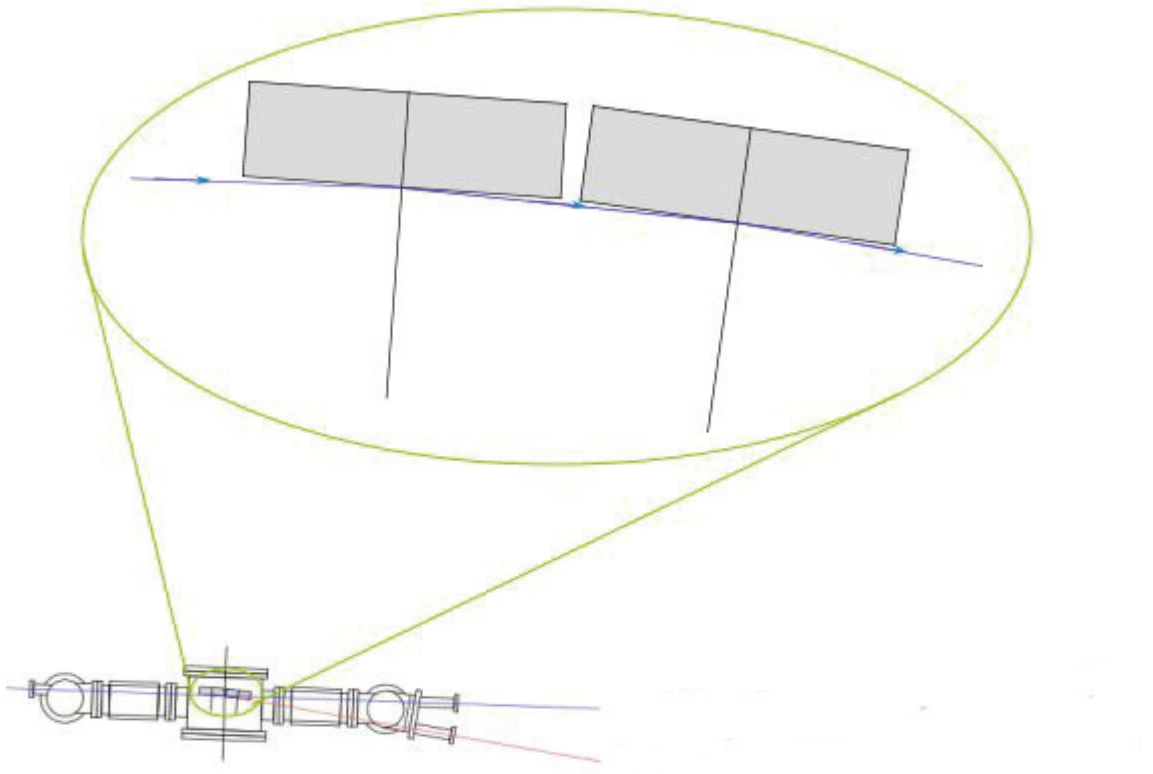

Figure 3: optical scheme of the switching mirror

**Gas absorber**

The request of intensity variation of FEL radiation at sample could be fulfilled by a beam attenuator on the optical path. This is the role of a gas absorber placed before entering into the monochromatizing section. It consists of, at least, 100 m long vacuum pipeline that can be filled with controlled rare gas flow in order to reach a proper pressure (up to $10^{-2}$ mbar) and regulate the transmission of the filter.

The attenuator may operate in flat mode, that means an attenuation of the beam almost constant over the range of interest, or in a selective mode in which the inserted gas absorbs radiation with different attenuation factors in the spectral regions of interest. The selective mode of operation is used when small attenuation is desired on the first harmonic FEL photon beam and the intensity ratio between first and higher harmonics need to be enhanced. The flat mode fulfills the needs of global intensity variation over the full spectral range of radiation, as requested in radiation damage experiments. The use of such a gas filter has some advantages summarized in the following:

- its response is almost insensitive to the incoming intensity.
- the attenuation factor can be easily varied in a very wide range (1- $10^{-8}$) by changing the gas pressure at fixed gas cell length.
- this attenuation method preserves the beam attributes such as coherence and pulse length.

The principal drawback is that the gas cell cannot be closed by windows and an efficient differential pumping system is required for a safe use of the device. A suitable instrumentation is placed at the end of the gas filter in order to measure the radiation intensity before the monochromator section. This could be a calibrated gas ionization.

**Focusing optics**

The focusing optics consists of different mirrors to change the spot size, then the energy density, on the focal point. Several optical configurations will be tested in detail during the design phase, such as a single toroidal mirror, two spherical mirrors in the so-called Kirkpatrick-Baez configuration or two bendable mirrors. The coatings of the optical elements will be chosen to maximize the reflectivity in the whole spectral region of operation. Elements working at fixed angle of incidence, e.g. the focussing mirrors, should be set according to the shortest wavelength to be reflected. The angle of incidence on the elements operating at variable angle, e.g. the monochromator, can be varied depending on the wavelength of operation: the longer the wavelength, the smaller the incidence angle which gives yet high reflectivity. The monochromator, when needed, depends strongly on the energy range we want to handle and will even deliver the full beam in non-monochromatic mode.

**End stations**

Ultrafast time and spin resolved spectroscopies proposed allows a direct access to both the relevant energy (1-100 meV) and time (1ps-10 fs) domain. The time resolved nature of the experimental methods directly will take full advantage pulses properties: coherence, brightness, and temporal structures.

Specific photon, ion and electron detectors will be designed. In this context some of the proposers (within the MIUR PIK2012 projects) are already actives for experimental tools that will be dedicated in FELs facilities.

**Ancillary equipment and estimation of cost**

The main ancillary device is an IR and UV-Vis lasers to excite samples in pump-probe experiments. It is necessary to have a wavelength tuneable system providing pulse durations in the order of few fs and pulse energy up to 1 mJ at 800 nm. Synchronisation to the x-ray beam in the order of the pulse duration or pulse-to-pulse measurements of the jitter is requested. Different wavelengths for the pump laser have to be provided, starting from the fundamental wavelength at ~800 nm up to the higher harmonic. High energy of 1-3 mJ should be also provided for fundamental and high harmonics asking for a large tunability by optical parametric tuning. A contrast ratio larger than $10^5$ is requested and the repetition rate has to allow following FEL pulses. It is likely that a trade-off between the laser pulse energy and the repetition rate will have to be applied. For pump-probe experiments a pulse duration better than 50 fs and a synchronisation with the FEL beam in the order of ~10 fs are required, otherwise suitable diagnostic tools to measure the time delay to a similar accuracy are necessary.

One beam line with such characteristic will cost about 4 Meuro, to this we must add the cost of the devices of the experimental end-station that could be of the order of 4-5 Meuro depending by the experiments and the energy range.

### 4.6. X-ray detectors

The parameters of the FEL X-ray source and the foreseen scientific activities lead to specific and challenging requirements with respect to X-ray instrumentation and in particular to X-ray detectors. These requirements ask for detection systems covering a wide energy range, from soft to hard X-ray energies, with specifications in some cases exceeding the existing technologies. Differently from high-energy/nuclear physics communities, FEL and synchrotron users are not intended to develop novel detection systems (the experiment is the *sample*, not the *detector*). Therefore a successful exploitation of the unprecedented features of the X-ray FEL radiation that will be available at IRIDE calls for a targeted and timely detector R&D program. "Baseline" X-ray detectors are therefore a building block of the facility and their development must start in parallel (and in synergy) with the other components of the FEL.

The main scientific challenges in developing X-ray detectors for FEL sources are briefly reviewed in the following. First of all the wide energy range of the X-rays – approximately from 100 eV up to 12 keV (corresponding to wavelengths ranging from 0.1 nm to 10 nm) - cannot be covered by a single detector or by a single technology but different detection systems optimized for a given energy range need to be developed. As an example, in case silicon is used as the detecting material (direct detection), at X-ray energies below 700 eV the attenuation length of photons becomes so small (<1 μm) that the entrance window for radiation on the detector surface has to be specially tailored in order to minimize photon absorption in the insensitive surface layer (Fig.1). Moving towards harder X-rays the problem of the entrance window is no more a critical point but detection efficiency starts dropping (e.g. below 90% at 10 keV for a 300 μm-thick silicon wafer) and radiation hardness of the interfaces on the opposite detector surface or of the ASIC may become an issue. Furthermore the increase of signal charge with energy has a severe impact on the dynamic range and on the spatial resolution.

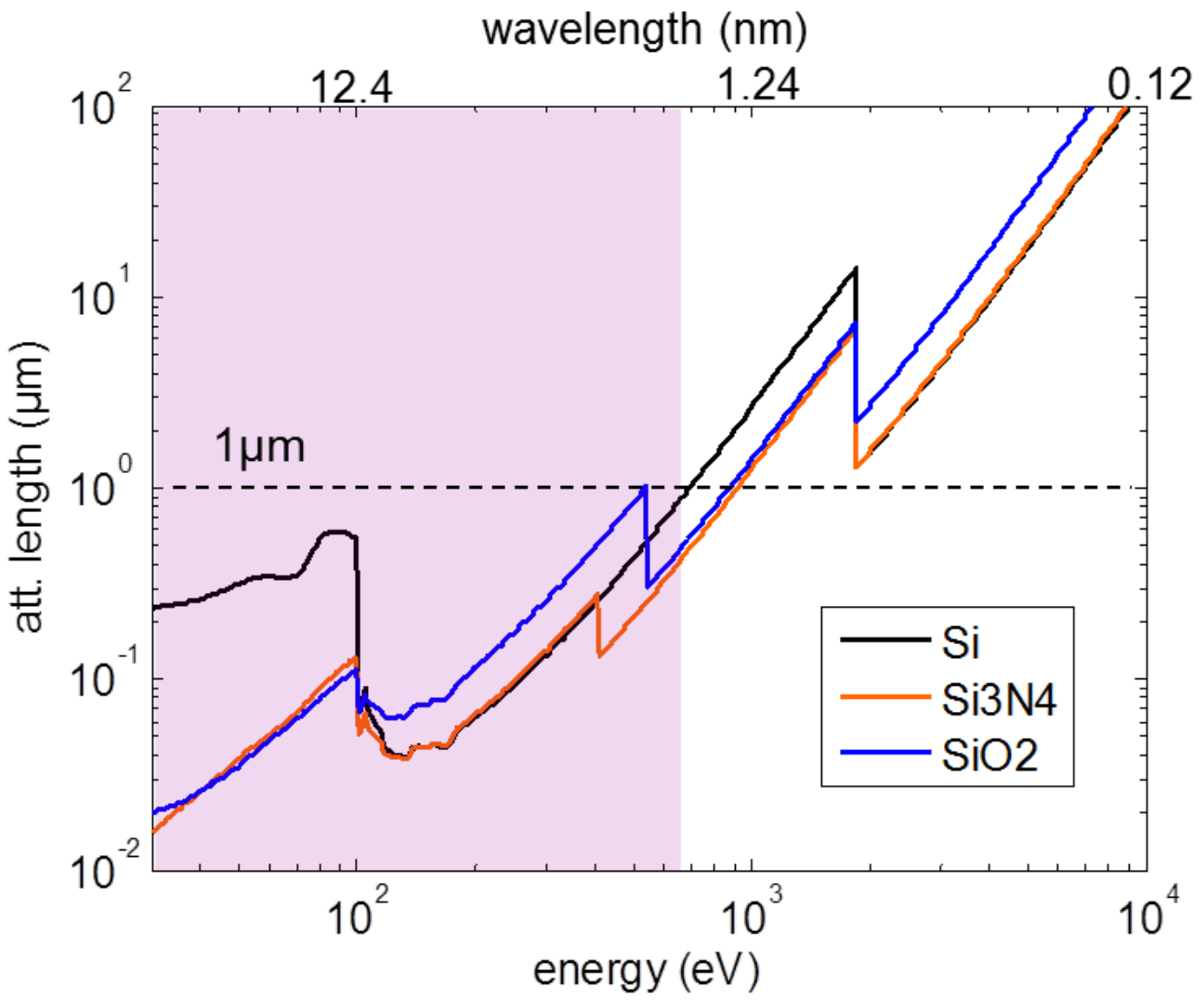

**Figure 1:** Attenuation length of photons in Si, $Si_3N_4$, $SiO_2$ (from X-ray database www.cxro.lbl.gov).

A further specific challenge for detector development at IRIDE-FEL is in the time structure of the X-ray beam. As every X-ray pulse can be regarded as a new imaging experiment, the imaging detector must be able to readout, digitize and store the data before the arrival of the next pulse. As a reference, the Linear Coherent Light Source (LCLS) in US, the Spring-8 Compact SASE source (SCSS) in Japan or FERMI@Elettra in Italy, using non-superconducting accelerator technology, currently feature evenly spaced pulses at rates from 120 Hz to 10 Hz respectively, a non-critical operating condition for state-of-the-art detection systems. The European XFEL facility at Hamburg will be able to produce much higher average pulse rate, about 30,000 pulses per second, but these pulses are delivered in bunch trains of about 3,000 pulses with a very challenging and critical time separation of 220 ns only, followed by a long silent gap (99.4 ms) which can be used to perform off-chip data transfer unfeasible during the bunch train. Therefore the time structure of the beam (pulsed or CW) at IRIDE should be carefully chosen. The option of the CW operation of the LINAC might deliver evenly spaced pulses at a rate up to e.g. 2 MHz - providing nearly two orders of magnitude more luminosity than the European XFEL – but leaving only 500 ns to readout each frame without any time gap which is an even more critical requirement.

Although every experiment may require a specific detection system, the detector needs for FEL experiments are divided into two major areas:

- **Detectors for spectroscopy (0-D or 1-D detectors)**

Time-resolved spectroscopy experiments will typically require energy-dispersive detectors with state-of-the-art energy resolution with possible position sensitivity for angular-dispersive experiments. The main challenges are the energy resolution, the quantum efficiency in the considered energy range and the expected photon rate, as conventional energy-dispersive detectors may suffer from excessive pile-up, and radiation hardness. Therefore also new and dedicated signal processing electronics must be developed to overcome these limitations. Silicon Drift Detectors (SDD) with integrated JFET optimized for X–ray spectroscopy (Fig.2) and, more recently, with external CMOS front-end, is the state-of-the-art technology that proved Fano-limited energy resolution, high count rate capability (up to MHz) and a quantum efficiency allowing operation from 100 eV to 10 keV [1]. High Z materials offer larger efficiencies than silicon and can extend the sensitivity to the harder X-ray spectrum. As they do not feature a native oxide layer they should also be extremely radiation hard in a high X-ray flux environment. Cryogenic detectors can achieve one order of magnitude better energy resolution than semiconductor detectors though their slow count rate may limit their applicability. For their exceptional energy-resolution they have been already employed at synchrotron facilities for X-Ray Fluorescence (XRF) and fluorescence-detected X-ray Absorption Spectroscopy (XAS) [2].

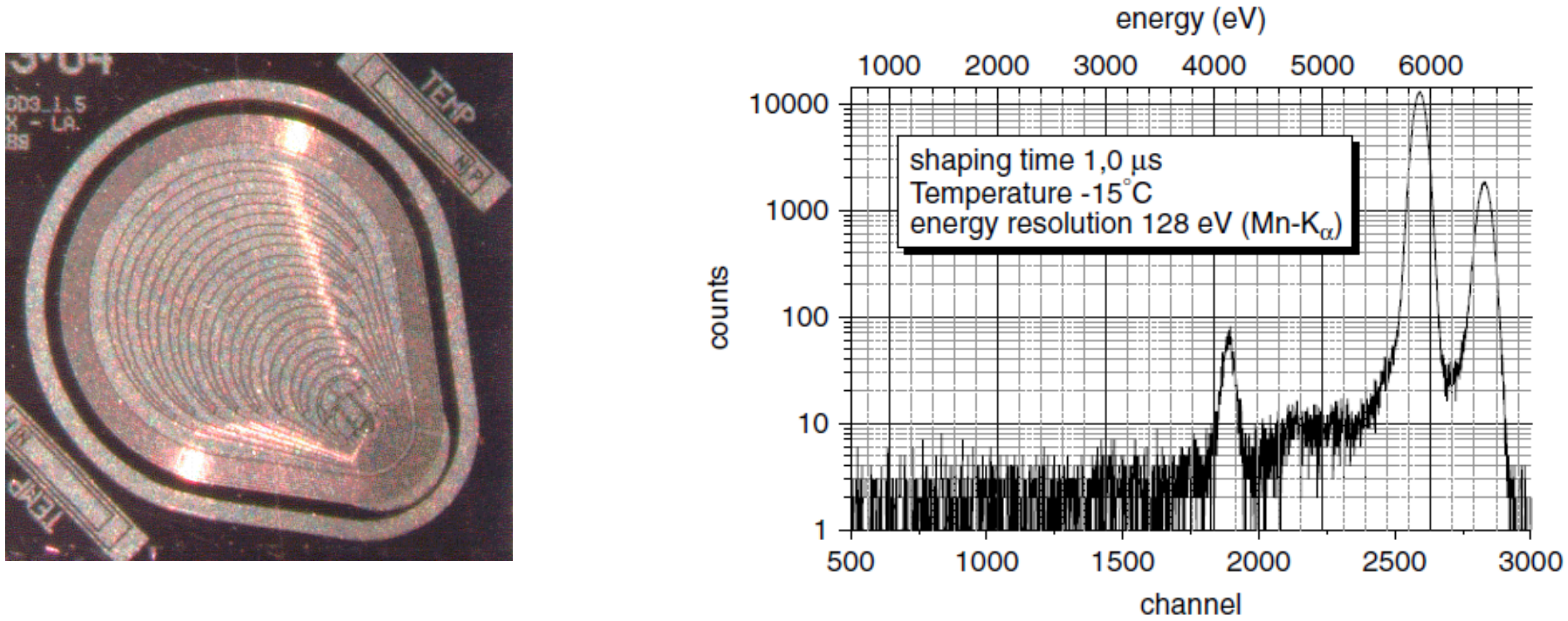

**Figure 2:** Photograph of a silicon drift droplet detector chip and recorded spectrum of Fe-55 source [1].

- **Detectors for imaging experiments (2-D detectors)**

Imaging experiments at FEL will require as baseline detectors a new generation of 2D detectors which demand novel and substantial research efforts. Several international developments are currently on-going. It must be pointed out that the available peak brilliance of the FEL X-rays will result in a large dynamic range at every detector element yet the detector must be able to discriminate the difference between zero and one photon (single photon sensitivity) in the low intensity regions.
Another issue is the signal generated per absorbed photon. In silicon, direct detection of a 50 eV photon will generate 14 electron-hole pairs which requires an exceptionally low noise performance of the system therefore investigation on avalanche multiplication structures and/or peculiar readout strategies in order to achieve single photon sensitivity should be evaluated (Fig. 3). A 12 keV photon will instead generate 3315 electron-hole pairs which instead poses a major limitation to the useable dynamic range and to the required full well capacity. Moreover such high local photon rate on the detector will produce a very high level of charge generation in the detector, leading to the known formation of a local electron-hole plasma which impacts on the charge collection process and on the point spread function (PSF), more significantly for the higher energy X-rays. This effect should be carefully taken into account as it broadens signal image with respect to the incident photon distribution and increases collection times.

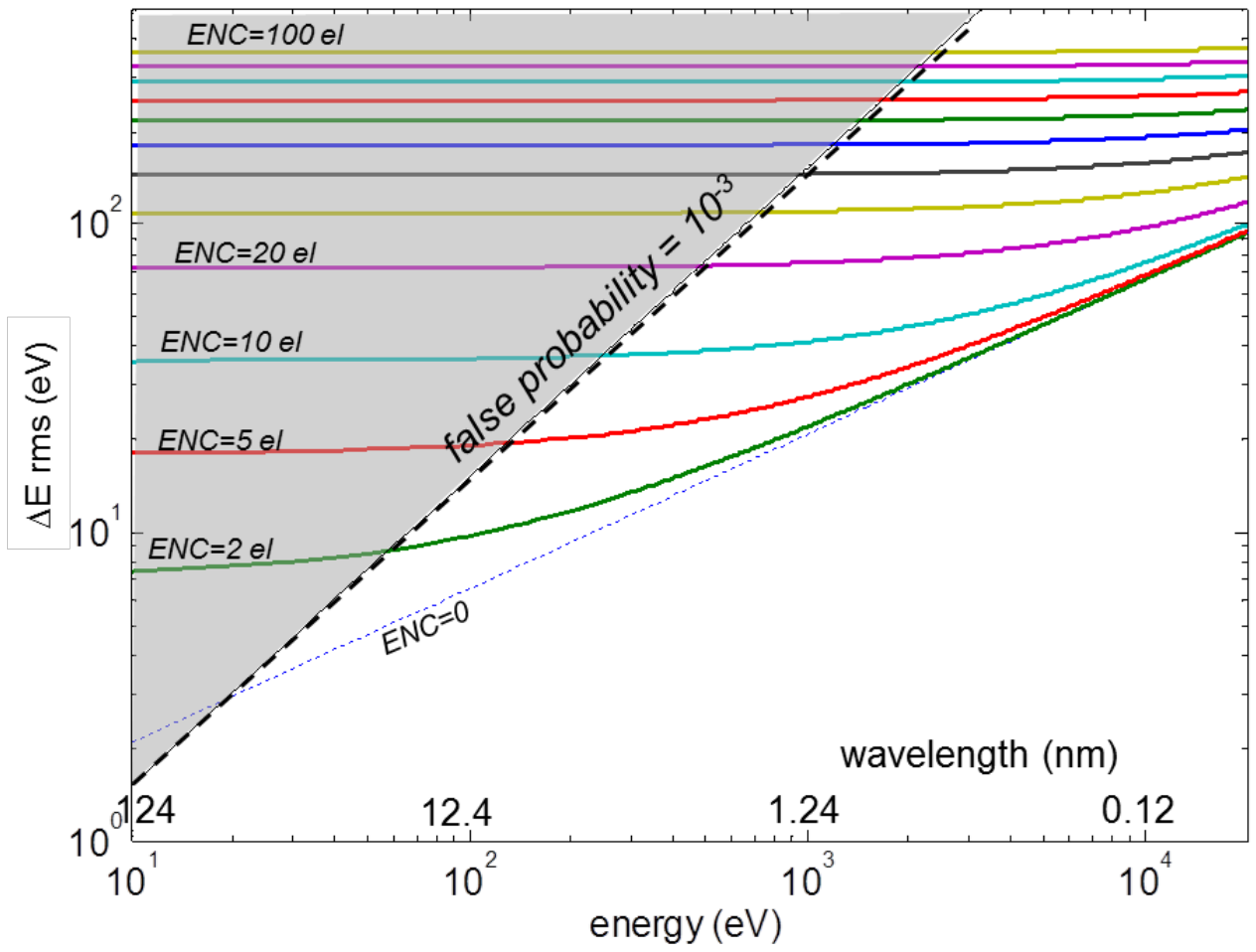

**Figure 3:** Energy resolution as a function of photon energy for different values of the electronic noise (equivalent noise charge, ENC). The dashed line is where the false probability in single-photon discrimination equals $10^{-3}$.

A non-exhaustive list of state-of-the-art 2D imaging detectors and technologies, either available or still in the development phase, should include charge-coupled devices (CCD), hybrid pixel detectors (HPD), monolithic active pixel sensors (MAPS), linear silicon drift detectors (LSDD).

Relevant examples of state-of-the art and ongoing detector developments for FEL experiments are given in the following (not a full review).
The CAMP chamber, developed by the Max Planck Advanced Study Group to meet the challenge of novel VUV and X-ray FEL sources, houses the world's largest pn-CCD chips. These pn-CCDs operate with a frame readout rate of up to 200 Hz and are thus fully capable of meeting the requirements at LCLS (up to 120 Hz) and SCSS (up to 60 Hz). The pn-CCD of the CAMP camera is built on fully-depleted high resistivity 450 µm-thick silicon and features a pixel size of 75 x 75 $\mu m^2$ and a read-out noise of 2.5 electrons (rms) at an operating temperature of -50°C [3].

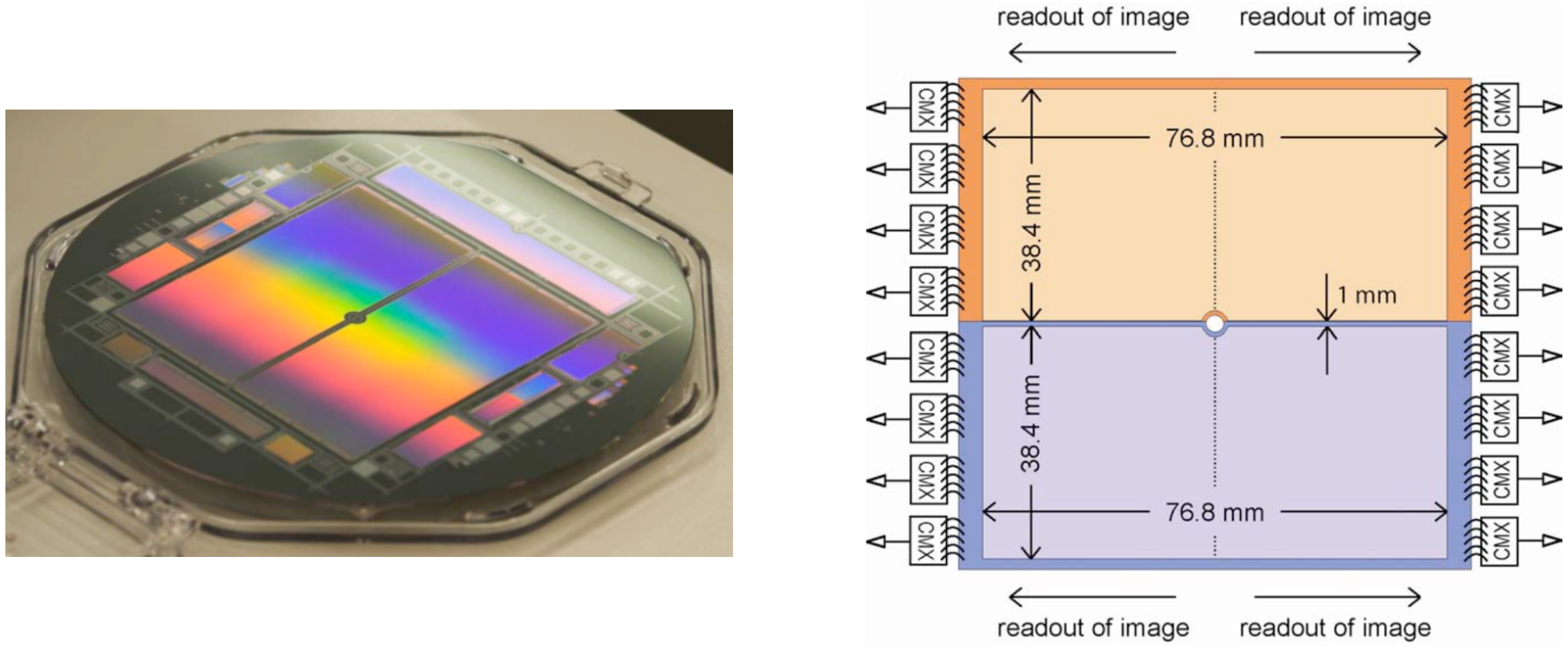

**Figure 4:** Photograph and architecture of the fully-depleted pnCCD of the CAMP chamber [3].

The Linear Collider Flavour Identification collaboration has developed MOS-CCDs with fast column parallel readout, bump-bonded to the readout ASIC, which can provide a pixel size of 20 µm or even smaller and aim at a fast readout frequency (50 MHz). It has also demonstrated the concept of the In-situ Image Storage Sensor (ISIS), a CCD in which signals are stored in-pixel during a burst train and can be read out in the inter-burst gap. Similar pixel size is obtained with conventional MOS-CCDs, capable of high-speed readout due to multiple output ports, fabricated on thick, high-resistivity silicon with tailored entrance window providing simultaneously optical sensitivity with excellent blue and red quantum efficiencies, sensitivity to X rays from the VUV to 10 keV, and to low energy electrons [4].

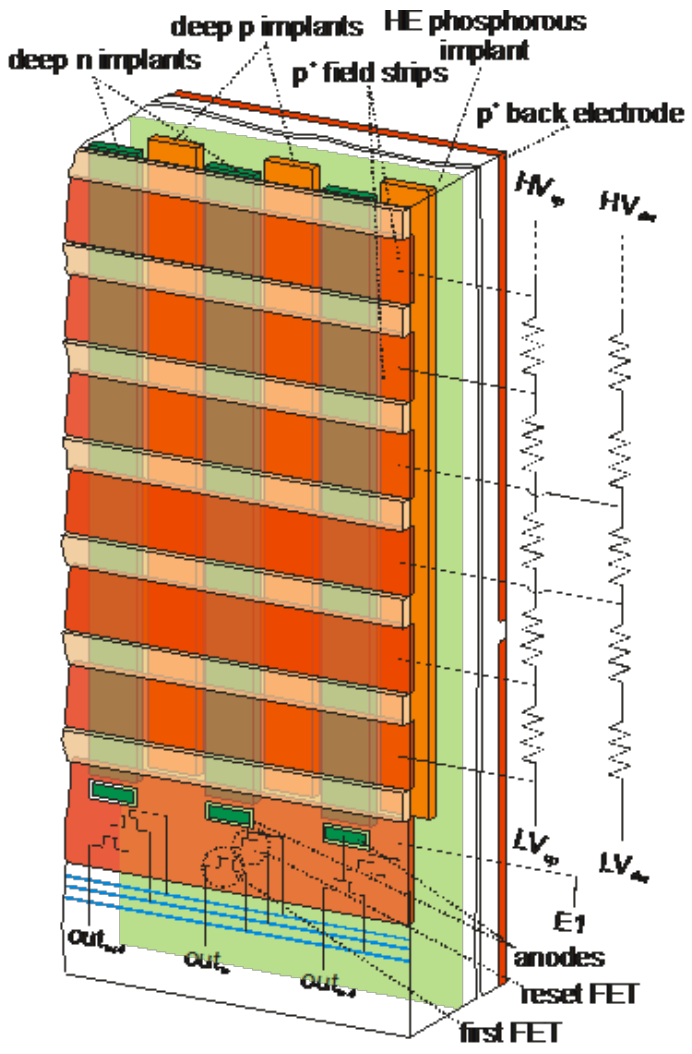

**Figure 5:** Architecture of a multi-linear silicon drift detector with on-chip JFETs [5]

Linear silicon drift detector architectures for position sensing and X-ray spectroscopy allow combining a fast readout speed, by continuously drifting the charges along the column by an electrostatic field, with very low noise leading to excellent spectroscopic performance. By keeping drift lengths below 1 cm, frame rates of 1 MHz can in principle be achieved. The drift detector principle for position sensing has been implemented several times in detectors for high-energy physics as well as for space research and can be designed in large areas. Its features include suppression of inter-column broadening of high signal charges and on-detector integrated electronics to reach low-noise spectroscopic performance [5].

In 2006 the European XFEL launched a call for proposals for 2D detectors to fulfill the challenge of the fast burst mode (up to 4.5 MHz burst frequency) and of the large dynamic range and three selected projects are currently under development [6]. The Adaptive Gain Integrating Pixel Detector (AGIPD) is a hybrid silicon pixel array bump-bonded to the CMOS ASIC mainly targeted to imaging applications for the hard X-rays [7]. The ASIC design uses dynamic gain switching to cover the large dynamic range combined with an analog pipeline to store the recorded frames during the 0.6 ms-long bunch train, that are subsequently digitized and stored off-chip during the 99.4 ms gap. The design goal for the ASIC is 200 µm x 200µm pixels containing more than 200 storage capacitors. The Large Pixel Detector (LPD) [8] attempts to cover the large dynamic range by implementing three different gain settings in parallel (multi-level gain), each followed by an analog pipeline with 500 storage cells per pixel which leads to larger pixel size of 500 µm. The DePMOS Sensor with Signal Compression (DSSC) is one of the few HPD developments addressing the low energy range (<6 keV) [9]. DSSC uses analog signal compression in the active pixel to accommodate the high dynamic range thanks to a specially designed non-linear response of the DePMOS transistor embedded in the silicon pixel. The DePMOS readout proved very low noise performance as the sensing node, buried in the silicon substrate, is coincident with the internal gate of the DePMOS transistor. DSSC foresees hexagonal pixels of 200 µm with a large SRAM that can store 640 words (9 bit) per bunch train. The ASIC, bump-bonded to the silicon sensor, performs readout, digitization and storage of the samples within the available 220 ns between the pulses.

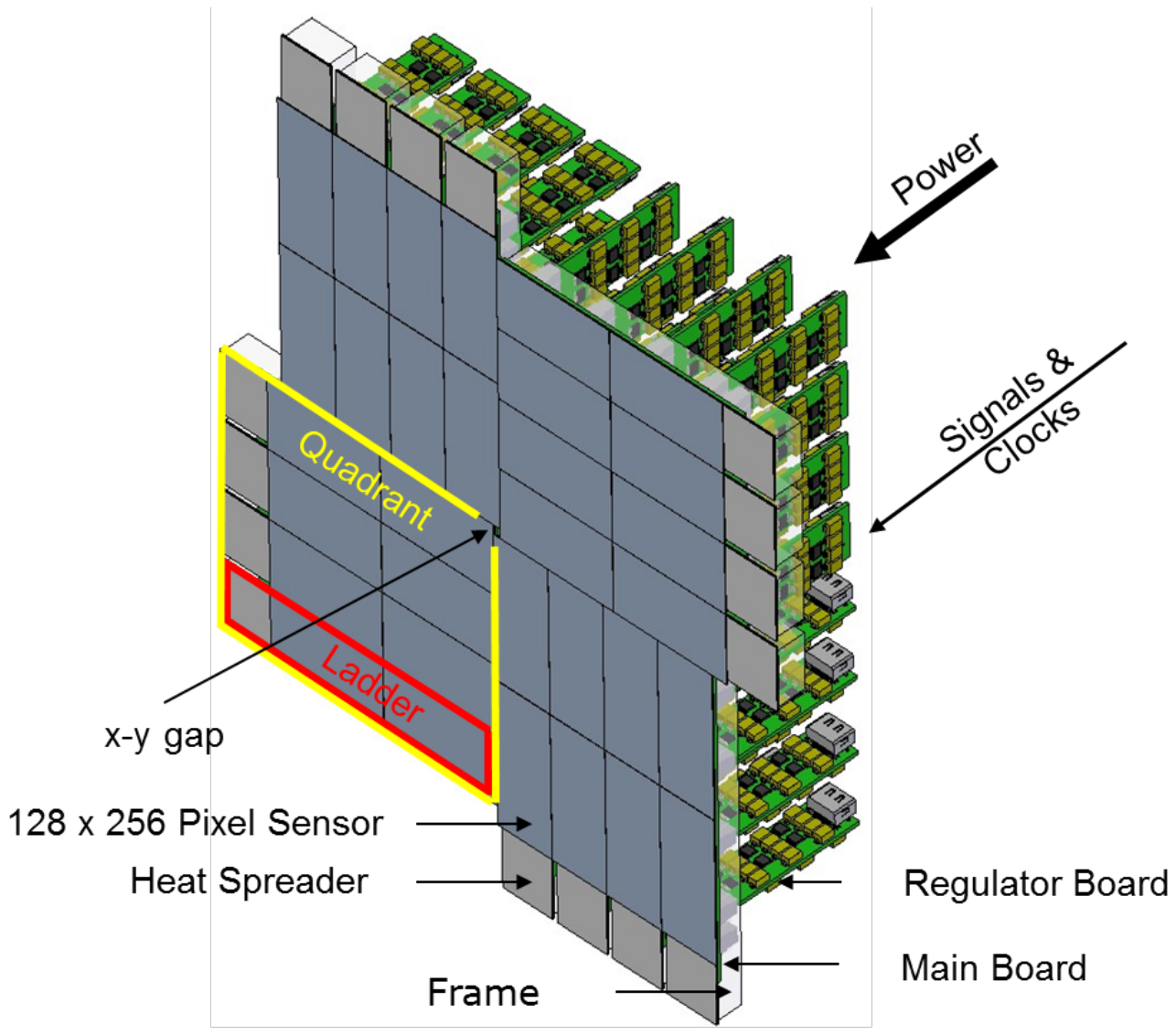

**Figure 5:** View of the full camera head of the DePMOS Sensor with Signal Compression (DSSC) presently under development for the European XFEL [10].

The X-ray Active Pixel Sensor (XAMPS) is a HPD designed for protein crystallography with synchrotron radiation and now tailored for the Linear Coherent Light Source (LCLS). The new design was specifically developed for the X-ray Pump Probe instrument at LCLS [10]. The XAMPS is made on high resistivity n-type silicon and consists of a pixel array detector with integrated JFET switches. During charge integration the switches are open and then selectively closed to allow charge flow to the readout lines. Pixels are arranged in a matrix fashion which allow parallel readout of all the pixels in the same row. XAMPS is meant to provide millisecond readout, pixel size of 90 µm, input dynamic range on the order of $10^4$ photons of 8 keV and a resolution of half a photon FWHM.

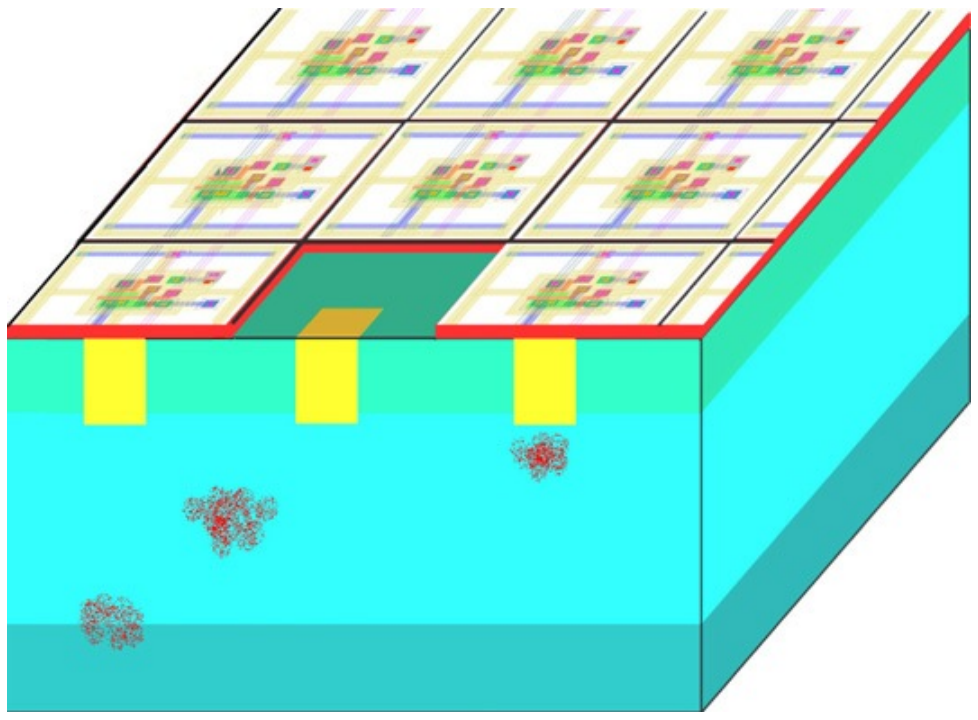

**Figure 6:** Sketch of principle of a CMOS sensor [12].

In the last years a number of developments of Monolithic Active Pixel Sensors (MAPS) for scientific applications have been designed and tested [11]. Given the thin active silicon layer, coupling with a scintillation layer is required in order to reach acceptable efficiency for higher X-ray energies. Small pixel size can be achieved (e.g. 20 µm or smaller) though spatial resolution (PSF) is typically larger due to charge broadening. MAPS developments specifically targeted to FEL and synchrotron facilities are presently under development (e.g. PERSIVAL, Pixelated Energy Resolving CMOS Imager, Versatile And Large). Readout frequencies can reach 1 kframe/s but development aiming at 10 kframe/s are on-going (e.g. MIMOSA-26) [12]. More recently some CMOS process offer the option of a high-resistivity epitaxial layer of some 10 µm thickness which is attractive to enhance X-ray detection efficiency for direct detection above 1 keV [13].
Several other open issues like radiation hardness, medium term stability (e.g. temperature, bias voltage and charge effects), as well as calibration and data acquisition, not analyzed here for brevity, should be careful taken into account at an early stage as they are part of any detector research effort.

## Conclusive remarks

Detectors for day-one operation should be decided both on the base of the available performance achievable with proven technologies and with novel development programs where a step in sensor performance is necessary to fulfill the requirements of the experimental cases. A balanced detector development program should follow a matrix approach between the different technologies and the foreseen experiments in order to highlight common needs and avoid unnecessary duplications.
The technological requirements discussed above lead to the development of a minimum of two baseline detection systems: a 2D X-ray imaging instrument, to be tailored for a given energy range (either the soft or the hard X-rays), and an energy-sensitive detection system for high-resolution X–ray spectroscopy. The corresponding cost estimates are given in Table 1, where the development efforts related to mechanics and DAQ systems are foreseen as common to both detection systems.

**Table 1**: Cost estimate for X-ray detectors

| Components | M€ |
|---|---|
| | |
| 2D X-ray imaging camera head | 3.5 |
| **Total for 2D X-ray cameras** | **3.5** |
| | |
| Energy-sensitive X-ray camera | 1.5 |
| **Total for Energy sensitive X-ray camera** | **1.5** |
| | |
| Shared DAQ and back-end systems | 1 |
| Shared mechanics and beamline integration | 2 |
| **Total for shared DAQ and mechanics** | **3** |
| | |
| **GRAND TOTAL** | **8** |

It is should be stressed that an X-ray detector R&D program in IRIDE is both timely and strategic as it is in the frame of the present demand and competition in the FEL international community to develop advanced X-ray detection systems. Such a program will not only enhance the Italian scientific and technological leadership in this cutting-edge field but will also establish a strong link to other present and future FEL facilities in order to exchange detector technologies in both directions.

### 4.7. Conclusions

By merging competences and desiderata coming from different scientific backgrounds and aiming at different scientific objectives, we have given a wide, surely not exhaustive, panorama of the experimental possibilities that may use the extraordinary properties of radiation produced by a FEL. In the effort of providing a useful platform for discussion and further contributions, we have organized our analysis of the possible applications of the IRIDE-FEL light, into three main area, collecting in each of them, though sometimes with a bit of unavoidable arbitrariness, some examples and open issues according to their critical requirements in terms of the main characterizing features of FEL radiation. It is also clear that it is impossible to evaluate all the possible applications of this new source, because of the unprecedented extreme beam conditions. Science with FEL sources is in its infancy, the results obtained at the operating facilities and the advent of few new facilities in the period between the writing of this proposal and the realization of the IRIDE facility will ask for a revision of its contents in a few years' time. Nevertheless the experimental proposals presented in this Scientific Case meant to represent a possible road map for all future experiments at FEL@IRIDE.

The proponents have acquired a solid experience in spectroscopic activities in Synchrotrons and FEL facilities. Most of them are now involved in the developments of experiments design and experimental tools to be used existing FEL facility. The possibilities coming with the FEL@IRIDE will be a strong opportunity for further developments in the new fields that are now opening with this new sources.

## 5. THE THZ SOURCE

The interest in THz science is recognized since many years for the potential to advance research in many scientific fields. In addition, THz research has many industrial prospects, so that THz activities may offer potential spin-off not only associated to materials science researches, e.g., semiconductor or superconductors materials whose characterization may have a direct impact on many technologies but also in R&D of detectors or imaging. A great expectation for industry is the development of imaging for biomedical applications and security issues.

The development of high-power, sub-picoseconds pulsed radiation covering the spectral range from terahertz (THz) to mid-infrared (MIR) is rapidly growing, both as it is a powerful tool for investigating the dynamics of matter at low energy, and as it allows for a number of possible applications spanning from Biology and Medicine to Security screening.

THz radiation lies between the photonics band on one hand and the electronics one on the other hand of the electromagnetic spectrum. Effectively, THz frequencies extend from the higher limit of millimeter waves, i.e. 300 GHz, up to the lower limit of infrared (IR) radiation, i.e. 10 THz. It is non-ionizing and highly penetrating in a large variety of insulating materials, e.g. plastic, ceramics, paper, etc. Usually, the THz band is defined for frequencies ranging from 0.3 to 10 THz (Figure 1).

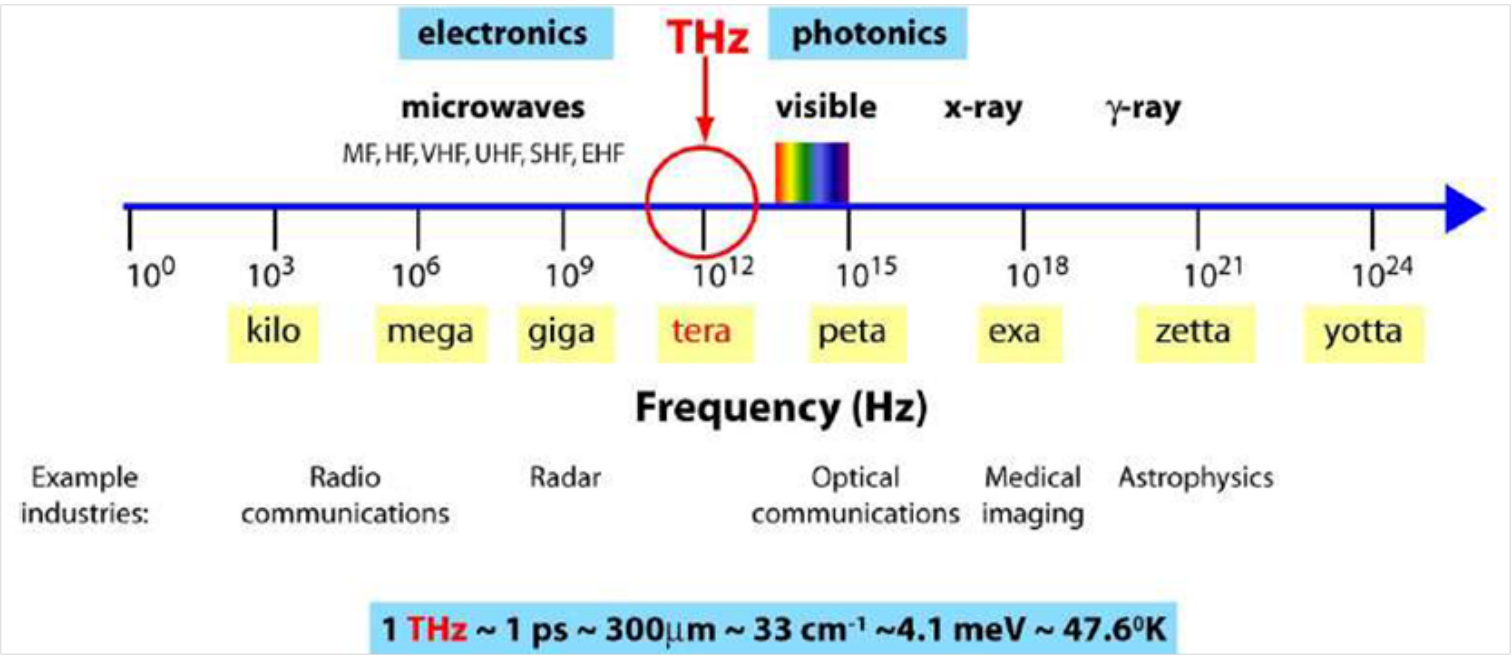

**Figure 1:** The electromagnetic spectrum, highlighting the THz region.

The THz part of the spectrum, though very hard to generate and detect, is energetically equivalent to many important physical, chemical and biological processes including superconducting gaps and protein dynamical processes. Therefore it is of great interest to facilitate experimental research in this region, which will enable new and largely unexplored ways of probing and manipulating matter. This has stimulated major steps in the past decade for filling this so-called THz gap of the electromagnetic spectrum, resulting in a great technological improvement. Notable examples are laser-based THz sources, e.g. THz emitters [1] (photoconductive antennas, optical rectification) and quantum cascade lasers (QCL) [2], as table-top sources. In addition, a new generation of sources, based on particle accelerators, allow to rise up the power, both average and peak, by many orders of magnitude, making this region routinely accessible for several applications. This great increment relies on two enhancement factors, namely relativistic electrons and super-radiance, which occurs at THz frequencies as the electron bunch gets shorter than the emitted wavelength [3]. Indeed, accelerator based photon sources are continuously extending their range of application by taking advantages of high brightness electron beams (HBEB). This allows the generation of very intense and short photon pulses in a wide range of the electromagnetic spectrum. The progress achieved in the production and optimization of ultra-HBEBs, as those needed for X-ray SASE FELs, readily put the basis for wider uses of their unique features, i.e. very low transverse normalized emittance (≈ mm mrad) and high peak current (≈ kA). Among several applications, the generation of high power, both broad-band and narrow-band, THz radiation is arousing a great deal of interest in many fields of science. Unlike conventional THz sources for time-resolved spectroscopy, a linac-driven THz source can deliver broadband THz pulses with femtosecond shaping, and with the possibility to store much more energy in a single pulse. In addition, taking advantage from electron beam manipulation techniques, high power, narrow-band THz radiation can be also generated [4]. Finally, HBEB also permits the possibility to extend the emission towards the MIR having an unique source covering three decades in wavelength from 1000 microns to 1 micron. This provides a unique chance to realize THz-pump/THz-MIR probe spectroscopy, a technique practically unexplored up to now.

5.1 THz source

Even if being still in a developing stage, THz spectroscopy has already displayed its potential in:

- **solid state physics**, to measure the dynamics of low-energy excitations like the Cooper pairs in superconductors [5, 6] and those of the charge and spin density waves in solids [7]. We can also cite phonon pumping, which is crucial for investigating the electron-phonon coupling is several materials.
- **electronics**, to directly probe the photo-excited carrier lifetime in a semiconductor, a topic of high relevance in photonics and nano-electronics;
- **biophysics**, to extract fs-resolved dynamical information on the different isomeric conformations of a given (bio-) molecule after a photo-excitation. This can be done in their native water environment [8, 9]. THz/MIR spectroscopy in liquids is extremely challenging, because of the strong absorption of water in this spectral range. Its role is nevertheless crucial, since the hydrogen bonds between hydration shells and the hydrophilic and hydrophobic tails determine the tertiary structure (fold) of the proteins;
- study of the **biological effects** of terahertz radiation [9];
- **medical imaging**, as a complementary technique to magnetic resonance;
- **security applications**, for the non-destructive detection of drugs and explosives.

The principle of femtosecond spectroscopy is that of pumping a system out of equilibrium by a pulse of sufficient fluence (up to several mJ per pulse). The photo-induced variation in the reflectivity and/or in the transmission of the system (often at a single frequency), is then tracked with a spectroscopic probe during the return to thermal equilibrium. The evolution of photo-induced charges, as well as their dynamical interactions with other microscopic excitations, is measured by delaying the pump and probe pulses with respect to each other using a high precision mechanical delay line. Pump-probe spectroscopy in the THz/MIR range is usually performed by three different experimental procedures:

a) **Optical pump-probe spectroscopy (OPPS)** is the usual technique: both the pump and the probe work in the visible or near-IR region of the electromagnetic spectrum where direct generation of femtosecond pulses from mode locked lasers is possible. Although many important results have been obtained by OPPS, this technique is affected by the relatively high-energy involved, which push the electrons, to high energy bands and thermalize, coming back to the Fermi energy ($E_F$), following complicated scattering processes. Therefore both the high-energy pulse as well as the high energy observations are affected by strong inelastic scattering effects that may mask the intrinsic electronic dynamics.

b) **Optical pump-THz/MIR probe spectroscopy**

A visible/near-IR laser emitting femtosecond pulses is used either to pump the sample, and to produce a THz/MIR pulse for probing it by in a THz/MIR set-up. In this case, inelastic effects in the electronic bath can be present due to the relatively high-energy involved, which push the electrons, to high energy bands and thermalize, coming back to $E_F$. Instead, the THz/MIR pulse directly probes the low-lying electronic excitations. Another advantage of OPTPS is that it can measure both the real and imaginary part of the optical conductivity using a self-synchronized EOS (Electro-Optics Sampling) technique. This, as discussed below, enables for instance the simultaneous measurement of the dynamics of quasi-particles and the superconducting condensate.

c) **THz/MIR pump-THz/MIR probe spectroscopy**

Pumping and probing the excitations in the same low-energy spectral region clearly opens new interesting scenarios in femtosecond time-resolved spectroscopy. However, only a pulse intensity on the order of 10-100 μJ/pulse (corresponding to an electric field of 1-10 MV/cm) can induce non-linear phenomena in physical systems. This task can be only accomplished using THz/MIR emission from a linac-based source, which provides both high intensity, short pulse and, therefore, a wide spectral coverage. On this ground, in the following sub-sections different research opportunities to be performed at a linac-based source will be presented and discussed.

### *5.1.1. Breaking Cooper pairs with THz photons*

Non-equilibrium superconductivity in both BCS and High-Tc superconductors has been widely studied by both pumping and probing the system in the near infrared.

Recently, synchrotron-based techniques [10] and the development of THz time-domain spectroscopy allowed a few groups to use a THz/MIR beam as a probe, thus directly addressing the photo-induced normal state Drude response and the superconducting gap $\Delta$. However all THz pump-probe experiments are still completely unexplored because of the low energy per pulse delivered by conventional THz sources.

It is believed that after the optical pump pulse, the high-energy photo-excited quasi-particles shortly relax through electron-electron and electron-phonon mechanism and settle to energies close to within less than 1 ps. It is only the subsequent dynamics occurring in the ps-timescale to be related to superconductivity via the so-called phonon-bottleneck effect. This means that because of the optical frequency of the pump pulse a complicated and poorly understood dynamics at frequencies much higher than the superconducting gap takes place before entering in the regime of the Cooper pair recombination. This has severe consequences on our understanding of the experiments. For instance, the number of excited quasi-particle cannot be accurately determined, since a sizable portion of the laser energy (maybe its largest portion) goes to the phonon bath.

Photo-induced Superconductor to Normal Phase Transitions (SNPT) have been observed in High-Tc superconductors for optical pump photon fluences of the order of 100 $\mu J/cm^2$ [11, 12]. The availability at a linac-based source of THz sub-ps pulses with energies in the 100 μJ scale, is therefore expected to break a sufficient number of Cooper pairs on a sample area large enough to record the Terahertz spectrum, and possibly to induce a SNPT in either BCS or High-Tc superconducting nano-wires and nano-bridges with minor heating of the phonon bath. The THz fluence threshold could then be determined experimentally and directly compared with available theoretical models as the T* and μ* models [13]. Moreover, the recombination of the Cooper pairs could then be studied in a system where the above gap photo-excited quasi-particles are not thermalized with phonons, with potentially interesting consequences on their dynamics.

Furthermore, the MV/cm electric field E associated with the THz pulse may induce a large supercurrent value (possible exceeding its critical value). In this case, the normal state can be reached without producing photo-excited charges. The recover to equilibrium of the superconductor may be investigated by a delayed THz pulse.

### *5.1.2. Inducing structural instabilities in solids*

In recent years huge interest has been attracted by materials in which ferroelectricity and ferromagnetism are coupled. These materials, known as multiferroics [14], are highly promising for novel device applications. The technological use of multiferroics may enable the control of charges by applied magnetic field, and spins by applied voltages.

Strongly correlated electron materials are known to display strong variations of their electronic properties as a consequence of small variations in their control parameters as the lattice constants or bond angles. The possibility of driving an electronic phase transition through a mode-selective THz/MIR excitation has been recently demonstrated in manganites (MnO) [15], with the help of 1 mJ pump pulses at 17 THz, which is the actual low-energy limit in THz laser based sources. This topic is extremely interesting not only from a fundamental point of view, but also for its potentials in ultrafast (>GHz) optical data storage or optical switching. The THz/MIR radiation from a linac-based source (which can be easily extended below 20 THz), could be used to directly pump the normal modes associated with structural transitions (soft modes, Jahn-Teller phonons), thereby inducing the phase transition. The internal degrees of freedom of composite quasi-particle like large and small polarons can be excited as well [16]. The same THz/MIR pulse can also be used as a probe to monitor the induced structural rearrangement or the eventual occurrence of an insulator to metal transition.

### *5.1.3. Collective Excitations in Charge-Density, Spin-Density Wave Materials and Charge-Ordered Materials*

A Charge-Density Wave (Spin-Density Wave) is a modulation of the electronic density (spin density) in a metal which, below a given temperature, opens a gap at the Fermi energy changing the metal into an insulator. The low-energy Drude behavior of the metallic state disappears while a single-particle gap in the MIR opens. Moreover, the CDW retains excitations corresponding to phase (phason) and amplitude (amplitudon) collective modes, which fall in the THz range [7]. Their out-of-equilibrium dynamics as well as their intrinsically non-linear (non-ohmic) behavior have been scarcely studied. For instance, inducing a depinning of the phason, might provide the possibility to study the dynamical (Froehlich) metallic state associated with the free motion of the phason collective mode. Moreover, this kind of investigation could easily be extended to other charge-ordered states, for example observed in manganites, cuprates as well as in transitional metal oxides. The availability at a linac based source of high electric field THz/MIR radiation could be used to directly depin those collective modes and investigate their sub-ps dynamical metallic state.

### *5.1.4. Magnetism at sub-ps time resolution*

The low-energy dynamical properties of magnetic materials are studied by different techniques spanning from Inelastic Neutron Scattering (INS), to steady-state Electron-Spin-Resonance (ESR). Recently, steady-state ESR has been extended to high (THz) frequencies, using Fourier-Transform spectroscopy coupled to infrared synchrotron radiation (IRSR) [17]. In this case, broad band infrared synchrotron radiation induces transitions between states which are split by a static magnetic field applied onto the sample. On this ground we can cite the full characterization of the far-IR antiferromagnetic resonances in manganites as well as in other quantum antiferromagnets [17, 18].

A very intriguing scenario in magnetic materials may be open using the magnetic field B (up to 1 T) accompanying the high power THz/MIR photons which would be produced by a linac-based source. Indeed, due to the intrinsically large B-field related to those photons, a magnetic excitation, even at zero external field, might be induced in a material. Its dynamic return to the equilibrium can be then followed by a THz/MIR probe, obtaining information on the magnetic structure at sub-ps temporal scale.

### *5.1.5. Torsional dynamics in biomolecules*

The importance of THz/MIR spectroscopy in Chemistry and Biology is steadily growing since the key role of the THz energy scale has been acknowledged in the study of weak rotational and torsional potentials. Prominent examples are those occurring around the C=C bonds in the extended molecular chains of polycarbonates.

Bio-molecules as those containing the retinol group undergo light-induced cis-trans isomerisation processes which do indeed involve rotations around specific chemical bonds [19]. These mechanisms allow nature to store energy or detect light through a conformational change. An accurate knowledge of these processes - which lie at the core of vision and photosynthesis - is of primary importance, not only for our understanding of biology. Photo-isomerisation holds indeed also promising applications for optical switches, data storage, and nano-machinery.

All optical pump and probe absorbance and fluorescence experiments on the femtosecond time scale give access to important information on the dynamical phenomena associated with photoreactions. Optical pump -THz/MIR probe experiments with table-top THz/MIR -TDS set-up have been envisaged, but still not performed, to our knowledge. These experiments are highly promising since they should allow to directly probe with sub-ps resolution the rotational motion associated with the isomerisation process. One major difficulty is related to the need of obtaining a sufficient THz/MIR intensity, while keeping a low repetition-rate in order to avoid sample damage by heating. The THz/MIR beamline at a linac-based source could overcome these issues since it can deliver 100 μJ per pulse at a variable repetition rate. The high power of the THz/MIR pulses at a linac-based source should allow one to go further.

The THz/MIR pulses may be used to induce strong (unharmonic) distortions in the low energy modes involved in the cis-trans isomerization. The effect of the THz/MIR induced twisting could then be probed both in the THz/MIR range and in the visible (absorbance/fluorescence). The results can provide important information on the configuration energy landscape of the chromophores, and more in general, on the folding mechanisms of proteins [20, 21].

5.2. Beam requirements and performances

An electromagnetic source can be characterized in terms of field strength, pulse shape, bandwidth and frequency, and those requirements depend on the class of experiments. An effective THz/MIR source should have higher peak fields, from 100 kV/cm to 10 MV/cm, the coverage of a spectral range up to a frequency equal to 10 THz and a full pulse shaping.
The source at IRIDE will be designed to achieve these requirements.
In particular THz/MIR radiation will be generated at IRIDE as coherent radiation from ultra-short, i.e. ≈ $10^2$ fs, electron bunches. Diffraction and undulator radiation are the chosen mechanisms for broad band and narrow band emission, respectively, with the advantage of being non-disruptive for the electron beam.

Table 1 summarizes the performances of THz/MIR coherent radiation source from the undulator. Beam and undulator parameters also reported.

**Table 2: Parameters of the THz/MIR coherent undulator radiation (CUR) source.**

| Electron beam parameters | |
|---|---|
| Electron energy (GeV) | 0.1 |
| Charge/bunch (pC/bunch) | 500 |
| RMS bunch length (μm) | 60 |
| Normalized emittance (mm mrad) | 1 |
| Undulator | |
| Period (cm) | 40 |
| Number of periods | 10 |
| Magnetic field (T) | 0.1 -1 |
| Coherent Radiation parameters | |
| Wavelength (μm) | 100 (with K = 6, i.e. B ≈ 0.2 T)-10 (K=1.4, i.e. B≈ 0.04T) |
| Peak power (MW) | > 100 |
| Micropulse energy (mJ) | ≈ 10 |
| Micropulse duration (fs) | 200 |

Table 2 summarizes the performances of THz/MIR coherent radiation source from a diffraction radiation (DR) target. Beam parameters also reported.

**Table 3: Performances of coherent diffraction radiation (CDR) source.**

| Electron beam parameters | |
|---|---|
| Electron energy (GeV) | 1.5 |
| Charge/bunch (pC/bunch) | 250 |
| RMS bunch length (μm) | 60 |
| Normalized emittance (mm mrad) | 1 |
| Coherent Radiation parameters | |
| Wavelength (μm) | > 50 |
| Peak power (MW) | > 100 |
| Micropulse energy (μJ) | ≈ 100 |
| Micropulse duration (fs) | 200 |

5.3. User beam line

Two beam lines are considered in the first phase of the IRIDE project, both after compression stages, one at low energy, i.e. 100 MeV, and the second one at higher energy, i.e. 1.5 GeV, in proximity of the FEL extraction site. The first one, consisting in an infrared undulator able to emit quasi-monochromatic radiation from THz to MIR (Coherent Undulator Radiation, CUR), will be optimized for experiments involving high peak power, narrow band THz radiation, while the second one will combine the Coherent Diffraction Radiation, emitted from a rectangular slit in a metallic screen, to the FEL radiation to perform THz pump X-ray probe experiments.

Coherent radiation produced by the electron beam in this undulator strongly depends on the bunch profile and charge. Indeed, the energy radiated by the electron bunch is given by [22]

$$E_{Coh} = \frac{\pi e^2 {A_{jj}}^2 {\omega_0}^2 K^2}{c(1+\frac{K^2}{2})} \left[N + N(N-1)F(\omega)\right] \ ,$$

$N$ is the number of electrons per bunch, $F(\omega)$ is the square of the Fourier transform of the bunch profile function, $\omega_0$ is the resonant frequency of the first undulator harmonic, $A_{jj}=[J_0(Q)-J_1(Q)]$, with $J_0$ and $J_1$ the Bessel functions and $Q = K^2/[4(1+K^2/2)]$, $K$ is the undulator adimensional parameter. At wavelengths longer than the bunch length the longitudinal form factor tends to the unity, thus leading to an enhancement, by a factor of $N$, of the radiation power with respect to the power of incoherent radiation.

Coherent Diffraction Radiation [23, 24] is emitted when an ultra-short (≈ sub-ps) electron beam travels through a slit cut on a metallic screen, due to the interaction between the electro-magnetic (EM) field of the travelling beam and the target metal surface. The metallic target is placed inside the ultrahigh vacuum beam pipe of the linac.

The extension of the EM field of a relativistic particle is a flat circle of radius $\gamma\lambda/2\pi$, where $\gamma$ is the Lorentz factor and $\lambda$ the emitted wavelength. In case of an infinite long slit of aperture $a$, DR is emitted if $\gamma\lambda/2\pi$ is comparable with $a$. Depending on the slit aperture and the beam energy the emitted wavelength is in the range of THz/MIR/FIR. Since the beam goes through the slit, DR is a non-intercepting tool for the generation of THz/MIR radiation source and excellent to be used parasitically.

The THz/MIR radiation, either diffraction or undulator, is coupled out through a diamond window and transported to the experimental hall through an evacuated tube equipped with proper light optics. THz and MIR radiation [25, 26] will be collected by a mirror, under broad acceptance angles (100 - 150 mrad, due to the wide emission cone at millimeter wavelengths). The large emission angles in the THz/MIR region may imply diffraction effects. Moreover, the pulse time structure can be deteriorated by multiple reflections along the line. The beamline optics will be calculated by using different ray-tracing codes (SRW, Shadow, etc.) and by exploiting the experience accumulated in designing the infrared beamline SISSI@Elettra and the THz facility at SPARC. Both facilities efficiently working since 2006 (SISSI) and 2009 (THz source at SPARC) [27, 28].

Considering the very large wavelengths in the THz/MIR range, no special requirements on flatness, nor on local roughness, is demanded on the mirror surface. This strongly reduces the overall cost of the optics. The extraction mirror, due to the large acceptance angle, should be mounted directly inside the vacuum chamber and will be remotely controlled. The others will be located in the laboratory and are therefore directly accessible. In order to avoid water vapor absorption, the entire beamline will be kept in low vacuum conditions ($10^{-3}$ mbar). To this aim, a diamond window will separate the ultra-high vacuum chamber from the beamline which transfer the radiation to the experimental station under low vacuum environment.

The experimental station of the THz/MIR beamline will be equipped with several spectrometers to be developed in house: Martin-Puplett and Michelson interferometers, a grating monochromator, and an Electro-Optic Sampling set-up.

A Martin-Puplett interferometer is basically a Michelson interferometer where the beam splitter is replaced by a wire-grid polarizer. Thus the two polarized beams from the two arms recombine at the exit ideally with no loss, into a beam elliptically polarized as a consequence of the path difference between the two arms. A wire-grid polarizer is finally used to go back to a plane polarized beam whose amplitude varies periodically with the path difference (interferogram) and is measured by a bolometer.

Grating monochromators may also be useful at the THz/MIR beamline, due to the recent developement of linear arrays of pyroelectric detectors. As demonstrated in Ref. [29], the design and construction of the monochromator and related diffraction gratings is rather straightforward. By spatially dispersing the radiation at various wavelengths, the monochromator can be used not only as a spectrometer, but also as a tool for tailoring the spectral content of the radiation beam. The rotation of the diffraction grating can therefore permit to continuously tune the frequency of the pulse to be used as "pump" beam.

Electro-Optic sampling is a widely used technique for THz/MIR detection in THz-TDS systems and can be extended to the detection of pulses from a high brightness photo-injector. EOS consists in probing the THz/MIR-induced optical anisotropy in an optically active crystal (typically ZnTe) by using an external polarized femtosecond laser pulse [30]. The modification induced by the pulse on the polarization state of the fs-laser pulse is proportional to the electric field associated to the pulse. The advantage of EOS with respect to the above mentioned spectroscopic techniques is that both the amplitude and the phase of the electric field can be probed as a function of time. The simultaneous measurement of these quantities allows one to determine both the real ($n(\omega)$) and the imaginary ($k(\omega)$) part of the refractive index: $\tilde{n}(\omega) = n(\omega) + ik(\omega)$. A cartoon of the EOS setup is shown in Figure 2.

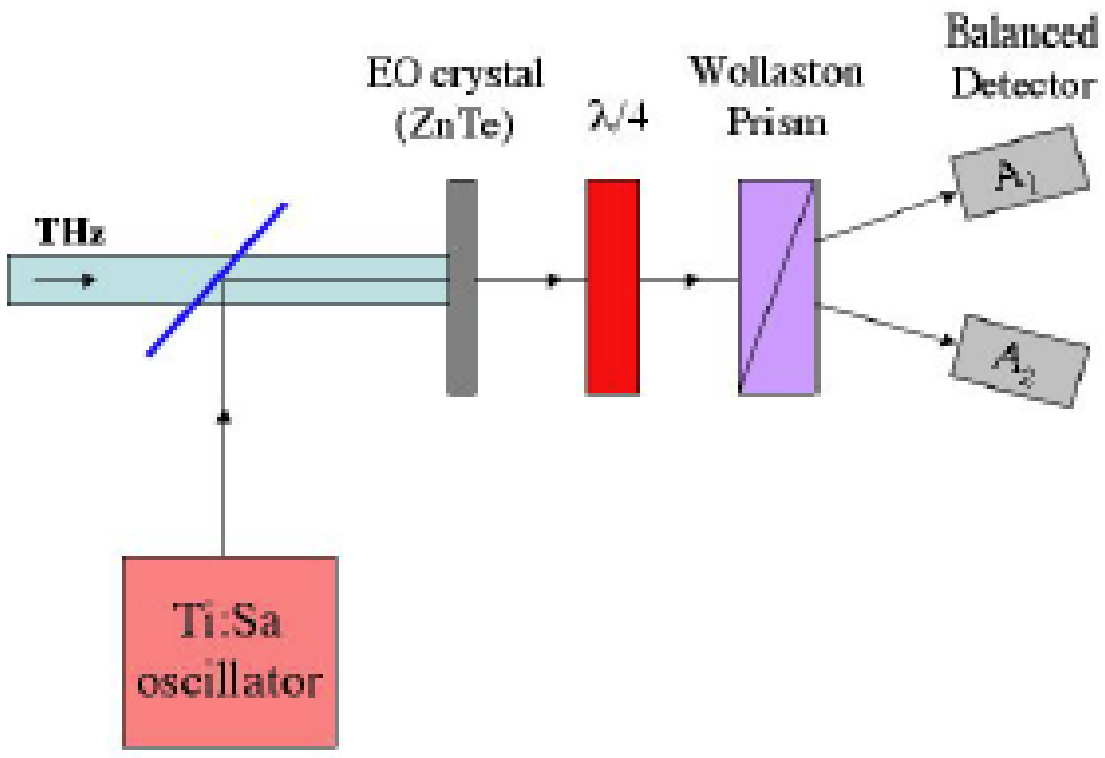

**Figure 2:** Schematic layout of the EOS set-up.

The THz/MIR and optical beams hit simultaneously the non-linear crystal. The optical laser pulse experiences the birefringence induced by the THz/MIR field in the crystal through Pockels effect. The out-coming optical laser beam is then sent to a quarter-wave plate, where its elliptical polarization is changed into linear. A Wollaston prism separates then the two orthogonal polarization components and guides them to the balanced detector. The difference signal is then proportional to the sine of the retardation parameter, Γ: $|A_1|^2 - |A_2|^2 \propto \sin\Gamma$ . Since it is directly proportional to the actual value of the THz/MIR electric field ($E_{THz/MIR}$), its measurement for several delays, between the THz/MIR and the laser beam, permits to sample the $E_{THz/MIR}(t)$ value of the THz/MIR pulse. By Fourier Transform, one can finally reconstruct the THz/MIR spectral shape.

5.4. Cost analysis

**Table 3: Cost analysis in keuro (VAT is not included).**

| | |
|---|---|
| 2 Chambers (including actuators, motors, control, etc.) | 100 |
| Diagnostics (interferometers, spectrometers, EOS, THz camera, THz detectors, etc.) | 300 |
| Ancillaries (breadboards, Cryostat…) | 100 |
| Electromagnetic undulator | 700 |
| Power supply | 400 |
| 2 Diamond Extraction windows | 100 |
| Guiding optics | 200 |
| Beamlines (mechanics, vacuum, controls for a 20 m length) | 1000 |
| Vacuum system (pipe, valves, …) | 100 |
| **TOTAL in Meuro** | **3** |

# 6. THE NEUTRON SOURCE

## 6.1. Introduction

Neutrons represent a unique probe for studying matter on the molecular scale, thus opening a wide range of applications: from material science to life science, from engineering and industrial applications to fundamental physics experiments. Highest intensity neutron sources in the world are based essentially on three different production mechanisms: some fission processes (such as that of $^{235}U$) in nuclear reactors, proton-driven spallation sources, and electron-driven (photo-production) sources. Reactor sources usually deliver a continuous flux of neutrons (but it is possible to have a pulsed reactor source, like in JINR, Russia, or in Triga reactors like the one in Pavia), while accelerator driven sources are generally pulsed (with some exception, like SINQ at PSI, Switzerland). Pulsed sources provide the additional information of the time-stamp of the neutron bunches that, together with techniques determining the time of arrival of neutrons at the detector, enable the use of time-of-flight for energy measurement. The pulsed sources generally provide fast neutron bursts a fraction of μs long ("short" pulses), that generally will be moderated to hot and thermal energies of tens-hundreds of μs, while the new European facility in Lund, Sweden (ESS) will produce "long" pulses of 1 ms.

The limitations of reactor technology are well know, so that the next generation neutron source in Europe will be based on accelerators. Of course spallation is a more efficient approach with respect to production from an electron beam, so that is the choice for the most important facilities: SNS at Oak Ridge and LANSCE in Los Alamos, USA, JSNS at Jaeri, Japan, CSNS facility in Dongguan, China; while in Europe SINQ at PSI, and ISIS at RAL, UK, will be joined by the new ESS facility. A separated chapter is represented by the CERN facility, n-TOF, mainly dedicated to nuclear reactions cross-section measurements, driven by a 20 GeV proton beam extracted from the SPS and producing a broad spectrum, ranging from thermal to 1 GeV neutrons.

However, with a large community using neutrons and due to the high specialization of instruments for the different classes of neutron experiments, photo-production sources demonstrated to be very useful, as the GELINA facility based on a 70-140 MeV electron beam in Geel, Belgium. Another interesting example is the n-Elbe source, driven by a 40 MeV electron beam, at FZ Dresden-Rossendorf. Indeed, photo-production facilities can be more cost-effective than spallation sources for neutron fluxes up to $10^{15}$-$10^{16}$ n/s at the target, even though the neutron yield per primary electron is (depending on the primary beam energy) at least 10-20 times lower with respect to proton-induced spallation.

Some of the most significant examples of neutron accelerator-driven sources, together with their main parameters, and summary of the main applications, are listed in Table 6.1

**Table 6.1 Representative selection of existing neutron sources and their characteristics**

| Facility (Country) | Main parameters | Neutron rate [n/s] | Applications |
|---|---|---|---|
| **photoproduction** | | | |
| **HUNS** (Japan) | $E_e$=35MeV, $I_{average}$=30 μA Frequency=50-100 Hz | $1.6\times10^{12}$ | Imaging, small angle neutron scattering (SANS), device development, nuclear data |
| **ORELA** (USA) | $E_e$= 180 MeV $I_{peak}$= 20 A Frequency=525 Hz | $10^{14}$ | Neutron capture for nuclear physics and nuclear astrophysics |
| **GELINA** (Belgium) | $E_e$= 100 MeV $I_{average}$= 100 μA Frequency=800 Hz | $3.4\times10^{13}$ | Neutron resonant capture analysis (NRCA), neutron data, nuclear structure |
| **n-ELBE** (Germany) | $E_e$=: 40 MeV $I_{average}$= 1 mA Frequency=1.5 MHz | $2.7\times10^{13}$ | Neutron cross section data or fast fission reactors physics; Long-lived to short lived nuclei transmutation processes |
| **spallation** | | | |
| **SNS** (USA) | $E_p$=1 GeV $I_{average}$= 2 mA, Frequency= 60Hz, | $10^{17}$ (estimated) | Diffraction, spectroscopy, reflectometry |
| **ISIS** (UK) | $E_p$= 800 MeV $I_{average}$= 200 μA, Frequency=40Hz(10 Hz) TS-I (TS-II) | $10^{16}$ (estimated) | Diffraction, spectroscopy, reflectometry |
| **J-PARC** (Japan) | $E_p$= 3 GeV $I_{av}$= 333 μA Frequency= 25 Hz | $3\text{-}5\times10^{16}$ | Diffraction, spectroscopy, spin-echo, imaging |
| **SINQ-PSI** (Switzerland) | $E_p$: 590 MeV $I_{av}$= 1 mA frequency;. Continuous | $3\times10^{16}$ | Diffraction, spectroscopy, imaging, reflectometry |

As far as reactor sources are concerned, with the exception of the main European facility in Grenoble (ILL), many countries in the Mediterranean region lack nuclear reactors or are considering decommissioning. Adding that the main accelerator-driven sources are concentrated in central-northern Europe, this leaves room for a medium neutron source in southern Europe, even considering that it could be operative in 3-4 years from approval. Indeed, concentrating on neutron scattering, in the last decade (2000-2009) approximately 3,000 experiments were carried out in European facilities, using about 20,000 beam days; the 6,000 European scientists published more than 2,000 papers/year [Report from the ILL Associates' Working Group on Neutrons in Europe for 2025 (2011)].

Finally, the impact of a neutron facility relies on many factors beyond the pure source power, such as the availability, uniqueness and quality of the instrumentations, the operational support, cost and ease of use, support of the academic community, etc.

The opportunity of using an intense, high-energy, electron beam – produced in the case of the IRIDE project by superconducting high average current linacs – opens the possibility of having a new photo-production neutron source (as well as of charged hadrons, as described in the following), that can have a significant place in the European panorama in the coming decade and can be a very useful facility. A source in the range of **$2\times10^{14}/1\times10^{15}$ n/s** at the production target (for the pulsed/continuous options, respectively, as described in the following), would indeed not only enlarge the access capacity for the large community of European scientists, but also could provide a complementary and original approach with respect to the existing neutron facilities, thus enriching the experimental and application opportunities, and profiting of the large wealth of expertise in the field.

The time structure of the electron beam from the IRIDE linac(s) is of course connected with the requirements of the other components of this interdisciplinary project, ranging from the X-FEL, to the Compton back-scattering source, from the electron-electron and gamma-gamma collider, to laser-plasma research. The baseline option for the neutron source is to have a "short" (referred to accelerator-driven neutron sources) electron pulse of the order 1 μs, of course micro-bunched with the radio-frequency of the superconducting linac (we assumed 0.5 μs bursts, bunched at 1.3 GHz, of 1 nC charge). At a repetition rate of 50 Hz and a beam loading of 20% this would translate in $2\times10^{14}$ n/s at the production target, almost an order of magnitude more with respect to GELINA and n-ELBE, and obviously still very far from a spallation source. The alternative option is to exploit the possibility of continuous wave operation of the super-conducting linac. As described in the following sections detailing the scientific opportunities, this would limit the range of possible experiments and beam-lines (as further discussed in Section 5), but of course would translate in a much higher neutron flux of $1\times10^{15}$ n/s, and practically with no influence on the operations of the other IRIDE components.

Apart beam power, the neutron yield depends (weakly) on the electron beam energy, while the target material and geometry are very important parameters. The target optimisation and some considerations on the target station and biological shielding are discussed in Section 4: the baseline figures of $2\times10^{14}/1\times10^{15}$ n/s have been computed with a conservative assumption of a tungsten target struck by a 1.5 GeV electron beam of 32.5 mA/1 mA peak current in the pulsed/continuous options. A factor 3 higher yield can be roughly expected just by using an Uranium target.

In addition to the different research areas that can be covered using this kind of neutron source (briefly described in Section 2), a number of industrial applications can be envisaged for such a neutron facility: a wide range of applications can be grouped under the definition of non-destructive analysis, while another interesting area is neutron irradiation. Section 3 summaries the concrete interest that could be found in industries and more generally in the applied field.

Additional scientific opportunities are related to the production of charged hadrons, in addition to neutrons, and are summarized in Section 6.

Finally, a neutron facility of the kind we are proposing, can have a positive impact under other two important points: training and education of young scientists, and development of new neutron technique, including detection techniques. Indeed, in addition to synergy and complementarity with the main larger infrastructures, it can also play a role as replacement of small (mainly reactor-based) facilities. Not only a lower neutron intensity can be sufficient for some classes of experiments, but it can be beneficial for different aspects:, designing and testing new experimental and data acquisition techniques, development of new or challenging instruments, radiation safety and protection problems; more generally, developing knowledge and expertise to be then exploited to larger, higher-power sources, where the pressure for access and operational costs are significantly higher. Another interesting example in this respect is the field of neutron detectors, that has gained importance in the last years due to the growing problem of $^{3}$He-replacement, and can clearly benefit of an easier access and usability, lower intensity facility.

### 6.2. Scientific Opportunities

This section summarizes the techniques in which neutrons are used for fundamental and applied research and that can be developed on a neutron source as the one proposed in this project. There are other techniques that will not be mentioned here. It is important to underline that the italian neutron user community is formed by more than 500 users who performed at least one experiment in the last 10 years. Among them there are about 340 frequent users who did at least 5 publications involving the use of neutrons in the last 10 years. The scientific productivity of the neutron user community is anyway fairly good since the total number of publication since 1990 is about 3700 (2012) with an average of 167 publications per year while the average of the last 5 years is 219. The Italian users community is represented by the Società Italiana di Spettroscopia Neutronica (SISN) who can count about 250 associates in the last 10 years.
The development of a national neutron source could give a strong pulse to the growth of the community since it could also work as training facility and a closer and important place to perform neutron scattering experiments.

- *Neutron Activation*

Neutron Activation is a nuclear process used for determining the concentrations of elements in a large amount of materials. The method is based on neutron activation of a sample and the analysis of the energy of the gamma rays emitted by the decay of the activated nuclides. The sample is bombarded with neutrons, causing the elements to form radioactive isotopes and since the radioactive decay path for each element is known, it is possible to detect the origin nuclide and then to perform an absolute count of such an isotope. Using this information, it is possible to determine the concentrations of the elements within the irradiated sample. A particular advantage of this technique is that it does not destroy the sample, and then it can be used for analysis of works of art and historical artifacts. The accuracy of INAA is in the region of 5%, and the relative precision is usually better than 0.1%. There are actually two methods used at the moment: the Instrumental Neutron Activation Analysis (INAA) in which the analysis is performed at the end of the irradiation process. It needs an irradiation facility inside the neutron source where a the irradiation is monitored, and a Ge detector equipped chamber, far enough from the neutron source so that the gamma background is negligible, to measure the gamma spectrum of the activated sample. The second method implies the simultaneous neutron irradiation and gamma emission measurement of the sample and it is then called Prompt Gamma Activation Analysis (PGAA).

*- Neutron Diffraction*

Neutron diffraction is the application of neutron scattering to the determination of the atomic and/or magnetic structure of a material. A sample to be examined is placed in a beam of thermal or cold neutrons to obtain a diffraction pattern that provides information of the structure of the material. The technique is similar to X-ray diffraction but due to their different scattering properties, neutrons and X-rays provide complementary information. The technique requires a source of neutrons. Neutrons are produced in a nuclear reactor or spallation source. At a research reactor, other components are needed, including a crystal monochromator as well as filters to select the desired neutron wavelength. In order to change the neutron wavelength selected for the measurement, some parts of the setup may be movable. At a spallation source, the time of flight technique is used to sort the energies of the incident neutrons (higher energy neutrons are faster), so no monochromator is needed, but rather a series of chopping elements, synchronized with the neutron burst produced by the spallation phenomenon, to filter neutron pulses with the desired wavelength range.

The technique is most commonly performed as powder diffraction, which only requires a polycrystalline powder. For single crystal analysis, the crystals must be much larger than those used in X-ray crystallography. It is common to use crystals that are about 1 $mm^3$.

Although neutrons are uncharged, they carry a spin, and therefore interact with magnetic moments, including those arising from the electron cloud around an atom. Neutron diffraction can therefore reveal the microscopic magnetic structure of a material (Izyumov 1991). Magnetic scattering does require an atomic form factor as it is caused by the much larger electron cloud around the tiny nucleus. The intensity of the magnetic contribution to the diffraction peaks will therefore reduce towards higher angles.

Neutron diffraction can be used to determine the static structure factor of gases, liquids or amorphous solids. Most experiments, however, aim at the structure of crystalline solids, making neutron diffraction an important tool of crystallography.

Neutron diffraction is closely related to X-ray powder diffraction (Ibberson 2002). Because the data is typically a 1D powder diffractogram they are usually processed using Rietveld refinement. In fact this method was born in neutron diffraction (at Petten in the Netherlands) and was later extended for use in X-ray diffraction.

One practical application of elastic neutron scattering/diffraction is that the lattice constant of metals and other crystalline materials can be very accurately measured. Together with an accurately aligned micropositioner, a map of the lattice constant through the metal can be derived. This can easily be converted to the stress field experienced by the material. This has been used to analyse stresses in aerospace and automotive components to give just two examples. This technique has led to the development of dedicated stress diffractometers, such as the ENGIN-X instrument (Santisteban 2006) at the ISIS neutron source.

At the moment there are about 60 neutron diffraction beamlines in Europe with open access related to the proposal submission system but even if the number might appear high there is a ratio 2:1 between the number of experiments proposed with respect to the ones approved to be performed in the facilities because of the high level of requests.

*- Small Angle Neutron Scattering*

Small angle neutron scattering (SANS) (Kostorz 1982, Feigin 1987) allows the determination of structural features, such as size and volume fraction, of matrix inhomogeneities in a huge variety of materials.
Typical applications include: (i) Biology: organization of biomolecular complexes in solution; mechanisms and pathways for protein folding; (ii) Polymers: conformation of polymer molecules in solution; structure of microphase for separated block polymers; (iii) Chemistry: structure and interactions in colloid suspensions, mechanisms of molecular self-assembly in solutions; (iv) Materials Science: precipitation mechanisms and kinetics as functions of thermal treatments (e.g. ageing); cavitation induced by thermomechanical processes (e.g. welding); crystalline structure investigations.
The order of magnitude of the size objects that can be detected is in the approximate range 1-$10^3$ nm. It is usually considered as a complementary technique to Transmission Electron Microscopy (TEM). In fact, a correct use of SANS is often based on some information that can be obtained from TEM, such for instance the particle shape. On the other hand, SANS/SAXS are performed on cross-sectional areas of some $mm^2$, or even $cm^2$, so that their results are much more significant, from a statistical point of view, than the ones obtained by TEM (few $\mu m^2$).
Information about the gyration radius Rg and the specific surface of the particles can be obtained from the behaviours at low (Guinier's approximation) and high (Porod's approximation) transferred momentum Q, respectively.
In the case of polydispersion, the experimental data can be fitted to determine the size distribution N(R). According to the system under investigation, Gaussian, log-normal, Weibull, Pearson or other distributions can be used for N(R). If no particular distribution shape can be assumed a priori, or also when more than one particle families could be present, then a common method is to write N(R) as a linear combination of β-spline functions.

- *Elastic Incoherent Scattering*

In order to characterize on a nanoscopic scale the flexibility and the resilience of material systems, through the evaluation of the atomic Mean Square Displacement (MSD) and its temperature dependence, Zaccai (Zaccai 2000) formulated a successful approach based on the determination of Elastic Incoherent Neutron Scattering (EINS) intensity as a function of temperature, at a fixed instrumental energy resolution. Such an approach has been applied for investigating a wide range of systems of both scientific and industrial interest, this latter including biomedical, pharmaceutical, cosmetic and food products. From the pure research point of view, this method has been successfully employed to investigate the mean macromolecular dynamics of many biological systems of increasing complexity, such as, just to exemplify, extremophiles (Tehei 2004), which are organisms living in extreme conditions, such as very low (psychrophiles) and very high temperatures (thermophiles and hyperthermophiles), by revealing that the mean macromolecular resilience increased with physiological temperature, as pointed out by the increasing values of the effective mean force constant, so allowing to conclude that larger resilience allows macromolecular stability at high temperatures, while maintaining flexibility within acceptable limits for biological activity.

In a complementary way, the Resolution Elastic Neutron Scattering (RENS) approach [3], based on the measure of the elastic scattering law as a function of the instrumental energy resolution, allows to characterize the same system properties, at a fixed temperature value. As an example, through the scattered intensity trend versus the logarithm of instrumental energy resolution, which is described by an increasing sigmoid, the system relaxation time can be extracted without the use of any model fitting function.
Both the approaches can take advantage from the high intensity of elastically scattered intensity in comparison to quasi elastic neutron intensity at low energy exchange, which is an important properties in dealing with exotic systems, and from the direct access to the system properties.

- *Chip irradiation*

A beam line dedicated to e study of neutron-induced Single Event Effects (SEE) in electronics (single chips as well as complex architectures such as Field Programmable Gate Array or devices under operation like PCs) has to provide a neutron spectrum that resembles the atmospheric one as close as possible with an intensity as higher as possible to speed up SEE tests. Pencil beam and open beam configurations has to be taken into consideration to allow for single-to large electronics arrays to be irradiated. As a matter of fact, SEE studies in the atmospheric neutron field using balloons, or at altitude (in mountain) are long lasting measurements (in the order of several months). Fig. 2.1 reports on a single plots te available neutron spectral fluence rates for the atmospheric field (scaled for a factor 1E7 and 1E8) and at different large scale facilities where chip irradiation is performed [Andreani_2008].

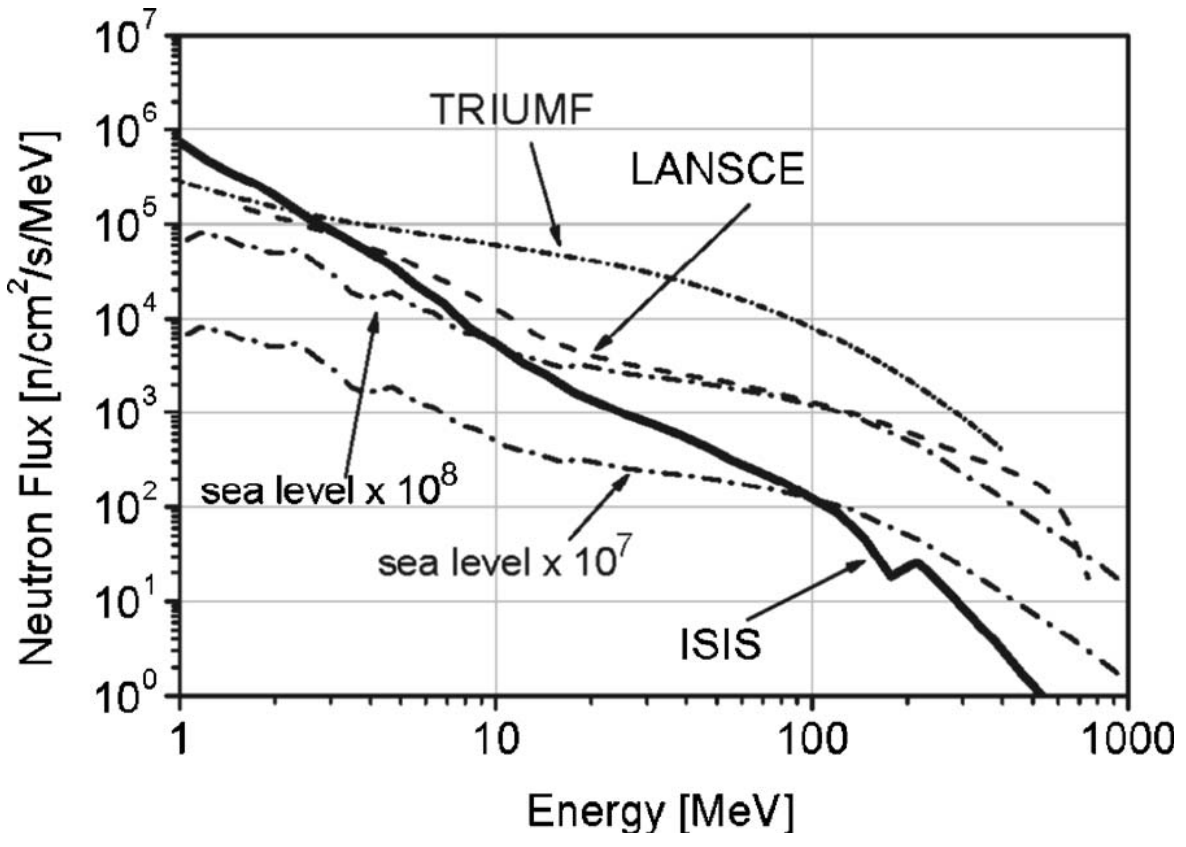

Figure 6.1: Neutron energy spectrum provided by the ISIS spallation neutron source at about 12 m from a water. Also shown forcomparison are the neutron energy spectra at LANSCE and TRIUMF facilities,as well as the terrestrial one at sea level, multiplied by 107 and 108.

The IRIDE neutron source is expected to provide about two order of magnitude less spectral fluence rate with respect to ISIS and extending up to 100-200 MeV thanks to the use of 1 GeV electrons beam for neutron photoproduction.
The neutron spectrum from the target should be tailored properly to resemble as close as possible the atmospheric profile. The possible use of filters is envisaged to allow the selection of restricted energy region in the thermal ad epithermal eergy region to study the effect of neutron energy related dose effects. With a fluence rate of about 1E4 n/s/cm$^2$ a the irradiation point, one year of data taking in atmosphere should be simulated within tens of minutes.

*- Imaging*

Neutron imaging is the process of making an image with neutrons. The resulting image is based on the neutron attenuation properties of the imaged object. The resulting images have much in common with X-ray images, but since the image is based on neutron attenuation properties of the material instead of X-ray attenuation ones, some features easily visible with neutron imaging may be very challenging or impossible to see with X-ray imaging techniques (and vice versa).
X-rays are attenuated mainly basing on material density. High density materials will stop more X-rays. Concerning neutrons, the capacity of a material to attenuate a neutron beam is not related to density. Some light materials such as boron will absorb neutrons while hydrogen will generally scatter neutrons, and many commonly used metals allow most neutrons to pass through them. This can make neutron imaging better suited in many instances than X-ray imaging; for example, looking at hydrogen rich components or complex metal artifacts.
The first commercial facilities came on-line in the late sixties, mostly in the USA and France, and eventually in many other countries including Canada, Japan, South Africa, Germany, and Switzerland.
To produce a neutron image the following components are necessary: a source of neutrons, a collimator to shape the emitted neutrons into a fairly mono-directional beam, an object to be imaged, and a detection system. The latter is composed by a scintillator plate to convert neutrons into light and by an optical device to focus the light emitted into a CCD digital camera to record the pictures.
At the moment there are 5 beamlines available in Europe for neutron imaging and there is a very high request with respect to the available beam-time (more than 3:1 ratio).

*- NRCA*

Several elements in the periodic table show in the absorption cross section intense and relatively narrow peaks (resonances), each resonance being peculiar of the absorbing nuclide. Different isotopes of the same element may have different resonance energy. By using pulsed epithermal neutrons (roughly below 10 keV down to 1 eV) resonances are identified as peaks in the time of flight spectrum. Each resonance is the fingerprint of a nuclear specie (isotopical recognition) thus allowing for the elemental material analysis (qualitative and quantitative) especially on metallic samples (e.g. cultural heritages). The main requirement for such measurements is the use ot the TOF technique with epithermal neutrons.

*- Bragg Edge Transmission*

Most neutron tomography stations work with a ‘white’ spectrum where the neutron wavelength range from 0.5 up to 10 Angstrom. The standard transmission analysis integrates the attenuation coefficients over all the full energy range of neutrons so that the energy dependent features are mostly lost in the averaged data. Therefore, the exploitation of energy dependent attenuation effects is an important development in the field of neutron imaging. Strong changes in the total neutron cross sections of crystalline materials are related to Bragg edges (Fermi 1947), that are variations in the transmission data related to the removal of specific wavelength neutrons from the primary beam due to Bragg diffraction. Bragg edges are most pronounced and best separated in the cold neutron energy range which covers the low-indexed lattice planes of crystals. Energy-selective transmission studies require a high energy resolution for Bragg edge transmission studies or for selection of narrow energy (wavelength) intervals in order to use the Bragg edge contrast effects. When a polychromatic neutron beam passes through a crystalline material, neutrons of different energies are attenuated differently due to this effect. As a result, the energy spectrum of the neutron beam changes according to the sample. A careful analysis of the intensity ratio between the transmitted and incident beams can provide information about the microstructure of the sample. Fig. 1 shows four typical transmission spectra, all of them from specimens made of Cu-based alloys (Kockelmann 2007). Fig. 1a shows the transmission of an imperfect single crystal, fig. 1b represents the transmission of an archaeological bronze made of relatively large grains, fig. 1c shows the transmission of a fine-grained isotropic Cu specimen, and fig. 1d also corresponds to a fine-grained material, but in this case the material is textured.
There are no operating beam-lines at the moment offering Bragg edge analysis capabilities as a standard technique but there are possibilities for skilled users to apply for them in a few of the neutron imaging or neutron diffraction beam-lines.

*- Metrology*

One of the most significant features of neutron fields is the very wide range of possible neutron energies. In the nuclear industry, for example, neutrons occur with energies few meV to 10 MeV, while for cosmic ray dosimetry the energy range extends into the GeV region. This enormous range sets a challenge for designing measuring devices and a parallel challenge of developing measurement standards for characterizing these devices according to the requirements of the International Organization for Standardization, ISO (ISO 2000).
The Italian National Institute of Ionization radiation Metrology (INMRI) is interested in having in Italy, and especially in Roma area, a high energy neutron source in order to develop primary standards for neutron emission rate and energy spectrum calibration. The ENEA-INMR can currently dispose of primary standards for neutrons up to the Am-Be neutron source (few MeV). IRIDE, due to a higher energy electron beam compared to the other photo-production sources, could allow an extension of such standards to hundreds of MeV.

*- Detector development*

The research and development activity on He-free neutron detectors is of strategic importance for neutron applications in science and technology. As a matter of fact, the lack of $^3$He is triggering an interesting and stimulating technological effort in finding out effective substitutes of the He-gas tubes typically operating at both reactor and/or accelerator driven neutron sources worldwide. The most important request for the new detectors is to provide a high detection efficiency (above 50%), high rate capability (MHz/cm$^2$ or more) and large area covering (several m$^2$ just to give an order of magnitude). Further details will be given in Sec. 5

*- Electron Driven Subcritical Reactor for nuclear research (AdS)*

Among the various issues arising from the nuclear question, the one which has by far the deepest impact on public opinion is certainly the safe treatment of nuclear waste from energy production plants, and more generally of radioactive waste from industry, research and medicine.

Fission reactors or Accelerator-driven System (AdS) are good candidates to offer high rates of incinerating the Plutonium, Minor Actinides (MA) and Fission Products by transmutation (Stanculescu 2013). This goal is best achieved if a fast neutron spectrum is produced, owing to their larger excess neutron fraction and less neutron radiative captures, hence less MA generation. While critical Fast Reactors look more efficient and neutron cost-effective, AdSs present advantages in safety and flexibility. More specifically:

- power control in an AdS is possible up to a certain extent via the beam current
- even burn-up compensation could be done via the accelerator if properly dimensioned
- level of sub-criticality can be chosen, within the technological limits set by the accelerator, larger than βeff which is beneficial to power control and safety.

Spallation neutron sources, though very effective in neutron production, are large, expensive and presently would involve certain difficulties in their operation (e.g. beam trips). The use of an external neutron source driven by an electron accelerator has been considered in the past (Ridikas 2002), being an electron driver a rather low cost and compact machine that might bring many advantages in terms of reliability. An overall comparison between the electron-neutron converter and the spallation process was done for an electron driver coupled to 1) a nearly critical core, 2) a well subcritical core. In both cases the required beam power from the electron accelerator was huge, $5 \div 8$ MW, for an average flux of $\sim 10^{14}$ n/sec/cm$^2$ moreover the overall performances are strongly limited by the rather complicated and unsafe criticality control system in case 1) and the small active volume in case 2).

So, if an operating electron-driven AdS doesn't appear a viable option for P&T strategies, an e-driven 'classical' research reactor for AdS studies is however quite feasible and seems attractive, also because of its broad potentiality in other research fields (ultracold neutrons). Measurements of coupling efficiency, reactivity, multiplication, source-driven transients, neutron source importance in subcritical external source-driven assemblies, the development of benchmarks, validation of computer codes, etc are typical research goals for AdS studies that can be well carried out in a moderately (keff = $0.95 \div 0.98$) subcritical reactor coupled to a medium power photoneutron source, such as the one from the IRIDE SC Linac.

A significant experience in this sector was done within the RACE (Reactor Accelerator Coupling Experiments) project, a series of AdS studies which was meant as a part of the US DoE Advanced Fuel Cycle Initiative(Carta 2006). A comparison of the performance of an electron accelerator-driven TRIGA core experiment and of the proton driven TRADE originally planned at ENEA-Casaccia was done, focusing mainly at the dynamic power response of the core following reactivity and/or source transients. Preliminary results showed that transient system response at a level of subcriticality keff ≥ 0.98 have to be investigated at a minimum core power of 50 kW, which is attainable only with an Uranium target in the central position of the TRIGA reactor for an impinging beam power of 25 kW.

While more calculations and simulations are needed to get at least a preliminary estimate of expected performances, the realization of a research reactor with a classical lattice structure, made of low-enriched (< 20%) U-Al fuel, light water-cooled, surrounded by a graphite reflector and driven by a 150 kW electron beam impinging on a U-target (equivalent to a neutron source strength of ~ $5 \bullet 10^{14}$ n/sec) appears quite feasible in the actual IRIDE framework (see Figure). If we consider that the expected neutron flux is ~ $7 \bullet 10^{13}$ n/sec/cm$^2$ for the RACE U-target (hollow cylinder geometry) scheme with keff = 0.978 and reactor power 48 kWth, driven by a 25 kW e-beam, we just can expect a flux of ~ $5 \bullet 10^{14}$ n/sec/cm2 in a volume much bigger than then one described in (Ridikas 2002).

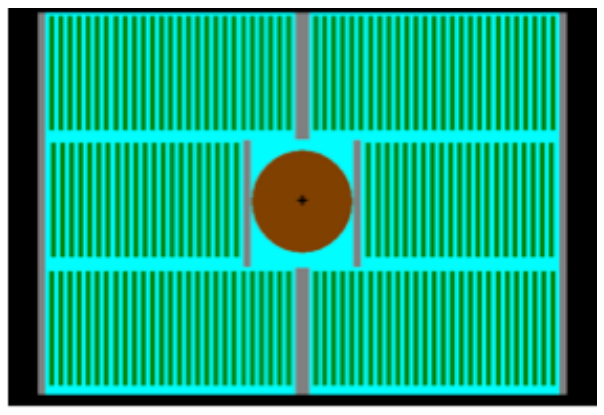

Figure. 6.2: A possible core design of an electron-driven reactor structure (Beller 2004), which shows a cross section of an MCNPX model of fuel trays with unequal tray sizes to produce a rectangular core. Color code: water is blue, target is brown, aluminum is gray, graphite is black, and fuel plates are green.

This system appears simpler, more flexible and of practical use as compared to the previously shown designs. The neutron producing target might be the only critical point for this structure, to keep thermal power at a reasonable level (< 500 W/cm$^3$).

### 6.3. Neutron oscillations

This section is devoted to fundamental physics studies on neutron oscillations feasible at the IRIDE neutron source.

**Neutron-antineutron oscillations** are very interesting since they would allow precision testing of the fundamental CPT-symmetry, where C is the charge conjugation, P is the inversion of space and T is the time-reversal, very closely connected to the quantum field theory through the CPT-theorem.

After the discovery of the neutron, Majorana considered it to be a Majorana particle, i.e. intrinsically neutral spinor with the Majorana mass. This idea did not work, since the neutron is electrically neutral but it has baryon number which protects its stability in nuclei, so the neutron must be a massive Dirac particle. On the other hand, nowadays understanding of particle physics and cosmology indicates that the baryon number *must* be violated, otherwise the existence of the matter in the Universe would be impossible: as was first suggested in a prominent work of Sakharov [Sakharov 1967], the generation of baryon asymmetry in the primordial universe requires violation of baryon number.

The possibility of neutron-antineutron oscillation, a baryon-number violating process with $\Delta B = 2$, was first suggested by Kuzmin in the context of the primordial baryogenesis.

This in fact means that the neutron, apart of the Dirac mass term, has also a Majorana type mass term, analogous to that of the neutrinos. The first particle mechanism of n→nbar oscillation in the context quark model based of unified gauge theories was suggested by Mohapatra and Marshak [Mohapatra 1980] while many other models were suggested in later times. In particular, n→nbar oscillation could naturally emerge in the context of supersymmetry from the R-parity breaking terms $u^c\ d^c\ d^c$ [Zwirner 1983].

More recently it was concluded that these oscillations would represent one of the most accurate test of the CPT symmetry [Abov 1984].

The baryon number violation implied by neutron-antineutron oscillation has to be compatible with large wealth of measurements on the stability of matter and with cosmological observations. On the other hand, the focus on this kind of experiment has been for a long time the proton decay, mainly through the process $p \to e^+ p^0$. In the framework of the seesaw mechanism, this is not the best suited process since it respects the new U(1) symmetry, the B-L quantum number of the grand-unified theory.  On the contrary, neutron-antineutron transition violates B-L by 2, exactly as in the neutrino masses seesaw mechanism.

The question is whether all physics beyond the standard model is encapsulated in grand unified theories, with neutrino masses being also a signal of the same GUT scale, or whether the neutrino masses indicate a rich new layer of physics at intermediate scales that involves low B-L breaking [R. Mohapatra 2009].

From the phenomenological point of view, we can consider a simple 2×2 Hamiltonian for the evolution of an initial beam of slow moving neutrons: the probability of finding a nbar in this beam after a time t is:

$$i\hbar\frac{\partial}{\partial t}\begin{pmatrix} n \\ \bar{n} \end{pmatrix} = \begin{pmatrix} E_n & \delta m \\ \delta m & E_{\bar{n}} \end{pmatrix} \begin{pmatrix} n \\ \bar{n} \end{pmatrix}$$

where δm expresses the n-nbar mixing, thus yielding the following formula for the probability of finding an antineutron at time t, given that at t = 0, $P_n(0) = 1$ and $P_{nbar}(0) = 0$:

$$P_{\bar{n}}(t) \ = \ \frac{4\delta m^2}{\Delta E^2 + 4\delta m^2} sin^2(\sqrt{\Delta E^2 + 4\delta m^2})t;$$

where $\Delta E = E_n - E_{nbar}$. Neutron decay does not affect this formula, as long as the flight path is much smaller than the lifetime.

In the case of neutrons inside matter, the difference of nuclear potential is so high (100 MeV), ΔE·t>>1, and that the probability expression simplifies to:

$$P_{n\to\bar{n}} \sim \frac{1}{2}\left(\frac{\delta m}{2\Delta E_n}\right)^2$$

where $P_{n\rightarrow nbar}$ is the oscillation probability in vacuum; oscillation probability inside the nucleus will be proportional to the vacuum probability.
The off-diagonal 2×2 matrix element, dm, is related to the oscillation time:

$$\tau_{n-\bar{n}} = \frac{h}{2\pi\delta m}.$$

For free neutron, the opposite limit is true: $\Delta E \cdot t \ll 1$, so that the oscillation probability simplifies to:

$$P_{n\rightarrow\bar{n}} \sim (\delta m \cdot t)^2 \equiv \left(\frac{t}{\tau_{n-\bar{n}}}\right)^2$$

The limit $\tau_{n\text{-}nbar} > 0.86\times10^8$ s obtained in the experiment by Baldo Ceolin et al [Baldo Ceolin 1994] at the Institut-Laue-Lengevin (ILL), Grenoble, remains a strongest direct limit on n-nbar oscillation until today. On the other hand, comparable or even somewhat stronger limits were obtained from the nuclear stability bounds [Beringer 2012]. It should be noted, however, that these limits are subject of the nuclear matrix elements which have rather big theoretical uncertainties. This value translates into a limit on $\delta m < 6\times10^{-33}$ GeV, so that the nucleus instability lifetime can be inferred, taking also into account nuclear effects [Friedman 2008], at the level of $3\times10^{32}$ years. Similar limits are obtained with nuclear decay searches at SuperKamiokande.

If discovered, n→nbar oscillation will establish a new force of nature and a new phenomenon leading to the physics beyond the SM at the energy scale above TeV. In addition, will help to provide understanding of matter-antimatter asymmetry and origin of neutrino mass.

If not discovered, the oscillation will however set new limits on the stability of matter, exceeding the sensitivity of large nucleon decay experiments, and will place constraints on large class of R-parity conserving super-symmetric models.

The figure of merit of an experiment probing n→nbar oscillation is thus the product of the neutron flux times $(t/\tau)^2$. The idea is to have an intense cold-neutron beam, and then let it pass through a vacuum and de-magnetized decay tunnel, and then search for antineutron annihilation signal in the detector, i.e. a multi-pion event with an energy of about 2 GeV.

The de-magnetization requirements comes from the neutron-antineutron energy splitting due to the 0.5 G terrestrial magnetic field: $\mu_n B \sim 3\times10^{-21}$ GeV, which, with a typical flight time of order 1 second, largely exceeds the free neutron limit.

Since the two crucial factors are the flux and the flight-time, the slower the neutrons, the longer the time on a given path, thus increasing the sensitivity of the experiment. Thus, a cold neutron source would be desirable for such kind of experiments. It will be of reasonable interest to perform such an experiment using cold neutrons generated at IRIDE, propagating them in >100 m distance in good vacuum conditions and magnetic field suppressed below 0.1 mG, essentially as it was done in the experiment [5], however with larger luminosity of neutrons. In addition, pulsed shape also can be used for discriminating the background.

1 meV neutrons travel at v=437 m/s, the time of flight over a length of ≈250 m will be t≈0.57 s. If we can have a cold moderator, e.g. solid methane or solid para-hydrogen, we can expect to reach a sensitivity depending on many parameters: in addition to the neutron energy spectrum, the distance between neutron source and moderator, the emission area of the moderator, the distance between moderator and annihilation target, diameter of the annihilation target, the type, geometry and details of the mirrors/reflectors/guides, etc.

Once a consistent neutron flux is available, the cost of such kind of experiment would be driven by the decay tunnel, both from the point of view of vacuum and of the magnetic shield. Considering an approximate maximum decay tunnel length of 250 m long experiment, a cost of 2.5+2.5 millions can be roughly envisaged (apart from excavation), to which the cost of the cold moderator and reflector (5 millions) and of the detector should be added. The beam-line should take into account the not negligible deflection of such a slow neutron under gravity: about 1.6 m for t=0.57 s.

A similar proposal has been put forward by the NnbarX Collaboration at Fermilab Project-X, and recently the PAC has approved the Expression of Interest letter. Of course, this experiment requires the realization of a spallation source in the framework of the Project-X complex will be necessary.

Another process which can be tested at IRIDE is the **neutron oscillation into mirror neutron**, which is the neutron's mass degenerated partner from hypothetical parallel mirror world. Such a parallel sector can represent dark matter and its neutrinos can be natural candidates for sterile neutrinos existence of which is required by some present experimental data on neutrino oscillations (for a review, see e.g. [Berezhiani 2008]). The neutron oscillation n-n' into its sterile mirror partner is a baryon number violating process with $\Delta B=1$. Surprisingly, neither present experimental data nor astrophysical and cosmological limits exclude the possibility that n-n' oscillation is very fast, with an oscillation time $\tau_{nn'}\approx 1$s [Berezhiani 2009], i.e. faster than the neutron decay. This fact attracted the interest of the experimentalist and in last years several experiments were performed for the search of n-n', mainly at the ILL, Grenoble [Ban 2009, Serebrov 2009]. These experiments searched for anomalous losses of ultra-cold-neutrons and interestingly, the data of last experiment [Serebrov 2009] indicate to anomalous dependence of the neutron losses on the applied magnetic field, with more than 5σ away from the null hypothesis [Berezhiani 2012] which can be a signal for n-n' transition.

Oscillation $n\to n'$ can be searched by the same setup which can be used for the search of neutron-antineutron oscillation: propagation of cold neutrons at distances >100 m in a horizontal tube with good vacuum conditions and controlled magnetic field. Namely, one can place a neutron absorber in the middle of the propagation baseline and study the neutron regeneration process (walking through the wall) due to $n\to n'\to n$ conversion.

This would be a direct prove of n-n' oscillation phenomenon.

### 6.4. Applications outside research

The characteristics of the neutrons that make them of interest for applied research can of course be used in industrial research. Examples of industrial field with known applications with neutrons are:

- Efficient and cost-effective fabrication of a variety of **advanced materials**. Neutron scattering techniques can be used for the development of novel transformation-induced plasticity steels or precipitation-hardening Al and Ni alloys, to be used, for example, in aeronautics.
- **Pharmaceutical products**. The development of new drugs and drug delivery systems, which is strictly related to the detailed understanding of the mechanisms of disease [Whitten 2008, Christie 2012, Hofmann 2010, Sanson 2011], as well as the improvement of the product shelf-life can be carried out by the employment of neutron techniques.
- **Thermoelectric materials**. Here, neutrons allows to identify efficient and non-pollutant systems for the development of innovative thermoelectric devices combining low thermal conductivity with high electrical conductivity [Keppens 1998], to be employed for waste-heat recovery and in the refrigeration industry.
- **Renewable energy** sources. In such a field, more and more effective engines, materials for lower heat loss and less energy spill and greener processes for industry are requested. Novel materials for solar and fuel cells [Lee 2011; Parnell 2010;Magraso 2012; Gebel 2011], as well as hydrogen storage materials [Yan 2010;Brown 2009] can be developed thanks to neutrons.
- **Agro-food** systems. Plant strategies and metabolism in resistance to drought [Wood 2007; Pieper 2008;Oswald 2008] can be characterized by neutron methods.

The development and improvement of advanced materials is important in many fields, such as aeronautics, e.g. Al and Ni super alloys and/or materials for turbines. Such systems requires to combine high efficiency and low energy consumption. This can be achieved by employing neutron techniques, furthermore, thanks to the highly penetrating nature of neutrons, the control of technological materials can be performed. Domain-structure for the design of new devices can be characterized by Small Angle Neutron Scattering (SANS). In addition, new sol-gel surface-treatment formulations for the bonded repair of aircrafts and the laminated structure and toughening mechanism of abalone shell can be developed starting from neutron scattering findings.
The method of chip irradiation by neutrons helps in overcoming the serious problems created by the chip malfunctioning induced by Soft Errors (SE), corrupting one or several bits on the chip, in safety-critical systems, such as the electronic systems on board of airplanes and satellites.

The investigation of engineering materials the study of the properties of multi-component devices can be performed by neutron techniques thanks to their non-destructivity, penetration, high precision and sensitivity, this latter property be relevant for allowing to distinguish among isotopes, such as hydrogen and deuterium. In the non-destructive testing of engineering materials, neutron imaging and SANS can be effectively employed to improve, for example, materials for construction, fuel cells, soot sedimentation and hydrogen storage and to study strains, textures, porosity, structural defects and inhomogeneity and microstructures deep in metal components and thin-film facing metals, allowing investigation of crucial processes, such as crack growth. The integrity and performances of these materials can be investigated by neutron imaging also under processes, such as production conditions and hydrogen content variation. Furthermore, neutrons can be used also to characterize the water diffusion in polyamide engineering plastics [Laurati 2012].

The improving of the sample environment make neutrons a very effective probe for investigating materials of smaller and smaller dimensions and of increasing complexity also under realistic processing conditions, therefore novel systems, showing, for example, high temperature superconductivity, new light-emission properties or negative thermal expansion, which are temperature and pressure dependent properties, can be characterized by neutron scattering. Furthermore, neutron techniques allow to characterize the interaction of local and acoustic phonon modes which affect the thermal conductivity [Schweika 2007; Christensen 2008] as well as the cage-like host structures, so providing information on thermoelectric materials that can be used in thermoelectric devices for waste-heat recovery.

Quasi Elastic Neutron Scattering (QENS) is a powerful technique for investigating processes involving systems of pharmaceutical interest, such as the diffusion of lidocaine through a polymer film as a function of hydration [Padula 2010], while the stabilizing effect of natural bioprotectant systems on labile biomolecules has been demonstrated by QENS and Elastic Incoherent Neutron Scattering (EINS) [Magazu' 2004], so improving the applications of these systems in food, cosmetic and pharmaceutical industry.

In medical field, the Neutron Stimulated Emission Computed Tomography (NSECT) technique, which uses prompt gamma emission at higher neutron energies to provide many of the advantages and accuracy of neutron activation analysis (NAA) and allowing a rapid *in vivo* elemental analysis, can be employed to evaluate the concentration of some elements which are related to human diseases, such as breast cancer.

To get insight into catalysis, in particular in the process steps and the specific catalytic activity, neutrons are very effective in locating hydrogen atoms and in determining the chemical state [Albers 2000; Lennon 2012; Stepanov 1998], so elucidating the structure and function of catalysts. Furthermore they contribute to design newly-developed photocatalytic materials and surfaces for the purposes to environmental clean-up applications as well as hydrogen generation and storage.

In the field of hydrogen storage, safety and efficiency are fundamental conditions to be realized. Here, QENS and EINS [Salles 2009] can be used to determine the amount of unbound hydrogen and dihydrogen [Georgiev 2006] as well as the diffusion of hydrogen in systems to be employed in the hydrogen industry, such as hydrates, nanotubes and clathrates, while SANS allows to tools for studying the confinement of small molecules in the reservoirs [Clarkson 2012; Melnichenko 2012]. On the other hand, ion exchange and lithium and hydrogen conductivity in systems for energy storage to be applied for example for backup energy supply or batteries can be investigated by neutron reflectivity and imaging [Nishimura 2008; Sharma 2010; Kardjilov 2005; Senyshyn 2012], while the diffusion of lithium ions can be characterized by QENS.

The industrial applications of polymeric systems are very well known and a lot of neutron scattering investigations are addressed to understand their interactions and phase transition. To exemplify, polymer networks with a hierarchical configuration [D. J. Waters, K. Engberg, R. Parke-Houben, et al. Structure and mechanism of strength enhancement in interpenetrating polymer network hydrogels. Macromolecules, 44, 5776-5787, 2011] can be investigated in a wide Q-range by SANS, while the dynamic properties of many polymeric nano-composite materials driving the material functions and the polymer segmental dynamics in grafted silica-polymer composites coupled to the aggregation of nanoparticles [P. Akcora, S. K. Kumar, V. Garcia Sakai, et al. Segmental dynamics in PMMA-grafted nanoparticle composites. Macromolecules, 43, 8275-8281, 2010] are characterized by QENS. This elucidates the structure-property relationship of nano-composite materials.

On the other hand, neutron reflectometry is very effective in the study of polymeric thin films, by combining the deuterium labeling. The material integration, as developed by neutron scattering findings, produces a synergistic effect and allows the development and production of non-conventional materials exhibiting new functionalities.

Research on polymeric systems is also addressed to the understanding of their behavior in relation with their industrial applications in sensors, polymer batteries and organic photovoltaic systems. Here, since at the neutron instruments it is possible to duplicate industrial processing conditions, neutrons can represent an useful probe for characterizing the collective dynamics and self-assembly as well as the surface and interface interactions in polymeric systems. In organic photovoltaic devices it is important to investigate the structure of the active layer [K. H. Lee, P. E. Schwenn, A. R. G. Smith, et al. Morphology of all-solution-processed bilayer organic solar cells. Advanced Materials, 23(6):766-770, 2011] in order to enhance their performances and this is possible by neutron reflectometry. In addition, in order to optimize the energy efficiency of fuel cells, neutron imaging, covering an extended spatial domain, provides information about the location and motion of light atoms in fuel cells, so allowing to identify novel candidate materials [Magraso 2012; Ahmed 2008] and the failures in the components [Gebel 2011;Strobl 2009].

SANS instruments can be also used to investigate the mechanisms of the hydration process of various waste management materials, so allowing to develop durable barriers for waste management. Furthermore, structural details on various length scales in materials like rocks, sands and soil can be determined by neutron imaging and NAA, providing useful information on water economy and transport as well as on soil characteristics and to improve the knowledge of natural resources. Finally, the internal composition of artifacts of archaeological and cultural interest can be investigated without any damage by neutron imaging, NAA and SANS, which require small sample sizes and no chemical preparation, by connecting macroscopic and microscopic structures.

Among neutron techniques, QENS, EINS, SANS, reflectometry, imaging (radiography and tomography), NAA/NSECT and chip irradiation are therefore the most proper choice for industrial and technological applications, since a wide range of length scales can be explored using high temporal resolution. Let us discuss some instruments that could be realized for IRIDE for industrial investigations.

The characterization of the molecular processes can be performed by **EINS**, by means of the so called "fixed-windows" method, taking advantage from the fact that, besides to a relatively low number of fitting parameters, in respect to the quasi-elastic contribution at low energy transfer, the elastic one is often a factor 100-1000 higher and then for obtaining good quality data in reasonable times, due to the usually limited neutron fluxes, one can cope with a relatively small amount samples, with small sized samples or with strongly absorbing. In this frame, the Resolution Elastic Neutron Scattering (RENS) approach [Magazu' 2011] provides information which are independent on the used model functions for fitting spectra.
A **SANS** instrument based on multiple detector banks and covering a Q-range from $10^{-3}$ $Å^{-1}$ to 0.1 $Å^{-1}$ needs a high resolution and bulky sample environment in order to analyze industrial processes, while polarized neutrons with good Q resolution will allow to study correlated structures.

Neutron **reflectometry** instruments can have different employments: in a horizontal reflectometer, an inclined beam allows to explore a wide Q range without moving the sample and to measure horizontal surfaces, such as liquid-air interfaces, while solid interfaces can be investigated with high resolution by a vertical reflectometer. In general terms, this technique can be addressed to the study of surface and interface structures in the Å-μm range.

Neutron **imaging** instruments need to be flexible in order to adapt bandwidth and wavelength resolution, with a spatial resolution possibly down to the micron range, in order to investigate a wide range of systems of engineering, magnetic, energy and cultural heritage interest.

Since **NAA** and **NSECT**, which are used to determine the concentration of elements in different matrices, based on the gamma rays emitted when radioactive nuclides, hit by neutrons, decay, with energies characteristic for each nuclide, use low intensity neutrons for stimulating gamma emission and, in order to get the better performances, a high-voltage power supply, a spectroscopy amplifier, an analog-to-digital converter and a multi-channel analyzer should be combined into a single module, the instrumentation having relatively low costs in comparison with many analytical techniques.

Finally, in **chip irradiation** instrumentations, intense fluxes and atmospheric-like neutron spectrum can be achieved by means of an optimized choice of the moderator material and geometry.

**Declarations of interest from companies**
During a Workshop held in the Laboratori Nazionali di Frascati on June 10-11th 2013 (http://agenda.infn.it/event/IRIDE_Neutroni) representative industries were invited to express their interest in such a facility.

The **LFOUNDRY/MIT** (http://www.lfoundry.com/) leading company in analog and mixed-signal process technology, is potentially interested in the IRIDE neutron source to perform ageing tests with chip irradiation for generic applications or specifically for aero-spatial, avionics, bio-medical applications or for high energy physics detectors R&D. It would also envisage radiation damage studies on alternative semiconducting materials (like borated silicon wavers or hydrogen-enriched materials) and nuclear activation analyses for the identification of contaminations in the raw materials like silicon wafers. Finally, LFOUNDRY/MIT stresses that neutron irradiation could allow its suppliers to produce high quality n-type wafers by neutron transmutation doping.

The **Meniki s.r.l.**, leading company in the production of steam vacuum cleaning devices is interested in using the imagin properties of neutrons to investigate the formation and diffusion of steam inside their products in order to optimize them.

The **Selex-ES**, Finmeccanica group, leading company on electronics systems is interested on the possibility to use IRIDE facility to investigate neutron induced damages in electronics.

The **Esercito Italiano** (the italian Army) is interested in using neutron beams for non-destructive material inspection inside World War II un-exploded devices.

The **Aeronautica Militare** (the italian Air Force) is interested in non-destructive diagnostics to be used on avionics components damaged in accidents. These parts are usually made of aluminum or carbon fiber composites. This diagnostic activity could be carried out using neutron beams impinging on components to be analyzed, obtaining stress maps, creeks detections, or presence of water inside the honeycomb. The use of neutrons beam allows to probe the materials in depth where the X-rays cannot penetrate.

The **Alenia, Augusta, e Avio**, Italian companies on avionics branch, are interested on IRIDE neutron source to carry out tests on new materials and construction procedures. One of the issues is the real time monitoring of resins distribution inside the mold: the curing procedures are typically made under pressure and the resin paths should be designed carefully to obtain a uniform distributions. Avionics companies world wide already use neutron beams to characterize materials ( e.g. for plane wings and reactors).

### 6.5. Neutron target and shieldings

*- Photo-neutron Source by high Energy Electrons on high Z targets*

Beside reactors, the most powerful neutron sources are accelerators, where neutrons can be produced by spallation or by photo-production, sending high energy particles (typically protons in spallation and electrons in photo-production) to impinge on a suitable target. Several studies (Ridikas 2002) have shown that, when neutron fluxes higher than 1E+16 n/s are required for applications, the spallation sources are preferred, while for lower neutron fluxes the photonuclear process will be more convenient in term of investment costs.

The physics that underlies the photo-neutron production by means of high energy electrons can be briefly described as follows: when high energy electrons interact with a bulk of high Z target, they loose energy mainly by bremsstrahlung (more than 80%), producing an electromagnetic shower cascade. The photons of the shower can excite the nuclei of the target with which they interact and these excited nuclei go back into the fundamental state by emitting one or more nucleons. The most probable reaction is ($\gamma$,n), in which only a neutron is boiled off the nucleus, but also ($\gamma$, 2n), ($\gamma$, p) and ($\gamma$, $\pi$) reactions are possible even if less probable than the first one (in particular, the charged particle photo-production is repressed by the Columbian barrier in case of high Z material). Typically high Z materials have much higher photo-neutron cross sections than the medium and light nuclei. It is important to point out that the neutron photo-production is a threshold reaction since a neutron can be delivered from a nucleus only if the energy transferred by the photon overcomes the average binding energy of its nucleons. Again, the threshold is lower in heavy nuclei (high atomic number) than in light nuclei: 5 to 7 MeV in the first case and 15 to 20 MeV in the second case.

The main mechanisms according to which neutrons can be delivered from the excited nuclei depends on the primary photon energy impinging on it and they are reported, in order of increasing primary photon energy required, hereafter: the Giant Dipole Resonance ($E_{\gamma}$ < 20 MeV), the Quasi Deuteron Resonance ($E_{\gamma}$ > 30 MeV) and the intranuclear cascade from heavier particles, i.e pions, ($E_{\gamma}$> 140 MEV).

The direct “electro-nuclear” reactions give significant contribution only at very high electron energy (above 100 GeV), and it is quite well negligible respect to the “gamma-nuclear” ones in the energy range of interest for IRIDE (1-2 GeV).

*- Choice of Materials and Geometry for the Target*

The materials with the largest photo-neutron cross sections are Tantalum, Tungsten and Uranium, the last one overcoming the others by a factor three. Current studies have concentrated with the first two since there is less experience, due to radioprotection issues, with Uranium.

The geometrical dimensions and configuration of the target for the IRIDE facility have to be optimized by Monte Carlo simulations, in order to maximize the number of neutrons leaving the target, but taking into account, in addition to the neutronic aspects, also other important constraints related to the thermal heat transfer, the material embrittlement and, last but not least, the material activation and corrosion.

Anyway, the case of Uranium at present has been not furthermore investigated, even if it is well known that could be an interesting solution for the new neutron IRIDE Facility. It will be certainly included as possible material in the study of the best optimized target for the this project.

The neutron yield [n/s] per unit power [kW] has been estimated, respectively, for Tantalum, Tungsten and Lead, using the Monte Carlo code FLUKA. The predictions have been obtained for 500 MeV electrons on thick targets (i.e, about 10 radiation length thickness) and have been compared to the corresponding values provided by Swanson's empirical correlations (Swanson 1977), for benchmarking purpose. Swanson operated at SLAC at the end of 1970s, producing an important literature that still today constitutes the main reference for the design guidelines of shields around electron accelerators.

The results reported in Table 6.2 show that Tungsten have higher neutron yield respect to the other examined materials.

**Table 6.2: Neutron yield for enough thick targets (≈10 X0) and for Ee- = 500 MeV**

| Material | Swanson [n/kWs] E+12 | Fluka [n/kWs] E+12 |
|---|---|---|
| **Pb** | 1.98 | 2.06 |
| **Ta** | 2.13 | 2.37 |
| **W** | 2.42 | 2.67 |

Moreover, Tungsten offers several important thermo-mechanical advantages: it has a better thermal diffusivity and higher melting temperature respect to the other examined materials, as it is shown in Table 6.3.

**Table 6.3: Thermophysical properties of several high Z materials**

| Properties (T= 300K) | Ta | W | Pb | Unat |
|---|---|---|---|---|
| **density(g/cm$^3$)** | 16.69 | 19.25 | 10.66 | 19.1 |
| **Z** | 73 | 74 | 82 | 92 |
| **P.M (g/mol)** | 180.95 | 183.84 | 207.2 | 238.03 |
| **Rad Length [cm]** | 0.41 | 0.35 | 0.56 | 0.32 |
| **K (thermal-cond) [W/m K]** | 57.5 | 173 | 35.3 | 27.5 |
| **E(young) [GPa]** | 186 | 411 | 16 | 208 |
| **Poisson Ratio** | 0.34 | 0.28 | 0.44 | 0.23 |
| **alpha µm/m K** | 6.3 | 4.5 | 28.9 | 13.9 |
| **T(melting point)** | 3290 | 3695 | 606 | 1405 |

| **[k]** | | | | |
|---|---|---|---|---|

For this reason, at the state of the art of the project, we have mainly focused on the Tungsten as possible choice for the target of IRIDE: the estimated rate emissions of neutrons and other secondary particles, that are described in more details in the following sections, have been obtained for a Tungsten cylindrical target with 7 cm diameter and 6 cm height.

*- MC Prediction Validation of the Source Term: the n@BTF Experimental Results*

In order to predict, with a certain confidence level, the expected yields and energy spectra of neutrons from a suitable target, several Monte Carlo (MC) simulations have been performed using the major MC codes for nuclear and high energy physics (FLUK A, MCNPX and GEANT4). Photonuclear calculations are typically complex and heavy from the computational point of view, since they require extensive libraries of isotopic data, high statistics simulations and, very often, adequate biasing techniques (due to the low photo-neutron cross sections).

In this context, it is easy to understand the importance of validating the Monte Carlo simulation results respect to experimental data and/or by multiple code benchmarking.

The estimations for the neutron yield expected for IRIDE could take advantage of the successful comparison between the MC predictions and experimental measurements carried on at the DAΦNE Beam Test Facility (Buonuom 2008), in the frame of the n@BTF experiment (Bedogni 2011)

A low intensity pulsed neutron source has been realized at the DAΦNE Beam Test Facility in 2005 BTF (Bedogni 2011, Quintieri 2011): neutrons are produced in BTF sending high energy (510 MeV) electrons to impinge on an optimized Tungsten target (the "n@BTF" target). This kind of source produces neutrons with an energy spectrum that spans over more than 9 decades of energy (from few meV up to hundred of MeV), even if most of them have energy around 1 MeV, since in that energy range the GDR is the predominant production mechanism.

In Fig. 6.3 the BTF neutron spectrum, in equilethargic representation, is shown together with the spectra calculated with MCNPX and FLUKA in the energy interval, from thermal up to 20 MeV neutrons. The neutron spectrum was measured with a well-established Bonner Sphere Spectrometer (BSS) equipped with dysprosium activation foils (less sensitive to the harsh $\gamma$ background, typical of this kind of source).

The experimental results are well in agreement with the simulations: the majority of produced neutrons belong to the energy range from 10 KeV to 20 MeV and the calculated neutron spectrum has a Maxwellian shape with average around 0.77 MeV (that is the nuclear equilibrium temperature for Tungsten).

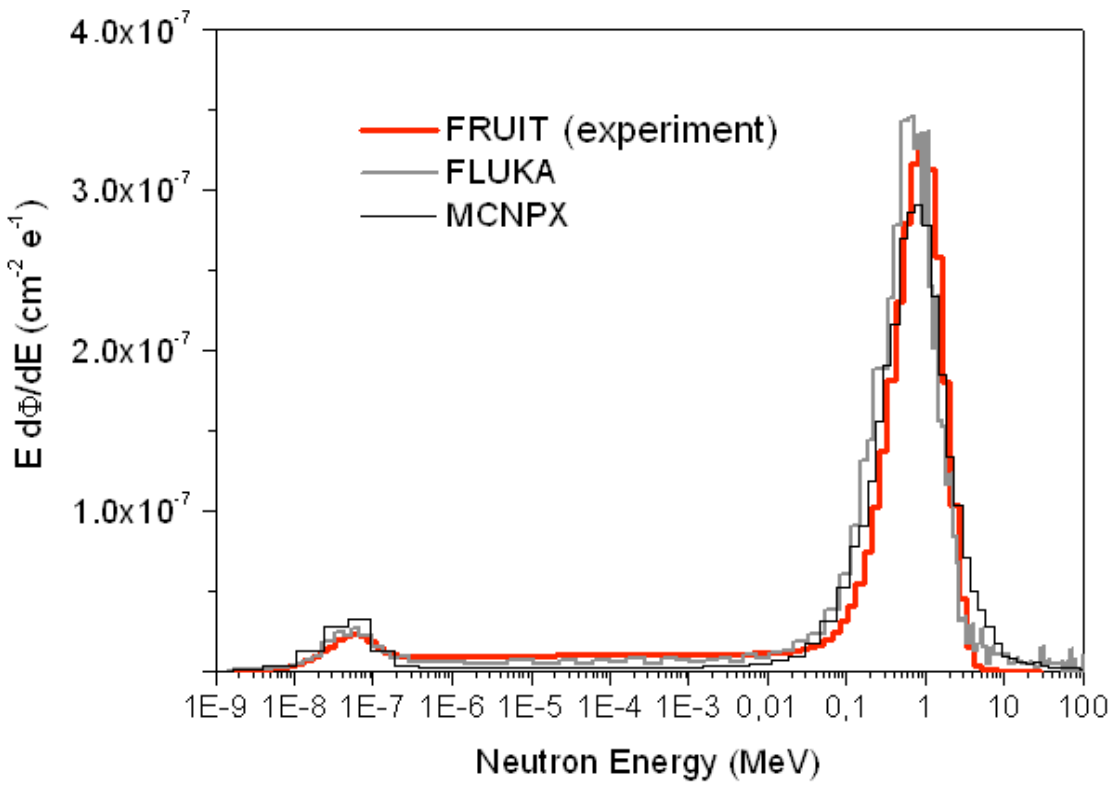

*Figure 6.3: Comparison between the simulated and experimental neutron spectra over the whole energy range. (The uncertainty of the experimental fluence <3.1%. MC Statistical errors <4%)*

The experimental and calculated total neutron fluences, collected over a sphere of 10 cm diameter, at the reference point, located 1.5 m apart from the target along the neutron beam extraction line, are reported in the Table 6.4, confirming the good agreement between experimental and predicted values

**Table 6.4: Total neutron fluence per primary: comparison between experimental measurements and predictions at a well defined reference point around the target.**

| **Measurements (BSS) $1/cm^2/pr$** | **MCNPX $1/cm^2/pr$** | **Fluka $1/cm^2/pr$** |
|---|---|---|
| 8.04E-7 ±3% | 8.06E-7±4% | 8.12E-7±5% |

The maximum yield of neutron leaving the optimized Tungsten target has been estimated to be 0.212 neutrons per primary (it refers to the DaΦne electron beam parameters), integrated on whole energy spectrum and all the solid angle. The actual yield of neutrons produced per primary, instead, has been estimated to be 0.2189, showing that only about 3% of the produced neutrons are actually absorbed in the optimized n@BTF W target.

The Monte Carlo estimations and the experimental results obtained for the n@BTF experiment are considered, in this context, as a validated reference from which it derive, with a high level of confidence, the preliminary estimations of the neutron yields by photo-production as a function of the primary energy electrons, using in the simulations the n@BTF W target (cylinder of natural Tungsten with 7 cm Diameter and 6 cm Height),.

Starting from the neutron yield pre primary, the maximum emission rates are computed using the maximum electron beam currents that the Superconducting Linac foresees to provide on the target, assuming correctly that, in the range of energy of interest for IRIDE (1-2 GeV), quite all the energy of the primary electrons is deposited inside the target itself.

*- Photo-neutron Spectra as a function of the Primary Electron Energy*

The experimental results are well in agreement with the simulations: the majority of produced neutrons belong to the energy range from 10 KeV to 20 MeV and the calculated neutron spectrum has a Maxwellian shape with average around 0.77 MeV (that is the nuclear equilibrium temperature for Tungsten).

The preliminary estimations of the secondary neutrons and hadrons for IRIDE have been obtained for a target having the same material and dimension of the n@BTF target, that is a cylinder made of Tungsten with 7 cm diameter and 6 cm height. The neutron yields have been estimated as a function of the impinging electron energy, using the validated FLUKA code. The results for 1 GeV primary electrons have been also compared to the ones obtained with MCNPX and GEANT4, respectively. The benchmarking with other codes is of great importance especially for the estimation of the high energy component of the generated neutrons. In fact, it is worthwhile to stress that the FLUKA predictions at very high energy (above 20 MeV) have not yet been validated: the energy spectrum in the n@BTF experiment had been measured with a BSS spectrometer that did not allow to cover the higher part of the energy neutron spectrum.

The photo-neutron yield, that can be obtained by high energy electron beam on a bulk target, depends mainly on the following factors:

• Thickness of the target

• Material of which the target is made

• Energy of the primary beam

The photo-neutron energy spectra that can be obtained for IRIDE have been estimated considering a primary beam with 0 divergence, an energy spread less than 0.1% and supposing a circular spot size of the beam on the target of about 2mm diameter. The results are shown in figure 6.4

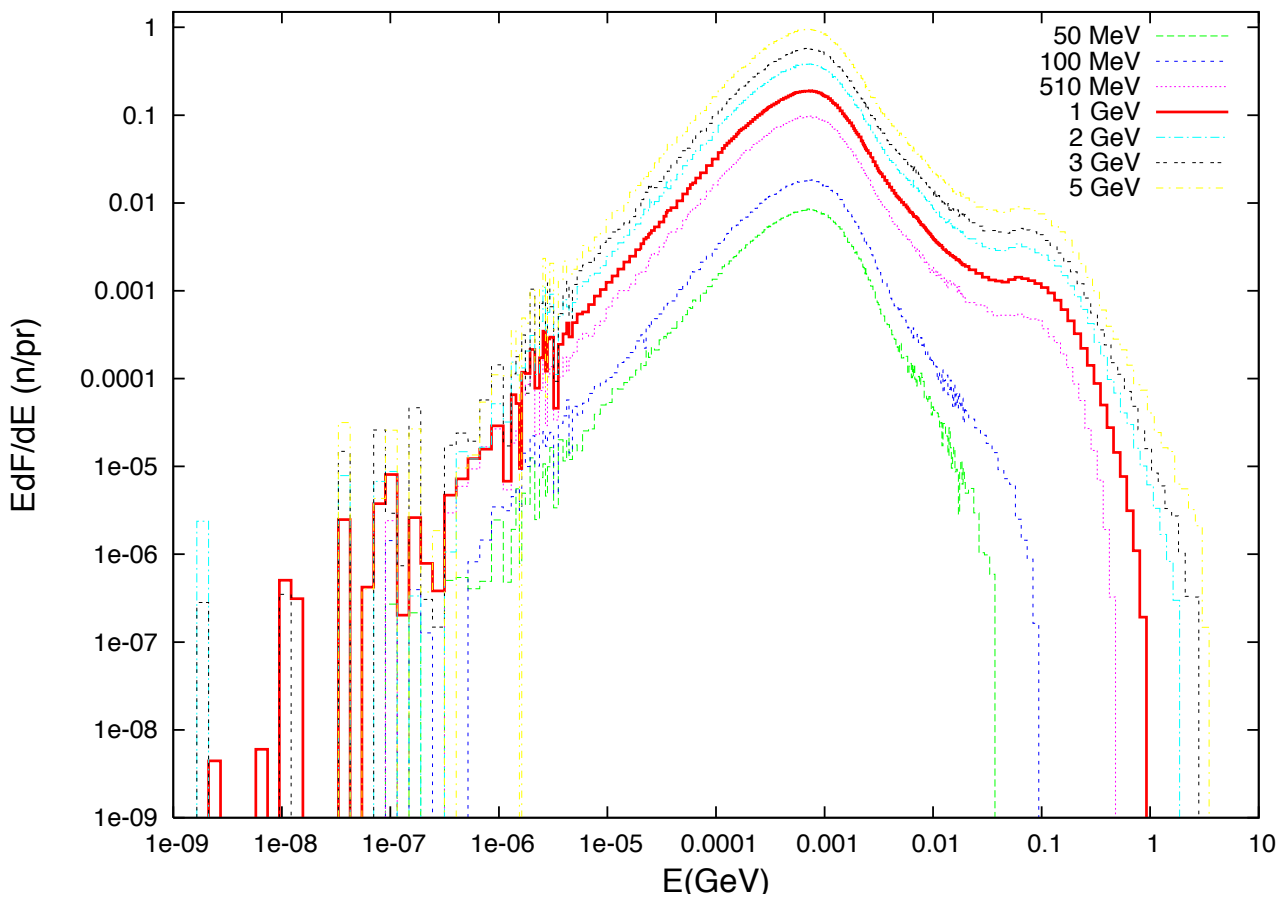

*Figure 6.4: Energy Spectra of neutron leaving the target parameterized with the electron energy*

It can be seen that for the lower energies ($E_e$-< 100 MeV) the only relevant mechanism of production is the GDR (Giant Dipole Resonance): neutrons are emitted from the nuclei mainly for evaporative process with average energy around 1 MeV. The tail at higher energies is no longer present. The main contribution in the spectra at higher energies comes from the Quasi Deuteron Effect and from the pion decay intranuclear cascade. The bump around 140 MeV is due to the pion re-absorption (the π mass being 139 MeV).

The FLUKA predictions of the neutrons per primary electron escaping from the W target (integrated all over the solid angle and on the whole energy spectrum) have been reported as function of the primary electron energy in Fig. 6.5. The estimated neutron yield in case of 1 GeV is 0.45 n/pr.

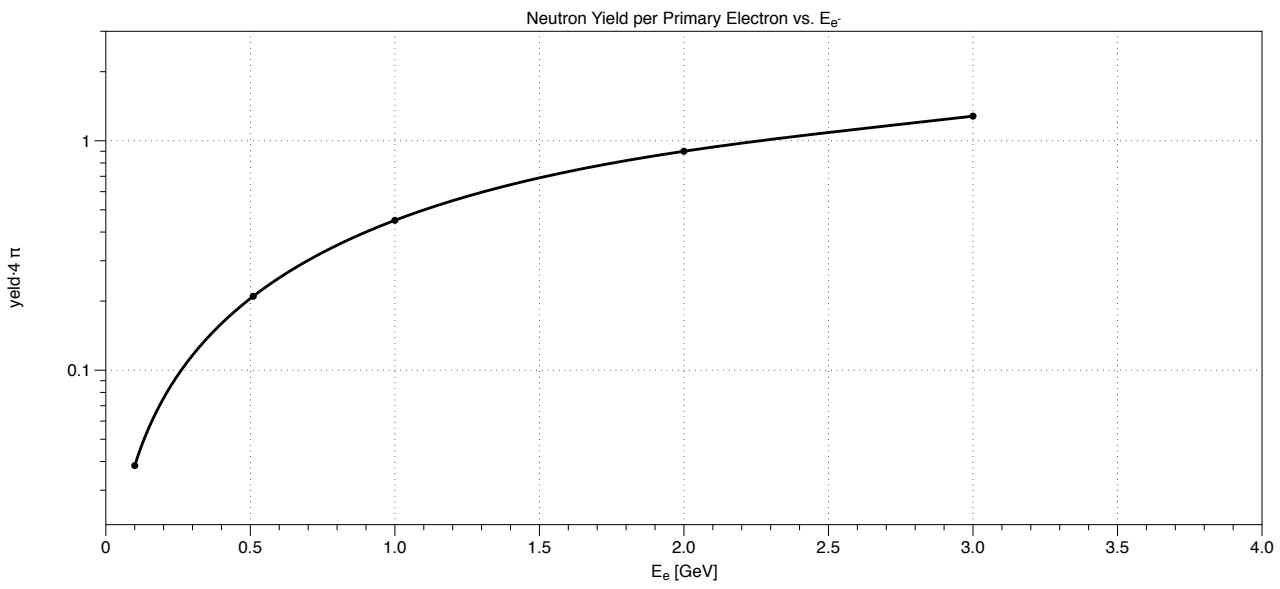

*Figure 6.5: Neutron yield form the optimized W target as a function of the impinging electrons. Errors are below 1%*

Figure 6.6 shows the comparison of the neutron energy spectrum in case of 1 GeV electrons as obtained, respectively with FLUKA and GEANT4.

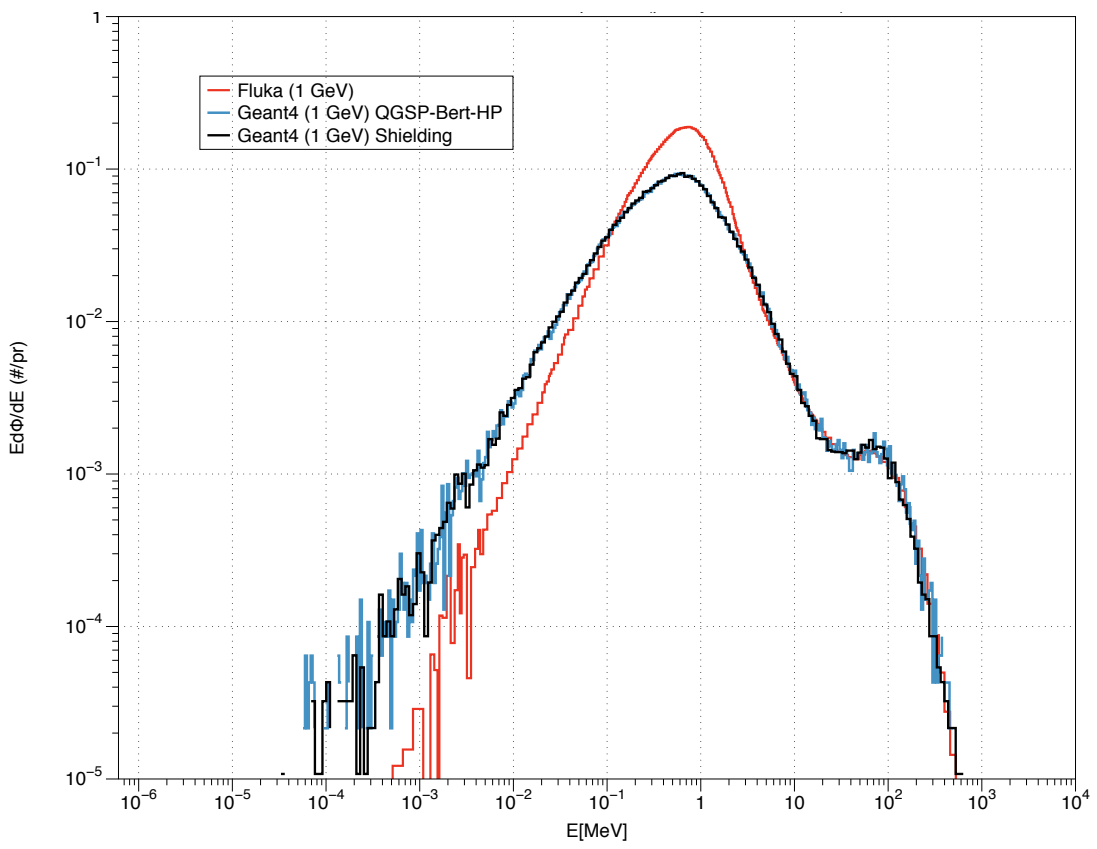

*Figure 6.6: Comparison between FLUKA and GEANT4 at 1 GeV*

In spite of the excellent agreement at the higher energies, the two codes give different predictions below 1 MeV that are still not well understood. Moreover, the neutron yield predicted by GEANT4 (0.28 n/pr) is a factor 1.6 less than the one predicted by FLUKA (4.5 n/pr). Detailed investigations are in progress to check and explain the discrepancies found between the codes.

*- Neutron Source Strength expected at IRIDE for 1 GeV Electron on the base of the Power deposited on the Target*

The power that could be deposited on the target is foreseen to go from a minimum of 32 kW (in a pulsed scenario) to a maximum of 1MW (in the continuous injection scheme), requiring a suitable thermo-mechanical design of the target able to preserve as much as possible the maximum neutron yield.

According the predicted conversion yield “n/e- “at 1 GeV ( 0.45 +/- 0.7 %), a preliminary estimation of the neutron emission rate at the W target (n@BTF like) can be obtained as function of the deposited power:

**Table 6.5: Preliminary estimation of the maximum neutron emission rate that can be obtained with a W thick target (DaΦne BTF like) with 1 GeV electrons as a function of the power deposited in the target**

| Power [kW] | Neutron emission rate (integrated over the whole energy spectrum and on 4π) [n/s] |
|---|---|
| 32 | 1 E+14 |
| 250 | 7 E+14 |
| 1000 | 3 E+15 |

At energies above 1 GeV and for a given thick target (i.e, able to enclose the whole electromagnetic cascade), the neutron production rate depends exclusively on the electron rate, i.e. to the ratio between power and electron energy, This because the yield per primary goes toward saturation.

These preliminary estimations allow to figure on that the new neutron Facility we are proposing could reach quite an order of magnitude of emission rate higher than GELINA and nELB, that are at the present the most powerful photo-neutron sources in Europe.

In Tab. 6.6 the IRIDE facility is compared with the major neutron accelerator–driven sources, nowadays working all over Europe. In terms of the strength of the source, IRIDE should be in between the photo-production facilities, that have an order of magnitude less than the value that it wants to realize, and the spallation source as ISIS, that, on the contrary, have a source intensity at least an order of magnitude higher. Finally, it is expected to have with IRIDE the same emission rate as with nTOF at CERN.

**Table 6.6: Preliminary estimation of the maximum neutron emission rate that can be obtained with a W thick target (DaΦne BTF likle) with 1 GeV electrons as a function of the power deposited in the target**

| Facility Parameters | nElbe* | Gelina** | nToF | ISIS | IRIDE |
|---|---|---|---|---|---|
| **Source** | SC e-Linac | e-Linac | p spallation | p spallation | SC e- Linac |
| **Part E (MeV)** | 40 | 120 | 20000 | 800 | 1000 (2000) |
| **Max Power (kW)** | 18 | 11 | 45 | 160 | 32/1000 (pulsed/continuous) |
| **Neutrons/s** | 3.4E+13 | 3.2E+13 | 8.1E+14 | 1E+16 | 7E+14/3E+15 |

*- The thermal Design of the Neutron Target*

The thermal constraints could imply some technical choices (i.e . segmentation and rotation of the target, use of heavy,cool down system, etc) that might affect in a significant way, both from the quantitative (modifying the spectrum shape) and qualitative points of view, the rate and the shape of the photo-produced neutrons.

In order to have an idea of how severe could be the thermal problem, a first simplified analytical approach has been followed to estimate the maximum temperature that could be reached inside the target. As starting simplified configuration, the energy has been supposed to be deposited through the whole cylindrical volume of the W target (R=3.5 cm and H=6 cm). This is a non-conservative analysis since it extends the energy deposition region to the whole target volume (in a more realistic scenario the real beam spot size as it arrives on the target has to be considered).

Assuming to deliver 250 kW uniformly through the volume of the target, with a volume of 170 $cm^3$ (cylinder radius= 3cm and height 6 cm), the power density, q''', in the target is q''' = 1.5E+9 W $m^{-3}$. For Tungsten the thermal conductivity $\lambda$ is 173 $Wm^{-1}$ $K^{-1}$, and supposing to solve a stationary problem with a fixed external surface temperature $T_0$= 300 K, $T_{max}$ is obtained by the equation (Beek 2000) $Tmax=T0+ q'''R^2/4\lambda$.

Substituting the value in the formula, a maximum temperature of 2500 K is calculated along the cylinder axis. This value can increase by a factor 2 if we consider to deposit the energy with a more realistic beam spot size of 1.5 cm: it this would be the case, the melting temperature of Tungsten will be highly exceeded (T=3700 K).

To obtain a more realistic estimation of the temperature profile inside the target, the real energy deposition profile has to be taken into account. In fact it is not uniform along the cylindrical axis (z in the plot), but it has a peak in the first 3 cm, as it is shown in figure 6.7.

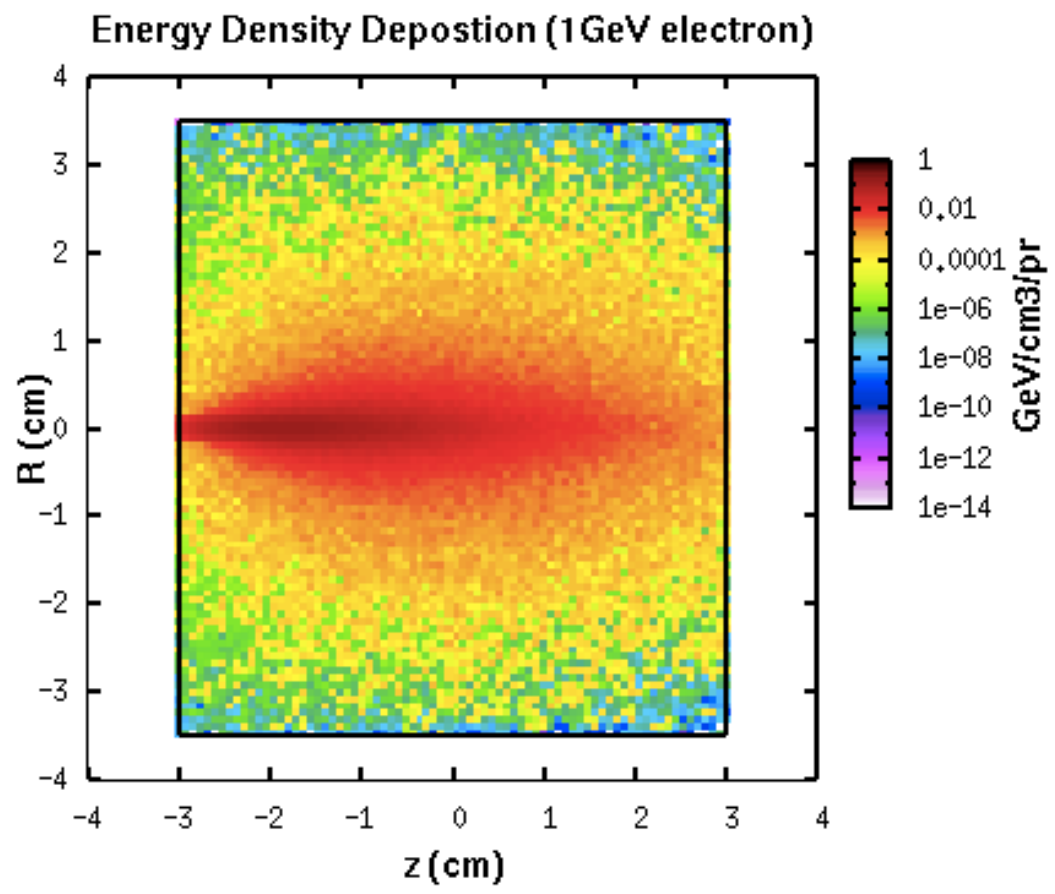

*Figure 6.7: Energy density deposition inside the target (1 GeV electron)*

The possibility of segmenting the W target into plates perpendicular to the axis has to be properly investigated. Such a segmentation would create cooling channels in between the plates, through which a cooling liquid can flow. The choice of the coolant has to be done also on the base of neutronic considerations: in principle water can affect more deeply the shape of the energy spectrum, due to its moderation properties, contrarily to heavy metals, that beside having a lower moderating power, also allow to have much higher cooling power density (resulting for this more effective to save room in between the target slices).

In case of very high energy power deposition (1 MW) , also a rotating solution for the target should be investigated.

Accurate finite element analysis (by using the ANSYS code) has been planned in order to properly define the thermal design of the target (transient analysis in multiphysics context), taking into account the real duty cycle and Linac working regime (continuous one or pulsed).

*- Preliminary Design of the Shielding for Target Station*

In order to have a preliminary estimation of the shielding thickness around the neutron source in the target station, some simplified but conservative simulations have been performed by FLUKA code, keeping in mind that the complete design of the biological shielding has to take into account a series of complex factors not yet well defined at this point of the project.

We have assumed a conservative scenario and configuration to evaluate the dose rate in a controlled zone just behind the shield . The assumptions are the following ones:

1) the target is located close to the shield: only about 20 cm apart
2) a continuous injection scenario has been reproduced (Ie= 1 mA that is 10^16 e-/s striking against the target)
3) the shield thickness has to guarantee that the environmental equivalent dose in the close controlled zone just behind the shield has never to be higher than 1 μSv /h (the law imposes for the class A controlled zones to not overcome the annual limit of 6mSv/y)

Two preliminary scenarios have been simulated: one with an iron shield of 1 m thickness and the second one with 3 m thickness. The results of simulations are reported in figure 6.8: the ambient dose equivalent rate has been estimated in several regions around the target: the middle rectangle on the left is the target station, where the target is located and irradiated. The middle rectangle in the center is the Iron shield and the middle rectangle on the right is the controlled zone, where the dose rate must be kept below 1 μSv/h. the horizontal upper and lower rectangle represent the lateral concrete walls (2 m thick).

The simulations shown that using a 3m Iron thickness this constraints is well respected in the controlled zone, also summing the equivalent dose contributions from neutrons and photons

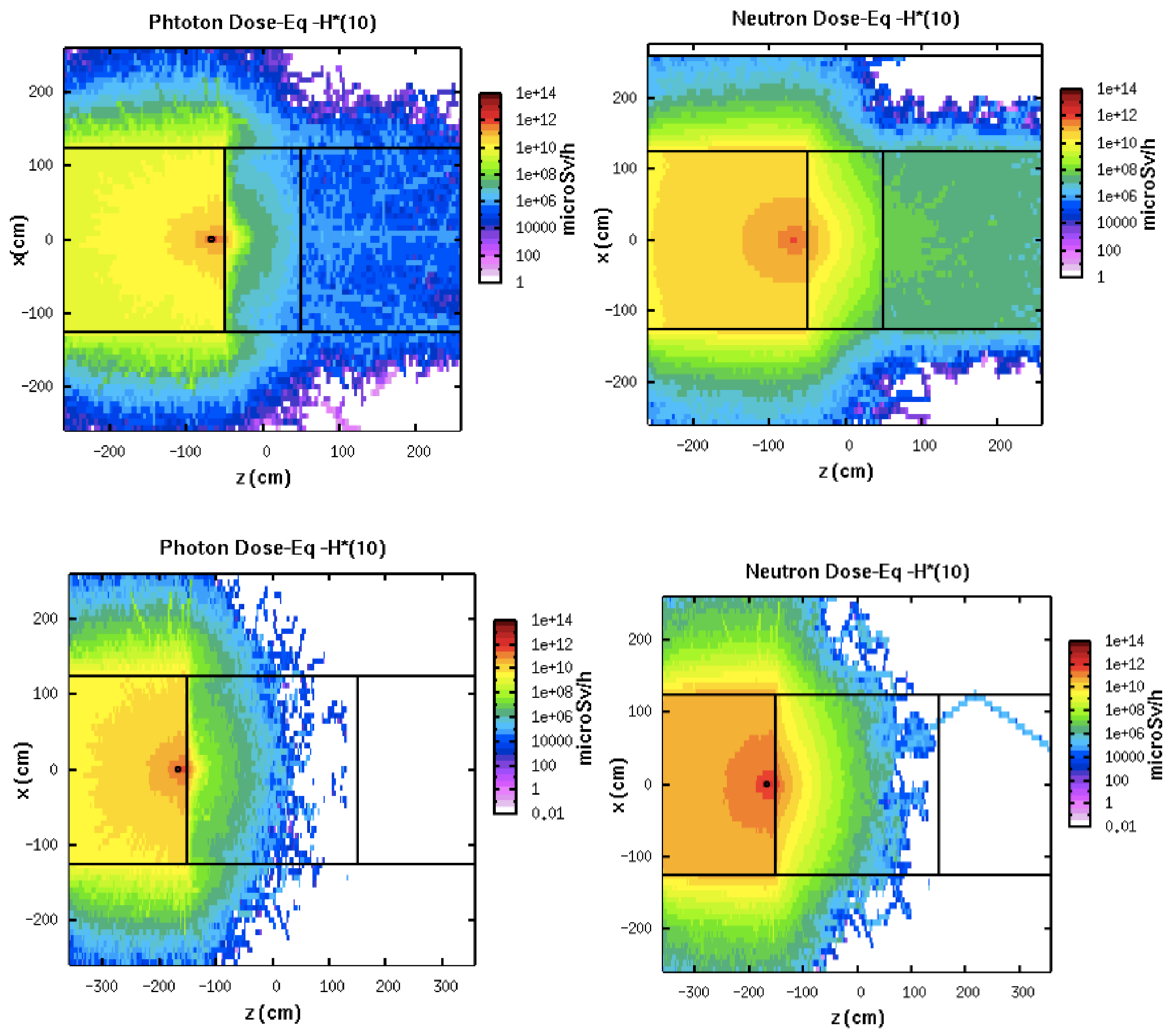

*Figure 6.8: Ambient Dose Equivalent rate: up 1 m Iron shield; down 3 m Iron shield*

## 6.6. User Beamlines And Detectors

In the following are schematically reported the beamline requirements and characteristics for the different techniques discussed in section "Scientific opportunities". The pulsed and continuous beam cases are discussed when applicable.

### *A.* Imaging beamline

An imaging beamline is composed by a **vacuum flight tube**, a **pin hole selector** permitting to select different diameters for beam divergence definition, a **sample position** including a rotating stage, a **scintillator screen** for neutron conversion into light, and a digital **image acquisition device**. An exampleis reported in Fig. 6.9

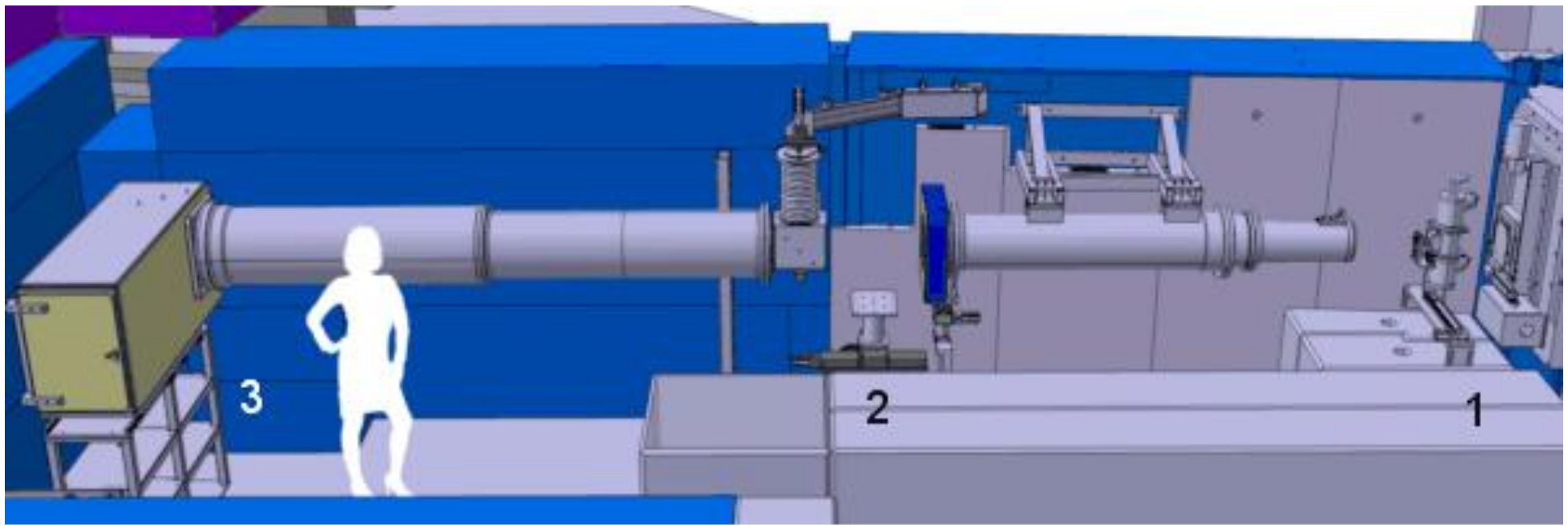

*Fig. 6.9: scheme of the NEUTRAimaging beamline at Paul Scherrer Institut.*

The cost of an imaging beamline can be about 1M€.

### B. Neutron diffraction beamline

**Continuous source**

The typical configuration for a neutron diffraction beamline on a continuous source is based on the selection of a constant neutron wavelength to irradiate the sample and use angular dispersion for d-spacing determination. The beamline is then arranged in the following way: **flight tube** or **guide** from the **moderator** to a **monochromator** usually made of pyrolitic graphite or copper to select the wavelength of interest for the diffraction measurement. A second **flight tube** or **guide** to bring neutrons to the **sample position** and a **serie of detectors** or a **single movable detector** placed at constant distance covering the widest possible scattering angle around the sample. A scheme is shown in Fig. 6.10

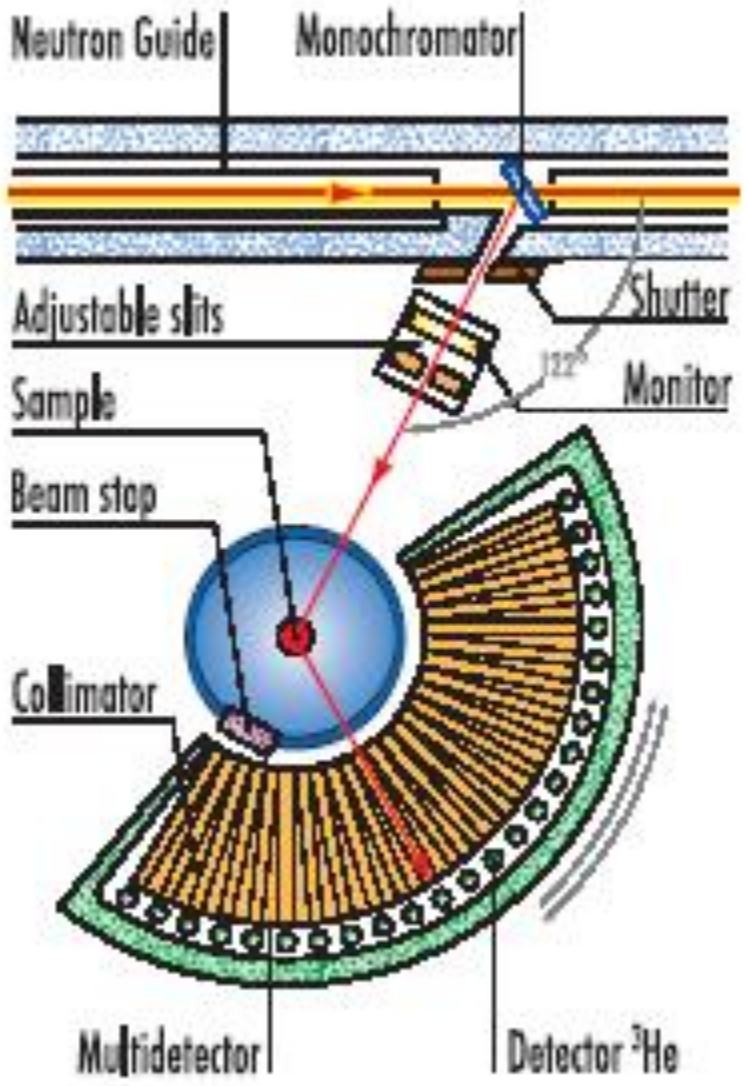

*Fig. 6.10: scheme of a constant walvelength neutron diffractometer*

**Table 6.7: : technical specifications for a neutron diffraction beamline at a continuous source**

| | |
|---|---|
| **Detectors** | $^3$He counters |
| **Moderator** | water or liquid methane |
| **Neutron incident wavelength** | change according to the monochromater (copper or graphite), ranging from 0.8 to 3.0 Å |
| **Neutron distribution** | beam size: 30 x 50mm |
| **Provisional cost** | 3000 kEuro |

**Pulsed source**

A diffractometer working on a pulsed source exploits the whole wavelength available range since there is a linear distribution of the neutron wavelength with time. In this way the full white spectrum is used to analyze the sample and the only experimental requirement is the tuning of the neutron pulse production instant with the detection time on the detector to identify the neutron wavelength. The beamline is then composed by a **flight tube** or **guide** to bring neutrons to the **sample** and one or more **detector banks** for neutron counting. Fig. 6.11 shows a scheme of a time of flight neutron diffractometer.

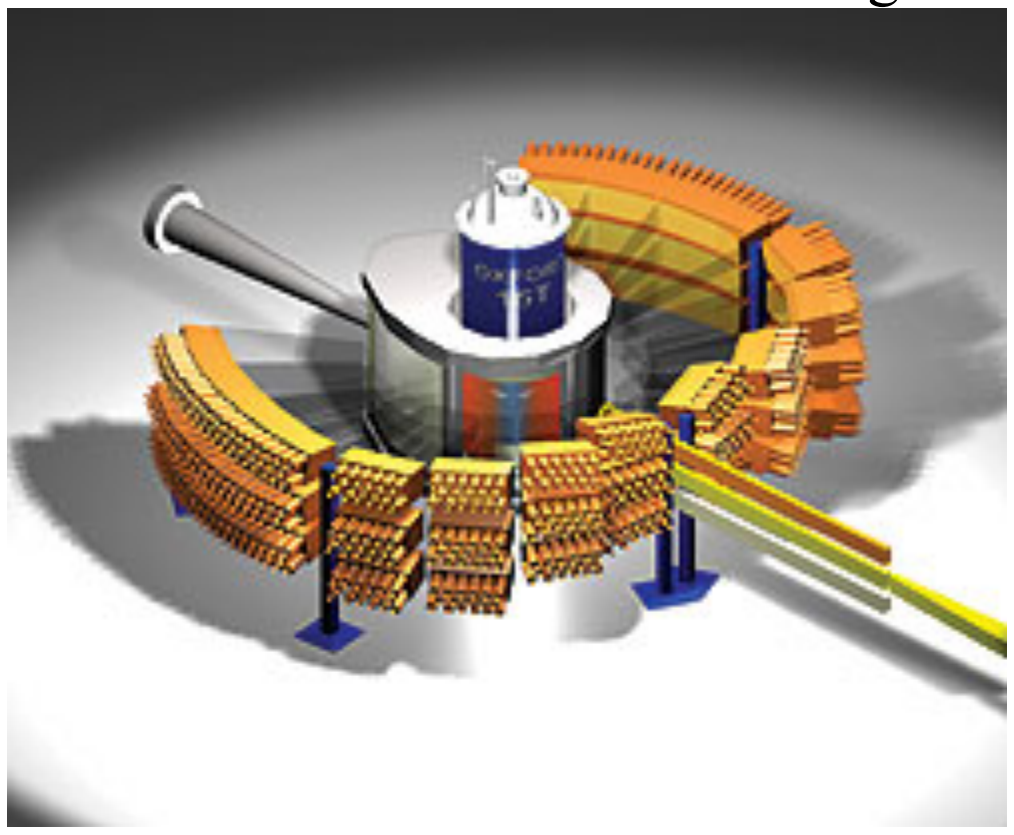

*Fig. 6.11: scheme of a ToF diffractometer. The multiple detector banks reveal a different d-spacing range each and have different resolution.*

**Table 6.8: technical specifications for a neutron diffraction beamline at a pulsed source**

| | |
|---|---|
| **Detectors** | $^3$He counters |
| **Moderator** | water or liquid methane |
| **Neutron incident wavelength** | change according to the moderator, ranging from 0.3 to 6.0 Å |
| **Neutron distribution** | beam size: 30 x 50mm |
| **Provisional cost** | 3000 kEuro |

*C.* Small angle neutron scattering beamline

**Continuous source**

A possible configuration for a continuous source could be similar to the one of the SANS instrument at the SINQ spallation source at PSI - Villigen (CH) (fig.1) [**Kohlbrecher_2000, Kurpsi_2013**] as schematically shown in Fig. 6.11. Nevertheless in the past years there have been significant evolutions that will allow different solutions.

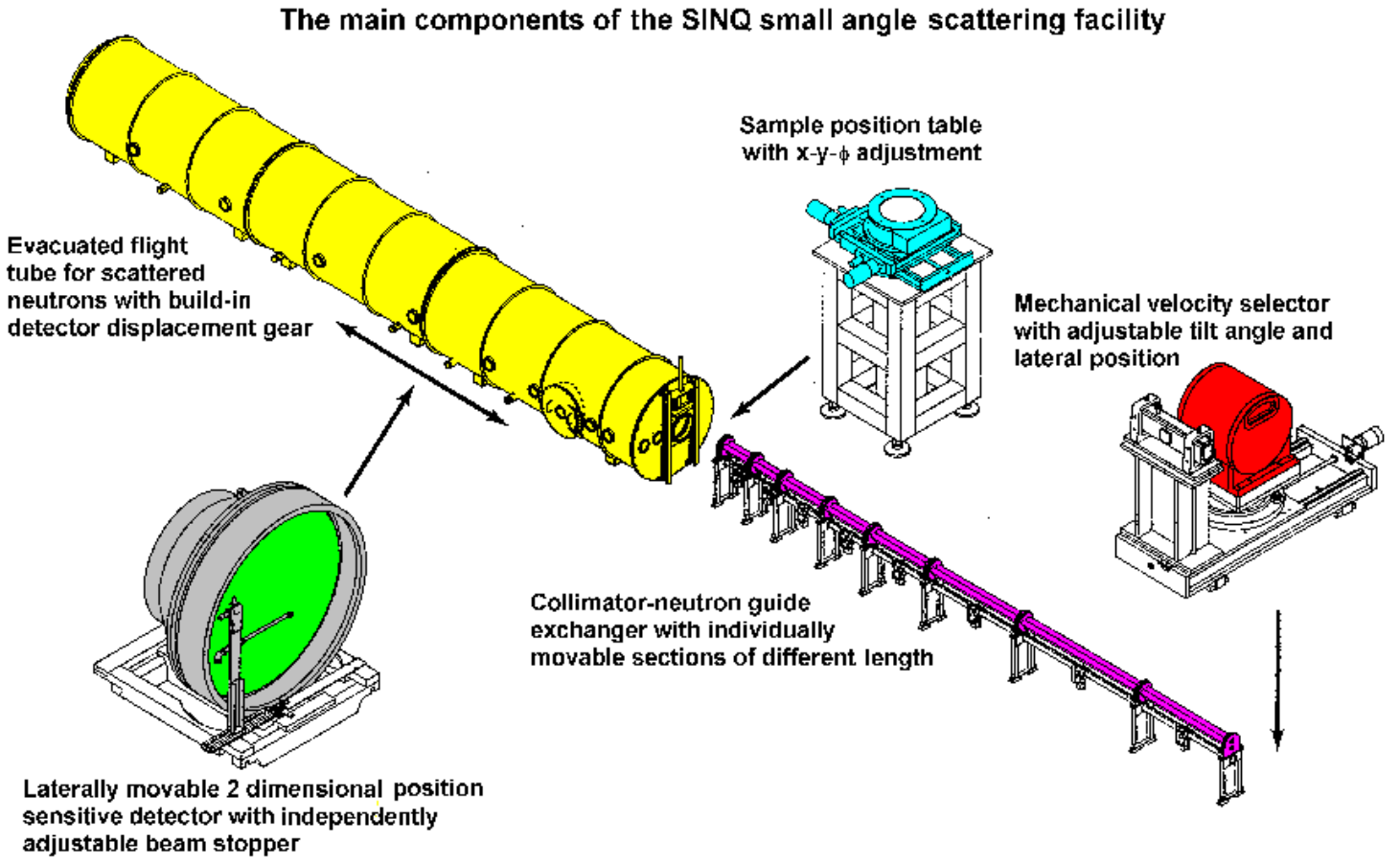

*Fig.6.12 - Layout of the SANS instrument at PSI-SINQ [1].*

**Table 6.9: - Technical specifications of the PSI-SINQ SANS spectrometer**

| | |
|---|---|
| **Collimation, Detector, Beamstop** | • Curved neutron guide to filter epithermal and higher energy neutrons. Incoming neutrons monochromatized by mechanical velocity selector and collimated on variable length from 1 to 18 m.<br>• 2-D 3He multiwire proportional counter, with a sensitive area of 96 x 96 cm2 and 128 x 128 detection elements of 7.5 x 7.5 mm2 each; mounted on a rail-guided trolley such that it may be positioned at any distance (L) between 1.50 m and 20 m from the sample. Optional lateral displacement of the detector up to 50 cm, to increase Q-range at any L, combined with a rotation around the central vertical axis of the detector to minimize parallax effects.<br>• Beamstop of B4C-plates mounted on a thin-walled aluminium tube, moveable in vertical and horizontal direction, to prevent detector damage due to the direct beam. Four beamstops of dimensions 40x40 mm2, 70x70 mm2, 85x85 mm2 and 100x100 mm2 are available and may be exchanged by remote control in the vacuum. |
| **Moderator** | D2 (cold neutrons) |
| **Neutron incident wavelength and Q-range** | Cold neutrons: ; resolution: ≈10% FWHM. Maximum Q-range from 6 x $10^{-3}$ $nm^{-1}$ to 5.4 $nm^{-1}$ (up to 10.5 $nm^{-1}$ with the detector displaced laterally by 50 cm) The realization of a cold neutrons moderator is a big challenge. |
| **Neutron flux at sample position** | See fig.2 |
| **Provisional cost** | Around 6,000 kEuro |

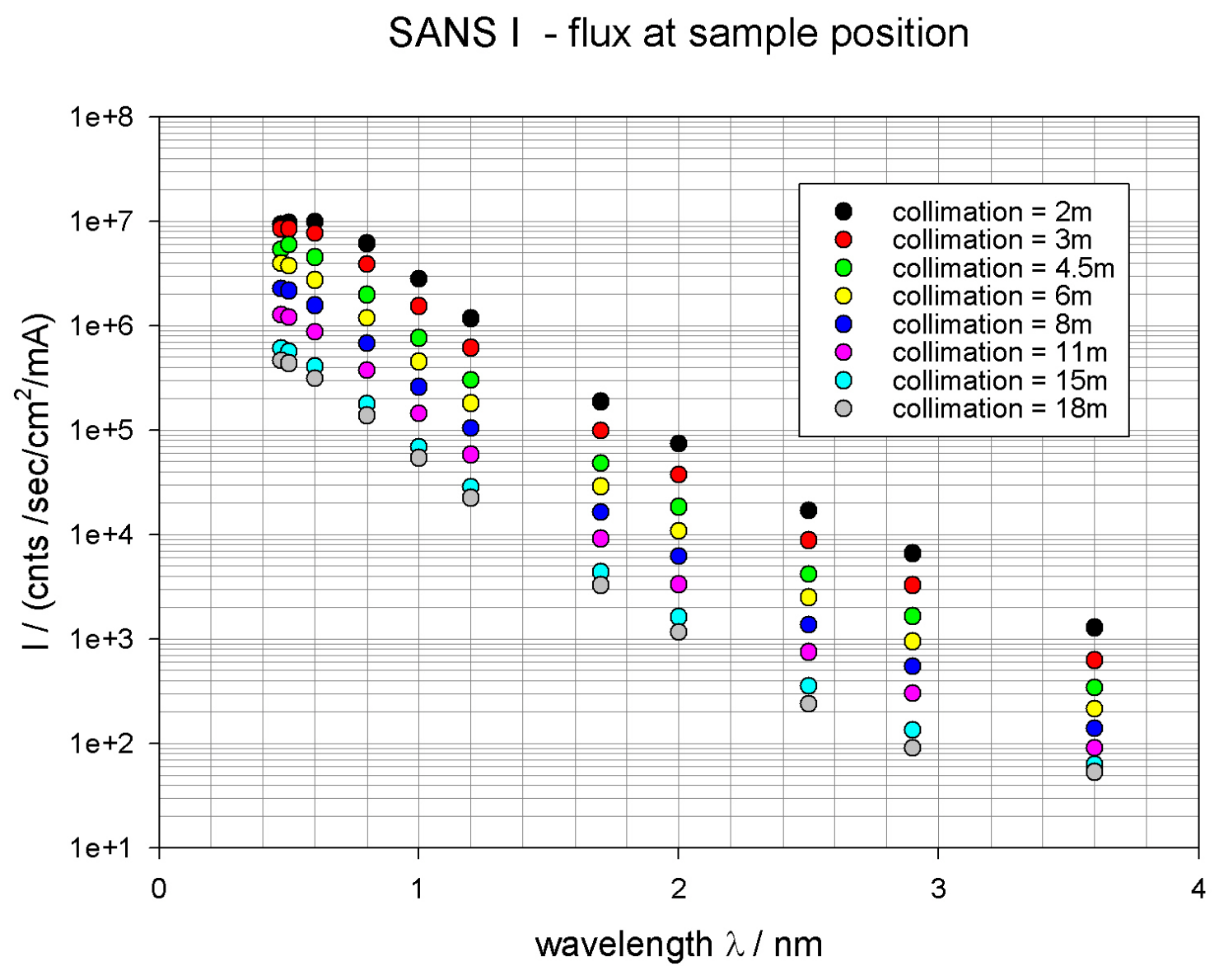

*Fig.6.12 - Neutron flux at sample position of the PSI-SINQ SANS spectrometer.*

### Pulsed source (ToF technique)

As a possible configuration, Fig.13 and Tab.II show the layout and the main features of the TOF SANS spectrometer presently under design for the CPHS pulsed neutron source at Tsinghua (China).

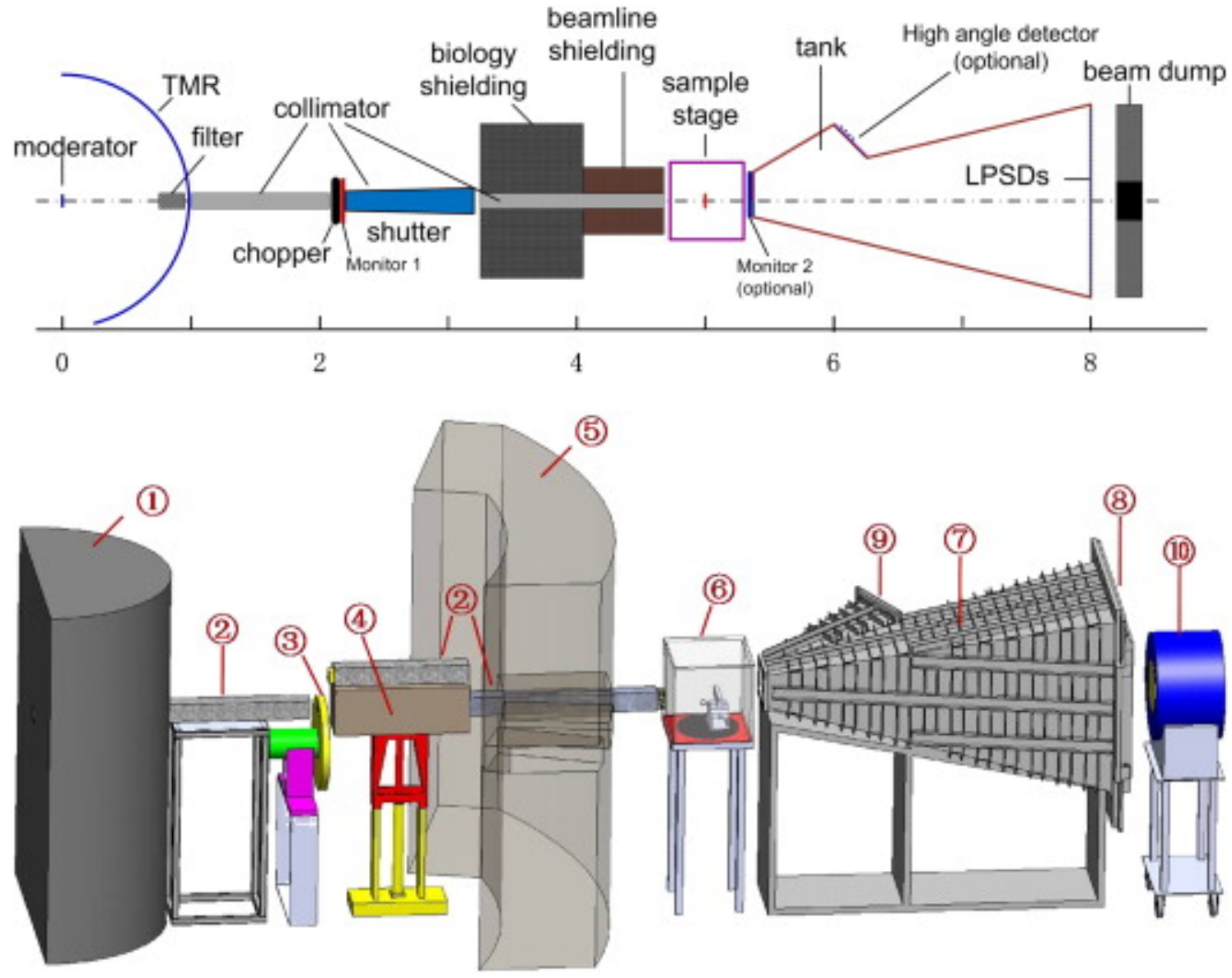

*Fig.6.13 - Schematic diagram of the ToF SANS beamline at the CPHS pulsed neutron source (Tsinghua - China) [3]. (1) TMR assembly, (2) collimator, (3) bandwidth chopper, (4) shutter, (5) biological shielding, (6) sample chamber, (7) scattering tank, (8) $^3$He LPSDs, (9) high angle detectors (reserved), (10) beam dump.*

***Tab.6.10*** - **Technical specifications of the TOF SANS spectrometer designed for the CPHS pulsed neutron source at Tsinghua (China) [Huang_2012].**

Design parameters of the SANS instrument.

| Parameter | Design value |
|---|---|
| Source frequency | 50 Hz |
| Wavelength range | 1–10 Å |
| Source-to-sample distance | 5 m |
| Sample-to-detector distance | 3 m |
| Collimation | Pinhole collimation |
| Sample size | 1–2 cm diameter |
| Area detector | $^3$He LPSD Array |
| Active area | $1 \times 1$ m$^2$ |
| Pixel size | 12 mm |
| Q-range | 0.007–1 Å$^{-1}$ |
| Q-resolution | 2–30% |
| Flux at sample position | $\sim 10^4$ n/cm$^2$/s |

*D.* Elastic Incoherent Neuron Scattering- Resolution Elastic Neutron Scattering beamline

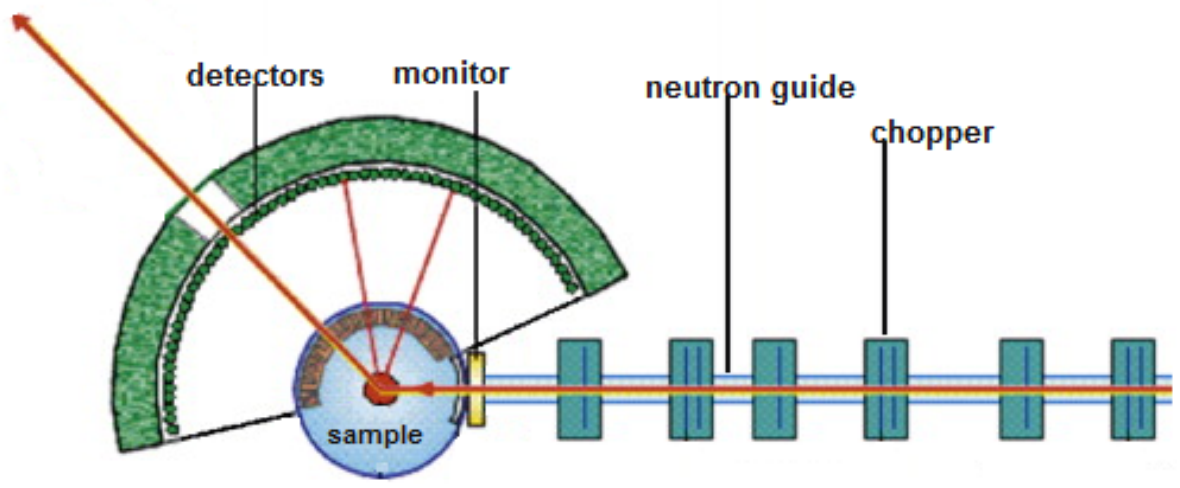

*Figure 6.4: Schematic layout of the EINS-RENS spectrometer*

**Table 6.11: Technical specifications of the EINS-RENS spectrometer**

| | |
|---|---|
| **Detectors** | $^3$He counters |
| **Moderator** | coupled solid methane/ liquid $^2$D @ T=20K |
| **Neutron incident energy** | 1÷20 meV (energy resolution: 5÷200 μeV) |
| **Neutron distribution** | high "peak" flux (beam size: 30 x 50mm) |
| **Provisional cost** | 2500-3000 kEuro |

The instrument configuration is for pulsed and continuous sources: **1)** in the pulsed configuration, the chopper is used to define the incident neutron energy cutting away unwanted neutrons that are background source; **2)** in the continuous configuration he chopper is used to allow for time of flight technique to be used.

*E.* **Neutron Resonance Capture Analysis beamline (pulsed only)**

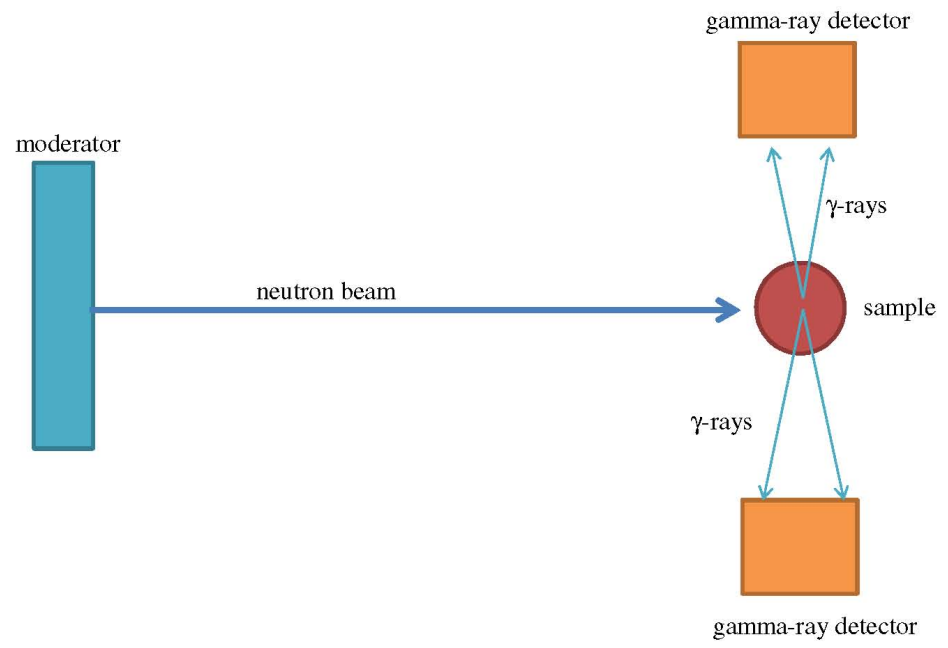

*Figure 6.15: Schematic layout of NRCA station*

**Table 6.12: Technical specifications of a NRCA station**

| | |
|---|---|
| **Detectors** | Array of scintillators |
| **Moderator** | coupled Water moderator @ T=295K, |
| **Neutron incident energy** | Full spectrum, used up to 1 keV effectively |
| **Provisional cost** | 500-1000 kEuro |

This instrument maybe integrated on a diffractometer when used to study structure and elemental composition of cultural heritage metallic artifacts [**Pietropaolo_2011**].

*F.* Bragg Edge Transmission (pulsed only)

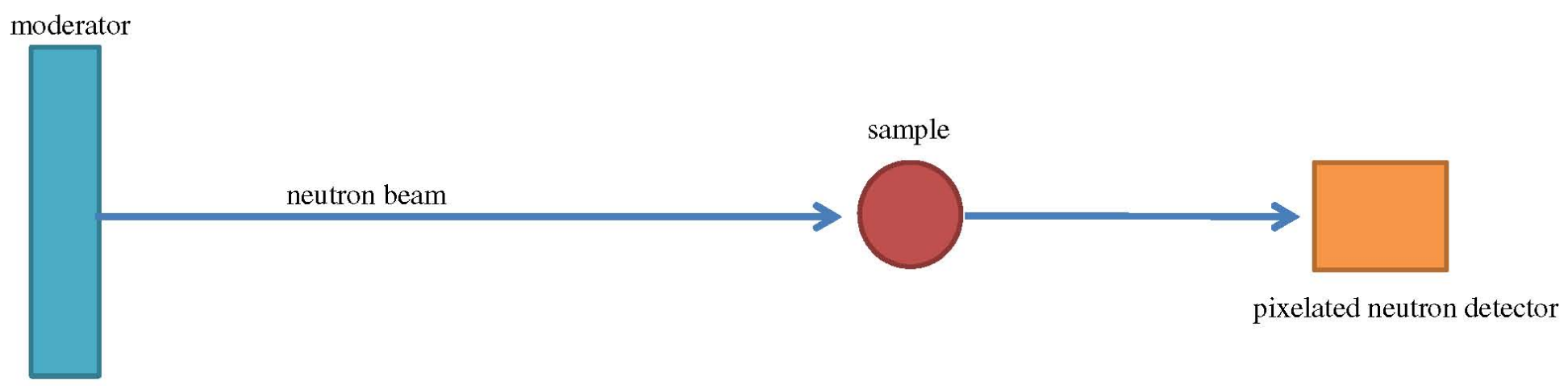

*Figure 6.16: schematic layout for a BET experimental station*

**Table 6.13: Technical specifications of a BET station**

| | |
|---|---|
| **Detectors** | Pixelated detector, sub-mm resolution |
| **Moderator** | coupled Water moderator @ T=295K, |
| **Neutron incident energy** | Full spectrum |
| **Provisional cost** | 500 kEuro |

This instrument maybe integrated on a diffractometer, together with NRCA, when used to study structure and elemental composition of cultural heritage metallic artifacts [**Pietropaolo_2011**].

*G.* **Neutron Activation Analysis beamline (pulsed and continuous)**

A beamline for NAA consists of an irradiation area coupled to an external laboratory where gamma-ray spectroscopy at high resolution can be performed. An important issue for the bamline is incident beam monitor to measure precisely the incident neutron flux onto the irradiated sample. Provisional cost is around 500 kEuro.

*H.* **Chip irradiation beamline (pulsed and continuous)**

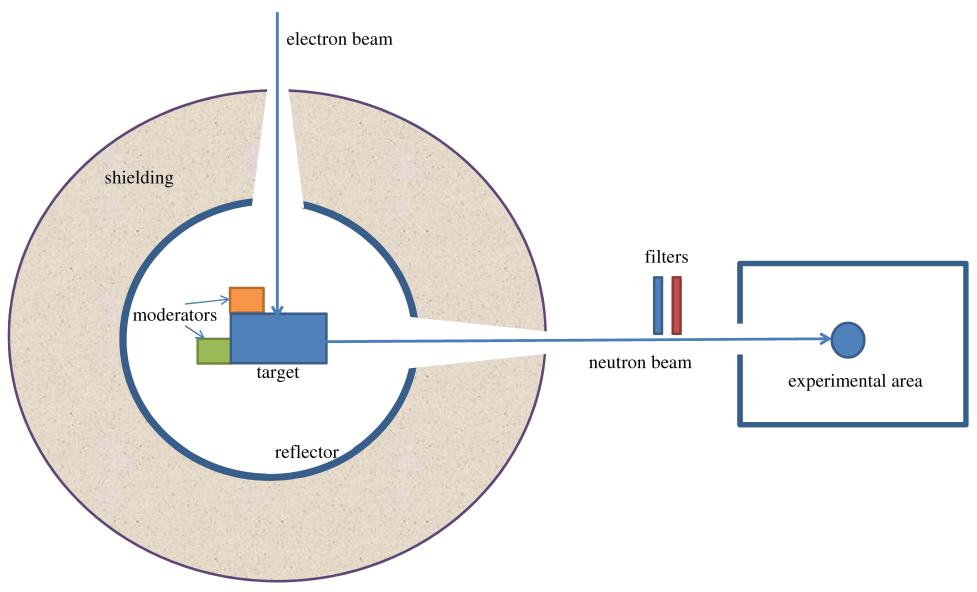

*Figure 6.17: Schematic layout for a chip irradiation beamline.*

**Table 6.14: Technical specifications of a Chip Irradiation station**

| | |
|---|---|
| **Detectors** | thermal-to-fast neutron monitors |
| **Moderator** | To be studied to resemble terrestrial spectrum. Filters |
| **Neutron incident energy** | Full energy spectrum |
| **Beam configuration** | Pencil and large neutron beams |
| **Provisional cost** | 1000 kEuro |

One important issue to be addressed for this beamline is the precise measurement of the neutron fluence within the beam spot as well as localized measurements within the spot (mm2 resolution) to have a measure of the fluence as close as possible to the irradiated sensitive area of a chip. Indeed the precise measurements of SEE cross sections of a device is strictly related to the precision obtainable in fluence determination **[Violante_2007]**.

## *I.* Detectors R&D

The research and development activity on He-free neutron detectors is of strategic importance for neutron applications in science and technology. As a matter of fact, the lack of $^{3}$He is triggering an interesting and stimulating technological effort in finding out effective substitutes of the He-gas tubes typically operating at both reactor and/or accelerator driven neutron sources worldwide. The most important request for the new detectors is to provide a high detection efficiency (above 50%), high rate capability (MHz/cm$^{2}$) and large area covering (1 m$^{2}$ just to give an order of magnitude).

Boron technology seems to be an effective way as indicated by a series of experimental and simulation studies [**Kouzes_2009**, **Lintereur-2011, Klein_2011,Tremsin_2005]** although other approaches, for example based on radiative capture, exploiting the (n,γ) reactions in Cd, are also being investigated [**Festa_2011**].

Possible solutions may be the use of gaseous detectors like GEM (Gas Electron Multipliers) that can provide large area detection areas, or solid state detectors like MEDIPIX or hybrid combinations of the two. These strategy are currently subject to experimental investigation [**Murtas_2012,Pietropaolo_2013**].

One important issue to be addressed in boron technology approach is the deposition of film of 1-2 mm thickness with good mechanical quality to provide stability against aging effects. This issue may stimulate synergic activities among research centers and small/medium enterprises working on film depositions for different industrial applications.

In what follows are briefly described, as examples, the approaches based on GEM and radiative capture.

### *1)* GEM detectors

Fast neutron beams available at large scale facilities are becoming strategic for industrial applications, especially in relation to the assessment of radiation hardness of silicon-based nano-sized electronic chips, but also for non-destructive analysis of materials and components. A stringent request for neutron beam lines dedicated to chip irradiation is the possibility to monitor and characterize the neutron beam above 1 MeV (the more concerning energy region of the spectrum) with a spatial resolution in the millimeter range or below. The triple GEM detector can be considered as a good candidate for the purpose, thanks to its constructive characteristics and physical mechanisms upon which the detection mechanism relies on [**Alfonsi_2005, Bonivento_2002,Murtas_2010**] .

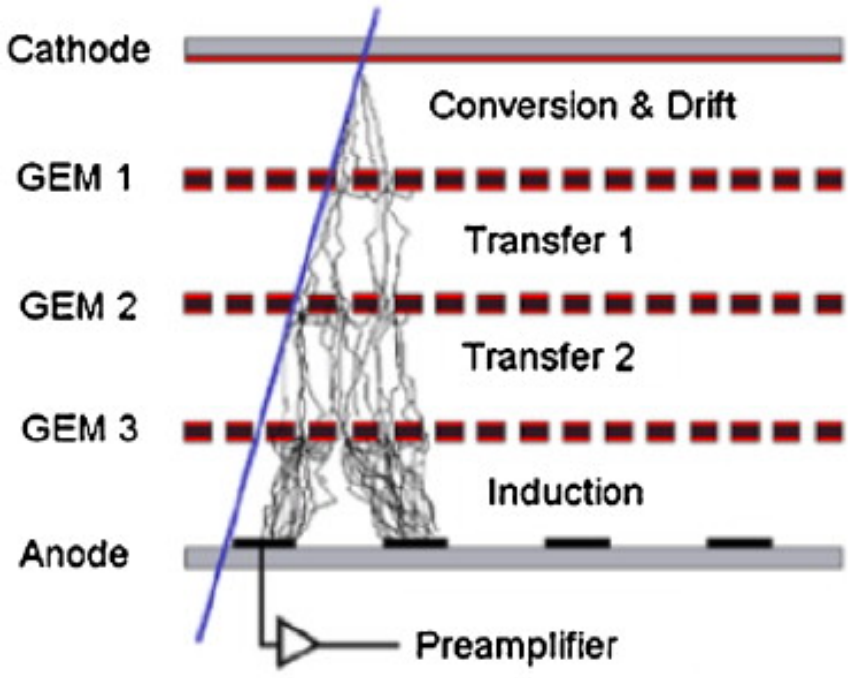

*Figure 6.18: The triple GEM detector*

**Fast neutron beam monitor**

A triple Gas Electron Multiplier (GEM) detector was developed as a fast neutron beam monitor for the ISIS spallation neutron source in UK. The spatial distribution of the neutrons was measured in real time in the energy region between 2 and 800 MeV achieving a spatial resolution of a few millimeters thanks to the patterned readout of the detector.

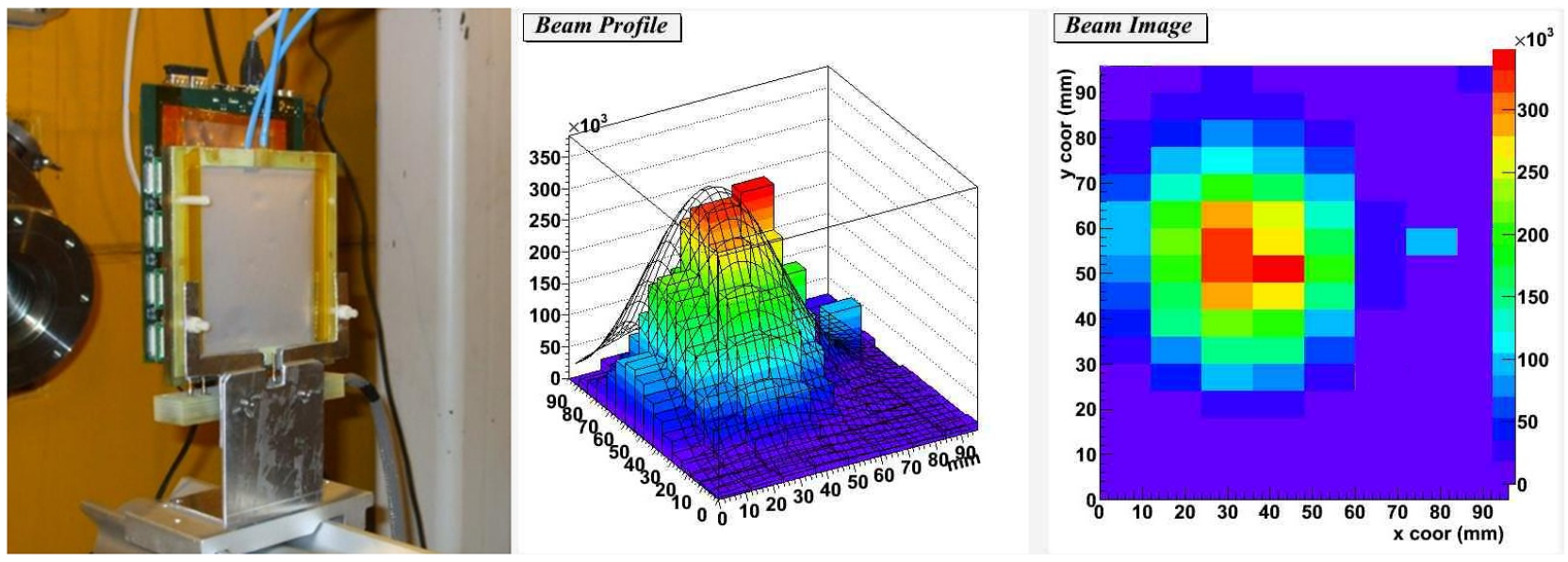

*Figure 6.18: On the left the experimental setup; on the right the online beam spot monitor*

In this setup also a time scan was performed to measure the time spectrum of the fast neutrons as shown in figure 6.19. The low operating bias allowed for an almost complete rejection of the gamma background as shown in figure LLLL.

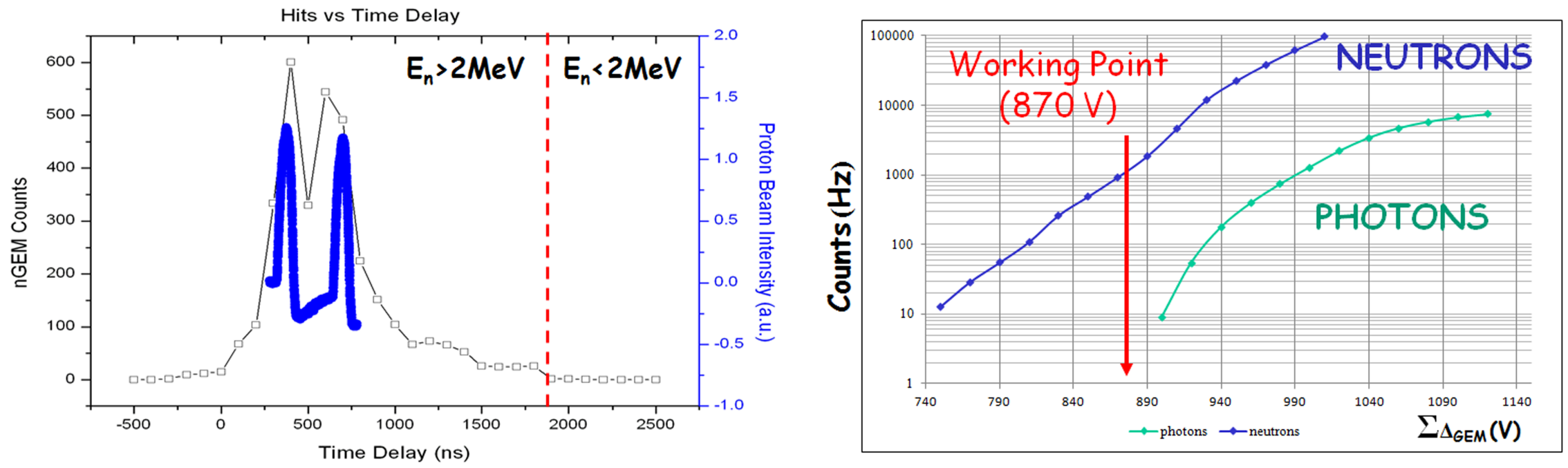

*Figure 6.19: On the left the comparison between the neutron flux and the two proton bunches producing neutron via spallation reactions; on the right the particles detected vs the HV settings.*

**Thermal neutron detector**

A different approach was used to build a first protityoe of a high efficiency neutron detector to be used in scattering geometry. Fig. 6.20 shows the inner section of the detector and a schematic of the neutron detection mechanism. This detector assembly was tested both at ISIS and ENEA-Casaccia (TRIGA reactor) to assess response linearity and efficiency [**Pietropaolo_2013**]. Tests were perfomed also at the n-TOF facility at CERN [**Murtas2_2012, Claps:2012**]

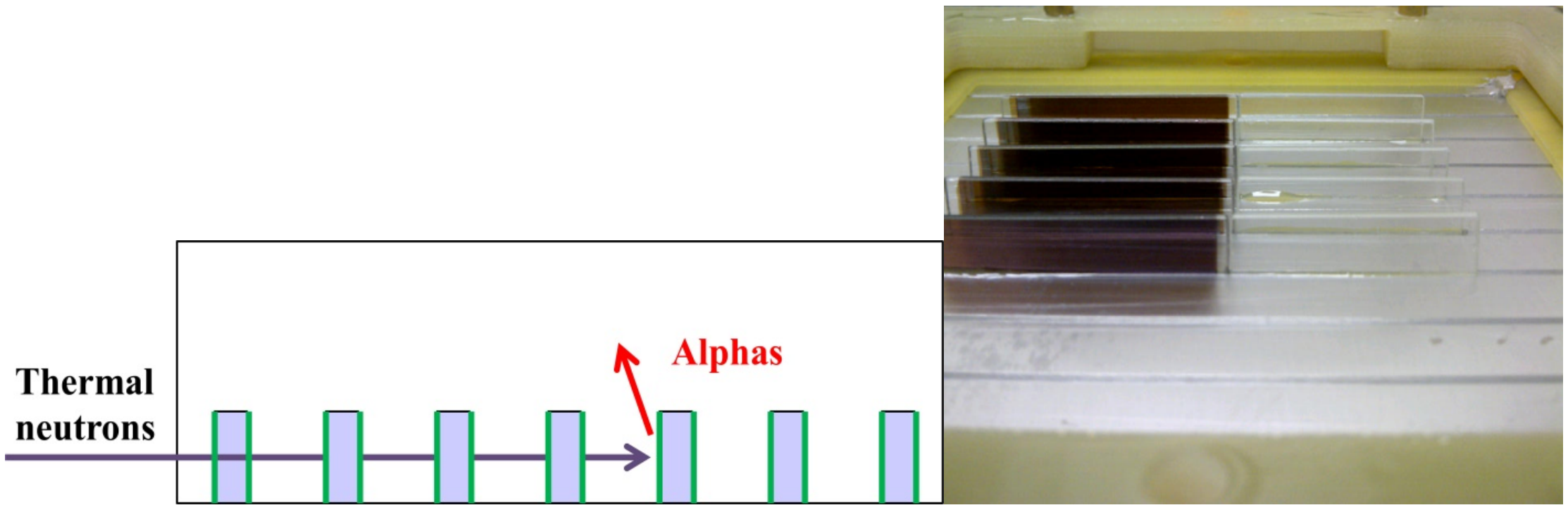

*Figure 6.20: (right)View of the cathode with borated glass (left) and simple glass (right); (left) Conversion mechanism in the borated sheets*

*2)* **Radiative capture detectors**

Although boron and most generally neutron-charged particles converters widely investigated, a different approach that is subject of investigation is the use of thermal neutron radiative capture [**Festa_2011**] in cadmium. Indeed high neutron absorption efficiency (close to 100%) cab be reached in small thickness of Cd, an in principle geometrical efficiency close to 100% can be obtained by a proper Cd-gamma detector setup. Typically inorganic scintillators have to be used and in this respect an important issue to be addressed is the improvement of the signal o background ratio that can be optimized by properly chosen discrimination thresholds [**Pietropaolo_2006,**

**Tardocchi_2004**]. Preliminary tests on the use of radiative capture for thermal neutron counting and an investigation of detector cross talk was performed [**Pietropaolo_2013**] and encouraging results were obtained. Figure 1 shows the detector setup for the measuements and the diffraction patterns (in the time of flight domain) of a metallic sample recorded by two different detection systems: a 1 mm-thick Cadmium sheet (2 x 10 cm$^2$) faced to a Yttrium-Aluminum-Perovskite YAP) inorganic scintillator, and a standard squashed $^3$He tube at 10 bar pressure (2 x 10 cm$^2$ effective area).

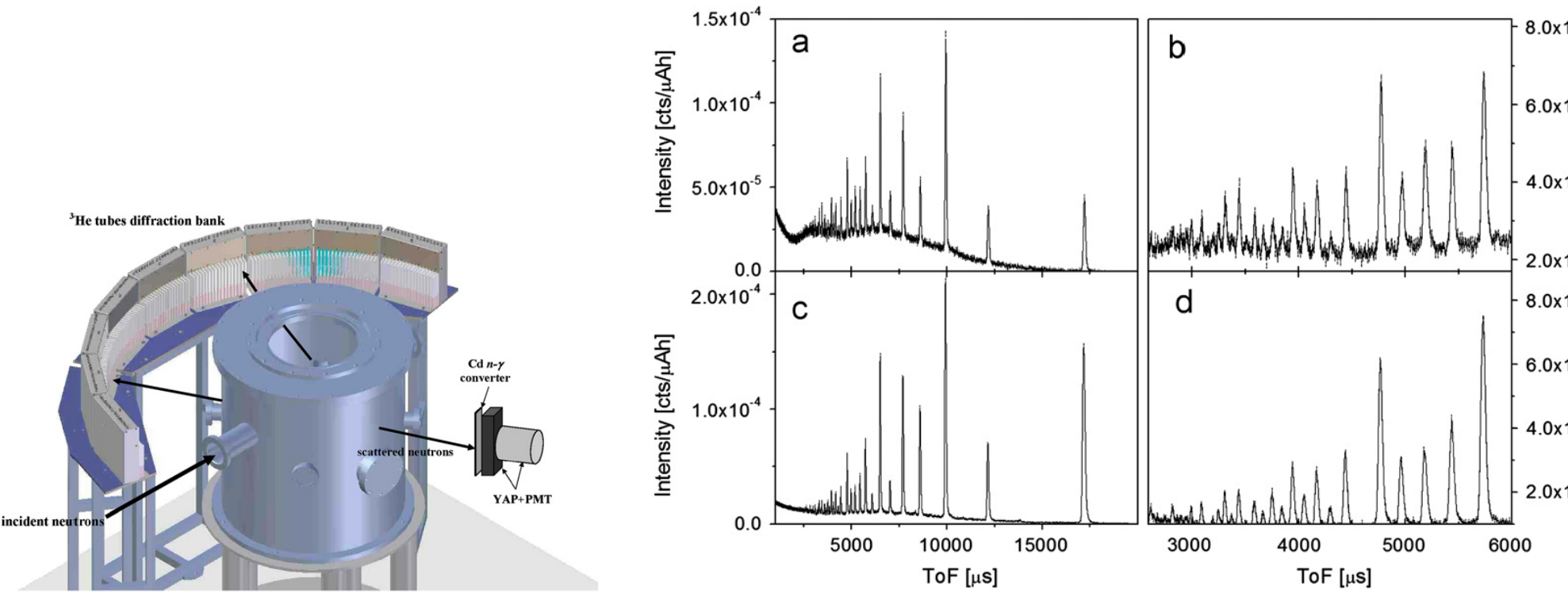

*Figure 6.21: left: Schematic layout of the INES powder diffraction diffractometer at ISIS during the experiment. The instrument was equipped with the new YAP–Cd device for non-standard diffraction measurements. Right: (a) Time of flight diffraction pattern recorded by the YAP–Cd counting device in the whole thermal neutron energy region; (b) blow up in the ToF region between 2500 and 6000 ms. Panels (c) and (d) show the same ToF diffraction patterns of panels (a) and (b), respectively, but recorded by the 3He tube at a similar scattering angle of the YAP–Cd*

### 3) Real-time neutron spectrometry in the eV-GeV energy interval

The measurement of neutron spectra over more than ten orders of magnitudes in energy (from thermal up to GeV) represents a key-issue for the moderated or un-moderated (chip irradiation) neutron beams to be produced at IRIDE. The knowledge of the energy distribution of the neutron field impinging on a sample, or on an electronic device under test, may provide important additional information with respect to a simple beam-monitor system.

A desirable added feature would be the directional spectrometry: if a real-time directional neutron spectrometer existed, this would allow separating the target-based neutron component from that originated in the scattering with the walls, the objects and the structures present in the irradiation room.

A solution to these problems is represented by the spectrometers CYSP and SP$^2$ developed within the NESCOFI@BTF project (INFN – CSN V). They adopt a similar philosophy as the Bonner sphere spectrometer, but using multiple thermal neutron detectors (TND) in a single moderator instead of multiple moderators. NESCOFI@BTF showed that such a new spectrometer has similar performances as the Bonner spheres, and in addition it can work as a real-time instrument and is well suited for monitoring applications (machine diagnostics as well as radiation protection dosimetry).

Two types of spectrometers with different number of TND and different moderating geometries have been designed and tested within NESCOFI@BTF, namely CYSP and $SP^2$:

- CYSP spectrometer [Bedogni1] has cylindrical geometry and shows a sharply directional response, thus allowing very accurate directional spectrometry. It includes seven active thermal neutron detectors (TND) working in parallel at different depths in a polyethylene cylindrical moderator.

- $SP^2$ spectrometer [Bedogni2] has spherical geometry and isotopic response. It includes thirty-one TNDs symmetrically arranged along the three perpendicular axes of a polyethylene-lead sphere (25 cm in diameter). It is suited for measuring the neutron spectrum disregarding its directional distribution. This device is optimal for radiation protection monitoring of the areas surrounding the neutron beam-lines .

Suitable low-cost, miniaturized thermal neutron detectors have been developed to be embedded in the CYSP and $SP^2$ spectrometers

The so called **"thermal neutron rate detector" TNRD** [Bedogni3] produces a DC voltage level that is proportional to the thermal neutron flux. Adequate photon rejection is achieved through an intrinsic compensation effect. The lowest measurable thermal neutron flux is <100 $cm^{-2}s^{-1}$. Its overall dimensions are approx. 1.5 cm x 1 cm x 0.4 cm. This signal is amplified in a low-voltage electronics module especially developed by the project team. The amplifier output is sent to an ADC controlled by a PC through a LabView application. The detector and its typical time-dependent output are shown in Figs 6.22.

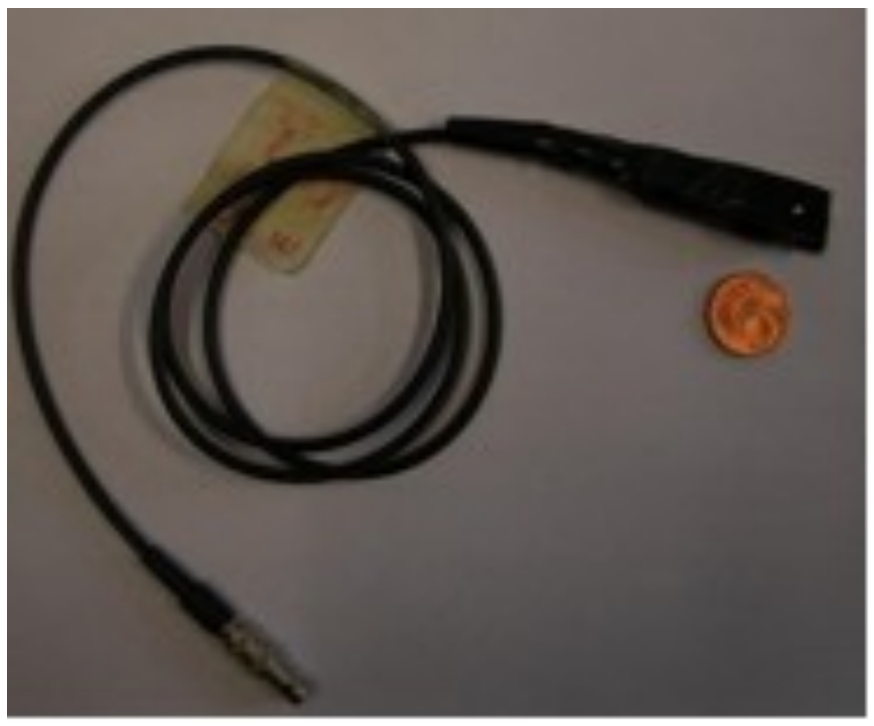

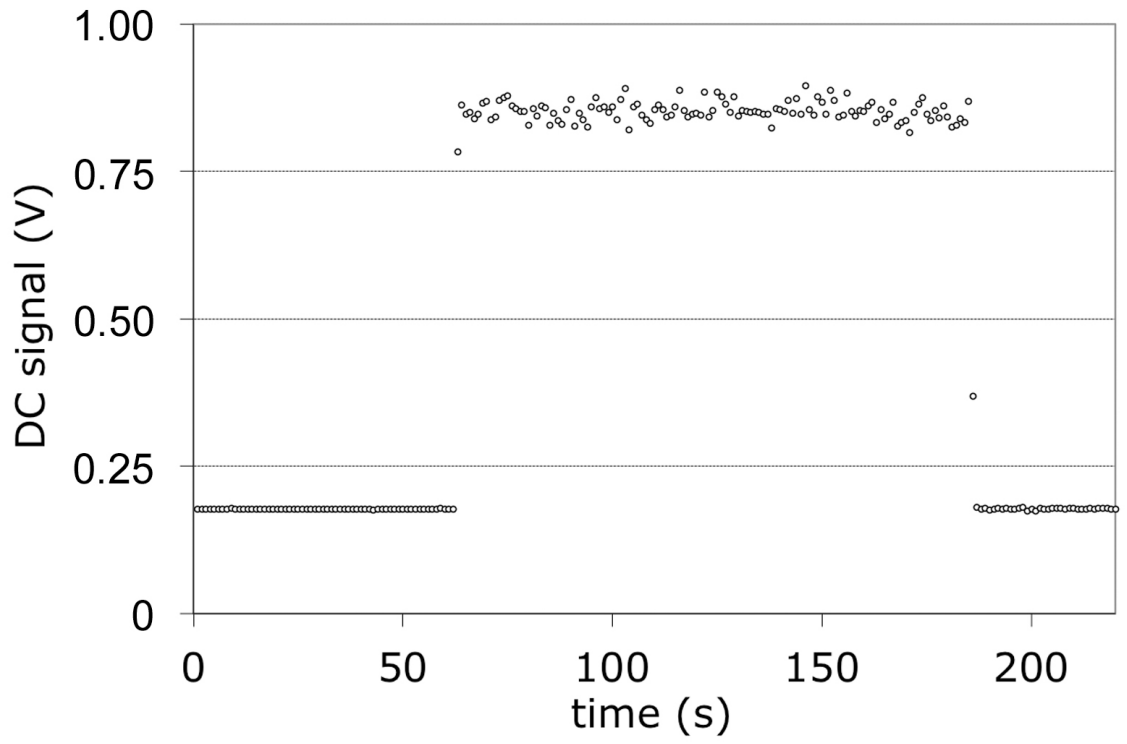

**Figure 6.22**. *Left: The thermal neutron rate detector (TNRD). Right: Time-dependent output of the TNRD when exposed in an ex-core thermal neutron beam from the ENEA Casaccia TRIGA reactor at power 46 kW (thermal flux 6E+4 $cm^{-2}s^{-1}$).*

TNRDs are individually calibrated in a reference moderating assembly at INFN-LNF. Typical calibration factor (thermal flux per unit output voltage) is (96±3) $cm^{-2}s^{-1}$ $mV^{-1}$. To estimate the reproducibility of the manufacturing process, the response of ten TNRDs with nominally identical fabrication characteristics was compared, and its variability is 5% (one s.d.). The response to photon radiation, measured in a reference 137Cs field, is (0.51±0.02) mGy $h^{-1}$ $mV^{-1}$.

**CYSP** is a cylindrical spectrometer embedding 7 active TNDs and showing a prominent directional response. See Fig. 6.23 for details.

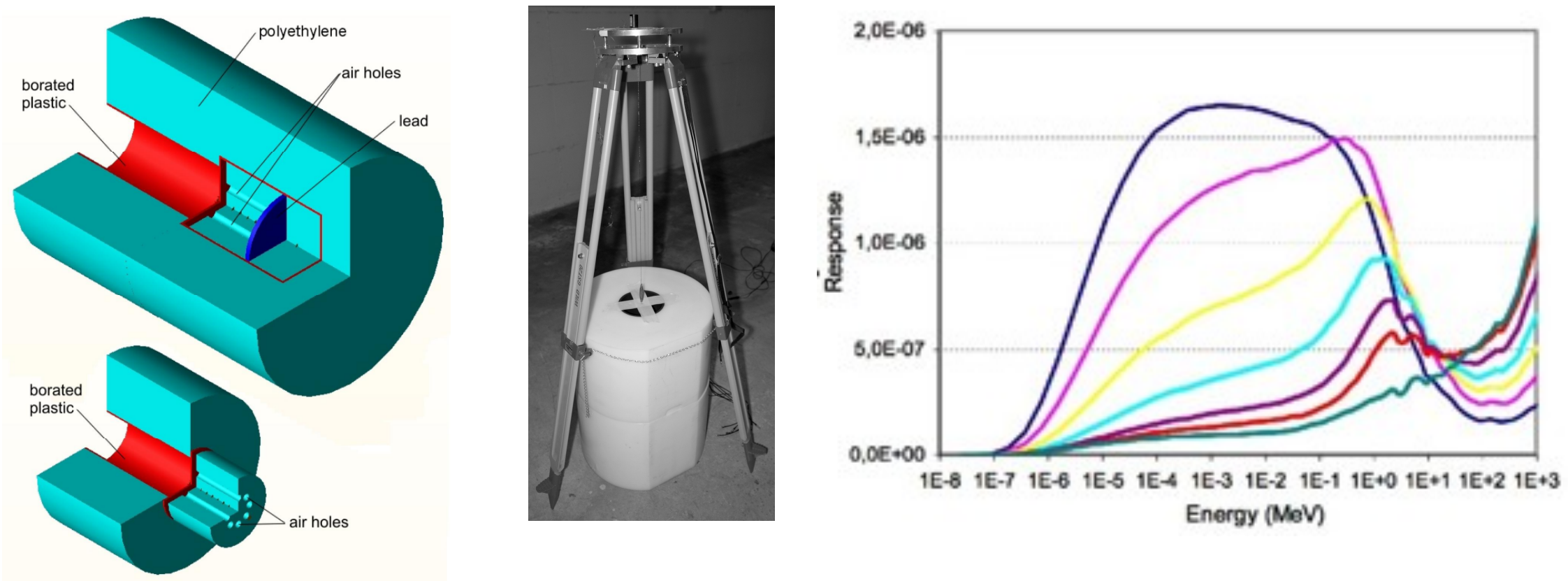

**Figure 6.23.** *Left: 3D image of the CYSP with internal parts. Center: Picture of the CYSP*. *Right: matrix of the CYSP (MCNPX simulations).*

The cylindrical moderator is built in high-density polyethylene (blue) and has overall dimension 50 cm (diameter) x 65 cm (height). The collimating aperture (16 cm in diameter, 30 cm in length) on the left defines the acceptance solid angle. The seven detectors (internal cavities) are located at increasing depth in a polyethylene internal capsule (called “detector capsule”). These detectors are covered by different thicknesses of moderating material, thus producing response variation as a function the initial neutron energy and of the detector position. In this device the different detection positions play the same role as the spheres of different diameter in the Bonner sphere spectrometer.

The 0.5 cm borated plastic layer (in red) plus the large external polyethylene layer protect the detector capsule against neutrons coming from different directions than the one defined by the collimator. In fact neutron from unwanted directions are moderated / absorbed in the lateral polyethylene, and residual slow neutrons are absorbed in the borated plastic layer. Neutrons above 20 MeV undergo inelastic (n,xn) reactions in the lead layer (1 cm thick), thus producing secondary low-energy neutrons that are detected from detectors in the vicinity of the lead. Figure 6.23 also shows the CYSP response matrix, i.e. the reading of the different thermal neutron detectors per unit incident neutron fluence, as a function of the neutron energy. The irradiation geometry is plane parallel.

The calculations have been performed with MCNPX. From thermal neutrons up to 10 MeV, the response decreases as the detector depth increases (The shallowest detector, n. 1, is in blue. The deepest detector, n. 7, is in green). Above 10 MeV the effect of the lead converter is evident, since the response of detectors n. 5, 6 and 7 are higher than detectors n. 1 to 4.

**The SP$^2$ spectrometer** consists of 31 thermal neutron detectors arranged along three perpendicular axes at 5 radial distances (5.5, 7.5, 9.5, 11 and 12.5 cm) and at the center of a polyethylene sphere of diameter 25 cm. An internal 1 cm thick lead shell between 3.5 and 4.5 works as an energy converter via (n,xn) reactions thus enhancing the response above 20 MeV, either for the central detector and for those located at 5.5 cm and 9.5 cm.

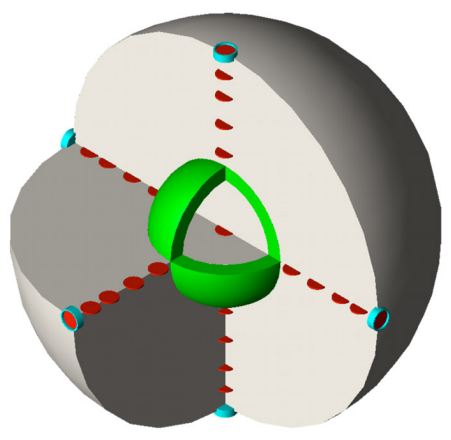

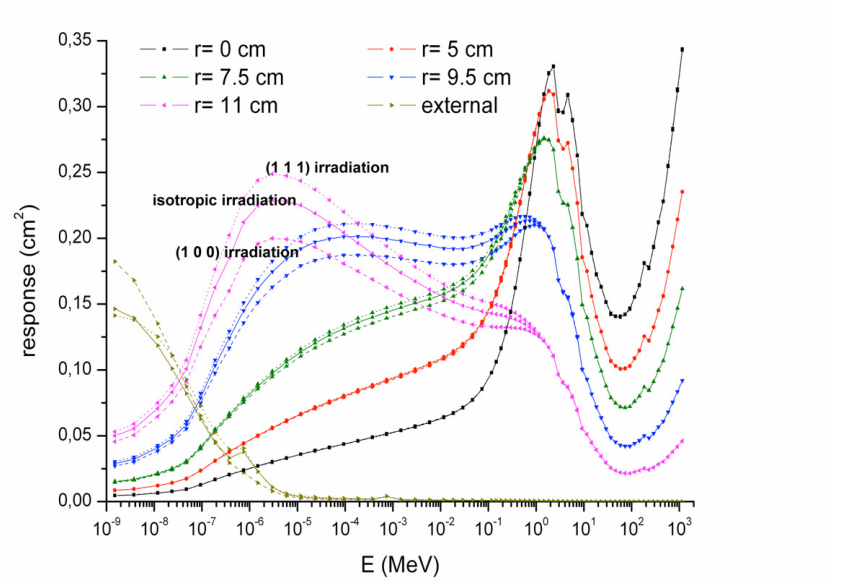

**Figure 6.24.** *Left: 3D image of the SP$^2$ with measuring positions in evidence. Right: Response matrix of the SP$^2$. Dependendece on the irradiation geometry can be appreciated by comparing the result for (1 0 0), (1 1 1) and isotropic irradiation.*

The SP$^2$ response matrix for neutron energis energies from $10^{-9}$ to $10^3$ MeV is ploted in Fig. 6.24 for three irradiation geometries: an isotropic irradiation and a parallel beam in two different incidence directions, namely (1 0 0) and (1 1 1) referred to the three perpendicular axis of detectors (Fig. 23). As it can be seen, the lower and higher response values are obtained for the (1 0 0) and (1 1 1) irradiation, respectively and this effect is of practical importance only for the shallow detectors (d=11 cm and d=9.5 cm) and for energies in the keV region and below. Thus and although the response of a single TND in a given location should not be isotropic, a nearly isotropic response is obtained by averaging the readout of detectors located at the same radial response.

Because the three geometries give almost coincident response functions above 1 keV, the spectrometer response is practically isotropic in this energy domain. Moreover, the response functions above 10 MeV are significantly enhanced by the presence of the lead layer due to the (n,xn) reactions in lead, not only for the central detector but also for those located between 5.5 and 9.5 cm. As it can be seen, the response above 1 MeV for fast and high energy neutrons is completely isotropic. According to the response matrix displayed in Fig. 6.24, a good spectrometric performance is expected above 1 keV. Moreover, the response matrix has been validated using the quasi-monoenergetic reference neutron fields of 0.147, 0.565, 1.2, 5.0 and 14.8 MeV available at PTB Braunschweig and is overall uncertainty was evaluated as ±3%.

### 6.7. Charged Particles Production

Albeit the main goal of a photo-production source is to produce neutrons, the interactions of electrons with a target generate also charged particle.
Provided a suitable target, the Bremsstrahlung of electrons up to 3GeV in energy can produce secondary positron, pion and muon beams either via pair production (γ+Nucleus → $e^+e^-/\mu^+\mu^-/\pi^+\pi^-$ +Nucleus) or via nuclear evaporation (γ+Nucleus → hadrons including n,p or π). The pions produced can in turn decay into muons or electrons.

As an alternative, also possible within IRIDE, the original photon instead of being produced by Bremsstrahlung, could be generated by Compton backscattering of a high power laser on the electron beam. This process could allow the production of up to 1 GeV monochromatic photons that could produce charged particles by pair production. This production mechanism allows to have a different spectrum on the emitted secondary particles, as studied for the positron production [Broggi 2013].

The characteristics of the target depend on the radiation of interest. Here we explore two cases: the production of beams of positrons, muons, pions and protons, and the production of muons from pions at rest.

The **production of collimated beams of charged particles** is favoured when using the Compton backscattering radiation because of its collimation. In any case the targets need to be thinner than for neutron production in order to reduce the scattering.
A preliminary evaluation of the production of the secondary particles from the above reactions, has been carried out by using the FLUKA Monte Carlo code [Fassò 2005,Battistoni 2007]. We have considered targets made of one radiation length of Tungsten (3.5 mm) and Lead (5.6 mm), hit by 1 GeV photon beam and 3 GeV electron beams.
The results obtained with $10^6$ primary particles are summarized in the following table. Here the total number of secondary particles (positron, protons etc.)/primary particles is reported. The number after the slash is the number of secondary particle (positron, protons etc.)/primary particles emitted in a forward cone of 10° aperture.

**Table 6.15: Charged particles production rates**

| Case | e+ | p | $\pi^+$ | $\pi^-$ |
|---|---|---|---|---|
| γ on W | 0.75/0.44 | 3.27E-4/ 2.81E-6 | 5.23E-5/ 5.93E-7 | 6.72E-5/ 1.76E-6 |
| γ on Pb | 0.75/0.43 | 2.68E-4/ 3.42E-6 | 4.12E-5/ 1.06E-6 | 5.43E-5/ 1.2E-6 |
| Electrons on W | 1.30/0.55 | 4.00E-4/ 4.59E-6 | 5.69E-5/ 7.00E-7 | 8.01E-5/ 1.09E-6 |
| Electrons on Pb | 1.30/0.55 | 3.62E-4/ 5.26E-6 | 4.35E-5/ 7.87E-7 | 6.46E-5/ 7.98E-7 |

Muon production is not considered because the energy is not enough to obtain a beam-like structure.

The fluencies of the particles from the cases listed above are shown in the Fig. 6.25. The primary beam comes from the left and the target is located at (0.0). While positrons have a collimated structure, protons (as well as pions albeit not shown here) have an almost isotropical distribution. The resulting energy spectra are instead similar among the particle types, as shown in Fig. 6.26.

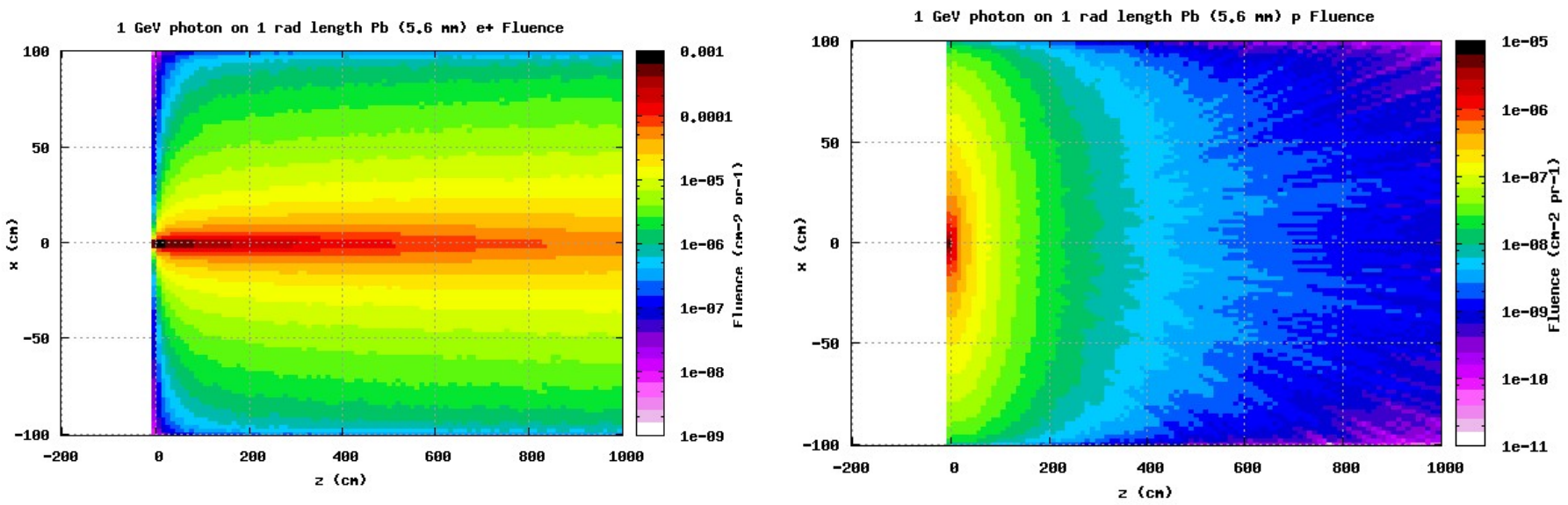

*Fig. 6.25 Positron (Left) and proton (Right) fluence from 1 GeV photons on Pb target*

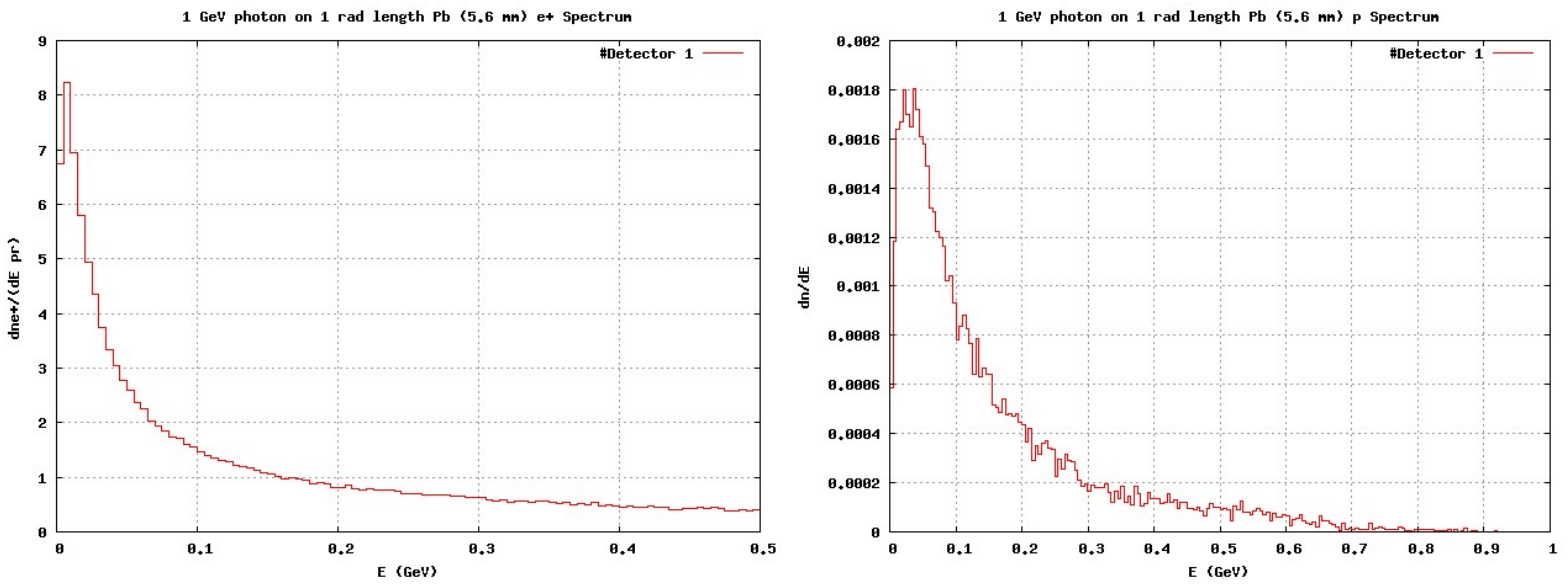

Fig. 6.26 *Positron (Left) and proton (Right) energy spectra from 1 GeV photons on Pb target*

The production of **muons from pions at rest** is needed by $\mu \rightarrow e\gamma$ experiments like MEG [Adam 2013]. Also this application requires a relatively thin target because only the muons produced on the surface can escape the target. To increase the range of the muons of interest also a target made of carbon was considered. To estimate the expected production rate a simulation based on GEANT (Allison 2006) was performed with a cylindrical target and its radius and length was varied. Also the impinging electron beam energy was scanned.

The optimal surface muon rate was obtained with 3GeV electrons on a 35cm long target of graphite, in which case $5x10^{-5}$ $\mu^+$ of interest per incident electron are produced. It is to be noted that the $\mu^-$ rate is negligible since $\pi^-$ interact before decaying. Tungsten targets produce an order of magnitude less muons. For an electron energy of 1GeV the rate drops to $6x10^{-6}$ muons per electron, and therefore high energy impinging beams are preferable.

Not all muons are of interest because they need to be stopped downstream in a thin target. The MEG experience established that the muons of interest are those with a muon momentum between 25 and 30 MeV. From the IRIDE spectrum in Fig. 6.27, where a high energy tail has been neglected, considering the nominal electron rate for IRIDE of $10^{15}$ electros per second, the expected muon rate is $5\ 10^9\ \mu^+$ per second. Although further losses downstream are to be expected, this result is extremely encouraging for an upgrade of the Lepton Flavor Violation searches, for instance if we compare it with the PSI max beam rate of $10^8\ \mu^+$ per second.

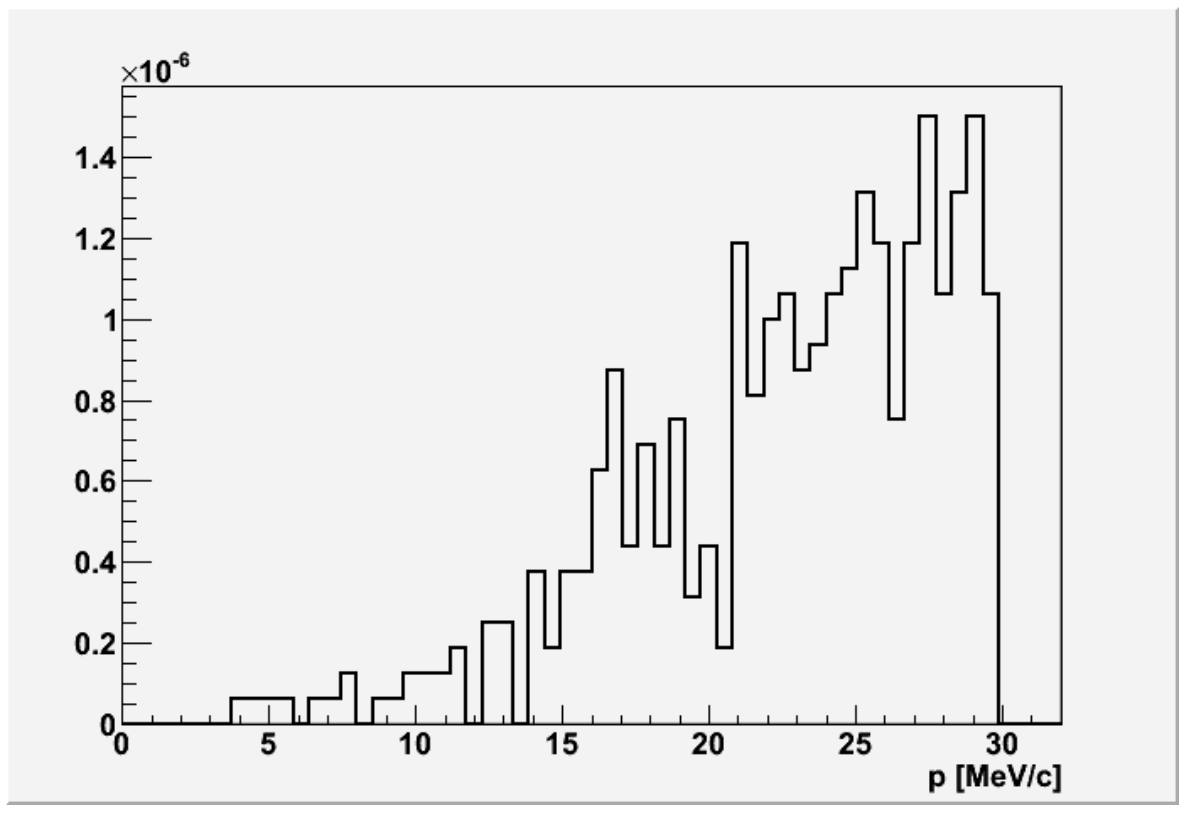

Fig. 6.27 *Surface muon energy spectrum (in a.u.) from 1 GeV electrons on carbon target*

### 6.8. Cost Analysis

This section will summarize the costs of the neutron source. The electron beam production and the infrastructures (buildings, supplies,…) are not considered in this budget.

The larger costs are due to the target area. Tab. 6.16 reports a preliminary estimation of the costs, based upon the design described in Sec. 6.5. The largest cost comes from the shielding. Here we have assumed to use the same design of ISIS although the neutron rate is expected to be smaller at IRIDE. On the other side the photon rate is larger at IRIDE and this might cause some adjustment. At ISIS the target area required 6,000 tons of steel and 1,000 tons of concrete. At the current prize of ~700Euro/ton for steel and ~50Euro/ton of concrete, the material costs of the shielding amount to ~3MEuro.

**Table 6.16: Estimated material costs of the neutron target area, not including design, development, manufacture and qualification.**

| **Item** | **Overall costs [MEuro]** |
|---|---|
| Shielding | 4 |
| diagnostics | 0.5 |
| Target Rotating support and Cooling system | 1.5 |
| Target+Moderator | 1.5 |
| Reflector | 0.5 |
| **Total Target Area** | **8** |

Separate costs have to be addressed for each beamline. They include the shielding, the beam transport, the diagnostics and the measurement equipment that needs to be provided for the beamline to be effective. The less expensive lines are those for Chip Irradiation and Imaging since they are short and not require $^3$He detectors. SANS moderator more expensive (D2 and cooling) .Diffraction, NRCA and BET could be done in a single beamline. Detectors make instead a difference for the other lines, and in addition SANS requires a long vacuum pipe. An approximate evaluation is detailed in Tab. 6.7. The costs of the surrounding buildings are not considered here.

**Table 6.17: Estimated costs of each type of neutron beamline. All costs are in MEuro**

| Beamline | Chip Irradiation | SANS | Imaging | Diffraction/ NRCA/BET | EINS |
|---|---|---|---|---|---|
| Shielding | 0.5 | 0.3 | 0.3 | 0.3 | 0.3 |
| Beam transport | 0.3 | 2.2 | 0.2 | 0.3 | 0.3 |
| Detectors and diagnostics | 0.2 | 3.5 | 0.5 | 2.4 | 2.4 |
| **Total (MEuro)** | **1** | **6** | **1** | **3** | **3** |

## 6.9. Conclusions

The neutron source at an electron facility of the type that is envisaged for the IRIDE project can deliver a neutron rate ranging from $1.6 \times 10^{14}$ n/s at the target with a pulsed structure to $5.0 \times 10^{15}$ n/s without time structure.

In the scenario without time structure, which is the easiest to implement and to integrate within the IRIDE project, the facility will allow Chip Irradiation, Neutron Radio/Tomography, Nuclear Activation Analysis and Small Angle Neutron Scattering, techniques routinely used in applied, biomedical, condensed matter, and industrial research. For such applications only 30% of the requests in Europe are satisfied by the existing facilities.

In the case the time structure, and therefore a measurement of the neutron's wavelenght with a Time Of Flight technique, can be implemented, the potentialities will broaden to include the Neutron Resonance Capture Analysis, the Bragg Edge Trasmission, and the Neutron Diffraction. For such applications only 50% of the requests in Europe are satisfied by the existing facilities.

As far as particle physics research is concerned, the high neutron flux can be exploited for the search for neutron-antineutron oscillations. We are also exploring the charged particle production, in particular muons.

The costs of the neutron source, provided the electron beam and the buildings, are contained, about 3.5MEuro for the target area and less than 500kEuro for each beamline.

The project is clearly modular and thus it can be pursued in steps and with different funding sources.

Finally, there is a strong interest in a neutron source both in the scientific community, composed of ~500 scientists that publish ~200 papers per year, and the industrial and aerospace community. There is infact no other dedicated facility in Italy and, as mentioned above, the availability of beam-time in Europpe is significantly below the request.

# 7. COMPTON SOURCE

## 7.1. Laser system and performances for Compton Source

In order to convert the electron beam generated by the IRIDE Linac into a bright beam of gamma photons we need to collide the electron beam with an intense optical laser pulse [7.1], that may be generated with three possible techniques:

A) Intra-cavity FEL pumped optical cavity (typically 10 MHz)

B) Fabry-Perot externally coupled optical cavity (typically 100 MHz)

C) Amplified High Power Lasers coupled to Laser Recirculators (10 kHz)

All of them aim at maximizing the flux $N_\gamma^{bw}$ of gamma ray beam generated in the Compton back-scattering: this quantity scales like the product between the number of electrons in the colliding bunch times the laser intensity in the collision point,

$$N_\gamma^{bw} \approx Q[pC] I_{las}\left[\frac{W}{cm^2}\right]$$

In case of FEL pumped cavities (case A) the upper limit is about 100 kW of average FEL radiation focused down to 150 $\mu m$ at a rep rate of 100 $\mu m$ (ref. Krafft): maximum intensity reached in the interaction point $I_{las} = 5 \cdot 10^8 \frac{W}{cm^2}$ . Fabry-Perot externally pumped cavities have achieved up to 100 kW of stored laser power at focal spot sizes as small as 50 $\mu m$, implying $I_{las} = 4 \cdot 10^9 \frac{W}{cm^2}$. The newly developed technique of high power amplified lasers coupled to laser recirculators will allow to reach 100 W x 100 round trips = 10 kW effective power at the interaction point, with spot sizes much smaller than those of optical cavities, *i.e.* 10 $\mu m$, implying $I_{las} = 10^{10} \frac{W}{cm^2}$ .

In order to evaluate the maximum flux achievable with each technique we need to evaluate the max charge in the electron bunch compatible with the maximum average current available of 300 $\mu A$. This is 30 pC for case A, 3 pC for case B and 1 nC for case C (limited by injector performances more than by average beam current constraints). We then obtain: $N_\gamma^{bw} \approx 10^{10}$ for case A and B, while $N_\gamma^{bw} \approx 10^{13}$ for case C. Therefore the amplified laser technique is more performant as far as the photon flux is concerned, mainly because of the better focusability of the laser pulse in the interaction region. If we consider now the scaling for the luminosity of a gamma-gamma and an electron-gamma colliders, we found an even larger difference between the three techniques. In fact: $L_{e-\gamma} \approx Q[pC] N_\gamma^{bw}$ for the electron-gamma, implying $L_{e-\gamma} \approx 4.5 \cdot 10^{11}$ for case A, $L_{e-\gamma} \approx 3.6 \cdot 10^{10}$ for case B and $L_{e-\gamma} \approx 10^{16}$ for case C. For the gamma-gamma collider we have $L_{\gamma-\gamma} \approx \left(N_\gamma^{bw}\right)^2 / f$ , implying $L_{\gamma-\gamma} \approx 2 \cdot 10^{13}$ for case A, $L_{\gamma-\gamma} \approx 1.4 \cdot 10^{12}$ for case B and $L_{\gamma-\gamma} \approx 10^{22}$ for case C. In conclusion, the technique based on amplified lasers coupled to laser recirculators is more performant by about 3 orders of magnitude than FEL based or Fabry-Perot based optical cavities as far as gamma photon flux is concerned, while it is at least 5 orders of magnitude more performant as far as the luminosity of electron-gamma colliders is concerned, and at least 8 orders of magnitude more performant as fas as the luminosity of a gamma-gamma collider is concerned. It is interesting noticing that these considerations are almost independent on the gamma photon energy, for the energy range we are considering (1-60 MeV), where the Compton recoil effects in the back-scattering are small. The best candidate for this technique is nowadays the Yb:Yag laser technology: this was recently proven to deliver laser pulses carrying 1 J of energy per pulse at 100 Hz, at 1.03 $\mu m$ wavelength. Since the laser pulse after the interaction with the electron beam is almost unaltered (as the number of scattered photons is smaller by orders of magnitudes than the number of photons carried in the laser pulse) the laser pulse itself can be recirculated in a wrapped-up transport line, called the recirculator, and taken back to the interaction point again, after a delay of about 10-20 nsec, to collide with a new fresh bunch of electrons. A maximum number of round trips up to 100 can be conceived (a present lay-out for the ELI-NP-GBS proposal [7.2] foresees 32 round-trips), creating an effective 100x100=10 kHz repetition rate of the interaction, or, in other word, an increase by a factor 100 of the effetive laser power delivered at the interaction point.

### 7.2. Cost analysis

A 1 J 100 Hz Yb:Yag laser system costs up to 6 M€ : for the gamma-gamma collider two of these systems are needed. The two laser recirculators have been evaluated to cost 1.5 M€ each. To this one should add the cost of laser clean rooms, transport lines and laser instrumentation and diagnostics, which may be on overall evaluated up to 3 M€ (ref.ELI-NP).

# 8. NUCLEAR PHOTONICS

## 8.1. Scientific opportunities

In nuclear physics there is large interest at present for the neutron-rich systems. On the one hand, existing and planned radioactive beams facilities aim to locate the position of the neutron and proton drip-lines (i.e., the limits defining whether a nuclear system is bound), and to study the properties of the isotopes in which the neutron/proton ratio differs from the values that characterize stable nuclei. On the other hand, another example of nuclear matter under extreme conditions is the matter that compose the compact astrophysical objects like the neutron stars. From a general point of view, an understanding of these "exotic" nuclear systems cannot be reached without a better assessment of the isoscalar (T = 0) and isovector (T = 1) components of the nuclear Hamiltonian. Even in stable nuclei the isovector properties are poorly known: the neutron radii, the systematics of isovector collective states, the pairing interaction in the T = 0 and T = 1 channel - to name only a few - are still object of strong debate. The possibility of new experiments that clarify these issues is consequently of great importance for the development of nuclear science.
Experiments with real photons offer quite obvious advantages: the electromagnetic probes excite selectively the isovector states, and the electromagnetic interaction is well known.
A first clear goal is the study of isovector collective states, namely isovector giant resonances. A nuclear Giant Resonance (GR) is a collective vibration of the nucleus that is made up with the coherent contribution of many particle-hole (p-h) excitations and exhausts a large fraction of the appropriate sum rules [8.1, 8.2]. The Giant Dipole Resonance (GDR) is the first resonance that has been experimentally investigated, due to the availability of real photon beams [8.3]. Much less experimental information is available in the case of the isovector GRs having different multipolarity. Hadronic probes have been used in the past to try to locate them, but (i) the lack of selectivity between spin-flip and non spin-flip GRs, (ii) the presence in the same energy region of different multipole excitations and of background processes, and (iii) the uncertainties associated with the hadronic reactions in general, have always hindered a precise determination of the IV monopole, quadrupole etc. Recently, using real photons available at the HIgS facility, Duke University (USA), S.S. Henshaw *et al.* have measured the excitation strength function of the IV giant quadrupole resonance (IVGQR) in 209Bi [8.4]. The relevant operator for this mode is

$$Q = \sum_{i=1}^{A} r_i^2 Y_{2\mu}(\hat{r}_i) t_z(i).$$

The IVGQR peak shows up at an excitation energy consistent with the law 135 $A^{-1/3}$. If one wishes to extend the knowledge of the IVGQR to lighter nuclei, because of this scaling, one needs an energy range which is accessible for IRIDE.
As already mentioned in [8.4], the interest of the study of the IVGQR stems also from the possibility of obtaining new information about the higly debated nuclear symmetry energy. Along a similar line as in our previous discussion, one usually distinguishes an isoscalar and isovector part of the EoS, namely one writes

$$\frac{E}{A}(\rho,\delta) = \frac{E}{A}(\rho,\delta = 0) + S(\rho)\delta^2,$$

in terms of the total density and of the proton-neutron asymetry d = ($r_n$-$r_p$)/r. The first term in the previous equation is the EoS of symmetric nuclear matter and, although still uncertain at high density, this part is certainly much better known than the second term especially around saturation. The second term defines the so-called symmetry energy S. The symmetry energy is basically unconstrained. At the same time, it is a quantity of paramount importance as it controls not only the nuclear stability but also the properties of neutron stars, the way they are formed through supernova explosion and the way they cool. Cf. [8.6] and references therein.

There are other physics issues that cannot be accessed easily without an intense photon beam at large enough energy, and on which the contribution of IRIDE would be highly welcome.

Although a number of experimental data and theoretical studies have been cumulated, the question still exists whether we can access only the inclusive properties of the GRs (energy and fraction of exhausted sum rule), or more exclusive properties associated with the wave function of the GR. Ultimately, we can say that we still miss an unambiguous confirmation of the microscopic picture of this collective motion. Giant resonances have a finite lifetime. Being as a first approximation described by coherent superposition of p-h configurations, their most probable damping mechanism is their coupling to progressively more complicated states of 2p-2h … *n*p-*n*h character (up to the eventual compound nucleus state). The associated contribution to the total width, the so-called spreading width, is the dominant one. The decay width associated with the emission of one nucleon in the continuum (escape width or $G^-$) is of some relevance in light nuclei but much less important in heavy nuclei. The gamma-decay width is a small fraction ($\approx 10^{-3}$) of the total width. Despite this, the study of the gamma-decay decay of GRs has been considered a valuable tool for about 30 years. The basic question is whether the decay will enlight some fine structure of the IVGDR. Already in the pioneering experiment of Ref. [8.5], it was pointed out that the non-statistical component of the neutron decay of the IVGDR in $^{208}$Pb, depends strongly on the IVGDR excitation energy and final state energy. The reasons for this fine structure, the details of the GR wavefunction that emerge from the measurement of its decay – that can be performed by exciting the GR and measuring the electromagnetic decay either to ground state and excited states – are another issue for IRIDE.

IRIDE can also allow the study of multi-phonon states, namely multiple giant resonances. This is related to fundamental questions like how large is the phonon-phonon interaction and how large are the deviations of the atomic nucleus from the harmonic picture. Coulomb excitation data exist for the excitation of the double IVGDR in $^{136}$Xe and $^{208}$Pb, yet with low energy resolution. For the double ISGQR there exist hadronic data characterized by large backgrounds. All these data have not answered unambiguously the question if the emerging picture is harmonic or not. Scanning the high energy region (20-30 MeV) for a double-GDR search with good energy resolution and without uncertainties related to the reasons process, would be very beneficial.

## 8.2. Measurement of γ-rays from collective states

IRIDE is a facility which will provide an intense, highly monochromatic and polarized beam of g-rays in a very large energy window (from few hundreds of keV up to at least 30-40 MeV). The IRIDE beam can excite one or two phonon states in collective excitation of different multipolarity (i.e. IVGDR, PDR, IVGQR).

The IRIDE facility will allow NRF (Nuclear Resonance Fluorescence) experiments which requires the production of a large number of collectively excited nuclei and the measurement of their gamma, neutron or combined decay to the ground state. This is a unique feature of this facility especially because of the extremely reduced bandwidth ($s_E$ < 0.3-0.2%) and beam intensity which is expected to be very large in terms of g-rays/sec ( > $10^9$) but small in terms of g-rays/shots/10ns ( < $10^3$). It is important, in fact, that the electromagnetic flash produced by the IRIDE g beam hitting the target should not blind the gamma and neutron detectors.

The IRIDE facility should in addition provide a very accurate 'shot by shot' measurement of the beam energy ($E_g << s_E$), an accurate time signal ($s_T$ < 100 ps) and a time-separation between the different shots of some ns (depending on the time resolution of the used detectors, see following discussion). In this way, using time coincidences, the IRIDE g-ray detector array can univoquely associate the measured g-ray to a defined IRIDE shot of precisely known energy. It would be additionally useful to shield each detector from cosmic rays (with some VETO system) and from scattered gamma and neutrons (with some lateral shielding).

A typical NRF experiment requires the measurement high energy g-rays (0.5-40 MeV) and fast neutrons. A g-ray detector is usually described in terms of full energy peak efficiency (defined by the effective atomic number, density and detector size) time and energy resolution. Features like position sensitivity or radiation hardness are aspects which are less important. One aspect, which will be discussed at the end of this section, is the capability to identify fast neutron induced events and measure their kinetic energy.

From the g-detection point of view it is useful to divide a NRF experiment into two different scenarios, which could be performed at the same time, indeed: i) the elastic scattering scenario, namely the measurement of one single high energy g-ray (of the same energy of the incident IRIDE g-rays) transition to the ground state (one step decay), ii) the inelastic scattering scenario, namely a transition to the ground state divided into two steps. The first step could be either a high energy g-ray or a neutron, the second is a g-ray from a discrete low lying state (two step decay). Such kind of multiple decay it is however expected to be weaker than the previous one.

In the case of the one step decay the detectors are supposed to measure one single high energy g-ray. Its energy could range between 5 to 40 MeV and high efficient (namely large volume) g-ray detectors would be preferred. In such a scenario, the energy of the measured elastically scattered g-ray would be exactly the same of the IRIDE beam therefore the bandwidth of IRIDE will define the energy resolution of the measurements. A g-ray detector should only be capable to clearly separate the Full Energy Peak (FEP) from the First Escape one (1EP). In such a situation both solid state HPGe and scintillator detectors like $LaBr_3$:Ce [8.7-8.10] would be suitable at least up to g-rays of 25-30 MeV (see figure 8.1). For larger g-ray energies probably only HpGe detectors are capable to distinguish between FEP and 1EP. However, $LaBr_3$:Ce scintillators, if compared to HpGe, benefit of much larger available volumes and 10 times better time resolution, which will strongly boost its efficiency.

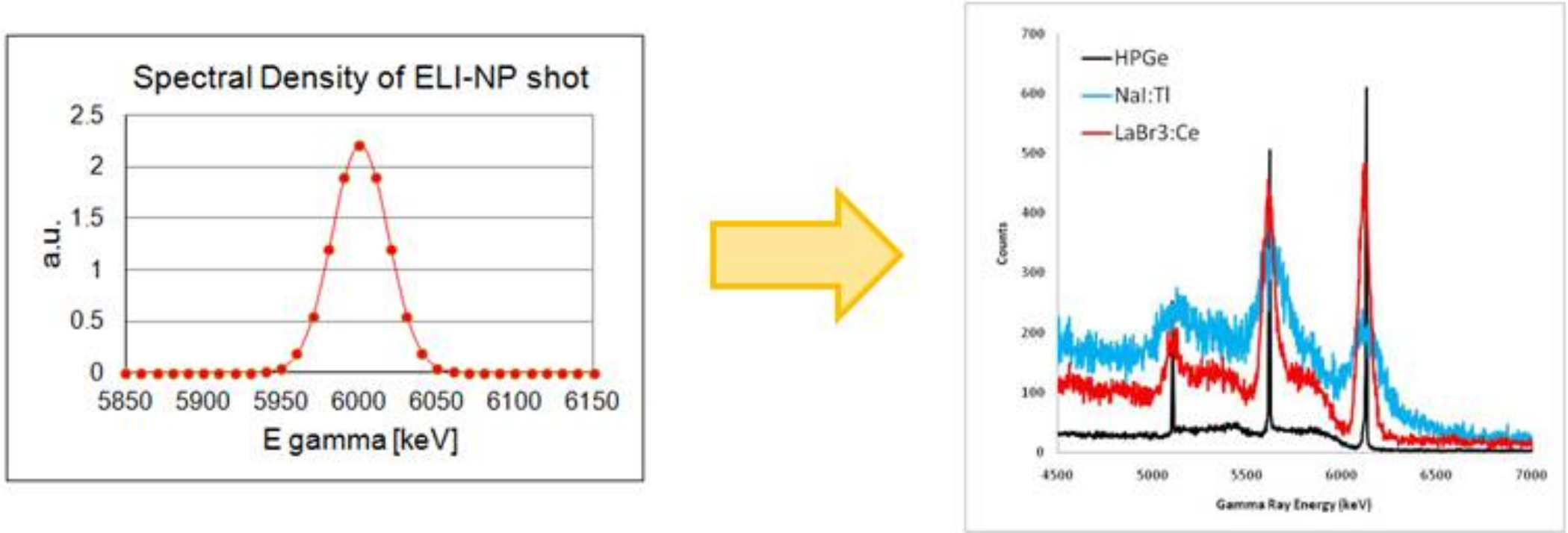

Figure 8.1: Left panel: the spectral density of a 6 MeV IRIDE beam (supposing sE/E ≈ 0.3%). Right Panel: the measured spectra in NaI, $LaBr_3$:Ce and HpGe in case of monochromatic 6.13 MeV g-rays. In all detectors the FEP is clearly separated from the 1EP. In an NRF experiment with IRIDE the HPGe peak FWHM will be very similar to that of LaBr3:Ce because of the intrinsic spectral density of IRIDE. In case of 15 MeV g-rays NaI energy resolution is not good enough to distinguish the FEP from the 1EP therefore only LaBr3:Ce and HpGe detectors should be used.

In the case of the two step decay scenario the detectors are supposed to measure in coincidence one high and one low energy g-rays. The sum of their energy should match that of the IRIDE therefore the detector sE/E should match $2^{-1/2}$ times the bandwidth of IRIDE. In such a situation, solid state HpGe are certainly superior to scintillator detectors. However, like in the previous case the much better time resolution and larger efficiency of $LaBr_3$:Ce could make the use of LaBr3:Ce preferable because of efficiency and background rejection.
The measurement of fast neutrons emitted in the decay of collective states is an experimental aspect which requires a dedicated discussion. At the moment there is not an efficient neutron detector capable to efficiently identify and stop neutrons and, at the same time, measure the neutron kinetic energy. Time of flight techniques would require very thin detectors and this will destroy the efficiency of the apparatus. However, recently, some new materials has appeared and among them seams particularly interesting the CLYC scintillator [8.11-8.13]. It is a brand new scintillator material and only very small samples are at the moment available. The CLYC scintillator presents very fast rise time (namely excellent time resolutions), good energy resolution (FWHM/E < 5% for g-rays of 661 keV), excellent neutron identification using PSA and the capability to stop fast neutrons and measure their kinetic energy through the reaction $^{35}$Cl(n,p)$^{35}$S. At the moment the R&D related to CLYC scintillator is only at the very beginning and crystals are still too small (at least a 3”x3” crystal would be needed) for comprehensive tests. However, as CLYC could efficiently identify and measure both g-rays and neutrons, the perspective that such detector provide are extremely encouraging.

### 8.3. Collimation system

The Compton source gamma beam is obtained by collimating the photons emerging from the interaction region. A careful design of the collimation system is required to reach the design brilliance together with a precise energy calibration of the gamma beam and a continuous monitoring of the parameter stability. A characterization system providing fast feedback on the energy spectrum, intensity, space and time profile of the beam is also essential for the commissioning and development of the machine.
The most challenging parameter to be achieved is the gamma energy bandwidth that is defined as the ratio of the r.m.s to the maximum of the beam energy spectrum. As mentioned above the desired range is 0.2 ÷ 0.3% and it results in a quasi-monochromatic beam. Starting from the white spectrum of the Compton radiation distribution the desired high energy contribution can be selected by means of an extraction beam line equipped with a collimator arrangement able to filter out the low energy photons. In table 1 the necessary r.m.s divergences and relative collimation aperture diameters are reported that provide a 0.3 % bandwidth for three example gamma beam energies.

**Table 8.1. Collimation parameters for selected gamma beam energies**

| | **Low Energy** | **High Energy** | **High Energy** |
|---|---|---|---|
| Energy (MeV) | 4.60 | 9.72 | 18.50 |
| Source rms divergence (μrad) | 48 | 36 | 24 |
| Collimator aperture (μm) | 892.8 (@ 9.3 m) | 309.6 (@ 4.3 m) | 206.4 (@ 4.3 m) |

The collimation system efficiency relies on the main following requirements:

1. Very low transmission of gamma photons (high density and high atomic number material);
2. Continuously adjustable aperture, in order to reach the correct energy bandwidth in the whole energy range;
3. Avoid production of secondary radiation (electromagnetic, neutrons…)

For the collimators a possible choice can be made of a dual slit collimator similar to the one already designed and assembled at INFN-Ferrara for a high energy X-ray experiment [8.14]. As reported in Fig. 8.2 it consists of an aluminum framework and two 20 mm-thick tungsten edges, where the slit width can be adjusted by translating the tungsten edges. A system of several collimators of the same type at different angle can provide a pin-hole collimator with a small enough angular acceptance and a significant thickness.
As an example for a 1-5 MeV gamma beam a possible collimator arrangement could foresee a 12 slit system with a relative 30° rotation between each other as shown in Fig. 8.3.

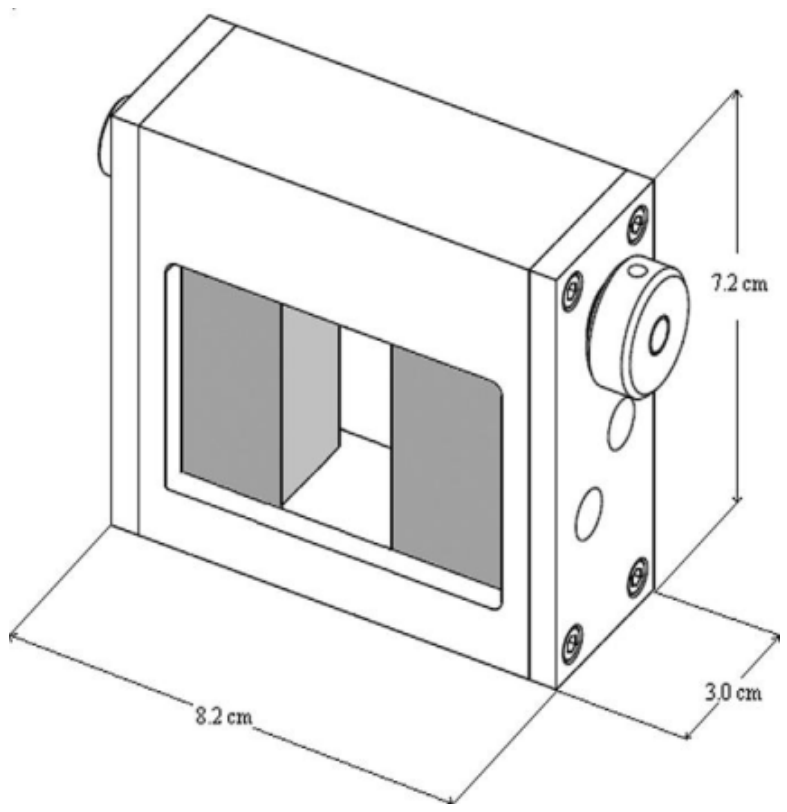

Fig. 8.2 Dual slit collimator design,

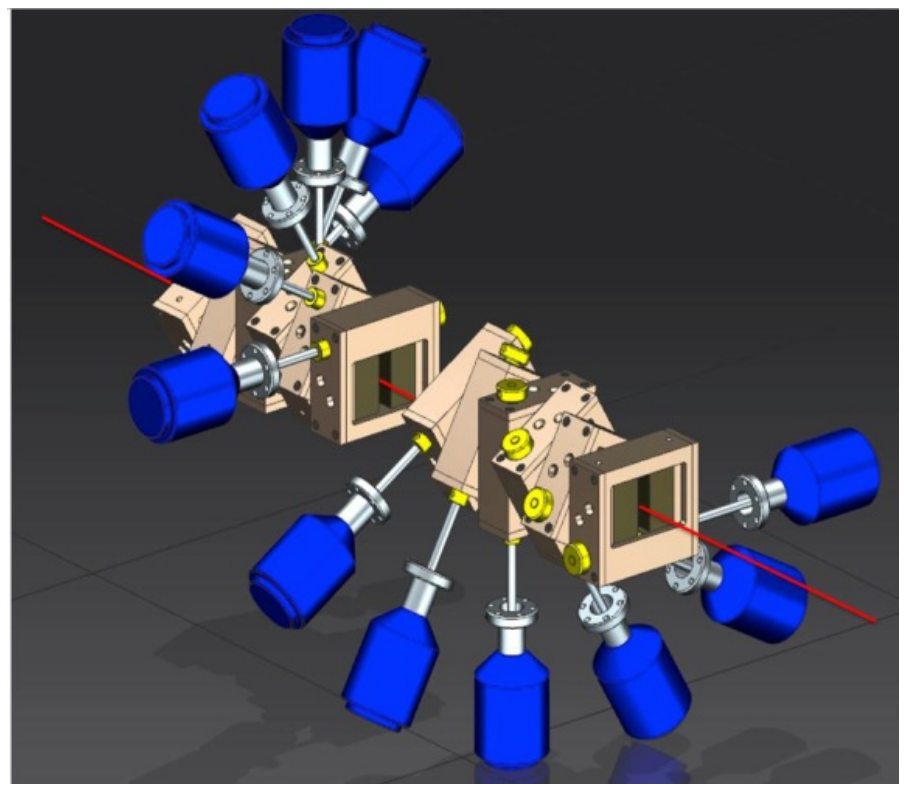

Fig. 8.3 Drawing of the configuration of low energy collimator made up of 12 tungsten adjustable slits with a relative 30° rotation each.

The transmission efficiency setup has been evaluated by a Monte Carlo simulation of the collimator assembly using the MCNPX code. In table 2 the results are reported as obtained considering only the direct (source) contributions, i.e. neglecting all scattering events, and a point source located 10 m upstream. Each slit is in this case composed by two blocks of tungsten of size 3.0 x 3.0 x 2.0 $cm^3$ and a 0.96 mm aperture. In Fig. 8.4 the number of photon transmitted by the collimation system are presented as a function of position in grey levels or as a 3-d plot. It is possible to see the pattern due to the superimposition of different thicknesses of tungsten on a logaritmic scale.

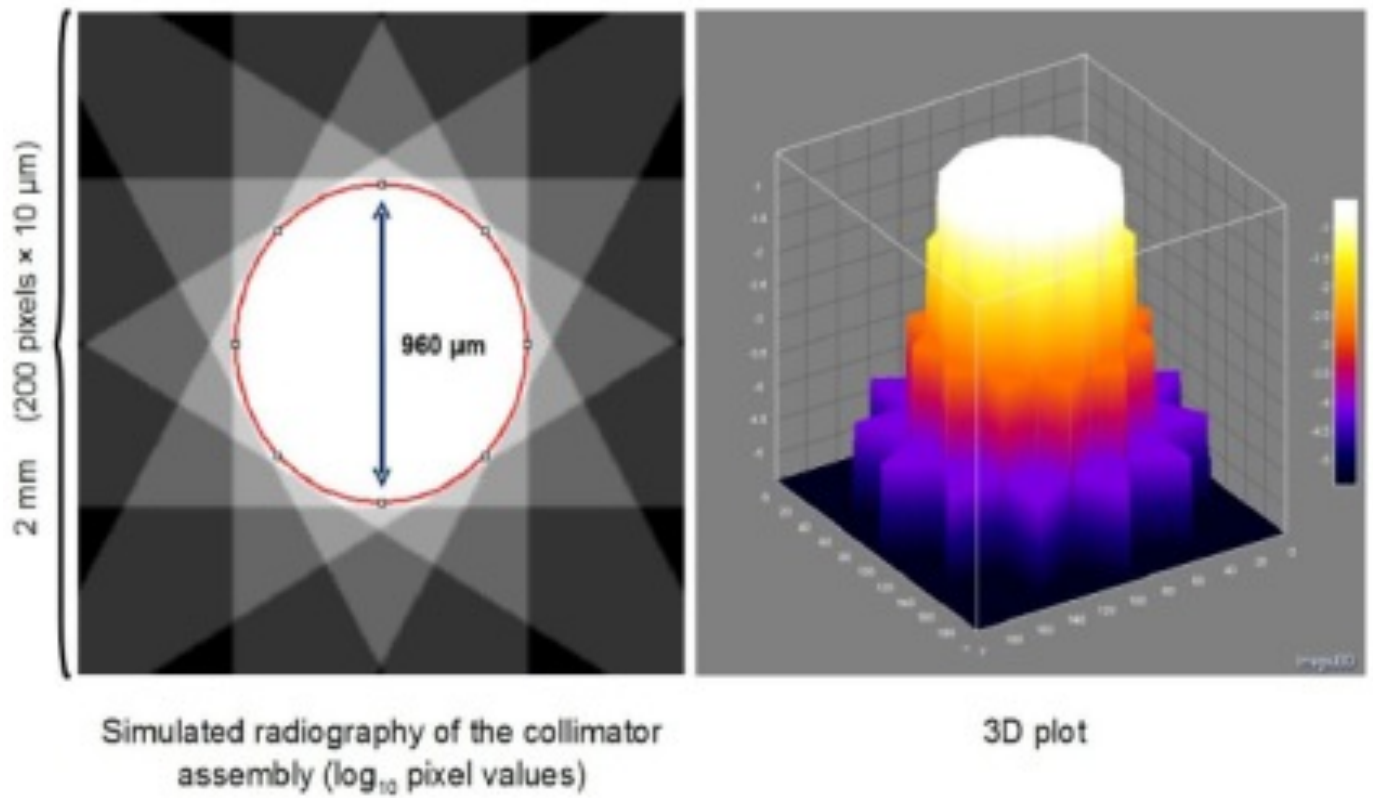

Fig. 8.4 Results of MC simulation for a 4.6 MeV gamma beam

For the whole gamma beam energy range the system setup has to be optimized by means of different arrangements of the type and number of the slits and their relative orientation. Furthermore to provide an adequate collimation system stability each one of the slits has to be equipped with mechanical adjustment structures able to set and optimize distance and orientation between the frames with the needed precision.

Two kinds of backgrounds are also expected downstream the collimation system: one from the photon beam itself and the other one from the photon beam losses along the collimation system. After the Compton scattering some of the photons at the IP can have a significant divergence (> 10 mrad) and crossing the beam pipe without interacting the can reach the downstream experimental zone. On the other hand most of the photons will be stopped by the collimation system and will generate secondary particles such as gamma and neutron. Both sources of particles should be stopped before reaching the experimental zone area by means of proper concrete block arrangement.

Finally in order to assess the machine performance and monitor its operation the beam brilliance has to be measured reconstructing its intensity and energy spectrum. The required resolution on the energy measurement should not be worse than the expected bandwidth of 0.2 ÷ 0.3% in the whole photon energy range. The capability of assessing the time dependence of the beam parameters within a macro-pulse, other than their long-term stability, is also a desirable property of the system that would allow providing feedback on the tuning of the laser recirculation system and of the interaction region. To meet these requirements for the intensity monitoring a luminometer can be used based on thin detectors exposed to the beam, providing the integrated beam flux on two different time scales: fast (ns scale) to resolve single pulses and slow (ms scale) to monitor the integrated macro-pulse flux.

The space profile of the beam can be monitored with high resolution and minimal beam attenuation using thin scintillator screens readout by CCD cameras.

For measuring the gamma energy, the main difficulty comes from the very short length of the intense pulses, preventing to easily disentangle the response of single photons to any detector directly exposed to the beam line. The energy of the gamma-beam can be evaluated indirectly, by measuring the energy and position of electrons and scattered photons produced by Compton interaction in a suitable micrometric target. The advantage of this technique is the minimal interference with the beam operation, making it an ideal tool for beam energy monitoring. From a sufficiently large amount of such measurements, the shape of energy spectrum can be precisely determined. The optimization of this method for very high energies is anyway under study.

### 8.4. A possible application: γ-induced positron spectroscopy (GiPS)

Classical slow positron spectroscopy techniques use implantation of low energy positron (up to few tens of keV) inside the target and the collection of the annihilation radiation. A new technique uses the sample material itself to produce positron by the irradiation of gamma rays. The positron can be used for annihilation spectroscopy. Positron Annihilation Spectroscopy is a unique and powerful tool to study crystal lattice defects like vacancies, dislocations or cluster and open-volume cavities at the nanometers scale. It is a non-destructive method with negligible impact on the sample material [8.15].
The production of positrons inside the sample volume by means of high-energy photons offers the unique possibility of investigating the entire sample volume (several $cm^3$). Annihilation radiation from the sample can be detected using time and energy sensitive detectors in order to realize standard positron techniques. Positron generation inside the sample takes place within the short time of the g pulse so the accelerator signal for electron bunch generation is used as sharp start signal for positron lifetime measurements. Positron lifetime spectroscopy is realized by measuring the time difference between the accelerator bunch pulse and the annihilation quanta detected by time sensitive detectors. This is just realizable because of the short bunch lengths that are negligible compared to estimated positron lifetimes (≈100 ps to ns range).
A multi detector system can register annihilation photons emitted from the sample. High-Purity Germanium detectors can be used to measure the energy of gamma radiation; time sensitive Barium Fluoride detectors can measuring the time of photon detection in order to realize positron lifetime measurements (the typical timing resolution is 180 ps FWHM). HPGe detectors feature a relative efficiency of 100% (IEEE 325-1996 standard), typical energy resolutions of around 2.4 keV FWHM (at 1.33 MeV), and they can be equipped with escape-suppression shields (ESS) made from bismuth germanate (BGO) scintillation detectors. These shields efficiently reduce the background in energy spectra stemming from incompletely absorbed photons inside the HPGe (e.g. pair production or Compton scattering events).
The setup allows for Coincidence Doppler Broadening Spectroscopy (CDBS) using two HPGe detectors placed face-to-face; an HPGe detector paired with a $BaF_2$ detector in coincidence works as an independent spectrometer for Age–Momentum Correlation (AMOC). In this coincidence setup, lifetime and Doppler spectra are recorded in parallel enabling to construct time-correlated energy spectra.

# 9. PARTICLE PHYSICS OPPORTUNITIES WITH IRIDE

## 9.1. Overview

It is commonly accepted that the Standard Model (SM) of elementary particles interactions is the model, which describes the visible part of the nature and of the Universe. Recently the experimental results from the Large Hadron Collider at CERN have provided us with very important information on the mass of the SM higgs-like particle. However, the existence of this particle with a given mass does not solve, by itself, all the long-standing puzzles of the SM, such as a problem of the SM hierarchy [9.1], the naturalness of the higgs boson and the electroweak (EW) symmetry breaking. Even though all the SM parameters are now measured to a high accuracy, the necessity of the New Physics (NP) existence for explaining the SM puzzles is still an open question. From a theoretical point of view, precise and complicated calculations are required to answer these questions, and high-precision input information on the SM parameters is a must. Due to the intrinsic complexity of the calculations, as one needs to study the running of the non-abelian gauge theory parameters over a dozen of orders of magnitude up to the Planck scale, even small experimental uncertainties in the SM parameters have a drastic impact on the conclusions, which can be drawn from such computations. The implications affect our understanding of the fundamental issues of the "conspiracy" between the SM couplings, the EW phase transition, Universe inflation, the cosmological constant, and also the nature of the Dark Matter (DM).

*It is important to stress that the precise values of the SM parameters, due to the renormalization group evolution, can be obtained only by simultaneous studies at high-energy and low-energy scales.* The former point highly motivates the International Linear Collider (ILC) initiative, while the IRIDE project can pursue the latter one and serve as an accelerator-technology test installation and a research facility. The latter point motivates the possible use of the IRIDE facility as a precision tool for the SM exploration at low- and medium-energy scales, with a high priority on the information about the EW couplings of SM, which drives the evolution of the electromagnetic running coupling and the squared sine of the weak angle. Also a rich hadron phenomenology is accessible at these scales, which allows to study issues of the QCD confinement, where the ordinary perturbation theory approaches fail to work. While the technological aspects are discussed in another Chapter of this White Book, the present Chapter deals with an overview of the fundamental particle physics opportunities of the IRIDE project.

It is anticipated that the construction of the IRIDE facility will be realized step-by-step. We review the particle physics goals of the full accelerator complex according to the order in which one can launch the various steps of the facility. We start with the physics program that can be pursued with an electron beam on target, further we investigate that of the electron-photon collider, of photon-photon collisions and finally of the electron-positron and electron-electron collider. It is important to stress that a synergy of all the proposed measurements can lead to a very reliable and cross-checked experimental exploration of the SM.

The **electron-on-target** physics program makes IRIDE a discovery and also a precision physics machine. Among the searched candidates there are the hypothetical particles, like the very-weakly interacting massive U(1) gauge boson (U-boson) as a DM particle candidate and the non-hypothetical, well investigated theoretically, but yet undiscovered, "true muonium" states (TM), which are the bound states of muon and anti-muon with the lifetime of an order of a picosecond. Utilizing the polarized electron beam dumped onto the proton target, one can measure the left-right parity violating asymmetry of electron-proton scattering at the per cent level, and thereby extract precisely the electroweak mixing angle.

The **electron-photon collider** allows to utilize the elementary Primakoff process to produce the light pseudoscalar (and scalar) mesons in order to precisely measure their two-photon decay widths and thus to tackle the triangle anomaly of QCD. In addition, one can perform the U-boson search in the lepton triplet production channel. A special feature here is the availability of the highly-polarized photon beam. This allows to use the lepton triplet production at IRIDE as a research laboratory for development of the methods of polarimetry to be used in astrophysics to measure the polarization directions of incoming high energy γ-rays. Finally, triple Compton effect can be used to study the properties of entangled states (see Ref. [9.2]). *These measurements, which provide important tests of the SM, are not possible at present electron-photon colliders due to the low photon intensities of the machines.*

Low-energy **photon-photon collisions** give a direct view into the vacuum properties of Quantum Electrodynamics (QED), allowing for precision tests of QED in the MeV range, and more generally of Quantum Field Theory (QFT). *The IRIDE accelerator complex can generate for the first time colliding photon-photon beams by Compton backscattering, and this opens the fascinating field of low-energy photon-photon physics.* The technology needed to carry out a photon-photon physics program at energies close to 1 MeV would disclose new developments at higher energies, where a photon-photon Higgs factory could be a nearly ideal discovery machine.

The high-luminosity **electron-positron** and **electron-electron collider** with variable energy would be an extremely useful tool for the study of hadronic vacuum polarisation effects, measurements of the effective electroweak mixing angle and contributing to the description of the muon anomalous magnetic moment and the running QED coupling constant by providing the hadronic cross sections with high accuracy. In addition, these measurements can contribute to the extraction of the light quark masses, flavour symmetry breaking pattern in the light meson sector and allow to study precisely the meson mixing phenomenology through the various meson decays produced with high statistics in lepton collisions. The gamma-gamma fusion sub-processes in the positron/electron-electron inelastic scattering gives us the opportunity to investigate the two-photon couplings and form-factors of the various hadronic resonances (and also the many-particles states, like $\pi^+\pi^-$ or $\pi^0\pi^0$), which is important for the understanding of the quark contents of these resonances, of hadron phenomenology and for improvement in the estimate of the hadronic light-by-light scattering contribution to the anomalous magnetic moment of the muon. *The LHC, or a future $e^+e^-$ International Linear Collider (ILC), will answer already many questions. However, their discovery potential may be substantially improved if combined with more precise low energy tests of the SM. In this framework an electron-positron collider such as IRIDE with luminosity of $10^{32}$ $cm^{-2}s^{-1}$* *with centre of mass energy ranging from the mass of the ϕ-resonance (1 GeV) up to ~ 3.0 GeV, would complement high-energy experiments at the LHC and a future linear collider (ILC).* The direct competing project is VEPP-2000 at Novosibirsk which will cover the center-of-mass energy range between 1 and 2 GeV with two experiments. This collider has started first operations in 2009 and is expected to provide a luminosity ranging between $10^{31}$ $cm^{-2}$ $s^{-1}$ at 1 GeV and $10^{32}$ $cm^{-2}$ $s^{-1}$ at 2 GeV. Other "indirect" competitors are the higher energy $e^+ e^-$ colliders (τ-charm and B-factories) that can cover the low energy region of interest by means of radiative return (ISR). However, due to the photon emission the "equivalent" luminosity produced by these machines in the region between 1 and 3 GeV is much less than the one expected in the collider discussed here.

## 9.2. Electron beam-on-target experiments

### 9.2.1. *Electron beam dump experiment: dark forces*

Hidden photons (γ') with masses in the MeV to GeV range are being searched for by many experiments in the world, both at colliders and at fixed target facilities. In particular, electron beam dump experiments are particularly well suited for probing low masses ($m_{\gamma'}$) and very low kinetic mixing values ε. In this kind of experiment, hidden photons are produced by directing an intense electron beam on a high-Z material target, in a process similar to ordinary bremsstrahlung, but with a cross section suppressed with respect to the latter by a factor $\varepsilon^2$. On the other hand, since the lifetime of the hidden particle is inversely proportional to the fourth power of the mixing parameter, for very small ε it can become relatively long, allowing the detection of its decay products much downstream the dump, where all of the "ordinary" physics signals are absorbed.

Results (derived by old experiments) have been obtained by the authors of references [9.3 – 9.5], where the potentials and limitations of this type of technique applied to possible future facilities are also discussed. Figure 9.1, taken from reference [9.5] shows the limits obtained so far by the aforementioned experiments in the plan $\varepsilon$-$m_{\gamma'}$ (note that the mixing parameter is there dubbed as χ), together with those obtained by detectors at colliders or by fixed target experiments on a thin target. As previously mentioned, beam dump results cover the leftmost part of the plot, while the others reach much lower sensitivity of the coupling while covering a much larger interval of the mass. Among the beam dump experiments, in particular, SLAC E137 was able to achieve limits down to $\varepsilon \sim 2\text{-}3 \times 10^{-8}$, thanks essentially to the big amount of dumped current, ~ 30 C.

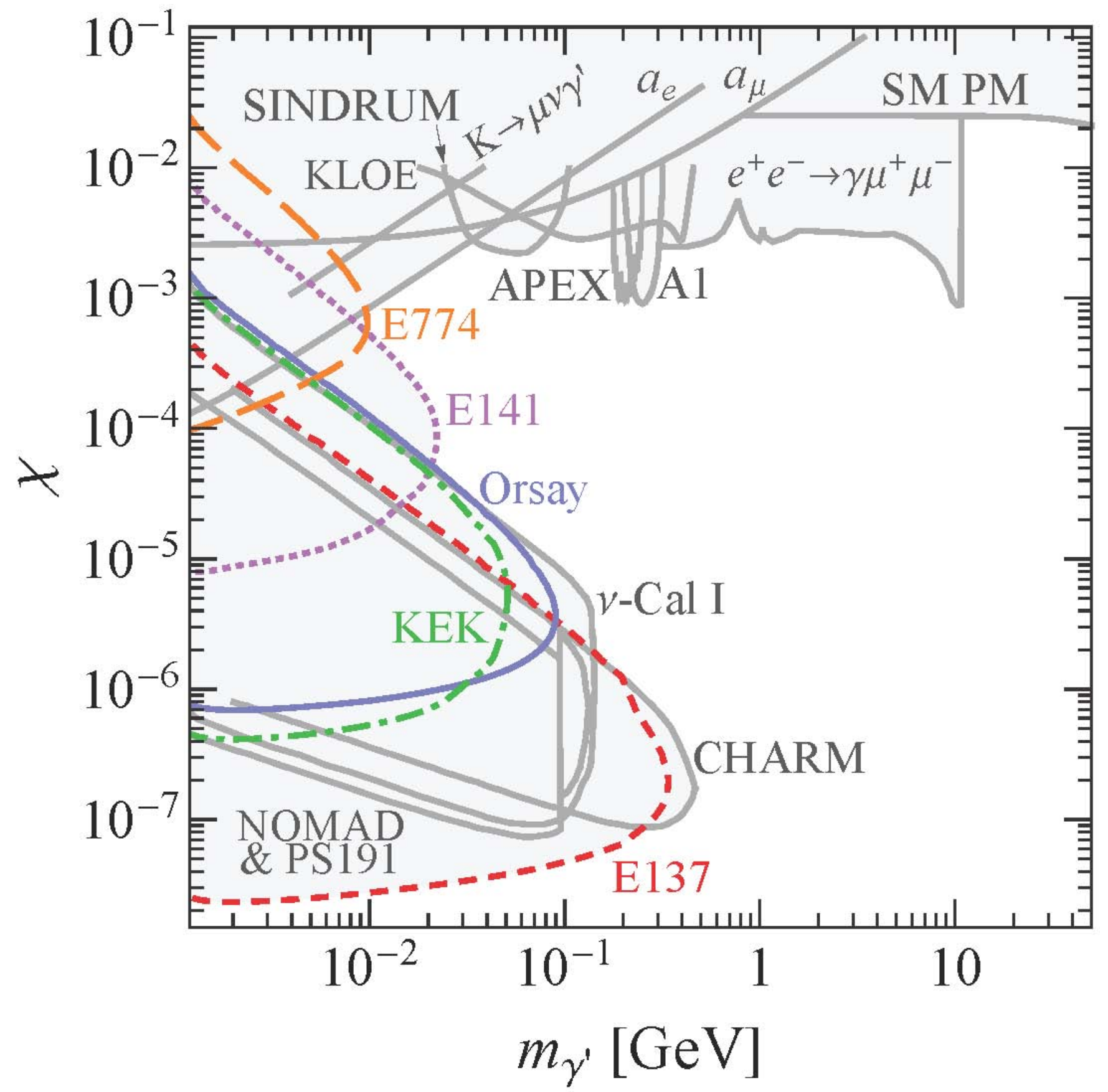

**Figure 9.1 -** Limits on the hidden photon mass and kinetic mixing parameter ($\chi$ in the figure, $\varepsilon$ in the text) obtained by past and current experiments [9.5].

New fixed target activities are in a construction phase, both at Mainz and Jefferson Lab, and are expected to deliver their first results within a few years. All of them plan to perform experiments using thin targets, thus being sensitive to the region of $\varepsilon > 10^{-5}$ and $100 < m_{\gamma'} <\sim 500$ MeV.

It is conceivable to exploit the high intensity beam of IRIDE to perform a beam dump experiment. In principle this could allow one to be sensitive to very low effective couplings. Unfortunately, however, the sensitivity to ε scales only with the fourth power of the dumped current, therefore the improvement with respect to E137 would be only marginal, even operating with IRIDE at the highest possible currents. On the other hand, a relatively short dump, combined with the high beam intensity could allow one to probe the region of moderately long lifetimes. Using formula (3.14) of reference [9.5] one can see for instance that for $\varepsilon \sim 10^{-6}$ and $m_{\gamma'} \sim 200$ MeV, at IRIDE energies the mean decay path of a dark photon is of ~ 3 m. Note that the energy of the IRIDE beam is comparable to the one used for the Orsay experiment, but in the latter the dumped current was only ~ 3 mC. Moreover in the Orsay experiment the dump was about 1 m long, a value which probably can be consistently decreased. The usage of a short dump poses the problem of the containment of the backgrounds induced by ordinary e.m. processes, as well as technical problems connected to heat dissipation and radiation control. Still this solution looks very promising, allowing one to consistently improve with respect to present limits in a region of the parameter space not covered (or only partially covered) by planned experiments. More detailed simulation studies are planned for the near future.

### 9.2.2. *Elastic electron scattering: proton weak charge*

In the SM, the weak charge of the proton, $Q^p_W$, at low momentum transfer, $Q^2$, is related to the sine of the weak mixing-angle by the relation [9.6, 9.7]:

$$Q_W^p = 1 - \sin^2\theta_W \approx 0.0716\,.$$

While for the neutron this quantity is -1, the proton coupling to the weak-force mediator is suppressed, making the measurement of $Q^p_W$ sensitive to contributions from new physics. The SM connects the value of $Q^p_W$ measured at low $Q^2$ with the one measured at the Z pole by running the weak mixing-angle from $M_Z$ down to low energies. Combining the uncertainty contribution of the extrapolation, 0.00028, with that of the Z-pole best fit, 0.0005, the total error is 0.00057 (0.8% relative error).

New physics such as new neutral heavy bosons (called Z') could affect Z-pole and low-energy measurements in a different way [9.8]. The former are mostly affected by Z-Z' mixing, which lowers the mass of the Z relative to the SM prediction and modifies the Zff vertices. The latter are affected also by Z' exchange. Direct searches at LHC with 20 $fb^{-1}$ already exclude such bosons with masses below 2.5 TeV and a factor two improvement is expected in the first months of 2015 when the machine will run at a higher center of mass energy. Contributions from other models such as supersymmetry, dark photons etc., may as well be tested [9.6].

Precision measurement of the proton weak-charge at low $Q^2$ are performed by measuring the cross-section asymmetry of longitudinally polarized electrons scattering from the proton:

$$A = \frac{1}{P}\frac{\sigma^+ - \sigma^-}{\sigma^+ + \sigma^-}$$

where $P$ is the longitudinal polarization of the electron beam. For small scattering angles this relation reduces to:

$$A = -\frac{1}{P}\frac{G_F Q^2}{4\sqrt{2}\pi\alpha}\left[Q_W^p + Q^2 B\left(Q^2\right)\right]$$

where $G_F$ is the Fermi coupling and $\alpha$ the fine structure constant. $B(Q^2)$ is the contribution coming from the finite size of the nucleon. For $Q^2 \sim 0.03\ \mathrm{GeV}^2$ it contributes to about 30% of the asymmetry. Its value is obtained by extrapolating existing elastic electron-proton parity-violating asymmetry data at higher $Q^2$. Further relevant corrections arise from box diagrams involving exchange of a photon and a Z boson.

The Qweak Experiment at JLab [9.9,9.10] collected, during 2011 and 2012, 2,500 hours of collisions of a 150 μA electron-beam with an 85% longitudinal polarization incident with energy of 1.16 GeV on a $LH_2$ target. They are now analyzing data and aim at a relative precision on the proton weak charge of 4%, corresponding to a parity-violating new physics at a scale of 2.3 TeV. At these $Q^2$ and energy, the hadronic and box corrections contribute with a 1.5% to the final error. Beam polarization is measured at percent level both with Compton and Moller scattering. In the latter case a polarized electron target is used.

The P2 Project [9.11] proposes a new measurement at the MESA accelerator in Mainz, reaching 2% relative precision on $Q_W^P$. They plan to operate at a lower $Q^2$, about 0.003 $GeV^2$, in order to reduce the hadronic contributions to a negligible level. Moreover, choosing a beam energy $E$ = 200 MeV the γ-Z box correction to $Q_W^P$ is also strongly suppressed. The proposed experimental conditions foresee a 150 μA electron-beam with an 85% longitudinal polarization incident with an energy of 200 MeV on a $LH_2$ target almost a factor 2 thicker (60 cm) than the one used at JLab. They plan to take data for 10,000 hours. Beam polarization is expected to be measured at 0.5% level. The MESA commissioning is expected by the end of 2017.

Given the difficulty of such measurement, the comparison of results from experiments with different systematics is mandatory. A future experiment at IRIDE should take into account the contribution of theoretical and experimental errors and choose the best working point in term of beam energy and $Q^2$. For instance, MESA is limited to 200 MeV while IRIDE could choose a higher energy. While this choice would increase the contribution of hadronic and box corrections, below 300 MeV polarization measurement with Compton scattering is difficult if not impossible [9.12].

### 9.3. Electron-photon collider option

A high brightness linac like IRIDE could be used, in a stand alone operational mode, to drive a Compton γ–ray source. In this way γ–rays with energies of 1-100 MeV can be produced and directed head-on against electrons of 100-1000 MeV. A reach physics program can be studied [9.13], which includes –among others- the precise measurement of the $\pi^0$ width through the process $e\gamma \rightarrow \pi^0 e$ (*Primakoff effect*), and the search for light dark bosons in the energy region of a few to hundreds MeV. These measurements, which provide important tests of the SM, are not possible at present electron-photon colliders due to the low photon intensities of the machines.

#### 9.3.1. *$\pi^0$ width measurement*

The axial anomaly of Adler, Bell and Jackiw (non-conservation of the axial vector current) is responsible for the decay of the neutral pion into two photons. It bridges the strong dynamics of infrared physics at low energies (pions) with the perturbative description in terms of quarks and gluons at high energies. The anomaly allows to gain insights into the strong interaction dynamics of QCD and has received great attention from theorists over many years. Due to the recent advances, the $\pi^0$ decay width is now predicted with a 1.4 % accuracy [9.14]. The major experimental information on this decay comes from the photo-production of pions on a nuclear target via the Primakoff effect [9.15]. The PrimEx Collaboration, using a Primakoff effect experiment at JLab, has recently achieved a 2.8 % precision [9.16]. This improved the PDG average for the width and mean lifetime of the pion. Nevertheless, the uncertainty of the width average is still inflated (a scale factor 1.2 at the moment), which gives an additional motivation for a new precise measurement of the pion lifetime and the two-photon decay width. The Primakoff effect-based experiments on a nucleus target suffer from model dependence due to the contamination by the coherent and incoherent conversions in the strong field of a nucleus [9.17]. Therefore, a measurement using a different method is highly desirable.

The electron-photon collider option is in a given sense similar to the traditional Primakoff effect setup, but provides a much cleaner environment from the theory and data analysis perspective. The proposed measurement of the pion width will have impact on the evaluation of the SM prediction of the anomalous magnetic moment of the muon, $a_\mu$.

The precision of the theoretical value of $a_\mu$ is currently limited by uncertainties of the hadronic vacuum polarization and the hadronic light-by-light (HLbL) scattering contribution. The value of the latter is currently obtained using hadronic models and leads to an uncertainty in $a_\mu$ which is almost as large as the one from hadronic vacuum polarization. In view of the proposed new g-2 experiments at Fermilab and JPARC with a high precision, the HLbL contribution needs to be controlled much better, in order to fully profit from these new experiments, to test the Standard Model and to constrain New Physics. According to model calculations, the exchange of neutral pions yields the dominant contribution to HLbL scattering. The evaluation of the pion-exchange contribution has to use the normalization value (the pion decay constant) $f_\pi$. This is the same quantity, which fixes the two-photon decay width of the neutral pion. Therefore, the precision of the $\pi^0$ width controls directly the precision of the HLbL contribution to the muon *g-2*.

At IRIDE we propose to use electron-photon collisions as a source of $\pi^0$ mesons produced via the Primakoff effect having the electron as a "target" instead of a nucleus, see Fig. 9.2.

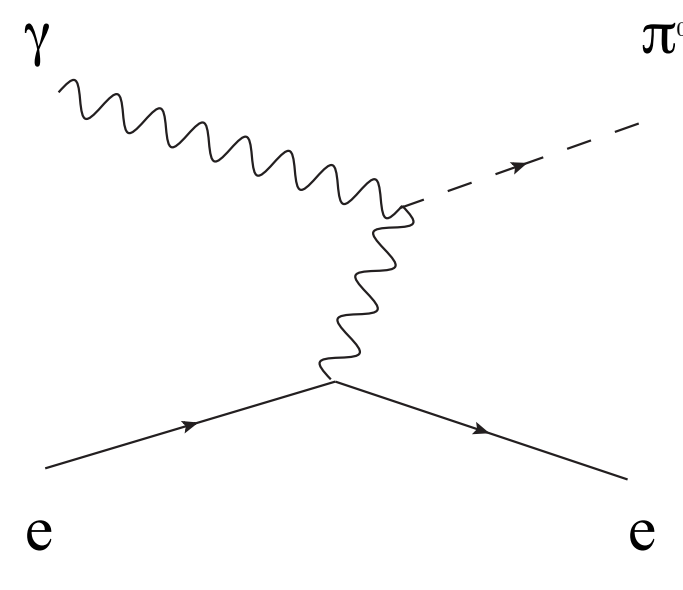

**Figure 9.2** - $e^- \gamma \rightarrow \pi^0 e^-$ diagram (*Primakoff effect*).

On the one hand, in this way one looses the $Z^2$ factor enhancement of the event rate; and thus a high intensity of the photon and electron beams is required. On the other hand, the photon-electron collision is a much cleaner environment as compared to the photon-nucleus case. A Monte Carlo generator for the $\pi^0$ production has already been developed and we have started the simulation studies, taking into account the possible beam spread (at a level of 0.1% for the electron and 10% for the photon beam) and looking for suitable event selection criteria. Our preliminary calculation shows that by colliding photons of 20 MeV photons against 750 MeV electrons a cross section of 1-2 nb for the process $e\gamma \rightarrow \pi^0\gamma$ can be expected. Our preliminary calculation shows that by colliding photons of 20 MeV photons against 750 MeV electrons a cross section of 1-2 nb for the process $e\gamma \rightarrow \pi^0\gamma$ can be expected. This means that with a Luminosity of $10^{30}$ $cm^{-2}$ $sec^{-1}$ for the $e\gamma$ collider (well within reach of this proposal) a measurement of 1% is possible. This measurement would be better than the current experimental world average and the theory prediction, and competitive with the planned measurement at KLOE-2, using two-photons channel [9.18].

### 9.3.2. *$\eta$ and $\eta$' width measurement*

There are many theoretical and experimental motivations for a precise measurement of $\eta$ and $\eta$' widths:

- *The role of the $\eta$ and $\eta'$ widths to constrain the hadronic light-by-light scattering contribution to the anomalous magnetic moment of the muon*

The anomalous magnetic moment of the muon is one of the most accurately measured quantities in particle physics. Any deviation from its prediction in the Standard Model of particle physics is a very promising signal of new physics. The present world average experimental value of its deviation from the Dirac value, i.e., $a_\mu = (g_\mu-2)/2$, is given by $a_\mu$ = 116 592 08.9(6.3) x $10^{-10}$, Ref. [9.19]. This impressive result is still limited by statistical errors, and a proposal to measure $a_\mu$ to a precision of 1.6 x $10^{-10}$ has recently been submitted to FNAL [9.20].

At the level of the experimental accuracy, the QED contributions to $a_\mu$ from photons and leptons alone are very well known. Recently the calculation has been completed up to the fifth order of the fine-structure constant $\alpha_{em}$, giving the result 11658471.885(4) x $10^{-10}$ [9.21].

The main uncertainties at present in the Standard Model calculation for $a_\mu$ originate from the hadronic vacuum polarization (HVP) as well as from the hadronic light-by-light scattering (HLBL) (see also the discussion in the $e^+e^-$ section). We show the present estimates and their uncertainties for the QED, HVP, HLBL, and the electroweak (EW) corrections in Table 9.1 below. The existing discrepancy between the experimental value for $a_\mu$ and its Standard Model prediction stands at about $3\sigma$.

In order to interpret the upcoming new experiment at FNAL, with an anticipated precision of 1.6 x $10^{-10}$, there is an urgent need to improve both on the HVP as well as the HLBL contributions. The accuracy of the HVP contribution depends on the statistical error of the experimental data for the $e^+e^-$ annihilation cross-section into hadrons. With future experiments, in particular at BES-III [9.26], and the possibility to measure $e^+e^- \rightarrow e^+e^- \gamma^*\gamma^* \rightarrow e^+e^-X$ (see photon-photon physics section) one foresees this error to quantitatively decrease. The HLBL, $a_\mu^{HLBL}$ cannot be directly related to any measurable cross section however, and requires the knowledge of nonperturbative QCD contributions at all energy scales. Since this is not known yet, one needs to rely on hadronic models. Such models introduce systematic errors which are difficult to quantify. Using large $N_c$ [9.27,9.28] and also the chiral perturbation calculations, it was proposed in [9.29] to split the diagram of Fig. 9.3 into a set of different contributions where the numerically dominant contribution arises from the pseudo-scalar exchange diagram shown in Fig. 9.4 [9.30].

**Table 9.1**

| Contribution | Result in $10^{-10}$ units | Ref. |
|---|---|---|
| QED (leptons) | 11658471.885±0.004 | [9.21] |
| HVP (leading order) | 692.3±4.2 | [9.22] |
| HVP (higher order) | -9.84±0.07 | [9.23] |
| HLBL | 11.6±4.0 | [9.24] |
| EW | 15.4±0.2 | [9.25] |
| **Total** | 11659181.3±5.8 | |

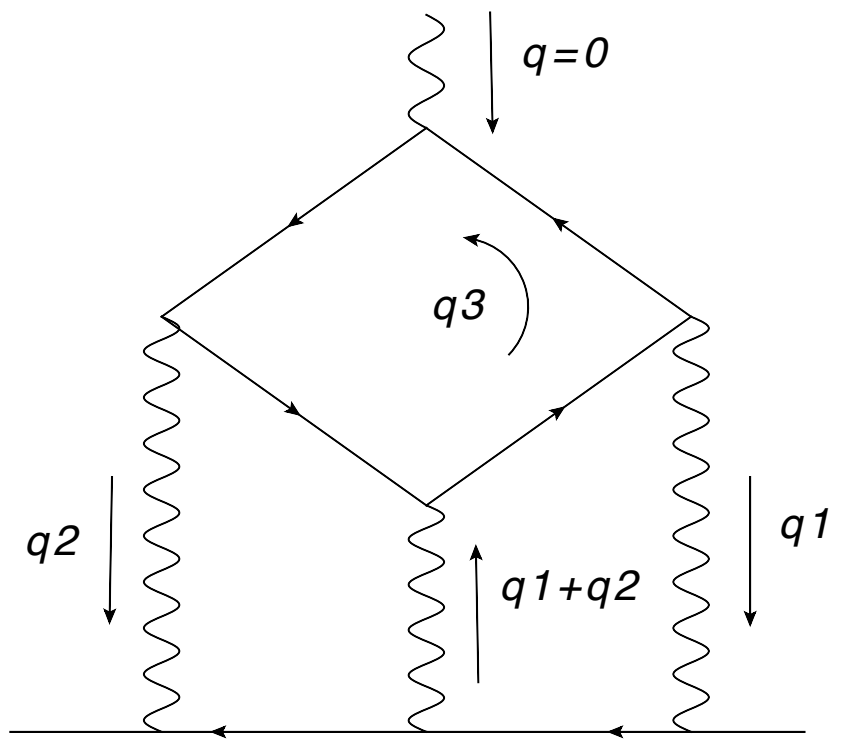

**Figure 9.3 -** Quark loop diagram with running quark mass responsible for the $a_\mu^{HLBL}$ contribution.

The large $N_c$ approach however has two shortcomings: firstly, the assumption of pion-exchange dominance implies that the remaining pieces are small enough to justify their omission. Although this seems reasonable [9.30], it might lead to an underestimation of the error. Secondly, calculations carried out in the large $N_c$ limit demand an infinite set of resonances for computing any quantity. As such sum is not known in practice, one ends up truncating the spectral function in a resonance saturation scheme, the so-called

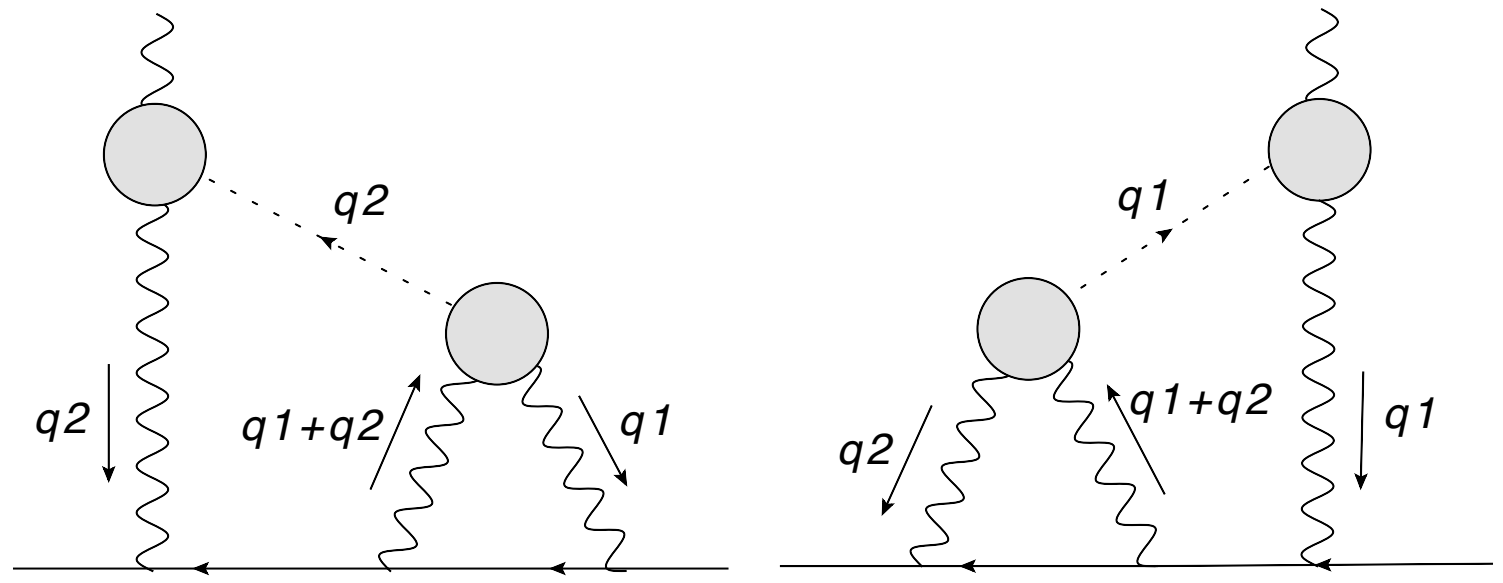

**Figure 9.4 -** Pseudoscalar-exchange contribution to $a_\mu^{HLBL}$ scattering. The shaded blobs represent the form factor $F_{P^*\gamma^*\gamma^*}(p^2,q_1^2,q_2^2)$.

Minimal Hadronic Approximation (MHA) [9.31]. The resonance masses used in each calculation are then taken as the physical ones from PDG [9.32] instead of the corresponding masses in the large $N_c$ limit. Both problems might lead to large systematic errors not included so far [9.33,9.34]. It was pointed out in Ref. [9.33] that, in the large $N_c$ framework, the MHA can be understood in the framework of Padé Approximants (PA) to meromorphic functions. Obeying the rules from this method, one can compute the desired quantities in a model-independent way and is even be able to ascribe a systematic error to the approach [9.35].

To account for the $a_\mu^{HLBL}$ contribution, one should use, instead of a hadronic model for the transition Form Factors (FF)-blobs in Fig.9.4, a sequence of Padé Approximants [9.33] in two variables (called Chrisholm Approximants) build up from the low-energy expansion of the pion FF obtained in [9.40] after a fit to the available space-like experimental data, to minimize a model dependence. The FF should be considered to be off-shell (see Ref. [9.24] where this point is addressed). Its influence, which nevertheless needs further investigation, was found to be rather small [9.24]. A description of the whole procedure can be found in Ref. [9.36]. One, then has to deal with the $F_{P\gamma^*\gamma^*}(q_1^2,q_2^2)$. Since the FF is not known, and only data for $F_{P\gamma\gamma^*}(q_1^2,0)$ is available, thanks to measurements using the single-tag method, one then should construct a hadronic model and constrain it using the low-energy parameters obtained from that data. We notice in Ref. [9.36] and [9.37] the important role of precise measurements of such low-energy parameters in order to correctly constraint hadronic models. The impact of such precise measurements is two-fold, both with clear relevance for IRIDE.

First, all the hadronic models should be normalized as the FF, i.e., all the models should correctly reproduce the decay of the pseudoscalars to two photons (which is the particular kinematic point in the complex plane of the FF with two photons off-shell that coincides with the one when both photons are real, on-shell $F_{p\gamma\gamma}(0,0)$). This measurement, then, yields, the normalization of the integrals involved in the calculation of the pseudoscalar contribution to $a_\mu^{HLB}$. We notice that the contribution of the lightest pseudoscalars account for $a_\mu^{HLBL} \sim 10 \times 10^{-10}$ (number that might oscillate depending on the particular model used, see Ref. [9.24]), with $a_\mu^{HLBL} \sim 7 \times 10^{-10}$ coming from the $\pi^0$ contribution and $a_\mu^{HLBL} \sim 1.5 \times 10^{-10}$ from η and η' each. Notice, on top, that even though the η' meson has a mass which is almost the double of the η meson, the contribution to $a_\mu^{HLBL}$ is similar, due to the different decays of these mesons into two photons. Recalling that the forthcoming experiment at FermiLab should reach a precision of $1.6 \times 10^{-10}$, the correct estimation of both η and η' is crucial to understand the current discrepancy between SM and experiment.

Second, and to our understanding even more important, is the synergy between the experiment and theory proposed in Ref. [9.34] where it is shown that a precise determination of the η and η' width could also provide a precise and model-independent extraction of the low-energy parameters governing the FF entering the diagrams of Fig. 9.4. With such synergy, one can build up a model-independent approximation to the FF (without relaying on a particular model) and account for the $a_\mu^{HLBL}$.

Precise measurements, feasible at IRIDE, of both η and η' meson decay widths (or just decays into two photons), are crucial to compete with the forthcoming experimental precision on the $a_\mu$ and reduce the SM prediction on one of its weaker points, the HLBL.

- *Physics of the η- η′ system and symmetry structures of QCD at low energies*

The question of η- η′ mixing, i.e., how their wave functions are composed of $SU(3)_{fl}$ singlet and octet quark pair components is a long standing problem, going back to the beginning of the quark model era. With the advent of QCD it became even more intriguing, due to the possibility of this system to mix with a purely gluonic component. Since in general all components compatible with the quantum numbers of a state can appear in that state's wave function, there is no *a priori* reason why the η and η′ wave functions could not contain such configurations. This mixing has been actively searched for, since it would establish for the first time that gluons play an independent role also in hadronic spectroscopy.

Information on the mixing angle has been looked for in a variety of processes, mostly radiative vector and pseudoscalar meson decays and decays into two photons or production in γγ collisions. At present, the different determinations of the η—η′ mixing are generally consistent, but the message concerning the gluon content remains ambivalent. By analyzing the latter in several experiments, it is possible to show that one major contributions to the mixing angle comes from the uncertainty of the total width $\Gamma_{\eta'}$ [9.38,9.39]. The total width $\Gamma_{\eta'}$ extracted by PDG and the value of the partial width $\Gamma(\eta' \to \gamma\gamma)$ are strongly correlated, which may create difficulties when the total and the partial width are used at the same time, as in the case of the mixing angle extraction. It goes without saying how important is to refine the measurement of the partial width $\Gamma(\eta' \to \gamma\gamma)$.

The system of $\pi^0$, η, η' mesons represent a laboratory to study symmetry structure of QCD at low energies. The importance to improve the knowledge of mesons parameters comes from the limits they induce on many important investigations. We start from the impact on η–η' mixing. A not exhaustive list of other items, which will profit from the improvements on η and η' mesons, are the investigation of quark mass differences, isospin breaking in QCD, and Chiral Anomalies, in the partial widths $\Gamma(\pi^0/\eta/\eta' \to \gamma\gamma)$ and $\Gamma(\eta/\eta' \to \pi\pi\gamma)$.

The decay $\eta \to 3\pi$ violates isospin invariance. Electromagnetic contributions to the process are very small [9.38] and the decay is induced dominantly by the strong interaction via the u, d mass difference. The transition is therefore very sensitive to the quark mass difference. KLOE has studied the slopes of Dalitz plot [9.40], which represent an important information on the decay amplitude. To provide important constraints for the light quark mass difference we need to improve the accuracy of the width, from which the $\Gamma(\eta \rightarrow 3\pi)$ is evaluated. The decay widths of experiment and $\chi$PT prediction presently disagree: $\Gamma_{i(LO)} = 40$ eV, $\Gamma_{i(NLO)} = 160 \pm 50$ eV, while $\Gamma_{i(EXP)} = 295 \pm 16$ eV.

The Chiral Anomaly play an important role in the $\chi$PT: the Wess–Zumino–Witten (WZW) term in the ChPT Lagrangian accounts for anomalous decays involving an odd number of pseudoscalar mesons. In particular the triangle anomaly is responsible for the two-photon decays of the $\pi^0/\eta/\eta'$ mesons and box anomalies should contribute to the $\eta/\eta' \rightarrow \pi^+\pi^-\gamma$ decays. Two experiments have investigated $\eta-\eta'$ mixing: KLOE/KLOE-2 and WASA @ COSY. These two setups work in different ways: while KLOE studies $\eta$ and $\eta'$ decays, looking at the $\phi$ radiative decays $\phi \rightarrow \eta\gamma/\eta'\gamma$, at COSY the $\eta$ and $\eta'$ mesons are produced in pp → ppX and pd → $^3$He X decays. The total experimental uncertainties on the measurement of $\eta$ and $\eta'$ full widths are 5% [9.41] and 10% [9.42], respectively. Another experiment focused on pseudoscalar mesons is Primex/GlueX at Jlab. The $\gamma-e$ collider at IRIDE represents a complementary setup to all the existing ones, because it is based on the Primakoff effect without a nucleus, then there is no contamination due to the strong field of a nucleus. In addition it would be possible to measure the full width with ~ 1% accuracy which would be a dramatic improvement on the existing uncertainties.

PDG shows no direct measurement of $\Gamma_\eta$, while $\eta'$ has been recently measured at COSY-11 facility [9.42]. The $\eta$ width is determined from the partial decay $\Gamma(\eta \rightarrow \gamma\gamma)$ divided by the fitted branching ratio for that mode [9.41]. In 1994, a note on decay width $\Gamma(\eta \rightarrow \gamma\gamma)$ by Roe [9.43] stressed that results from two-photon production disagree with those from Primakoff production and that subsequent two-photon measurements are consistent with previous one. In experiments based on the Primakoff effect $\eta$ mesons are produced by interaction of a real photon with a virtual photon in the Coulomb field of the nucleus: there is a coherent background from the strong production of $\eta$ mesons in the nuclear hadronic field, and interference between the strong and Primakoff production amplitudes. The main problem is due to the evaluation of this background. Experiments in 1974 [9.44] suggested that background terms should be taken into account and reanalyzing old data of Ref. [9.45] finds consistent results. Disagremeent between two-photon and Primakoff results still remain but in 1994 the Primakoff is still used. In the last PDG (2012) data from Primakoff effect disappear from the world average.

The experimental scenario for the $\eta'$ meson is different: a direct measurement from COSY-11 is combined with others to produce the PDG mean value, which is obtained from the fit of 51 measurement of partial widths, branching ratios and combinations of partial widths obtained from integrated cross sections. The results of the fit is strongly correlated with the value of the. In the next future Primex and Gluex at Jlab, will give new data still via Primakoff effects.

At IRIDE the $\eta$ and $\eta'$ mesons can be produced via Primakoff scattering $e\,\gamma \rightarrow e\,\eta/\eta'$ with directing a 2 GeV electron beam on 65 MeV $\gamma$ (for $\eta$) and/or a 2.5 GeV electron beam on 100 MeV $\gamma$ (for $\eta'$). A preliminary estimate of the cross section to 2 photon final states gives ~ 0.6 nb for $\eta$ and 0.04 nb for $\eta'$. In order to achieve a 1% error on the $\gamma\gamma$ width of $\eta$ and $\eta'$ an integrated luminosity of ~ 40 $pb^{-1}$ for $\eta$ and 500 $pb^{-1}$ for $\eta'$ is requested. This corresponds to 1 year of data taking with $L \sim 4 \times 10^{30}$ $cm^{-2}$ $sec^{-1}$ for $\eta$ and $L \sim 5 \times 10^{31}$ $cm^{-2}$ $sec^{-1}$ for $\eta'$. While for $\eta$ the required luminosity is within reach for IRIDE, for η' an improvement of about one order of magnitude in luminosity is expected. This upgrade in luminosity is certainly not easy but well motivated.

### 9.3.3. *$f_0$ width measurement*

IRIDE could also contribute to exotic spectroscopy, and precisely to measurements related to $f_0(980)$, an isospin singlet with a mass of (980 ± 10) MeV, and a full width between 40 MeV and 100 MeV, which reflects the fact that the width determination is very model-dependent.In the literature, the hadronic structure of the $f_0(980)$ state has been discussed for decades and there are many different interpretations, from the conventional quark-antiquark picture to tetraquark states. It is an important question to explore its hadronic structure. This particle has been explored at hadronic high energy facilities (see e.g. Ref. [9.46]), but reactions induced by electromagnetic probes at lower energies may provide additional information.

### 9.3.4. *Search for dark forces*

Several puzzling astrophysical observations (PAMELA abundance of positrons, ATIC excess, WMAP haze, INTEGRAL signal) could be explained on a common ground by existence of a new, beyond-standard-model (BSM), weakly interacting boson "U", see, e.g. Ref. [9.47]. The mass of the U boson is expected to be at the MeV or GeV scale. Such a particle would be an elusive dark matter candidate and, technically, a gauge boson of a "hidden sector" abelian symmetry group U(1), see Ref. [9.48]. The U boson does not interact directly with the Standard Model (SM) particles, however, it couples with the SM photon due to kinetic mixing with a tiny mixing parameter factor ε, see Ref. [9.49]. Therefore the SM observables mediated by a photon, would receive a contribution from the hidden sector. The existing precision tests of SM to a high level constrain the possible range of the U boson mass and ε. Some of the SM inconsistencies can be explained by the existence of the U boson. Among the most important problems, a 3.6 sigma deviation of the muon (g-2) experiment from the SM prediction can be explained by U boson contribution, see Ref. [9.50]. A bulk of the parameter space for U is already excluded by various experiments, see for example a recent update of the big picture of Endo et al., [9.51]. However, there are numerous experimental searches for the U boson going on and planned [9.52].

At IRIDE we can search for U boson via the lepton triplet production process in the electron-photon collision. The main QED process of the lepton triplet production is through *u* channel exchange ("BH diagram") and the *t* channel exchange ("VCS diagram"), see Fig. 9.5.

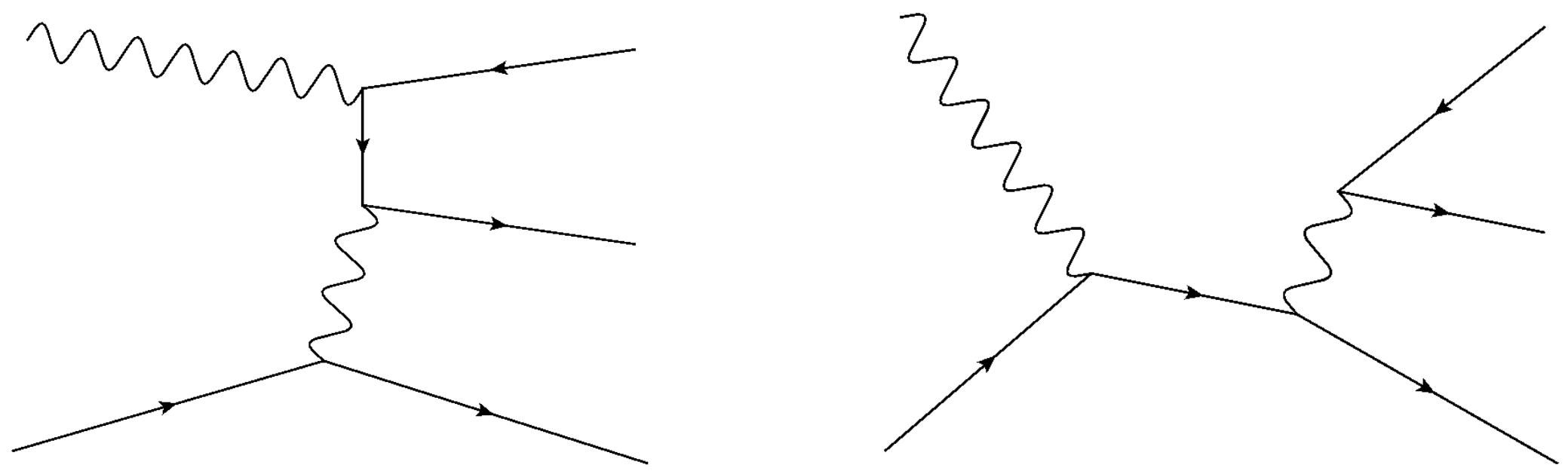

**Figure 9.5 -** QED process of the lepton triplet production: t-channel (left) and s-channel (right) contributions to the amplitude.

The U-boson contribution is included as the *t*-channel part with the photon line modified by the mixing with the U, see Fig. 9.6.

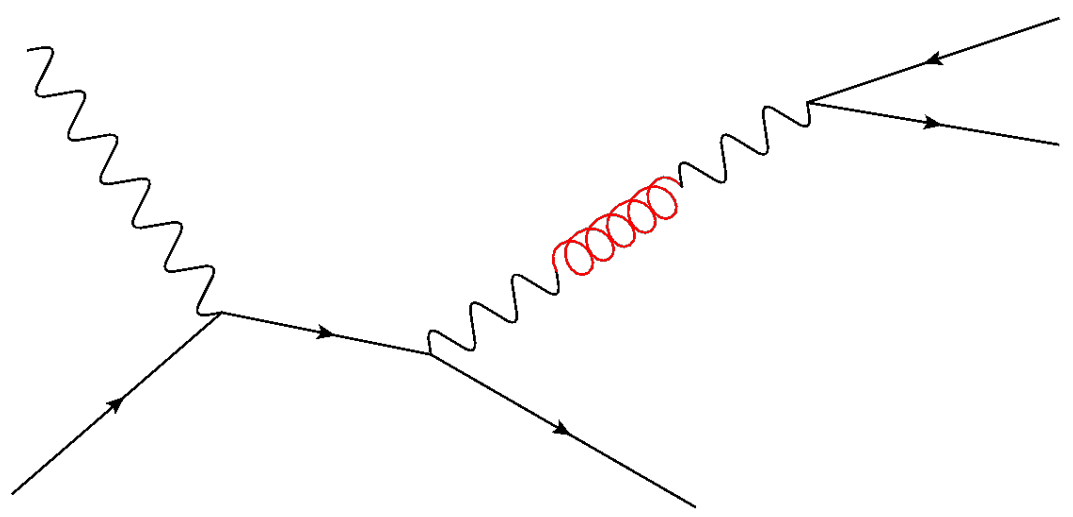

**Figure 9.6 -** U-boson production (in red) via s-channel process.

Therefore, the VCS part of the QED process is of the most importance, as it is intimately related to the U-boson production, while the BH contribution must be reduced as much as possible. This can be done by specific angular criteria for the event selection. Namely, it is foreseen that the detected electrons originated from the BH process mainly populate the region of small polar angles with respect to the beam collision axis.

We have studied the cross section integrated over the whole angular region of the final particles. The results are shown in Fig. 9.7 for different energies of colliding beams. As one can see, the ratio of VCS to BH contributions is decreasing when increasing the initial particle energies. In other words, the low-energy configuration is more suitable for these searches, although the U-boson mass window is reduced.

In order to find the optimized e-γ beam configuration for U boson searches, a detailed Monte Carlo simulation of the signal process and the possible QED, hadronic and machine background is in progress. To estimate the sensitivity for the U boson contribution to the cross section, one can rely on a useful relation between the QED and the U boson mediated cross section, integrated in the vicinity of the U boson resonance peak as derived by Bjorken et al. [9.3]. For such evaluation, one need to realistically account for the experimental resolution, which requires a detailed understanding of the detector equipment and of the analysis procedure. We have assumed a conservative detector resolution for the U mass of 5 MeV, and the QED process fully dominated by VCS contribution (i.e. we consider the optimistic case in which BH contribution can be neglected). The corresponding cross section for the U boson with $\varepsilon \sim 10^{-3}$ is also shown in Fig. 9.7, which makes this search promising in the low U-mass range. In particular our results indicate an increase of sensitivity for energy-beam configuration with photons and electrons of lower energies (of the order of 1 MeV for photons and 100 MeV for electrons), as shown in Fig. 9.8. In this case to reach values of $\varepsilon < 10^{-3}$ a luminosity larger than $10^{30}$ $cm^{-2}$ $sec^{-1}$ is needed.

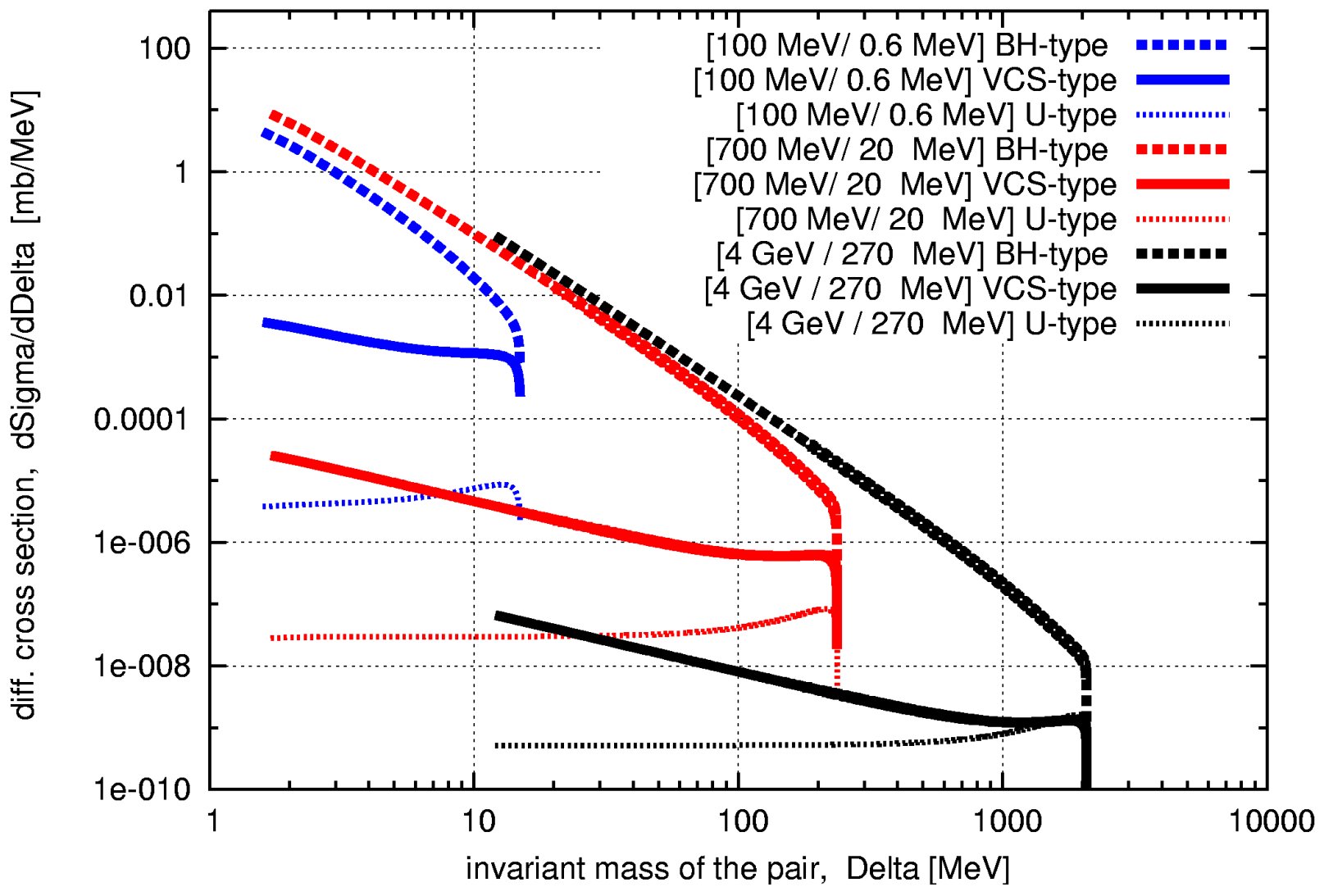

**Figure 9.7 -** Differential cross section vs ee invariant mass for the process $e^- \gamma \rightarrow e^- e^+ e^-$ at different e-γ energies.

Besides the precision measurement of the $\pi^0$ width and the searches for U-bosons, other physics motivations for the e-γ interaction are: the triple Compton scattering for studies of entangled states [9.2] and the production of μμ and ππ at rest for tests of Coulomb and strong effects for bound states.

Although still detailed studies must be carried on, the detector required for the e-gamma interaction will consist mainly by a tracking system, a calorimeter and possibly a spectrometer in the forward direction (to detect the outcoming electrons emitted at low angle).

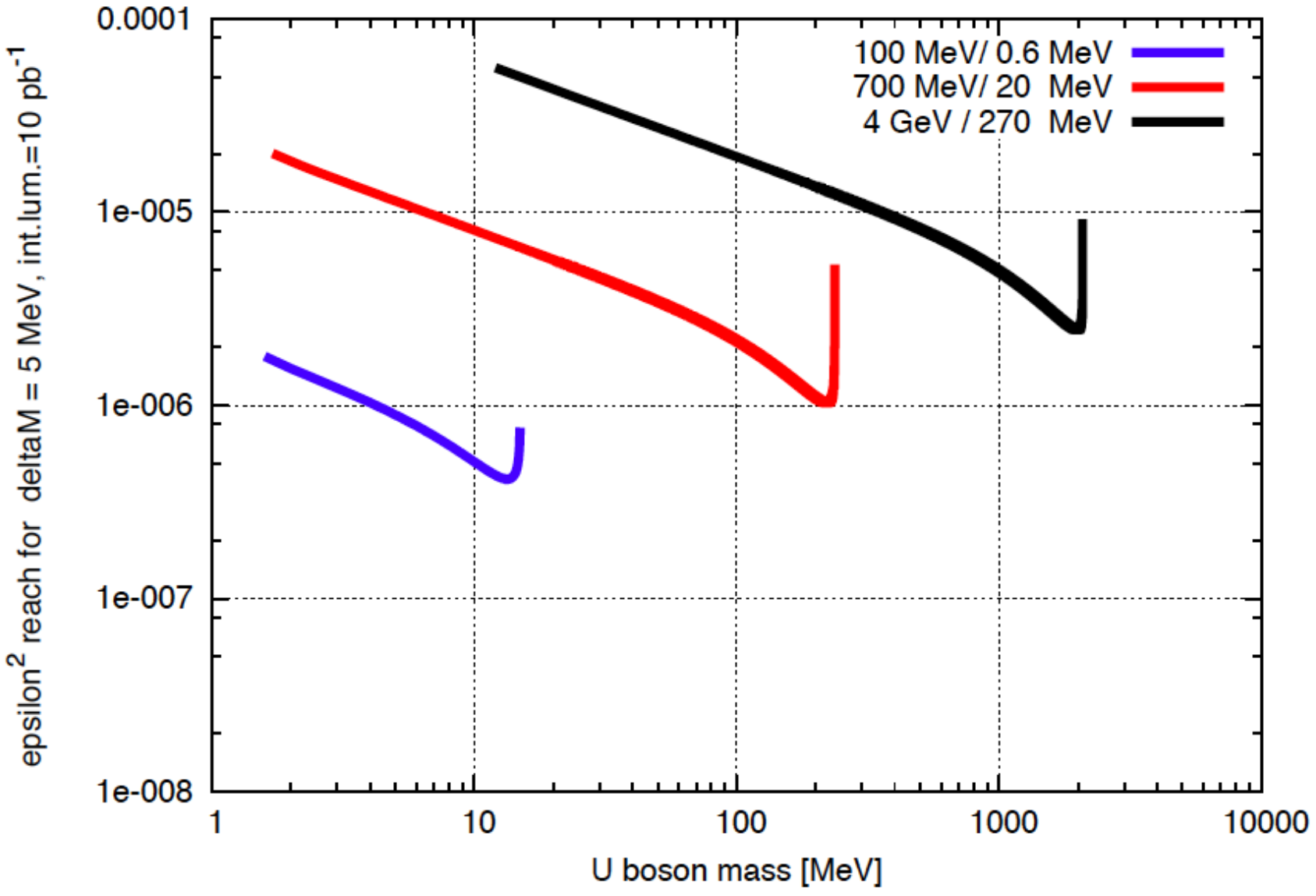

**Figure 9.8 -** Sensitivity on $\varepsilon^2$ as function of the U boson invariant mass for the process $e^- \gamma \rightarrow e^- e^+ e^-$ at different e-γ energies for an integrated luminosity of 10 $pb^{-1}$.

Depending on the physics case, U boson searches or the measurement of $\pi^0$, η, η’ widths, the detector design could be oriented towards the one used in γγ or $e^+e^-$ physics programs.

## 9.4. Photon-photon collider option

### 9.4.1. *Introduction*

The IRIDE accelerator complex can generate colliding photon-photon beams by Compton backscattering, and this opens up the fascinating field of low-energy photon-photon physics. A good hold on the technology needed to carry out a photon-photon physics program at energies close to 1 MeV would disclose new developments at higher energies [9.53], where a photon-photon Higgs factory could be a nearly ideal discovery machine [9.54]. However, even at low energy, such a collider would provide plenty of interesting physics, especially as a testing ground of Quantum Electrodynamics (QED).

Indeed, QED, and more generally the quantum theories of fields, rank among the most spectacularly successful theories of physics, and yet they display evident shortcomings that show that we still do not understand some basic elements of the physical world. The most striking failing is the huge mismatch between the measured energy density of vacuum and the energy density of the ground level of the fundamental fields [9.55 – 9.57] which are distant by something like 120 orders of magnitude. This is the so called “cosmological constant problem”, and several potential solutions have been devised to get rid of it, but at the moment none of them can be considered the final one [9.55].

In its essence, the cosmological constant problem can be stated as follows [9.58]. Consider first the number of modes per unit volume of the electromagnetic (EM) field at a given frequency, which yields the energy density per unit frequency

$$u(\omega) = \frac{\hbar}{2\pi^2 c^3}\omega^3$$

then, integrating over all frequencies we find a badly divergent integral. We can select a suitable cutoff frequency, but even a mild choice like the GZK cutoff[2] leads to huge values for the integrated energy density: in the case of the GZK cutoff the energy density of the EM vacuum turns out to be about 1.6 x $10^{40}$ GeV/fm$^3$, which is approximately $10^{41}$ times larger than the nuclear density. Clearly, taking the Planck energy as a more natural cutoff, the energy density of the EM vacuum reaches the huge value 1.7 x $10^{76}$ GeV/fm$^3$, which is about 120 orders of magnitude larger than the critical energy density of the universe (i.e., the energy density of a flat universe with Minkowski metric). When we include other fields, and we note that the elementary fermionic degrees of freedom (21) give a contribution with sign opposite to that of elementary bosonic degrees of freedom (27), we find that the problem becomes even worse. In this context Zumino noted many years ago [9.59] that supersymmetry – where each fermionic degree of freedom is balanced by a corresponding bosonic degree of freedom – might cure the problem, if only the threshold of supersymmetry breaking were sufficiently low. Unfortunately, we do already know that this threshold, if it exists, is certainly much higher than required for the solution of the cosmological constant problem.

All this means that any experimental information that we may glean about the QFT vacuum is invaluable, and could bring us closer to the solution of this problem. There are indeed a few processes that afford a direct view into QED vacuum, and photon-photon scattering is one of them. This process has no tree-level graph, and the first order contribution comes from the one-loop graph shown in Figure 9.10.

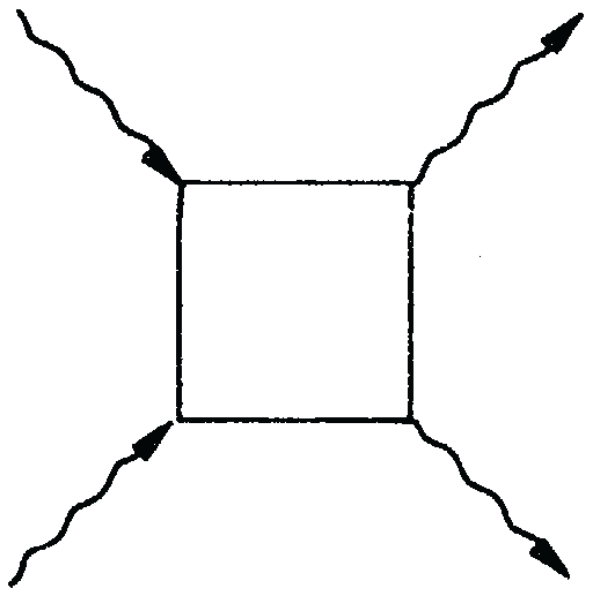

**Figure 9.10** - The basic diagram

The photon-photon scattering cross-section was first considered by Halpern [9.61], and the low-energy cross section was first calculated by Euler and Kockel [9.62]. The complete cross-section was first calculated by Akhieser et al [9.63]. Later, it was studied in depth in a series of papers by Karplus and Neuman [9.64], and by De Tollis [9.65]. It turns out that a quantity that plays a very important role is the so called *vacuum polarization tensor* $G_{\mu\nu\lambda\sigma}^{(\kappa)}$, which depends on the photon four-momenta:

$$G_{\mu\nu\lambda\sigma}^{(\kappa)}\left(k^{(1)},k^{(2)},k^{(3)},k^{(4)}\right)=G_{\mu\nu\lambda\sigma}^{(\kappa)}\left(-k^{(1)},-k^{(2)},-k^{(3)},-k^{(4)}\right)$$

[2] Since we observe photons with energies as large as the GZK cutoff, the corresponding cutoff in the integral must be equal or greater than that.

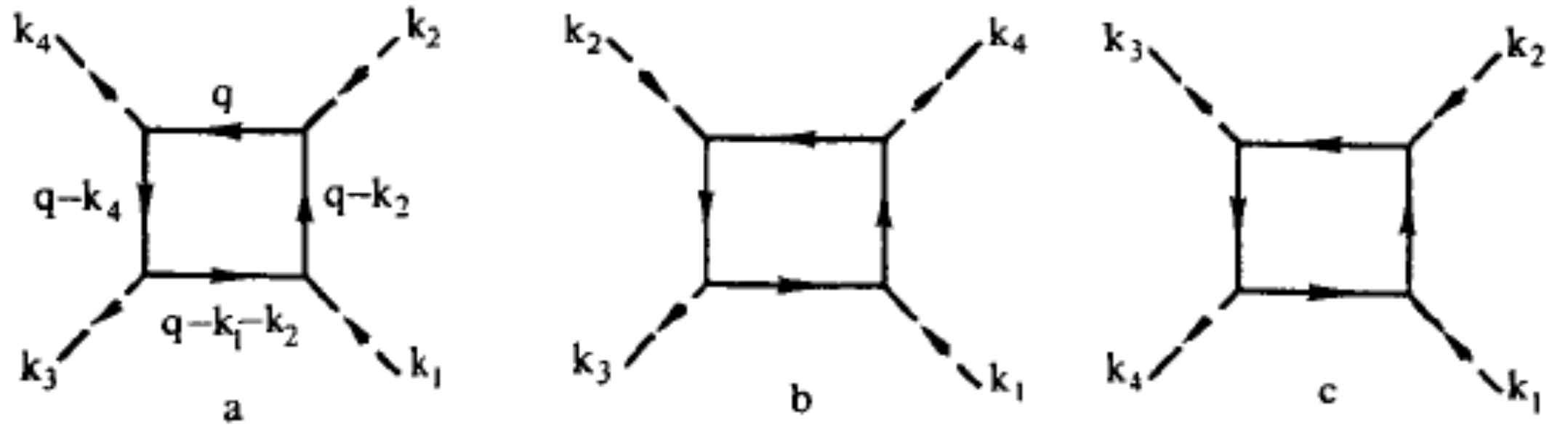

**Figure 9.11** - To compute all the scattering amplitudes we must sum over graphs with different photon indexes.

Here the upper index denotes photons, and conventionally 1 and 2 denote the incoming photons; $\kappa = mc^2/\hbar c$ is the reciprocal length related to the mass of the fermion (electron) in the loops shown in Figures 9.10 and 9.11.

In order to obtain meaningful results, the vacuum polarization tensor must be regularized

$$G_{\mu\nu\lambda\sigma} = \lim_{M\to\infty}\left[G_{\mu\nu\lambda\sigma}^{(\kappa)} - G_{\mu\nu\lambda\sigma}^{(M)}\right] \tag{9.1}$$

The differential cross-section can be computed directly from the vacuum polarization tensor, and in the CM reference frame it is

$$\frac{d\sigma}{d\Omega}=\frac{\alpha^2 r_e^2}{4\pi^2 k^2}|M|^2$$

where $r_e$ is the classical electron radius; here $M$ represents generically the polarization dependent amplitudes that are directly related to the vacuum polarization tensor

$$M_{\lambda_1\lambda_2\lambda_3\lambda_4}(\Omega)=\frac{1}{4}e_\mu^{\lambda_1}e_\nu^{\lambda_2}e_\lambda^{\lambda_3}e_\sigma^{\lambda_4}G_{\mu\nu\lambda\sigma}(\mathbf{p},\omega,-\mathbf{p},\omega,\mathbf{q},\omega,-\mathbf{q},\omega)$$

When we consider the scattering of unpolarized photons, we must replace $|M|^2$ with its average over all the incoming photon polarizations and sum over all outgoing photon polarizations. At energies less than about 0.7 MeV, the CM differential cross-section – averaged over initial polarizations and summed over final polarizations – is

$$\frac{d\sigma}{d\Omega}\approx\frac{139\alpha^4}{(180\pi)^2}\frac{(\hbar\omega)^6}{\left(mc^2\right)^8}\left(3+\cos^2\theta\right)^2 \tag{9.2}$$

It is important to remark that the cross-section depends directly on the vacuum polarization tensor and that the regularization procedure plays an all-important role. Indeed, if one forgoes the regularization step (9.1), the cross-section turns out wrong: this was recently evidenced by a couple of preprints in arXiv [9.65] and by a paper that rectified the wrong claims in the preprints [9.66]. The relevance of regularization here involves some deep issues, as evidenced by Jackiw [9.67], who distinguished three classes of radiative corrections: those that diverge and do absolutely need regularization, those that do not diverge and yield definite results without regularization, and those that do not diverge, but lead to undefined results unless they are regularized. Jackiw conjectured that "... if the form of the radiative correction is such that inserting it into the bare Lagrangian would interfere with symmetries of the model or would spoil renormalizability, then the radiative result will be finite and uniquely fixed. Alternatively, if modifying the bare Lagrangian by the radiative correction preserves renormalizability and retains the symmetries of the theory, then the radiative calculation will not produce a definite result – it is as if the term in question is already present in the bare Lagrangian with an undetermined strength, and the radiative correction adds a further undetermined contribution. With this criterion, the radiatively induced $g - 2$ Pauli term in QED and the photon mass in the Schwinger model are unique ...". These considerations place the tests of QED based on $g - 2$ [9.68] in a different class with respect to photon-photon scattering and indicate that the measurement of the photon-photon cross-section is not just a test of QED as it is, but a deep test of the regularization/renormalization procedure as well.

The approximate expression (9.2) shows that the CM cross-section is a very fast-growing function of photon energy, and close to 1 MeV, it is about 36 orders of magnitude higher than in the visible photon range. This consideration alone places photon-photon scattering in a different category with respect to the PVLAS experiment, which tests instead photon-photon scattering at very low energy [9.69].

It is also important to note that the ability of setting the initial photon polarization gives a partial access to the individual polarization dependent amplitudes. This is illustrated in Figures 9.12 and 9.13 which show the cross-section for unpolarized photons, as well the cross-section for photons that are polarized parallel and perpendicular to the scattering plane.

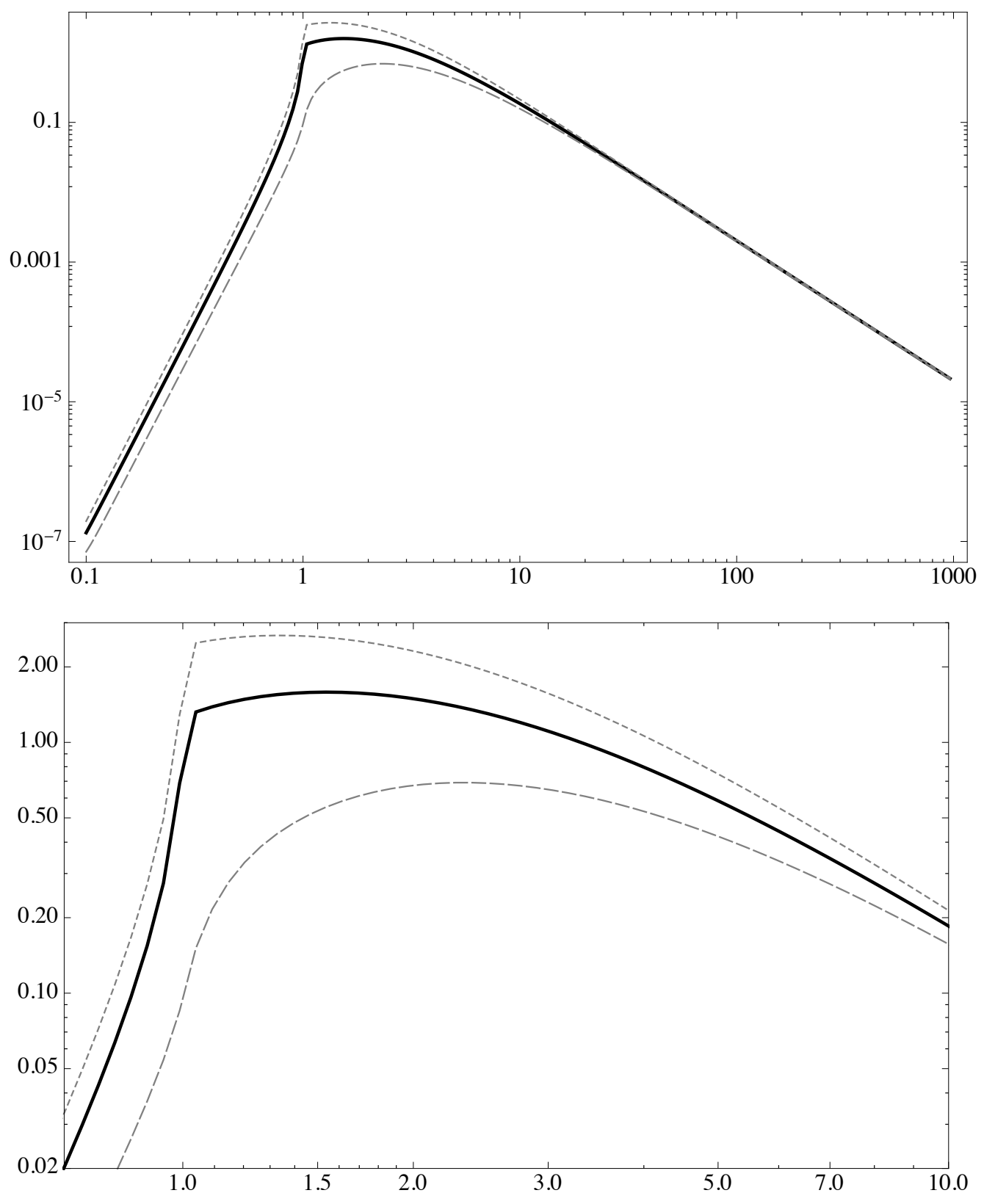

**Figure 9.12 -** Upper panel: total cross-section (μbar) vs. CM energy (MeV). Solid line: cross-section averaged over initial photon polarizations. Dotted line: incoming photons have the same circular polarization. Dashed line: incoming photons have opposite circular polarization. Lower panel: zoomed version of the curves in the upper panel, close to the peak.

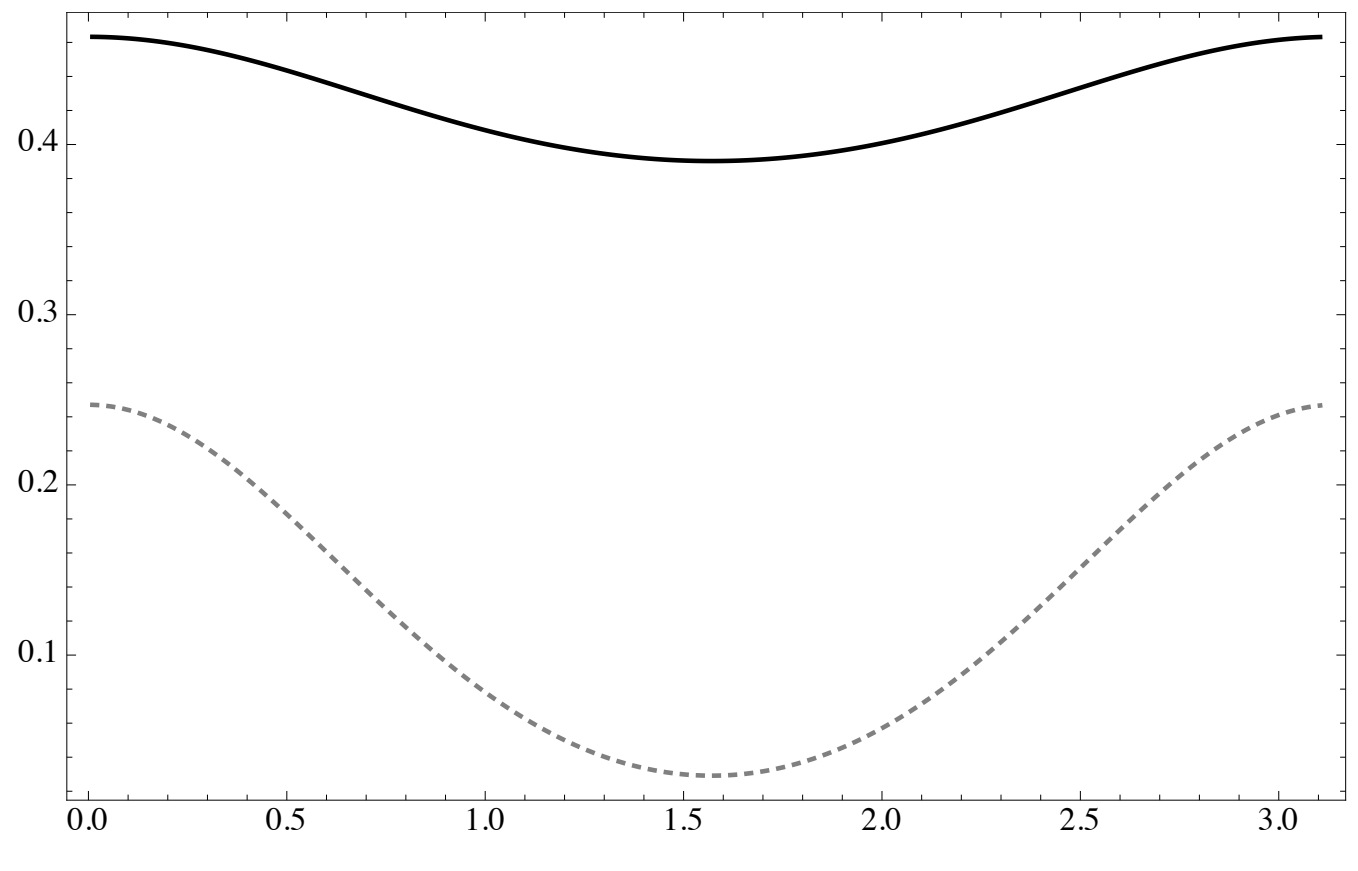

**Figure 9.13 -** Differential cross-section $d\sigma/d\theta$ (μbarn/rad) vs. $\theta$ (radians) close to the peak of the total cross-section (1.6 MeV CM energy). Solid line: incoming photons with the same circular polarization. Dotted line: incoming photons with opposite circular polarizations.

The photon-photon cross-section is directly related to the fermion loops that polarize vacuum (Figure 9.12), and therefore a photon-photon scattering experiment with photon energies in the 0.5-0.8 MeV range – where the cross-section is reasonably large – would be an important test of our understanding of the QED vacuum. Furthermore, the considerations of Jackiw, and the structure of the regularization equation (9.1), suggest that the chiral anomaly plays a role as well in the exact determination of the cross-section (9.2), and therefore an accurate measurement of the photon-photon cross-section would test the Hamiltonian part as well as the regularization/renormalization procedure that so much disturbed Dirac at the end of his life[3] [9.70].

It is important to note that although a few processes involving the diagram of Figure 9.12 have long been observed – the first detection of Delbrück scattering dates back to 1973 [9.72] – the sheer detection of photon-photon scattering in vacuum would be an important confirmation of our views of quantum vacuum.

However, besides detection, we also need to verify the correctness of the differential cross-section (9.2). Consider the situation at low energy: there we know that photon-photon interactions can be effectively described by adding the Euler-Heisenberg-Weisskopf (EHW) Lagrangian correction to the usual EM Lagrangian density [9.75]

$$L_{\mathrm{EHW}} = \frac{A_e}{\mu_0}\left[\left(\frac{E^2}{c^2} - B^2\right)^2 + 7\left(\frac{\mathbf{E}}{c}\cdot\mathbf{B}\right)^2\right]$$

and the generic functional form of this Lagrangian satisfies the requirements of Lorentz invariance, parity invariance, gauge invariance and locality constraint (i.e., it has no derivatives of the fields of order greater than 1). In order to take into account all the possible variants, it is interesting to generalize the non-linear EM Lagrangian correction by introducing two dimensionless free parameters depending on the model, $\eta_1$ and $\eta_2$ [9.76]:

$$L_{\mathrm{pM}} = \frac{\xi}{2\mu_0}\left[\eta_1\left(\frac{E^2}{c^2} - B^2\right)^2 + 4\eta_2\left(\frac{\mathbf{E}}{c}\cdot\mathbf{B}\right)^2\right] \tag{9.3}$$

where ,

$$\xi = 1/B_{\mathrm{crit}}^2 = \left(\frac{e\hbar}{m_e^2c^2}\right)^2$$

---

[3] Dirac was disturbed by the *ad hoc* character of the regularization/renormalization procedure, which he viewed as a breach in the elegance of the theory [9.70], and as a violation of the concepts of beauty which he cherished so much [9.71]. Schwinger and Feynman also expressed criticisms, although based on different grounds. Currently, most theorist are quite happy with the renormalization procedure, which has found a mathematically pleasing expression in the renormalization group approach. Unfortunately, the procedure is still *ad hoc*, and divergences still plague the theory, as evidenced by the cosmological constant problem.

and $B_{\text{crit}}$ is the critical magnetic field. We find that in general all the vacua described by these Lagrangian corrections become birefringent when a magnetic field is present – and this corresponds to the scattering of photons off the virtual photons of the magnetic field. This birefringence is related to the coefficients introduced above by the equation

$$\Delta n = 2\xi\left(\eta_2 - \eta_1\right)B_{\text{ext}}^2$$

and we see that it vanishes if $\eta_2 = \eta_1$; this very special condition occurs for the modified Lagrangian

$$L_{BI} \approx \frac{1}{2}\left\{\left(\mathbf{E}^2 - \mathbf{B}^2\right) + \frac{1}{4b^2}\left[\left(\mathbf{E}^2 - \mathbf{B}^2\right)^2 + 4\left(\mathbf{E}\cdot\mathbf{B}\right)^2\right]\right\}$$

(where $\hbar = c = 1$), which was originally introduced by Born and Infeld to remove the classical divergence of the fields produced by pointlike charges. It is remarkable that the Born-Infeld Lagrangian also occurs as a low-energy limiting Lagrangian in superstring theories.

Thus we see that photon-photon scattering – which at low energy is related to the specific form of the parameterized Lagrangians (9.3), and effectively produces the birefringence of the magnetically loaded vacuum – occurs in different vacua, and this means that it is important to observe the shape of the differential cross-section (9.2) to provide a compelling confirmation that vacuum has the structure predicted by QED. This in turn leads to the estimate of the minimum integrated luminosity required to carry out a meaningful test.

### 9.4.2. *Luminosity and beam requirements*

Photon-photon scattering has never been observed, and it is one of the main goals of projects like ELI [9.75]. Current proposals such as ELI utilize extremely powerful lasers in the near-visible range to search for elusive photon-photon scattering events [9.75]. However the total cross-section grows fast – see equation (2) – and near its peak it is close to 2 μbarn (see Figures 9.12 and 9.16), about 36 orders of magnitude higher than in the visible region, and this underscores the importance of moving to higher energy.

Figure 9.14 shows the partially integrated differential cross-sections

$$I_{++}\left(\theta_1,\theta_2\right) = \int_0^{2\pi} d\varphi \int_{\theta_1}^{\theta_2} \sin\theta d\theta \frac{d\sigma_{++}}{d\Omega}$$

and

$$I_{+-}\left(\theta_1,\theta_2\right) = \int_0^{2\pi} d\varphi \int_{\theta_1}^{\theta_2} \sin\theta d\theta \frac{d\sigma_{+-}}{d\Omega}$$

close to the peak of the total cross-section (1.6 MeV CM energy). Their ratios are shown in figure 9.15.

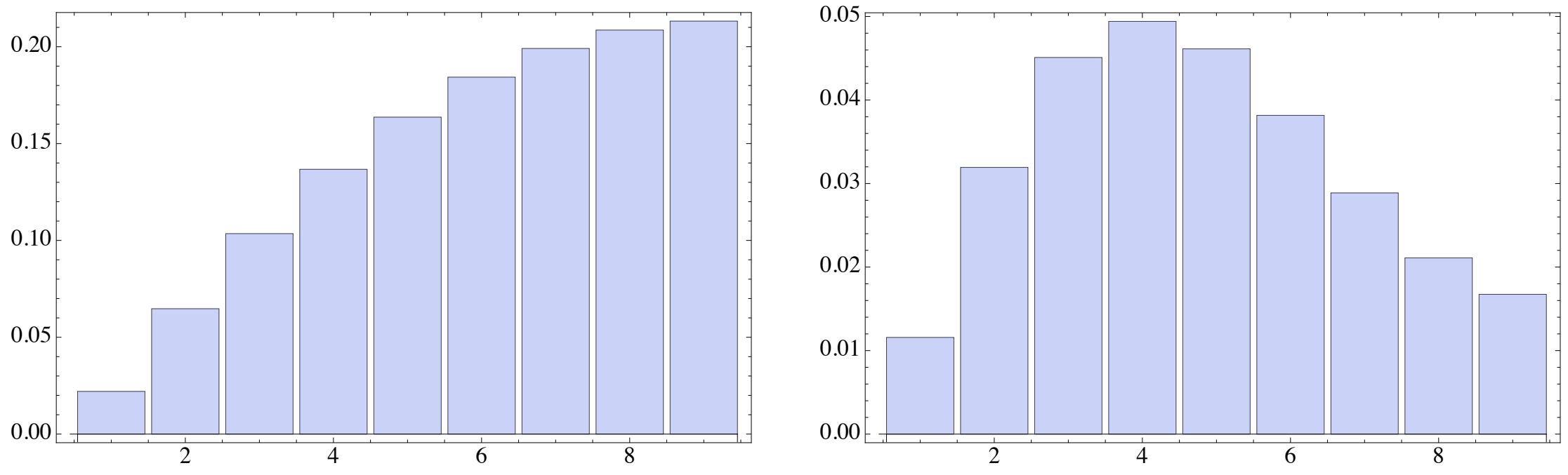

**Figure 9.14 -** Partially integrated differential cross-sections (μbarn) close to the peak of the total cross-section (1.6 MeV CM energy). Each bin corresponds to a 10° region for θ starting from θ = 0° up to θ = 90°. Left panel: partially integrated cross-section; right $I_{++}$ panel: partially integrated cross-section $I_{+-}$.

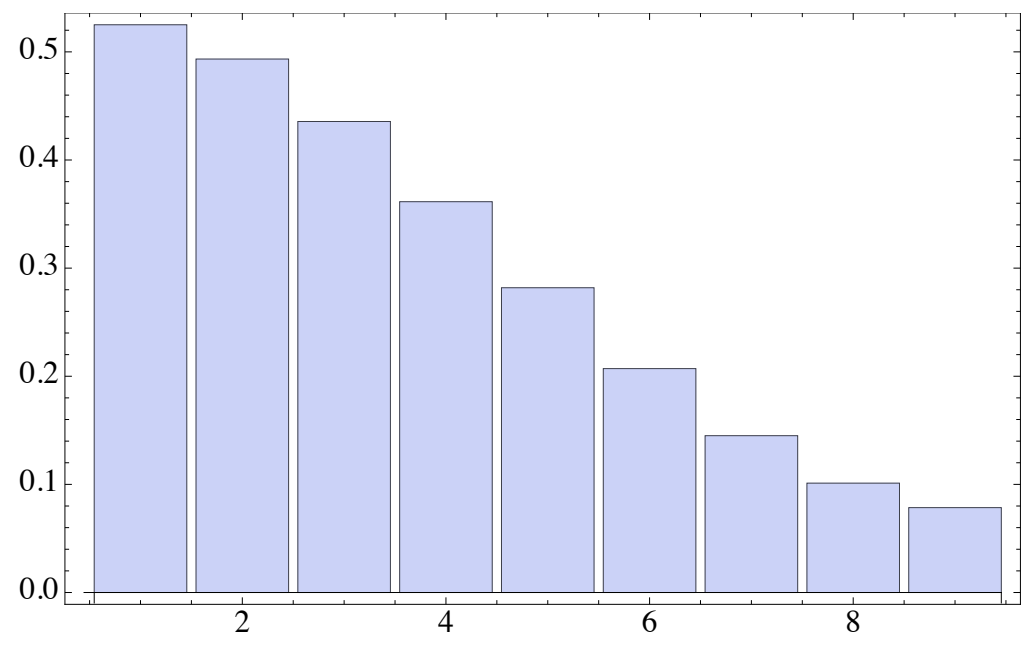

**Figure 9.15 –** Ratio $I_{+-}/I_{++}$ of the partially integrated differential cross-sections shown in figure 5.

The smallest value of the ratio $I_{+-}/I_{++}$ is about 8%, and it occurs in the bin that spans the 80°-90° interval. This means that in order to obtain a signal-to-noise ratio of 1 in this bin, we must have an uncertainty smaller than about 8% in the determination of the number of scattering events with incoming photons that have the same circular polarization, i.e., about 170 events in the ++ channel and about 13 events in the +− channel. Since the partially integrated cross-section $I_{+-}$ is about 0.017 μbarn in this bin, the integrated luminosity must be

$$\mathcal{L}\,\Delta t \geq \frac{13}{17\,\text{nbarn}} \approx 0.8\ \text{nbarn}^{-1}$$

### 9.4.3. *Background processes*

Although the photon-photon cross-section is much larger around its peak than at low energy, it is still small when compared to other processes which contribute to the background, and this makes the experiment even more challenging. Here the main background processes are due to electron pair production (Breit-Wheeler process) γγ → $e^+e^-$, to the Bethe-Heitler process eγ → $ee^+e^-$, and to the Landau-Lifshitz process $e^+e^- \to e^+e^- e^+e^-$.

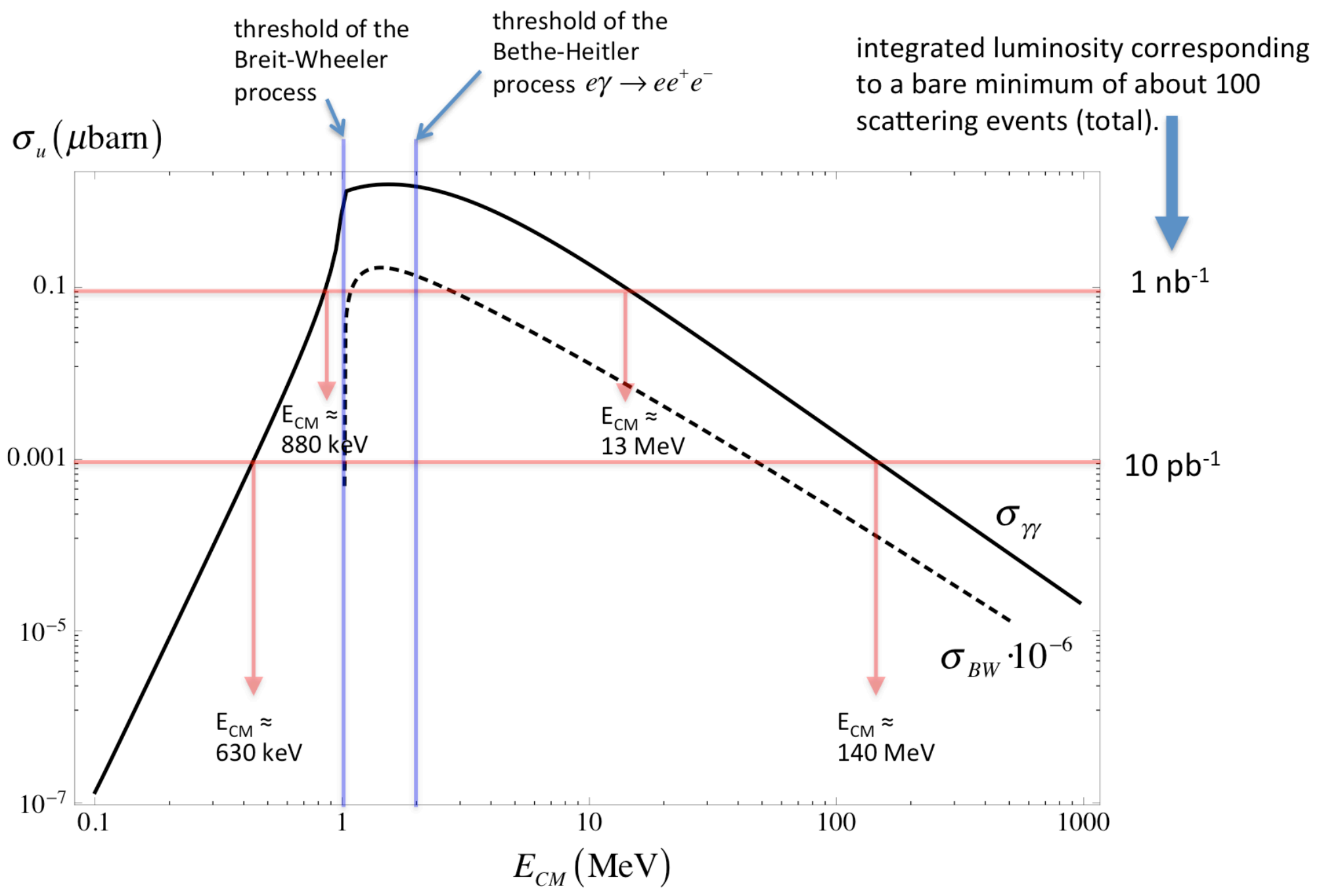

**Figure 9.16 -** Unpolarized photon-photon ($\sigma_{\gamma\gamma}$) and pair-production ($\sigma_{BW}$) total cross-sections with constraints from gamma-gamma luminosity.

We assume from the outset that electrons are swept away from the interaction region after producing high energy photons by Compton backscattering, and therefore we neglect the Landau-Lifshitz process. Moreover we take $E_{CM}$ < 2 MeV, so that the Bethe-Heitler process can also be neglected.

Then the only important remaining background process is the Breit-Wheeler electron pair production, which has the differential cross-section (see, e.g., [9.76] and [9.77]):

$$\frac{d\sigma_{BW}}{d\cos\theta_{CM}} = \pi r_e^2 \frac{\beta}{\gamma^2}\left\{\left[\frac{1}{2}\frac{1+\beta^2\cos^2\theta_{CM}}{1-\beta^2\cos^2\theta_{CM}} + \frac{1-\beta^2}{1-\beta^2\cos^2\theta_{CM}} - \frac{\left(1-\beta^2\right)^2}{\left(1-\beta^2\cos^2\theta_{CM}\right)^2}\right]\right.$$
$$\left. + \frac{h}{2}\left[\frac{1+\beta^2\cos^2\theta_{CM}}{1-\beta^2\cos^2\theta_{CM}}\left(1-2\frac{1-\beta^2}{1-\beta^2\cos^2\theta_{CM}}\right)\right]\right\}$$

where $E_\gamma$ is the initial photon energy, $\theta_{CM}$ is the scattering angle in the CM of the final electron pair,

$$\beta = \frac{\sqrt{E_\gamma^2 - \left(mc^2\right)^2}}{E_\gamma}$$

and $h$ is the product of the initial photon helicities. The corresponding total cross-section is plotted in Figure 9.17. The partially integrated cross-section (in 10° bins like those of Section 9.4.2) is listed in Table 9.2, and the corresponding background event rate is given in Table 9.3 for a machine luminosity $10^{28}$ cm$^{-2}$ s$^{-1}$.

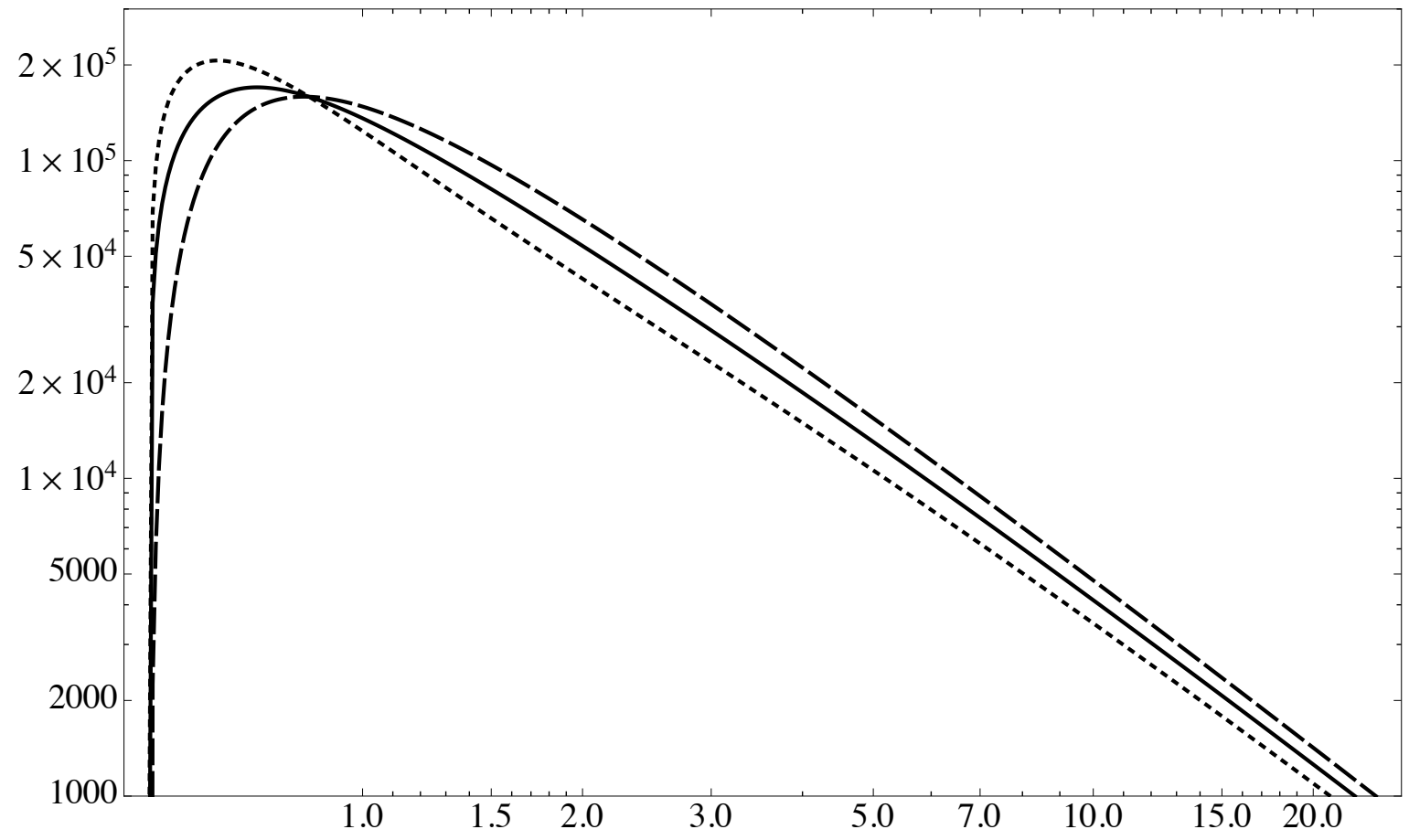

**Figure 9.17 -** Total cross-section $\sigma_{BW}$ (μbarn) vs. $E_\gamma$ (MeV) for the Breit-Wheeler process. Solid line: unpolarized initial photons; dashed line, same helicity, $h=1$; dotted line, opposite helicity, $h=-1$.

**Table 9.2 -** Integrated cross-section (μbarn) for unpolarized initial photons. Here the differential cross-section is integrated in bins that are 10° wide. This can be used to estimate the number of background events for a given machine luminosity.

| $E_\gamma$ (MeV) | 0°-10° | 10°-20° | 20°-30° | 30°-40° | 40°-50° | 50°-60° | 60°-70° | 70°-80° | 80°-90° |
|---|---|---|---|---|---|---|---|---|---|
| 0.55 | 8555.32 | 8335.09 | 7888.85 | 7209.44 | 6295.64 | 5159.83 | 3832.15 | 2358.91 | 796.279 |
| 0.6 | 11545. | 11368.6 | 10976.9 | 10301.1 | 9263.81 | 7809.09 | 5935.5 | 3711.82 | 1262.76 |
| 0.65 | 12233.9 | 12175.6 | 11996.4 | 11575.7 | 10753. | 9368.59 | 7330.62 | 4681.85 | 1610.09 |
| 0.7 | 11887.8 | 11944.1 | 11985.6 | 11867.2 | 11372.5 | 10244.1 | 8269.01 | 5408.83 | 1884.02 |
| 0.75 | 11103.4 | 11248.6 | 11470.4 | 11621.1 | 11459.5 | 10657.4 | 8877.27 | 5956.48 | 2104.42 |
| 0.8 | 10171.9 | 10378.9 | 10732.5 | 11095.8 | 11226.6 | 10757.1 | 9241.01 | 6365.93 | 2283.68 |
| 0.85 | 9232.3 | 9479. | 9921.82 | 10441.1 | 10809.8 | 10646.7 | 9422.63 | 6666.79 | 2430.49 |
| 0.9 | 8347.29 | 8616.93 | 9115.59 | 9742.92 | 10295.4 | 10398.9 | 9468.79 | 6881.31 | 2551.27 |
| 0.95 | 7541.25 | 7821.92 | 8351.71 | 9049.15 | 9737.69 | 10064.3 | 9414.77 | 7026.68 | 2650.91 |
| 1. | 6819.83 | 7103.29 | 7646.54 | 8385.3 | 9170.46 | 9678.48 | 9287.37 | 7116.31 | 2733.17 |

The differential cross-section of the Breit-Wheeler process is peaked in the forward-backward direction, and it is plotted in Figures 9.18 and 9.19 for two different values of the initial photon energy.

**Table 9.3 -** Background events (Hz) from the Breit-Wheeler process (events) with unpolarized initial photons in the same angular bins specified in Table 9.2. The number of background events has been estimated for a machine luminosity $10^{28}$ $cm^{-2}$ $s^{-1}$. Further (azimuthal) detector segmentation can be used to reduce these background rates.

| $E_\gamma$ (MeV) | 0°-10° | 10°-20° | 20°-30° | 30°-40° | 40°-50° | 50°-60° | 60°-70° | 70°-80° | 80°-90° |
|---|---|---|---|---|---|---|---|---|---|
| 0.55 | 7.963 | 23.59 | 38.32 | 51.60 | 62.96 | 72.09 | 78.89 | 83.35 | 85.55 |
| 0.60 | 12.63 | 37.12 | 59.35 | 78.09 | 92.64 | 103.0 | 109.8 | 113.7 | 115.5 |
| 0.65 | 16.10 | 46.82 | 73.31 | 93.69 | 107.5 | 115.8 | 120.0 | 121.8 | 122.3 |
| 0.70 | 18.84 | 54.09 | 82.69 | 102.4 | 113.7 | 118.7 | 119.9 | 119.4 | 118.9 |
| 0.75 | 21.04 | 59.56 | 88.77 | 106.6 | 114.6 | 116.2 | 114.7 | 112.5 | 111.0 |
| 0.80 | 22.84 | 63.66 | 92.41 | 107.6 | 112.3 | 111.0 | 107.3 | 103.8 | 101.7 |
| 0.85 | 24.30 | 66.67 | 94.23 | 106.5 | 108.1 | 104.4 | 99.22 | 94.79 | 92.32 |
| 0.90 | 25.51 | 68.81 | 94.69 | 104.0 | 103.0 | 97.43 | 91.16 | 86.17 | 83.47 |
| 0.95 | 26.51 | 70.27 | 94.15 | 100.6 | 97.38 | 90.49 | 83.52 | 78.22 | 75.41 |
| 1.0 | 27.33 | 71.16 | 92.87 | 96.78 | 91.70 | 83.85 | 76.47 | 71.03 | 68.20 |

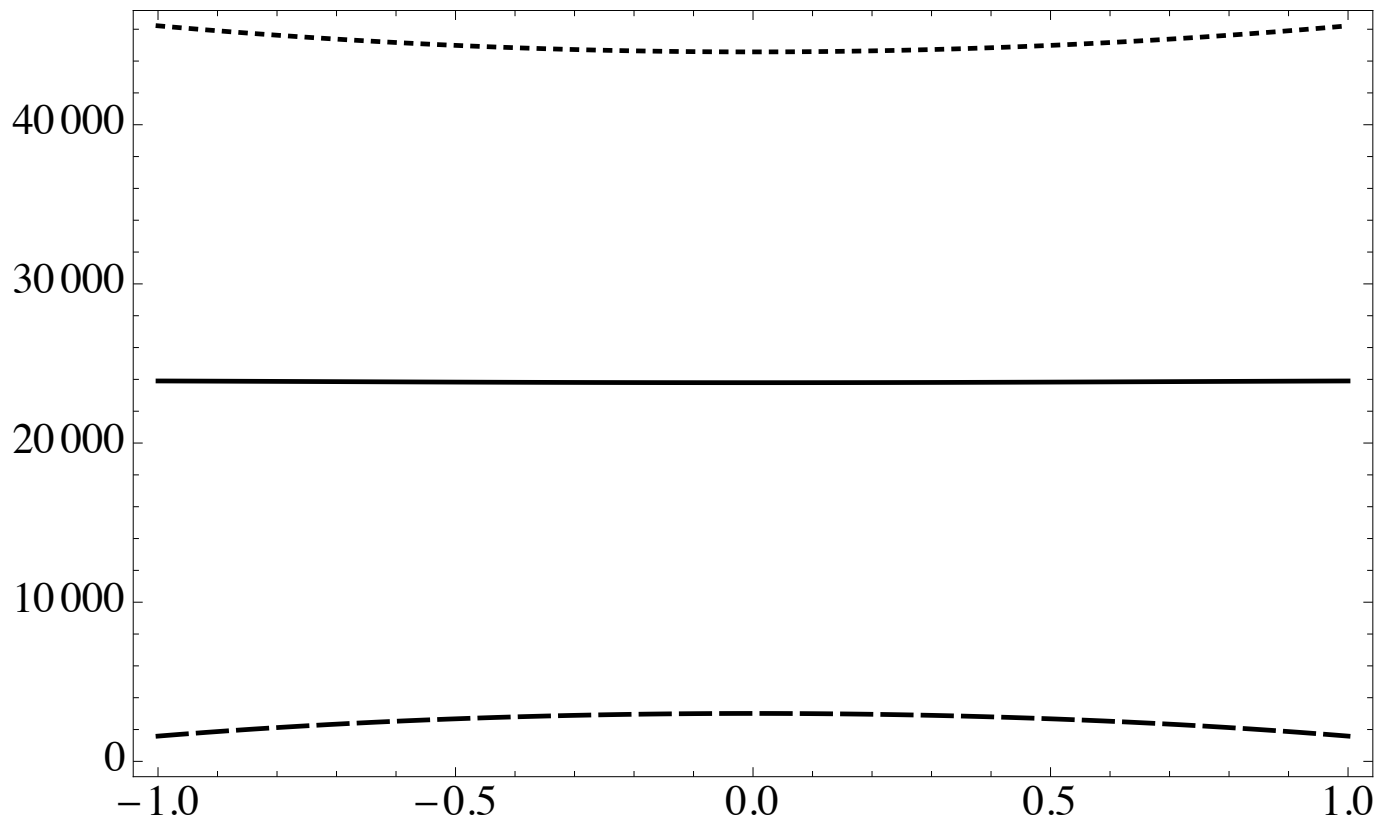

**Figure 9.18 -** Differential cross section $d\sigma_{BW}/d\cos\theta_{CM}$ (μbarn) vs. $\cos\theta_{CM}$ for the Breit-Wheeler process, close to the production threshold ($E_\gamma$ = 520 keV). Solid line: unpolarized initial photons; dashed line, same helicity, $h = 1$; dotted line, opposite helicity, $h = -1$.

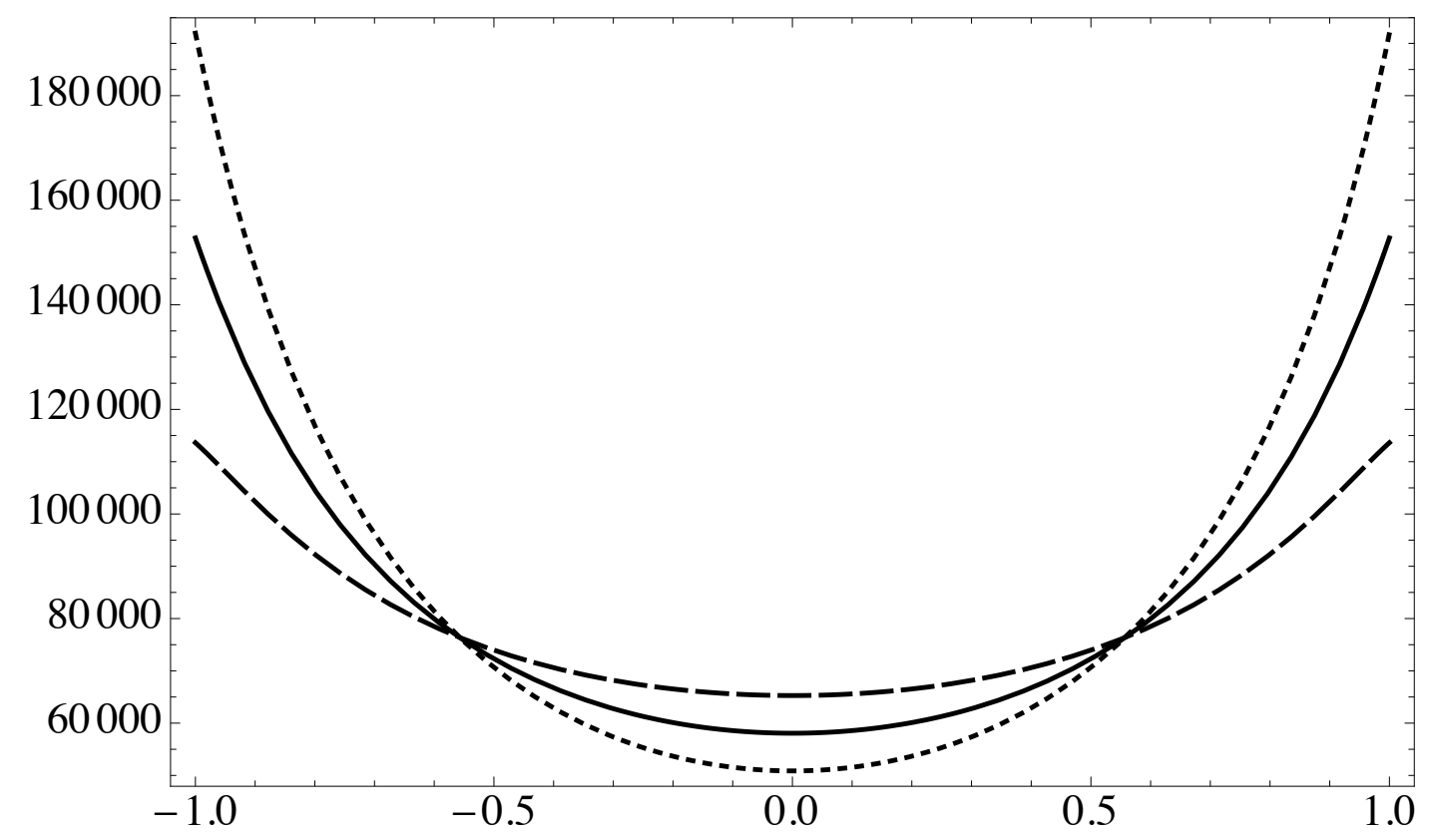

**Figure 9.19** – As Figure 9.18 for $E_\gamma = 800$ keV.

A few comments are in order:

- these curves point to the need of a detector that discriminates between photons and charged particles, to get rid of the considerable background from this process;
- the threshold effects in the differential cross-section are important and cannot be neglected;
- the Breit-Wheeler process has never been observed to date [9.78,9.79], and its first observation and measurement would itself be an interesting achievement.

### 9.4.4. *Interaction region and detector*

Given the energy of the incoming and of the scattered photons – in a range close to the electron rest energy – the detector can in principle be very similar to those used in PET systems, where annihilation photons are emitted back to back and detected in arrays of scintillating crystals, using photomultipliers or solid-state detectors like APD's. To date there are many good reviews of detector design in PET, see e.g., [9.80], [9.81].

The design of the detector in the present context is complicated by its interaction with the need of a sweeping magnet and of the (low-energy) photon beam optics, and at the time of writing the project is not yet mature enough to allow a detailed description of the components, although we can already specify a few important features:

- the electron beam intensity and the optical laser intensity are not sufficient to achieve the required photon-photon luminosity, and this requires an ELI-like optical recirculator scheme [9.82];
- the optical recirculator complicates the interaction region, however a measurement of the photon-photon cross-section is still feasible with the main detector lying on the optical table that hosts the optical recirculator;

- since some photons of the scattered photon-photon pair are lost, it is important to measure with good precision the energy of the single photons. In this way we can identify single photons from scatterings and avoid wasting the available solid angle;
- the detectors must be able to discriminate electrons, and this can be achieved either with tracking detectors like those described in Ref. [9.83], or with very fine-grained calorimeters, so that photons release all their energy in single pixels (while electrons release their energy over long paths, i.e., over many pixels).

Figure 9.20 shows a schematic picture of the interaction region, taken from the first of Ref. [9.53]: polarized photons from a powerful laser are Compton-backscattered by the counterpropagating electron beams. The resulting high-energy photons are emitted in narrow cone with aperture $\sim 1/\gamma$ , where $\gamma$ refers to the electron beam; higher energy photons are closer to the center of the cone, and are surrounded by a cloud of lower energy photons. Thus the geometry of the interaction region also defines the energy distribution at the interaction point.

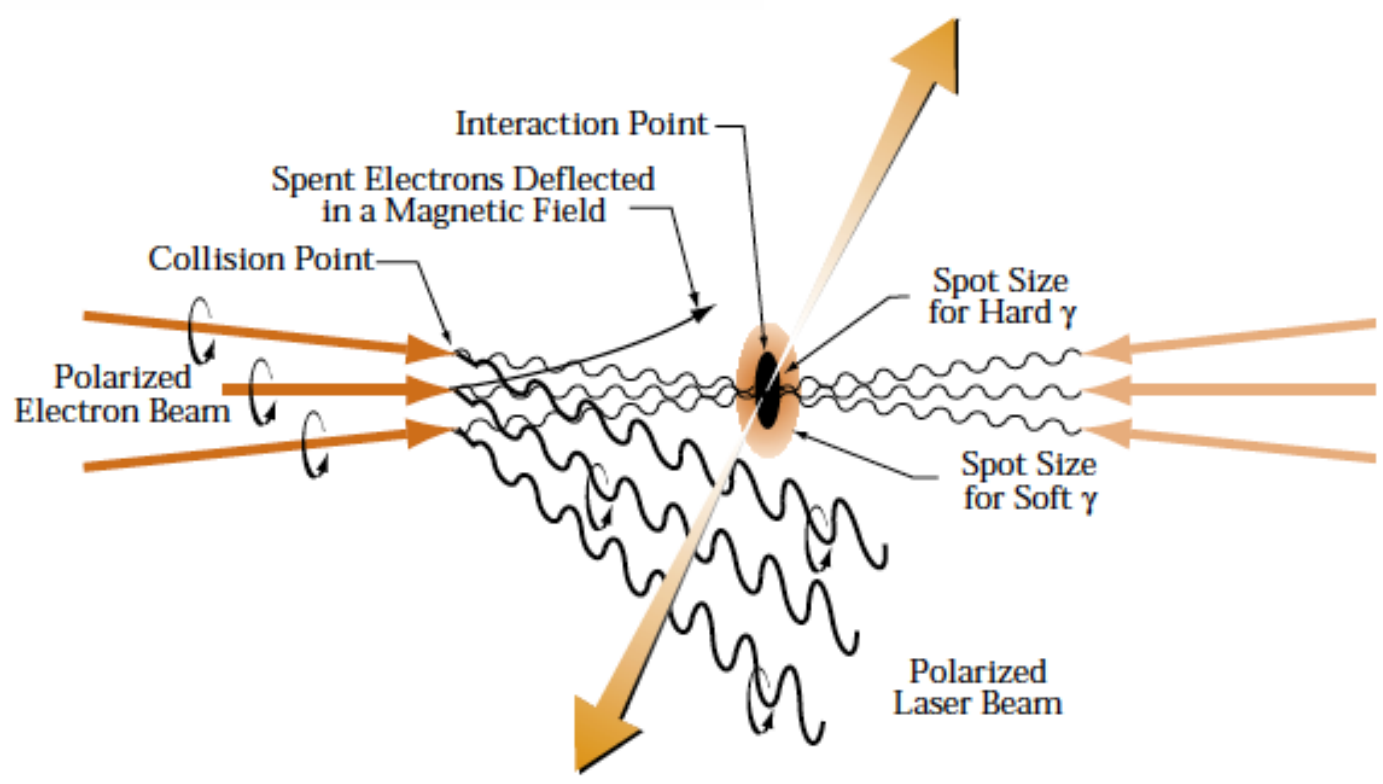

**Figure 9.20 -** A sketch of the interaction region [9.53]. See text for a description.

This already complex interaction region must be embedded in the optical recirculator: a schematic view of the ELI recirculator is shown in Figures 9.21 and 9.22 (from Ref. [9.84]), and such a scheme could be replicated here.

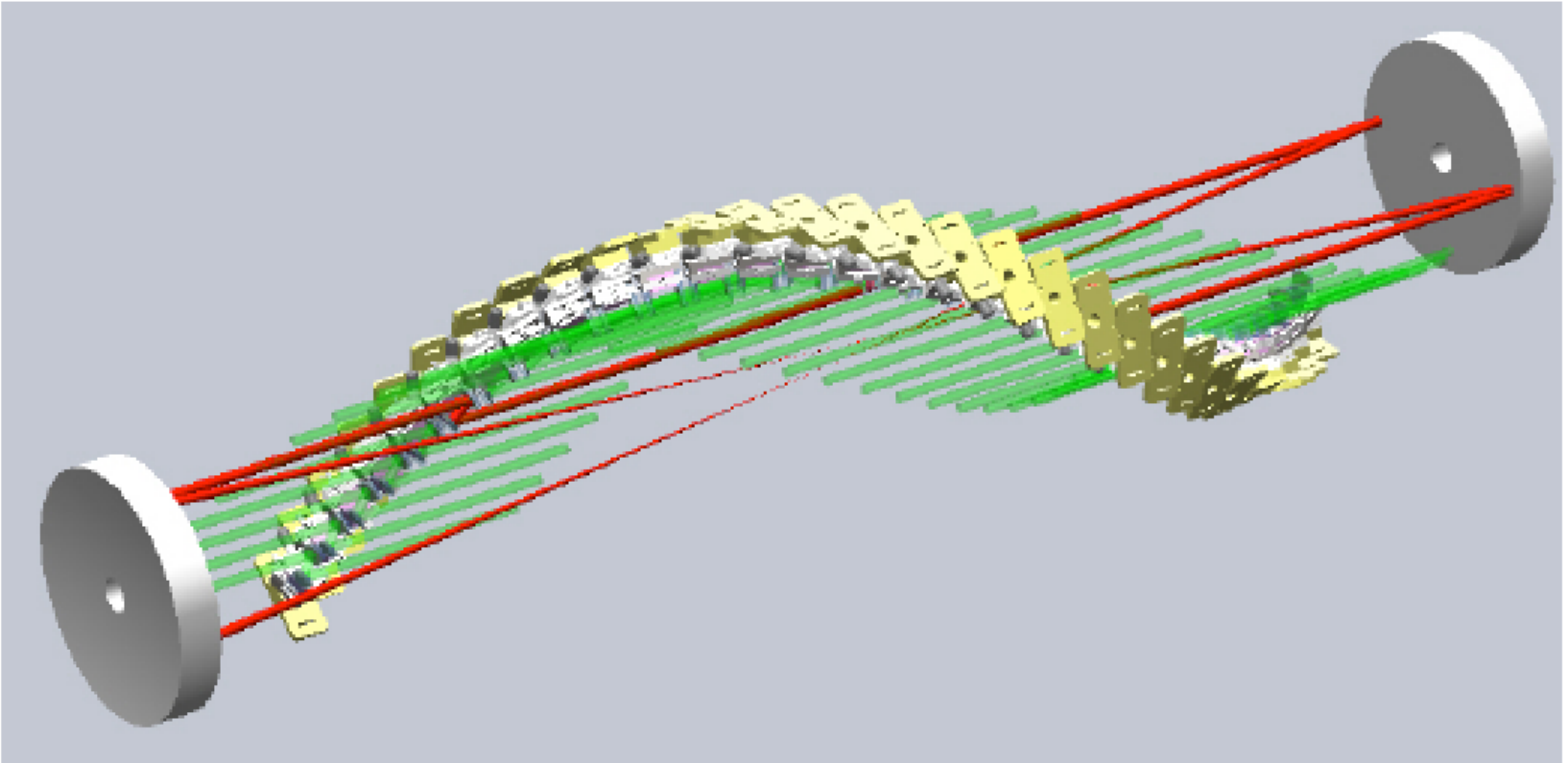

**Figure 9.21.** Schematic view of the optical recirculator [9.84]. The laser beam is repeatedly reflected by the end mirrors, like in a conventional multipass cavity. However, in order to achieve multiple passes at the very center of the cavity, the recirculator needs an array of auxiliary mirror pairs (in a conventional multipass cavity the main mirrors are astigmatic, and the beam fills a large spread-out region, rather than pass through a given point many times). The electron beams pass through the holes in the main mirrors.

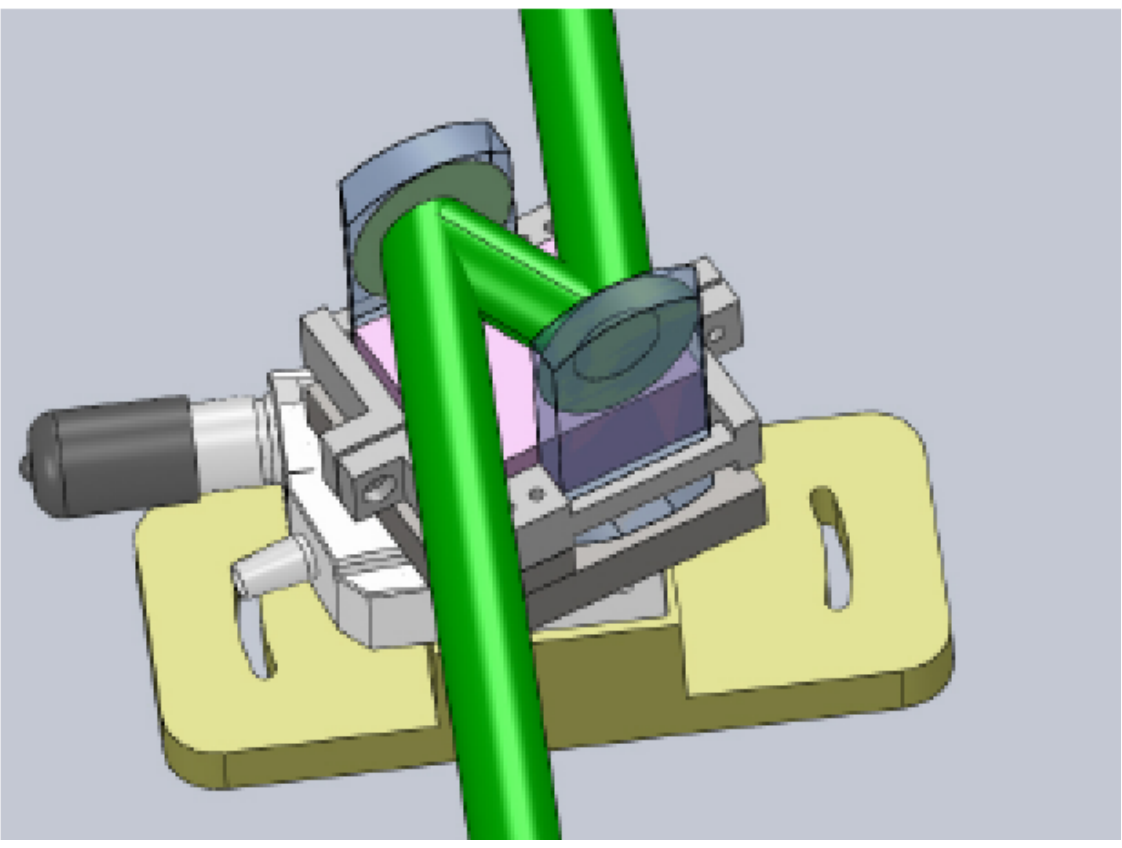

**Figure 9.22.** Arrangement of the mirror pairs used in the recirculator [9.84].

Figure 9.23 shows how the detector could fit with this complex interaction region and the recirculator. In this arrangement part of the solid angle is lost: this means that in some cases the detected scattered photons are single, and they must be identified by their energy, because there is no collinear photon to help in the event selection.

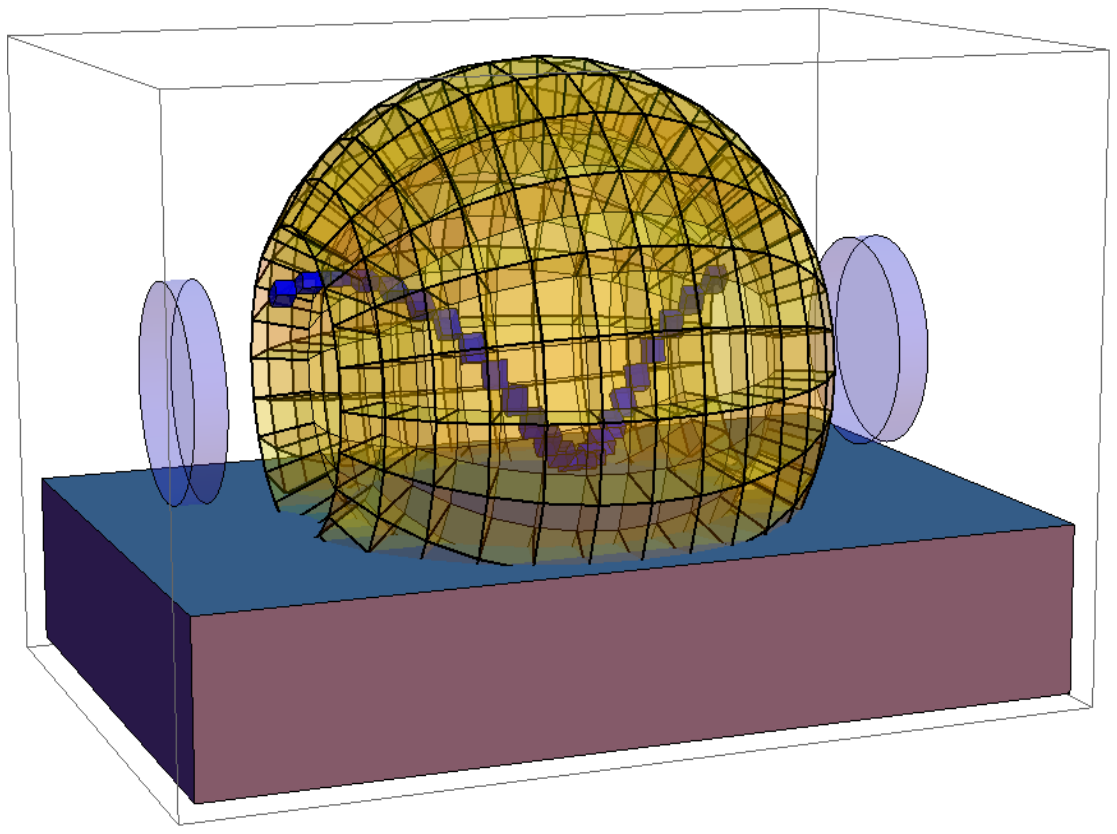

**Figure 9.23.** Schematic arrangement of the detector on the optical table that supports the recirculator.

This setup leaves many open questions, that can only be solved with a careful iterative design work, involving several steps

- design of Compton backscattering scheme that takes into account the photon directions determined by the optical recirculator;
- design of the photon polarization dynamics, which is determined by the initial polarization as well as by the many reflections in the recirculator;
- choice of detector technology;
- optimization of detector shape and positioning by means of extensive Monte Carlo simulations.

### 9.5. Physics at electron-positron collider

The systematic comparison of Standard Model (SM) predictions with precise experimental data served, in the last decades, as an invaluable tool to test this theory at the quantum level. It has also provided stringent constraints on "new physics" scenarios. The (so far) remarkable agreement between the measurements of the electroweak observables and their SM predictions is a striking experimental confirmation of the theory, even if there are a few cases where the agreement is not so satisfactory. On the other hand there are clear phenomenological facts (dark matter, matter-antimatter asymmetry in the universe) as well as strong theoretical arguments hinting at the presence of physics beyond the SM. The LHC, or a future $e^+e^-$ International Linear Collider (ILC), will hopefully answer many questions. However, their discovery potential may be substantially improved if combined with more precise low energy tests of the SM.

In this framework an electron-positron collider with luminosity of $10^{32}\,cm^{-2}s^{-1}$ with centre of mass energy ranging from the mass of the $\phi$-resonance (1 GeV) up to ~ 3.0 GeV, would complement high-energy experiments at the LHC and future linear collider (ILC). Such a machine can easily collect an integrated luminosity of about 5 $fb^{-1}$ in a few years of data taking, a statistics much larger than that collected at any previous machine in this energy range. This will allow one to measure the $e^+e^-$ cross section to hadrons with a total fractional accuracy of 1%, a level of knowledge that has relevant implications for the determination of SM observables, like, the *g*-2 of the muon and the effective fine-structure constant at the $M_Z$ scale. The latter are, through quantum effects, sensitive to possible beyond SM physics at scales of the order of hundred GeV or TeV.

The only direct competitor project is VEPP-2000 at Novosibirsk which will cover the center-of-mass energy range between 1 and 2 GeV with two experiments. This collider has started first operations in 2009 and is expected to provide a luminosity ranging between $10^{31}$ $cm^{-2}$ $s^{-1}$ at 1 GeV and $10^{32}$ $cm^{-2}$ $s^{-1}$ at 2 GeV. Other "indirect'' competitors are the higher energy $e^+e^-$ colliders (τ-charm and B-factories) that can cover the low energy region of interest only by means of radiative return (ISR). However, due to the photon emission, as we will show later, the "equivalent" luminosity produced by these machines in the region between 1 and 3 GeV is much less than the one expected in the collider here discussed.

In the following we will give a description of the most relevant physics issues that can be explored at this machine.

### 9.5.1. *The effective fine-structure constant at the scale $M_Z$*

Precision tests of the Standard Model require the appropriate inclusion of higher-order effects and the knowledge of the input parameters to the best possible accuracy. One of the basic input parameters is the fine-structure constant $\alpha$, determined from the anomalous magnetic moment of the electron with an impressive accuracy of 0.37 parts per billion (ppb) [9.85], relying on the validity of perturbative QED [9.86]. However, physics at non-zero squared momentum transfer $q^2$ is actually described by the effective electromagnetic coupling, the 'running' $\alpha_{em}(q^2)$, rather than by the low-energy constant $\alpha$ itself. The evolution of the fine-structure constant from the Thomson limit to higher energies involves low-energy non-perturbative hadronic effects, which spoil the precision of the determination of $\alpha$ at high $q^2$. In particular, the effective electromagnetic coupling at the scale of the Z boson, $\alpha_{em}(M_Z)$, is a key ingredient in the global electroweak fits of the SM. Its uncertainty affects the indirect determination of the Higgs mass already at present, and the limitations will become more severe as other parameters (like the top quark mass) are determined ever more precisely. As measurements of the effective electroweak mixing angle $\theta_W$ at a future linear collider may improve its precision by one order of magnitude, a much better accuracy of $\alpha_{em}(M_Z)$ is required. One can write:

$$\alpha(M_Z{}^2)=\frac{\alpha}{1-\Delta\alpha(M_Z{}^2)}$$

with $\Delta\alpha(M_Z^2) = \Delta\alpha_{lep}(M_Z^2) + \Delta\alpha_{had}^{(5)}(M_Z^2)$. The leptonic contribution is calculable in perturbation theory and known up to three-loop accuracy: $\Delta\alpha_{lep}(M_Z^2) = 3149.7686 \times 10^{-5}$ [9.87]. The hadronic contribution $\Delta\alpha_{had}^{(5)}(M_Z^2)$ of the five light quarks (u, d, s, c, b) can be computed from hadronic annihilation data via the dispersion relation [9.88]

$$\Delta\alpha_{had}^{(5)}(M_Z^2) = -\left(\frac{\alpha M_Z^2}{3\pi}\right)\mathrm{Re}\int_{m_\pi^2}^{\infty} ds \frac{R(s)}{s(s - M_Z^2 - i\varepsilon)} \qquad (9.4)$$

where $R(s) = \sigma^0_{\text{had}}(s)/(4\pi\alpha^2/3s)$ and $\sigma^0_{\text{had}}(s)$ is the total cross section for $e^+e^-$ annihilation into any hadronic states, with vacuum polarisation and initial state QED corrections subtracted. The current accuracy of this dispersion integral is of the order of 1 %, dominated by the error of the hadronic cross section measurements in the energy region below a few GeV (see Ref. [9.24]). An uncertainty $\delta\Delta\alpha_{\text{had}}^{(5)} \sim 5 \times 10^{-5}$, needed for precision physics at a future linear collider, requires the measurement of the hadronic cross section with a precision of 1 % from threshold up to the Y peak.

The dispersion integral in Eq. (9.4) can be calculated in a different and more precise way [9.24]: it is sufficient to calculate the function $\Delta\alpha_{\text{had}}^{(5)}(s)$ not directly at $s = M_Z^2$, but at some much lower scale $s_0 = -M_0^2$ in the Euclidean region, chosen in a way such that the difference $\Delta^{(5)} = \Delta\alpha_{\text{had}}^{(5)}(M_Z^2) - \Delta\alpha_{\text{had}}^{(5)}(-M_0^2)$ can be reliably calculated using perturbative QCD (pQCD). In Eq. (9.4) pQCD is used to compute the high energy tail, including some perturbative windows at intermediate energies. An extended use of pQCD is possible by monitoring the validity of pQCD via the Adler function, essentially the derivative of the function $\Delta\alpha_{\text{had}}^{(5)}(s)$ evaluated in the spacelike region:

$$\frac{D(Q^2)}{Q^2} = -\frac{3\pi}{\alpha}\frac{d}{dq^2}\Delta\alpha_{had}^{(5)}\Big|_{q^2=-Q^2}$$

Using a state-of-the-art pQCD prediction for the Adler function one finds that $\Delta^{(5)}$ can be neatly calculated from the predicted Adler function for $M_0 \sim 2.5$ GeV as a conservative choice. Also the small missing $\Delta\alpha_{\text{had}}^{(5)}(M_Z^2) - \Delta\alpha_{\text{had}}^{(5)}(-M_Z^2)$ terms can safely be calculated in pQCD. The crucial point is that pQCD is used in a fully controlled manner, away from thresholds and resonances. This strategy allows a more precise determination $\Delta\alpha_{\text{had}}^{(5)}(M_Z^2)$ than the direct method based on Eq. (2) but it requires a very precise QCD calculation and relies on a very precise determination of the QCD parameters $\alpha_s$, $m_c$ and $m_b$. Most importantly, as shown in the lower panel of Fig. 9.23, the method relies mainly on a precise cross section measurement in the region below 2.5 GeV, which at the same time is most important for reducing the uncertainty of the prediction of the muon *g-2*.

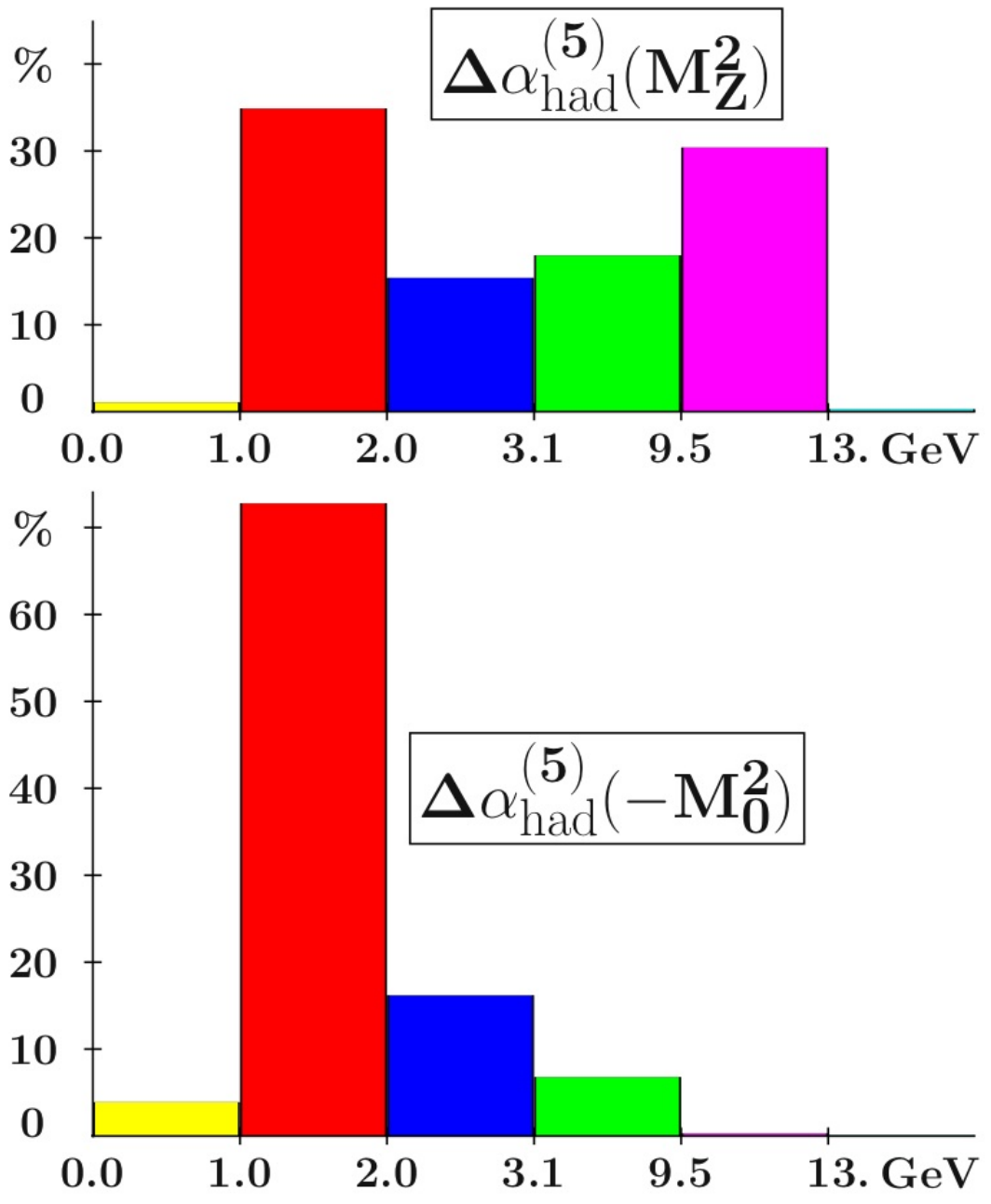

**Figure 9.23** - Present error profiles for: *(upper panel)* $\Delta\alpha_{had}^{(5)}$ $(M_Z^2)$ (standard integration); *(lower panel)* $\Delta\alpha_{had}^{(5)}$ $(-M_0^2)$ (Adler function). As it can be seen with this second method about 70 % of the total error comes from the region below 2 GeV.

### 9.5.2. *The muon (g-2)*

Like the effective fine-structure constant at the scale $M_Z$, SM determination of the anomalous magnetic moment of the muon $a_\mu$ is presently limited by the evaluation of the hadronic vacuum polarisation effects, which cannot be computed perturbatively at low energies. However, using analyticity and unitarity, it can be shown [9.89] that this term can be computed from hadronic $e^+e^-$ annihilation data via the dispersion integral:

$$a_\mu^{HLO} = \frac{1}{4\pi^3}\int_{m_\pi^2}^{\infty} ds K(s)\frac{R(s)}{s}$$

where the kernel function *K(s)* decreases monotonically with increasing *s*. This integral is similar to the one entering the evaluation of the hadronic contribution $\Delta\alpha_{had}^{(5)}$ $(M_Z^2)$ (see Eq. (9.4)). Here, however, the kernel function in the integrand gives a stronger weight to low-energy data. The contributions to $a_\mu^{HLO}$ and to its uncertainty $\delta a_\mu^{HLO}$ from different energy regions are shown in Fig. 9.24. The region below 2.0 GeV accounts for about 95% of the squared uncertainty $(\delta a_\mu^{HLO})^2$, 55% of which comes from the region 1-2 GeV [9.24].

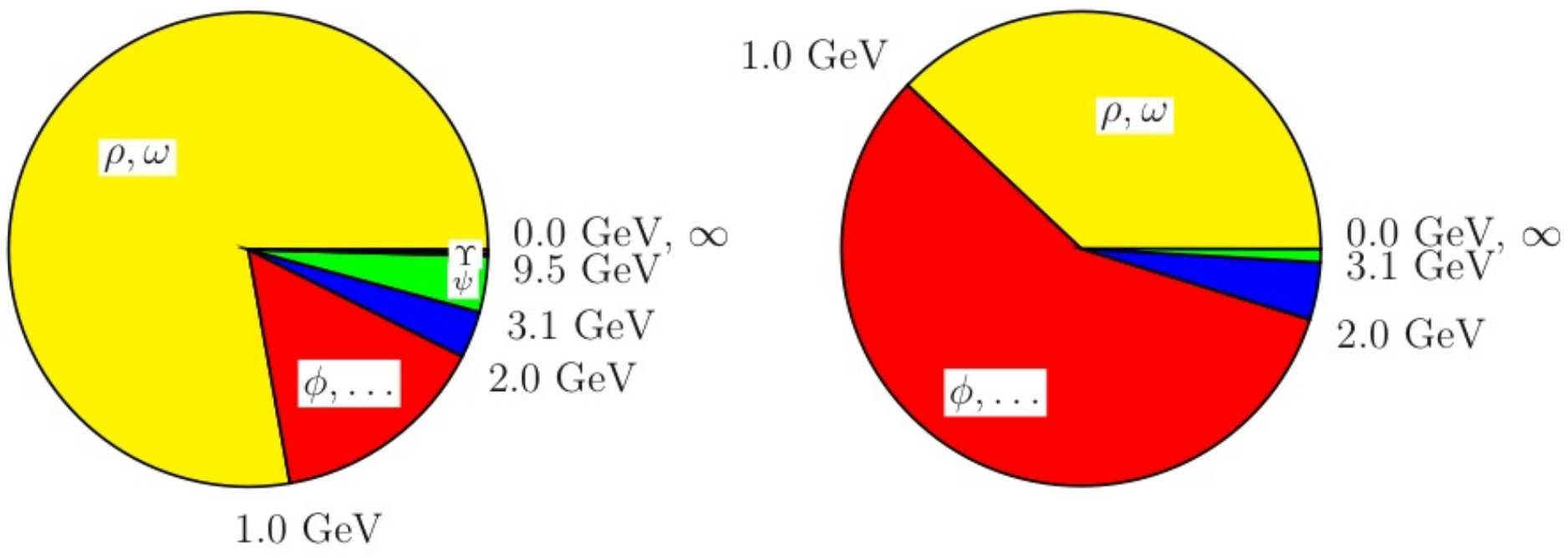

**Figure 9.24** - The contributions of the different energy regions to $a_\mu^{HLO}$ *(left)* and $(\delta a_\mu^{HLO})^2$ *(right)*. The error of *i*th contribution is $\delta^2_{i,tot}/\Sigma_i\, \delta^2_{i,tot}$ in %. The total error combines statistical and systematic errors in quadrature.

In the last few years several papers have been published aiming to determine the SM value $a_\mu^{SM}$, including the evaluation of the $a_\mu^{HLO}$ term based on new measurements of the $e^+e^-$ hadronic cross-sections at low energy (particularly at VEPP-2M, DAFNE, BEPC, PEP-II and KEKB). The resulting estimates (for a recent review see Ref. [9.90]) are systematically lower than the experimental result $a_\mu^{EXP} = 11659\ 2089\ (63) \times 10^{-11}$ [9.19] by an amount (depending on the estimate used for $a_\mu^{HLO}$) between 3.2 and 3.6 standard deviations

$$\Delta a_\mu = a_\mu^{EXP} - a_\mu^{SM} = 287(80) \times 10^{-11}.$$

As widely discussed in the literature, this result could well be the first indirect signal of physics beyond the SM. Deviations of this size are indeed expected in several realistic NP-scenarios, such as the minimal supersymmetric extention of the SM.

As for the error on $a_\mu^{SM}$, the more recent estimate [9.24] of main contributions to it are shown in Table 9.4. The two dominant contributions to the uncertainty, namely the lowest order hadronic term (HLO) and the so called hadronic Light-by-Light scattering term (LbL), are shown separately. In order to clarify the nature of the observed discrepancy between theory and experiment, and eventually reinforce its statistical significance, new direct measurements of the muon *g-2* with a fourfold improvement in accuracy have been proposed at Fermilab [9.20] and J-PARC [9.91]. With these experiments the uncertainty of the difference $\Delta a_\mu$ between the experimental and the theoretical value of $a_\mu$ will be dominated by the uncertainty of the hadronic cross sections at low energies, unless new experimental efforts at low energy are undertaken. The last column of Table 9.5 shows a future scenario based on realistic improvements in the $e^+e^-$ → hadrons cross-sections measurements. Such improvements could be obtained by reducing the uncertainties of the hadronic cross-sections from 0.7 % to 0.4% in the region below 1 GeV, and from 6% to 2% in the region between 1 and 2 GeV as shown in Table 9.5.

**Table 9.4** - Estimated uncertainties $\delta a_\mu$ (in units of $10^{-11}$) according to Ref. [9.24] and (last column) prospects in case of improved precision in the $e^+e^-$ hadronic cross-section measurement (the prospect on $\delta a_\mu^{LbL}$ is an *educated guess*). In the last row is reported the uncertainty on $\Delta a_\mu$ assuming the present experimental error [9.19] of 63 (second column) and of 16 (last column) as planned by the future experiments [9.20, 9.91].

| $\delta a_\mu$ | Estimate | Prospect |
|---|---|---|
| SM | 65 | 35 |
| HLO | 53 | 26 |
| LbL | 39 | 25 |
| SM - EXP | 88 | 40 |

**Table 9.5** - Overall uncertainty of the cross-section measurement required to get the reduction of uncertainty on $a_\mu$ in units $10^{-11}$ for three regions of $\sqrt{s}$.

| | $\delta\sigma/\sigma$ (present) | $\delta a_\mu$ (present) | $\delta\sigma/\sigma$ (prospect) | $\delta a_\mu$ (prospect) |
|---|---|---|---|---|
| $\sqrt{s} < 1$ GeV | 0.7 % | 33 | 0.4 % | 19 |
| $1 < \sqrt{s} < 2$ GeV | 6 % | 39 | 2 % | 13 |
| $\sqrt{s} > 2$ GeV | | 12 | | 12 |
| Total | | 53 | | 26 |

In this scenario the overall uncertainty on $\Delta a_\mu$ could be reduced by a factor 2. In case the central value would remain the same, the statistical significance would become 7-8 standard deviations, as it can be seen in Fig. 9.25.

The effort needed to reduce the uncertainties of the $e^+e^- \rightarrow$ hadrons cross-sections according to Table 9.5 is challenging but possible, and certainly well motivated by the excellent opportunity the muon *g-2* is providing us to unveil (or constrain) "new physics" effects. Once again, a long-term program of hadronic cross section measurements at low energies is clearly warranted.

### 9.5.3. *Measurement of the hadronic cross-sections below 2.5 GeV*

As discussed above, in the last years the improved precision reached in the measurement of $e^+e^-$ annihilation cross-sections in the energy range below a few GeV has led to a substantial reduction in the hadronic uncertainty on $\Delta\alpha_{had}^{(5)}(M_Z^2)$ and $a_\mu^{HLO}$. However, while below 1 GeV the error of the two-pion channel, which dominates the cross section in this energy range is below 1%, the region between 1 and 2 GeV is still poorly known, with a fractional accuracy of ~ 6% (see Table 2). Since, as shown in Figs. 9.23 and 9.24, this region contributes about 40% to the total error on $\Delta\alpha_{had}^{(5)}(M_Z^2)$ (up to ~ 70% by using the Adler function), and ~ 50% to the error on $a_\mu^{HLO}$, it is evident how desirable an improvement on this region is.

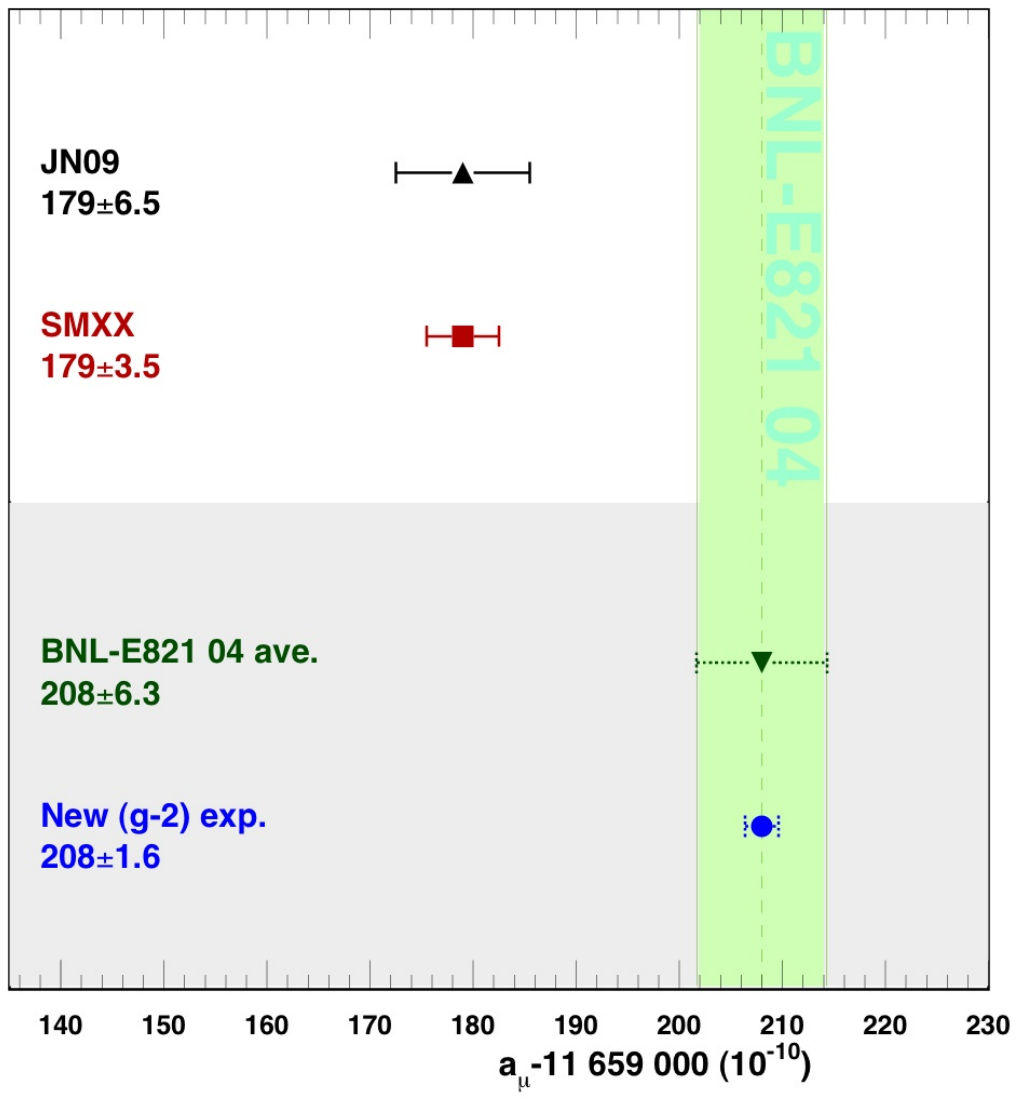

**Figure 9.25** - Comparison between $a_\mu^{SM}$ and $a_\mu^{EXP}$. "JN09'' is the current evalution of $a_\mu^{SM}$ using [9.24]; "SMXX'' is the same central value with a reduced error as expected by the improvement on the hadronic cross section measurement at the collider here proposed (see text); "BNL-E821 04 ave.'' is the current experimental value of $a_\mu$; "New (g-2) exp.'' is the same central value with a fourfold improved accuracy as planned by the future (g-2) experiments at Fermilab and J-PARC.

A variable energy $e^+e^-$ collider with a luminosity of $10^{32}$ $cm^{-2}$ $s^{-1}$, can perform a scan in the region from 1 to 2.5 GeV, collecting an integrated luminosity of 20 $pb^{-1}$ (corresponding to a few days of data taking) per point. Assuming an energy step of 25 MeV, the whole region would be scanned in one year of data taking. As shown in Fig. 9.26 the statistical yield will be one order of magnitude higher than with 1 $ab^{-1}$ at BaBar, and significantly better than at BES-III.

Figure 9.27 shows the statistical error for the channels $\pi^+\pi^-\pi^0$, $2\pi^+$ $2\pi^-$ and $\pi+\pi^-$ $K^+K^-$, which can be achieved by an energy scan with 20 $pb^{-1}$ per point, compared with BaBar with published (89 $fb^{-1}$), and tenfold (890 $fb^{-1}$) statistics. As can be seen, an energy scan allows to reach a statistical accuracy of the order of 1% for most of the energy points. (Of course, one can benefit from the high machine luminosity to use radiative return as well). Comparison of exclusive vs inclusive measurements can be performed as well.

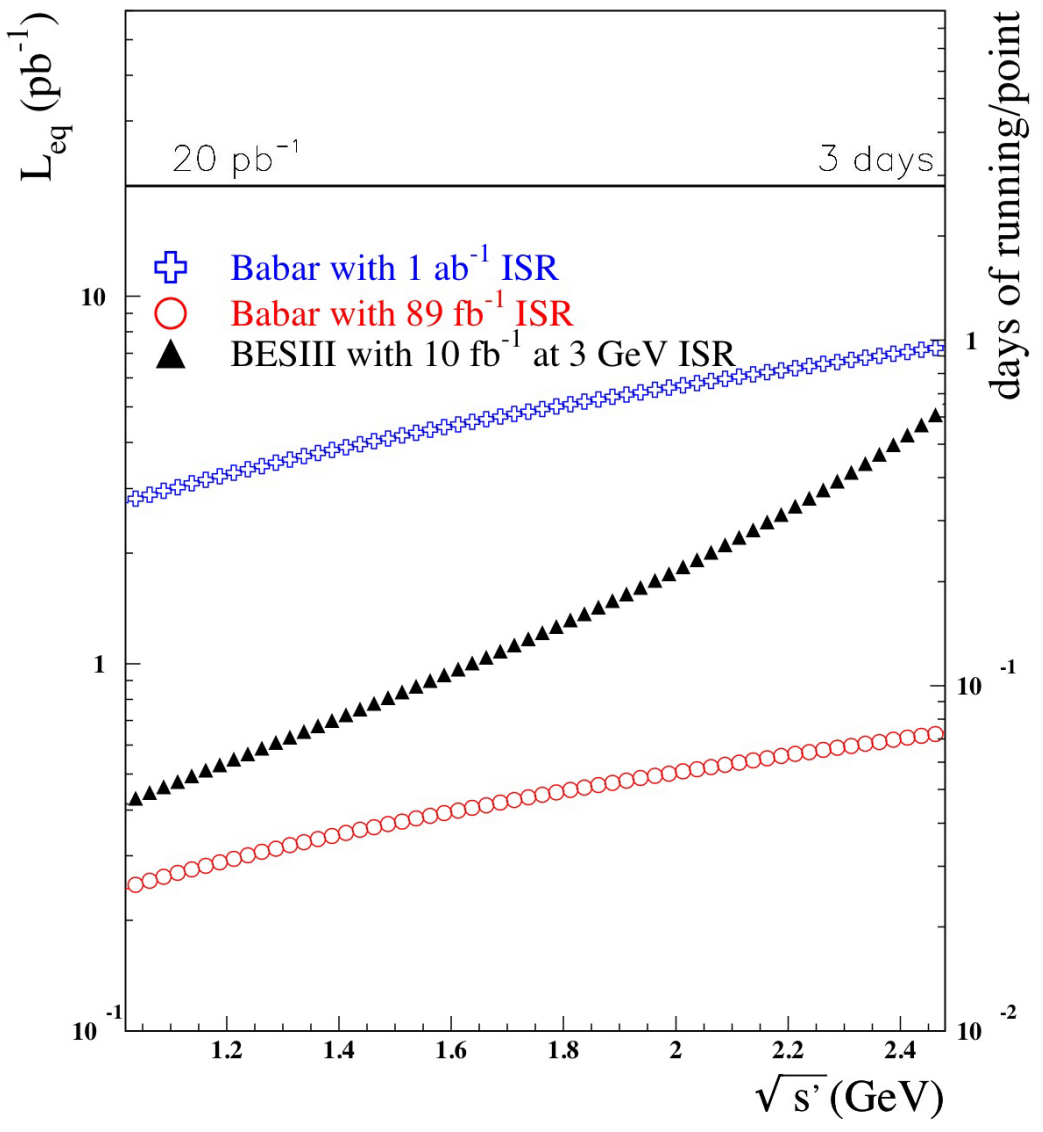

**Figure 9.26** - Equivalent luminosity for: BaBar with 1 ab$^{-1}$ (+); BaBar with 89 fb$^{-1}$ (o); BES-III with 10 fb$^{-1}$, using the ISR at 3 GeV (▲). For the solid line, see text. A photon polar angle larger than 20 degrees is assumed.

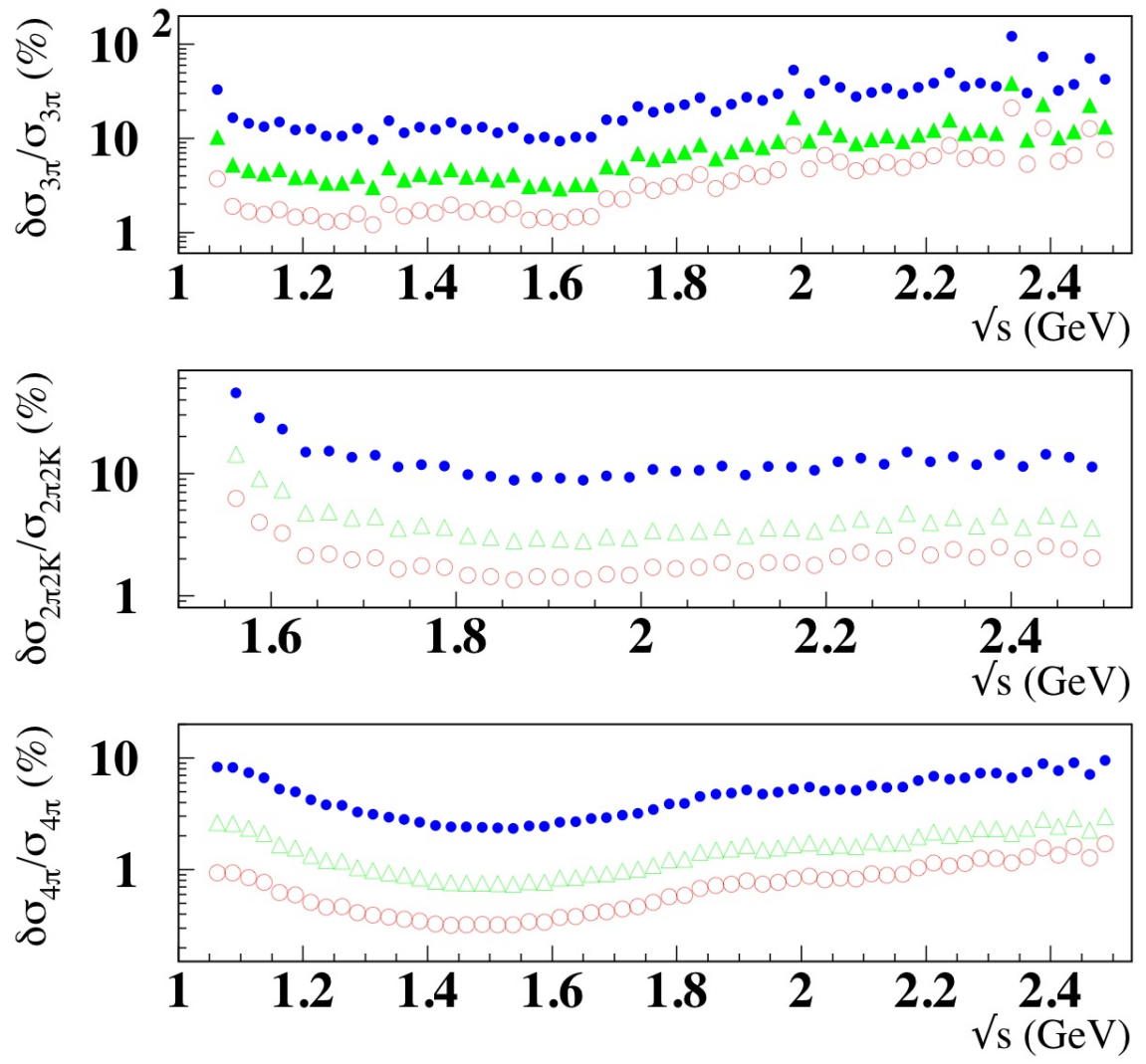

**Figure 9.27** - Comparison of the statistical accuracy in the cross-section between the proposed collider with an energy scan with 20 pb$^{-1}$ per point (o); published BaBar results (●), BaBar with 890 pb$^{-1}$ statistics (Δ) for $\pi^+\pi^-\pi^0$ (top), $\pi+\pi^-K^+K^-$ (middle) and $2\pi^+ 2\pi^-$ (down) channels. An energy step of 25 MeV is assumed.

### 9.5.4. *Other physics motivations*

#### 9.5.4.1. *Photon-photon physics*

The two-photon *(γγ)* physics program (i.e., the study of the process $e^+e^- \rightarrow e^+e^- \gamma^*\gamma^* \rightarrow e^+e^-X$) gives the opportunity to investigate many aspects of the low-energy regime of QCD.

Figure 9.28 shows the γγ flux function emphasizing the thresholds for producing systems of interest, as a function of the γγ invariant mass, and for three different total energy of the beams.

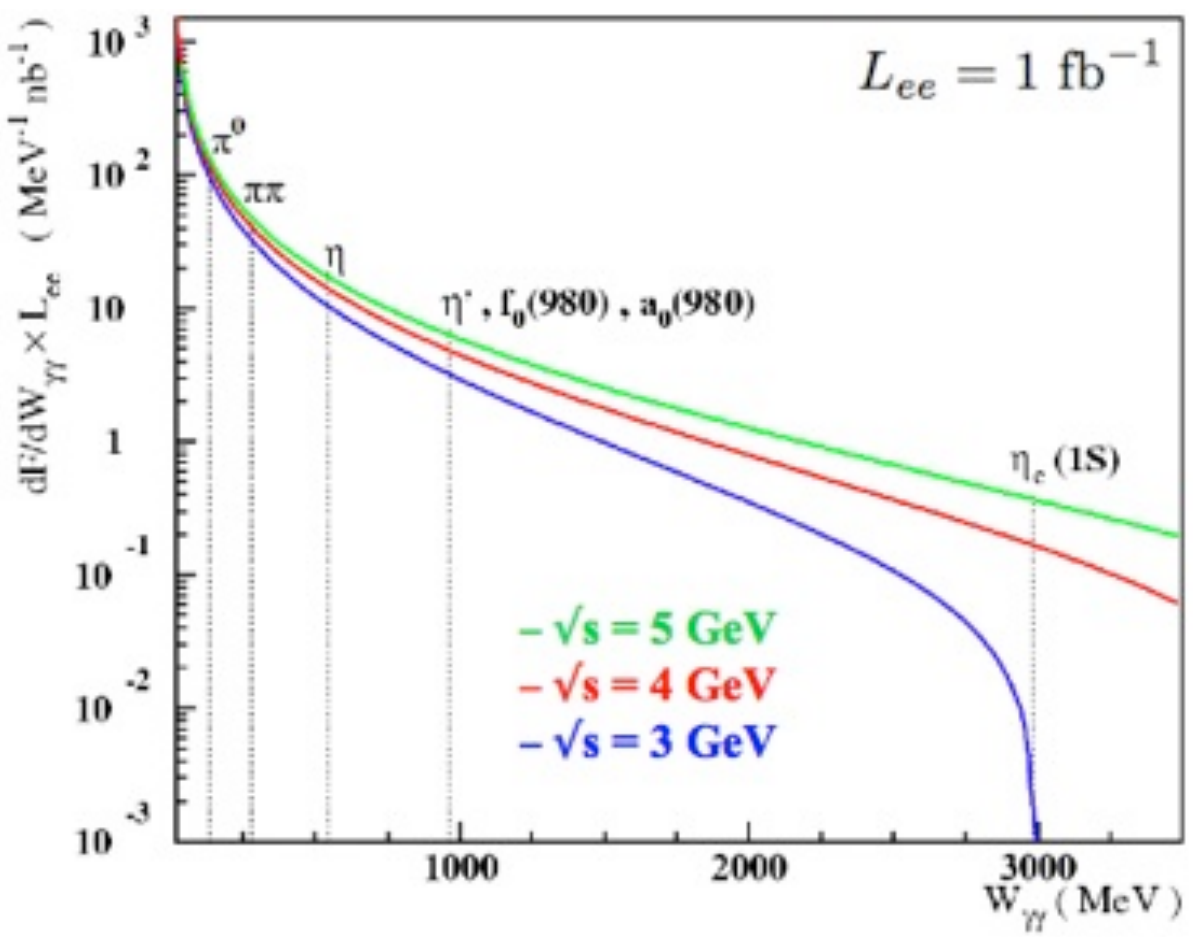

**Figure 9.28** - γγ flux function obtained using the Weizsäcker-Williams approximation.

For example, studying the process of two-pion production by two photons (X = ππ) can significantly contribute to the solution of various open questions in low-energy hadron physics. This process is a clean probe to investigate the nature of the scalar resonances. The nature of the isoscalar scalars seen in ππ scattering below 1.6 GeV, namely the $f_0(600)$ or σ, $f_0(980)$, $f_0(1370)$ and $f_0(1510)$ mesons, is still controversial. Various models have been proposed in which some of these states are qbar-q, some qbar-qbar-q-q, sometimes one is a Kbar-K molecule, and one a glueball, but definitive statements cannot be drawn. Their two-photon couplings will help unraveling this enigma.

Single pseudoscalar (X= $\pi^0$, η or η') production is also accessible and would improve the determination of the two-photon decay widths of these mesons, relevant for the measurement of the pseudoscalar mixing angle $\varphi_P$. For example, the η width $\Gamma_{\eta\gamma\gamma}$ has been recently measured to a precision of 5%, at the DAFNE $e^+e^-$ collider, using an integrated luminosity of 240 $pb^{-1}$. Collecting few $fb^{-1}$, it is possible to achieve a better precision. The high √s value allows also to carry out the precise measurements of the η' cross section produced in γγ, in the dominant decay channels. A combined fit of the different channel results may lead to a precision determination of the $\Gamma_{\eta'\gamma\gamma}$, from which one can infer the valence gluon content in the η' wave function.

Moreover, the study of the same processes gives access to the transition form factors $F_{X\gamma*\gamma*}(M_X^2,Q_1^2,Q_2^2)$ at spacelike photon momentum transfer, which is relevant for the hadronic LbL scattering contributions to the (*g-2*) of the muon. (We remark that the pseudoscalar (P) form factors can also be studied in the process. $e^+e^- \rightarrow \gamma^* \rightarrow P\gamma$). By detecting one electron at large angle with respect to the beams, the transition form factor $F_{X\gamma*\gamma*}(M_X^2,Q^2,0)$ with one quasi-real and one virtual spacelike photon ($Q^2 = -q^2$) can be measured. Experimental investigations of these form factors have been done by the CELLO, CLEO and, recently, BaBar collaborations in the range 1 GeV$^2$ < $Q^2$ < 40 GeV$^2$ using single-tagged samples. The region of very low $Q^2$ (less than 0.5 GeV$^2$, the more important for the HLbL contributions), is devoid of experimental data and is only accessible at the DAΦNE $\phi$-factory in Frascati.

By increasing the energy of the machine, new higher-mass states will become accessible: pseudoscalars (like the η'), scalars (like the $f_0$) and axial-vector (like the $a_1$) mesons. A measurement of transitions form factors for these mesons will be of fundamental importance to reduce the uncertainties that currently affect the estimates of the hadronic LbL scattering.

In addition, the precision measurements of $\gamma\gamma \rightarrow e^+e^-$, $\mu^+\mu^-$ differential distributions may hint to a possible exchange of new bosons at the GeV scale, such as those predicted in the Dark Hidden Sector (see below).

Finally, it's worth to notice that the two-photon physics program can also be pursued with an electron-electron collider (see the corresponding section). Formulae and figures are the same for both the $e^+e^-$ and the $e^-e^-$ collider mode, with the most relevant notice that background processes from annihilation are absent in the latter configuration.

#### 9.5.4.2. *Spectroscopy and Baryon Form Factors*

Cross sections of exclusive final states are also important for spectroscopy of vector mesons, whose properties provide fundamental information on interactions of light quarks. PDG lists the following vectors between 1 and 2 GeV: ω(1420), ρ(1450), ω(1650), φ(1680), and ρ(1700). However, even their basic parameters ($M, G, \Gamma_{ee}$) are badly known. In addition many states still need a confirmed identification. As discussed in [9.92] there are still many unsolved points; some progress can be achieved in ISR studies at BaBar and Belle, but such analyses are statistically limited, and a real breakthrough can be expected only at the dedicated colliders like the proposed one.

Finally, above a center-of-mass energy of $\sqrt{s} = 2m_N$ = 1.88 GeV, proton-antiproton and neutron anti-neutron pairs are produced and can be detected. The measurement of the cross-section for nucleon-antinucleon pairs allows to extract the nucleon time-like form factors. While the proton time-like form factors have been extensively measured in a wide $q^2$ region, the neutron time-like form factors are poorly known. More precise information for both time-like form factors would certainly have an important impact on our understanding of the nucleon structure.

#### 9.5.4.3. *Test of CVC*

The hypothesis of the conserved vector current (CVC) and isospin symmetry relate the $e^+e^-$ annihilation into isovector hadronic states and the corresponding hadronic decays of the $\tau$ lepton [9.93]. Using experimental data on $e^+e^- \to$ hadrons with $I = 1$ one can compare the CVC predictions and $\tau$ lepton data both for decay spectra and branching ratios. A systematic check of these predictions showed that at the (5-10) % level they work rather well [9.94]. However new high-precision data on the $2\pi$ final state challenged this statement [9.95] and some evidence for a similar discrepancy is also observed in $e^+e^- \to \pi^+\pi^-2\pi^0$. A test of CVC with very high accuracy will require detailed measurements of the energy dependence of the relevant exclusive processes, like $\pi^+\pi^-$, $4\pi$ (2 final states), $6\pi$ (3 final states), $\eta\pi^+\pi^-$, $K_S\ K_L$ and $K^+K^-$, from threshold to $\tau$ lepton mass.

#### 9.5.4.4. *Searches for physics beyond the SM*

Low energy, high luminosity electron-positron colliders are an ideal tool to search for hypothetical U vector bosons weakly coupled to SM particles. These bosons are predicted in extensions of the SM recently appeared in the literature as a consequence of intriguing and, as yet not completely explained, astrophysical observations. In fact, KLOE, BaBar and BES-III have already started measurements in the field. There are several possible signatures to look at, such as $e^+e^- \to e^+e^-\ \gamma$, $e^+e^- \to \mu^+\mu^-\ \gamma$, $e^+e^- \to E_{missing} + \gamma$, $e^+e^- \to E_{missing} + e^+e^-$, or events with 4 or 6 leptons in the final state. A careful analysis of such reactions in the region of interest for this proposal would complement the above mentioned searches, particularly in the case of the channels with missing energy or multilepton jets. The cross sections for these processes are expected to be of the order of 10-100 fb, thus one could expect to observe a few hundred events.

## 9.6. Physics at electron-electron collider

### 9.6.1. *Photon-photon physics*

Electron-electron collider at $\sqrt{s} \sim$ 3-4 GeV with a Luminosity of $\sim 10^{32}$ cm$^{-2}$ sec$^{-1}$ is an ideal laboratory for two-photon physics due to the absence of the $J = 1^{--}$ annihilation channel (as present in $e^+e^-$ interactions). The absence of this background makes it possible, for example, to precisely measure the cross sections of the production of two charged pseudoscalars (like $\pi^+\pi^-$) or axial-vector particles by two photons which are very important to constrain the model dependence of the hadronic contribution to HLbL scattering of the *g-2* of the muon.

As discussed in the e-gamma section, the usual approach to classify the HLBL diagrams, is to consider both the chiral and the large Nc counting. We stated that the dominant piece in this calculations come, numerically, from the exchange of pseudoscalars. This is, indeed, a numerical fact and all the computation so far agreed on this fact. Strictly speaking, however, the leading-order piece of the combined chiral and large-$N_c$ framework comes from other processes, the pion and kaon loops. These processes represent the leading-order in the chiral perturbation calculations (the pseudoscalar exchange $O(p^2)$ is suppressed with respect to these ones) and behave as $O(N_c^0)$ in the large $N_c$ counting. Models for such pion and kaon loops have problems with the renormalization scale. At the end they are not predictive enough and produce results with large uncertainties. The state-of-the-art average is $(2 \pm 2) \times 10^{-10}$. Until now this supposed small quantity did not worry the community since both experimental and theoretical errors were much larger. Again, the forthcoming experimental precision demands better understanding of these contributions.

After the recent works of Engel et al [9.96] and Bijnens et al [9.97], the previous conclusion about the smallness of the pion and kaon loops is under discussion. Both references indicate an overestimation of such loop contributions due to underestimation of counterterms. One can relate the desired process with the $e^+e^- \to e^+e^- \gamma^*\gamma^* \to e^+e^-\pi\pi$ using the optical theorem. However, such a process has never been measured before. This, indeed, can be an interesting and promising breakthrough from IRIDE. Providing accurate experimental information for such processes would contribute to understand better the role of the leading-order ChPT piece of the HLbL of the muon.

### 9.6.2. *Determination of the effective QED running coupling and the hadronic contribution to $(g\text{-}2)_\mu$*

The running of $\alpha_{em}$ is driven by the vacuum polarization and, while the leptonic contributions are calculable to very high accuracy, the hadronic loops have to be evaluated by using a dispersion integral over the measured cross section of $e^+e^- \to$ hadrons at low energies [9.98].

This cross-section is however highly oscillating due to the contributions of several resonating states and particle production threshold effects. An alternative approach might be proposed [9.99] that uses the evaluation of the running of alpha occurs in the negative $t$ (space-like) region and, in particular, the running of $\alpha_{em}$ can be evaluated by extracting the angular dependence of the $e^-e^- \to e^-e^-$ scattering cross section.

A similar approach has been proposed for the small angle Bhabha scattering [9.100]. There, however, the unavoidable $s$-channel contribution, appearing as pure $s$ as well as interference contribution has to be subtracted. This is not the case for the $e^-e^- \to e^-e^-$ scattering being this a pure $t$-channel process.

A simple expression can be used to compute the LO hadronic contribution to the muon anomaly, $a_\mu^{HLO}$, from the vacuum polarization in the spacelike region [9.101].

Being the vacuum polarization a smooth function at negative $t$, the accuracy of the determination of the vacuum polarization function is thus only limited by the statistics and by the control of the systematics of the experiment. Being the cross section in the forward region extremely high, a precise determination of $\alpha_{em}(t)$ and of $a_\mu^{HLO}$ can be obtained in one single experiment at IRIDE, with an accuracy competitive of (or even better than) the one obtained by the standard approach which uses timelike data [9.102].

## 9.7. A further option: search for WISPs

The avalaibility of an intense laser source also offers the possibility to search for *very weakly interacting sub-eV* particles (WISPs). The discovery of these particles would be a major step in the comprehension of fundamental physics [9.103]: the solution of the Dark Matter (DM) problem could be their most important implication. Indeed, given the more and more constrained status of Supersymmetry searches at the LHC, it seems natural to look back to axion interpretations of DM.

One distinguishes between Peccei-Quinn axions and ALPs, Axion-Like-Particles. The former were introduced in QCD as an explanation of the "unnatural" smallness of the $\theta$ parameter, as suggested by measurements of the neutron electric dipole moment. The energy of the vacuum is found to be a function of $\theta$ and has an extremum when this parameter is equal to zero. At $\theta = 0$ also CP is conserved, as $\theta$ weights a CP-violating term in the lagrangian. On the other hand, it has been noted several times that points of higher symmetry are also stationary points. This observation would fit here if $\theta$ were a dynamical variable and not just some fixed parameter. Peccei-Quinn introduce $\theta$ a pseudoscalar field $a$ (the PQ axion) having a shift-symmetry useful to eliminate $\theta$ by a shift in $a$. The vacuum energy as a function of $\theta$ is now interpreted as the potential energy of the axion, having an extremum in $\theta = 0$.

The mass $m_a$ and the coupling to photons $\alpha/f_a$ of the PQ axions are related. This has helped in constraining them from astrophysical data, especially from white dwarf stars [9.102]. PQ axions could be the main constituents of Cold DM if $m_a \sim 1$ μeV and $f_a \sim 10^{12}$ GeV; white dwarfs put constraints in the meV range.

The relation between $m_a$ and $f_a$ (which for PQ axions is $m_a f_a \approx m_\pi f_\pi$) is lost for ALPs, generic pseudo-Goldstone bosons of other broken global symmetries [9.103, 9.5,9.105] (we will call $g$ their coupling to photons). This increases the space of experimental searches. On the other hand, in the case of ALPs, the coupling to photons, the main research channel, is not guaranteed. Indeed it is through the coupling with light that one can constrain ALPs searches studying stellar evolution (the stellar cooling could be altered by photon conversion into ALPs), searching for deviations in CMB spectra due to processes like $\gamma$ +... $\rightarrow$ ALP +..., studying gamma-ray bursts and devising experiments with laser light (like "shining-through-wall" experiments) or radiofrequency, like ADMX [9.106].

In particular, laboratory experiments promise to beat the astrophysical and cosmological bounds on ALPs which set the benchmark point to $m_a \lesssim 10^{-12} \div 10^{-9}$ eV and $g \lesssim 10^{-12} \div 10^{-11}$ GeV$^{-1}$. These are mostly "light-shining-through-wall" experiments: two cavities tuned in such a way to be resonant with each other, separated by a thick wall impeding the passage of anything but ALPs. The latter are produced by photon conversion in one of the two cavities, say cavity 1, and pass though the wall regenerating a photon in the second cavity by the inverse process to ALP production. Cavity 1 can be fed with laser light traveling back and forth in a magnetic field (inducing the Primakoff $a$-$\gamma$ conversion), whereas the magnetic field in cavity two is responsible for the reappearance of a photon in it. In normal cavities used, light can make $\approx 10^5$ passages. (Radiofrequency could be a step forward as $10^{11}$ passages can be sustained for microwaves in similar cavities: this largely increases the exposition of the microwaves to the magnetic field making the axion conversion more likely to occur. )

Experiments with polarized laser light are also very much sensible in ALPs searches. One can inject a Fabry-Perot cavity with laser light linealy polarized and switch on a magnetic field **B** orthogonal to the beam direction. The polarization plane of the injected light is prepared with an angle $\theta$ with respect to the direction of **B**, i.e., the photon beam has a polarization component along **B** and one orthogonal to it according to θ. The photons with a large component of their polarization along **B** are those most effectively converted into (pseudoscalar) ALPs. The latter escape from the cavity depleting the number of aligned photons: overall one should observe a rotation of the polarization plane, with a background-free discovery potential down to $g \sim 1/M_P$, where $M_P$ is the Planck mass. One can also seek for linear → elliptical polarization effects due to a phase shift induced by the external magnetic field between different polarization directions.

Another kind of WISPs are minicharged particles, like electrons with a very low mass and a tiny fraction of electric charge. They are usually associated to hidden photon sectors. Benchmark points for direct and indirect searches are in the range $m_\varepsilon \lesssim 1$ meV and $\varepsilon \sim 10^{-8} \div 10^{-9}$. Such particles might be produced and detected using some apparatus exploiting the Schwinger effect. Indeed the rate of pair production of electron-positron pairs due to a strong electric field **E** is $\Gamma \gtrsim (eE)^2/(4\pi)^3 \exp(-m_e^2/eE)$. The ratio $m_e^2/e$ requires electric fields $E \gtrsim 10^{16}$ V cm$^{-1}$ to get any non-negligible pair-production rate. Minicharged particles could allow these fields to be as low as $10^6$ V cm$^{-1}$ for $\varepsilon \sim 10^{-7}$.

## 9.8. Conclusions

The Particle Physics program offered by IRIDE is quite reach: it goes from precision test of the Standard Model through a precise determination of the anomalous magnetic moment of the muon and the effective fine-structure constant, as well as the the squared sine of the weak angle. It allows also to search for something new at low energy, like, e.g., elusive light bosons, and therefore firmly establish (or strongly constrain) new physics effects.

IRIDE will complement high-energy activities at the LHC and future linear colliders through its ability to improve the determination of precision observables of the SM, like, e.g., *(g-2)* of the muon, which are, through quantum effects, sensitive to possible BSM physics at high scales of the order of hundred GeV or TeV.

Also a rich hadron phenomenology is accessible at these scales, which allows to study issues of the QCD confinement and test effective field theory of strong interaction, where the ordinary perturbation theory approaches fail to work and one has to rely on models. Low-energy photon-photon collisions give a direct view into the vacuum properties of Quantum Electrodynamics (QED), allowing for precision tests of QED in the MeV range, and more generally of Quantum Field Theory (QFT).

"classical electron radius" $r_0$ as in the case of, e.g., Compton scattering. However, pair production by two real photons has not been observed experimentally since it is difficult to prepare two colliding beams of high-energy photons."

# 10. ADVANCED ACCELERATOR CONCEPTS

## 10.1. Introduction

In this section we outline the case for the construction of IRIDE-PW, a Petawatt Laser Installation at IRIDE aimed at delivering advanced accelerator techniques based on the exploitation of the unique properties of plasmas.

The need for the development of new concepts for particle acceleration originated from the limit of conventional RF accelerating cavities in using higher accelerating fields due to breakdown. This limits the gradient of RF cavities to 50-100 MV/m. TeV scale linear colliders require, with such a technology, tens of km to accelerate the beams up to their final energy. Future accelerators with ultra-high fields will not be based on conventional metallic resonant cavities, also because the power necessary to excite such structures becomes excessive. This leads to consideration of shorter wavelength linac structures. Existing linacs, which work in the 1 to 10 cm wavelength range, would be naturally scaled down to mm and sub-mm wavelengths for GV/m operation, ideally in the THz region, $\lambda \sim 0.3$ mm where, however, limitations come from the lack of a suitable power source to achieve GV/m fields. A variety of advances acceleration techniques are therefore being explored and plasma based schemes are considered a very promising alternative to RF devices for future accelerators.

Indeed, advanced acceleration is the building block of the European Extreme Light Infrastructure (ELI), a European project selected in 2007 by the European Strategy Forum for Research Infrastructures. ELI foresees the use of ultraintense lasers, also combined with conventional accelerators, to enter new domains of research in the intense field physics which are expected to deliver benefits to society with more practical exploitation of intense lasers and new discoveries in fundamental science. Prompted by this initiative, a number of new high power laser installations have been established or are currently being designed across Europe, with unprecedented investments in high power laser installations for civilian applications. A similar path is being followed in the rest of the world, with new installations coming on line in the US, Japan, Canada, Cina, S.Korea, just to mention a few. These installations will allow plasma accelerators schemes to be scaled to higher particle energy and will enable further exploration of possible future scheme for high energy linear colliders for particle physics [10.1].

Key to the success of these initiatives is the cooperation between laser, plasma and accelerator physics communities. Examples of such cooperation are already emerging worldwide at some existing accelerator facilities and are starting to demonstrate the great potential of advanced acceleration techniques. An installation like IRIDE is the ideal location where these communities can cooperate and build the expertise needed for such cross-field research. The international scenario indeed indicates that our proposal of establishing a high power laser system at IRIDE is timely. The construction and running cost are expected to be a small fraction of the costs of the entire installation.

In details, there are two main strong reasons to pursuit innovative acceleration schemes in the framework of IRIDE project. On one end, plasma acceleration could be used [10.2] to further increase the energy of the electrons of the main linac using plasma based acceleration stages, either laser driven or particle driven. On the other end, we foresee the possibility to use a world-record powerful laser to explore all the novel “all-optical” schemes based of laser-driven acceleration based on Laser Wakefield Accelerator (LWFA) [10.1] in which electrons are accelerated from the plasma itself (self-injection) in the so-called bubble regime [10.3]. In addition to these two main electron acceleration schemes, a number of laser-plasma interaction schemes exist which can be used to accelerate ions [10.4]or to produce positrons and neutral beams of major interest for astrophysics [10.5].

Currently all these schemes are being explored at different facilities and physics models are being developed using the most advanced numerical techniques available nowadays. The SPARC_LAB facility at LNF was indeed established to start developing these studies over a wide range. It features an ultraintense laser system (FLAME) which was designed to deliver the laser pulses energy, power and intensity required to access the plasma acceleration regimes introduced above. The knowledge already acquired by our community in using FLAME, as well as other facilities world-wide, including for example, Astra-Gemini at RAL, the LOA at Palaiseau, UHI-100 at Saclay, J-Karen at KPSI, Japan, make us confident in conceiving and planning the next *IRIDE–PW* installation were scaling of advanced acceleration techniques at higher laser intensities will be studied and established.

## 10.2. Laser-driven plasma acceleration

In view of the above, a plasma can be considered ideal medium as it is immune from further breakdown (it is already ionized) and can naturally support electron plasma waves characterized by very high longitudinally electric fields, up to tens of GV/m. Depending on the density of particles, the phase velocity of the wave can be very close to the speed of light and electrons trapped in the wave can experience high accelerating fields for sufficiently long distances to reach relativistic energies.

| GAS-JET | GAS CELL | CAPILLARY DISCHARGE | HOLLOW FIBER | SOLID TARGET(μm) | SOLID TARGET (<<μm) | … |
|---|---|---|---|---|---|---|
| 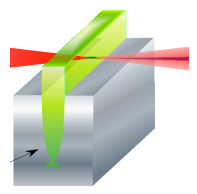 | 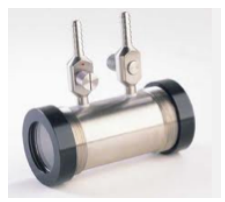 | 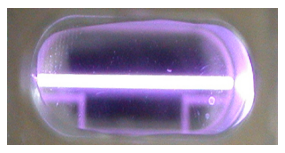 | 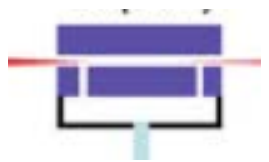 | 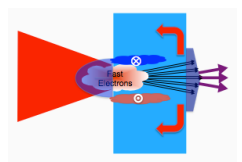 | 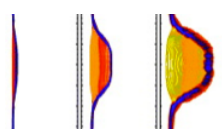 | … |
| LWFA (non linear) Self-injection | LWFA (non linear) Self-injection | LWFA (linear) Self-injection | PWFA Particle driven with COMB-like structure | TNSA Ion acceleration by fast electrons | RPA Pondero-motive Ion accelera- tion | … |
| High density | High density | Medium density | Low to medium density | Solid density | Solid density | … |

**Table10.1**. A number of configurations of target plasmas for acceleration of particles. Each target enables plasma geometrical and physical properties ideal for different acceleration schemes.

These simple principles are behind a worldwide intensive R&D to develop alternative methods to accelerate electron beams. The currently explored plasma acceleration methods mainly differ from the specific target plasma configuration used and for the plasma wave driver configuration. A non-exhaustive list of targets is summarized in the Table 10.1 showing different techniques to achieve the various plasma parameters needed for the specific acceleration mechanism. Depending on the background electron density of the plasma, different laser guiding schemes have been proposed and investigated experimentally as shown above. For relatively high densities, relativistic self-focusing (RSF) can occur provided that the laser power exceeds the critical power. Increase of the propagation length due to RFS was demonstrated in a LWA of self-injected electron [10.6]. Alternative guiding techniques include capillary discharge [10.7] and gas filled hollow waveguide [10.8]. As discussed in the next section, guiding techniques have been demonstrated to be effective tools in controlling beam propagation in plasmas and offer effective how-to approaches to the possibility of multi-staging of laser-driven electron accelerators.

In addition to electron acceleration, plasmas can also accelerate protons and ions. In this case, acceleration relies of the onset of a quasi-static accelerating electric field at the rear of an irradiated solid target. More advanced schemes based upon the effect of the radiation pressure will be accessible with the next generation of laser line IRIDE-PW. In the following we focus our attention the most promising acceleration processes involving plasma also exploiting dielectric structures. Also, we focus briefly, at the end of this chapter, on an interesting astrophysics application of the intense laser pulse used for plasma acceleration

### *10.2.1. Laser Wakefield*

As outlined above, several schemes can be considered in order to produce high energy particle beams using a plasma accelerator. Among these, laser driven plasma accelerators are currently developing rapidly thanks to the impressive development of ultrashort, high power laser technology based upon Chirped Pulse Amplification (CPA) [10.9]. When a single, very intense, short laser pulse is focused inside a plasma, due to its strong ponderomotive force, it can generate a quasi-resonant electron plasma wave along the laser propagation path. This approach leads to a substantial reduction of the complexity of the experimental set up compared to previously used techniques like the beat-wave.

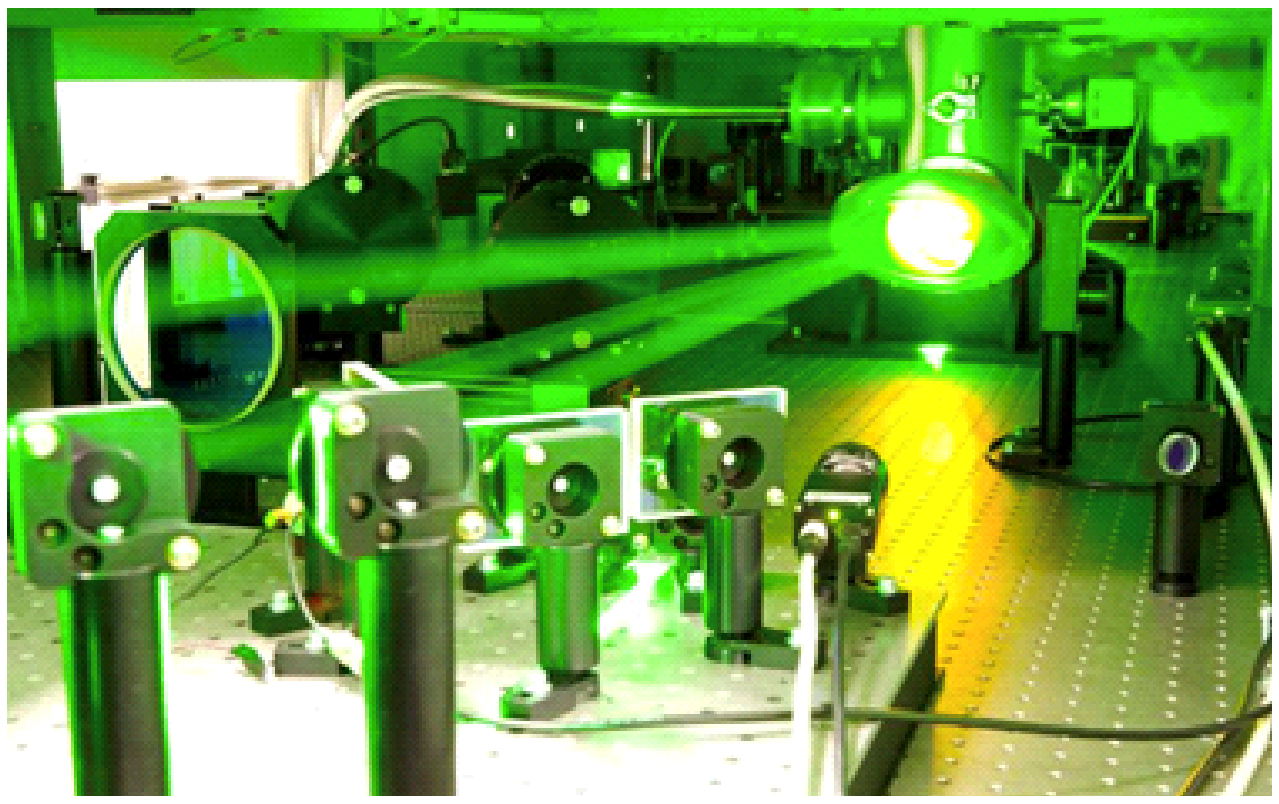

**Figure 10.1** – The final amplifier of a petawatt-scale laser system. Titanium sapphire amplifying media, pumped by frequency doubled Ne:YAG lasers can amplify laser pulses up to tens of Joule which can be then compressed to a few tens of femtoseconds, leading to Petawatt optical power.

If the longitudinal size of the laser pulse is about a half (or less) of the wavelength of the natural oscillation mode of the plasma, a high amplitude plasma wave develops quasi-resonantly on the wake of the pulse, excited by the action of the longitudinal ponderomotive force. Indeed, this so-called Laser Wakefield Acceleration (LWFA) [10.10] is currently being explored as a promising technique for the acceleration of electron bunches that takes advantage of the high longitudinal electric fields supported by electron plasma waves. Numerical simulations show that at sufficiently high laser intensities, the wake formation becomes nonlinear and the accelerating structure exhibits unique properties, including very strong longitudinal fields, focusing transverse fields and self-trapping capabilities [10.11]. In this self-injection scheme of LWFA, electrons are trapped in the plasma wave that has a phase velocity very close to the speed of light. Trapped electrons experience an accelerating field that can be as high as tens of GeV/m and can gain energy as long as they are in phase with the accelerating region of the field. Experimental campaigns carried out in several laboratories worldwide [10.12] have shown that quasi-monoenergetic electron bunches can be accelerated from the background electron plasma population up to high (>100 MeV) energies. More recently, the same schemes were used to demonstrate GeV acceleration of electrons [10.13] and multi GeV peak energy (e.g. GIST, S.Korea and University of Texas).

External injection of electrons in a laser wakefield is also being pursued as this scheme is likely to preserve the quality of electron bunches produced by conventional accelerators, while offering potential for energy gain in relatively compact schemes. The idea here is to inject electron bunches produced by an external linac in the electron plasma wave in such a way that the electron bunch experiences an accelerating field and its energy is thus increased. External injection was initially pursued using the so-called laser beat-wave driven relativistic plasma wave. However, experiments carried out so far using intense, ultrashort CPA pulses have demonstrated acceleration gradients which are up to two orders of magnitude larger than those achieved in laser beat-wave experiments and suggest a perspective of much more compact accelerators, provided that schemes for relatively long interaction length are demonstrated. In fact, since most interaction schemes are based upon propagation is a gas, ionization of the medium is carried out by the laser pulse itself. Due to the typical radial distribution of the laser intensity, a radial gradient of the refractive index in the plasma is created due to the stronger laser field on the axis, which leads to a defocusing of the beam. This is a major issue in this field and beat wave experiments are producing very interesting results which demonstrate the possibility of extending the interaction length via control of ionization induced defocusing (IID) [10.14].

### *10.2.2. Proton acceleration*

As for protons and ion acceleration by lasers, this is mainly based upon the interaction of intense laser pulses with solid targets. Laser-driven proton and ion beams with high brilliance, ultrashort (picosecond) duration and Multi-MeV energies are already being used successfully for ultrafast radiography [10.15] and for applications in nuclear and warm dense matter physics [10.16]. The next generation of laser systems, yielding powers exceeding 10 PW (such as the APOLLON laser in France, delivering 150J in 15fs), will allow to access the regime of relativistic ions (>1GeV/nucleon), suitable for HEP applications such as, e.g, the production of low energy neutrinos for CP violation studies [10.17]. Theoretical predictions in such regime have been based on the radiation pressure dominant acceleration mechanism, of which experimental signatures have been recently observed [10.18]. In addition, multi-PW lasers will allow the generation of relativistic collisionless shocks which also may both provide a mechanism for acceleration of high-quality proton beams [10.19] and a test bed for astrophysical acceleration on a laboratory scale [10.20].

### *10.2.3. Plasma Wakefield Accelerator*

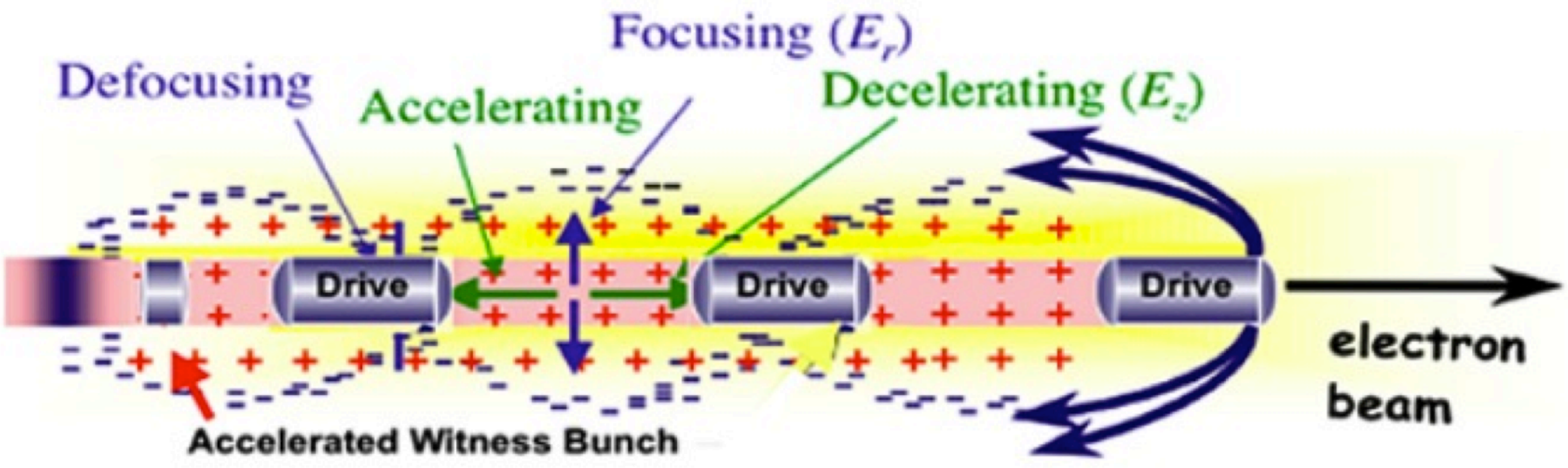

**Figure 10.2** : Operation principle of a comb like beam plasma acceleration

The basic concept of the plasma wakefield accelerator (PWFA) [10.21] is the same as the LWFA except that the plasma wave is excited by an electron (or proton) beam, called driver. The plasma can be preformed by ionizing a gas with a laser [10.22], using a high current discharge or, under some conditions, directly by the driver itself through field-ionization by Coulomb field [10.23]. Another important difference is that, once an electron bunch is somehow injected into the plasma wave, it will remain “in phase” with the accelerating field, since the perturbation travels with the driver's velocity, which is usually much higher than that of LWFA. Moreover, guiding devices are not required because the driver beta function can be tuned, in principle, to any desired length.

Several experiments are planned worldwide, mainly at FACET (SLAC) [10.24] and at FLASH (DESY) [10.25] in order to explore this mechanism and to produce beams with small transverse emittance of such a quality that it is possible to foresee their use in the framework of a future linear colliders. The gradient achievable in such a system is strongly dependent on the plasma density and the incoming beams characteristics. The charge, the pulse length, the transverse dimensions are all parameters that severely affect the resulting beam quality. A gradient in excess of 1 GeV/m, and up to 10 GeV/m, is feasible.

In the IRIDE framework we can consider to use such an acceleration scheme to at least double the beam energy. IRIDE layout has two different stages of bunch compressors. With a fast kicker we can select a single bunch of the long train to bypass the magnetic chicanes, preserving the bunch length. Using a mask in a dispersive arm [10.26] we can produce a comb-like longitudinal structure (drivers + witness), and use it in the plasma accelerator stage. In an alternative, using a second photoinjector, it is possible to inject a comb-like already produced at the photocathode.

### 10.3. Dielectric based acceleration

Another interesting class of very compact acceleration makes use of dielectric structures. In a dielectric wakefield accelerator [10.27], electromagnetic power is radiated by an ultra-short, intense “driving” electron bunch propagating in a high impedance environment formed by a hollow dielectric fiber. This power is then used to accelerate another “witness” bunch just as in the case of the plasma wakefield accelerator.

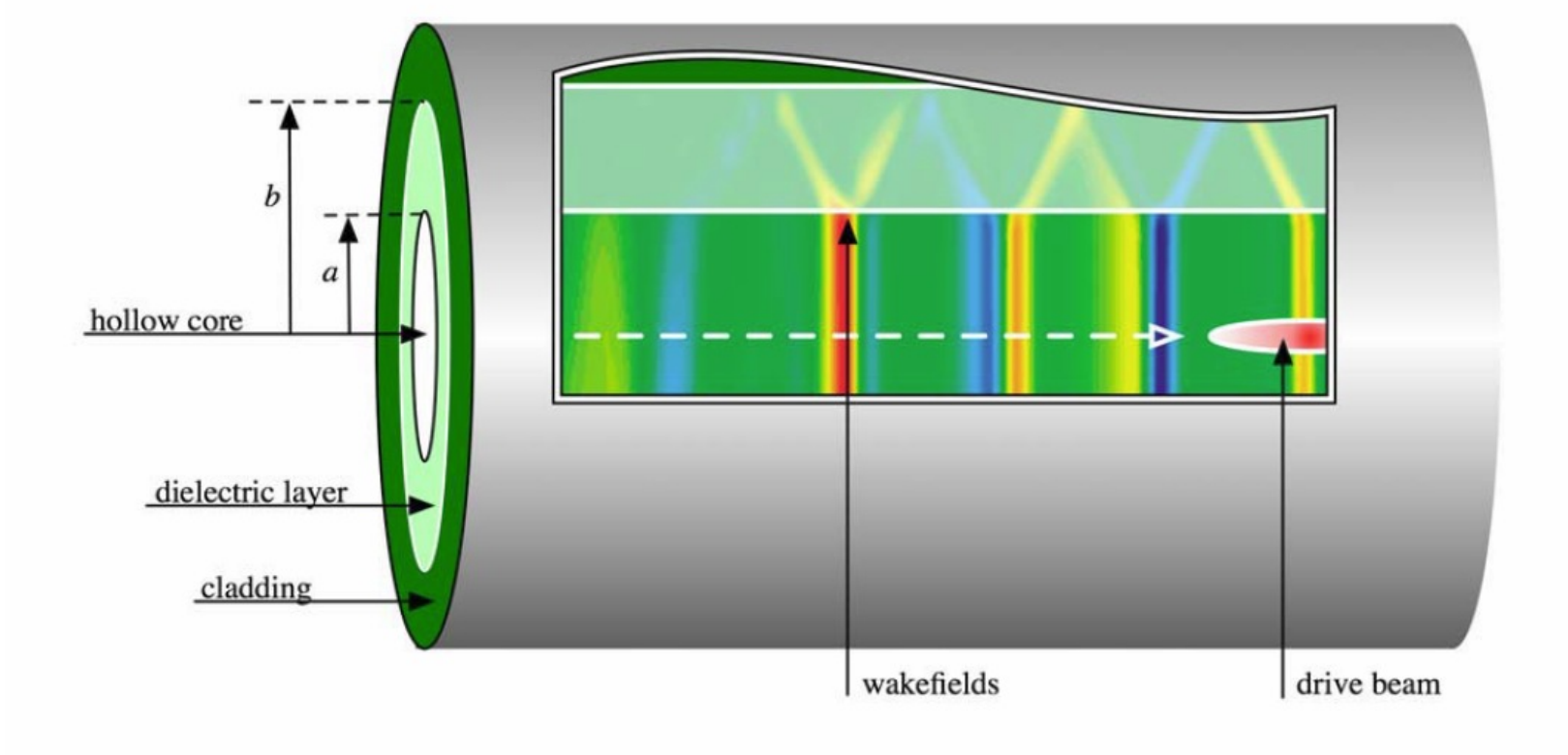

**Figure 10.3** : Conceptual drawing of the dielectric wakefield accelerator. A “drive” beam excites wakefields in the tube, while a subsequent “witness” beam (not shown) would be accelerated by the reflected wakefields (bands of color).

The Dielectric Wakefield Accelerator (DWFA) based on a hollow dielectric tube is one such approach (Figure 10.3). A short (<1 ps) drive-bunch traversing the tube creates Cerenkov wakefields that propagate towards the dielectric boundary at the Cerenkov angle, and are reflected back towards the center axis, where a second bunch then is accelerated [10.28].

The dielectric wakefield accelerator solves the THz-power problem by using radiated fields from short electron bunches, leveraging high-precision fabrication technology from developments in fiber optics, and provides a straightforward means of producing large on-axis accelerating fields. These fields may be used most straightforwardly by accelerating a trailing on-axis bunch, or by directing the radiated fields to an off-axis, higher impedance structure (step-up transformer) [10.29]. Note that these mechanisms work equally well with electrons or positrons. Gradients up to 5 GeV/m were recently obtained [10.30]

In the IRIDE framework we can apply here the same scheme already proposed for PWFA, producing a comb-like beam in the same way, tailoring the longitudinal structure for this particular application. The achievable final energy strongly depends by the accelerating gradient set by the structure (up to GV/m) and by the total energy available in the driver beam. Doing this, we can easily foresee an energy doubling for one properly dimensioned DWFA accelerating stage.

In the Dielectric Laser Acceleration (DLA), there is even no need for a driving bunch. This method relays on lithographically produced photonic structures [10.31] which are pumped by a high power laser pulse. The laser pulse transverse electric field is converted by the structure itself in a longitudinal field that is then used to accelerate a particle beam.

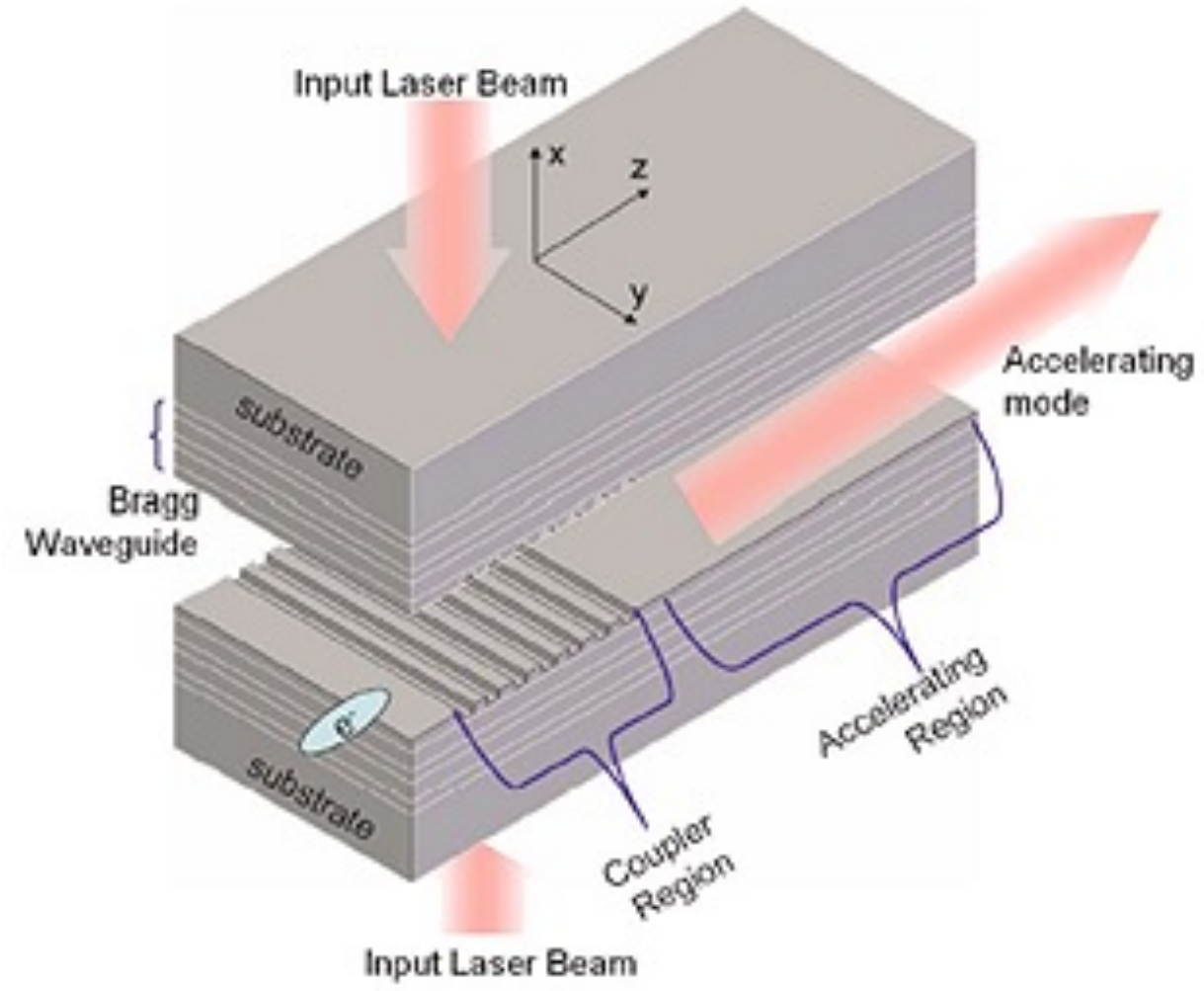

**Figure 10.4**: example of side pumped photonic structure

The dielectric laser acceleration (DLA) concept leverages well-established industrial fabrication capabilities and the commercial availability of tabletop lasers to reduce cost, while offering significantly higher accelerating gradients, and therefore a smaller footprint. In contrast to other novel acceleration schemes, desirable luminosities would be obtained by operating with very low charge per bunch but at extremely high repetition rates. This particular feature makes this scheme really appealing in the IRIDE project, profiting on the natural high repetition rate of such a machine.

Laser-powered acceleration of charged particles promises extraordinary acceleration gradients due to the immense power available from lasers. This possibility has led to many proposed structures for laser acceleration; however, the field of vacuum laser acceleration is almost wholly unexplored experimentally. Recent advances in laser efficiency and phase synchronization have made feasible a practical laser accelerator and are a strong motivation for this experimental program.

### 10.4. Laboratory Astrophysics and radiation physics

Advanced radiation and particle sources can access physics processes that can be of great relevance to astrophysics and can support astrophysical observations. This is particularly true when dealing with processes that are known to be scalable. In these circumstances, by choosing the appropriate experimental conditions in the laboratory, experiments can be made whose results can be scaled to the astrophysical domain. Plasma processes driven by radiation and magnetic fields are common to a wide range of astrophysical problems for which scaled models can be conceived and studied in the laboratory. Supernova explosions and accretion discs are just examples of key astrophysics systems that can be studied with laboratory investigations. In this scenario, intense, high energy laser pulses are proving to be very important tools that can allow radiation driven hydrodynamic conditions to be achieved and studied directly in plasmas [10.32].

More recently, secondary sources generated by ultraintense lasers, have been considered as tools for the investigation of non-local processes such as propagation of high energy particle beams and high energy photons. In fact, multi-PW laser systems will soon allow to break the current record of laser intensity ($10^{21}$ W/cm$^2$) and access a regime of extremely high fields and ultra-relativistic electron dynamics, which may become dominated by radiation friction effects and be characterized by a copious production of positrons, resulting in the laboratory production of pair plasmas, a subject of great interest for astrophysics. Such plasmas are known to be created and ejected at ultra-relativistic velocities from compact and energetic objects such as pulsars and black-holes [10.33]. These jets strongly interact with the low-density interstellar plasma and is subject to instabilities such as the filamentation, Weibel, and hosing instabilities [10.34]. A clear understanding of such processes may help to understand many exotic astrophysical phenomena, such as Active Galactic Nuclei, micro-quasars [10.35] and pulsar winds. Ultraintense lasers offer the possibility of producing quasi-neutral electron-positron beams in the laboratory. These circumstances would open the prospect for characterizing propagation of neutral beams in controlled conditions in a much more similar way to what observed in astrophysics.

At the highest foreseeable intensities, probing of nonlinear QED effects also becomes accessible. Hence this regime is highly relevant both to applications, such as the development of coherent gamma-ray sources, and to tests of fundamental physics. These latter may be based both on all-optical experiments and on the exploitation of laser interaction with linac-produced electrons. See [10.36] for a recent review.

In this scenario, direct signatures of Non-Linear Quantum Electro-Dynamics are finally within reach with state-of-the-art and near-future laser facilities and experiments have are starting. At the same time electron-laser interactions have already demonstrated pair production using large scale conventional accelerators. In fact, Compton scattering of an electron beam in an intense laser field generates a bright source of collimated gamma-ray beams with a tunable energy per photons in the regime of a few to tens of MeV. This could be an ideal secondary source for a wide range of nuclear studies in IRIDE.

## 10.5. Advanced Positron source

Positron beam intensity and emittance are strongly related to the methods of production and to the collection system. Positron generation by electromagnetic interaction or nuclear β decay are well known and under intense studies in various world accelerator centers. However this processes are complicated by thermic processes in the target. The knowledge of the positron production rate is insufficient to calculate the actual beam intensity. Typically one of the main issue is the beam emittance. Present accelerator systems (storage rings and colliders) require small emittance values.

Positron sources used in present day accelerators are generated by a linac electron beam. The increasing positron yield with electron energy leads to the use of very energetic electron beams, and linear colliders are now constructed or planned with multi-GeV incident electron beams on the converter. Instead of using very high energy electron beams on thick targets, which require a remarkable beam power, we could use photons to generate positrons by pair production in the target. One of most promising photon source is an undulator, which is characterized by narrow forward directed radiation, that can provide small source emittance. Additionally also electron channeling in crystals becomes an attractive method to produce high intensity positron fluxes.

Instead of generating high energy photons by means of strong magnetic field in an undulators, the electrons,  propagating in aligned crystals, experience strong transverse electric fields generated by crystallographic planes, which form deep potential wells. The electrons that cross the crystal with a quasi rectilinearly motion near fixed planes or axes become trapped by transverse attractive potential that results in high frequency transverse oscillating trajectory within defined potential wells. Incident particle trajectories in the crystal are very similar to those in a magnetic wiggler with a periodicity several times the atom separation distance. This atomic wiggler, called crystal undulator, presents high levels of photon production, which can be used to generate positrons via pair creation in an amorphous target.

Extensive simulation analysis using GEANT code was undertaken aiming to develop a positron source with light crystals such as Si or Ge and an amorphous target. Results showed that a large amount of soft photons – much higher than with classical bremsstrahlung (BS) – could be created. These photons are more interesting for positron production leading to yield enhancement up to a factor 5 in the energy range 2-20 GeV, for targets having the same overall thickness. Such results could be obtained also at lower energies for W crystals, for which the energy threshold, to overcome Bethe-Heitler cross section, is significantly lower than for germanium.

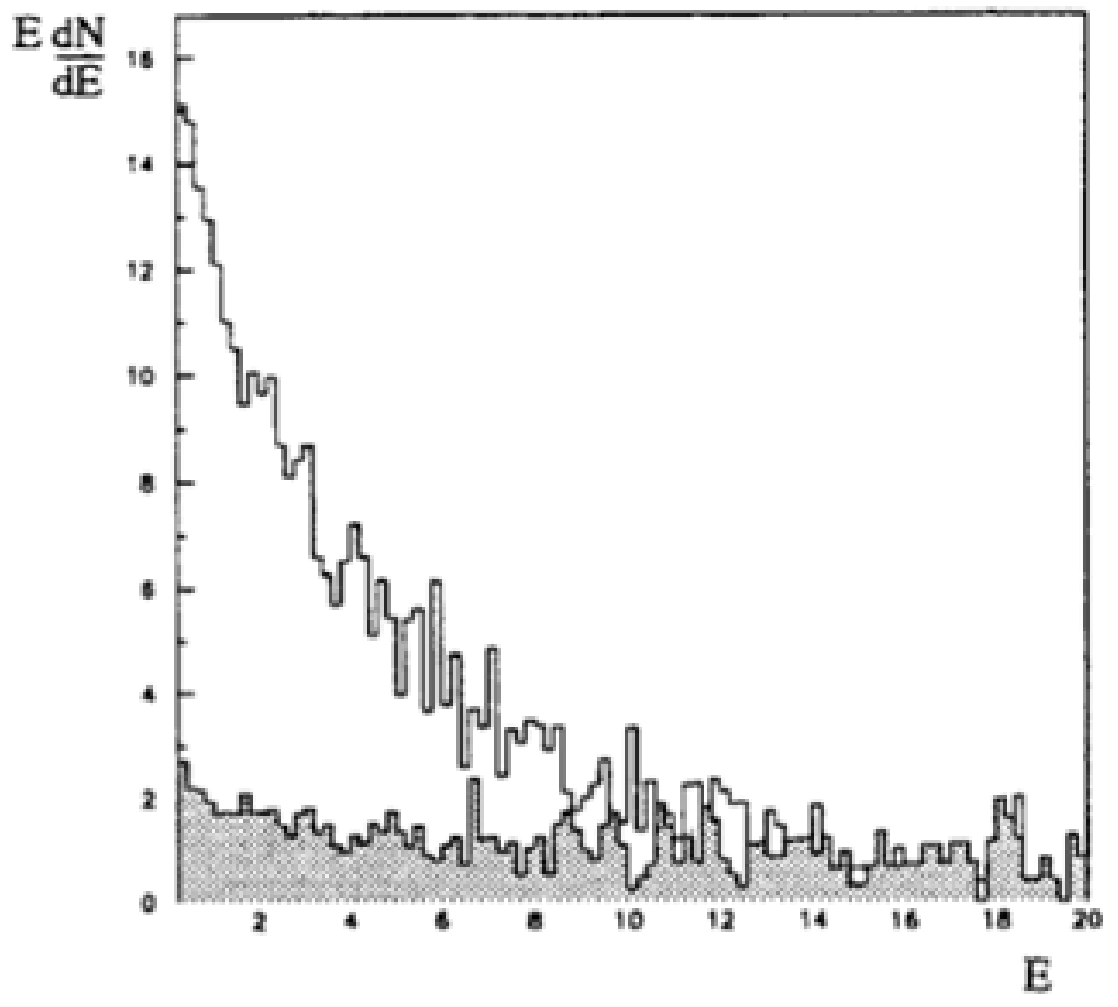

**Figure 10.5** Photon spectrum for incident electron energy of 20 GeV. Tungsten on <111> axis – 1 mm thickness

A new experiment called NTA-SL-POSSO has been recently started at LNF to study the features of moderate-energies (0.1-1 and up to 3 GeV) electron beam channeling in various crystals [10.37]. The project aims to create in SPARC_LAB a group with knowledge and experience for using the features of the passage of high energy beam through the crystals in order to to shape the beams (beam bending, collimation) as well as to generate a powerful x-ray and γ-radiation source (coherent bremsstrahlung, channeling radiation, parametric x-ray radiation) [10.38,10.39].

Electron channeling, namely channeling radiation by ultrarelativistic electrons (~ 1 GeV) in crystals, is a rather promising technique for getting high brilliant positron beams for the future $e^-/e^+$ colliders. The main advantage of crystalline targets consists in a higher photon rate at relatively low photon energies due to the generation of channeling radiation (CR). The two possible schemes are depicted in the figure 10.6. In the first case (a), generation of CR and BS, as well as radiation conversion, proceeds simultaneously in one single crystal of a heavy element (W). At the second (hybrid) setup (b), radiation and positron production take place in separate targets, a crystalline radiator (C, Si, Ge, W) and an amorphous convertor. In the single-target configurations the main issue is the rather large heat load from the primary beam. Using the hybrid scheme, the primary beam is deflected behind the crystal to reduce the heat load into the convertor. Promising positron yields using crystalline targets have already been obtained experimentally (see in [10.40]).

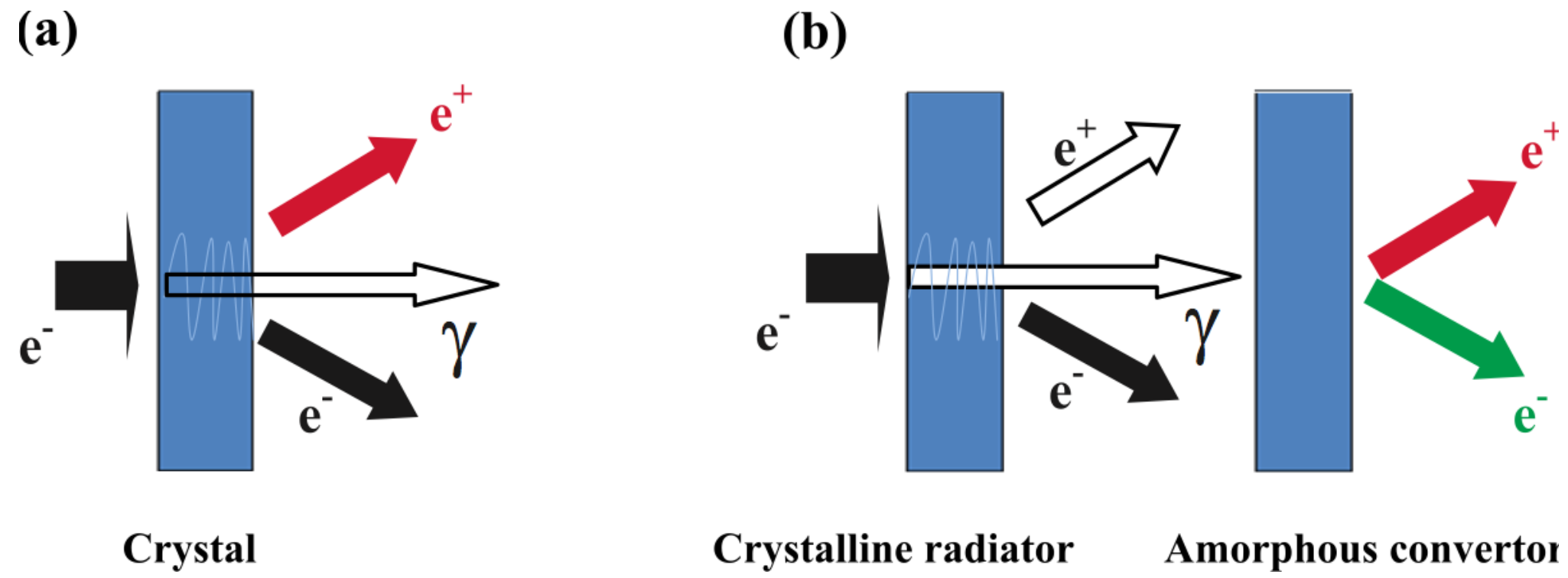

**Figure 10.6** Schemes of non-conventional positron sources. a) One single crystal. b) Crystalline target combined with an amorphous convertor.

With the increase of electron energy the radiation loss due to the beam channeling becomes essential; for instance, for the electron energy change from 150 MeV to 200 MeV the radiation intensity in Si (110) increases two times, while for 800 MeV electrons the radiation flux becomes one order of magnitude higher. Keeping in mind that the depth of a Si (110) potential well (~20 eV) is much less deep than the one for a W <100> (~800 eV), we can expect extremely high channeling radiation flux in W to be emitted within a cone of $1/\gamma \sim 10^{-3}$ around the particle trajectory [10.38,10.39].

The spectra of planar CR reveal the change of the CR energy distribution with increasing angle of incidence of the electrons with respect to the channeling plane. Herewith, both the total yield as well as the energy distribution of CR influence the residual positron yield. The main reason is the contribution from quasi-channeling which leads to emission of coherent bremsstrahlung (CB). Since a real particle beam has some emittance, the optimum positron yield will, therefore, depend on the beam divergence.

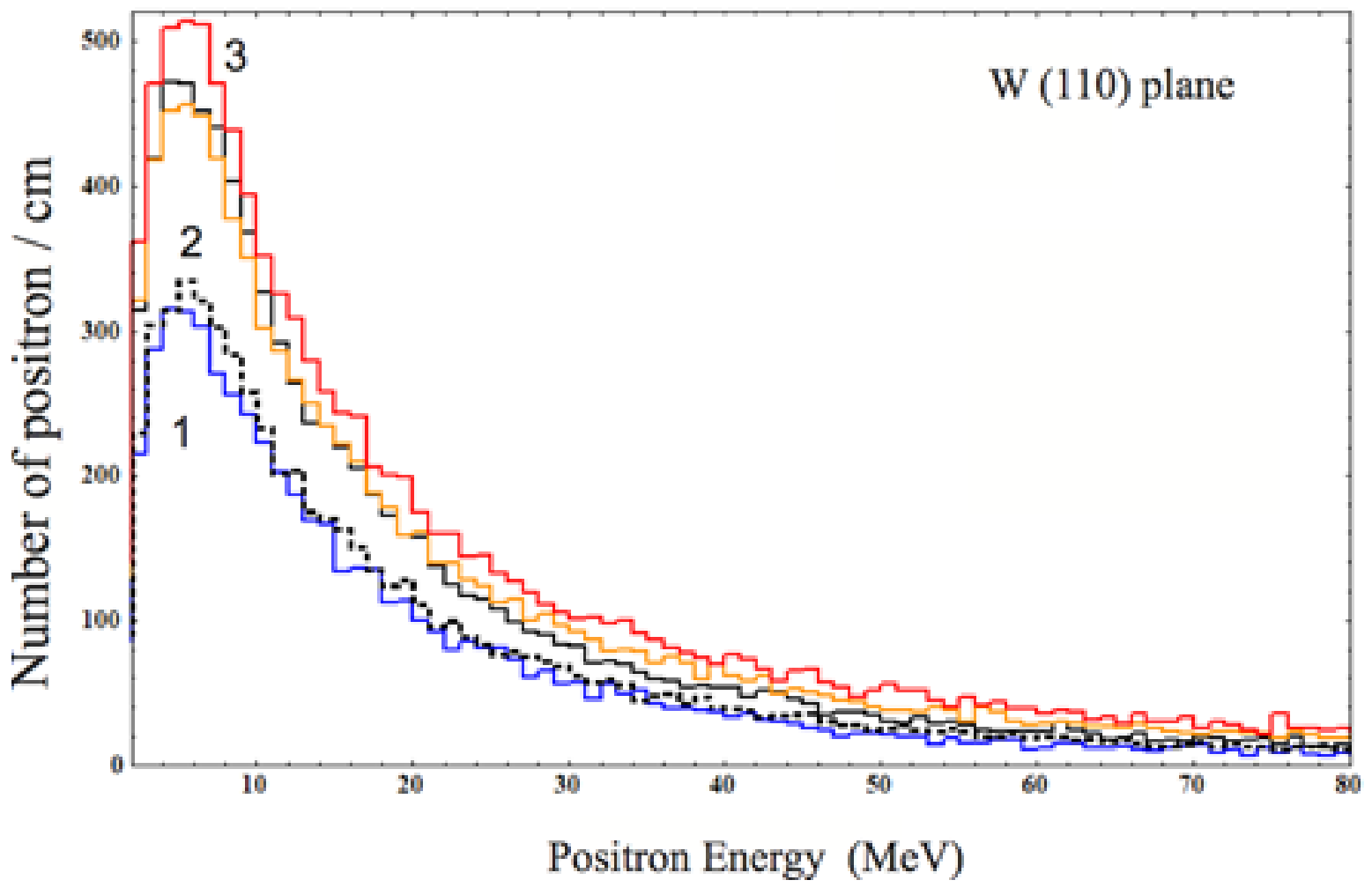

**Figure 10.7** Positron energy distributions behind the W convertor simulated for the CR spectra at planar channeling. (1) incident angle is a half of critical angle for channeling; (2) dashed line incident angle equals to the critical angle; (3) incident angle is twice more than the critical one when we have strong contribution of CB

The process of scattering under channeling regime is presently under evaluation. Our expectations are based also on positive results got by positron production Hiroshima experiment on the use of 1.2 GeV electron channeling in W <100> crystal of 1.2 mm thickness that shows the 2-3 times yield with respect to amorphous targets [10.41].

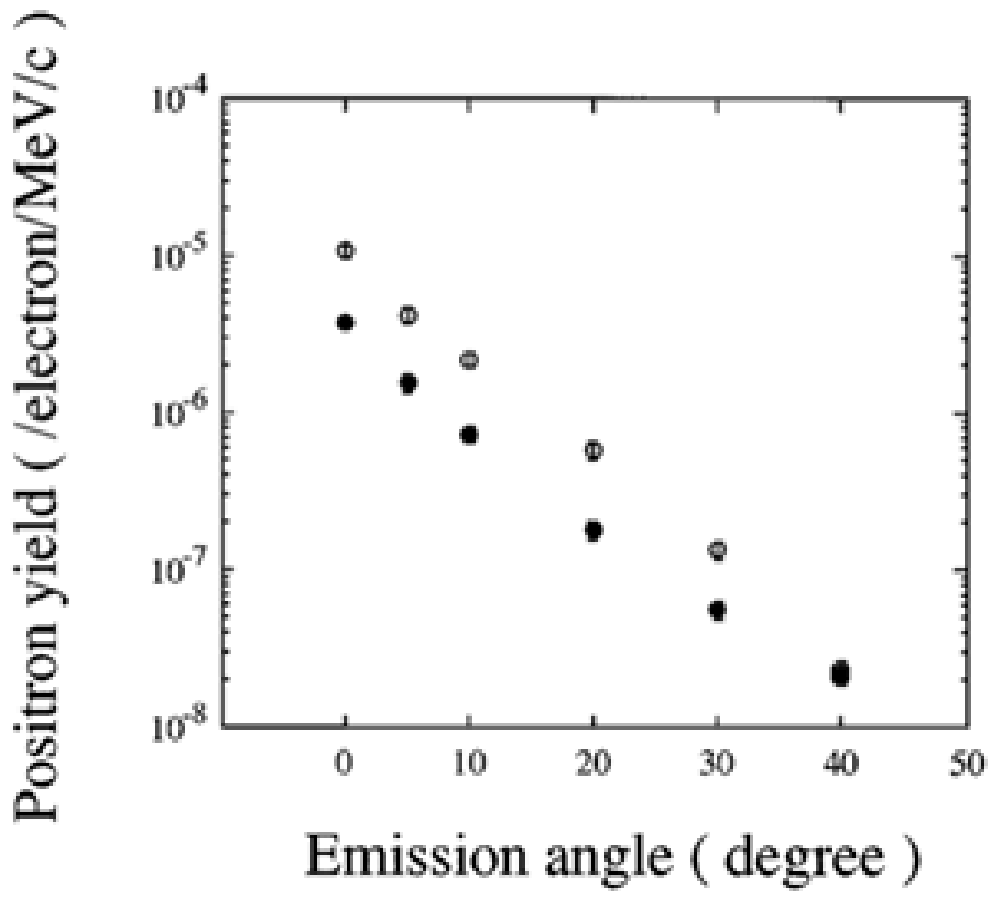

**Figure 10.8** Angular distribution of positrons with a momentum of 20 MeV/c from the 1.2-mm-thick W <100> crystal at channeling (open circles) and off channeling (solid circles)

Our theoretical studies as well as analysis of the results obtained by other groups proves the feasibility of ~1 GeV electron channeling method to produce high brilliant positron source for a new Frascati advanced accelerator facility.

# 11. RADIATION SAFETY

## 11.1. Introduction

High-energy electron accelerators are complex devices containing many components. All facilities contain the same basic systems:

Accelerators structures
RF power components
Vacuum system
Magnetic system associated with steering and focusing the beam
Water-cooling
Air ventilation

Prompt radiation and radioactivity induced by particle nuclear interaction in beam line elements and shielding structures represents the main radiation hazard of high energy accelerators.
The accelerator's design parameters are of crucial importance in the determination of the nature and magnitude of radiation source. The most important parameters are:

Type of accelerated particle
Particle energy
Beam power
Target material
Work load
Beam losses

## 11.2. Operating parameter

The IRIDE project consists in a large infrastructure for fundamental and applied physics research. Conceived as an innovative and evolutionary tool for multi-disciplinary investigations in a wide field of scientific, technological and industrial applications, it will be a high intensity "particle beams factory", Based on a combination of a high duty cycle radio-frequency superconducting electron linac (SC RF LINAC) and of high energy lasers. It will be able to produce a high flux of electrons, photons (from infrared to $\gamma$-rays), neutrons, protons and eventually positrons, that will be available for a wide national and international scientific community interested to take profit of the most worldwide advanced particle and radiation sources.
The electron energy, not completely fixed at the moment, can reach 4 GeV with an average current less of 300 $\mu$A

## 11.3. Machine protection

Even if the maximum power of the beam is quite low, an active machine protection system in under study to limit beam losses

## 11.4. Radiation Protection

11.4.1. *Shielding outlines*

The new machine general layout has been previously shown.
Using the previous operating parameters calculations of shielding are in progress. Because of a great number of the precautions introduced, the results should be a conservative approximation of the doses actually expected. During the commissioning phase, the reliability of the assumptions made will be verified and, if necessary, additional precautions will be made.

11.4.2. *Shielding Design Criteria*

The shielding design criteria have been base on the text of the Italian legislation (D.Lgs. 230/95); according to European Directives as well as the recent ICRP recommendations (ICRP 103) According previous documents the individual limits are 20mSv/y for radiation workers, and 1 mSv/y for the members of the public.
Moreover the definitions of controlled and supervised areas are useful as guidelines. A controlled area is every area where 3/10 of the limits recommended for radiation worker may be exceeded. A supervised area is one area where the overcoming of 1/10 of the previous limit may occur.
Taking into account the dose levels normally found around accelerators, the thickness of the shielding was calculated maintaining the doses, within the areas outside the shield frequented by the staff, 1mSv/year and 0.25 mSv/year within the areas outside the shield frequented by members of the "public".
A shifting from these values could at most change the radiation classification of some areas.
In normal working condition the dose rate outside shielding should not exceed a fraction of μSv/y

## 11.5. Source Term

For shielding evaluation purposes, three components of radiation field which are produced when an electron beam, with an energy of hundred of MeV both a vacuum chamber wall or a thick target have to be considered

11.5.1. *Bremsstrahlung*

Prompt photon fields produced by bremsstrahlung constitute the most important radiation hazard from electron machines with thin shielding. Bremsstrahlung yield is very forward peaked, and increasingly so with increasing energy.
The following equation describes this behaviour:

$$\theta_{1/2} = 100 / E_0$$

where $q_{1/2}$ is in the angle in degrees at which the intensity drops to one half of that at 0º, and $E_0$ is the energy of the initial electrons in MeV. In order to evacuate the shield thickness a "thick target", usually a target of sufficient thickness to maximize bremsstrahlung production, was considered. Photon yield from a thick target as a function of angle consists of two components: sharply varying forward component, described in equation , and a mildly varying wide-angle component. Forward (or zero-degree) bremsstrahlung contains the most energetic and penetrating photons, while bremsstrahlung at wide angles is much softer.
The source term (per unit beam power) for bremsstrahlung at 90º is independent of energy.

### 11.5.2. *Neutrons*

Photons have larger nuclear cross-sections than electrons, so neutrons and other particles resulting from inelastic nuclear reactions are produced by the bremsstrahlung radiation. Neutrons from photonuclear reactions are outnumbered by orders of magnitude by electrons and photons that form the electromagnetic shower. However, some of these neutrons constitute the most penetrating component determining factor for radiation fields behind thick shielding.

#### 11.5.2.1. *Giant resonance production*

The giant resonance production can be seen in two steps:
1) the excitation of the nucleus by photon absorption;
2) the subsequent de-excitation by neutron emission, where memory of the original photon direction has been lost.
The cross-section has large maximum around 20-23 MeV for light nuclei (mass number $A \leq 40$) and 13-18 MeV for heavier nuclei.
The angular yield of giant resonance neutrons is nearly isotropic.
The giant resonance is the dominant process of photoneutron production at electron accelerators at any electron energy.

#### 11.5.2.2. *Pseudo-deuteron production*

At photon energies beyond the giant resonance, the photon is more likely to interact with a neutron-proton pair rather than with all nucleons collectively. This mechanism is important in the energy interval of 30 to ~300 MeV, contributing to the high-energy end of the giant resonance spectrum. Because the cross-section is an order of magnitude lower than giant resonance, with the added weighting of bremsstrahlung spectra, this process never dominates.

#### 11.5.2.3. *Photo-pion production*

Above the threshold of ~140 MeV production of pions (and other particles) becomes energetically possible. These pions then generate secondary neutrons as byproduct of their interactions with nuclei. While substantially less numerous than giant resonance neutrons, the photopion neutrons are very penetrating and will be the component of the initial radiation field from a target (with the exception of muons at very high energies) that determines the radiation fields outside very thick shields.

### 11.5.3. *Muons*

Muon production is analogous to $e^+/e^-$ pair production by photons in the field of target nuclei when photon energy exceeds the threshold $2m_m c^2 \approx 211$ MeV.
Above a few GeV the muon yield per unit electron beam power is approximately proportional to electron energy $E_0$. Muon angular distribution is extremely forward-peaked, and this distribution narrows further with increasing energy. At energies of a few GeV adequate photon and neutron shielding will be also sufficient for muons. Calculation are in progress

### 11.5.4. *Induced Activity*

Personnel exposure from radioactive components in the beam line is of concern mainly around beam lines, collimators, slots, beam stopper or beam dump, where the entire beam or a large fraction of the beam is dissipated continuously, while unplanned beam losses result from beam mis-steering due to inaccurate orbit adjustment or devices failure. Beam losses induce activation in machine component as well as in

| | |
|---|---|
| the beam pipe | ($^{60}$Co, $^{54}$Mn, $^{51}$Cr, $^{46}$Sc, $^{22}$Na, $^{11}$C, $^{7}$Be) |
| the cooling water | ($^{3}$H, $^{7}$Be, $^{15}$O, $^{13}$N, $^{11}$C) |
| the air | ($^{15}$O, $^{13}$N, $^{38}$Cl, $^{41}$Ar) |
| the concrete walls | ($^{152}$Eu, $^{154}$Eu, $^{134}$Cs, $^{60}$Co, $^{54}$Mn, $^{22}$Na) |

The activation of soil as well as the groundwater by neutrons and other secondary particles can have an environmental impact but at electron accelerators the radioactivity levels are generally low.

Calculations should be made accordind with the final definition of the beam parameters, the beam losses and the characteristic of cooling water system, the air circulation system and the beam dump layout.

### 11.5.5. *Machine accesses*

During machine operation the linac tunnel will be an excluded area.
During no operation periods the linac tunnel will be a controlled area, due to the possible activation of the machine structure.
The technical areas behind the roof shield will be classified as controlled or supervised areas.
The experimental areas will be a free access area. Only areas close to the front ends or at the end of the beam line will be classified.
In order to protect workers in the experimental areas, the electron beam will be dumped below the floor after the FEL undulators. A deflection of 45º is effected by electromagnets.

For additional safety permanent magnets and active radiation detectors interlocked with the beam will be used.

### 11.5.6. *Beam line radiation shielding design*

For each shielding situation (insertion device white beam, radiation transport, monochromator, hutches etc.) the synchrotron radiation, the gas bremsstralhung, the high-energy bremsstralhung, from beam halo interactions with the structures of the machine, will be calculate for a representative geometry.

### 11.5.7. *The operational radiation safety program*

The purpose of the operational safety system program is to avoid life-threatening exposure and/or to minimize inadvertent, but potentially significant, exposure to personnel. A personnel protection system can be considered as divided into two main parts: an access control system and a radiation alarm system.
The access control system is intended to prevent any unauthorized or accidental entry into radiation areas.
The access control system is composed by physical barriers (doors, shields, hutches), signs, closed circuit TV, flashing lights, audible warning devices, including associated interlock system, and a body of administrative procedures that define conditions where entry is safe. The radiation alarm system includes radiation monitors, which measure radiation field directly giving an interlock signal when the alarm level is reached.

#### 11.5.7.1. *Interlock design and feature*

The objective of a safety interlock is to prevent injury or damage from radiation. To achieve this goal the interlock must operate with a high degree of reliability. All components should be of high grade for dependability, long life and radiation resistent. All circuits and component must be fail safe (relay technology preferably).
To reduce the likelihood of accidental damage or deliberate tampering all cables must run in separate conduits and all logic equipment must be mounted in locked racks.
Two independent chains of interlocks must be foreseen, each interlock consisting of two micro switches in series and each micro switches consisting of two contacts.
Emergency-off buttons must be clearly visible in the darkness and readily accessible.
The reset of emergency-off buttons must be done locally.
Emergency exit mechanisms must be provided at all doors.
Warning lights must be flashing and audible warning must be given inside radiation areas before the accelerator is turned on.
Before starting the accelerator a radiation area search must be initiated by the activation of a "search start" button. "Search confirmation" buttons mounted along the search path must also be provided. A "Search complete" button at the exit point must also be set.
Restarting of the accelerator must be avoided if the search is not performed in the right order or if time expires.
The interlock system must prevent beams from being turned on until the audible and visual warning cycle has ended.

Any violation of the radiation areas must cause the interlocks system to render the area safe.
Restarting must be impossible before a new search. Procedures to control and keep account of access to accelerator vaults or tunnels must be implemented.

### 11.6. Electron Beam Dump

The electron beam will be dumped below the floor at the end of the undulators in the linac tunnel. The beam deflection of about 45 degrees in made using permanents magnet for additional safety.
The layout of beam dump as well as the size and type of shielding materials is under study.